\documentclass[11pt,a4paper,cite]{report}
\usepackage[english]{babel}
\usepackage[T3,T1]{fontenc}
\usepackage[latin1]{inputenc}
\usepackage{graphicx} 
\usepackage{cite}
\usepackage{amsmath}
\usepackage{amstext}
\usepackage{amsfonts}
\usepackage{amssymb}
\usepackage{rotating}
\usepackage{amsthm}
\usepackage{color}
\usepackage{bm}
\usepackage{datetime}
\usepackage{xcolor}
\usepackage{multicol}
\usepackage{slashed}
\usepackage{extarrows}
\usepackage{vwcol}
\usepackage{stackengine}
\usepackage{hyperref}
\usepackage{tikz}
\usepackage{float}
\usepackage{upgreek}
\usepackage[a4paper, left=3cm, right=2.5cm, top=3cm, bottom=3cm]{geometry}
\makeatletter
\newcommand{\vast}{\bBigg@{4}}
\newcommand{\Vast}{\bBigg@{5}}
\makeatother
\newcommand{\eqdef}{\xlongequal{\!\!\!\!\!\!\!\eqref{E2.4.99}\!\!\!\!\!\!\!}}

\DeclareSymbolFont{tipa}{T3}{cmr}{m}{n}
\DeclareMathAccent{\invbreve}{\mathalpha}{tipa}{16}

\newcommand{\ze}{\kern 0.05em}

\newcommand{\insertplot}[5]{\begin{figure}
		\hfill\hbox to 0.05in{\vbox to #5in{\vfill
				\inputplot{#1}{#4}{#5}}\hfill}
		\hfill\vspace{-.1in}
		\caption{#2}\label{#3}
\end{figure}}
\newcommand{\inputplot}[3]{
	\special{ps: plotfile #1}
}
\newcounter{fig}

\usepackage{subfiles}
\usepackage{pdfpages}

\newdateformat{monthyeardate}{%
	\monthname[\THEMONTH], \THEYEAR}

\begin{document}



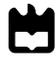
**Universidade de Aveiro**
**2022**

**Alexandre Mira Pombo**

**Buracos negros e objectos solitónicos com campos bosónicos**

**Black holes and solitonic objects with bosonic fields**



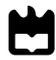
**Universidade de Aveiro**
**2022**

**Alexandre Mira Pombo**

**Buracos negros e objectos solitónicos com campos bosónicos**

**Black holes and solitonic objects with bosonic fields**

*"Not only is the Universe stranger than we think, it is stranger than we can think."*

— Werner Heisenberg



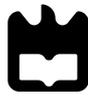

Universidade de Aveiro
2022

Alexandre Mira Pombo

**Buracos negros e objectos solitónicos com campos bosónicos**

**Black holes and solitonic objects with bosonic fields**

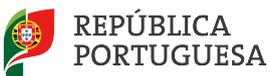
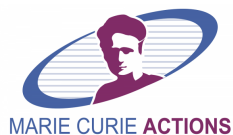
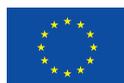



I dedicate this thesis to my family and my good friends, without which this journey would have been unbearable.



**o júri / the jury**

| | |
|---|---|
| presidente / president | Joaquim Arnaldo Carvalho Martins |
| | Professor Catedrático, Universidade de Aveiro |
| | |
| vogais / examiners committee | Daniela Doneva |
| | Head of Emmy Noether Researcher Group, Universidade de Tuebingen |
| | |
| | Yves Brihaye |
| | Professor Ordinário, Universidade de Mons |
| | |
| | Pedro Pina Avelino |
| | Professor Associado, Universidade do Porto |
| | |
| | Carolina Loureiro Benone |
| | Professora Adjunta C, Universidade Federal do Pará |
| | |
| | Carlos Alberto Ruivo Herdeiro |
| | Investigador Coordenador, Universidade de Aveiro |



**agradecimentos / acknowledgements**

First of all, I would like to offer my most profound gratitude to my supervisor Prof. Dr. Carlos Herdeiro for the opportunity to do a PhD under his supervision and for the patient and disponibility. I would also like to thank sincerely my co-supervisor, Dr. Eugen Radu, whose support and mentorship was essential. This thesis was only possible due to their guidance.

To Jorge Delgado, João Oliveira and Nuno Santos for their unconditional friendship. Without you, this would not be as fun. I would like to thank you for the countless hours of discussions and for sharing the happiness and hardships of life. The ability to do a thesis surrounded by such good friends and end up working with them was the best part of all.

For all the sympathy, patience and help to deal with the bureaucracies and documents regarding all types of topics, I want to give my acknowledgements to Cristina Grosso.

At last, I would like to thank my family, which have always supported me.



**Resumo**    Os buracos negros (BH) estão entre os objetos físicos mais incomuns e excitantes. Além de serem, simultaneamente, objetos relativistas e quânticos, o que permite a descoberta e/ou testes de nova física, os BHs apresentam também falta de "individualidade'. Se duas estrelas tiverem exactamente a mesma massa $M$ e momento angular $J$, nada os obriga a serem distribuídos da mesma forma nas duas estrelas, porém, dois BHs de origem distinta com os mesmos $M$ e $J$, serão indistinguíveis. Nos últimos anos, vários modelos alternativos de objetos ultracompactos surgiram para explicar o Universo. Entre eles, objetos feitos e/ou cercados por campos bosônicos estão no centro dos desenvolvimentos recentes.

Esta tese tem como objetivo construir e estudar tais objetos bosônicos. Em particular, os fenômenos de escalarização – onde um BH fica imerso em um campo escalar não trivial – e a construção de estrelas de bosões – configurações autogravitantes de campos bosônicos; bem como técnicas para estudá-los, a saber: a decomposição em harmônicos esféricos e o estudo "completo" da identidade virial através do argumento de Derrick. Nesse sentido, construímos um integrador Runge-Kutta de ordem 6(5) de passo adaptativo e paralelizado para resolver, com altíssima precisão, o conjunto de equações diferenciais ordinárias. No que diz respeito às equações diferenciais parciais, recorremos a um solver profissional (pacote de programas CADSOL/FIDISOL) devido à complexidade acrescida.



**Abstract**        Black holes (BH) are among the most unusual and exciting physical objects. Besides being simultaneously Relativistic and quantum mechanical objects, which allows the discovery and/or tests of new physics, BHs also present a lack of "individuality". As a recall, if two stars have the same mass $M$ and angular momentum $J$, nothing binds them to be distributed in the same way in both stars. However, two distinctly originated BH with the same $M$ and $J$, will be indistinguishable.

In recent years, several alternative models of ultra-compact objects have emerged to explain the unknown Universe. Between them, objects made and/or surrounded by bosonic fields are at the centre of recent developments.

This thesis aims to construct and study such bosonic objects. In particular, the scalarization phenomena – where a BH gets immersed in a non-trivial scalar field – and the construction of Boson Stars – self-gravitating configurations of bosonic fields; as well as techniques to study them, namely: decomposition into spherical harmonics and the "complete" study of the virial identity through Derrick's argument. In this regard, we have constructed a parallelized, adaptative-step 6(5) order Runge-Kutta integrator to solve, with very high accuracy, the set of ODEs. Concerning PDEs, we resort to a professional solver (CADSOL/FIDISOL program package) due to the added complexity.



\pagenumbering{roman}

\tableofcontents	
	
\listoffigures

\listoftables

%
\chapter{Introduction}\label{C1}
\pagenumbering{arabic}
%
		The newest results from LIGO collaboration~\cite{abbott2016ligo,abbott2021population,abbott2019tests,abbott2019gwtc,abbott2021gwtc} (gravitational waves detection) and event horizon telescope collaboration~\cite{event2019first,akiyama2019first,akiyama2019third,akiyama2019fourth,akiyama2019fifth,event2019six,akiyama2021seven,akiyama2021eight} (shadow image) indicate a population of extremely compact objects, known as black holes (BH). However, it remains uncertain if these BH candidates are depicted by general relativity, some alternative model of gravity (see \cite{damour1992tensor} for an extensive review), or even distinct compact objects without an event horizon. It is then the right time to emphasise or rule out possible candidates.

	In that regard, and in an endeavour to explain the observed universe, it is tentative to connect the dark matter and dark energy phenomena with the newly observational data. Either through the study of exotic ultra-compact objects or alternative BHs models. 
	
	One of the most straightforward approaches is to consider bosonic fields. Excluding the well-known vector particles present in the standard model, the recent result of the Higgs boson\cite{atlas2012observation,chatrchyan2013cms,higgs1966spontaneous,higgs1964broken} emphasises the existence of bosonic fields in nature. The simplest bosonic field is the scalar field (e.g. the Higgs boson). Scalar fields play a significant role as a plausible explanation for the high accelerated expansion of the early universe (inflation) and the non-vanishing cosmological constant at the current times. If bosonic fields existed in the early universe and are stable or long-lived, they can survive until the present and be part of the dark sector of the universe. 
		
	The interaction between a scalar field and the strong spacetime curvature of a BH originates a new exciting phenomenon: spontaneous scalarization~\cite{salgado1998spontaneous,dima2020spin,antoniou2018evasion,doneva2018new}. The latter endows BHs surrounded by a massless (massive) real bosonic field\footnote{The same phenomena can be generalized to higher-spin fields such as the vector field (spin-1), endowing the spontaneous vectorization phenomena~\cite{oliveira2021spontaneous,minamitsuji2020spontaneous,barton2021spontaneously}, and spin-2 particles, a spontaneous tensorization phenomena~\cite{ramazanouglu2019spontaneous,ramazanouglu2019generalized}.}. Such a phenomenon creates new, alternative BH models with interesting and distinct characteristics.
	
	In the absence of gravity and with proper self-interactions, the bosonic fields can generate self-stabilised solitonic solutions known as $Q$-balls~\cite{coleman1985q}. These are made of a massive, complex, self-interacting bosonic field and are everywhere regular solutions of the matter (Klein-Gordon for scalar~\cite{coleman1985q,volkov2002spinning,lee1989gauged} and Proca for vector~\cite{heeck2021proca,brihaye2017proca} bosons) equations, where the dispersiveness of the bosonic field frequency is enough to counteract the attractive self-interaction. 
	
	Taking backreaction into account relaxes the conditions for the self-interactions and self-gravitating configurations of bosonic fields (\textit{a.k.a. Boson Stars (BS)}~\cite{jetzer1992boson,herdeiro2017asymptotically,herdeiro2019asymptotically}) emerge. The latter can be made of scalar or vector fields. While the former is usually designated as Boson Star, the latter is dub as a Proca Star~\cite{brito2016proca}. To avoid conflict and distinguish between the several solutions, we will always denominate a generic self-gravitating bosonic field as a Boson Star; if the bosonic field is \textit{scalar}, we anoint it Scalar Boson Star (SBS)~\cite{liebling2017dynamical,schunck2003general,yoshida1997rotating}; if the boson is a \textit{vector}, we call it Proca Star (PS).

	The previous SBS can be in equilibrium with a spinning black hole (at its centre) if rotation is taken into account, endowing \textit{Kerr black holes with scalar hair (KBHsSH)} \cite{herdeiro2014kerr,herdeiro2015construction}. The intricacies of such configurations have long alienated further developments. However, if one decomposes such solutions into the spherical harmonics basis, not only do the angular and radial components separate, but one can also gain some insight into the structure and intricacies of the solutions -- a kind of spectroscopic analysis, hence a \textit{spectral decomposition (SD)}.
	
	At last, observe that all these configurations are not random. They have to obey certain conditions. In particular, the no-go theorems~\cite{JoaoThesis} and the ``no-hair'' conjecture~\cite{ruffini1971introducing} has imposed many restrictions on the existence of BHs with additional degrees of freedom (\textit{a.k.a. hair}). Some of these conjectures rely on viral identities -- the latter has been usually computed through Derrick's scaling argument~\cite{derrick1964comments} and never completely understood. The requirement of the addition of an additional term, the Gibbons-Hawking-York term, was missing from the literature. Such a result allows the computation for a generic $n$-dimensional metric. In particular, for $2$-dimensional, axially symmetric metric ansatze. 

	This chapter will introduce the main topics and concepts that we will delve into along the thesis, as well as notations. After a brief discussion on the observational evidence for BHs and the general relativity BH paradigm, we shall discuss the importance/interest of considering bosonic fields in nature. 
		
\bigskip
		
	As a final remark, we will always use index notation in a $4$-dimensional Schwarzschild-like coordinate system $(t,\, r,\, \theta,\, \varphi)$. Expressions and equations will come in geometrized units $\big(4\pi G=c=4\ze \pi \epsilon _0 =1\big)$ and we will consider the metric signatures to be $(-, +, +, +)$. Asymptotic flatness is assumed unless stated otherwise. The complex conjugate of a given function $X$ is denoted by an overhead bar $\bar{X}$. In all the plots, the mass and Noether charge is normalized by the field's mass, and we are assuming the following notation: a generic derivatives will be explicitly represented using the reduced notation $\partial _{y} X \equiv X_{,y}$; a derivative in order to the proper time $X_{,t} \equiv \dot{X}$; a derivative in order to the radial coordinate $X_{,r}\equiv X'$; a derivative in order to the matter field \textit{e.g.} $(\Phi )$ is $X_{,\Phi} \equiv \hat{X}$; and a derivative with respect to the $\theta$ coordinate is $X_{,\theta} \equiv \invbreve{X}$. In addition, after being first introduced, functions will be represented without the dependence argument $\big($\emph{e.g.} $X(\Phi) \equiv X\big)$. Besides, we introduce the notation $c^{\rm te}\equiv constant$. 

%
	\section{General relativistic black holes}\label{S1.1}
%
	When in $1784$, John Michell and Pierre-Simon Laplace~\cite{montgomery2009michell} suggested the existence of a celestial object whose gravitational force was strong enough that not even light could escape, it was considered fiction. The computations performed under Newton's gravitational theory paradigm result in an object whose escape velocity is the velocity of light ($\, c\,$).

	Almost $130$ years later, in $1915$, Albert Einstein developed his theory of General Relativity (GR). Months later, Karl Schwarzschild solved the Einstein field equations for the gravitational field of a spherically symmetric point mass. The latter became known as the \textit{Schwarzschild black hole}~\cite{Schwarzschild:1916ae}. At what is called the \textit{Schwarzschild radius ($\, r_S \ze$)}, the solution becomes singular, meaning that some term in the Einstein equations diverges. In the latter, the strength of the gravitational field makes it impossible for light to escape to infinity. In a beautiful demonstration of the connection between GR and Newton's gravitational theory, the formula reached by Michell and Laplace corresponds to the Schwarzschild radius.
	
	When an extremelly massive star reaches the end of its life, the core must continue to collapse: either to a new unknown ultra-dense state of matter halted by additional degeneracy  (like quark degeneracy~\cite{Ivanenko:1965dg}, electroweak degeneracy~\cite{Dai:2009br}, Preon degeneracy~\cite{hansson2005preon}, ...), or to a BH. Anyway, there must exist a mass at which the total collapse is unavoidable (or the collapsed object radius is smaller than $r_S$), turning the study of BH physically significant. 
	
    Black holes are one of the most unusual and exciting physical objects. Besides being one of the most extreme objects in the Universe, they are simultaneously GR and quantum mechanical objects, which allows the discovery and/or tests of new physics.

	In physical terms, a BH is a region of spacetime causally disconnected from the rest of the Universe. The boundary of this region is known as the \textit{event horizon}. More technically, a BH is a solution to Einstein's general relativity (or some generalization thereof) possessing an event horizon.
	
	Although initiated at the beginning of the last century, never in their one hundred years history, there has been a more exciting time to study these fascinating objects. A diversity of observational data delivers information with unprecedented accuracy on the strong gravity region. It is, therefore, timely to test the \textit{Kerr hypothesis}, on which most discussions of astrophysical BHs are based. This hypothesis can be explained by quoting the words of the great astrophysicist and BH theorist Subramanian Chandrasekhar~\cite{chandrasekhar2013truth} 

\bigskip

\textit{``In my entire scientific life, extending over forty-five years, the most shattering experience has been the realization that an exact solution of Einstein's field equations of general relativity, discovered by the New Zealand mathematician, Roy Kerr, provides the absolutely exact representation of untold numbers of massive black holes that populate the Universe.''}

\bigskip

	The Kerr hypothesis is the paradigm that, the Kerr metric, discovered in 1963 by Roy Kerr~\cite{kerr1963gravitational}, \textit{``provides the absolutely exact representation of untold numbers of massive black holes that populate the Universe.''} The establishment of this paradigm relies on two main steps. 

	Firstly, in the late $1960$s and early $1970$s, a set of mathematical theorems established that the most general physical BH solution in (electro-)vacuum GR is the Kerr(-Newman) solution~\cite{newman1965note,newman1965metric}. As stated by Brandon Carter, \textit{``the Kerr metrics represent ``the'' (rather than merely ``some possible'') exterior fields of BHs with the corresponding mass and angular momentum values.''}~\cite{Carter:1971zc}. We emphasize: the uniqueness theorems are a set of solid mathematical results. However, they apply only to (electro-)vacuum. 

	Secondly, inspired by the uniqueness theorems, a much more ambitious idea became widespread, known as the \textit{``no-hair conjecture''}, put forward by John Wheeler and Remo Ruffini~\cite{ruffini1971introducing}. According to this idea, \textit{regardless of the type of matter/energy one starts with}, the gravitational collapse leads to equilibrium BHs uniquely determined by their total mass, angular momentum (and eventually electric-type charges). All of which are asymptotically measured quantities subject to a Gauss law and no other independent characteristics (collectively referred to as ``hair''). This more ambitious concept suggests that generically ($i.e.$ not restricted to electro-vacuum), Kerr(-Newman) BHs are the only type of BH that can emerge \textit{dynamically}.

	If true, the ``no-hair'' conjecture would support the Kerr hypothesis. Assuming that an electric charge is inconsistent with astrophysical BHs\footnote{In a dynamical astrophysical environment, the presence of plasmas around the BH leads to prompt discharge. Alternatively, the neutralization can occur through Hawking charge evaporation~\cite{gibbons1975vacuum}.}, then all BHs in the Cosmos should be described by the Kerr metric since this is the only equilibrium solution that can emerge dynamically. 

	It should be emphasized that if accurate, the Kerr hypothesis makes BHs remarkably different from any other macroscopic object. For instance, two stars with the same total mass and total angular momentum can be quite distinct, as the mass and spin can be distributed differently within the star, a discrepancy that will manifest in the higher multipole moments of the gravitational field. Nevertheless, the Kerr hypothesis states that the resulting BHs must be \textit{exactly equal}. These multipoles are non-trivial but entirely defined by the total mass and angular momentum via an elegant formula derived by Hansen~\cite{Hansen:1974zz}. 

	On the theoretical side, testing the Kerr hypothesis is tantamount to keeping some healthy scepticism about the no-hair conjecture and asking whether there may be other good BH/compact object models beyond the Kerr metric. Three broad criteria~\cite{degollado2018effective} for a good model of a compact object include 1) appearing in a well-motivated and consistent physical model; 2) having a dynamical formation mechanism; 3) being sufficiently stable.
%
	\section{Spontaneous scalarization}\label{S1.2}
%
	Black holes have a surprisingly small number of macroscopic degrees of freedom in General Relativity \textit{and} electro-vacuum, where a remarkable uniqueness holds -- see $e.g.$~\cite{chrusciel2012stationary} for a review. In gravitational theories beyond GR or even GR with matter sources, however, ($i.e.$ beyond electro-vacuum), one finds a much richer landscape of BH solutions -- see $e.g.$ the reviews~\cite{herdeiro2015asymptotically,volkov2018hairy} for different types of non-Kerr BHs. These are often called ``hairy'' BHs since they have more macroscopic degrees of freedom. Then, the central question becomes if there are~\textit{dynamically} viable ``hairy'' BHs that could represent alternatives to the Kerr BH paradigm. 

	A dynamical mechanism that could lead to the formation of BHs that differ from the standard GR electro-vacuum BHs is~\textit{spontaneous scalarization}. First proposed by Damour and Esposito-Far\'ese~\cite{Damour:1992we,Damour:1993hw}\footnote{Reports of an earlier proposal by Zagaluer, back in $1992$, also exist; however, due to some artificial considerations, it has been criticized ever since~\cite{Zaglauer:1992bp}.}, scalarization occurs when, a non-trivial configuration of a scalar field with vanishing asymptotic behaviour is dynamically prefered. It is said to be spontaneous scalarized when such scalar configuration occurs without an inducing external perturbation (hence the name).
	
	In this context, the presence of non-conformally invariant matter (such as a neutron star) sources scalar field gradients due to the non-minimal coupling of the scalar field to the Ricci curvature. It turns out that BHs are immune to this tendency to scalarize because they are conformally invariant in scalar-tensor theories, as BH solutions in these theories, in general, coincide with the electro-vacuum solutions~\cite{hawking1972black,sotiriou2012black}. Thus, they do not source scalar field gradients and do not scalarize. Nevertheless, if the non-conformally invariant matter would surround the BHs, they should scalarize similarly, as suggested in~\cite{cardoso2013matter,cardoso2013black}. This sort of BH scalarization was confirmed in a set of concrete field theory models in~\cite{herdeiro2019black}\footnote{A similar phenomenon of ``tensorization'' (for neutron stars) was discussed in~\cite{ramazanouglu2017spontaneous}.}.
	
	Spontaneous scalarization is essential to distinguish between some alternative theories from GR. It can be achieved by (non-)minimally coupling the scalar field to gravity, producing a field-dependent gravitational constant.
		
	The phenomenon of BH spontaneous scalarization is largely inspired by the well-known phenomenon of spontaneous scalarization of neutron stars~\cite{Damour:1993hw}. However, a key difference is that in the latter, matter induces scalarization, whereas, in BH scalarization, the phenomenon is triggered by the strong spacetime curvature.

	This thesis will discuss two models wherein the Kerr hypothesis is challenged: the \textit{Einstein-Maxwell-Scalar (EMS)} model~\cite{herdeiro2018spontaneous} and the \textit{extended Scalar-Tensor-Gauss-Bonnet (eSTGB) gravity}~\cite{doneva2018new,antoniou2018evasion,silva2018spontaneous}. Despite not being directly related, these two are part of a larger class, which has recently attracted attention in the literature.

	When a BH becomes unstable and undergoes scalarization, there is a transfer of energy and charge to a  surrounding ``cloud'' of scalar particles. While in the eSTGB model, the scalar field has a non-minimal coupling to the Gauss-Bonnet curvature term, in the EMS, the scalar field is non-minimally coupled to the Maxwell tensor. Observe that, in both cases, the scalar field has a minimal coupling to the Ricci scalar.   

\bigskip

	The considerations in this thesis apply to a family of models generically described by the action: 
	\begin{equation}\label{E1.2.1}
	 \mathcal{S}= -\frac{1}{4}\int d^4 \ze x \, \sqrt{-g}\bigg[\frac{R}{4\pi G}-2\ze\phi _{,\mu}\phi ^{,\mu} - f(\phi)\ze \mathcal{I} (\psi, g)\bigg]\ ,
	\end{equation}		
	where $R$ is the Ricci scalar, $\phi$ is the real scalar field that is non-minimally coupled through the \textit{coupling function} $f(\phi)$ to the \textit{source term} $\mathcal{I}$. The latter depends only on the spacetime metric $g_{\mu \nu}$ (with $g$ the metric field determinant) or also on extra matter fields (collectively denoted by $\psi$). Variation of the action with respect to the scalar and metric fields gives the corresponding field equations, respectively,
	\begin{align}\label{E1.2.2}
	 & \Box \phi = \frac{\mathcal{I}}{4}\,\hat{f}\ ,\\
	 & E_{\mu \nu} = R_{\mu \nu}-\frac{1}{2}\ze g_{\mu \nu} = 8\pi G\ze T_{\mu \nu}\ .\label{E1.2.3}
	\end{align}
	With $T_{\mu \nu}$ the total stress-energy tensor. In this thesis we shall focus in two specific models within the family \eqref{E1.2.1}:
	\begin{itemize}
	\item[\textbf{EMS:}] a ``matter'' source: $\mathcal{I}=\mathcal{L} _M \equiv F_{\mu \nu} F^{\mu \nu}$, with $\psi=A_\mu$ the $4$-vector potential and $F_{\mu \nu} =  A_{\nu\, ,\mu} - A_{\mu\, , \nu}$ the Maxwell tensor.
	\item[\textbf{eSTGB:}] a geometric source: $\mathcal{I}=R_{GB} ^2 = \mathcal{L} _{GB} \equiv R^2 -4R_{\mu \nu}R^{\mu \nu}+R_{\mu \nu \rho \delta}R^{\mu \nu \rho \delta}$ the Gauss-Bonnet scalar. 
	\end{itemize}
	While for the latter model no additional extra matter fields are present ($\psi =0$); the former model's equations of motion are supplemented by the Maxwell equations for the electromagnetic field
	\begin{equation}\label{E1.2.4}
	 \big(\sqrt{-g} \ze f\ze F^{\mu \nu}\big) _{,\mu} = 0\ ,
	\end{equation}	 

	\bigskip
	
	Observe that the Reissner-Nordstr\"om (RN) BH is a solution of the action~\eqref{E1.2.1} with $f(\phi)=c^{\rm te}$, $\phi ={\rm c^{te}}$. For more general $f(\phi)$, however, the RN BH may or may not solve~\eqref{E1.2.2}-\eqref{E1.2.4}. This naturally leads to two classes of EMS models. (Note that, in this classification~\cite{astefanesei2019einstein}, we assume, without any loss of generality, that the scalar field vanishes asymptotically, $\lim\limits_{r\to + \infty} \phi = 0$.) 
	\begin{itemize}

	 \item[\textbf{I}] \textbf{dilatonic-type:} In this class of models $\phi = 0$ does $not$ solve the field equations\footnote{For the EMS model, there is an exceptional case: if $Q_e =P$, $\phi=0$ solves this class, so that the dyonic, equal charges RN BH is a solution.}. Then, the scalar field equation \eqref{E1.2.2} implies that
		\begin{equation}\label{E1.2.5}
		 \hat{f} (\phi =0) \equiv \frac{d\ze f (\phi)}{d \phi}\bigg |_{\phi=0} \neq 0\ .
		\end{equation}
	A representative example of a coupling for this class is the standard dilatonic coupling
		\begin{equation}\label{E1.2.6}
		 f (\phi)=e^{\alpha \phi} \ ,
		\end{equation}
	in which case we refer to $\phi$ is a dilaton field. The arbitrary non-zero constant $\alpha$ (\textit{a.k.a.} coupling constant) is taken to be positive without any loss of generality. Indeed, the solutions remain invariant under the simultaneous sign change $(\alpha,\phi) \to -(\alpha,\phi)$. Thus, flipping the sign of $\alpha$ simply corresponds to flipping the sign of $\phi$. The coupling~\eqref{E1.2.6} appears naturally in Kaluza-Klein (KK) models~\cite{dobiasch1982stationary,gibbons1986black} and supergravity/low-energy string theory models~\cite{gibbons1975vacuum}. Three reference values for the coupling constant $\alpha$ in \eqref{E1.2.6} are:
		\begin{equation}\label{E1.2.7}
		 \alpha=0 \ ,\ {\rm (EM \ theory)} \qquad \alpha=1 \ ,\ {\rm (low \ energy \ strings)} \qquad \alpha=\sqrt{3} \  \ {\rm (KK \ theory)} \ .
		\end{equation}
	Some exact, closed form BH solutions of \eqref{E1.2.2}-\eqref{E1.2.4} with \eqref{E1.2.6} are known and presented in Appendix~\ref{C}. Other exact solution examples in this class (with a non-dilatonic coupling) are given in~\cite{fan2015charged}.

	Once embedded in string theory, the dilaton $\phi$ controls the string coupling, which
 is related to the vacuum expectation of the asymptotic value of the dilaton, $g_s=e^{\langle\phi_{+\infty} \rangle}$. Therefore, a consistent analysis of hairy BHs in string theory should consider a dynamical dilaton whose asymptotic value can vary \cite{gibbons1986black} (see, also, \cite{astefanesei2019einstein} for a resolution of the appearance of the scalar charges in the first law of thermodynamics). This need, however, is mitigated by \textit{the attractor mechanism} \cite{ferrara1996supersymmetry,ferrara1996university,ferrara1995n}: the near horizon data (particularly, the entropy) of extremal BHs is independent of the asymptotic values of the moduli. The mechanism is based on a simple physical intuition; when the temperature vanishes, there is a symmetry enhanced near horizon geometry: $AdS_2\times S^2$. The infinite long throat of $AdS_2$ yields the decoupling between the physics at the boundary from the physics at the extremal horizon \cite{kallosh1992supersymmetry}. A similar decoupling plays a central role in the $AdS/CFT$ duality (see, $e.g.$  \cite{larsen2019nattractor,lin2019symmetries}). 
	\item[\textbf{II}] \textbf{scalarized-type:} In this class of models $\phi= 0$ solves the field equations. This demands that 
		\begin{equation}\label{E1.2.8}
		 \hat{f} (\phi =0) \equiv \frac{d\ze f (\phi)}{d \phi}\bigg |_{\phi=0}= 0 \ .
		\end{equation}
	This condition is naturally implemented, for instance, if one requires the model to be $\mathbb{Z}_2$-invariant under $\phi\rightarrow -\phi$. The BH solution, however, is (in general) not unique. These models may contain a second set of BH solutions, with a non-trivial scalar field profile -- {\it the scalarized BHs}. Such a second set of BH solutions may or may not continuously connect with the standard BHs. Such leads to two subclasses. Below, some conditions for this to occur are discussed.
		\begin{itemize}
		 \item[\textbf{II.A}] \textbf{scalarized-connected-type:} In this subclass of models, the scalarized BHs bifurcate from the standard BH, and reduce to the latter for $\phi=0$. This bifurcation moreover, may be associated to a tachyonic instability, against scalar perturbations, of the standard BH. Considering a small-$\phi$ expansion of the coupling function
			\begin{equation}\label{E1.2.9}
			 f(\phi)=f(0)+\frac{1}{2}\frac{d^2 f(\phi)}{d \phi^2}\bigg |_{\phi=0}\phi^2+\cdots \ ,
			\end{equation}
	equation \eqref{E1.2.2} linearized for small-$\phi$ reads:
			\begin{equation}\label{E1.2.10}
			 \big(\Box-\mu_{\rm eff}^2\big)\phi =0\ , \qquad {\rm where} \ \ 
\mu_{\rm eff}^2= \mathcal{I}\ze \frac{d^2 f(\phi)}{d \phi^2}\bigg |_{\phi=0}   \ .
			\end{equation}
	The instability arises if $\mu_{\rm eff}^2<0$, which in particular requires

			\begin{equation}
			 \hat{\hat{f}} (\phi =0) \equiv \frac{d^2 f(\phi)}{d \phi^2}\bigg |_{\phi=0} \neq 0 \ ,
			\end{equation}
	and with the opposite sign of $\mathcal{I}$. Two reference example of a coupling function in this subclass, which we consider in this work are
			\begin{equation}\label{E1.2.11}
			 f(\phi )= e^{\alpha \phi^2}\ \text{\cite{herdeiro2018spontaneous,silva2018spontaneous}}\ ,\qquad\qquad f(\phi) = \frac{\phi ^2}{2}\ \text{\cite{silva2018spontaneous}}\ .
			\end{equation}		
		With the EMS case still relevant in cosmology \cite{martin2008generation,maleknejad2013gauge}. Depending on the coupling, this subclass could also contain \textit{another} family of disconnected scalarized BHs, akin to the ones of class \textbf{II.B} below.
 
		\item[\textbf{II.B}] \textbf{scalarized-disconnected-type:} In this subclass, the scalarized BHs do not bifurcate from GR BHs and do not reduce to the latter for $\phi=0$. This is the case if there is no tachyonic instability, for which a sufficient (but not necessary) condition is that 
			\begin{equation}\label{E1.2.12}
			 \hat{\hat{f}} (\phi =0) \equiv \frac{d^2 f(\phi)}{d \phi^2}\bigg |_{\phi=0} = 0 \ .
			\end{equation}
	A representative coupling in this subclass, which we shall consider in this work (for EMS~\cite{blazquez2020einstein,blazquez2021quasinormal}) is
			\begin{equation}\label{E1.2.13}
			 f(\phi)=1+\alpha \ze\phi^4	\ .
			\end{equation}			 
		\end{itemize}
	\end{itemize}
	Condition~\eqref{E1.2.8} guarantees that the standard GR BH is a solution. Yet, it does not guarantee the existence of scalarized BHs. In the case of purely electric (or magnetic) BHs, two Bekenstein-type identities can be derived, which put some constraints on $f$ so that scalarized solutions exist. These will be derived in Ch.~\ref{C2}.  
	
	\bigskip
	
	At last, observe that action \eqref{E1.2.1} imposes the scalar field equation of motion
			\begin{equation}\label{E1.2.15}
			 \Box \phi =\hat{f} \ze\ze \frac{\mathcal{I}}{4}\ ,
			\end{equation}
	which, after being linearized around a scalar free solution, yields $\big(\Box -\mu _{\rm eff} ^2 \big)\delta \phi = 0\ze $, where
			\begin{equation}\label{E1.2.16}
		 	 \mu _{\rm eff} ^{ 2} =  \hat{\hat{f}} (0)\ze \frac{\mathcal{I}}{4},		
			\end{equation}			
	is the scalar field effective mass. In order for a tachyonic instability to settle in, we must have $ \mu _{\rm eff} ^{2} <0\ze $. As an example, let us consider the EMS model -- with $\mathcal{I}=Q_e ^2 \, r^{-4}$ (full computation can be seen in Ch.~\ref{C2}) --, this implies that we have now three possibilities:
			\begin{itemize}
			 \item For class \textbf{I} the $\mu _{\rm eff} ^2=\alpha ^2 Q_e ^2 r^{-4}$, and a tachyonic instability is impossible to settle in.	
		 	\item Class \textbf{II.A} has $\mu _{\rm eff} ^2 = -|\alpha|\ze\ze Q_e^2 \ze r^{-4} <0$, a tachyonic instability can settle and endow spontaneously scalarized BHs.
			 \item Class \textbf{II.B} has $\mu _{\rm eff} ^2 = 0$, there is no tachyonic instability but disconnected scalarized solutions are still possible.
			\end{itemize}		
	Let us focus on class \textbf{II.A}. Spherical symmetry allows a scalar field's decomposition in (real) spherical harmonics, 
			\begin{equation}\label{E3.3.26}
			 \phi=U_\ell(r) Y_{\ell}^ m(\theta,\varphi) \ ,
			\end{equation}
	where  $Y_{\ell m}$ are the real spherical harmonics and $\ell,m$ are the associated quantum numbers with the usual ranges, $\ell=0,\,1,\,\dots$ and $-\ell \leqslant m \leqslant \ell$. The scalar field equation simplifies to
		\begin{equation}\label{E1.2.17}
	 	 \frac{1 }{\sigma r^2} \Big( r^2 N  \sigma\ze U' _{\ze \ell} \Big) '-\left[\frac{\ell(\ell+1)}{r^2}+\mu _{\rm eff} ^{ 2} \right] U_\ell=0\ ,
		\end{equation}
	which is an eigenvalue problem: fixing the coupling $\alpha$, for a given $\ell$, requiring an asymptotically vanishing, smooth scalar field, selects a discrete set of BHs solutions, $i.e.$ RN solutions with a certain $Q_e /M = q$. These are the bifurcation points of the scalar-free solution. They are labelled by an integer $n\in \mathbb{N}_0\ze $; $n=0$ is the fundamental mode, whereas $n>1$ are excited states (overtones). One expects only the fundamental solutions to be stable~\cite{myung2019instability}. Focusing on the latter, solutions with a smaller (larger) $q$ are stable (unstable) against spherical scalar perturbations for that coupling. Where $\sigma =1$ and $N =1-2M/r+Q_e^2 / r^2$ are the two metric functions of an usual RN metric (see Sec.~\ref{S1.5}). Then, a scalarized solution can be dynamically induced by a scalar perturbation of the background, as long as the scalar-free RN solution is in the unstable regime.

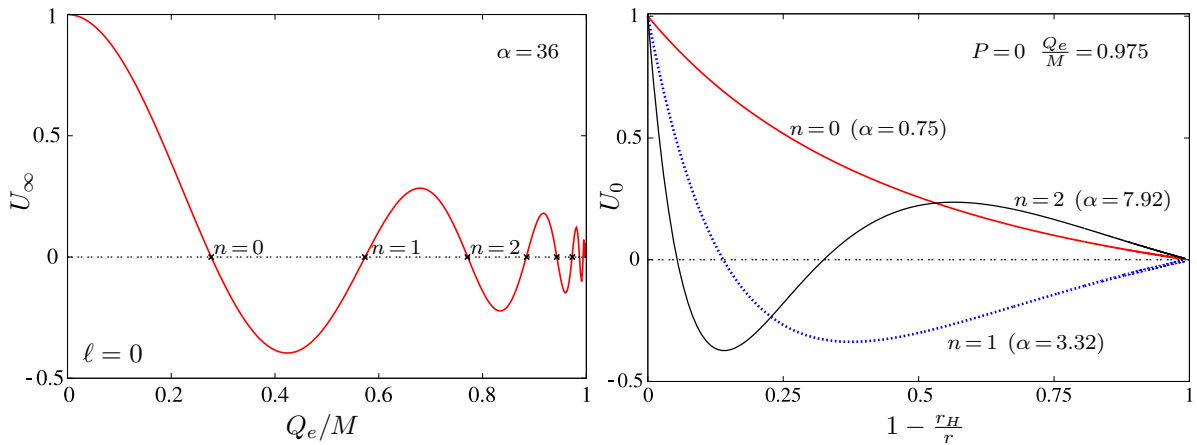
\begin{figure}[H]
			\begin{center}
				\begin{picture}(0,0)
		  		 \put(25,15){\small $\ell = 0$}
		  		 \put(180,130){$\scriptstyle \alpha\, =\, 36$}
		  		 \put(74,56){$\scriptstyle n\, =\, 0$}
		  		 \put(133,56){$\scriptstyle n\, =\, 1$}
		  		 \put(170,56){$\scriptstyle n\, =\, 2$}
		  		 \put(101,-12){\small $Q_e /M$}
		  		\put(-2,72){\begin{turn}{90}{\small $U_\infty$ }\end{turn}}
		   		\end{picture}
				 \includegraphics[scale=0.632]{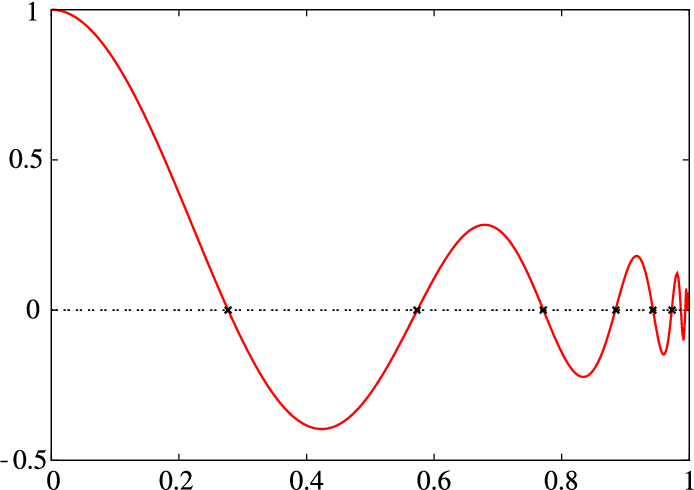}\hfill
				 \begin{picture}(0,0)
		  		 \put(140,130){$\scriptstyle P\,=\,0\ \, \frac{Q_e}{M}\,=\, 0.975$}
		  		 \put(72,101){$\scriptstyle n\, =\, 0\ (\alpha\, = \, 0.75)$}
		  		 \put(131,20){$\scriptstyle n\, =\, 1\ (\alpha\, = \, 3.32)$}
		  		 \put(156,74){$\scriptstyle n\, =\, 2\ (\alpha\, = \, 7.92)$}
		  		 \put(108,-12){\small$1-\frac{r_H}{r}$}
		  		\put(1,72){\begin{turn}{90}{\small $U_0$ }\end{turn}}
		   		\end{picture}
				 \includegraphics[scale=0.62]{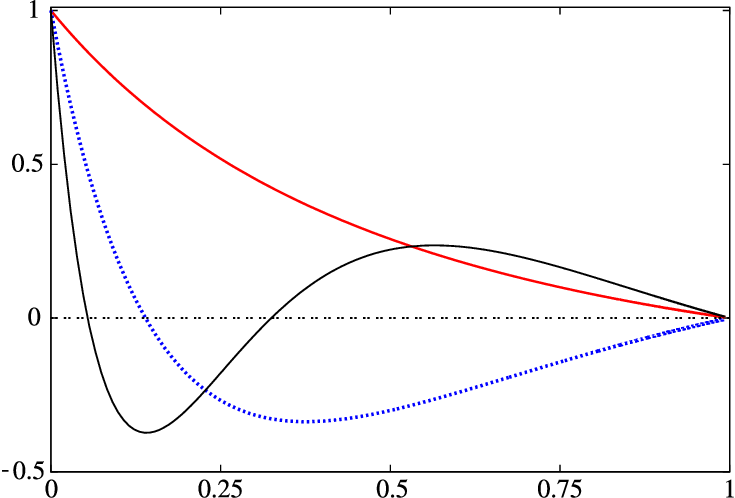}
				\end{center}
			 \caption{(Left panel) The asymptotic value $U_{\infty}$ of the zero-mode amplitude $U_0$ for $\alpha=36$ as a function of the charge to mass ratio of a RN BH. An infinite set of configurations with $U_\infty=0$ exist, labelled by $n$, the number of nodes of $U_0$. (Right panel) The profiles of three zero mode amplitudes $U_0$ with a different node number, for a given RN background.}
			\label{F1.1}
		\end{figure}

	As pointed out in~\cite{herdeiro2018spontaneous}, for $\ell=0$, one finds the following exact solution\footnote{No exact solution appears to exist for $\ell\geqslant 1$, and equation (\ref{E1.2.17}) has to be solved numerically. These modes, nonetheless, also possess non-linear continuations leading to static, non-spherically symmetric scalarized BHs \cite{herdeiro2018spontaneous}.}
		\begin{equation}\label{E1.2.18}
		 U_0 =P_u \left[1+\frac{2Q_e^2\ze (r-r_H)}{r_H^2-Q_e^2}\right]\ , \qquad {\rm where} \qquad u\equiv\frac{1}{2}\big( \sqrt{4\alpha-1}-1\big) \ ,
		\end{equation}
	and $P_u$ being a Legendre function. For generic parameters $(\alpha,\ze Q_e,\ze r_H)$, the function $U_0$ approaches a constant \textit{non-zero} value as $r\to +\infty$,
		\begin{equation}\label{E1.2.19}
		 U_0 \to {}_2F_1 \left[ \frac{1}{2}\big( 1-\sqrt{4\alpha+1}\big) ,\,\frac{1}{2}\big( 1+\sqrt{4\alpha+1}\big) ,\,1;\, \frac{x^2}{x^2-1}\right]+\mathcal{O}\left(\frac{1}{r}\right) \ ,
		\end{equation}
	where $x=Q_e/r_H$. Thus finding the $\ell=0$ unstable mode of the RN BH reduces to a study of the zeros of the hypergeometric function ${}_2F_1\ze $. Some values are given in Table~\ref{T2.1} and Fig.~\ref{F1.1}.

%
	\section{Solitonic objects}\label{S1.3}
%
	In quantum mechanics, every particle has an intrinsic property: their intrinsic angular momentum or \textit{spin}. Its value divides particles into two groups: the fermions, with half-integer spin, and the bosons, with integer spin. 
	
	While bosons follow the Bose-Einstein statistic, which does not restrict the occupation number of a given state, fermions follow the Fermi-Dirac statistic and obey the Pauli exclusion principle. There can only be one fermion in a given quantum state at a given time. For the boson case, since they can occupy the same quantum state, they can bunch together, creating lumps of coherent matter: \textit{Bose-Einstein condensates}.
	
	In particle physics, while the classical matter (quarks and leptons) are fermions, the force carriers (photons, gluons, ...), as well as the Higgs particle, are bosons. In the observable universe, every known astrophysical object is composed mainly of fermions. Hypothetical structures, on the other hand, composed of bosons have been theoretically suggested, the simplest of which are \textit{$Q$-balls} and \textit{boson stars}. These are compatible with current observations as long as they only interact weakly with electromagnetic radiation, therefore, dark matter objects.
	
	In this section, our main objective will be to investigate such hypothetical bosonic objects. At first, the $Q$-ball case will be considered. $Q$-balls consist of many bosons held together by non-linear self-interactions. In the second subsection, self-gravitating bosonic objects, \textit{boson stars} will be treated. 
	
	Described for the first time in $1834$ by John Scott Russel \cite{drazin1989solitons}, solitons play an important role in several branches of science, namely fibre optics, biology, hydrodynamics and models of high energy physics. In the latter, they have been suggested in the early $1960$s as a model of hadrons: the \textit{Skyrme model}~\cite{skyrme1994non}. In the $1980$s, Coleman observed that a single complex scalar field could yield solitonic solutions when appropriate (non-renormalizable) self-interactions are included. Such solutions of a non-linear scalar field theory on Minkowski spacetime became known as  \textit{$Q$-balls} \cite{coleman1985q}. In $1988$ Frieman \textit{et al}.~\cite{frieman1988primordial} and later in $1997$ Kusenko \textit{et al.}~\cite{kusenko1998supersymmetric}, speculated that dark matter might consist of $Q$-balls and that the latter plays a role in baryogenesis.
%
		\subsection{Q-balls}\label{S1.3.1}
%
		The familiar Klein-Gordon equation, with a simple mass term, on Minkowski spacetime, admits plane wave solutions. At any given time, a generic wave packet can be constructed as a superposition of these plane waves. However, the time evolution of such a wave packet is generically dispersive, as the different plane waves have different phase velocities. 

	The dispersiveness of the linear Klein-Gordon equation can be counter-balanced by introducing non-linear terms in the wave equation, \textit{i.e.}, self-interactions of the scalar field. When cancellation of non-linear and dispersive effects occurs, self-reinforcing solitary wave packets emerge: these are dubbed \textit{solitons}. Therefore, solitons are localized lumps of (in this case, scalar field) energy that are dynamically sufficiently stable. 

	$Q$-balls are non-topological solitons (\textit{i.e.} their existence does not rely on a non-trivial topological structure of the vacuum of the theory) made of a complex scalar field under a non-renormalizable self-interaction. These non-topological solitons circumvent the standard Derrick-type argument (see Ch. \ref{C7}) by having a time-dependent phase for the scalar field. $Q$-balls emerge in models with a global $\textbf{U}(1)$\footnote{Gauging the $\textbf{U}(1)$ symmetry, endows $Q$-balls with an electric charge~\cite{lee1989gauged,Gulamov:2015fya,Gulamov:2013cra}.} symmetry that leads to a conserved Noether charge $Q_S$, corresponding to the particle number. Such configurations have a rich structure and have several physically exciting applications.
		
	 They can be either spherically symmetric or spinning, gauged or ungauged. Some branches of $Q$-balls are stable. Dynamical properties of $Q$-balls have been study in \cite{battye2000q,axenides2000dynamics,bazeia2017split}. Stable $Q$-balls have the lowest energy per unit charge compared to any other configuration \cite{coleman1985q}. In particular, there are excited states, labelled by a quantum number $n$, with a discrete energy spectrum, for a fixed Noether charge. However, this thesis will focus on fundamental states ($n=0$).

	In the absence of interactions with fermions, there are two main types of $Q$-ball stability: the stability against (spontaneous) decay into $Q$-balls with a smaller charge (fission)\footnote{Where it is also included the stability concerning (spontaneous) decay into free particles.}; and classical stability (against small perturbations). A detailed study of stability for non-spinning solutions can be found in \cite{Gulamov:2015fya,Panin:2016ooo}.

	\bigskip	
	
	The action that describes a complex scalar field in a $(3+1)$-dimensional spacetime under a self-interaction potential $U(\Phi)$ and in the absence of gravity comes as
		\begin{equation}
		 \mathcal{S} _{\bar{\Phi}}=-\frac{1}{4\pi}\int d^4 \, x \sqrt{-g} \bigg[\frac{1}{2}g^{\mu \nu}\Big(\bar{\Phi} _{,\mu} \Phi _{,\nu} + \bar{\Phi}_{,\nu} \Phi _{,\mu}\bigg) + U\big( |\Phi|^2 \big)\bigg]\ .
		\end{equation}
	It is assumed that the potential $U$ has its global minimum at $\Phi=0$, where $U(0)=0$, while $U\to +\infty$ for $|\Phi|\to +\infty$. In addition, the potential must fulfil two additional criteria that we shall discuss below. 
	
	The global symmetry of the Lagrangian density, $\mathcal{L}$, under $\Phi \to \Phi\ze e^{i a}$ gives rise to the conserved Noether charge
		\begin{equation}
		 Q_S = -\textit{i} \int d^3 \, x \ze\Big[ \bar{\Phi} \dot{\Phi} -\Phi \dot{\bar{\Phi}}\Big]\ .
		\end{equation}
	The fundamental $Q$-ball solutions of the theory are minima of the energy for a given $Q_S$\cite{coleman1985q}. Since $\Phi$ must be time-dependent to have a non-vanishing $Q_S$, it is usual to assume an harmonic time-dependence. In the spherically symmetric case,
		\begin{equation}\label{E1.3.17}
		 \Phi = \phi (r) e^{i\omega t}\ ,
		\end{equation}
	where $\phi$ is a real function that describes the radial amplitude of the scalar field. The stress-energy tensor,
		\begin{equation}
		 T_{\mu \nu} = \Phi_{,\mu}\bar{\Phi}_{,\nu} +\Phi_{,\nu} \bar{\Phi}_{,\mu}-g_{\mu \nu} \mathcal{L}\ ,
		\end{equation}
	does not depend on time. The energy distribution\footnote{In a standard Minkowsky metric
	\begin{equation}
	 ds^2 = -dt^2 + dr^2 +r^2 \big(d\theta ^2 + \sin ^2 \theta d\varphi ^2 \big)\ , 
	\end{equation} with $\sqrt{-g}=r^2 \sin\theta$.} is therefore stationary, and the total energy is
		\begin{equation}
	 	 E=4\pi \int _0 ^{+\infty} dr\, r^2 \Big[ \omega ^2 \phi ^2 +\phi'^{\, 2}+U(\phi)\Big] \ , 
		\end{equation}
	The Klein-Gordon field equation
		\begin{equation}
	 	 \Box \Phi = -\hat{U}\ ,
		\end{equation}
	reduces to 
		\begin{equation}
		 \phi '' +\frac{2}{r}\phi ' -\hat{U}+\omega ^2 \phi =0\ ,
		\end{equation}
	This is equivalent to
		\begin{equation}
		 \frac{1}{2}\phi'^{\, 2} +\frac{1}{2}\omega ^2\phi ^2 -U = \mathcal{E}-2\int_0 ^{\tilde{r}} \frac{dr}{r}\phi'^{\, 2}\ .
		\end{equation}
	This effectively describes a particle moving with friction in the $1$-dimensional potential
		\begin{equation}
		 \mathcal{V}_{\rm eff}(\phi) = \frac{1}{2}\omega ^2 \phi ^2 -U\ .
		\end{equation}
	$\mathcal{E}$ is the integration constant playing the role of the total ``effective energy''. For a $Q$-ball to exist, since $\mathcal{V}_{\rm eff} ''(0)<0$, it follows that $\omega^2$ should not be too large:
		\begin{equation}
		 \omega ^2 < \omega _+ ^2 \equiv U''(0)\ ,
		\end{equation}
	On the other hand, $\omega ^2$ should not be too small either, since otherwise $\mathcal{V}_{\rm eff}$ will be always negative. $\mathcal{V}_{\rm eff}$ will become positive for some non-zero $\phi$, only if
		\begin{equation}
		 \omega ^2 > \omega_- ^2 \equiv \text{min} \left(2\ze\frac{U}{\phi^2}\right)_\phi\ ,
		\end{equation}
	where the minimum is taken over all values of $\phi\ze$. In addition, since we want a non-empty interval: $\omega_+> \omega _- \ze$. Observe that the only possible renormalizable self-interaction potential is $U = \frac{1}{2} \mu ^2 \phi ^2+\beta\ze \phi ^4$, does not obey this condition. Thus, non-renormalizable potentials must be considered.		
%
		\subsection{Boson stars}\label{S1.3.2}
%
	When backreaction is considered a \textit{Boson Stars (BS)} emerges. Unlike $Q$-balls, the dispersive behaviour of the scalar field is counter-balanced by the attractive gravitational pull. The established self-interaction conditions relax, and solutions with a single mass term are now possible. 

	BSs are speculative macroscopic Bose-Einstein condensates. They may be described as everywhere regular lumps (\textit{i.e.} self-gravitating solitons) of yet undetected ultra-light scalar~\cite{kaup1968klein,ruffini1969systems} or vector~\cite{brito2016proca} bosonic fields\footnote{BSs can also be gauged to generate charged~\cite{jetzer1989charged,jetzer1989stability} and obtained in the presence of a cosmological constant~\cite{hartmann2013compact,kichakova2014spinning,buchel2013boson,radu2012spinning}.} -- see \cite{colpi1986boson,lynn1989q,schunck1998rotating,yoshida1997rotating,schunck2003general,liebling2017dynamical,grandclement2014models,herdeiro2017asymptotically,alcubierre2018ell,herdeiro2019asymptotically,guerra2019axion,delgado2020rotating,minamitsuji2018vector,herdeiro2021multipolar,herdeiro2020asymptotically} for a more in-depth analysis. 

	Unlike ordinary stars, BSs would be transparent and invisible. In an environment with ordinary matter, they can be compact enough to bend light due to the gravitational pull, creating an empty region resembling a shadow of a BH event horizon. However, the absence of a horizon would cause the accreted matter to be visible in their interior.

	In a nutshell, BSs are localized solutions of the coupled system of Einstein and general relativistic Klein-Gordon (Proca) equations of a massive complex scalar (vector) $\Phi$ ($B_\mu$) field under a self-interaction potential $U$. 

	The action that describes a massive, complex matter field with spin $s=0,\, 1$, minimally coupled to Einstein gravity reads:
	\begin{equation}
	 \mathcal{S} _{s}=\int d^4 x \sqrt{-g} \left[ \frac{R}{16 \pi G}+\mathcal{L}_s \right]\ ,
	\end{equation}
where $R$ is the Ricci scalar of the spacetime represented by the metric $g_{\alpha \beta}$ with metric determinant $g$, $G$ is Newton's constant and the matter Lagrangians for the spin $0$ and spin $1$ fields are, respectively:
	\begin{equation}
	\mathcal{L}_0 = 
	-\frac{1}{2}g^{\alpha \beta} \big(\bar{\Phi} _{,\alpha} \Phi _{,\beta}+\bar{\Phi} _{,\beta}\Phi _{,\alpha} \big) - U_i(|\Phi |^2)\ , \qquad \qquad \mathcal{L}_1= -\frac{1}{4} G_{\alpha \beta}\bar{G} ^{\alpha \beta}-U (\textbf{B})\ .
	\end{equation}
	 The massive complex scalar field, $\Phi$, has a potential term $U_i(|\Phi |^2)$; the massive complex vector field, has a $4$-potential $B^\mu$ and is under a potential $U(\textbf{B})$. We have used the notation $\textbf{B}\equiv B_\mu \bar{B}^\mu$. 

	Variation of the action concerning the metric and matter fields leads to the following two sets of field equations, in the scalar and vector case, respectively
	\begin{align}
	 & E_{\mu \nu} = { 4\pi G} \Big[\bar{\Phi} _{,\mu} \Phi _{,\nu}+\bar{\Phi} _{,\nu}\Phi _{,\mu}-g_{\mu \nu} \mathcal{L}_0 \Big]\ , \qquad \Box \Phi = \hat{U_i}\ \Phi \ ,\\
	 & E_{\mu \nu} ={4\pi  G}\left[\frac{1}{2} \big(G_{\mu \delta} \bar{G}_{\nu \gamma} +\bar{G}_{\mu \delta} G_{\nu \gamma} \big) g^{\delta \gamma}+\hat{U}\big( B_\mu \bar{B}_\nu +\bar{B}_\mu B_\nu - g_{\mu \nu}\mathcal{L}_1 \big)\right] \ ,\nonumber\\
	 &\frac{1}{2} \nabla_\mu G^{\mu \nu} = \hat{U} B^\nu\ ,
	\end{align}
	with $E_{\mu \nu}$ the Einstein's tensor, $\Box$  $(\nabla)$ the covariant d'Alembertian (derivative) operator, $\hat{U_i}\equiv d U_i/d |\Phi |^2$ and $\hat{U}\equiv dU /d \textbf{B}$. 
%
	\section{Virial identity}\label{S1.4}
%
	The usefulness of the virial identity in physics is well known, mainly within the context of dynamical systems' equilibrium and stability properties. 
	
	In particle mechanics, the virial theorem is a \textit{statistical} result. It provides a useful relation between the averages over time of the total kinetic and potential energies for a stable system of $\mathcal{N}$ bound particles. The theorem reads \cite{goldestein} 
		\begin{equation}\label{E1.4.31}
		 \langle T \rangle =-\frac{1}{2}\sum_{i=1}^\mathcal{N} \langle \vec{F}_i\cdot \vec{r}_i \rangle \ ,
		\end{equation}
	where $T$ denotes the total kinetic energy and $\vec{F}_i$ the force over the $i^{\rm th}$ particle, which has position $\vec{r}_i$. The time averaging, denoted by $\langle \rangle$, amounts to a time integral, $\langle X \rangle \equiv (\Delta t)^{-1}\int_{t_i}^{t_f}X\, dt$, for  any quantity $X$.  Upon choosing appropriately an  integration interval $\Delta t\equiv t_f-t_i$, the theorem is, equivalently,
		\begin{equation}\label{E1.4.32}
		 \frac{1}{\Delta t}\int_{t_i}^{t_f} \left( T + \frac{1}{2}\sum_{i=1}^\mathcal{N}  \vec{F}_i\cdot \vec{r}_i \right)dt=0 \ .   \qquad \qquad {\rm {\bf [virial \  Clausius]}}
		\end{equation}
	Eq.~\eqref{E1.4.32} makes clear that the virial theorem amounts to an \textit{integral identity}. If the motion is periodic, choosing $\Delta t$ to be a multiple of the period, the integral exactly vanishes, and the $(\Delta t)^{-1}$ pre-factor is unnecessary. Nevertheless, even if the time integration is not exactly zero (for instance, if the motion is not periodic), for a system of bound stable particles, the integrand is bounded, and the \textit{lhs} of~\eqref{E1.4.32} can be made arbitrarily small by choosing a sufficiently large time interval. In either case, the virial theorem holds to arbitrary accuracy.

	If the forces are conservative, derivable from a total potential energy $U(\vec{r}_i)$, and if $U$ is a homogeneous function of degree $n$ of the particles' coordinates, then the virial theorem takes the form $\langle T\rangle =n \langle U\rangle/2$~\cite{goldestein}.

	For the particular case of inverse square law forces, $n=-1$, we recover the expected result that the average kinetic energy (in modulus) is one half of the average potential energy (which is negative)\footnote{Here, the integral is understood to be over a multiple of the period.}: 
		\begin{equation}\label{E1.4.33}
		 \int_{t_i}^{t_f} \left( T + \frac{U}{2}   \right)dt=0 \ .    \qquad \qquad {\rm {\bf [virial \  inverse \ square \ force \ law]}}
		\end{equation} 
	The virial identity~\eqref{E1.4.33} can be recovered by a \textit{scaling argument}. Consider the classical action of a particle, $\mathcal{S}=\int_{t_i}^{t_f}(T-U)dt$, where the kinetic energy is a homogeneous function of degree $2$ of the velocity, and the potential energy is assumed to be a homogeneous function of $\vec{r}$ of degree $n$. Consider that there is a solution of the classical equations of motion $\vec{r}=\vec{r}\ze (t)$.
	If one \textit{scales} this fiducial solution by a factor of $a$,  $\vec{r}(t)\rightarrow a\ze\vec{r}\ze (t)$, then $T\rightarrow a ^2\ze T$, while $U\rightarrow  a ^n \ze U$. The corresponding action\footnote{The action of the scaled solution becomes a \textit{function} of $a$, whereas it is a \textit{functional} of the particle's path.} $\mathcal{S}_a=\int_{t_i}^{t_f}\big(a^2 T-a^n U\big)\ze dt$ should be stationary at the original fiducial solution:
		\begin{equation}\label{E1.4.34}
		 \frac{\partial \mathcal{S}_a}{\partial a}\bigg|_{a=1}= 0 \ \ \stackrel{n\, =\, -1}{\stackrel{\  \Delta t\, =\, {\rm period}}{\Rightarrow}} \ \ \eqref{E1.4.33} \ .
		\end{equation}
	Note that $n=-1$ guarantees periodic motion, and choosing $\Delta t=$ period makes the above scaling a variational problem with periodic boundary conditions rather than fixed boundary conditions. The latter illustrates the derivation of a virial identity from a scaling argument. 

	Initially presented by R. Clausius in 1870~\cite{clausius1870xvi}, who dubbed the \textit{rhs} of~\eqref{E1.4.31} ``virial'', the virial theorem has found many applications in physics and mathematics. In the context of gravitation, for instance, F. Zwicky first deduced the existence of a gravitational anomaly and suggested the existence of  ``dark matter'' from an application of the virial theorem~\cite{Zwicky:1933gu}.

	In this thesis, we shall be interested in integral identities that are virial-like (and thus, following the literature,  will be referred to as ``virial identities''), but in field theory rather than particle mechanics, obtained from \textit{scaling} arguments.
%
	\section{Methods, notations and strategies}\label{S1.5}
%
	Let us assume that the computed solutions have a well-defined symmetry. In the majority of this work\footnote{In Ch.~\ref{C7} we will present and study several metric ansatze}, and unless stated otherwise, we use $(3+1)$-dimensional Shwarzschild-like spherical coordinates $(t,\, r,\, \theta,\, \varphi)$. The line element ansatz compatible with spherical symmetry comes as,
		\begin{equation}\label{E1.5.40}
	 	 ds^2 = -N(r) \sigma ^2 (r) dt^2 +\frac{dr^2}{N(r)}+r^2 \big( d\theta ^2 +\sin ^2 \theta \ d\varphi \big)\ ,\qquad \qquad N(r)=1-\frac{2\ze m(r)}{r}\ ,
		\end{equation}
	where $m(r)$ is the Misner-Sharp mass function~\cite{misner1964relativistic}. For axial symmetry, one considers a standard metric ansatz with four $\mathcal{F }_i (r,\, \theta)$ metric functions 
	\begin{equation}\label{E1.5.41}
	 ds^2 = -e^{2F_0}H(r) dt^2+ e^{2F_1}\left(\frac{dr^2}{H(r)}+r^2d\theta^2\right) + e^{2F_2}r^2\sin^2\theta\big( d\varphi-F_Wdt\big) ^2, \qquad  H(r)=1-\frac{r_H}{r} \ ,
	\end{equation}
	with $\mathcal{F}_i = \{ F_0,\, F_1,\, F_2,\, F_W\}$. 
	
	For the scalar field ansatz, one consider the following ansatz compatible with a complex, axialy symmetric scalar field with a harmonic time and angular dependence 
	\begin{equation}\label{E1.5.43}
	 \Phi (t,r,\theta,\varphi) = \phi (r,\theta) e^{\textit{i}\ze(m\varphi -\omega t)}\ ,
	\end{equation}			
	where $\omega $ is the scalar field frequency, $m$ is the azimuthal quantum number and $\phi $ is a real function that describes the amplitude of the scalar field. The spherically symmetric solution (Sec.~\ref{S1.3}) is recovered when $m=0$ and $\phi (r,\theta )\equiv \phi (r)$ -- see \eqref{E1.3.17}. Real scalar fields (Sec.~\ref{S1.2}) are recovered for $\omega = 0 = m$.

	In this work, we will solely deal with spherically symmetric vector fields, and hence the corresponding ansatz for a complex vector field comes as
	\begin{equation}\label{E1.5.44}
	 B_\mu = \big[ B_t(r) dt+\textit{i} B_r (r) dr \big] e^{-\textit{i}\ze\omega t} \ ,
	\end{equation}
	which reduces to a real vector field for $\omega=0 = B_r(r)$ (see Sec.~\ref{S2.5}).
	
	The Maxwell field in axial symmetry introduces two new functions, $V(r,\theta)$ and $A_\varphi (r,\theta)$, such that
	\begin{equation}\label{E1.5.45}
	 A_\mu = V dt +A_\varphi \sin \theta \big( d\varphi - F_W  dt\big)\ .
	\end{equation}
	At last, let us also define the effective Lagrangian formalism which is obtained by integrating the angular components of a radially dependent (besides the Jacobian $\sqrt{-g}$) Lagrangian density $\mathcal{L}$:
		\begin{equation}\label{E1.5.46}
		 \mathcal{L} ^{\rm eff} = \frac{1}{4\pi}\int _0 ^{2\pi} d\varphi \int_0 ^\pi  d\theta \,\mathcal{L}  \ .
		\end{equation}
	Besides, to simplify some relations we introduce the compact notation
	\begin{equation}
	\int d^3 x =\int _{r_H} ^{+\infty} dr\int _0 ^\pi d\theta \int _0 ^{2\pi} d\varphi \ .
	\end{equation}
\bigskip

	This work will deal with several field equations that we are unable to solved in a closed form. To surpass this, we will have to resort to numerical methods: all the studied spherically symmetric solutions are described by a set of coupled ordinary differential equations (ODEs). In that regard, we have developed a parallelized, adaptative-step $5(6)$ Runge-Kutta method (Appendix \ref{A}), which can impose proper boundary conditions either through a secant strategy or a bisection method (for the more unstable solutions).
	
	On the other hand, the axially symmetric solutions are described by coupled partial differential equations (PDEs). Even though we could, in principle, write our own numerical solver, the added complexity makes such an endeavour impractical. In that regard, we resorted to a professional solver, namely, the CADSOL/FIDISOL program package (Appendix \ref{B}). The latter is based on a finite difference method with a fixed step and a Newton's-Raphson method to implement the proper boundary conditions.
	
\bigskip

	To compare and test the obtained solutions, let us defined the relative error difference
	\begin{equation}
	err= 1-\frac{X}{Y}\ ,
	\end{equation}
	where $X$ and $Y$ are the two quantities to compare. As an example, the relative error difference between an analytical quantity $Y$ and a numerically obtained one $X$. This quantity will be highly significant in studying the validity of the spectral decomposition and virial identities.
%
	\section{Structure}\label{S1.6}
%
	Let us now briefly summarise the structure of this thesis. We dedicate Ch.~\ref{C2} to the Einstein-Maxwell model with a non-minimal coupling between a matter field (\textit{e.g.} scalar or vector field) and the Maxwell tensor (based on \cite{fernandes2019spontaneous,fernandes2019charged,blazquez2020einstein,blazquez2021quasinormal,astefanesei2019einstein,oliveira2021spontaneous}). We study their properties and stability. Some of the model's possible generalizations are also considered, namely: a massive scalar field, the addition of a magnetic charge and replacing the scalar field with a vector field. Ch.~\ref{C3} is also dedicated to the scalarization of charged BH. However, this time we consider an extended-scalar tensor model of gravity where the scalar field is non-minimally coupled with the Gauss-Bonnet invariant. While in the previous chapter, we have only considered spherical symmetry, in Ch.~\ref{C3} we introduce rotation, adding a new degree of complexity to the system (based on \cite{herdeiro2021aspects}).

	Then we proceed with the study of solitons as possible BH mimickers in Ch.~\ref{C4} (based on \cite{herdeiro2021imitation}). We study the possibility that either a scalar Boson Star or a Proca Star can mimick the shadow properties of a BH. Even though no stable BS can be ultra-compact, the presence of an ``innermost stable circular orbit'' like behaviour allows a PS to have the same scale as a Schwarzschild BH. A comparative study of the shadow of a BH and a PS configuration is performed.
	
	We follow by adding rotation to a BS and introducing a Kerr BH to its centre. The resulting Kerr BH with scalar hair is the test subject to the spectral decomposition procedure established in Ch.~\ref{C5}. We define the procedure and perform several numerical tests on the method -- all with a good indication of the power of such a technique (results to be published).
	
	In Ch.~\ref{C7} we study and generalize the virial identity computation through Derrick's scaling argument. We note that the presence of a boundary term, namely the Gibbons-Hawking-York term is tantamount to the complete computation of the virial identity for a generic metric. Several examples of the virial identity computation for both spherical (based on \cite{herdeiro2021virial}) and axial symmetry (results to be published) are presented. While in the former, we can make some conclusions from the identity (no-hair and no-go theorems), concerning the latter, the additional complexity makes it hard to have such a debate.
	
	We finish the main section of the thesis in Ch.~\ref{C8} with some general comments and remarks about the work covered throughout the whole thesis.
	
	To finish, in Appendix~\ref{A} we present the ODE solver developed throughout the PhD that resulted in this thesis. It consists of a parallelized, adaptative-step 6(5) order Runge-Kutta integrator with a shooting strategy that guarantees the agreement with the imposed boundary conditions. We also cover the professional CADSOL/FIDISOL program package that we used to solve the coupled of PDEs in Appendix~\ref{B}.  
%
\chapter{The Einstein-Maxwell-Matter model}\label{C2}
%
	As already pointed out (Sec.~\ref{S1.2}), in what concerns the BH spontaneous scalarization phenomenon, the eSTGB model belongs to a broader universality class that also contains the \textit{Einstein-Maxwell-Scalar} (EMS) models. In these models, scalarization occurs for electrically charged BHs, triggered by a large enough charge to mass ratio, $q=Q_e/M$. 

	EMS theories have allowed a deeper insight into the  BH spontaneous scalarization phenomena. This technically more straightforward model allowed a more accessible study of the domain of existence of solutions, in particular beyond the spherical sector, and it also allowed carrying out fully non-linear dynamical evolutions establishing that the instability of the scalar-free solution terminates in the scalarized BHs of the model~\cite{herdeiro2018spontaneous}. 

	The fundamental, spherical, scalarized solutions have been shown to be stable against generic perturbations (rather than only spherical)~\cite{myung2021scalarized} - see also~\cite{myung2019instability,bovskovic2019axionic} for additional work on related models. It is, therefore, relevant to ask how much the physics of the EMS depends on the coupling function. 

	This chapter is organized as follows. In Sec.~\ref{S2.1}  (based on \cite{fernandes2019spontaneous,blazquez2020einstein,blazquez2021quasinormal}) we present the EMS models with a single electric charge and propose six different coupling functions (Sec.~\ref{S2.1.1}) based on the classification of the BH solutions presented in Sec.~\ref{S1.2} -- one coupling of class \textbf{I}, four coupling functions belonging to class \textbf{II.A} and one to class \textbf{II.B}. In Sec.~\ref{S2.1.2} we show the solution's profile for an exemplar $(\alpha,\,q)$ configuration and observe that all solutions have the same behaviour. The zero mode of the RN BHs for the models that allow BH scalarization was derived in Sec.~\ref{S1.2}. We use the latter to construct the domain of existence for the six different couplings in Sec.~\ref{S2.1.3}. Follows a thermodynamical study of all the solutions in Sec.~\ref{S2.1.4}, a perturbative analysis in Sec.~\ref{S2.1.5} and a dynamical study of a coupling function of class \textbf{II.A} (the only one able to endow scalarized configurations from a hairless RN BH)\footnote{An oral presentation about this section can be seen at~\cite{ScalaVid}.}.
	
	With the base model well established and its properties understood, we follow by generalizing the model. In Sec.~\ref{S2.2} we perform a preliminary study of the previously referred EMS model by replacing the massless scalar field with a massive one. Follows the addition of a magnetic charge in Sec.~\ref{S2.3} and the study of a solution of class \textbf{I} and one from class \textbf{II.A}. The presence of a magnetic charge introduced an attractor mechanism that is studied in Sec.~\ref{S2.3.4} (based on \cite{astefanesei2019einstein}).
	
	In Sec.~\ref{S2.4} we introduce the Hodge dual and, with it, an axionic coupling (based on \cite{fernandes2019charged}). In this section, two solutions are presented, one of class \textbf{I} and one of class \textbf{II.A}. Both with a scalar field non-minimally coupled to the Hodge dual and with a scalar field minimally coupled to the Maxwell tensor. The same study on the solution's profile, the domain of existence, thermodynamical properties, perturbative stability and dynamical preference is performed for the latter. 
	
	We finish the generalization of the Einstein-Maxwell-Matter model by replacing the scalar field with a vector field (Sec.~\ref{S2.5}). For this Einstein-Maxwell-Vector model, only a coupling function of class \textbf{II.A} is presented  (based on \cite{oliveira2021spontaneous}). Some vector theorems are presented in Sec.~\ref{S2.5.1}, which restrain the shape of the ansatz of the vector field. We expose the bifurcation line and the full non-linear model in Sec.~\ref{S2.5.2}; being followed by the study of the solution's profile, the domain of existence, and thermodynamical stability in Sec.~\ref{S2.5.3}. A brief comparison between the scalarized BH solutions and the vectorized ones is performed at the end of the latter section. We conclude in Sec.~\ref{S2.6} with a discussion and some further remarks.
%
	\section{The EMS models}\label{S2.1}
%
	The EMS model describes a real scalar field $\phi $ minimally coupled to Einstein's gravity and non-minimally coupled to Maxwell's electromagnetism. The model is described by the action \eqref{E1.2.1} with $\mathcal{I}\equiv F_{\mu \nu}F^{\mu \nu}$ the ``source term'':
		\begin{equation}\label{E2.1.1}
		 \mathcal{S}_{EMS}= -\frac{1}{4}\int d^4 \ze x \, \sqrt{-g}\Big[R-2\ze\phi _{,\mu}\phi ^{,\mu} - f_i (\phi) F_{\mu \nu}F^{\mu \nu}\Big]\ ,
		\end{equation}
	where $F_{\mu \nu}$ is the usual Maxwell tensor that is non-minimally coupled to the real scalar field $\phi$ through the coupling function $f_i (\phi )$; the subscript index $i$ will be used to label the various coupling choices, as specified below. The generic, spherically symmetric line element \eqref{E1.5.40} will be used to describe both a scalar-free and a scalarized BH solution.
	
	In the absence of a magnetic charge, the $4$-vector potential is purely electroestatic, $A_\mu=V dt$ $\big( A_\varphi =0$ in \eqref{E1.5.45}\big). The scalar field is solely radial dependent $\phi(r) $.
	 
	This allows us to define an effective Lagrangian from which the equations of motion can be derived as
		\begin{equation}\label{E2.1.2}
		 \mathcal{L}_{EMS} ^{\rm eff} = \sigma \ze m'-\frac{1}{2}\ze \sigma \ze r^2 N\ze \phi'^{\, 2}+\frac{f_i}{2\sigma} \ze r^2 V'^{\, 2}\ .
		\end{equation}
	The resulting equations of motions are
		\begin{align}\label{E2.1.3}
	 	 & m' = \frac{1}{2}\ze r^2 N \phi'^{\, 2} +\frac{f_i}{2\ze\sigma ^2} r^2 V'^{\, 2}\ , \qquad \sigma ' = r \sigma \phi'^{\, 2}\ , \\
		 & \bigg( \frac{f_i }{\sigma}\ze\ze r^2\ze V' \bigg) ' = 0\ , \qquad \big( \sigma r^2 N\ze \phi ' \big) ' = -\frac{1}{2} \frac{\hat{f}_i}{\sigma} \ze\ze r^2\ze V'^{\, 2}\ ,\label{E2.1.4}
		\end{align}
	The equation for the electric potential  yields the first integral 
		\begin{equation}\label{E2.1.5}
		 V' = -\frac{Q_e \, \sigma}{r^2 \ze f_i}\ ,
		\end{equation}		
	with the integration constant, $Q_e$, being interpreted as the electric charge. Replacing the expression of the $1^{\rm st}$ integral \eqref{E2.1.5} in the mass and scalar field functions gives the simpler form
		\begin{equation}\label{E2.1.6}
		 m'=\frac{1}{2} r^2 N \phi'^{\, 2}+\frac{Q_e^2}{2r^2 f_i} \ , \qquad	\phi''+\frac{1+N}{rN}\phi'+\frac{Q_e^2}{r^3N f_i}\left(\phi'-\frac{\hat{ f}_i}{2\ze\ze r f_i }\right)=0\ .
 		\end{equation}		
	Observe that, in \eqref{E2.1.5}, the quantity $\varepsilon _\phi = f_i \ze\ze\sigma ^{-1}$ can be thought of as a relative electric permittivity that is caused by the non-minimal coupling between the scalar and Maxwell fields. 
		
	To solve the set of ODEs \eqref{E2.1.3}-\eqref{E2.1.5}, we have to implement suitable boundary conditions for the desired functions $\big( m,\, \sigma,\, \phi,\, V \big) $ and corresponding derivatives. Near the BH event horizon, located at $r=r_H >0$\ze , one can assume that the solutions possess a power series expansion in $(r-r_H)$, with
		\begin{align}\label{E2.1.7}
	 	 & m\ze (r) = \frac{r_H}{2}+\sum_{k\geqslant 1}m_{k}(r-r_H)^k,\qquad \qquad  \sigma (r) = \sigma _0 +\sum_{k\geqslant 1}\sigma_{k}(r-r_H)^k,\nonumber\\
	  	 & \phi \ze (r) = \phi _0 +\sum_{k\geqslant 1}\phi_{k}(r-r_H)^k\ ,\qquad \qquad V\ze (r) = \sum_{k\geqslant 1}{\rm v}_{k}(r-r_H)^k\ . 
		\end{align}
	The above expressions are replaced in \eqref{E2.1.3}-\eqref{E2.1.5}, which are solved order by order in $(r-r_H)$. It turns out that, at least to sixth order, the coefficients $m_{k},\,\sigma_{k},\,\phi_{k}$ and ${\rm v}_{k}$ are determined by the essential parameters $\phi _0$ and $\sigma_0$.

	The lowest order coefficients come as\footnote{Similar expressions can be written for higher-order coefficients. However, we have not been able to find a pattern or recurrence relations.}:
		\begin{equation}\label{E2.1.8}
 	 	 m_1 = \frac{ Q_e ^{\ze 2}}{2\ze\ze f_i (\phi _0)\ze\ze r_H^{\ze 2}}\ , \quad 
 	 \phi_1=\frac{\hat{f} _i(\phi_0)}{2r_H f_i(\phi_0)}\frac{Q_e^2}{\big(Q_e^2-r_H^2 f_i(\phi_0)\big) } \ , 
\quad
	 	\sigma _1 = -\phi _1 ^{\ze \ze 2}\ze \ze r_H\ , \quad 
	 	{\rm v}_1 = -\frac{\sigma _0 Q_e}{r_H^ 2 f_i \ze (\phi _0)}\ .
		\end{equation}
	The horizon data fixes the values of the Hawking temperature, and horizon area, 
		\begin{equation}\label{E2.1.8.5}
		 T_H = \frac{1}{4\pi} N'(r_H) \sigma _0\ , \qquad \qquad A_H = 4\pi r_H ^2\ .
		\end{equation}
	The expression of the Kretschmann scalar, $K\equiv R_{\mu\nu\alpha\beta} R^{\mu\nu\alpha\beta}$, and the energy density $\rho=-T_t^t$ at the horizon are also of interest
		\begin{equation}\label{E2.1.9}
		 K(r_H)=\frac{4}{r_H^4} \left[3-\frac{6 Q_e^2}{r_H^2 f_i(\phi_0)}+\frac{5 Q_e^4}{r_H^4 f_i^2(\phi_0)}
\right ]\ , \qquad \rho(r_H)=\frac{Q_e^2}{2\ze r_H^4 f_i(\phi_0)} \ ,
		\end{equation}
	while the Ricci scalar vanishes as $r\to r_H$. For future reference, observe that the energy density $\rho(r_H)$ vanishes when the coupling blows up and changes sign when the coupling changes sign.

	At infinity, the solutions can be asymptotically approximated by:
		\begin{align}\label{E2.1.10}
		 & m(r) = M-\frac{Q_e ^{\ze \ze 2}+Q_\phi ^{\ze\ze 2}}{2\ze\ze r}+\ze \cdots \ ,\qquad \qquad
	 	 \phi (r) = \frac{Q_\phi}{r}+\frac{Q_\phi\ze M}{r^2}+\ze \cdots \ , \nonumber\\
	 	 & V(r) = \Psi _e +\frac{Q_e}{r}+\ze \cdots \ , \qquad \qquad \qquad \quad \ 
	 	\sigma (r) =1+ \frac{Q_\phi ^{\ze\ze 2}}{2\ze\ze r^2}+\ze \cdots \ .
		\end{align}
	This expansion introduces another three constants: the ADM mass $M$, the electrostatic potential at infinity $\Psi _e$ and the scalar ``charge'' $Q_\phi$. Observe that, unlike the SBS (see Sec.~\ref{S1.3}), the scalar field is not under a conservation law and hence does not have a conserved Noether charge. The denomination of $Q_\phi$ comes as parallelism with the electric charge radial decay and is then a language abuse\footnote{Note that, even though we also denominate this as a scalar charge, we are using two different notations: a true Noether scalar charge is $Q_S$; while the asymptotic decay-like charge is $Q_\phi$.}. The full equations of motion can now be integrated with these asymptotic behaviours. 
	
\bigskip
	
	The solutions satisfy the \textit{virial identity}~\cite{herdeiro2018spontaneous,herdeiro2021virial} 
		\begin{equation}\label{E2.1.11}
	 	\int _{r_H} ^{+\infty} dr \left\{ \sigma \ze r^2 \ze \phi'^{\, 2} \left[1+\frac{2\ze r_H}{r}\Big(\frac{m}{r}-1 \Big)\right]\right\} =  \int _{r_H} ^{+\infty} dr \left[\frac{\sigma}{f_i}\left( 1-\frac{2\ze r_H}{r}\right) \frac{Q_e ^{\ze 2}}{r^2}\right] \ ,
		\end{equation}
	which is obtained via a scaling argument, \textit{cf.} Ch.~\ref{C7} for the full computation. The Smarr relation~\cite{bardeen1973four,smarr1973mass}, turns out to be unaffected by the scalar hair~\cite{herdeiro2018spontaneous}, 
		\begin{equation}\label{E2.1.12}
		 M=\frac{1}{2}\ze T_H A_H + \Psi_e   Q_e\ .
		\end{equation}
	The first law of BH thermodynamics is $dM = \frac{1}{4}T_H\ze dA_H +\Psi_e \, dQ_e$. The solutions satisfy also the following relation~\cite{herdeiro2018spontaneous}
		\begin{equation}\label{E2.1.13}
		 M^2 + Q_\phi ^{\ze\ze 2}= Q_e ^{\ze \ze 2} + \frac{1}{4}T_H^2 A_H^2\ .
		\end{equation}
	Remarkably, one can show that \eqref{E2.1.13}, dubbed \textit{non-linear Smarr} relation, holds for any $f_i \ze$ that behaves as $\phi \rightarrow Q_\phi /r$ asymptotically ($i.e.$ as $r\rightarrow +\infty $).
 
	\bigskip

	The solutions can be physically characterised by the following dimensionless quantities:
		\begin{align}\label{E2.1.15a}
	 	 & q=\frac{Q_e}{M}\ \, \text{(charge to mass ratio)}\ ,\\
	 	 & a_H= \frac{A_H}{16\pi M^2}=\frac{r_H ^2}{4M^2}\ \,\text{(reduced horizon area)}\ ,\\
	 	 & t_H = 8 \pi M T_H = 2M N'(r_H)\ze \sigma _0\ \,\text{(reduced horizon temperature)}\ .\label{E2.1.15c}
	 	 \end{align}
%
		\subsection{The coupling functions}\label{S2.1.1}
%
	Concerning class \textbf{II} (scalarized-type) -- see Sec.~\ref{S1.2} --, the coupling function $f_i $ must obey the following criteria: 1) accommodate non-scalarized solutions, which amounts to the condition $\hat{f}_i(0)=0$. This can be interpreted as implementing a $\mathbb{Z}_2$ symmetry $\phi\rightarrow -\phi\ze$; 2) the form of the coupling is  constrained by two Bekenstein type identities~\cite{bekenstein1972transcendence}, which require
			\begin{equation}\label{E2.1.14}
		 	 \hat{\hat{f}} _i > 0\ , \qquad \qquad \phi\ze\ze \hat{f} _i  > 0\ ,\\ 
			\end{equation}	 
	for some range of the radial coordinate; 3) obey $f_i(0)=1$, so that one recovers Maxwell's theory near spatial infinity. In this work we will consider six different forms for the coupling function consistent with the above requirements. Four couplings in class \textbf{II.A} (scalarized-connected): 
			\begin{itemize}
			 \item[\textbf{i)}] an exponential coupling, $f _E (\phi) = e^{\alpha\ze \phi ^2}$\cite{herdeiro2018spontaneous,fernandes2019spontaneous}; 
		 	 \item[\textbf{ii)}] a hyperbolic cosine coupling, $f _C (\phi) = \cosh\big( {\sqrt{|\alpha|}\, \phi}\big)$\cite{fernandes2019spontaneous}; 
		 	 \item[\textbf{iii)}] a power coupling, $f _P (\phi) = 1+\alpha\ze \phi ^2$\cite{bovskovic2019axionic};
		 		\item[\textbf{iv)}] a fractional coupling, $f _F (\phi) = \frac{1}{1-\alpha\ze \phi ^2 }$\cite{fernandes2019spontaneous}
			\end{itemize}
	and one in class \textbf{II.B} (scalarized-disconnected):
			\begin{itemize}
			 \item[\textbf{v)}] a quadratic coupling, $f _Q (\phi) = 1+\alpha \phi ^4$\cite{blazquez2020einstein,blazquez2021quasinormal}; 
			\end{itemize}
	In addition, and for sake of completion, let us also consider one example of the class \textbf{I} (dilatonic-type):
			\begin{itemize}
			 \item[\textbf{vi)}] a dilatonic coupling, $f _D (\phi) = e^{\alpha \ze \phi}$\cite{astefanesei2019einstein,gibbons1988black,garfinkle1992erratum}; 
			\end{itemize}	
	 The coupling constant $\alpha$ is a dimensionless constant in all cases, and, except for the hyperbolic function, the conditions on $f_i$ imply that $\alpha >0$ for a purely electric field, $i.e.$ $F_{\mu\nu}F^{\mu\nu} <0$. The $f_i$ candidates shall be specified by the subscript $i \in \{ E, \, C,\, P,\, F\,\, Q,\, D\}$, respectively. Concerning class \textbf{II.A} for $\alpha \ze \phi ^2 \ll 1\ze $ (and $\alpha>0$), $f_E$, $f_C$ and $f_F$ possess the same Taylor expansion to first order which coincides with the (exact) form of $f_P$:
		\begin{equation}
		 f_F \ze  \approx  f_C \ze  \approx f_E \ze  \approx 1+\alpha\ze\ze \phi ^2 +\mathcal{O}\ze (\phi ^4)\ .
		\label{E216}
		\end{equation}	 
	This observation implies, in particular, that the zero mode coincides for all cases in the spherical sector, fundamental branch, scalarized solutions. Thus, from~\cite{herdeiro2018spontaneous}, scalarized solutions exist in all cases for $\alpha >1/4\ze$.
	
	For $\alpha>1/4\ze$, scalarized solutions exist above a certain threshold for the charge to mass ratio $q$. From another perspective, there is an $\alpha$ minimum for each $q$ of a RN BH for scalarized solutions to exist. This minimum value corresponds to the branching point and is presented in Table~\ref{T2.1}  for some values of $q$. As the scalar field increases and non-linearities become relevant, the differences between the models with different couplings emerge. 
		\begin{table}[H]
 	 	 \centering
 	 	 \caption{Minimum value of $\alpha$ for class \textbf{II.A} scalarization of a RN BH with charge to mass ratio $q$.}
 	 	 \vspace{2mm}
 	 		\begin{tabular}{c|cccccccccc}

                         $q$ & $1.0$  & $0.9$  & $0.8$  & $0.7$ & $0.6$  & $0.5$  & $0.4$ & $0.3$ & $0.2$ & $0.1$\\ 	 
		 \hline
			 $\alpha$  & $0.25$ & $2.99$ & $5.12$  & $8.02$ & $12.37$  & $19.50$ & $32.56$ & $60.72$ & $141.00$ & $574.90$ 
			 \label{T2.1}
			\end{tabular}
 		 \end{table}
%
	
%
		\subsection{Solutions profile}\label{S2.1.2}
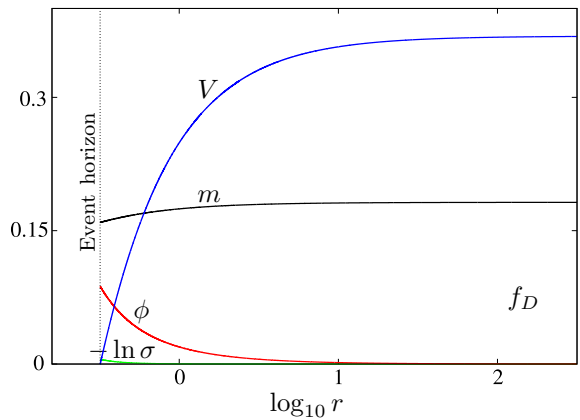
\begin{figure}[H]
		 \centering
			\begin{picture}(0,0)
		  		 \put(74,108){\small$V$}
		  		 \put(50,25){\small$\phi$}
		  		 \put(33,10){\small$-\ln \sigma $}		  		 
		  		 \put(74,69){\small$ m$}
		  		 \put(101,-10){\small$\log _{10} r$}
		  		\put(30,50){\begin{turn}{90}{$\scriptstyle{\rm Event\ horizon}$}\end{turn}}
		  	 \put(190,30){\small$f_D$}
	   		\end{picture}
	 	 \includegraphics[scale=0.62]{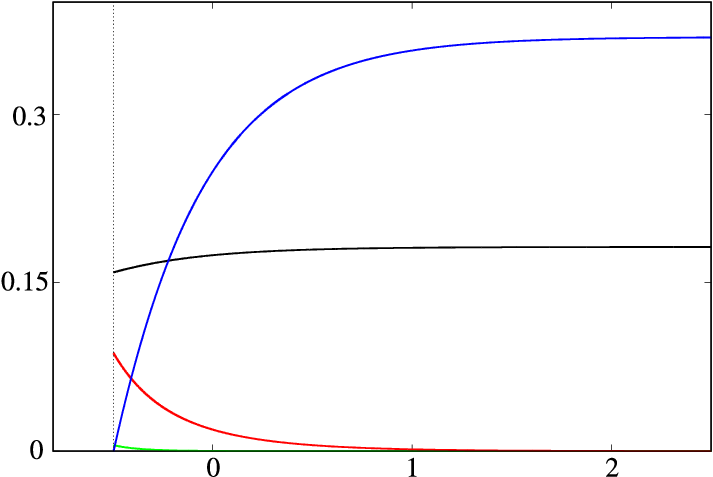}\\
	 	 \caption{Dilatonic, $f_D$, BH radial functions for $\alpha =10$ and $q=0.66\ze$.}
	 	 \label{F2.1}
		\end{figure}

%
	Let us start by exhibiting some typical solutions obtained from the numerical integration. Let us start with the dilatonic coupling (class \textbf{I}). In Fig.~\ref{F2.1} we perform a graphical representation of the radial profile of the metric functions $\sigma$ and $m$, the scalar field $\phi$ and electric potential $V$ for an exemplar coupling $\alpha = 10$ and charge to mass ratio $q \equiv Q_e /M =0.66$ for $f_D$.
	A universal feature of these nodeless solutions is that the scalar field is a monotonically decreasing function of the radius. Thus the scalar field value at the horizon, $\phi_0$, $cf.$~\eqref{E2.1.7}, is always the maximum of the scalar field. The scalar field vanishes asymptotically, $cf.$~\eqref{E2.1.10}. In fact, at far enough radius ($r>10 ^2$), all defining functions of the scalarized BHs converge to the ones of a comparable ($i.e.$ with the same global charges) RN BH.
	
\begin{figure}[H]
		 \centering
			\begin{picture}(0,0)
		  		 \put(74,108){\small$V$}
		  		 \put(50,25){\small$\phi$}
		  		 \put(34,10){\small$-\ln \sigma $}		  		 
		  		 \put(74,69){\small$ m$}
		  		 \put(101,-10){\small$\log _{10} r$}
		  		\put(30,50){\begin{turn}{90}{$\scriptstyle{\rm Event\ horizon}$}\end{turn}}
		  	 	\put(190,30){\small$f_E$}
	   		\end{picture}
	 	 \includegraphics[scale=0.62]{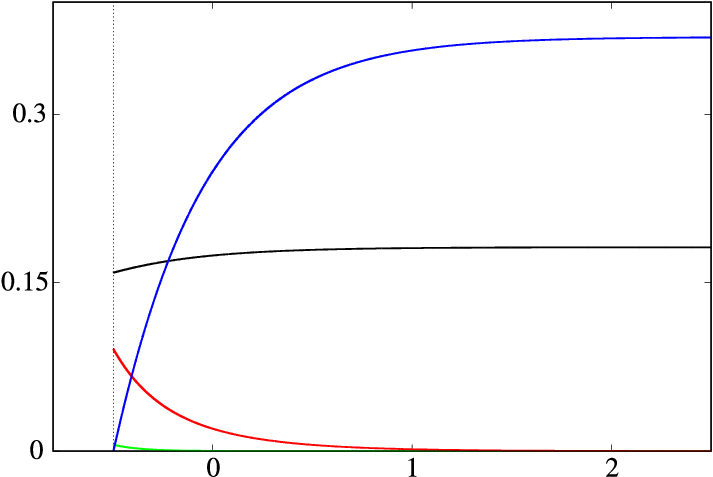}\hfill
	 	 		\begin{picture}(0,0)
		  		 \put(74,108){\small$ V$}
		  		 \put(50,25){\small$\phi$}
		  		 \put(34,10){\small$ -\ln \sigma $}		  		 
		  		 \put(74,69){\small$ m$}
		  		 \put(101,-10){\small$\log _{10} r$}
		  		\put(30,50){\begin{turn}{90}{$\scriptstyle{\rm Event\ horizon}$}\end{turn}}
		  	 	\put(190,30){\small$f_P$}
	   		\end{picture}
	 	 \includegraphics[scale=0.62]{Scalar/ProfileE}\vspace{6mm}
	 	 \includegraphics[scale=0.62]{Scalar/ProfileE}
	 	 	 	\begin{picture}(0,0)
		  		 \put(-148,108){\small$V$}
		  		 \put(-170,25){\small$\phi$}
		  		 \put(-185,10){\small$-\ln \sigma $}		  		 
		  		 \put(-150,68){\small$ m$}
		  		 \put(-119,-10){\small$\log _{10} r$}
		  		\put(-190,50){\begin{turn}{90}{$\scriptstyle{\rm Event\ horizon}$}\end{turn}}
		  	 	\put(-30,30){\small $f_F$}
	   		\end{picture}
	 	 \caption{Scalarized BH radial functions for $\alpha =10$ and $q=0.66\ze$. (Top left panel) $f_E$; (top right panel) $f_P$; (bottom panel) $f_E$.}
	 	 \label{F2.2}
		\end{figure}

	For class \textbf{II.A} observe Fig.~\ref{F2.2}. In the latter, the various radial functions defining the scalarized BHs are represented for an illustrative coupling of $\alpha=10$, charge to mass ratio $q=0.66$ and three different coupling choices.
	
\begin{figure}[H]
		 	 \centering
	 	 		\begin{picture}(0,0)
		  		 \put(40,15){\small$V$}
		  		 \put(47,108){\small$\phi$}
		  		 \put(72,15){\small$-\ln \sigma $}		  		 
		  		 \put(74,85){\small$m$}
		  		 \put(101,-10){\small $\log _{10} r$}
		  		 \put(100,125){$\scriptstyle \alpha \, =\, 5\ \ Q_e \, = \, 0.51\ \ r_H\, = \, 0.85$}
		  		\put(28,50){\begin{turn}{90}{$\scriptstyle{\rm Event\ horizon}$}\end{turn}}
		  	 	\put(190,30){$f_F$}
	   		\end{picture}
			 \includegraphics[scale=0.373]{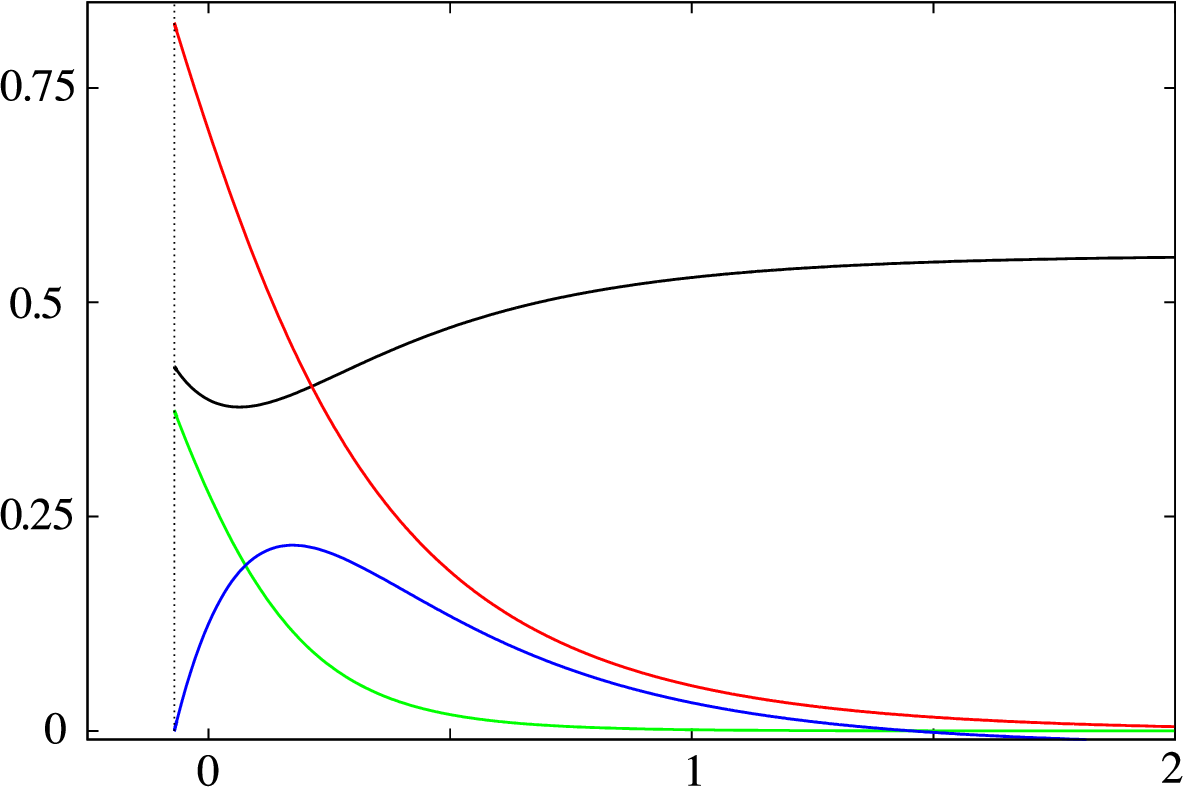}\hfill
			 	\begin{picture}(0,0)
		  		 \put(40,130){\small$\frac{K}{100}$}
		  		 \put(72,108){\small$R$}		  		 
		  		 \put(50,75){\small$\rho$}
		  		 \put(101,-10){\small $\log _{10} r$}
		  		 \put(170,132){$\scriptstyle f_F ^{-1}$}
		  		\put(28,50){\begin{turn}{90}{$\scriptstyle{\rm Event\ horizon}$}\end{turn}}
		  	 	\put(137,50){\small$\rho$}
	   		\end{picture}
 			 \includegraphics[scale=0.373]{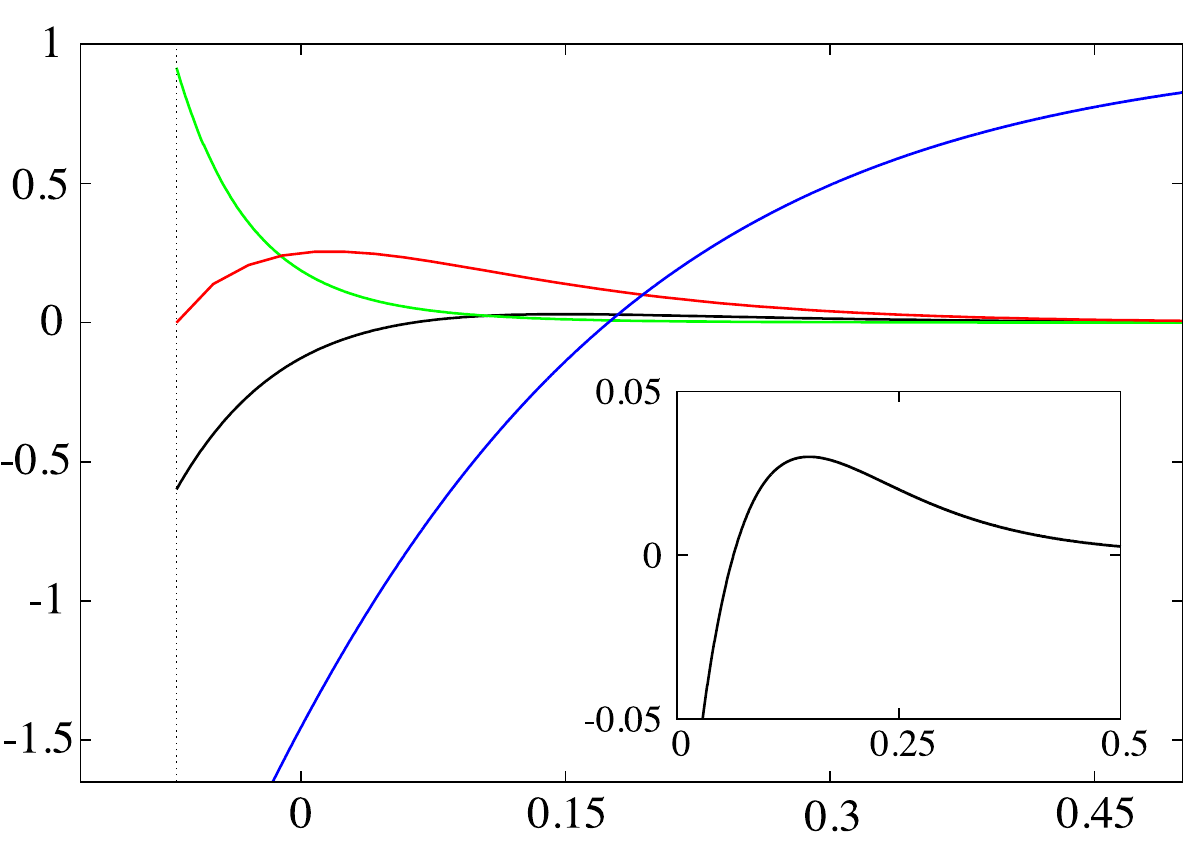}
	 	 	 \caption{A typical scalarized BH in an EMS model with the coupling function $f_F$, which possesses a region with negative energy density, $\rho<0\ze$. (Left panel) profiles of the metric and matter functions; (right panel) energy density (zoom-in presented in the inset), Ricci and Kretschmann scalars and the inverse of the coupling function $f_F$ which  changes sign at some finite $r$. This plot manifests that solutions with $\rho<0$ are smooth.}
			 \label{F2.3}
			\end{figure}
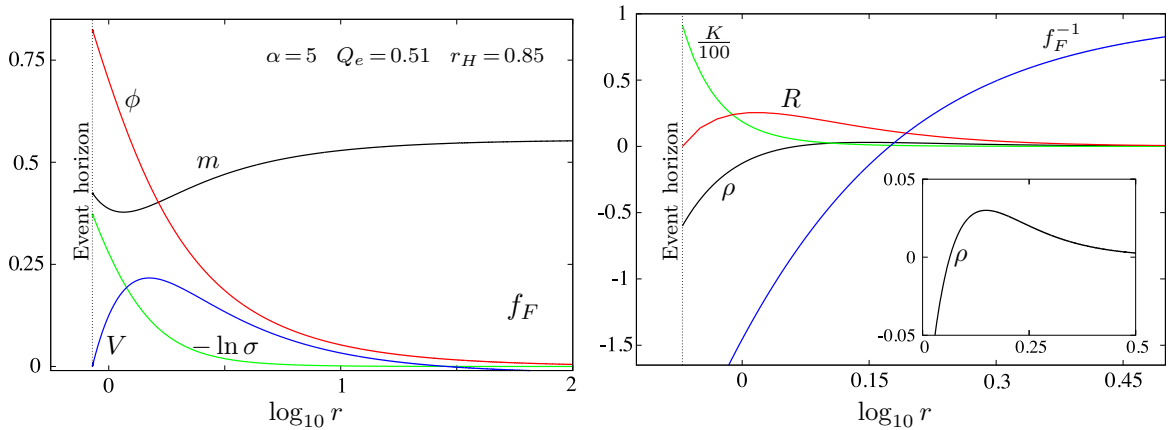

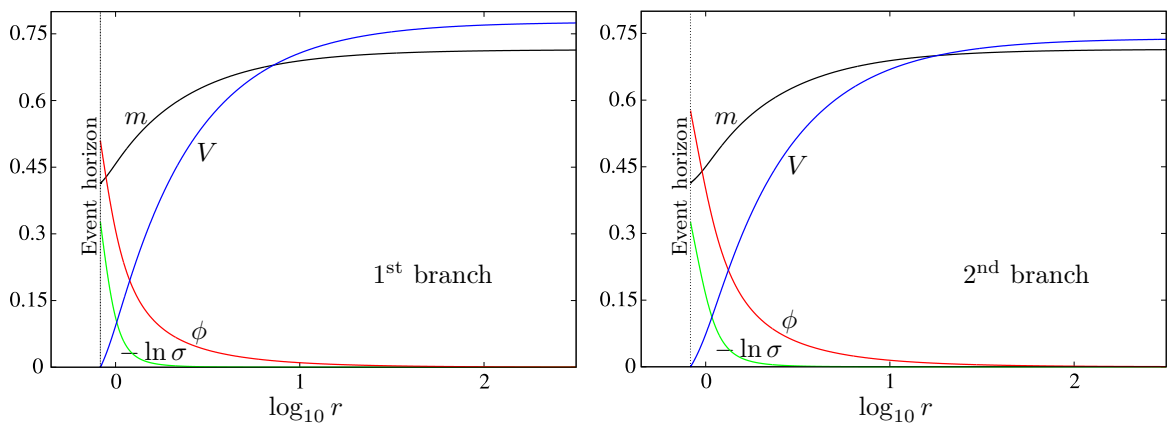
\begin{figure}[H]
		 	 \centering
				\begin{picture}(0,0)
		  	 	 \put(140,40){\small$1^{\rm st}\ \text{branch}$}
		  		 \put(45,11){\small$-\ln \sigma$}
		  		 \put(47,100){\small$m$}
		  		 \put(72,19){\small$\phi$}		  		 
		  		 \put(74,86){\small$V$}
		  		 \put(101,-10){\small $\log _{10} r$}
		  		\put(31,50){\begin{turn}{90}{$\scriptstyle{\rm Event\ horizon}$}\end{turn}}
	   			\end{picture}
	 	 	 \includegraphics[scale=0.62]{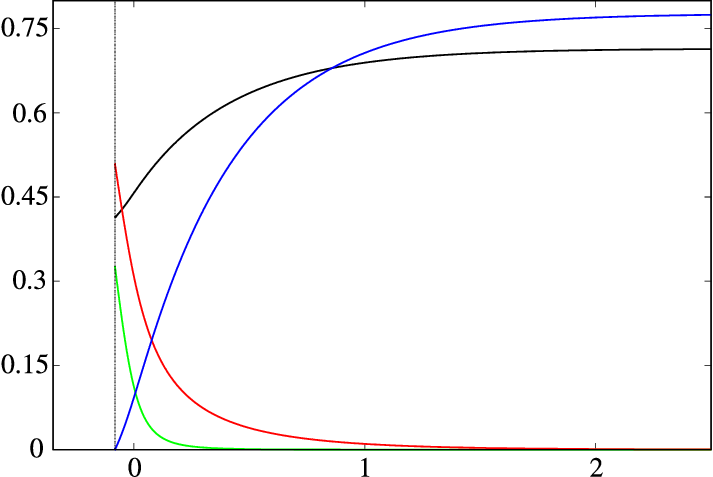}\hfill
	 	 	 	\begin{picture}(0,0)
		  	 	 \put(140,40){\small$2^{\rm nd}\ \text{branch}$}
		  		 \put(47,12){\small$-\ln \sigma$}
		  		 \put(47,100){\small$m$}
		  		 \put(72,23){\small$\phi$}		  		 
		  		 \put(74,81){\small$V$}
		  		 \put(101,-10){\small $\log _{10} r$}
		  		\put(31,50){\begin{turn}{90}{$\scriptstyle{\rm Event\ horizon}$}\end{turn}}
	   			\end{picture}
	 	 	 \includegraphics[scale=0.62]{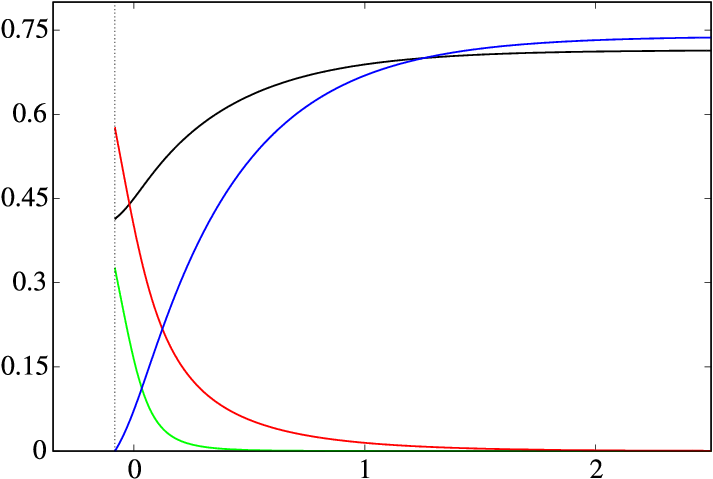}
	 	 	 \caption{Quartic scalarized BH radial functions for $\alpha =10$ and $q=0.98\ze$. (Left panel) first branch; (right panel) second branch.}
	 	 	 \label{F2.4}
			\end{figure}

	 Comparing between the presented couplings, one observes that the differences between the exponential and power-law are small\footnote{The same applies to the hyperbolic cosine coupling, thus not shown.} -- see Table~\ref{T2.2} for a full comparison and Fig.~\ref{F2.2} (top panel) -- and more pronounced for the fractional coupling (see Fig.~\ref{F2.2} bottom panel).
	We remark that these data is well within the numerical errors: our tests have exhibited a relative difference of $10 ^{-8}$ for the virial relation; $10^{-7}$ for the Smarr relation and $10^{-6}$ to the non-linear Smarr relation.
		\begin{table}[H]
 	 	 \centering
 	 	 \caption{Characteristic quantities for class \textbf{II.A} scalarized BH solutions with four choices of couplings,  $\alpha =10$ and $q=0.66\ze $.}
 	 	 \vspace{2mm}
 	 		\begin{tabular}{c|cccccc}
 	  		 $f_i (\phi )$  & $r_H$ & $M$ & $Q_\phi$ & $\Psi$ & $a_H$ & $T_H$\\
 	  			
			 \hline
			 $f_E$ & $0.318$ & $0.182$ & $0.017$ & $0.369$ & $0.766$ & $0.216$ \\
			 
			 $f_C$ & $0.318$ & $0.182$ & $0.013$ & $0.372$ & $0.766$ & $0.216$ \\
		
			 $f_P$ & $0.318$ & $0.182$ & $0.012$ & $0.373$ & $0.766$ & $0.215$ \\
			 	 
			  $f_F$ & $0.319$ & $0.182$ & $0.056$ & $0.285$ & $0.768$ & $\ 0.231$ 
			 \label{T2.2}
			\end{tabular}
 		 \end{table}
    For the particular case of $f_F$, however, a different type of solution that we call \textit{exotic} is possible. If $1-\alpha\ze \phi_0^2<0$, then the corresponding solutions have a region of negative energy density in the vicinity of the horizon, $cf.$~\eqref{E2.1.9} and Fig.~\ref{F2.3} (right panel). Moving away from the horizon, as the value of the scalar field decreases monotonically, $cf.$ Fig.~\ref{F2.3} (left panel), it passes through the point at which the coupling diverges. However, this divergence is benign, and the geometry is smooth therein. Such can be understood from the equations~\eqref{E2.1.9}, which contain $1/f_F$ terms but no divergencies. Moreover, beyond a critical radius, the energy density is again positive -- Fig.~\ref{F2.3} (right panel inset). The negative energy region in the vicinity of the horizon leads to a decrease in the mass function profile -- see Fig.~\ref{F2.3} (left panel). 
  	Concerning the quartic coupling, as we will see below (Sec.~\ref{S2.1.3}), there will be a degeneracy for a $q<1\ze $. The domain of existence has two coexisting solutions for the same $(\alpha,\, q)$ parameters. For completeness, let us plot in Fig.~\ref{F2.4} two profiles for an illustrative coupling $\alpha =10$ and $ q=0.98\ze$.
	In all cases, we see the same qualitative behaviour. 		
%
		\subsection{Domain of existence}\label{S2.1.3}
%
	Let us now focus on a comparative study of the domain of existence for the scalarized and dilatonic, fundamental, spherically symmetric solutions for the chosen couplings.
			\subsubsection*{Class \textbf{I} dilatonic}
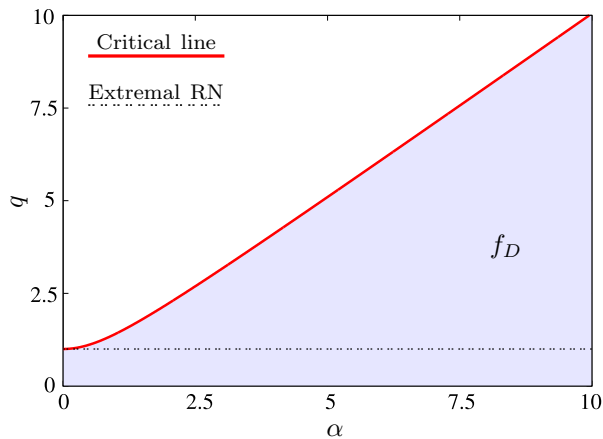
\begin{figure}[h!]
		 \centering
		 	 	 \begin{picture}(0,0)
		  		 \put(32,140){$\scriptstyle {\rm Critical\ line}$}
		  		 \put(29,121){$\scriptstyle {\rm Extremal\ RN}$}
		  		 \put(118,-6){\small $\alpha$}
		  		\put(0,80){\begin{turn}{90}{\small $q$}\end{turn}}
	   			\end{picture}
		 	\begin{tikzpicture}[scale=0.5]
\node at (0,0) { \includegraphics[scale=0.63]{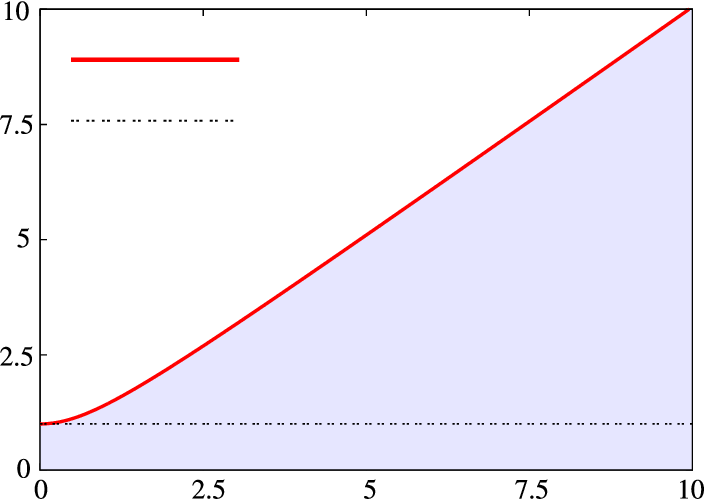}};
\node at (5,-1) {\small $f_D$};
		\end{tikzpicture}
	 	 \caption{Domain of existence of dilatonic BHs in EMS models (shaded blue regions). The domain of existence is always delimited by the Schwarzschild solution $q=0$ and the critical (red) line.}
	 	 \label{F2.5}
	\end{figure}

	Let us start with our reference class \textbf{I} solutions. The respective domain of existence can be seen in Fig.~\ref{F2.5}.
	The behaviour of the dilatonic BHs with any $\alpha >0$ are similar, albeit $\alpha = 1 $ is a somewhat particular point that separates the family into two subsets concerning the behaviour of some physical quantities. The latter can be seen from the study of the exact solutions in Appendix~\ref{C}. For a given $\alpha$, the branch of dilatonic BHs bifurcates from the Schwarzschild BH ($q=0$), rather than the RN BHs, and ends in a critical solution which is approached for a certain maximal $q$
				\begin{equation}
				 q^{\rm max} _D =\sqrt{1+\alpha ^2}\ ,
				\end{equation}
	The critical solution has, for any $\alpha >0$, a singular horizon, as one can see by evaluating the Kretschmann scalar~\eqref{E2.1.9}. At last, we would like to point out that along any branch with fixed $\alpha$, the ratio $q$ increases and becomes larger than unity at some stage. In this sense, \textit{overcharged} BHs are possible, in contrast with the RN family.
			\subsubsection*{Class \textbf{II.A} scalarized-connected}
	Concerning class \textbf{II.A} the solution of \eqref{E1.2.17} yields a RN BH surrounded by a vanishingly small scalar field. The full set of configurations creates the \textit{existence line} which, as discussed before in Sec.~\ref{S1.2}, is \textit{common} for all presented coupling functions of class \textbf{II.A} discussed herein, as they are identical for small $\phi$. The differences in the domain of existence of the four couplings emerge for larger values of $\phi$, wherein non-linearities become important.
    
\begin{figure}[h!]
		 \centering
		 	 \begin{picture}(0,0)
		  		 \put(25,137){$\scriptstyle {\rm Existence\ line}$}
		  		 \put(28,122.5){$\scriptstyle {\rm Critical\ line}$}
		  		 \put(25,108){$\scriptstyle {\rm Extremal\ RN}$}
		  		 \put(113,0){\small $\alpha$}
		  		\put(4,80){\begin{turn}{90}{\small $q$}\end{turn}}
	   			\end{picture}
	   			\begin{tikzpicture}[scale=0.5]
\node at (0,0) { \includegraphics[scale=0.35]{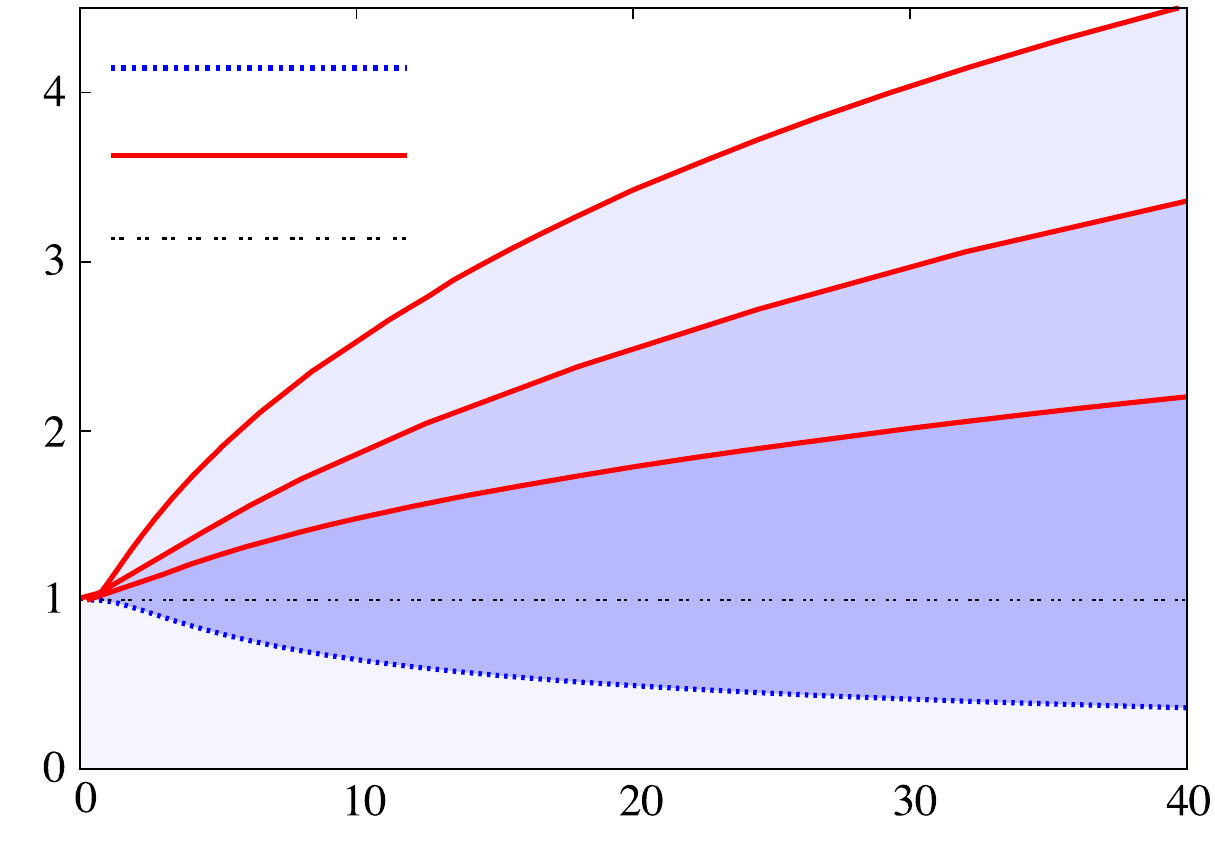}};
\node at (6,4) {\small $f_E$};
\node at (6,1.7) {\small $f_C$};
\node at (6,-0.6) {\small $f_P$};
\node at (-4.5,-3.5) {\small RN BHs};
				\end{tikzpicture}
				\begin{picture}(0,0)
		  		 \put(30,137){$\scriptstyle {\rm Extremal\ RN}$}
		  		 \put(122,137){$\scriptstyle {\rm Existence\ line}$}
		  		 \put(113,0){\small $\alpha$}
		  		\put(4,80){\begin{turn}{90}{\small $q$}\end{turn}}
	   			\end{picture}
				\begin{tikzpicture}[scale=0.5]
\node at (0,0) { \includegraphics[scale=0.35]{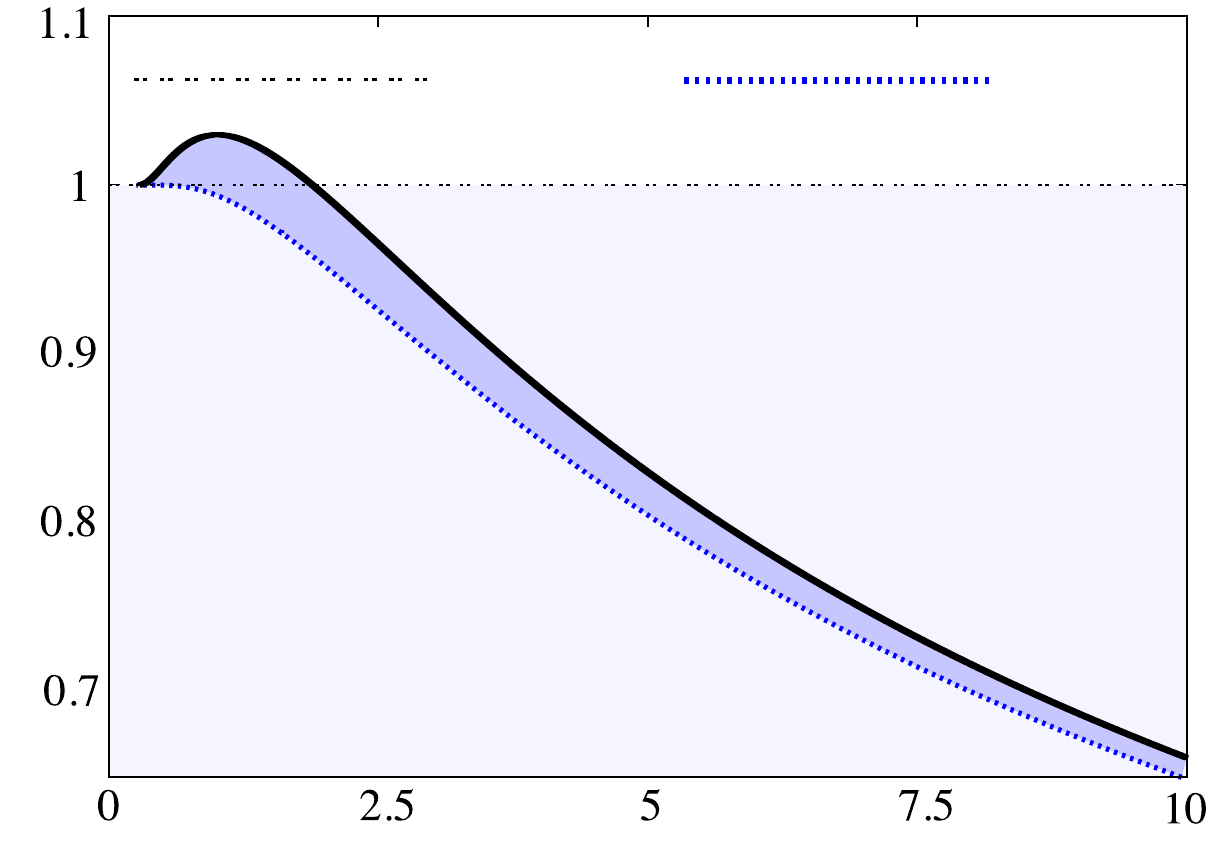}};
\node at (-4.5,3.1) {\small $f_F$};
\node at (-4.2,-3.5) {\small RN BHs};
\node at (4.5,-1.4) {$\scriptstyle {\rm Divergence\ line}$};
				\end{tikzpicture}
	 	 \caption{Domain of existence of scalarized BHs in EMS models (shaded blue regions). The domain of existence is always delimited by the existence line (dashed blue line) and the critical (red) line. (Left panel) $f_E\ze $, $f_C$ and $f_P$ couplings. (Right panel) $f_F$ coupling. Here we only exhibit the physical region, which is delimited by the existence line and the line at which the coupling function diverges at the horizon (\textit{a.k.a.} divergence line). The latter is the boundary of the physical region; above it, solutions have a negative energy density in the vicinity of the horizon.}
	 	 \label{F2.6}
	\end{figure}
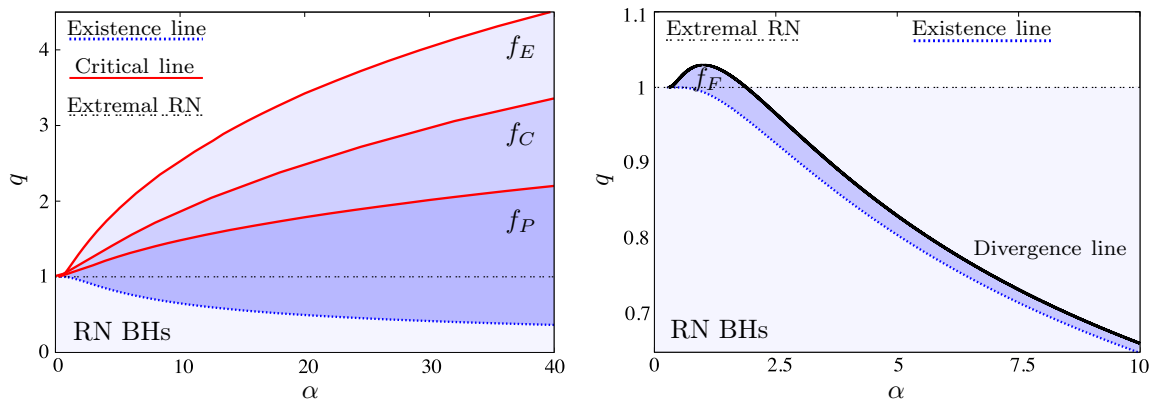

    The domains of existence for the scalarized BHs with the $f_E$, $f_C$, $f_P$ couplings are exhibited in Fig.~\ref{F2.6} (left panel). They are delimited by the \textit{existence line} -- (dashed blue) on which the RN BHs that support the zero-mode exist -- and a \textit{critical line} -- (solid red) which corresponds to a singular scalarized BH configuration. In between (shaded blue regions: dark for $f_P$, dark+medium for $f_C$, dark+medium+light for $f_E$), scalarized BHs exist. In particular, for $q\leqslant 1$, the usual RN BH and the scalarized solutions co-exist with the same global charges. In this region, there is non-uniqueness. The scalarized solutions are always entropically favoured (see Sec.~\ref{S2.1.4}). These spherical scalarized BHs are candidate endpoints of the spherical evolution (if adiabatic) of the linearly unstable RN BHs in the EMS model.
	At the critical line, numerics suggest $T_H,\ K\rightarrow +\infty\ze$; $A_H\rightarrow 0\ze$, while $M$ and $Q_\phi$ remain finite. As another feature, along  $\alpha =c^{\rm te}$ branches, $q$ increases beyond unity: therefore, scalarized BHs can be overcharged~\cite{herdeiro2018spontaneous}.

	Comparing the domain of existence of the exponential, $\cosh$ and power-law couplings (Fig.~\ref{F2.6}, left panel), we see that they are qualitatively similar. However, the critical set $\alpha = c^{\rm te}$ occurs at the smallest value of $q$ for the power-law coupling, an intermediate value for the hyperbolic coupling, and the largest value of $q$ for the exponential coupling. So, the exponential coupling allows \textit{maximising} the possibility of overcharging the BH and, in this sense, of maximising the differences with the RN BH case. Moreover, as seen before, $cf.$ Fig.~\ref{F2.2}, scalarization is ``stronger'' for the $f_E$ coupling than for $f_P$ (with an intermediate value for $f_C$). We also remark that for a given $\alpha $, as $q$ increases, so does the scalar field's initial amplitude $\phi _0 \ze $. As already mentioned, the scalar field profile is always such that the scalar field is monotonically decreasing. Thus, the global maximum of the scalar field occurs at the BH horizon and increases, for fixed $\alpha$, with $q$, and one can take $\phi_0$ as a measure of $q$ and vice-versa.

	The domain of existence of the $f_F$ coupling function (Fig.~\ref{F2.6} right panel) can be divided into two parts. For $\alpha = c^{\rm te}$,  $\phi_0$ grows from the existence line until it reaches $\phi_0^2=1/\alpha$ at the \textit{divergence line}, corresponding to the pole of the coupling. These solutions span the physical region wherein solutions have a positive energy density. Beyond the divergence line, solutions have $\phi_0^2>1/\alpha$ and thus a negative energy density region near the horizon extending up to a critical radius at which $\rho=0$ -- see Fig.~\ref{F2.3}. Beyond this point, the energy density is again positive. Solutions in the exotic region appear to be smooth, exhibiting no other obvious pathologies apart from the negative energy density. The physical region of the domain of existence will tend to thin down to zero as $\alpha$ increases. 
	
   Unlike the other studied couplings, for a model with $f_F$, the scalarized BH can only be overcharged and in the physical region if the coupling constant is in a compact interval: $\alpha \in [1/4\ze ,\ze 1.891]$, with a maximum of $q=1.030$ for $\alpha =1.012\ze $ -- $cf.$ Fig.~\ref{F2.6} right panel. 
			\subsubsection*{Class \textbf{II.B} scalarized-disconnected}
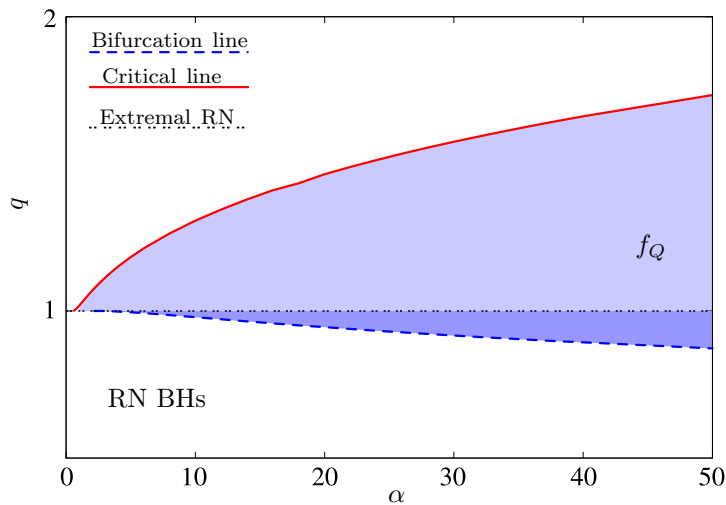
\begin{figure}[h!]
			 \centering
			 	\begin{picture}(0,0)
		  		 \put(28,140.5){$\scriptstyle {\rm Extremal\ RN}$}
		  		 \put(29,156){$\scriptstyle {\rm Critical\ line}$}
		  		 \put(25,169.5){$\scriptstyle {\rm Bifurcation\ line}$}
		  		 \put(136,-2){\small $\alpha$}
		  		\put(-5,108){\begin{turn}{90}{\small $q$}\end{turn}}
	   			\end{picture}
	   			\begin{tikzpicture}
\node at (0,0) { \includegraphics[scale=0.75]{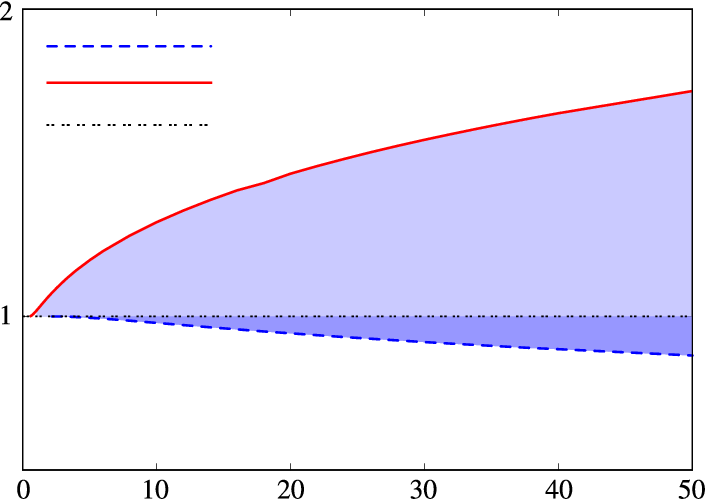}};
\node at (3.5,0) {\small $f_Q$};
\node at (-3,-2) {\small RN BHs};
				\end{tikzpicture}
	 		 \caption{Domain of existence of fundamental BH solutions in EMS models with a quartic coupling (blue shaded region) in a $q$ $vs.$ $\alpha$ diagram. The domain is bounded by the (solid red) critical line and the (dashed blue) bifurcation line. In the dark shaded region there are two distinct scalarized BH solutions and a RN solution. In the light shaded region there is only one scalarized BH solution (hot BHs).}
	 		 \label{F2.7}
	\end{figure}

	The domain of existence of the fundamental BH solutions of model~\eqref{E2.1.1} with the quartic coupling $f_Q = 1+\alpha\ze \phi ^4$, in a charge to mass ratio $vs.$ coupling constant diagram, is presented in Fig.~\ref{F2.7}.
	Fixing a generic value of $\alpha$, one observes the following behaviour: a first sequence of (cold) scalarized BHs \textit{starts} from the extremal RN BH. They form a first branch of solutions with  monotonically decreasing $q$ until a minimum value, $q_{\rm min}$, is attained. This minimum value is sub-extremal ($q<1$) and depends on $\alpha$, $q_{\rm min}=q_{\rm min}(\alpha)$. In Fig.~\ref{F2.7} the curve $q_{\rm min}(\alpha)$ is the blue dashed branch bifurcation line, which exists for $\alpha\gtrsim 2.185$. At $q_{\rm min}(\alpha)$ a second branch of (hot) scalarized solutions emerges. This branch has monotonically increasing $q$, extending into the over-extremal  regime $(q>1$). The second branch ends at a critical (singular) configuration wherein $a_H=0$, $t_H>0$ and $q=q_{\rm max}>0$. Again, $q_{\rm max}=q_{\rm max}(\alpha)$. The corresponding curve in Fig.~\ref{F2.7} is the red critical line, which exists for $\alpha\gtrsim 0.536$.

	It is worth emphasising the existence of a new sort of non-uniqueness amongst EMS models in this case. In the dark shaded blue region of Fig.~\ref{F2.7}, there are three different solutions for the same $q$, two scalarized ones (cold and hot) and the standard RN. Such is qualitatively different from the previous EMS models studied, where at most two solutions with the same $q$ were found in regions of non-uniqueness, corresponding to one scalarized BH and one RN BH. 
%
		\subsection{Entropic preference}\label{S2.1.4}
%
	In the EMS scalar model, the Bekenstein-Hawking BH entropy~\cite{hawking1975particle,bekenstein2020black,majumdar1998black} formula holds. Thus, the entropy analysis reduces to the analysis of the horizon area. 
			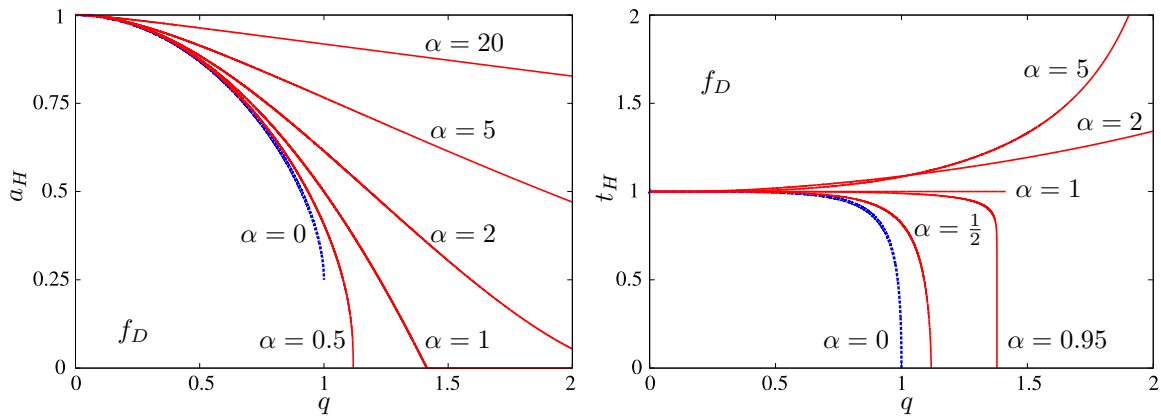
\begin{figure}[H]
	 \centering
	 			 	\begin{picture}(0,0)
	 			 \put(85,60){\small $\alpha =0$}
	 			 \put(92,20){\small $\alpha =0.5$}	 			 
	 			 \put(154,20){\small $\alpha =1$}
	 			 \put(156,60){\small $\alpha =2$}
	 			 \put(156,99){\small $\alpha =5$}	 			 
	 			 \put(154,132){\small $\alpha =20$}
		  		 \put(114,-3){\small $q$}
		  		\put(-1,75){\begin{turn}{90}{\small $a_H$}\end{turn}}
	   			\end{picture}
	 	   			\begin{tikzpicture}[scale=0.5]
\node at (0,0) { \includegraphics[scale=0.6]{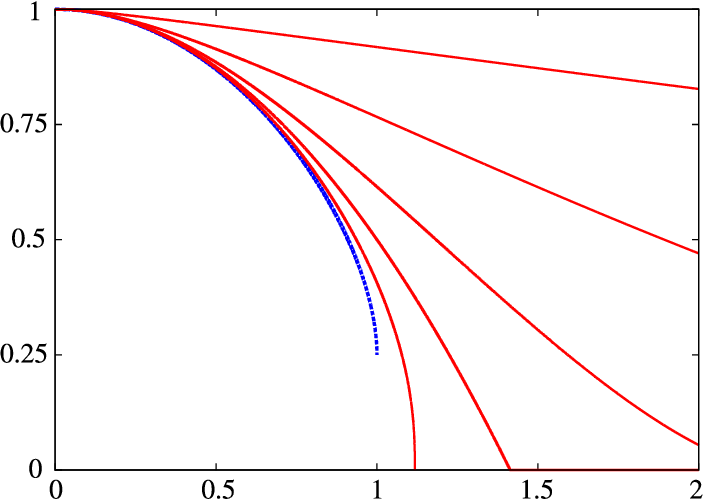}};
\node at (-4.5,-3.5) {\small $f_D$};
				\end{tikzpicture}
				\begin{picture}(0,0)
	 			 \put(85,20){\small $\alpha =0$}
	 			 \put(119,62){\small $\alpha =\frac{1}{2}$}	 	
	 			 \put(157,77){\small $\alpha =1$}	 			 
	 			 \put(154,20){\small $\alpha =0.95$}
	 			 \put(180,102){\small $\alpha =2$}	 			 
	 			 \put(160,122){\small $\alpha =5$}
		  		 \put(114,-3){\small $q$}
		  		\put(0,75){\begin{turn}{90}{\small $t_H$}\end{turn}}
	   			\end{picture}
				\begin{tikzpicture}[scale=0.5]
\node at (0,0) { \includegraphics[scale=0.6]{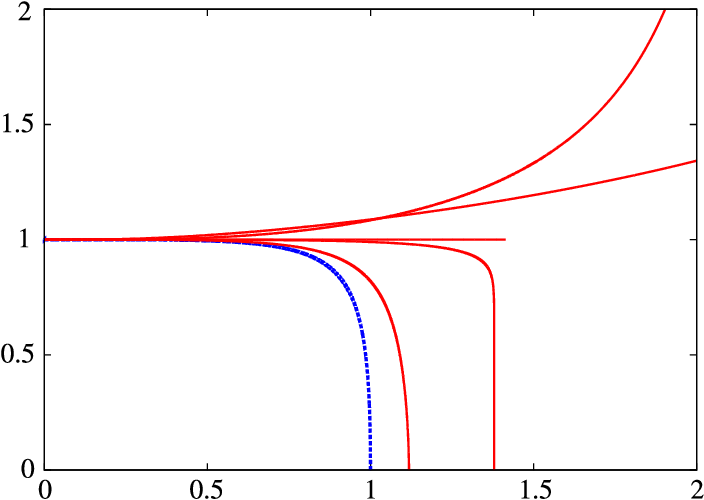}};
\node at (-4.5,3.1) {\small $f_D$};
				\end{tikzpicture}
	  \caption{Reduced area $a_H$ (left panel) and reduced temperature $t_H$ (right panel) $vs.$ the charge to mass ratio $q$ for dilatonic solutions. The blue lines are the set of RN BHs ($\phi =0$); the red lines are sequences of BHs with a non-trivial scalar field for a given $\alpha$. Different sequences are presented, for a range of values of $\alpha$.}
	 \label{F2.8}
	\end{figure}
	It is convenient to use the already introduced reduced event horizon area $[a_H \equiv A_H/(16\pi M^2)]$. Starting with the dilatonic coupling $f_D$ (Fig.~\ref{F2.8}), observe that the picture here is generic for $\alpha>0$, with some special features for $\alpha=1$ (this can be seen from the study of the exact solutions in Appendix~\ref{C}).

	From Fig.~\ref{F2.8} it is clear that the dilatonic solutions (solid red) bifurcate from the Schwarzschild solution $q=0$ and are parallel to the RN solutions. Even though the dilatonic solutions have higher entropy than the RN ones, such solutions are not related. Concerning the reduced temperature $t_H$, one observes that it goes to zero for $\alpha  < 1$ and diverges for $\alpha > 1$. The solutions with $\alpha = 1$ have $t_H = 1$. 

\bigskip

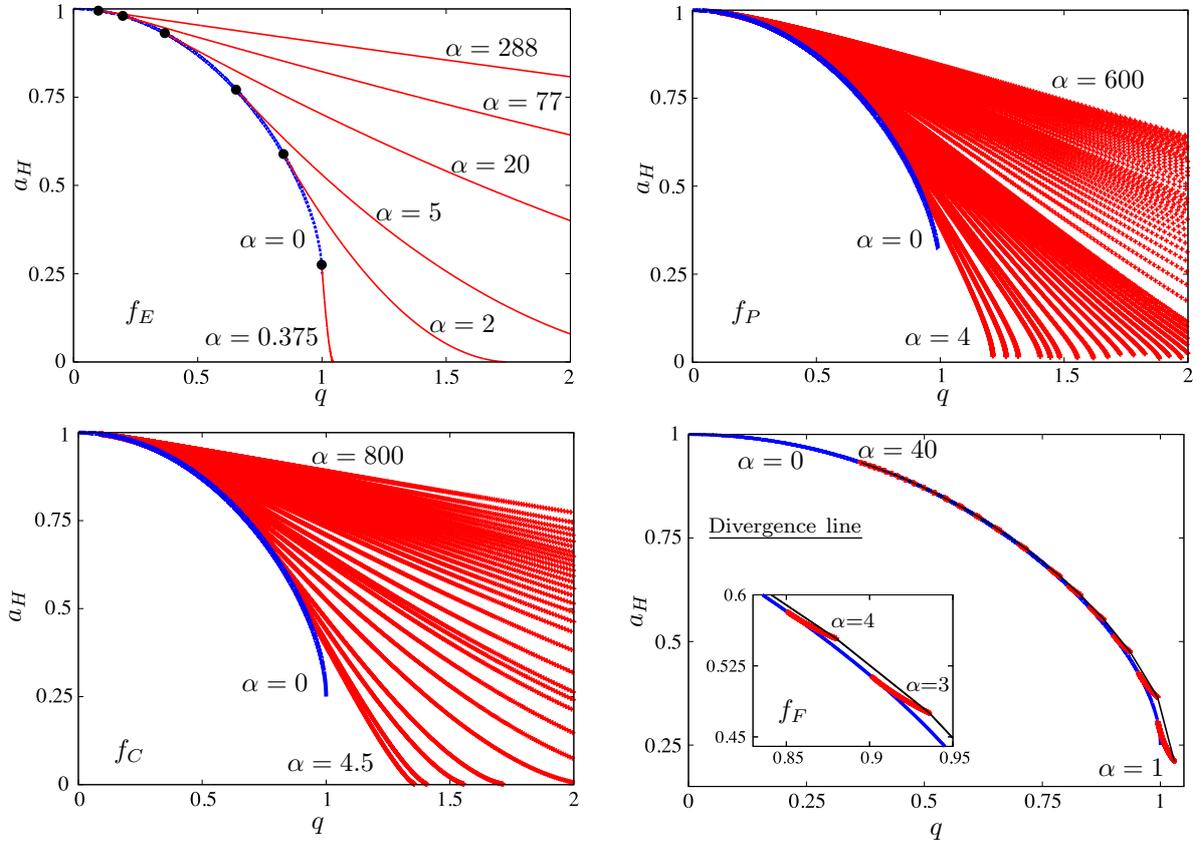
\begin{figure}[H]
			 \centering
	 			\begin{picture}(0,0)
	 			 \put(83,52){\small $\alpha =0$}
	 			 \put(70,15){\small $\alpha =0.375$}
	 			 \put(134,62){\small $\alpha =5$}	 	
	 			 \put(162,80){\small $\alpha =20$}	 			 
	 			 \put(154,20){\small $\alpha =2$}
	 			 \put(174,104){\small $\alpha =77$}	 			 
	 			 \put(160,124){\small $\alpha =288$}
		  		 \put(111,-6){\small $q$}
		  		\put(-1,73){\begin{turn}{90}{\small $a_H$}\end{turn}}
		  	 	 \put(40,25){\small $f_E$}
	   			\end{picture}
	  \includegraphics[scale=0.6]{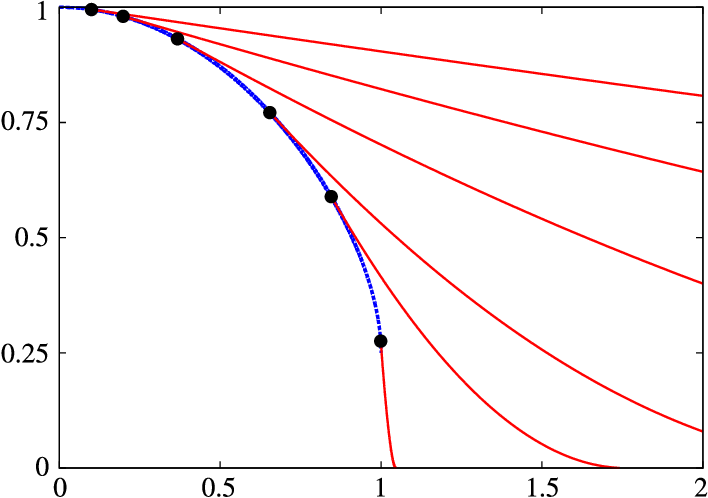}\hfill
	  	 			\begin{picture}(0,0)
	 			 \put(87,52){\small $\alpha =0$}
	 			 \put(105,15){\small $\alpha =4$}		 
	 			 \put(160,112){\small $\alpha =600$}
		  		 \put(117,-6){\small $q$}
		  		\put(5,72){\begin{turn}{90}{\small $a_H$}\end{turn}}
		  	 	 \put(40,25){\small $f_P$}
	   			\end{picture}
	  \includegraphics[scale=0.362]{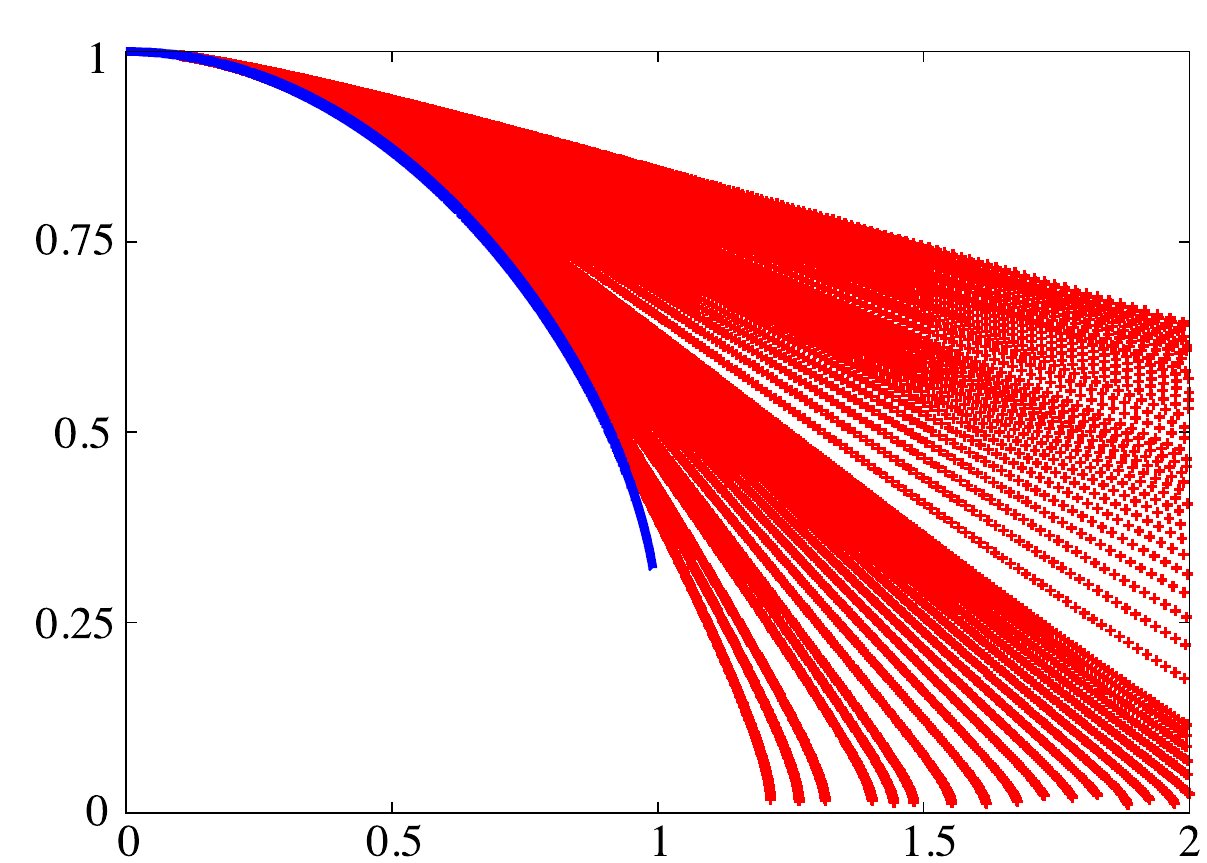}\vspace{5mm}
	    \includegraphics[scale=0.362]{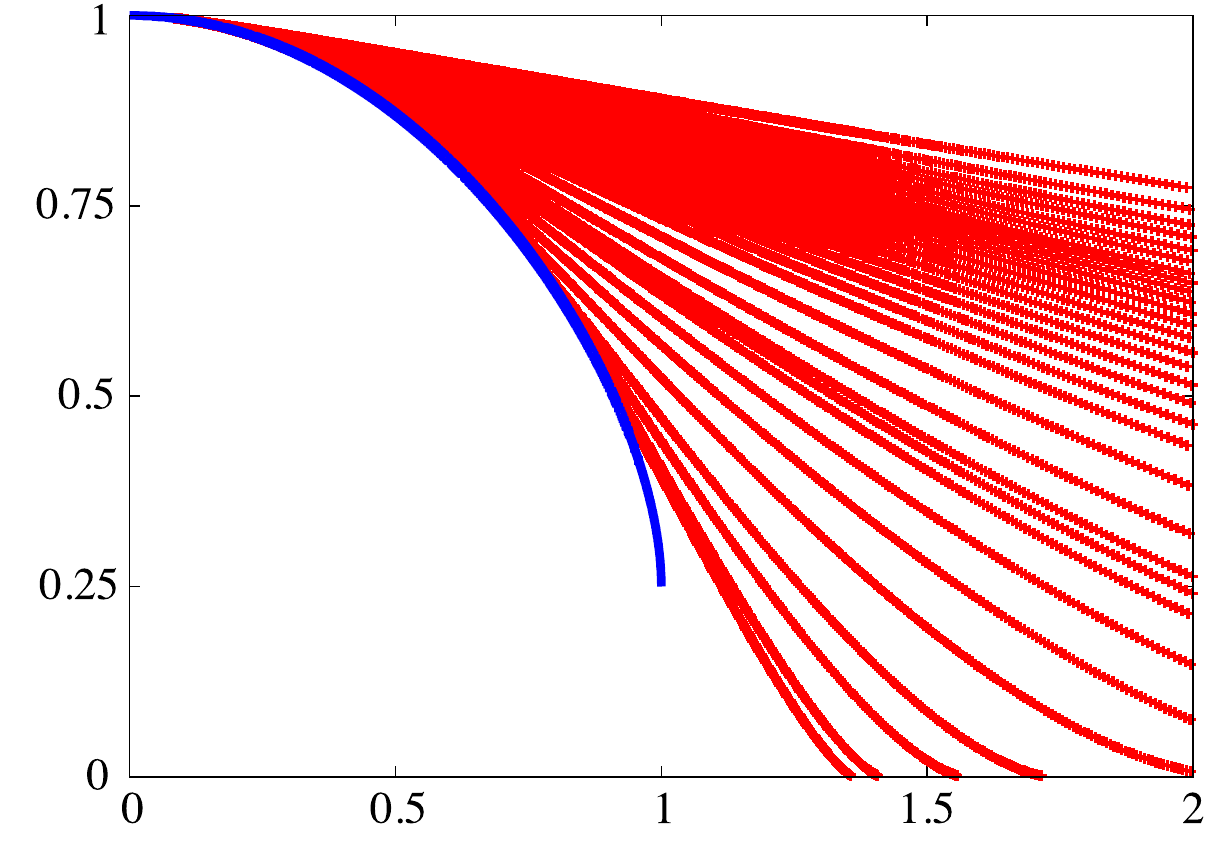}\hfill
	    	  	 \begin{picture}(0,0)
		  		 \put(-111,136){\small $\alpha = 800$}
		  		 \put(-120,20){\small $\alpha = 4.5$}
		  		 \put(-137,50){\small $\alpha = 0$}
		  		 \put(-111,-2){\small $q$}
		  		 \put(-224,76){\begin{turn}{90}{\small $a_H$}\end{turn}}
		  	 	 \put(-185,25){\small $f_C$}
	   			\end{picture}
	   		\begin{picture}(0,0)
	 			 \put(180,18){\small $\alpha =1$}	
	 			 \put(80,74){$\scriptstyle \alpha =4$}		
	 			 \put(108,50){$\scriptstyle \alpha =3$}	
	 			 \put(45,134){\small$\alpha =0$}
	 			 \put(34,110){$\scriptstyle {\rm Divergence\ line}$}		 
	 			 \put(90,138){\small $\alpha =40$}
		  		 \put(117,-4){\small $q$}
		  		\put(5,74){\begin{turn}{90}{\small $a_H$}\end{turn}}
		  	 	 \put(60,40){\small $f_F$}
	   			\end{picture}
	  \includegraphics[scale=0.362]{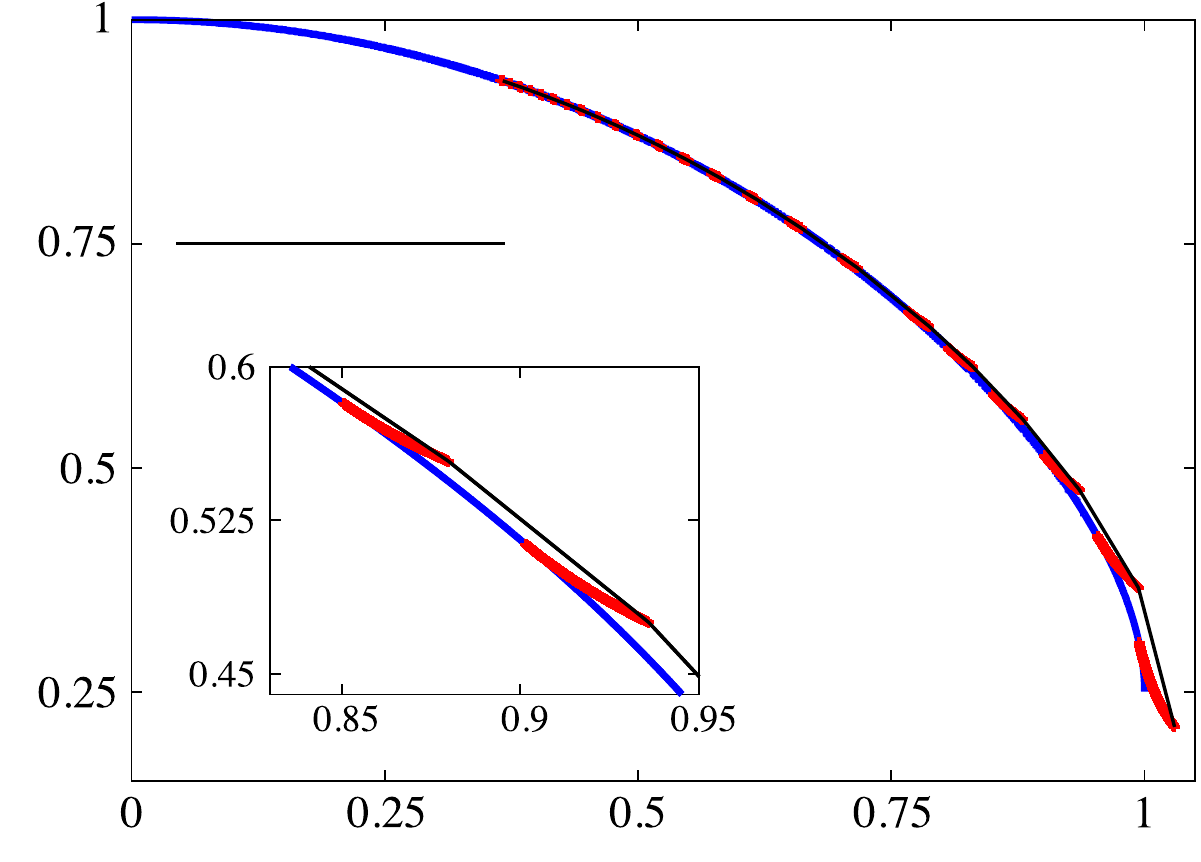}
	  \caption{Reduced area $a_H$ $vs.$ $q$ for: (top left panel) $f_E\ze$; (top right panel) $f_P \ze $; (bottom left panel) $f_C \ze$; (bottom right panel) $f_F \ze$. The blue lines are the sequence of non-scalarized RN BHs. The red lines are sequences of (numerical data points representing) scalarized BHs for a given $\alpha$. Black dots (top left panel) represent the bifurcation points of scalarized solutions from the RN BH. Different sequences are presented, for a range of values of $\alpha\ze $. The solid black line shows the sequence of solutions along the boundary of the physical region for the $f_F$ model.}
	 \label{F2.9}
	\end{figure}

	Considering the scalarized solutions, a true entropic preference can be achieved. In the region where the RN BH and scalarized BHs co-exist -- the non-uniqueness region -- for the same $q$, the scalarized
solutions are always \textit{entropically preferred}. Observe Fig.~\ref{F2.9} for all class \textbf{II.A} coupling functions studied herein. 
	Analysis of the $a_H \ vs.\ q$ plot (Fig.~\ref{F2.9}) shows that the spherically symmetric scalarized BHs bifurcate from the corresponding RN BH (with the bifurcation points represented by a black dot in Fig.~\ref{F2.9} top left), with a given $q = q(\alpha ) \neq 0$, as discussed above. For $\alpha = c^{\rm te}$, this branch has a finite extent, ending again in a critical configuration. This limiting solution possesses a singular horizon, as found when evaluating the Kretschmann scalar \eqref{E2.1.9}. The horizon area tends to zero as the critical solution approaches and the temperature diverges, while the mass and scalar charges remain finite. This behaviour parallels the dilatonic solutions with $\alpha > 1$. In the region of the parameter space wherein scalarized and RN BHs co-exist, one also observes that, for the same $q$, $a_H$ increases with the growth of $\alpha$.

\begin{figure}[H]
	 		\centering
	   		\begin{picture}(0,0)
	 			 \put(180,93){\small $\alpha =288$}	
	 			 \put(180,117){\small $\alpha =20$}	
	 			 \put(190,146){\small $\alpha =5$}			
	 			 \put(92,50){\small $\alpha =0$}	
	 			 \put(128,40){\small $\alpha =0.375$}			 
	 			 \put(138,125){\begin{turn}{70}{\small $\alpha =2$}\end{turn}}
		  		 \put(118,-8){\small $q$}
		  		 \put(-2,79){\begin{turn}{90}{\small $t_H$}\end{turn}}
		  	 	 \put(200,20){\small $f_E$}
	   			\end{picture}
	  \includegraphics[scale=0.66]{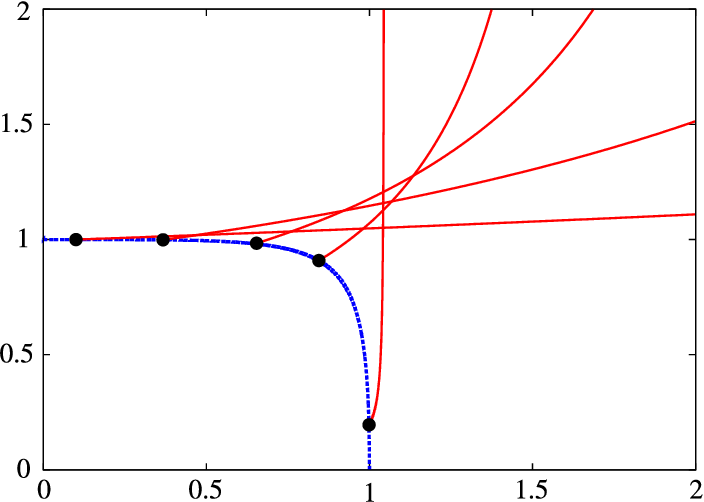}
	  \caption{Reduced temperature $t_H$ $vs.$ $q$ for $f_E\ze$. The blue line represent a sequence of non-scalarized RN BHs; The red lines are sequences of (numerical data points representing) scalarized BHs for a given $\alpha\ze$; black dots represent the bifurcation points of scalarized solutions from the RN BH. Different sequences are presented, for a range of values of $\alpha\ze $.}
	 \label{F2.10}
	\end{figure}
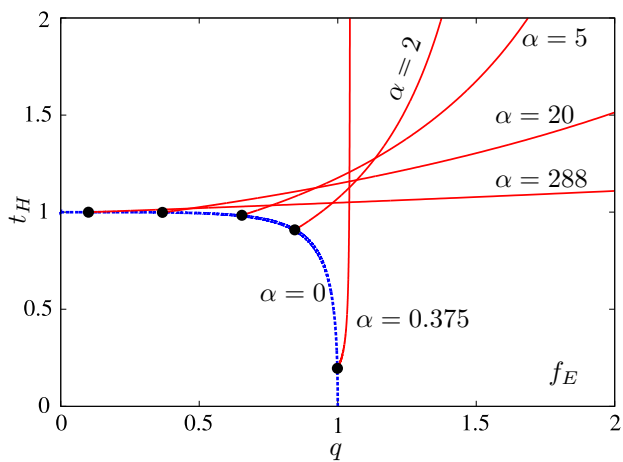

	Let us also quickly analyse the horizon reduced temperature as a function of the charge to mass ratio $t_H \ vs.\ q$ for an exemplary coupling, see $f_E$ in Fig.~\ref{F2.10}. We can observe that the solutions always have a temperature higher than a corresponding RN black hole.
	In addition, one can also observe that such scalarized solutions are never extremal $t_H=0$, having an increasing temperature with $q$ for a fixed $\alpha\ze$.
	
\bigskip
	
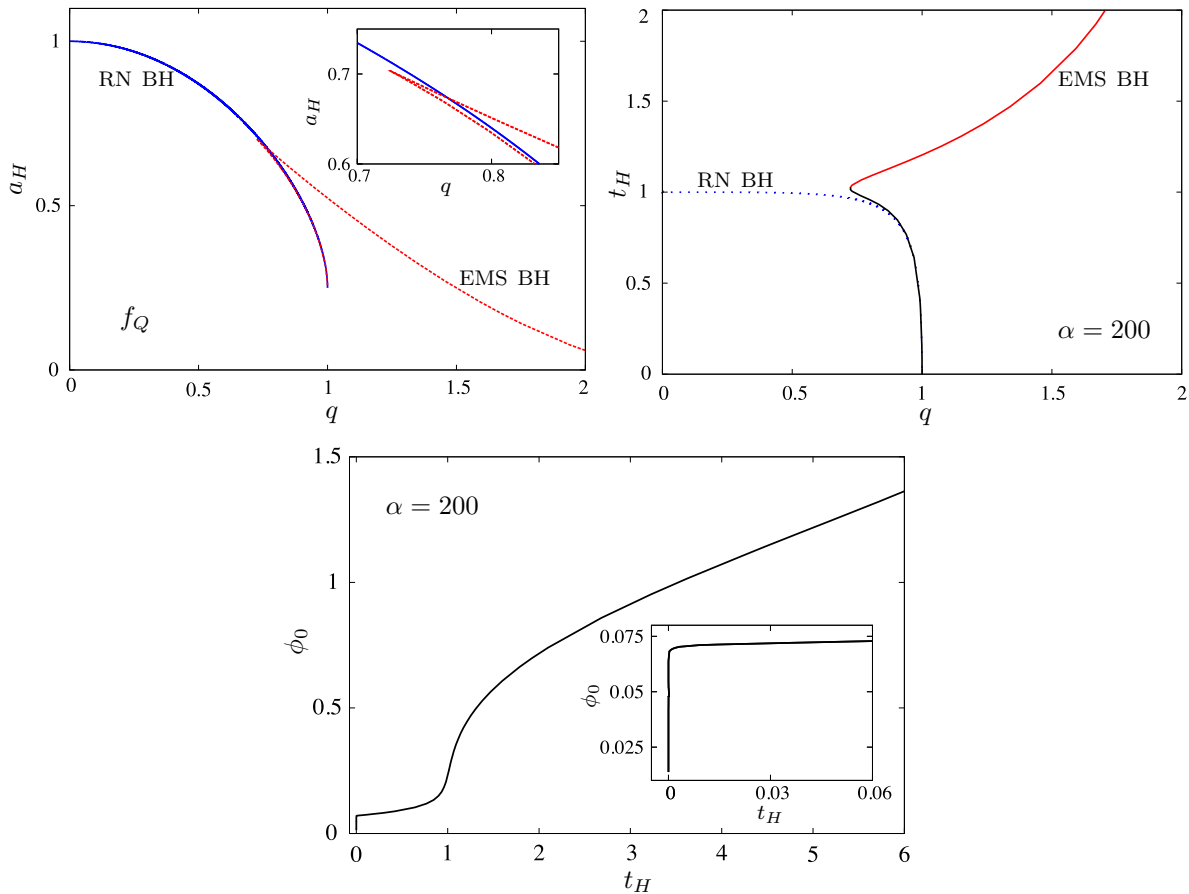
\begin{figure}[H]
	 		\centering
	   		\begin{picture}(0,0)	
	 			 \put(167,45){$\scriptstyle {\rm EMS\ BH}$}	
	 			 \put(32,120){$\scriptstyle {\rm RN\ BH}$}			 
		  		 \put(117,-5){\small $q$}
		  		 \put(-1,79){\begin{turn}{90}{\small $a_H$}\end{turn}}
		  	 	 \put(40,30){\small $f_Q$}
		  		 \put(110,105){\begin{turn}{90}{$\scriptstyle a_H$}\end{turn}}
		  	 	 \put(160,80){$\scriptstyle q$}
	   			\end{picture}
	  \includegraphics[scale=0.58]{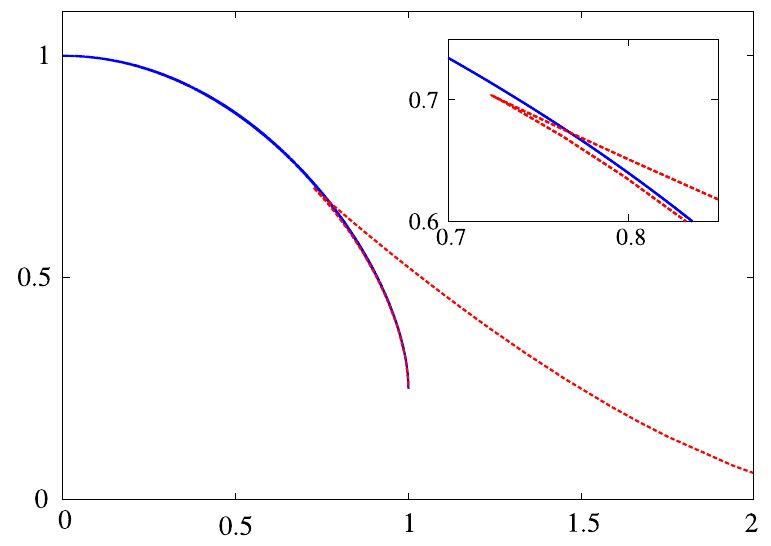}\hfill
	  	   		\begin{picture}(0,0)	
	 			 \put(167,120){$\scriptstyle {\rm EMS\ BH}$}	
	 			 \put(167,25){\small$\alpha=200$}	
	 			 \put(32,82){$\scriptstyle {\rm RN\ BH}$}			 
		  		 \put(115,-6){\small $q$}
		  		 \put(-1,78){\begin{turn}{90}{\small $t_H$}\end{turn}}
	   			\end{picture}
	  \includegraphics[scale=0.59]{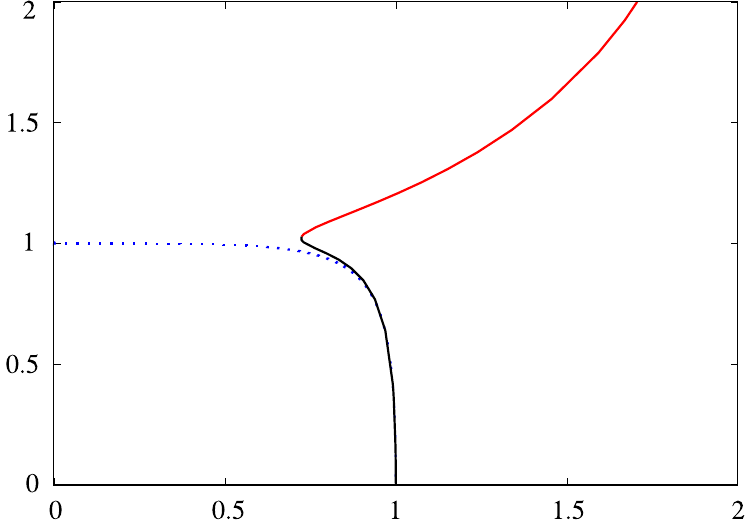}\vspace{5mm}\\
	  	  	  	   \begin{picture}(0,0)
	 			 \put(40,138){\small$\alpha=200$}			 
		  		 \put(129,-3){\small $t_H$}
		  		 \put(179,23){$\scriptstyle t_H$}
		  		 \put(2,85){\begin{turn}{90}{\small $\phi _0$}\end{turn}}
		  		 \put(113,65){\begin{turn}{90}{$\scriptstyle \phi _0$}\end{turn}}
	   			\end{picture}
	  \includegraphics[scale=0.4]{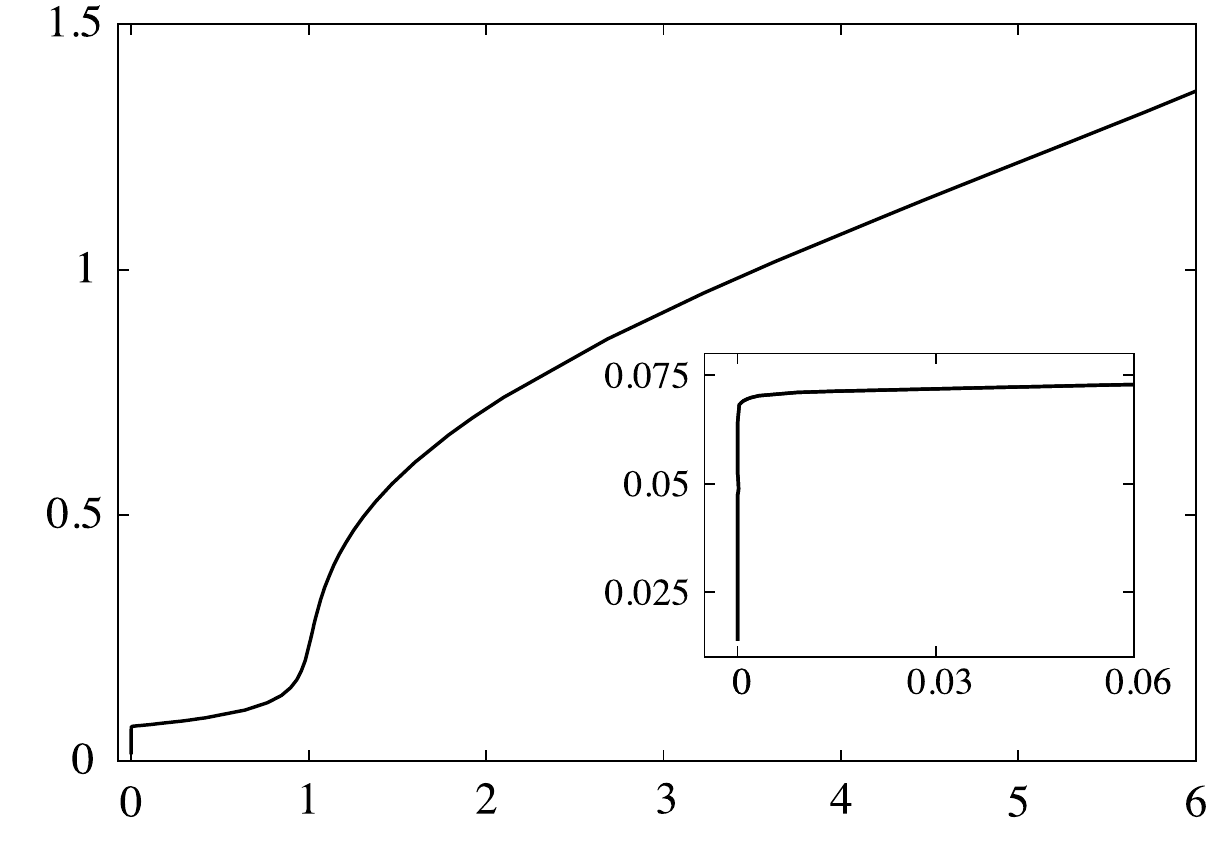}
	  \caption{Entropic quantities for a quartic coupling scalarized BH solution with $\alpha = 200\ze$. (Left panel) branches of scalarized BH solutions (red curve) and RN BHs (blue curve) in a $a_H\ vs. \ q$ diagram. The inset shows a zoom of the main panel wherein the two scalarized branches meet. (Right panel) $t_H\ vs. \ q$ diagram. Scalarized BHs are cold (black curve) in the first branch and hot (red curve) in the second branch. (Bottom panel) scalar field at the horizon $vs.$ reduced temperature. One may observe the fast surpression of $\phi _0$ as $t_H$ vanishes, of which the inset provides a zoom-in, showing the scalarized BHs continuously connected to the extremal RN solution.}
	 \label{F2.11}
	\end{figure}

	At last, let us analyse class \textbf{II.B}. In Fig.~\ref{F2.11} one can observe the entropy $a_H\ vs.\ q$  (left panel) and temperature $t_H\ vs. \ q$ (right panel) dependences for a quartic coupling. In these plots, one can appreciate the two branch structure of the scalarized BHs. Along the first (cold) branch, emerging from extremal RN, $q$ decreases and $a_H$ increases. Along the second (hot) branch, emerging from the solution with $a_H >0,\ t_H > 0$ and $q=q_{\rm min}$, $q$ increases and $a_H$ decreases.
	
	Fig.~\ref{F2.11} (left panel) also makes clear the vanishing of the reduced horizon area at the critical solution. Fig.~\ref{F2.11} bottom, which shows the value of the scalar field at the horizon $vs.$ the reduced BH temperature, clarifies that the scalarized solutions in the cold branch continuously connect to the extremal RN, which has $a_H=0.25,\ t_H=0$ and $q=1$. Moreover, whereas in previous class \textbf{II.A} models, the scalarized BHs were always entropically favoured for the exact global charges, this is not the case here, as manifest from Fig.~\ref{F2.11} (left panel). In part of the second branch (Fig.~\ref{F2.11} left panel inset), the scalarized BHs have a larger area, and hence larger entropy, than the comparable RN BH, $i.e.$ with the same $q$.  
	Finally, Fig.~\ref{F2.11} (right panel) shows the reduced temperature $vs.$ charge. The first branch of scalarized solutions starts at zero temperature (black solid line). The horizon temperature increases monotonically along the first and second branches (red solid line). RN BHs correspond to the blue dotted line. They are cooler than the BHs in the second branch. This diagram justifies the terminology cold/hot for
the scalarized BHs in the first/second branch.
        \begin{table}[H]

                       \centering

                       \caption{Physical properties of three $q$ degenerate BH solutions for $f_Q$ with $\alpha =200$ and $q=0.9\ze$.}

                       \vspace{2mm}

                              \begin{tabular}{c|ccccccc}

                                $ $  & $Q_\phi$ & $\Psi _e$ & $a_H$ & $t_H$ & $\phi _0$\\                    

                               \hline

                         RN (bald)      & $0.000$ & $0.627$ & $0.515$ & $0.846 $ & $0.000$\\

                         $1^{\rm st}$ branch (cold) & $0.038$ & $0.624$ & $0.513$ & $0.856$ & $0.136$ \\

                         $2^{\rm nd}$ branch (hot) & $0.393$ & $0.366$ & $0.585$ & $1.145$ &\ $0.401$ 

                         \label{T2.3}
                        \end{tabular}

                   \end{table}
	In Table~\ref{T2.3} we compare the three degenerate solutions for a specific choice of $\alpha = 200$ and $q =0.9\ze$. One can confirm that the scalarized BH in the hot branch has the largest area (and hence
entropy), temperature, scalar charge and scalar field value at the horizon.
%
		\subsection{Perturbative stability}\label{S2.1.5}
%
	Such entropic considerations are, however, not sufficient to establish if the endpoint of the instability of a RN BH is the corresponding hairy BH with the same $q$. In~\cite{herdeiro2018spontaneous}, fully non-linear dynamical evolutions were performed that established that for $f_E $, and sufficiently small $q$, this is indeed the case, which is consistent with the observation above that the scalarized solutions for the exponential (and also power-law and hyperbolic) coupling are, generically, stable against spherical perturbations. Nevertheless, the endpoint of the instability can only be established once fully non-linear numerical evolutions are studied. Such evolutions will be addressed in the next section. 
	
	Two intriguing questions, however, exist. Concerning the quartic coupling, we have, for the same space of parameters $(q,\ze \alpha)$ two distinct scalarized solutions. One must question if both of them are stable, unstable or one stable and another unstable. 
	
	About the fractional coupling, for a given $\alpha$ there are RN BHs that are unstable against scalar perturbations above the existence line in Fig.~\ref{F2.6} (right panel). However, no scalarized BHs exist for that value of $q$ (because it is above the critical set), in the physical region of the domain of existence with positive energy density. Therefore, the endpoint of the instability of such RN BHs is an interesting question.
	
	Following a standard technique for studying perturbative stability against radial perturbations, we consider spherically symmetric, linear perturbations of our equilibrium solutions, keeping the metric ansatz (\ref{E1.5.40}), but allowing the functions $N$, $\sigma$, $\phi$ and $V$ to depend on $t$ as well as on $r$:
		\begin{align}\label{E2.1.21}
		 & ds^2=- \tilde N(r,t)\tilde \sigma ^2 (r,t) dt^2+\frac{dr^2}{\tilde N(r,t)}+r^2\big( d\theta^2+\sin^2 \theta d\varphi^2\big) \ ,\qquad \tilde \phi(r,t) \ , \qquad \tilde V (r,t) \ .
		\end{align}
	The time-dependence enters as a periodic perturbation with frequency $\Omega$, for each of these functions:
		\begin{align}\label{E2.1.22}
		 & \tilde N(r,t)=N(r)+\epsilon \ze N_1(r)\ze e^{-\textit{i}\ze\ze \Omega\ze t}\ , \qquad  \tilde \sigma(r,t)=\sigma(r)+\epsilon\ze \sigma_1(r)\ze e^{-\textit{i}\ze \ze\Omega\ze t} \ , \nonumber\\
		 & \tilde \phi(r,t)=\phi(r)+\epsilon\ze \phi_1(r)\ze e^{-\textit{i}\ze\ze \Omega\ze t}\ , \qquad \tilde V(r,t)=V(r)+\epsilon\ze V_1(r)\ze e^{-\textit{i}\ze \ze\Omega\ze t}\ ,
		\end{align}
		with $\epsilon$ a parameter that helps us keep track of the perturbation functions. From the linearized field equations around the background solution, the metric perturbations and $V_1$ can be expressed in terms of the scalar field perturbation,
		\begin{align}\label{E2.1.23}
		 N_1=-2\ze r N\phi' \phi_1 \ , \qquad \sigma_1=-2\int dr ~r\ze\phi' \phi_1' \sigma_1  \ , \qquad V_1= -V'\ze\big( \sigma_1  + \phi_1 \hat f_i\big) \ ,	
		\end{align}
	thus yielding a single perturbation equation for $\phi_1$. This equation can be written in the standard Sch\"odinger-like form for $\Uppsi (x)$:
		\begin{equation}\label{E2.1.24}
		 -\frac{d^2 \ze\Uppsi}{dx^2}+U_{\Omega} \Uppsi=\Omega^2\, \Uppsi \ ,
		\end{equation}
	where we have defined $\Uppsi\equiv r \phi_1$ and the `tortoise' coordinate $x$ by
		\begin{equation}\label{E2.1.25}
		 \frac{dx}{dr}=\frac{\sigma}{N}\ . 
		\end{equation}
	The perturbation potential $U_{\Omega}$ is defined as:
		\begin{equation}\label{E2.1.26}
		 U_{\Omega}\equiv \frac{\sigma ^2 N}{r^2}\left\{ 1-N-2r^2 \phi'^{\, 2}+\frac{Q_e^2}{2r^2 f_i ^2}\left[2f_i ^2\big(1-2r^2 \phi'^{\, 2}\big)-2\hat{f}_i^{\ze 2}+\hat{\hat{f}}_i+4r\phi' \hat{f}_i\right]\right\} \ .
		\end{equation}
	Observe that we have fixed the sign convention of the imaginary part of the frequency, $\Omega_I$, by choosing the time-dependence $e^{-i\ze\Omega\ze t}$. Then, unstable modes (that grow in time) have $\Omega_I>0$. Indeed,  $e^{-i\ze\Omega\ze t}=e^{-i\ze\Omega_R\ze t} e^{\Omega_I \ze t} \to +\infty$ as $t\to +\infty$, when $\Omega_I>0\ze$.

	At spatial infinity, $i.e.$ $x \to +\infty$, we have $\Uppsi = A_+ \ze e^{i\ze\Omega \ze x}$. This means that in the case of an unstable mode of the form $\Omega=i\ze\Omega_I$, with $\Omega_I>0$, $\Uppsi = A_+\ze e^{-\Omega_I x} \to 0$. At the horizon, $i.e.$ $x \to -\infty$, we have $\Uppsi = A_-\ze e^{-i\ze\Omega\ze x}$. An unstable mode satisfies $\Uppsi = A_- \ze e^{\Omega_I\ze x} \to 0$. Hence unstable perturbations satisfy $\Uppsi (r=r_H)=\Uppsi(r=-\infty)=0\ze$. 

	The potential $U_\Omega$ is not positive definite, but is regular in the entire range $-\infty<x<+\infty$. Also, it vanishes at the BH event horizon and at infinity. It follows from a standard result in quantum mechanics (see $e.g.$~\cite{messiah1961quantum,ferrari1984new,buell1995potentials,pani2013advanced}) that \eqref{E2.1.24} has no bound states if $U_\Omega$ is everywhere larger than the lowest of its two asymptotic values, $i.e.$, if it is positive.\footnote{A simple proof is as follows. Write \eqref{E2.1.24} in the equivalent form
		\begin{equation}
 		 \frac{d }{dx}\left(\Uppsi \ze\ze \frac{d\ze \Uppsi}{dx}\right)=\left(\frac{d\ze \Uppsi}{dx}\right)^2+\big(U_{\Omega}  -\Omega^2\big) \Uppsi^2~. 		 
		\end{equation}
	After integrating from the horizon to infinity it follows that
		\begin{equation}
		\int_{-\infty}^{+\infty}dx \left[ \left(\frac{d \Uppsi}{dx}\right)^2+  U_{\Omega} \Uppsi^2 \right]=\Omega^2  \int_{-\infty}^{+\infty}dx\ze \Uppsi^2\ ,
		\end{equation}
which for $ U_{\Omega} >0$ implies $\Omega^2 >0$.}

\begin{figure}[H]
	 \centering
	 	 	 		\begin{picture}(0,0)
		  		 \put(101,-5){\small $\log _{10} \frac{r}{r_H}$}
		  		 \put(110,140){$\scriptstyle \alpha \, =\, 10\ \ Q_e \, = \, 0.12\ \ r_H\, = \, 0.318$}
		  		\put(26,68){\begin{turn}{90}{$\scriptstyle{\rm Event\ horizon}$}\end{turn}}
		  		\put(-4,93){\begin{turn}{90}{\small$U_\Omega$}\end{turn}}
		  	 	\put(200,30){\small $f_F$}
	   		\end{picture}
	  \includegraphics[scale=0.4]{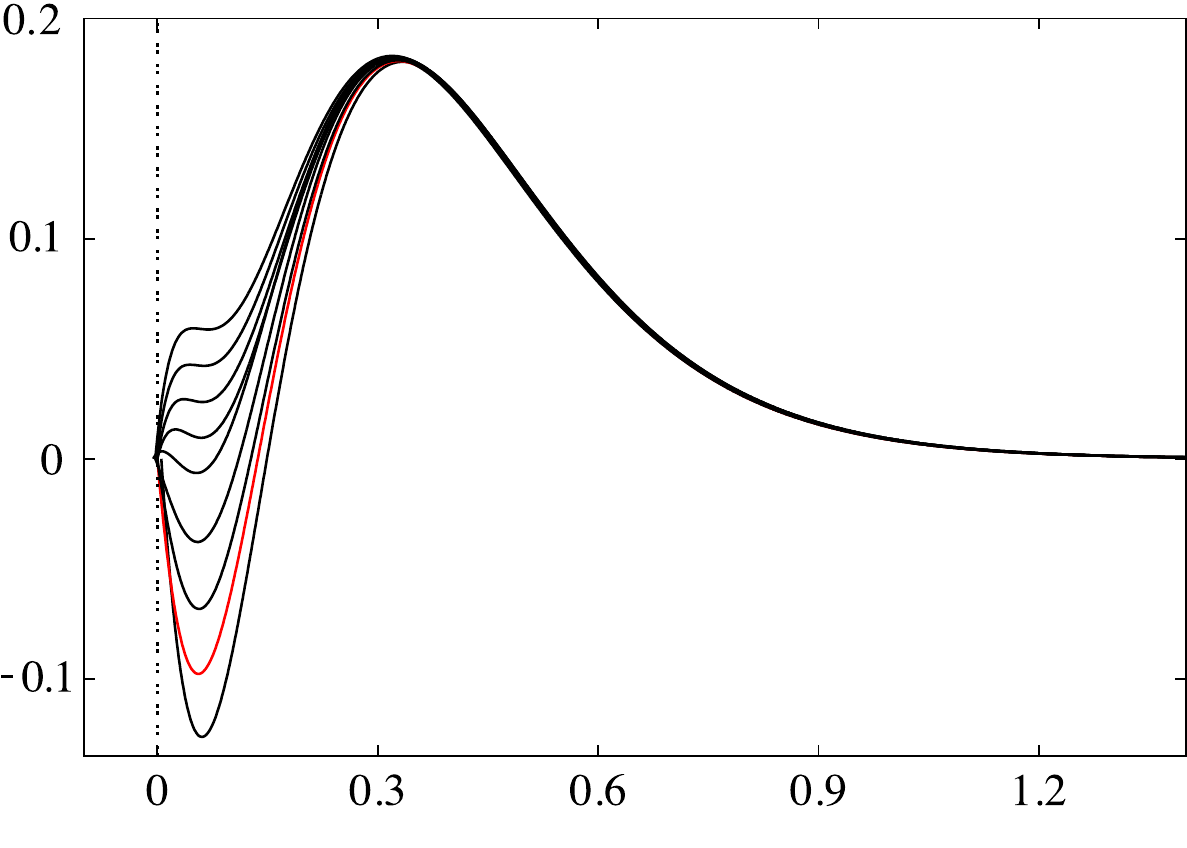}
	  \caption{Effective potential, $U_\Omega$, for a sequence of solution with the $f_F$, $\alpha=10$ and $Q_e=0.120\ze$. The solutions have $r_H=0.320$ ($q=0.658$) -- lowest curve -- up to $r_H=0.308$ ($q=0.676$) -- top curve. The curve in red corresponds to the $f_E$ solution in Fig.~\ref{F2.2} (bottom panel) with $r_H=0.318$ ($q=0.660$).}
	 \label{F2.12}
	\end{figure}
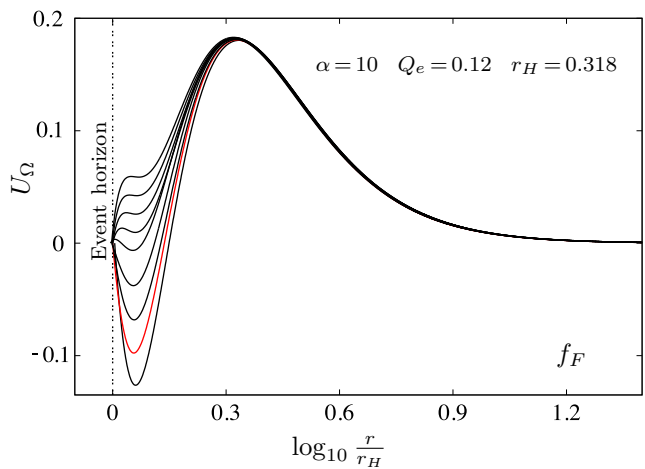

\begin{figure}[H]
		 \centering
		 	\begin{picture}(0,0)
		  		 \put(104,-176){\small $1-\frac{r}{r_H}$}
		  		\put(-7,-90){\begin{turn}{90}{\small$U_\Omega \times r_H ^2 $}\end{turn}}
		  	\put(160,-145){\small $\alpha = 200$}
		  	\put(60,-145){\small $q=0.8\ {\rm (cold)}$}
		   \put(110,-120){\small $q=0.8\ {\rm (hot)}$}
		   	 \put(120,-24){\small $q=1.2\ {\rm (hot)}$}
	   		\end{picture}
		 \includegraphics[width=0.39\linewidth,angle=-90]{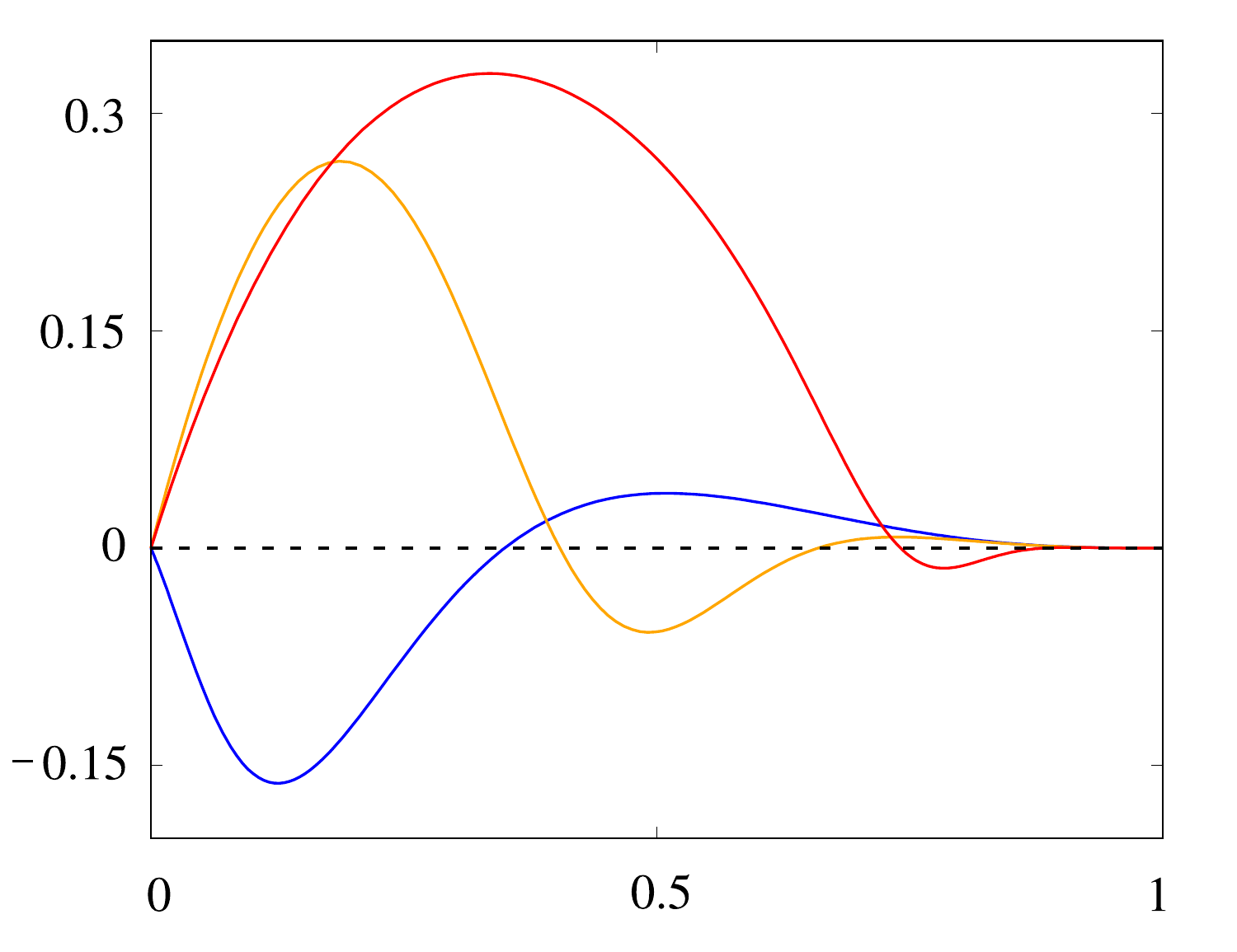}\hfill
		 \includegraphics[width=0.37\linewidth,angle=-90]{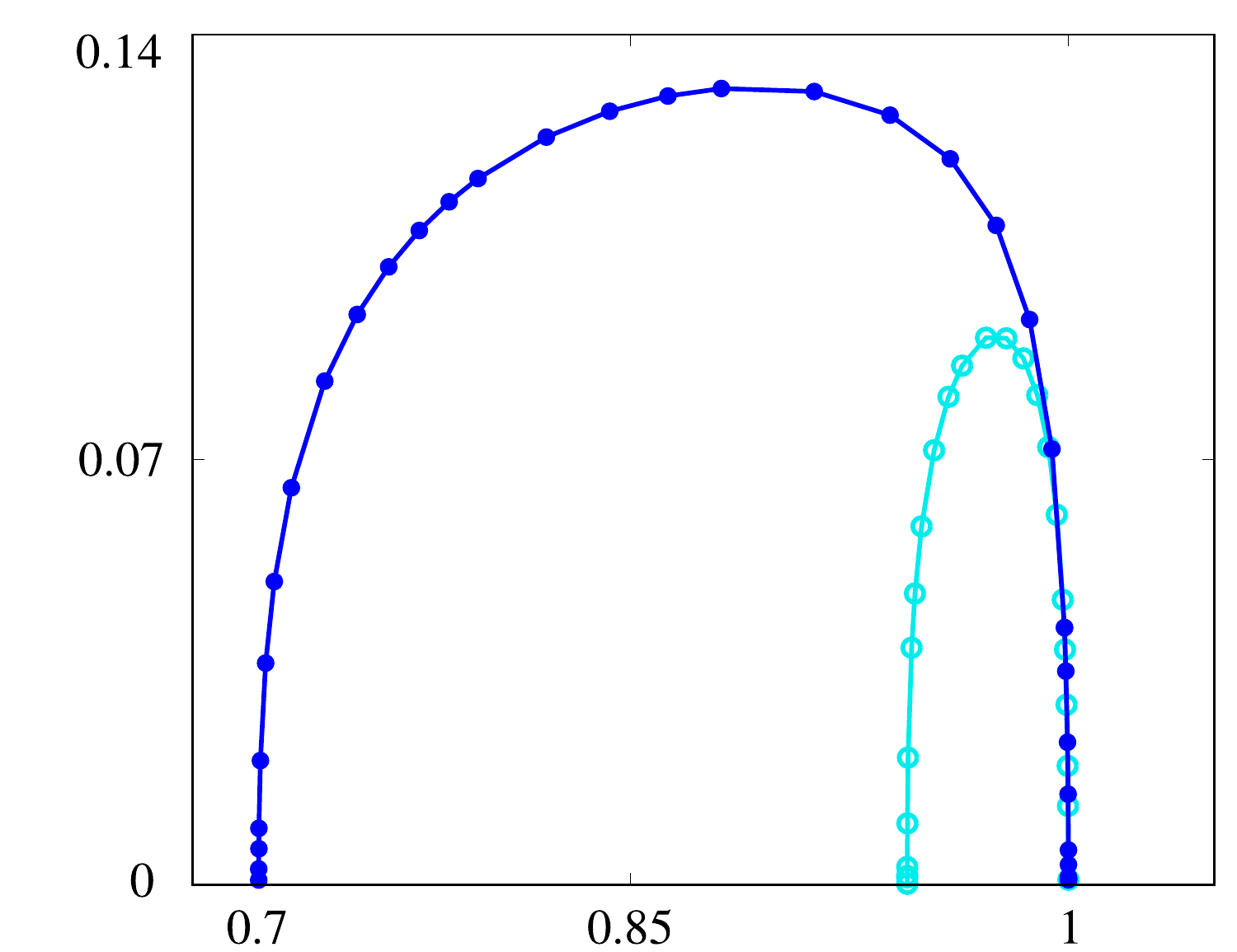}\vspace{2mm}
		 		 	   	\begin{picture}(0,0)
		  		 \put(125,5){\small $q$}
		  		 \put(60,70){\small $\alpha = 200$}
		  		 \put(130,50){\small $\alpha = 20$}
		  		 \put(110,140){$\scriptstyle {\rm cold\ branch}$}
		  		\put(10,80){\begin{turn}{90}{\small$\Omega _I \times M$}\end{turn}}
	   		\end{picture}
		 \caption{(Left panel) effective radial perturbations potential as a function of the radial compactified coordinate for one solution in the cold branch and two solutions in the hot branch.  (Right panel) scaled imaginary part of the mode as a function of $q$ for cold scalarized BHs with $\alpha=20,\ze 200\ze$.}
		\label{F2.13}
		\end{figure}
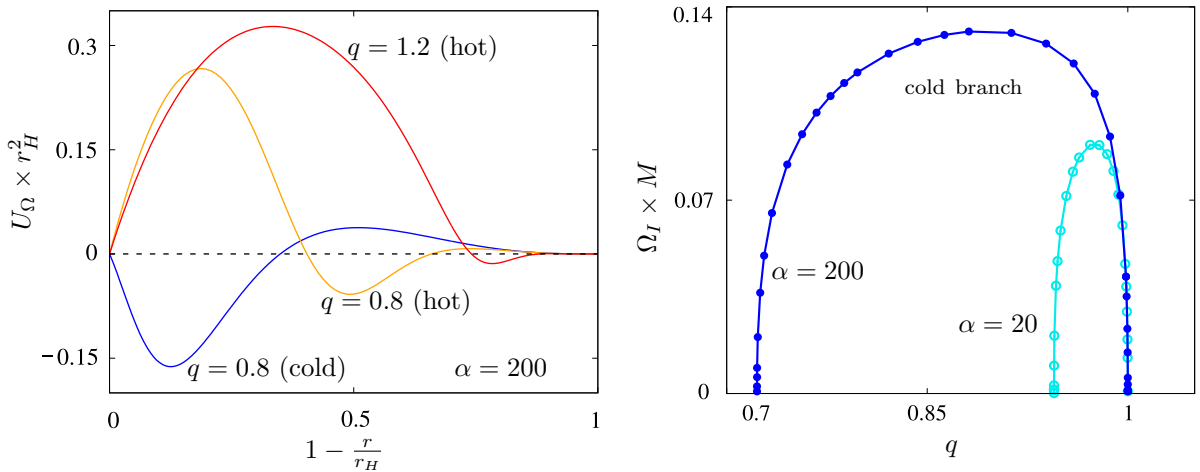

	For the case of the exponential, $\cosh$ and power-law coupling, the potential is, generically, everywhere positive for the vast majority of the solutions analysed, which are therefore free of instabilities -- see the related analysis in~\cite{herdeiro2018spontaneous,myung2021scalarized}. On the other hand, for the fractional coupling, there can be negative regions in the potential both for physical and exotic solutions. As an illustration,  in Fig.~\ref{F2.12} the potential is plotted for a sequence of solutions. One can see that the potential is smaller than zero in a small $q$-region close to the RN limit -- the RN BHs has the zero-mode at $q=0.649$ ($\alpha=10$). Then the potential becomes positive and remains so for arbitrary large $q$ along with the remaining $\alpha = c^{\rm te}$ branch. We emphasise that the existence of a negative potential region is a necessary but not sufficient condition for instability. It would be interesting to see if one can establish stability even in the presence of such negative regions, using, for instance, the $S$-deformation method~\cite{kimura2017simple,kimura2019stability,kimura2018robustness}\footnote{Unfortunatly, we did not pursue such a stability analysis for the $f_F$ coupling. However, an application of the $S$-deformation method was performed for $f_Q$ bellow.}.
	Concerning class \textbf{II.B}, it turns out that the potential always has a negative region. For solutions in the cold branch, the potential is strongly negative close to the horizon (see Fig.~\ref{F2.13} (left panel), blue curve for an exemplar solution). For solutions in the hot branch, on the other hand, the potential also has a negative part; however, this occurs away from the horizon. Moreover, this negative region of the potential becomes smaller along the hot branch when moving away from the bifurcation point, $i.e.$ for larger $q$. This is illustrated in Fig.~\ref{F2.13} (left panel), orange and red curves. The bottom line of these considerations is that the potential is not positive defined. Consequently, this analysis is inconclusive concerning radial stability.
	In order to assess the radial stability of the scalarized solutions, we resort to an explicit computation of possible unstable modes.

	Following the procedure used in other cases \cite{blazquez2018radial}, we have successfully obtained unstable modes of the previous potential, but \textit{only} for the scalarized solutions in the cold branch. The results are shown in Fig.~\ref{F2.13} (right panel), where we exhibit the positive imaginary part of the mode frequency, scaled by the mass, as a function of $q$. As the figure shows, cold scalarized BHs (first branch) have an unstable mode ($\Omega=i\ze\Omega_I$, with $\Omega_I>0$). The absolute value of $\Omega_I$ becomes very small at both end-points of this branch: close to extremality ($q \to 1$) and close to the bifurcation point with the hot branch. However, for hot scalarized BHs (second branch), we could not obtain numerically any unstable mode for any solution in this branch. This includes solutions for which the reduced area is lower than a comparable RN BH. Thus, we can conclude from this analysis that cold scalarized BHs are radially unstable, but nothing can be concluded about hot scalarized BHs.  
			\subsubsection*{$S-$deformation method}
   Since for hot BHs the potential is always negative in some region, unstable modes may exist, albeit we could not find them in the previous section. The $S$-deformation method~\cite{kimura2017simple,kimura2019stability,kimura2018robustness}, however, allows us to show that this is not the case and that all solutions in the hot branch are, in fact, radially stable. Let us first briefly explain the procedure and then prove this statement.

	Multiplying~\eqref{E2.1.24} by $\bar{\Uppsi}$, integrating from the horizon ($x=-\infty$), to $x=+\infty$, and then using partial integration, we obtain
			\begin{equation}\label{E2.1.29}
			 \int_{-\infty}^{+\infty} dx \Bigg[ 	\bigg|\frac{d\Uppsi}{x}\bigg|^2 + U_\Omega \ze |\Uppsi|^2\Bigg] = \Omega^2 \int_{-\infty}^{+\infty}{dx\ze |\Uppsi|^2} \ ,
			\end{equation}
	where we have used the boundary conditions on $\Uppsi$ for unstable perturbations. For unstable modes ($\Omega^2<0$), the right side is negative. This means that for unstable modes to exist, $U_\Omega$ cannot be strictly positive, a result already quoted above. 

	Next, we generalize \eqref{E2.1.29} by introducing the $S$-deformation function. To do this, first rewrite~\eqref{E2.1.29} by making use of the identity
			\begin{equation}\label{E2.1.30}
			 -\bar{\Uppsi}\ze\frac{d^2\ze\Uppsi}{dx^2} =-\frac{d}{dx}\bigg(\bar{\Uppsi}\ze\frac{d\ze\Uppsi}{dx} \bigg)+ \bigg| \frac{d\ze\Uppsi}{dx} \bigg|^2 \ .
			\end{equation}
	Then, introduce an arbitrary $S$ function (deformation function) into the wave equation
			\begin{equation}\label{E2.1.31}
			 -\bar{\Uppsi}\ze\frac{d^2\ze \Uppsi}{dx^2} + U_\Omega\ze |\Uppsi|^2 =-\frac{d}{dx}\bigg(\bar{\Uppsi}\ze\frac{d\ze\Uppsi}{dx} + S|\Uppsi|^2 \bigg)+ \bigg| \frac{d\ze\Uppsi}{dx} + S\Uppsi\bigg|^2 +  \bigg(U_\Omega - S^2 + \frac{dS}{dx}\bigg) |\Uppsi|^2= \Omega^2 |\Uppsi|^2 \ .
			\end{equation}
	Now repeat the integration. After using partial integration we get 
			\begin{equation}\label{E2.1.32}
			 -\bigg[\bar{\Uppsi}\ze\frac{d\ze\Uppsi}{dx} + S|\Uppsi|^2\bigg]_{-\infty}^{+\infty}
+\int_{-\infty}^{+\infty}{ dx\left[	\bigg|\frac{d\ze\Uppsi}{dx} + S\Uppsi\bigg|^2 + \left(U_\Omega - S^2 + \frac{dS}{dx} \right)|\Uppsi|^2	\right]}= \Omega^2 \int_{-\infty}^{+\infty}{dx\ze|\Uppsi|^2} \ .
			\end{equation}
	Next, we restrict the possible $S$ functions. We assume this function is smooth everywhere and does not diverge at the boundaries. Together with the boundary condition for unstable perturbations, these conditions make the first term of the previous expression vanish. We are left with
			\begin{equation}\label{E2.1.33}	
			 \int_{-\infty}^{+\infty}{ dx\ze \bigg|\frac{d\ze\Uppsi}{dx} + S\ze\Uppsi\bigg|^2}+\int_{-\infty}^{+\infty}{ dx\ze \bigg(U_\Omega+ \frac{dS}{dx} - S^2 \bigg)|\Uppsi|^2}= \Omega^2 \int_{-\infty}^{+\infty}{dx\ze |\Uppsi|^2} \ .
			\end{equation}
	The first term in the \textit{lhs} is positive. The \textit{rhs} is negative, if $\Omega^2<0$. Thus, if we show that the second term of the \textit{lhs} is positive or vanishes, we establish that no modes with $\Omega^2<0$ are possible for this potential. Observe that this does not require the potential $U_\Omega$ to be strictly positive anymore.

	In practice, the absence of unstable modes is established by defining the deformed potential $\tilde {U}_\Omega$~\cite{kimura2017simple,kimura2019stability,kimura2018robustness}:
			\begin{equation}\label{E2.1.34}
			 \tilde U_\Omega=U_\Omega + \frac{dS}{dx} - S^2 \ .
			\end{equation}
	Then, it is enough to show that it is possible to make $\tilde U_\Omega=0\ze$. This implies the original potential does not contain unstable modes. Such condition defines a Riccati-type differential equation
			\begin{equation}\label{E2.1.35}
			 \frac{dS}{dx}=S^2-U_\Omega \ ,
			\end{equation}
	or
			\begin{equation}\label{E2.1.36}
			 \frac{dS}{dy}=\frac{dr}{dy}\frac{dx}{dr}\big( S^2-U_\Omega\big) \ ,
			\end{equation}
	where $y=1-\frac{r_H}{r}$, $\frac{dr}{dy}=\frac{r_H}{(1-y)^2}$ and $\frac{dx}{dr}=\frac{1}{\sigma N}$. Since the potential is zero at $r=r_H$ ($y=0$) and $r=+\infty$ ($y=1$), a solution of \eqref{E2.1.36} has to satisfy $S(y=0)=S(y=1)=0\ze$.

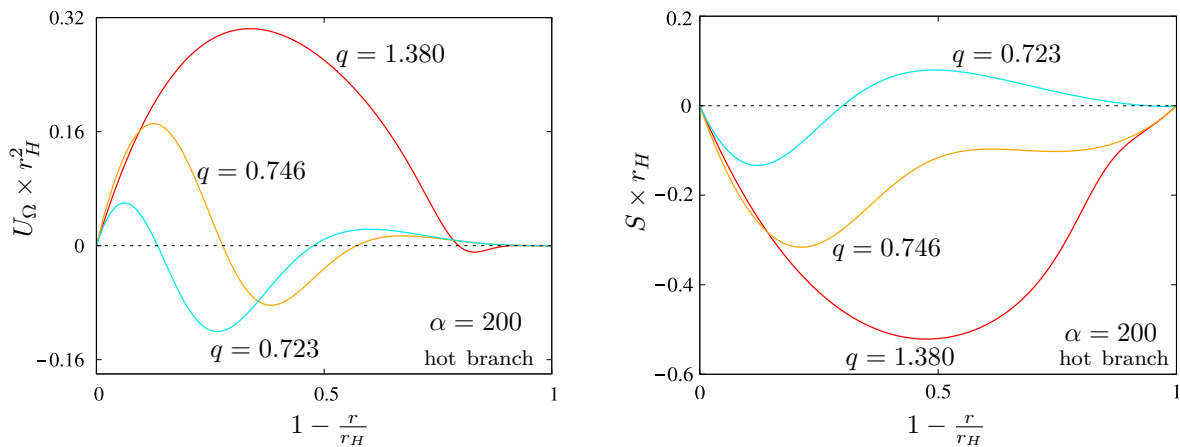
\begin{figure}[H]
			 \centering
			 \begin{picture}(0,0)
		  		 \put(105,-164){\small $1-\frac{r}{r_H}$}
		  		\put(0,-90){\begin{turn}{90}{\small$U_\Omega \times r_H ^2 $}\end{turn}}
		  	\put(155,-138){$\scriptstyle {\rm hot\ branch}$}
		  	\put(157,-125){\small $\alpha = 200$}
		  	\put(75,-135){\small $q=0.723$}
		   \put(70,-68){\small $q=0.746$}
		   	 \put(122,-24){\small $q=1.380$}
	   		\end{picture}
		 	 \includegraphics[width=0.37\linewidth,angle=-90]{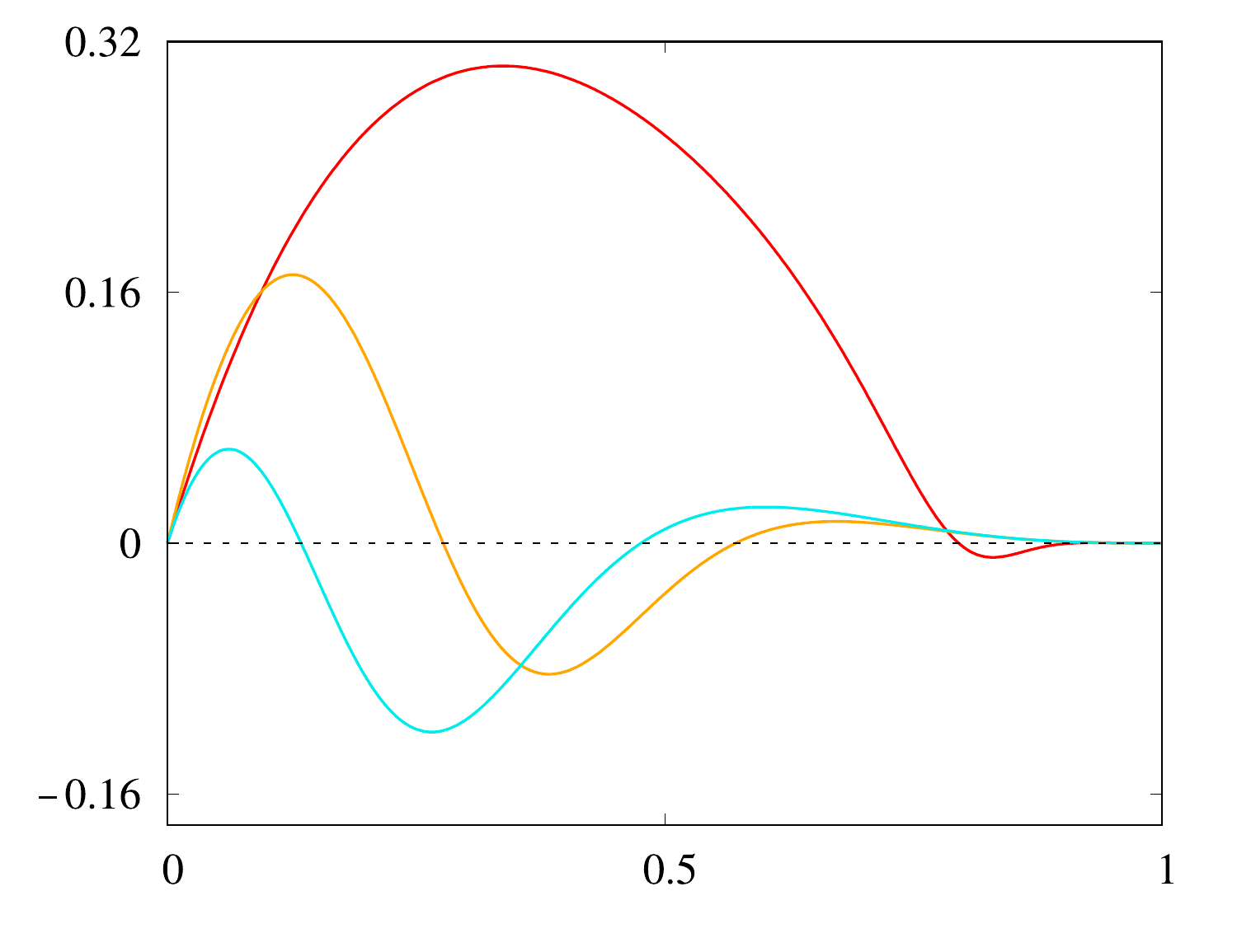}\hfill
		 	 	\begin{picture}(0,0)
		  		 \put(98,-164){\small $1-\frac{r}{r_H}$}
		  		\put(-7,-88){\begin{turn}{90}{\small$S \times r_H$}\end{turn}}
		  	\put(155,-138){$\scriptstyle {\rm hot\ branch}$}
		  	\put(157,-129){\small $\alpha = 200$}
		  	\put(75,-138){\small $q=1.380$}
		   \put(70,-97){\small $q=0.746$}
		   	 \put(115,-24){\small $q=0.723$}
	   		\end{picture}
			 \includegraphics[width=0.342\linewidth,angle=-90]{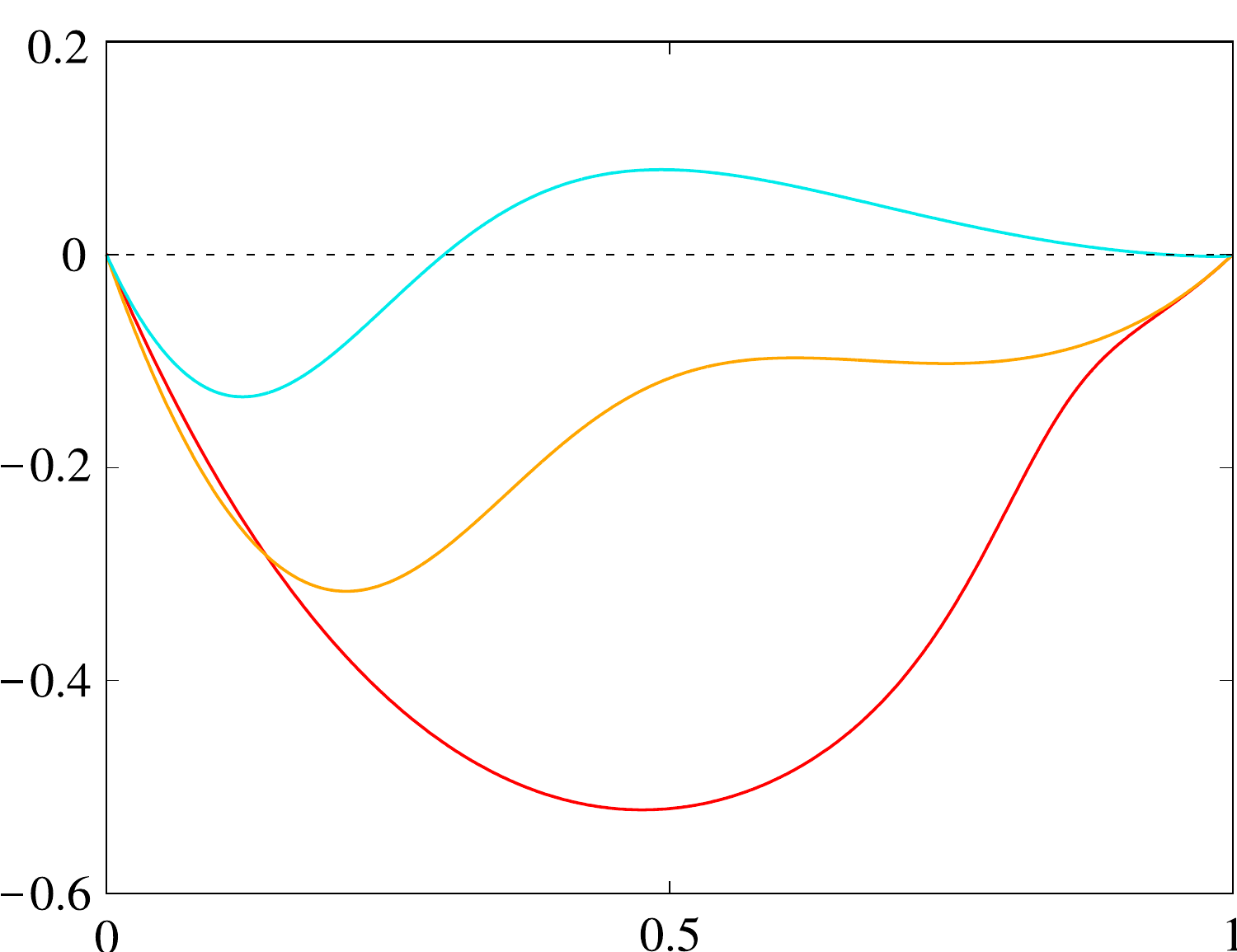}\vspace{2mm}		 	
			 \caption{(Left panel) Effective radial potential as a function of the radial compactified coordinate for several hot BH solutions. (Right panel) $S$-deformation function for the same solutions.}
			 \label{F2.14}
			\end{figure}

	We have numerically integrated \eqref{E2.1.36}, reading off the potential $U_\Omega$ from the numerical scalarized BH solutions. A few examples of the potential for hot scalarized BHs (interpolating between points with a cubic spline) are shown in Fig.~\ref{F2.14} (left panel). The differential equation is solved with boundary condition $S(0)=0$ in a domain $y\in[0,1]$. The result of the integration is that we were able to find solutions for $S(y)$ -- Fig.~\ref{F2.14} (right panel). The deformation function approaches zero on the right side. It was possible to obtain the deformation function for all hot BH solutions we have tackled. This includes solutions for which the reduced area is lower than that of a comparable RN BH -- $e.g.$ the blue curve in Fig.~\ref{F2.14}. Thus, since we have obtained a regular $S$ function for the BH solutions in the second branch, hot scalarized BHs are radially stable, even though the original potential is not strictly positive everywhere.

	As a final remark, if the same procedure is attempted for solutions in the first branch, one cannot integrate \eqref{E2.1.36}; the equation develops a singularity. The latter is consistent, of course, with the fact that we have numerically obtained unstable modes for cold BHs. Thus, it is clear that the $S$ function cannot exist\footnote{In \cite{blazquez2021quasinormal} we demonstrated (for the $f_Q$ coupling), through the full computation of the spectrum of quasinormal modes,  that the only unstable mode present is the radial scalar-led mode of the cold branch. Consequently, the bald RN and hot scalarized branches are both mode-stable. The non-trivial scalar field in the scalarized background solutions leads to the degeneracy between axial and polar modes present for RN solutions. This isospectrality is only slightly broken on the cold branch, but it is strongly broken on the hot branch.}. 
			
%
		\subsection{Dynamical preference}\label{S2.1.6}
%
	Following~\cite{herdeiro2018spontaneous}, with the numerical framework of~\cite{sanchis2016explosion,sanchis2016dynamical,hirschmann2018black,lim2018black}, we have performed fully non-linear evolutions of unstable RN BHs in the EMS class \textbf{II.A}\footnote{We recall that the only class of EMS solutions that can yield spontaneous scalarization from a perturbed RN BH is class \textbf{II.A}.} system under a small Gaussian scalar spherical perturbation, to assess the dynamical endpoint of the evolution\footnote{Dynamical evolution performed by Nicolas Sanchis-Gual.}. 

	We have also considered evolutions with a non-spherical perturbation using the freely available \texttt{Einstein Toolkit}~\cite{toolkit2012open,loffler2012einstein}. The scalar field initial data is 
			\begin{equation}\label{E2.1.37}
			 \phi(r,\theta) = p_{0}\,e^{-\frac{(r-r_0)^2}{\lambda^2}}\,Y^{0}_{\ell}(\theta)\ ,
			\end{equation}
	where $Y^{0}_{\ell}$ is the $\ell$-spherical harmonic with $m=0$ and $p_0,\, r_0$ two constants defining the amplitude and centre of the Gaussian radial profile of the scalar perturbation. A full description of the procedure can be seen in \cite{herdeiro2018spontaneous,fernandes2019spontaneous}. To perform the evolutions, we have used a numerical grid with $11$ refinement levels with 
			\begin{align*}
			&\Big\{ \big(192,\ze 96,\ze 48,\ze 24,\ze 12,\ze 6,\ze 3,\ze 1.5,\ze 0.75,\ze 0.375,\ze 0.1875\big),\qquad\\
			& \big( 6.4,\ze 3.2,\ze 1.6,\ze 0.8,\ze 0.4,\ze 0.2,\ze 0.1,\ze 0.05,\ze 0.025,\ze 0.0125,\ze 0.00625\big)\Big\} \ , 
			\end{align*}
	where the first set of numbers indicates the spatial domain of each level and the second set indicates the resolution. Due to the geometry of the spherical harmonics, we consider equatorial-plane symmetry and reflection symmetry concerning the $x$-$z$ plane for the $(\ell=2,\, m=0)$, but not for the $(\ell=1,\, m=0)$ mode, and reflection symmetry with respect to the positive values of $x$ and $y$ for both modes.
			\begin{figure}[h!]
	 		 \centering
	  		 \includegraphics[scale=0.46]{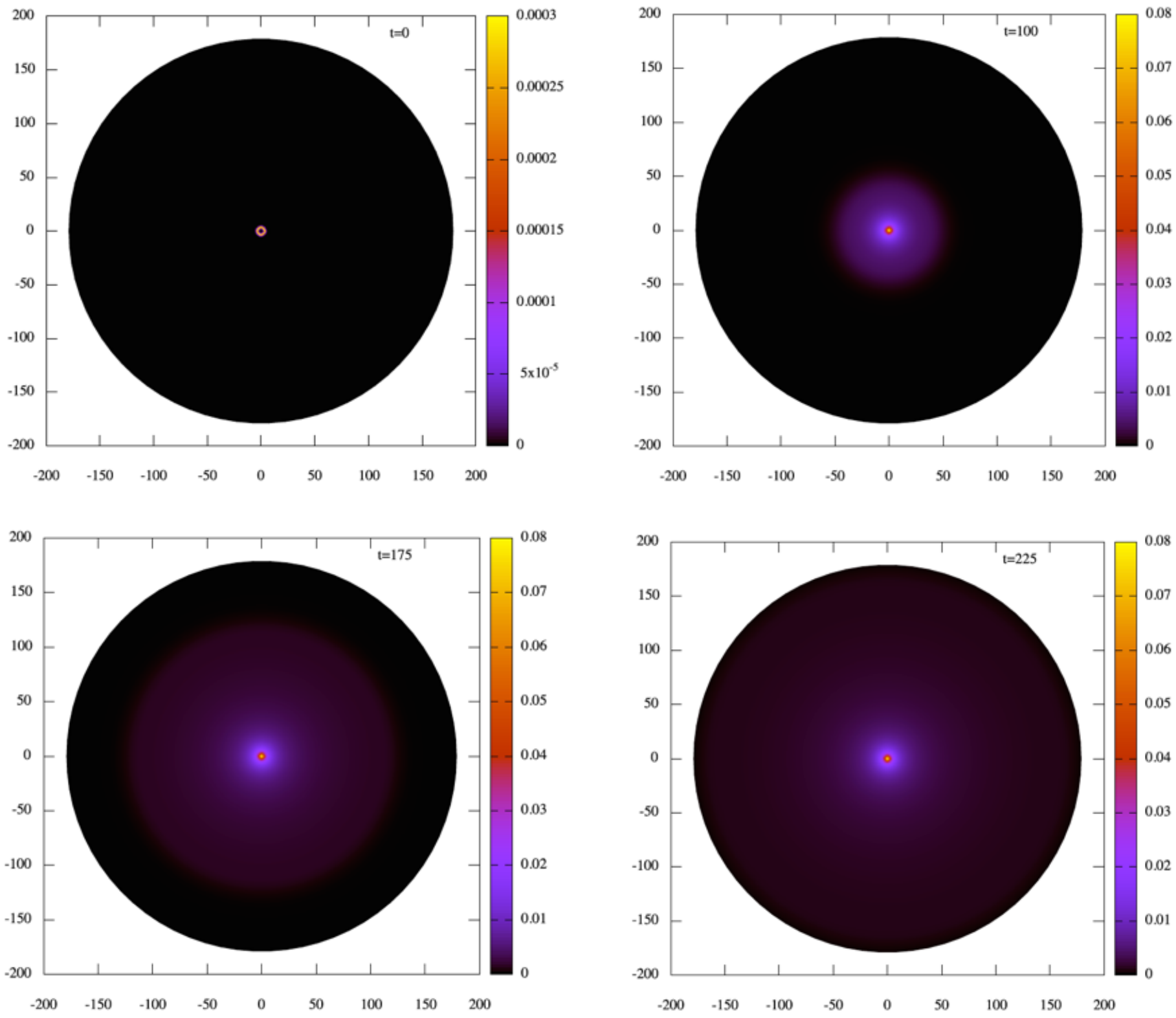}
	  		 \caption{Four snapshots of the time evolution of the scalar field around an unstable RN BH with $q=0.2$ in the EMS system, with the exponential coupling and $\alpha=400.979$. Full evolution can be seen at \cite{spontaneousVid1}.}
	 		 \label{F2.15}
			\end{figure}

	In~\cite{herdeiro2018spontaneous} the dynamical formation of scalarized BHs with the exponential coupling was established. The evolution of the process can be observed in Fig.~\ref{F2.15} (full evolution can be seen at \cite{spontaneousVid1}), wherein four snapshots, at times $t=0,\, 100,\, 175,\, 225$, are show for $f_E$, $q=0.2$ and $\alpha=400.979\ze$. The $\ell=0$ small Gaussian perturbation triggered the growth of a scalar cloud in the vicinity of the horizon that expands outwards and becomes a monotonically decreasing function of the radial coordinate. The energy transfer to the scalar field saturates by $t\sim 100$~\cite{herdeiro2018spontaneous}, and it reaches an equilibrium state, at least in the vicinity of the BH, around $t\sim 200$, albeit part of the more exterior scalar field distribution is still evolving outwards, settling down to the scalarized solution. The same qualitative pattern is observed for other class \textbf{II.A} couplings for which scalarization occurs. 
	    
	The endpoint of the evolution shown in Fig.~\ref{F2.15} is a scalarized BH with the same value of $q$. The aforementioned was established by comparing the value of the scalar field on the horizon obtained in the numerical evolution with one of the previously computed static scalarized solutions with the same coupling and $q$. As explained above, fixing $\alpha$ the value of $p_0\equiv \phi (r_H)$ serves as a measure of $q$. In Fig.~\ref{F2.16} (left panel), this comparison is made for various values of $\alpha$, fixing $q=0.2$ of the initial RN BH, for both the exponential coupling (data already shown in~\cite{herdeiro2018spontaneous}) and the power-law coupling. The crosses are from the numerical evolutions and the solid line from the static solutions. The agreement is quite good. As discussed above, the power-law coupling produces a weaker scalarization for the same coupling. 

	Fig.~\ref{F2.16} (right panel) performs a similar comparison, for $f_E$, but now exploring a larger range of values of $q$. Beyond $q\sim 0.4\ze$, the agreement between the value of the scalar field on the horizon obtained from the evolutions and that obtained from the static solutions with the same $q$, ceases to hold. In other words, the endpoint of the evolution of a RN BH with a certain value of $q$ is not a scalarized BH with the same value of $q$. Rather, the former matches a scalarized BH with a lower value of $q$. This is interpreted as a non-conservative evolution that ejects a larger fraction of electric charge than energy when forming the scalarized BH. 
			\begin{figure}[h!]
	 		 \centering
	 		\begin{picture}(0,0)
		  		 \put(120,-2){\small $\alpha$}
		  		\put(0,79){\begin{turn}{90}{\small$\phi _0$}\end{turn}}
	   		\end{picture}
	 	   			\begin{tikzpicture}[scale=0.5]
\node at (0,0) { \includegraphics[scale=0.27]{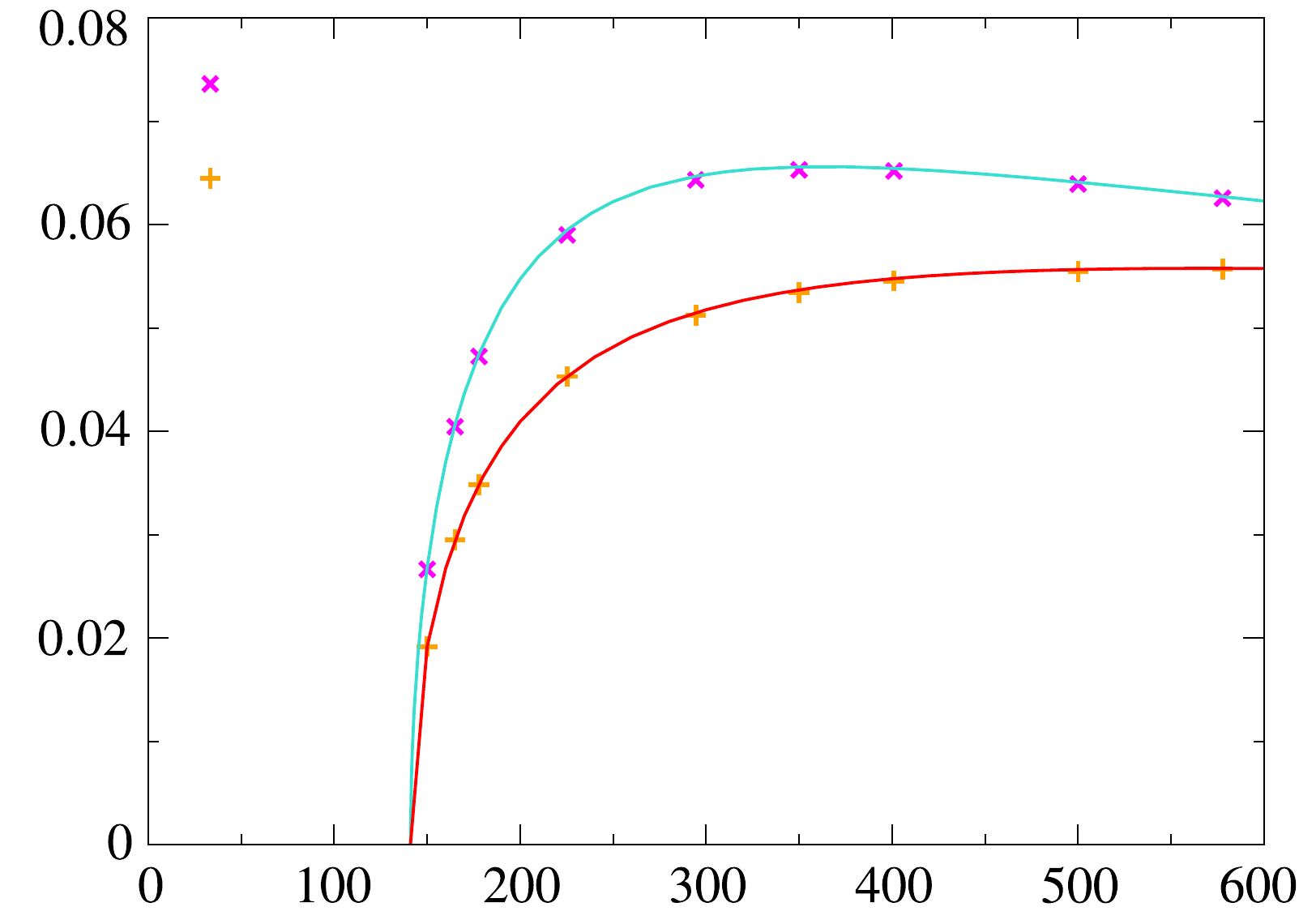}};
\node at (-4.1,4.25) {\small $f_E$};
\node at (-4.1,3.2) {\small $f_P$};
				\end{tikzpicture}
				\begin{picture}(0,0)
		  		 \put(120,-2){\small $\alpha$}
		  		 \put(30,140){$\scriptstyle q=0.2$}
		  		 \put(60,140){$\scriptstyle q=0.3$}
 		  		 \put(90,140){$\scriptstyle q=0.4$}
		  		 \put(120,140){$\scriptstyle q=0.5$}
		  		 \put(150,140){$\scriptstyle q=0.7$}
		  		 \put(180,140){$\scriptstyle q=0.8$}
		  		 \put(0,79){\begin{turn}{90}{\small$\phi _0$}\end{turn}}
	   		\end{picture}
	 	   			\begin{tikzpicture}[scale=0.5]
\node at (0,0) { \includegraphics[scale=0.27]{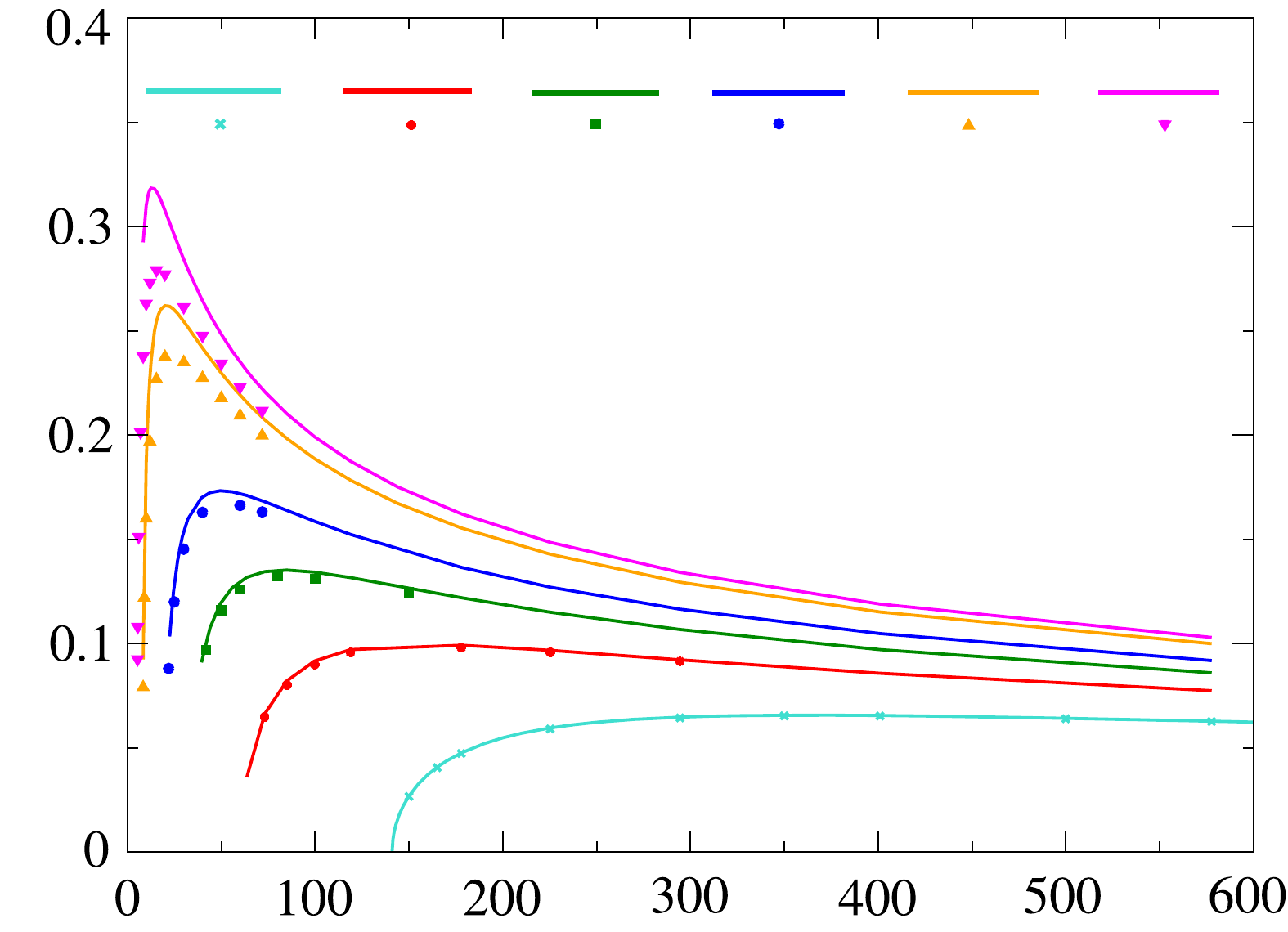}};
				\end{tikzpicture}
	  \caption{(Left panel) Scalar field value at the horizon for $q=0.2$ and a range of couplings $\alpha$, for the exponential and power-law coupling. The solid line is obtained from the static solutions. The crosses are the dynamically obtained value from the numerical simulations after saturation and equilibrium has been reached. The agreement is notorious. (Right panel) A similar study, for the exponential coupling, but for various values of $q$. The agreement between the points and the lines with the same $q$ is restricted to $q\lesssim 0.4$. For larger $q$, the evolution points match static solution lines with a smaller $q$.}
	 	 \label{F2.16}
		\end{figure}
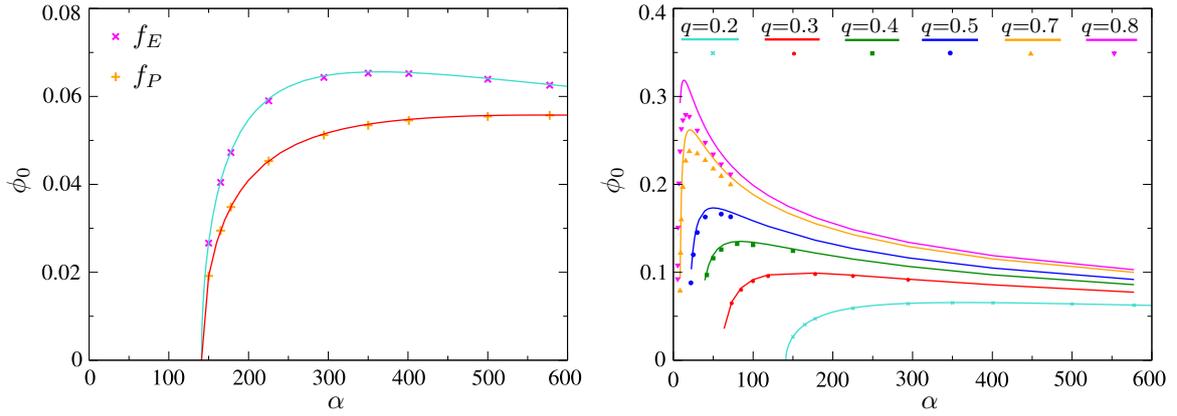
 		   \begin{figure}[H]
		 	\centering
	  		\includegraphics[scale=0.4]{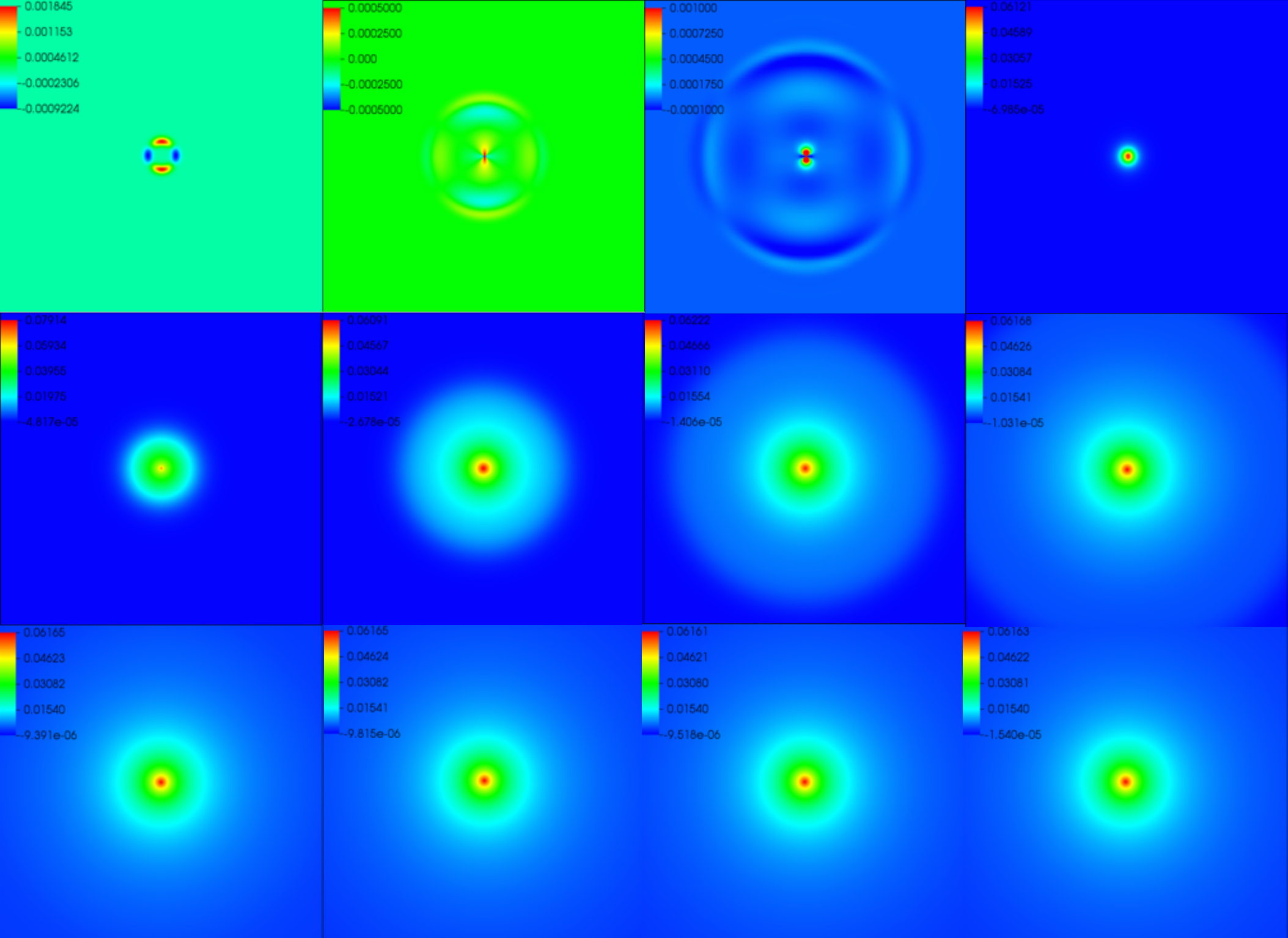}
	  		\caption{Twelve snapshots in the $x$-$z$ ($y=0$) plane of the time evolution of an unstable RN BH with $q=0.2$ in the EMS system, with the $f_E$ and  $\alpha=1200$ and an $\ell=2$, $m=0$ perturbation. The snapshots correspond to $t$ between $0$ and $140.8\ze$. The data for negative values of $x$ and $z$ are mirrored by the corresponding positive values, due to equatorial symmetry. Full evolution at \cite{spontaneousVid2,spontaneousVid3}.}
			\label{F2.17}
		\end{figure}
		
			An intriguing possibility raised in~\cite{herdeiro2018spontaneous} concerns the dynamical role of \textit{non-spherically symmetric} scalarized solutions. To address this issue, we have performed the evolution of an unstable RN BH under non-spherical perturbations, using~\eqref{E2.1.37} with $\ell=1,\, 2$. In Fig.~\ref{F2.17} we show snapshots of such an evolution for the $\ell=2$ case. It can be observed that, initially, the non-spherical mode is dissipated/absorbed; then scalarization proceeds much as in the case of a spherical perturbation. Similar results are obtained for the $\ell=1$ perturbation. Thus, scalarization is robust, even without imposing spherical symmetry, and we see no evidence of the formation of the non-spherical scalarized solutions described in~\cite{herdeiro2018spontaneous}. Hence, such solutions may be unstable.

%
	\section{Massive EMS}\label{S2.2}
%
   All the previous studies have been performed for massless fields. However, the known -- and several hypothetical -- bosonic particles (except the photon) have a mass and/or a self-interaction~\cite{fernandes2020einstein}. In this section, we will study the charge induced scalarization mechanism of a massive scalar field\footnote{All the results shown in this section were obtained by the author of this thesis before and independently from the results published in \cite{fernandes2020einstein}.}. Observe that this preliminary study can be seen as one of the simples generalizations of the EMS model (Sec.~\ref{S2.1}).

	The generalization of the action \eqref{E2.1.1} to contain a massive scalar field in an EMS model comes as:
		\begin{equation}\label{E2.2.34}
	 	 \mathcal{S}_{\mu EMS}=\frac{1}{4}\int d^4 x\ze \sqrt{-g}\ze\ze \Big[ R - 2\ze \phi _{\ze ,\ze \mu}\ze \phi ^{\ze ,\ze \mu} + f(\phi)\ze F_{\mu \nu} \ze F^{\mu \nu} -2\ze\mu _\phi ^2\ze \phi ^2 \Big]\ ,
		\end{equation}
	with $\mu _\phi$ the scalar field mass. The generic line element of a spherically symmetric is again given by \eqref{E1.5.40}. Considering a small scalar perturbation around a RN background, allows us to decomposed the scalar field in (real) spherical harmonics, $\delta \phi (r,\theta, \phi )=\sum _{\ell \, m}U_{\ze \ell} (r)\ze Y_{\ell}^m (\theta,\phi )$. The scalar field equations simplify to
		\begin{equation}\label{E2.2.35}
		 \frac{1}{r^2 \sigma } \left( r^2 N  \sigma \ze U' _{\ze \ell} \right)'-\left[\frac{\ell(\ell+1)}{r^2}+\mu _{\rm eff} ^2 +\mu _\phi ^{\ze 2} \right] U_\ell=0\ ,
		\end{equation}
	In this section, let us quickly consider the effect of a mass term on the spontaneous scalarization phenomena. For that, let us consider an exemplar coupling function, say $f_E (\phi)= e^{\alpha \phi ^2}$.

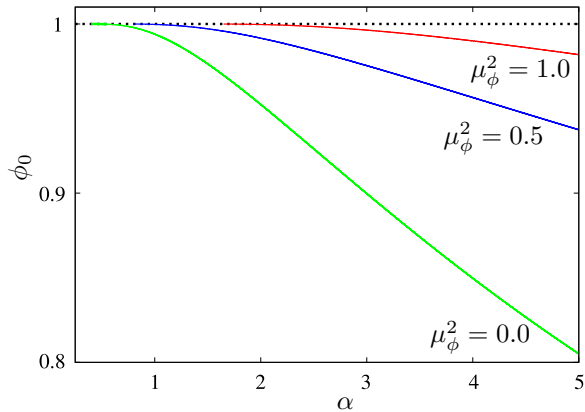
\begin{figure}[H]	
			 \centering
			\begin{picture}(0,0)
		  	 \put(165,118){\small $\mu^2 _\phi =1.0$}
		  	 \put(155,93){\small $\mu^2 _\phi =0.5$}
		  	 \put(150,18){\small $\mu ^2 _\phi =0.0$}
		  	 \put(115,-7){\small $\alpha$}
		  	 \put(-8,79){\begin{turn}{90}{\small$\phi _0$}\end{turn}}
	   		\end{picture}
	 	 \includegraphics[scale=0.6]{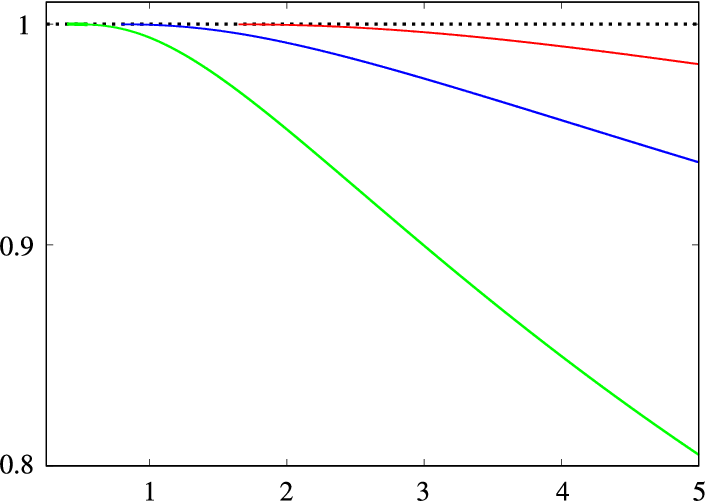}
	 	 \caption{Existence line $q\ vs. \ \alpha $ for an $\mu$EMS model with $f_E$ and (red line) $\mu _\phi ^2 = 1.0\ze$; (blue line) $\mu _\phi ^2 =0.5$ and (green line) $\mu _\phi ^2 =0.0\ze$. The dotted black line represents extremal RN BHs $q=1.0\ze$. The addition of the mass term atenuates the tachyonic instability and makes the onset of the instability move to higher $\alpha$ values.}
	 	 \label{F2.18}
		\end{figure}

  	Observe that one can redefine $\mu _{\rm eff} ^2 \equiv -\alpha\ze\ze Q_e ^2\ze \ze r^{-4} +\mu_\phi ^2 <0$ and \eqref{E2.2.35} reduces to \eqref{E1.2.17}. Hence, all the previously performed considerations (Sec.~\ref{S2.1}) can be generalized to the massive case. However, let us point out that $\mu _\phi ^2 >0$ and $Q_e ^2 \ze\ze r^{-4} <0\ze$. Since $\mu _{\rm eff} ^2$ must be negative for the spontaneous scalarization phenomena to occur, $Q_e ^2 \ze\ze r^{-4}>\mu _\phi^2$, one expects the addition of the mass term to attenuate the effects of the scalarization, requiring an higher coupling constant $\alpha$ to yield the same properties as in the absence of a mass term. In that regard, see Fig.~\ref{F2.18}, where we plot the existence line for three values of the scalar field mass: (red line) $\mu _\phi ^2 = 1.0\ze$; (blue line) $\mu _\phi ^2 = 0.5$ and (green line) $\mu _\phi ^2 = 0.0\ze$.
	In Fig.~\ref{F2.18} one can observe that, an increase in the scalar field's mass implies a shift to higher values of the coupling constant $\alpha$ to yield the same tachyonic instability -- compare $\mu _\phi ^2 =0.0 $ (green line) and $\mu _\phi ^2 =1.0$ (red line) cases. The mass term has a supressive contribution.
%
	\subsection{Numerical results}\label{S2.2.1}
%

	The system's effective Lagrangian density comes as:
	\begin{equation}
	 \mathcal{L} ^{\rm eff} _{\mu EMS}=\frac{1}{2\sigma}\left[2\ze m'+r\Big(-r\ze \mu_\phi ^2\ze \phi ^2+ \frac{f_E}{\sigma ^2}\ze\ze r \ze V'^{\, 2}-(r-2\ze m)\ze \phi'^{\, 2}\Big)\right]\ , 
	\end{equation}
	where the only difference from the case without the mass term \eqref{E2.1.2} is the presence of the mass term $-r\ze \mu _\phi ^2 \ze \phi ^2 $, thus by making $\mu_\phi = 0$ one recovers the previously obtained solution. The resulting Euler-Lagrange equations are:
	\begin{align}
	 & m' =  \frac{r}{2} \left[ \frac{Q_e ^{\ze \ze 2}}{f_E\ze r^3}+r\ze \mu _\phi ^2 \ze \phi ^2 + \big(-2\ze m +r ) \phi'^{\, 2}\right] \ ,\nonumber\\
	  & \phi '' = \frac{1}{r^3 \ze \big( -2\ze m+r\big)}\bigg[r^4 \ze \mu_\phi ^2 \ze \phi + r^2 \big( 2\ze m- 2\ze r + r^3 \ze \mu _\phi ^2 \ze \phi ^2 \big) \phi ' +\frac{ Q_e ^{\ze\ze 2}}{f_E}\big( \alpha \ze \phi +r\ze \phi ' \big) \bigg]\ ,\nonumber\\
	  & V' = -\frac{Q_e \ze\ze\sigma}{r^2 f_E}\ ,\qquad\qquad  \sigma ' = -r \ze \phi'^{\, 2} \sigma\ . 
	\end{align}
	Note that both the equations for $\sigma ' $ and $V'$ are explicitly independent of the mass term. However, this does not represent a lack of influence of the latter in the former. Since both $m'$ and $\phi ''$ are  explicitly dependent on $\mu _\phi$. 

	Due to the absence of a significant scalar field impact and the lack of an explicit influence in the $V'$ and $\sigma '$ equations, one obtains the same polynomial behaviour of the functions close to the horizon \eqref{E2.1.7}
	\begin{align}\label{E2.2.39}
	& m = \frac{r_H}{2}+\frac{e^{\alpha \ze \phi _0 ^2 }\ze\ze Q_e ^{\ze\ze 2}+\phi _0 ^{\ze\ze 2}\ze\ze r_H ^{\ze\ze 4}\ze\ze \mu_\phi ^2}{2\ze r_H ^{\ze\ze 2}}(r-r_H )\ +\cdots\  ,\nonumber\\
	 & \phi = \phi _0 - \ze\ze \frac{\phi _0\big(e^{\alpha \phi _0 ^{\ze\ze 2}}\ze\ze Q_ e ^{\ze\ze 2}\ze\ze \alpha+r_H ^{\ze\ze 4}\ze\ze \mu _\phi ^2 \big)}{e^{\alpha \ze\phi _0 ^{\ze\ze 2}} Q_e ^{\ze\ze 2}\ze\ze r_H -r_H ^{\ze\ze 3}+\phi _0 ^{\ze\ze 2}\ze\ze r_H ^{\ze\ze 5}\ze\ze \mu _\phi ^2}(r-r_H)+\cdots\ ,\nonumber\\
	 & V= -\frac{\sigma _ 0\ze\ze e^{\alpha \ze \ze \phi _0 ^{\ze\ze 2}} Q_e ^{\ze\ze 2}}{r_H ^{\ze\ze 2}} (r-r_H) +\cdots\ ,\nonumber\\
	 &\sigma = \sigma _0 -\sigma _0 \ze\phi _1 ^{\ze\ze 2} \ze\ze r_H (r-r_H ) +\cdots\ .
	\end{align}
	Concerning the asymptotic decay behaviour, we expect a more significant impact on the form decay due to the high contribution of the mass term. Considering that for negligible small $\mu _\phi $ we must recover the usual scalarized RN solution with $f_E$; for large $\mu_\phi$ and small $r_H$ we expect to obtain a solitonic-like solution. The region in between must be a mix of the two behaviours.
	
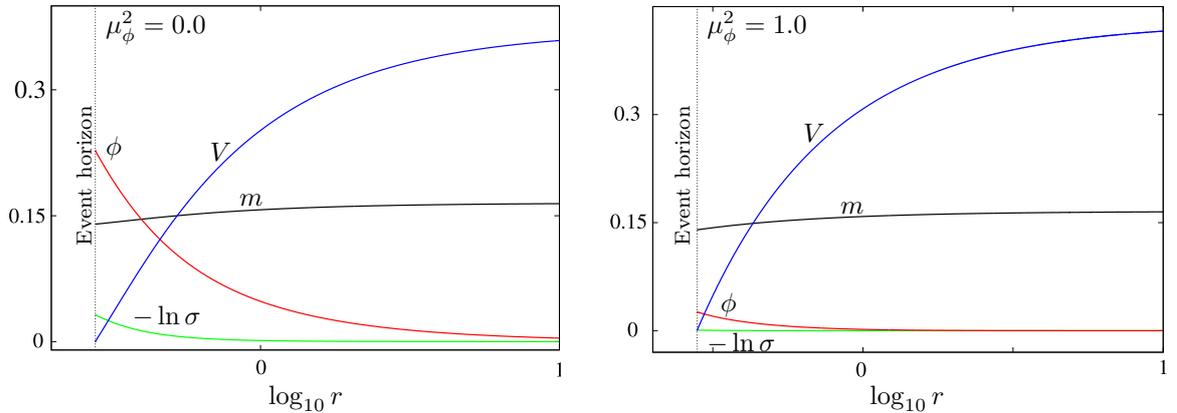
\begin{figure}[H]		 \centering
			\begin{picture}(0,0)
		  	 \put(40,130){\small$\mu^2 _\phi =0.0$}
		  		 \put(50,20){\small$-\ln \sigma$}
		  		 \put(90,65){\small$m$}
		  		 \put(40,84){\small$\phi$}		  		 
		  		 \put(79,82){\small$V$}
		  		 \put(101,-10){\small $\log _{10} r$}
		  		\put(29,50){\begin{turn}{90}{$\scriptstyle{\rm Event\ horizon}$}\end{turn}}
	   			\end{picture}
	 	 \includegraphics[scale=0.6]{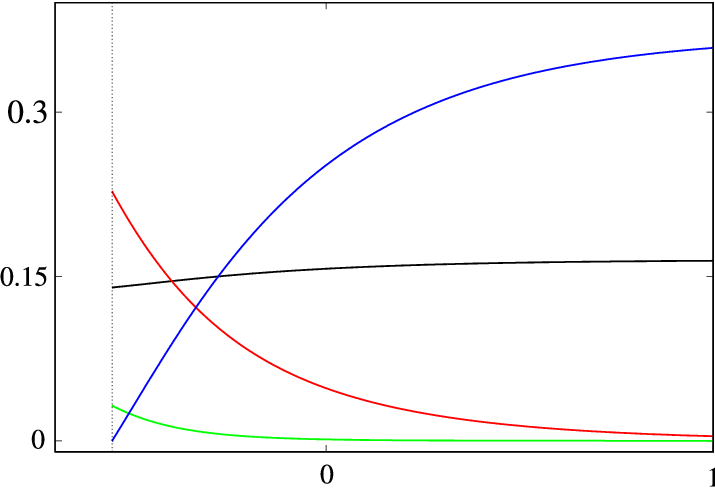}\hfill
	 	 			\begin{picture}(0,0)
		  	 \put(40,130){\small$\mu^2 _\phi =1.0$}
		  		 \put(40,10){\small$-\ln \sigma$}
		  		 \put(90,62){\small$m$}
		  		 \put(45,25){\small$\phi$}		  		 
		  		 \put(76,89){\small$V$}
		  		 \put(101,-10){\small $\log _{10} r$}
		  		\put(28,50){\begin{turn}{90}{$\scriptstyle{\rm Event\ horizon}$}\end{turn}}
	   			\end{picture}
	 	 \includegraphics[scale=0.615]{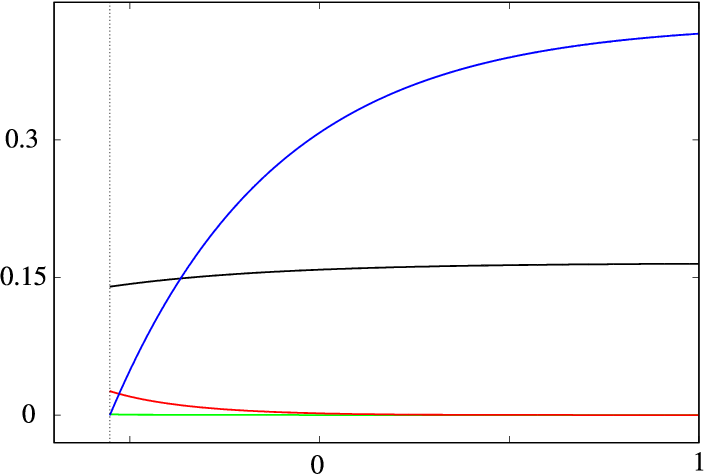}
	 	 \caption{Scalarized BH radial functions for $\alpha =10$ and $f_E$ model with $q=0.73\ze$, $r_H=0.28$ and $Q_e=0.12\ze $. (Left panel) $\mu _\phi ^2 =0.0$ (massless) and (right panel) $\mu_\phi =1.0$ (massive).}
	 	 \label{F2.19}
		\end{figure}

	For that reason, let us consider a modulation term that is proportional to $e^{-\mu_\phi r}$ for the scalar field equation. In this way as $\mu_\phi$ decreases, so does its contribution until we recover the expected scalarized RN solutions. As $r\rightarrow +\infty $ we expect: $m\rightarrow M$, $\phi \rightarrow 0$, $\sigma \rightarrow 1$ and $V \rightarrow \Psi _e$, as
	\begin{align}\label{E2.2.40}
	& m = M-\frac{Q_\phi ^{\ze\ze 2} \ze\ze \mu_\phi ^2}{2\ze r}+\cdots \ ,\qquad \sigma = 1+\frac{\sigma_3}{r^3}+\cdots\ ,\qquad \phi = \frac{Q_\phi}{r} e^{-\mu_\phi r}\ , \qquad V = \Psi _e-\frac{Q_e}{r}+\cdots\ . 
	\end{align}
	Comparing the solution's profile between two scalarized RN BH solutions -- Fig.~\ref{F2.19} -- both with an exponential coupling function $f_E$ and the same coupling constant value $\alpha=10$, electric charge $q=0.73$ and horizon radius $r_H=0.28$, however one has a massive scalar field $\mu_\phi ^2 =1.0$ (left panel) and the other has a massless scalar field $\mu_\phi ^2 =0.0$ (right panel).
	From Fig.~\ref{F2.19} it is clear that the massless solution (left panel) has a higher level of scalarization than the massive solution (right panel). A more detailed study of such solutions and domains of existence can be seen at \cite{fernandes2020einstein}.
	
%
	\section{Dyons}\label{S2.3}
%
	The scalarized BHs studied up to now contain only an electric charge. They possess no extremal limit. Rather, a critical solution is attained for the maximal charge a BH can support, which (numerical evidence suggests) is singular. This parallels the status of dilatonic BHs. For the latter, however, the introduction of an \textit{additional} magnetic charge leads to dyonic BHs with an extremal (non-singular) limit, which have been constructed for specific couplings~\cite{dobiasch1982stationary,gibbons1986black,kallosh1992supersymmetry}. Given the importance of extremal solutions, it is interesting to inquire which are the properties of the dyonic scalarized BHs family and, in particular, their extremal limit.

	The full Einstein-Maxwell (EM) theory, \textit{a.k.a.} electro-magnetic-vacuum, is the quintessential source-free, gravitational relativistic field theory. Its static physical BHs belong to the $3$-parameter RN family, described by mass $M$, electric $Q_e$ and magnetic $P$ charges. These BHs are perturbatively stable~\cite{moncrief1974odd,moncrief1974stability}, and, for a given $M$, can only sustain charges $(Q_e,\, P)$ if $\sqrt{Q_e ^2+P ^2}\leqslant M$. When equality holds, the extremal limit is attained. Extremal RN BHs are special. They are non-singular spacetimes, on and outside a degenerate and $C^\infty$ smooth event horizon, that: (i) have a vanishing Hawking temperature and are BPS states that possess Killing spinors when embedded in supergravity~\cite{gibbons1982bogomolny}; (ii) have a near horizon geometry which is, itself, a solution of the EM theory~\cite{gibbons1993vacuum} -- the Robinson-Bertotti ($AdS_2\times S^2$) vacuum~\cite{robinson1959solution,bertotti1959uniform}; and (iii) allow a no-force condition and a multi-BH generalization, described by the Majumdar-Papapetrou metrics~\cite{majumdar1947class,papapetrou1948einstein,hartle1972solutions}. 

	The considerations above suggest that a comparison between dilatonic and scalarized BHs can be instructive. The purpose of this section is to perform such a comparison for the canonical dilatonic coupling, and the reference model of scalarized solutions introduced in~\cite{herdeiro2018spontaneous}.

	The action of the EMS system with the presence of a magnetic charge is still \eqref{E2.1.1}, with the metric ansatz \eqref{E1.5.40}. The resulting field equations are given by \eqref{E1.2.2}-\eqref{E1.2.4}, being the only noticeable difference from the purely electrostatic case the $4$-vector potential ansatz that comes as
		\begin{equation}\label{E2.3.41}
		 A=V(r) dt+P \cos \theta d\varphi~,
		\end{equation}
	where $P=c^{\rm te}$ is the magnetic charge\footnote{Observe that this ansatz is equivalent to \eqref{E1.5.45} when $A_\varphi = P$ and $F_W =0$.}. The Maxwell equation's a first integral \eqref{E2.1.5} holds. One should point out that the equations of motion~\eqref{E1.2.2}-\eqref{E1.2.4} are invariant under the \textit{electro-magnetic duality} transformation
		\begin{equation}\label{E2.3.45}
		 \{P\to Q_e,~~Q_e\to P \}~~{\rm and}~~f(\phi)\to 1/f(\phi) \ .
		\end{equation}
	In what follows, we shall assume, without any loss of generality, that both $Q_e$ and $P$ are positive and that 
		\begin{equation}\label{E2.3.46}
		 Q_e\geqslant P \ ,
		\end{equation}
such that for scalarized BHs, the (electric) solutions in Sec.~\ref{S2.1} are recovered as $P\to 0\ze$.
%
		\subsection{Spontaneous scalarization of dyonic RN BHs: bifurcation line}\label{S2.3.1}
%
	Class \textbf{II.A} of EMS models is particularly interesting because it accommodates the dynamical phenomenon of spontaneous scalarization (Sec.~\ref{S2.2}). At the linear level this is manifest in the tachyonic instability~\eqref{E1.2.10}. For a dyonic RN BHs, $N(r)=1-2M/r+(Q_e^2 +P ^2)/r^2$ and  $F_{\mu\nu}F^{\mu\nu}=-2(Q_e^2-P^2)/r^4$. Thus, under the assumption~\eqref{E2.3.45} a tachyonic instability requires $\hat{\hat{f}}(0)>0 \ze$. Let us study this instability, generalizing the analysis in the previous Sec.~\ref{S2.2} for the dyonic RN case.

	Assuming separation of variables,
			\begin{equation}\label{E2.3.44}
			 \phi= U_\ell(r) Y_{\ell}^m (\theta,\varphi) \ ,
			\end{equation}
	the equation for the radial amplitude $U_\ell$ is \eqref{E1.2.17} with $ \mu^2_{\rm eff}=(P ^2- Q_e^ 2)\hat{\hat{f}}_i(0)/r^2$. For spherically symmetric perturbations $\ell=0\ze$, \eqref{E1.2.17} possesses an exact solution which is regular on and outside the horizon and vanishes at infinity
			\begin{equation}\label{E2.3.48}
			 U_\ell=P_u \left(1+\frac{2(Q_e^2-P^2)(r-r_H)}{r_H^2+P^2-Q_e^2} \right)\ , \qquad {\rm where} \ \ u=\frac{1}{2}\left(\sqrt{1-2\hat{\hat{f}}_i (0)}-1\right) \ ,
			\end{equation}
	For generic parameters $(\hat{\hat{f}}_i(0),\ze Q_e,\ze P,\ze r_H)$, the function $U_\ell$ approaches a constant \textit{non-zero} value as $r\to +\infty$,
			\begin{equation}\label{E2.3.49}
			 U_\ell \to U_\infty= {}_2F_1 \left[\frac{1-\sqrt{1-2\hat{\hat{f}}_ i(0)}}{2},\, \frac{1+\sqrt{1-2\hat{\hat{f}}_i (0)}}{2},\, 1;\, \frac{Q_e^2-P^2}{Q_e^2-P^2-r_H^2} \right]+\mathcal{O}\left(\frac{1}{r}\right)\ .
			\end{equation}
	Thus, finding the $\ell=0$ unstable mode of a RN BH with given $Q_e,\, P,\, M$ reduces, again, to a study of the hypergeometric function ${}_2F_1 $ roots, so that $U_\infty=0\ze $.

	The value of $U_\infty$ for $f_E$ and an illustrative value of $\alpha$ is shown in Fig.~\ref{F1.1} (left panel). To simplify the picture, the results in Fig.~\ref{F1.1} correspond to $P=0$;  a similar pattern holds also in the dyonic case.

	The solution \eqref{E2.3.48} yields a dyonic RN BH surrounded by a vanishingly small scalar field. The set of such RN BHs (varying $\alpha$) constitute the {\it dyonic existence line}.
%
		\subsection{Non-extremal black holes}\label{S2.3.2}
%
	Let us now construct, numerically, the non-linear BH solutions for both class \textbf{I} and \textbf{II.A}, starting with the non-extremal BHs.
	
	The equations of motion \eqref{E2.1.3}-(2.1.4), together with the first integral \eqref{E2.1.5} implies that the functions $m$, $\sigma,\phi$ solve the ordinary differential equations
			\begin{align}\label{E2.3.50}
			 &m'=\frac{1}{2} r^2 N \phi'^{\, 2}+\frac{1}{2r^2}\left(\frac{Q_e^2}{f_i}+f_i P^2 \right) \ ,\qquad \sigma'=-\sigma r \phi'^{\, 2}\ ,\nonumber\\ 
			& \big( \sigma r^2 N\phi'\big) '=-\frac{ \sigma}{2\ze r^2 f_i} \hat{f_i} \left(\frac{Q_e^2}{f_i}-f_i P^2 \right)=0 \ ,
			\end{align}
	which can also be derived from the effective Lagrangian:
			\begin{equation}\label{E2.3.51}
			 \mathcal{L}^{\rm eff} _{EMS}=\sigma m'-\frac{1}{2}\sigma r^2 N \phi'^{\, 2}+\frac{f_i}{2\ze \sigma}\left(r^2 V'^{\, 2}-\frac{\sigma ^2}{r^2}P^2\right) \ ,
			\end{equation}
	while $V'= Q_e /(r^2\varepsilon _\phi)$. To assess possible singular behaviours we remark that the expression of the Ricci and  Kretschmann scalars are only metric dependent and thus given by \eqref{E2.1.9}. 

\bigskip

	Close to the horizon, the solutions can be expanded as \eqref{E2.1.7}, with
			\begin{align}\label{E2.3.52}
			 & m_1=\frac{1}{2r_H^2} \left[\frac{Q_e^2}{f_i(\phi_0)} + f_i(\phi_0)P^2 \right] \ , \qquad \phi_1=\hat{f}_i(\phi_0)\frac{1}{2r_H} \frac{\frac{Q_e^2}{f_i(\phi_0)} - f_i(\phi_0)P^2 }{\frac{Q_e^2}{f_i(\phi_0)} + f_i(\phi_0)P^2 -r_H^2} \ , \nonumber\\
			 & \sigma_1=-r_H \sigma _0 \phi_1^2\ , \qquad {\rm v}_1=\frac{\sigma_0 Q}{r_H^2 f_i(\phi_0)} \ .
			\end{align}
	For large $r$, one finds the following asymptotic expansions:
			\begin{align}\label{E2.3.53}
			 & m(r)= M-\frac{Q_e^2+P^2+Q_\phi^2}{2\ze r}+\cdots\ ,\qquad \phi(r)=\frac{Q_\phi}{r}+\cdots\ ,\nonumber\\
			 & V(r)=\Psi_e-\frac{Q_e}{r}+\cdots\ ,\qquad \sigma(r) = 1+\frac{Q_\phi ^2}{2\ze r^2}+\cdots\ .
			\end{align}
	With one new essential parameter introduced in the expansion at infinity \eqref{E2.3.53}: the magnetic charge $P$.

	At the horizon the Ricci scalar vanishes while the Kretschmann scalar reads
			\begin{equation}\label{E2.3.54}
			 K=\frac{12}{r_H^4} \left\{ 1-\frac{2}{r_H^2}\left[\frac{Q_e^2}{f_i}+f_i P^2\right]+\frac{5}{3\ze r_H^4}\left[\frac{Q_e^2}{f_i}+f_i P^2\right]^2 \right\}+\mathcal{O}(r-r_H) \ .
			\end{equation}
	Both horizon physical quantities remain as $T_H=\frac{1}{4\pi}N'(r_H)\sigma_0$ and $A_H=4\pi r_H^2$. These, together with the horizon scalar field value $\phi_0$ compose the relevant horizon data.

	The Smarr-like relation~\cite{smarr1973mass} for this family of models turns out to have no \textit{explicit} imprint of the scalar hair,
			\begin{equation}\label{E2.3.55}
			 M=\frac{1}{2} T_H A_H+\Psi_e Q_e +\Psi_m P \ ,
			\end{equation}
	where we have defined a `magnetic' potential as $\Psi_m\equiv \int_{r_H}^{+\infty} dr \, \sigma f_i Pr^{-2}$.

	One can then compute the first law of BH thermodynamics for EMS BHs, that reads:
			\begin{equation}\label{E2.3.56}
			 dM=\frac{1}{4}T_H dA_H+\Psi_e dQ_e +\Psi_m dP \ .
			\end{equation}

	A non-linear Smarr relation ($i.e.$ mass formula) can also be established for this family of models,
			\begin{equation}\label{E2.3.57}
			 M^2+Q_\phi^2=Q_e^2+P^2+\frac{1}{4}A_H^2 T_H^2 \ ,
			\end{equation}

	Finally, the solutions satisfy the virial identity
			\begin{equation}\label{E2.3.58} 
			 \int_{r_H}^{+\infty} dr \left\{\sigma  \phi'^{\, 2} r^2 \left[1+\frac{2r_H}{r}\left(\frac{m}{r}-1\right) 
\right]\right\}=\int_{r_H}^{+\infty} dr \left\{\sigma\left(1-\frac{2r_H}{r}\right)\frac{1}{r^2}\left[\frac{Q_e^2}{f_i}+f_i P^2\right]\right\} \ .
			\end{equation}
	One can show that $1+\frac{2r_H}{r}(\frac{m}{r}-1) >0$, $i.e.$ the \textit{lhs} integrand, is strictly positive. Thus, the virial identity shows that a non-trivial scalar field requires a non-zero electric/magnetic charge so that the \textit{rhs} is non-zero.

	The model possesses the scaling symmetry
			\begin{equation}\label{E2.3.59a}
			 r \to \lambda\ze r \ , \qquad (Q_e,\, P)\to \lambda(Q_e,\, P) \ ,
			\end{equation}
	where $\lambda>0$ is a constant. Under this scaling symmetry, all other quantities change accordingly, $e.g.$, $M \to \lambda M$, while the coupling function $f$ is unchanged. Thus, for a physical discussion, we consider quantities that are invariant under the transformation \eqref{E2.3.59a}. Consequently, we introduce the new charge to mass ratio
			\begin{equation}\label{E2.3.60}
			 q\equiv \frac{\sqrt{Q_e^2+P^2}}{M}\ .
			\end{equation}
	Observe that dyonic RN BHs have closed expressions for $a_H,\, t_H$:
			\begin{equation}\label{E2.3.61}
			 a_H^{(\rm RN)}=\ \frac{1}{4}\big( 1+\sqrt{1-q^2}\,\big )^2\ , \qquad t_H^{(\rm RN)}=\frac{4\sqrt{1-q^2}}{\big(1+\sqrt{1-q^2}\,\big)^2} \ .
			\end{equation}
	In Appendix~\ref{C} we exhibit the corresponding expressions for other dilatonic BHs known in closed analytic form, which are class \textbf{I} solutions. The generic dilatonic dyonic solutions are not known in closed form, which holds for all scalarized BHs. These solutions are, again, found numerically.
			\subsubsection*{The BH solutions}
\begin{figure}[h!]
					\begin{center} 
					\begin{picture}(0,0)
		  		 		\put(30,130){\small$-500 \ln \sigma$}
				  		 \put(120,62){\small$V$}
				  		 \put(55,23){\small$\phi$}		  		 
		  				 \put(100,128){\small$m$}
				  		 \put(101,-10){\small $\log _{10} r$}
				  		 \put(150,35){$\scriptstyle {\rm P\,=\, 0.5 \ \, Q_e\, = \, 1.0 } $}
				  		 \put(155,22){$\scriptstyle {\rm \alpha\,=\, 4 \ \, r_H\, = \, 1.3 } $}
				  		\put(25,50){\begin{turn}{90}{$\scriptstyle{\rm Event\ horizon}$}\end{turn}}
	   			\end{picture}
					 \includegraphics[height=.33\textwidth, angle =0 ]{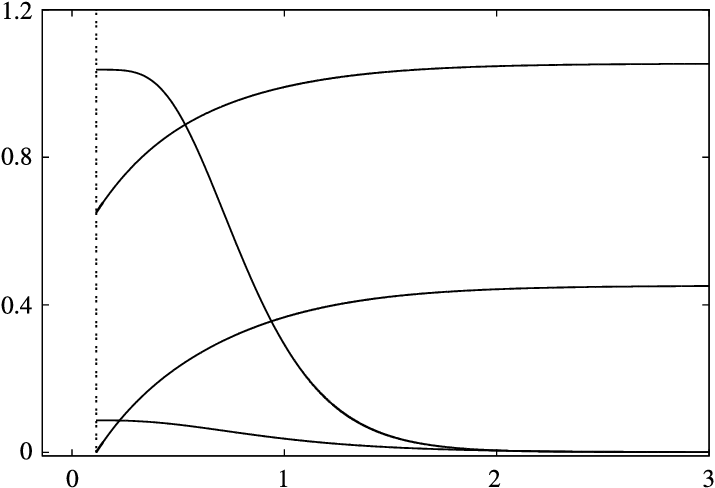}
					 \includegraphics[height=.33\textwidth, angle =0 ]{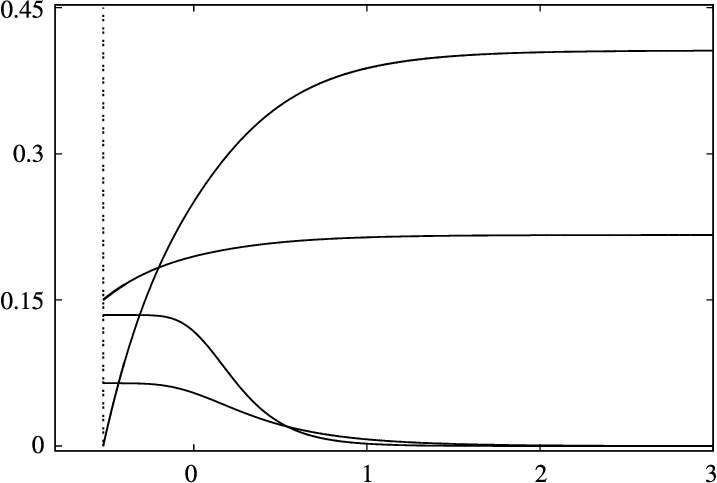}
					\begin{picture}(0,0)
		  		 		\put(-177,13){\small$-100 \ln \sigma$}
				  		 \put(-120,76){\small$m$}
				  		 \put(-152,40){\small$\phi$}		  		 
		  				 \put(-100,129){\small$V$}
				  		 \put(-101,-10){\small $\log _{10} r$}
				  		 \put(-100,35){$\scriptstyle {\rm P\,=\, 0.105 \ \, Q_e\, = \, 0.182 } $}
				  		 \put(-90,22){$\scriptstyle {\rm \alpha\,=\, 66 \ \, r_H\, = \, 0.3 } $}
				  		\put(-196,50){\begin{turn}{90}{$\scriptstyle{\rm Event\ horizon}$}\end{turn}}
	   			\end{picture}
					\end{center}
				 \caption{Examples of dyonic BHs radial profile functions for a dilatonic (left panel) and a scalarized (right panel) BH.}
				 \label{F2.20}
				\end{figure}
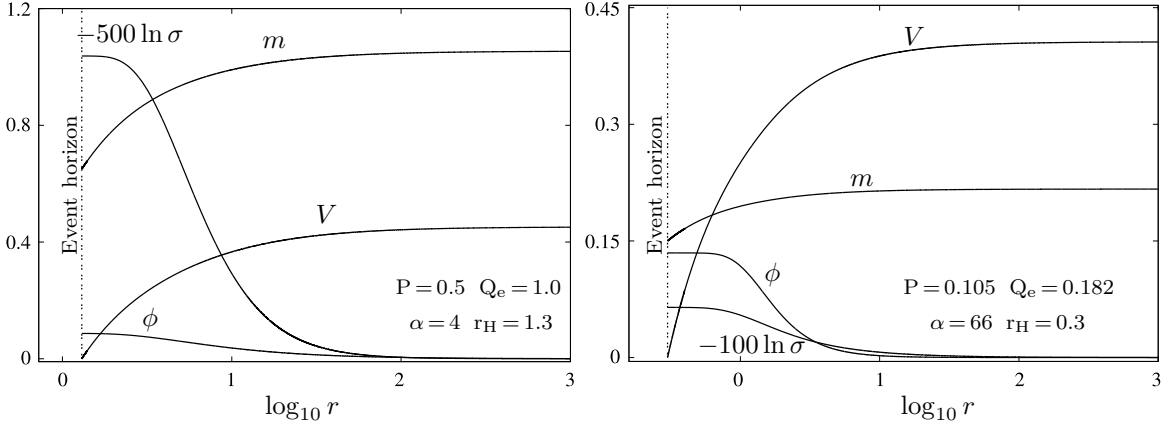

	The profile functions of illustrative dyonic BHs are shown in Fig.~\ref{F2.20}, for both the dilatonic and scalarized cases.

\begin{figure}[H]
			\centering
				\begin{picture}(0,0)
						\put(75,152){\small {\bf Class \textbf{I}} (dilatonic)}
				  		 \put(85,62){\small$\alpha =0$}		
				  		 \put(115,30){\small$\alpha =1$}		  		 
		  				 \put(100,118){\small$\alpha = 3$}
		  			     \put(175,130){$\scriptstyle {\rm \frac{P}{Q_e}\, =\, 0.1}$}
		  			     \put(125,11){$\scriptstyle {\rm Extremal}$}
		  				 \put(152,60){\begin{turn}{-60}{\small $\alpha = \sqrt{3}$}\end{turn}}
				  		 \put(116,-10){\small $q$}
				  		\put(-5,73){\begin{turn}{90}{$a_H$}\end{turn}}
	   			\end{picture}
				 \includegraphics[height=.34\textwidth, angle =0 ]{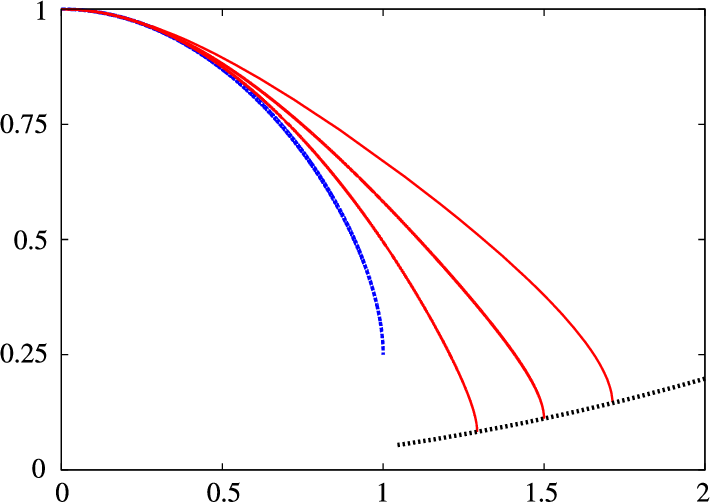}\hfill
				 	\begin{picture}(0,0)
						\put(75,152){\small {\bf Class \textbf{II.A}} (scalarized)}
				  		 \put(78,74){\small$\alpha =0$}		
				  		 \put(98,30){\small$\alpha =1$}		  		 
		  				 \put(100,118){\small$\alpha = 15$}
		  			     \put(172,130){$\scriptstyle {\rm \frac{P}{Q_e}\, =\, 0.1}$}
		  				 \put(144,60){\begin{turn}{-50}{\small $\alpha = 5$}\end{turn}}
		  				 \put(125,12){$\scriptstyle {\rm Extremal}$}
				  		 \put(116,-10){\small $q$}
				  		\put(-5,73){\begin{turn}{90}{$a_H$}\end{turn}}
	   			\end{picture}
				 \includegraphics[height=.34\textwidth, angle =0 ]{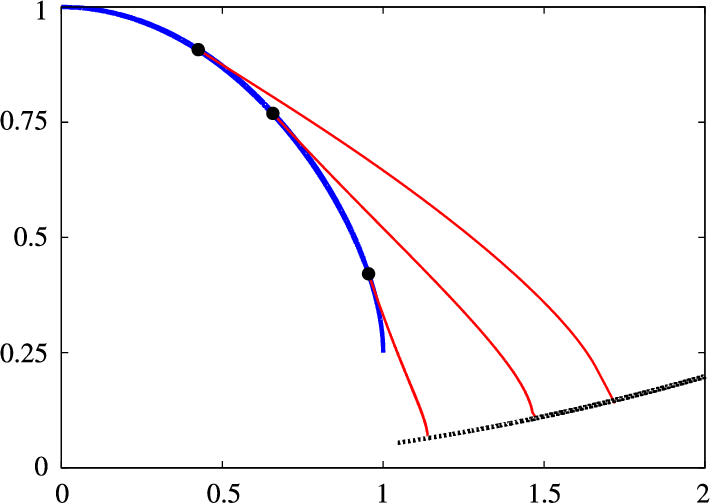}\vspace{6mm}\\
				 		\begin{picture}(0,0)
				  		 \put(85,74){\small$\alpha =0$}		
				  		 \put(118,30){\small$\alpha =1$}		  		 
		  				 \put(181,110){\small$\alpha = 3$}
		  			     \put(176,134){$\scriptstyle {\rm \frac{P}{Q_e}\, =\, 0.1}$}
		  				 \put(166,60){\begin{turn}{-90}{\small $\alpha = \sqrt{3}$}\end{turn}}
				  		 \put(116,-10){\small $q$}
				  		\put(-8,73){\begin{turn}{90}{$t_H$}\end{turn}}
	   			\end{picture}
				 \includegraphics[height=.335\textwidth, angle =0 ]{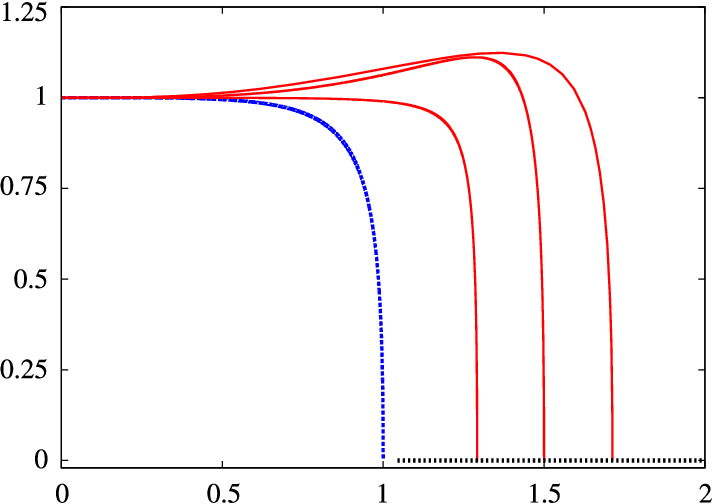}\hfill
				 		\begin{picture}(0,0)	
				  		 \put(87,30){\small$\alpha =0$}		  		 
		  				 \put(125,136){\small$\alpha = 5$}
		  				 \put(112,106){\small$\alpha = 1$}
		  				 \put(65,121){\small$\alpha = 15$}
		  			     \put(177,135){$\scriptstyle {\rm \frac{P}{Q_e}\, =\, 0.1}$}
				  		 \put(117,-10){\small $q$}
				  		\put(-5,72){\begin{turn}{90}{$t_H$}\end{turn}}
	   			\end{picture}
				 \includegraphics[height=.335\textwidth, angle =0 ]{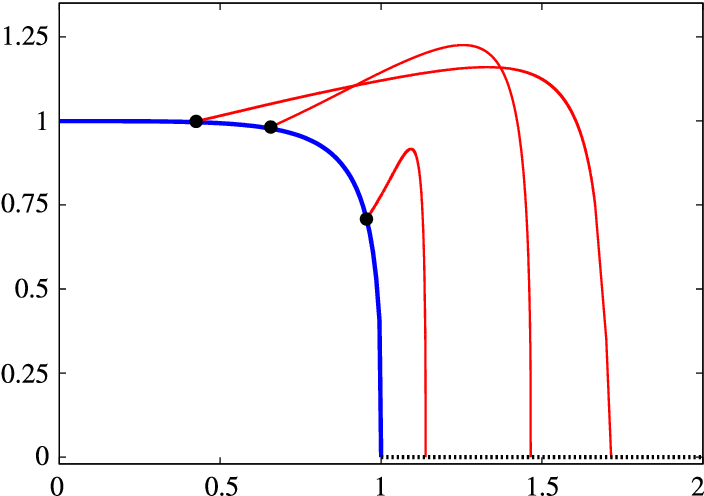}
			 \caption{Reduced area $a_H$ (top panels) and reduced temperature $t_H$ (bottom panels) $vs.$ charge to mass ratio $q$ for dilatonic, $f_D$ (left panels) and scalarized solutions with $f_E$ (right panels). All solutions have $P/Q_e=0.1\ze $. The blue lines are the set of RN BHs ($\phi = 0$). The red lines are sequences of BHs with a non-trivial scalar field for a given $\alpha$. Different sequences are presented, for a range of values of $\alpha$. The black dots indicate the RN solutions from which the scalarized BHs bifurcate. The dotted black line represents the extremal solutions.}
			 \label{F2.21}
			\end{figure}
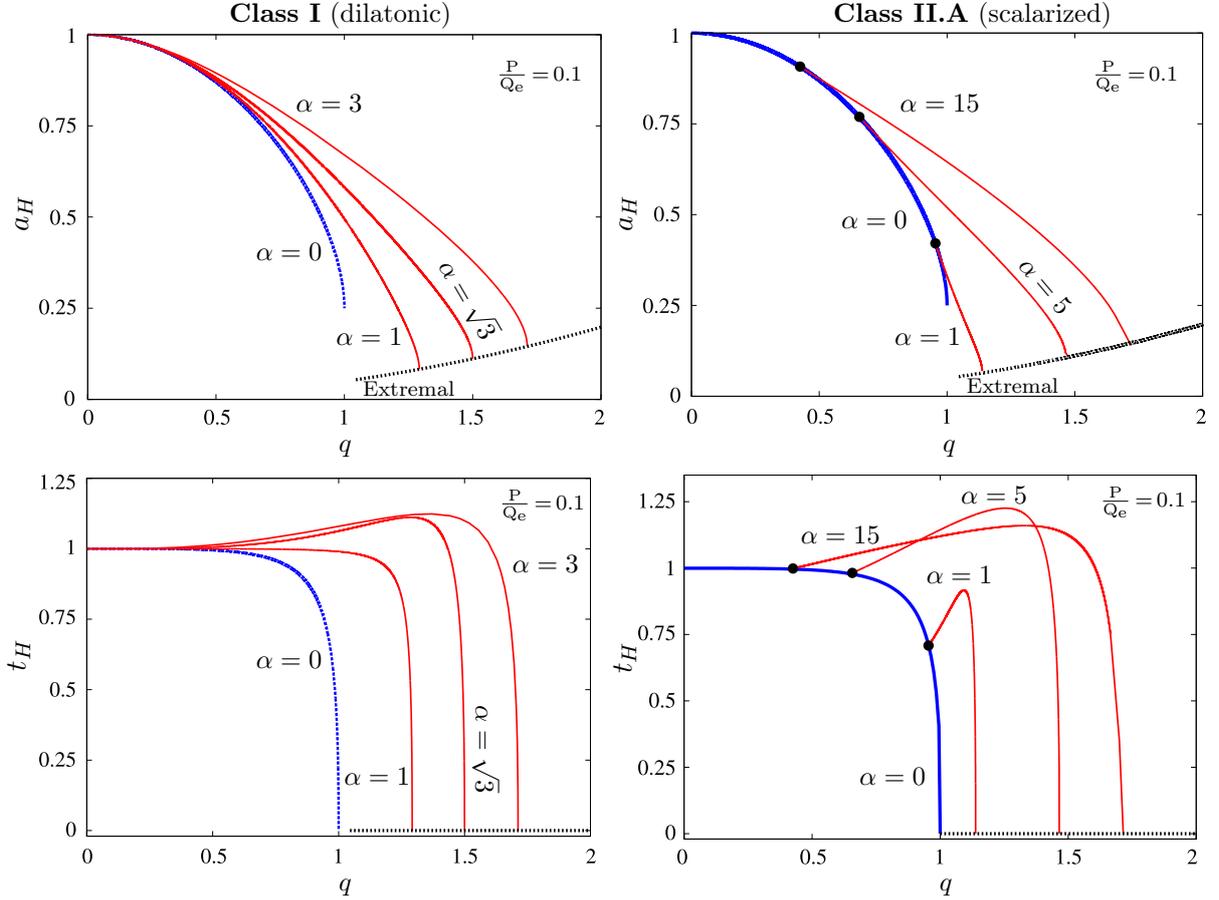

	Dyonic BHs preserve some, but not all, of the qualitative characteristics of the purely electric solutions (see Fig.~\ref{F2.1} and Fig.~\ref{F2.2}). In the dilatonic case, the branch of solutions with a given $\alpha$  starts again from the Schwarzschild limiting solution (which has $a_H=1\ze$, $t_H=1$ and $q=0$) and ends in a limiting configuration with  $a_H>0\ze $, $t_H=0$ and $q=q_{\rm max}>0$ -- Fig.~\ref{F2.21} (left panels). This limiting solution, however, is now an extremal BH (rather than a singular critical solution) and will be discussed in Sec.~\ref{S2.3.3}.	
	Unlike the dilatonic solutions, which exist for arbitrarily small $q$ for any $\alpha$,  scalarized BHs 
 with a given $\alpha$ exist for $q>q_{min}$ only. They bifurcate from a RN BH (with $q>0$) and form a branch ending again on an extremal solution with vanishing horizon temperature and non-zero horizon area -- Fig.~\ref{F2.21} (right panels). As for purely electric solutions, for the same global charges $M,\, Q_e,\, P$,  
the scalarized solutions are entropically preferred over the corresponding RN solution.

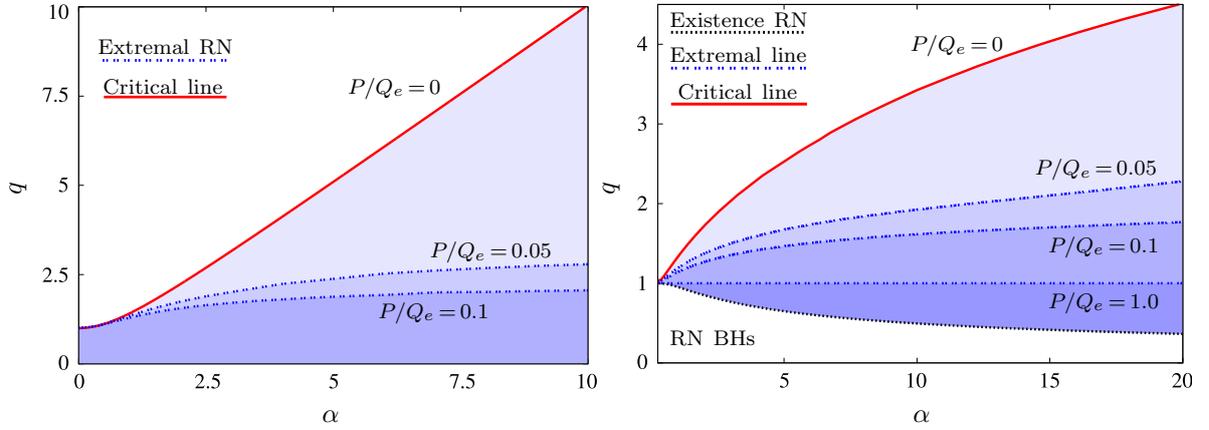
\begin{figure}[H]
				\centering
				\begin{picture}(0,0)
		  		 \put(28,130){$\scriptstyle {\rm Extremal\ RN}$}
		  		 \put(30,115){$\scriptstyle {\rm Critical\ line}$}
		  		 \put(118,115){ $\scriptstyle P/Q_e\, =\, 0$}
		  		 \put(112,-9){\small $\alpha$}
		  		\put(-5,77){\begin{turn}{90}{\small $q$}\end{turn}}
	   			\end{picture}
	   				 	\begin{tikzpicture}[scale=0.5]
\node at (0,0) {\includegraphics[height=.33\textwidth, angle =0 ]{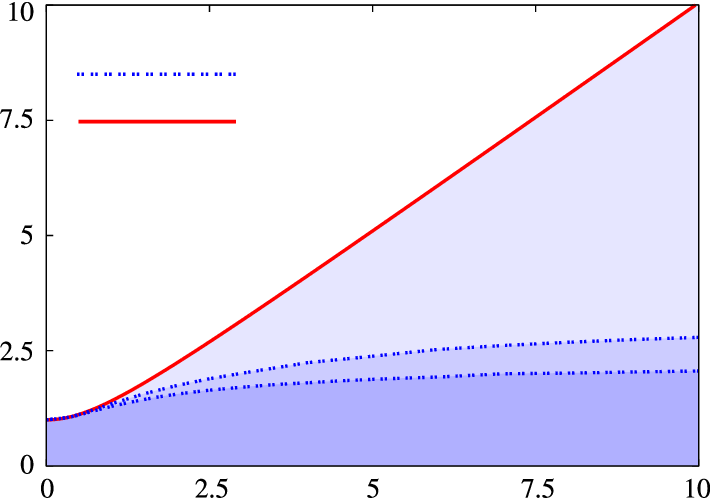}};
\node at (4.5,-1.5) { $\scriptstyle  P/Q_e\, = \, 0.05$};
\node at (3,-3.1) {$\scriptstyle  P/Q_e\, =\, 0.1$};
				\end{tikzpicture}
				\begin{picture}(0,0)
		  		 \put(20,140){$\scriptstyle {\rm Existence\ RN}$}
		  		 \put(20,127){$\scriptstyle {\rm Extremal\ line}$}
		  		 \put(23,113){$\scriptstyle {\rm Critical\ line}$}
		  		 \put(20,20){$\scriptstyle {\rm RN\ BHs}$}
		  		 \put(110,131){$\scriptstyle P/Q_e\, =\, 0$}
		  		 \put(111,-9){\small $\alpha$}
		  		\put(-5,77){\begin{turn}{90}{\small $q$}\end{turn}}
	   			\end{picture}
	   				 \begin{tikzpicture}[scale=0.5]
\node at (0,0) {\includegraphics[height=.33\textwidth, angle =0 ]{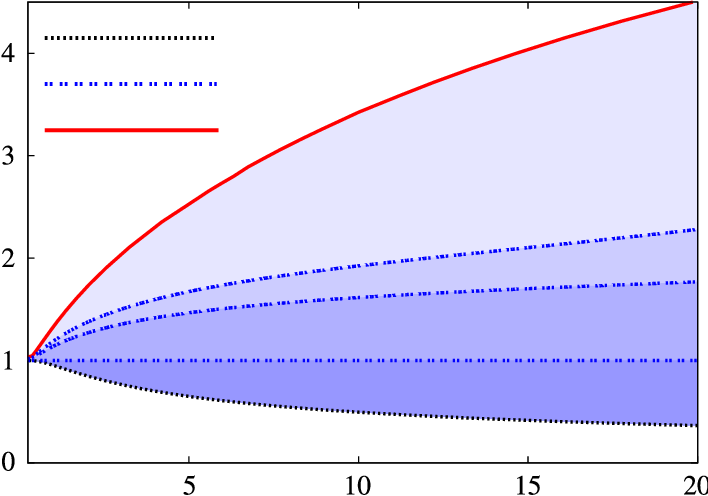}};
\node at (5,-1.35) { $\scriptstyle P/Q_e\, =\, 0.1$};
\node at (4.8,0.7){$\scriptstyle P/Q_e\, = \, 0.05$};
\node at (5,-2.83) {$\scriptstyle P/Q_e \,  =\, 1.0$};
				\end{tikzpicture}
	 		 \caption{Domain of existence of dilatonic BHs (left panel) and scalarized BHs (right panel) for several values of the ratio $P/Q_e$.}
			 \label{F2.22a}
			\end{figure}

	The domain of existence of dyonic BHs is shown in Fig.~\ref{F2.22a} for several values of the ratio $P/Q_e$ and both dilatonic and scalarized BHs. In particular, observe that in both cases, the maximal allowed value of $q$ for BHs with a given $\alpha$ decreases as the ratio $P/Q_e$ increases. In other words, the domain of existence shrinks as the magnetic charge increases, for fixed $Q_e$.
%

%
		\subsection{Extremal black holes}\label{S2.3.3}
%
	To address extremal BHs one needs to impose a different near-horizon expansion to that in~\eqref{E2.1.7}, which accounts for the degenerate horizon. The leading order terms of the appropriate expansion are:
			\begin{align}\label{E2.3.59}
			 & N(r)= N_2(r-r_H)^2+\cdots\ , \qquad \sigma(r)=\sigma_0+\sigma_1 (r-r_H)^{2k-1}+\cdots \ , \nonumber\\
			 & \phi(r)=\phi_0+\phi_c (r-r_H)^{k}+\cdots \ , \qquad V(r)= {\rm v}_1(r-r_H)+\cdots \ .
			\end{align}
	One can show that the next to leading order term  in the expression of $N$ is $\mathcal{O}(r-r_H)^3$. It is convenient to take $r_H$ and $\phi_0$ as essential parameters. Then the field equations imply
			\begin{equation}\label{E2.3.60}
			 Q_e =\frac{r_H\sqrt{f_i(\phi_0)}}{\sqrt{2}}\ , \qquad  P=\frac{r_H}{\sqrt{2f_i(\phi_0)}} \ , \qquad N_2=\frac{1}{r_H^2} \ .
			\end{equation}
	Consequently, given an expression of the coupling function $f_i$, one can express the value of the scalar field at the horizon $\phi_0$ as a function of $Q_e$ and $P$ by solving the equation
			\begin{equation}\label{E2.3.61}
			 f_i(\phi_0)=\frac{Q_e}{P}\ , \qquad {\rm while}\qquad r_H=\sqrt{2\ze P\ze Q_e}\ .
			\end{equation}

	The expansion \eqref{E2.3.59} contains two free parameter $\phi_c$ and $\sigma_0\ze$, which are fixed by numerics, while $\sigma_1,\, {\rm v}_1$ come as
			\begin{equation}\label{E2.3.62}
			 \sigma_1=-\frac{r_H\ze \phi_c^2\ze k^2}{2k-1}\ , \qquad  {\rm v}_1=\frac{\sigma\ze Q_e}{r_H^2 f_i(\phi_0)} \ .
			\end{equation}
	The power $k$ in \eqref{E2.3.59} is given by
			\begin{equation}\label{E2.3.63}
			 k=\frac{1}{2}\left[-1+\sqrt{1+2 \left(\frac{\hat{f}_i(\phi_0)}{f_i(\phi_0)} \right)^2} \ \right]>0 \ ,
			\end{equation}
	which, generically, takes non-integer values. However, a non-integer $k$ implies that a sufficiently higher-order derivative of the curvature tensor will diverge as $r\to r_H$. A minimal requirement for smoothness is that the metric functions $N,\, \sigma$ and their first and second derivatives are finite as $r\to r_H$; this yields the condition
			\begin{equation}\label{E2.3.64}
			 k >3/2 \ .
			\end{equation} 
	On the other hand, for analytic solutions on the horizon (as extremal RN), the power $k$ in the above 
near horizon expansion \eqref{E2.3.59} should be an integer. This imposes the condition 
			\begin{equation}\label{E2.3.65}
			 \frac{\hat{f}_i(\phi_0)}{f_i(\phi_0)}= \pm \sqrt{2\ze p\ze (p +1)}\ , \qquad {\rm with}~~p=1,\, 2,\dots \ .
			\end{equation}
	For the dilatonic case, condition~\eqref{E2.3.65} translates to \cite{gal2015triangular,zadora2018higher} (see also \cite{abishev2015dilatonic})
			\begin{equation}\label{E2.3.66}
			 \alpha=\sqrt{\frac{p\ze (p+1)}{2}} \ ,
			\end{equation}
	again with an integer $p$. For scalarized solutions with the coupling function $f_E$ the condition \eqref{E2.3.65} becomes
			\begin{equation}\label{E2.3.67}
			 \alpha= \frac{p\ze (p+1)}{4 \ln \frac{Q_e}{P}}\ .
			\end{equation}
	The extremal solutions share the far field asymptotics \eqref{E2.3.59} with the non-extremal ones; moreover, the relations \eqref{E2.3.52}-\eqref{E2.3.55} hold also for $T_H=0\ze$.

\begin{figure}[H]
				\centering
					 \begin{picture}(0,0)
		  		 \put(125,-10){\small $1-\frac{r_H}{r}$}
		  		 \put(100,14){\small $n=1$}
		  		 \put(137,75){\small $n=0$}
		  		 \put(68,42){\small $n=2$}
		  		 \put(135,145){$\scriptstyle {\rm P\, =\, 0.3\ \, Q_e \, = \, 0.35\ \, \alpha \, = \, 5}$}
		  		\put(35,60){\begin{turn}{90}{$\scriptstyle{\rm Event\ horizon}$}\end{turn}}
		  		\put(-8,85){\begin{turn}{90}{\small $\phi$}\end{turn}}
	   			\end{picture}
				 \includegraphics[scale=0.7]{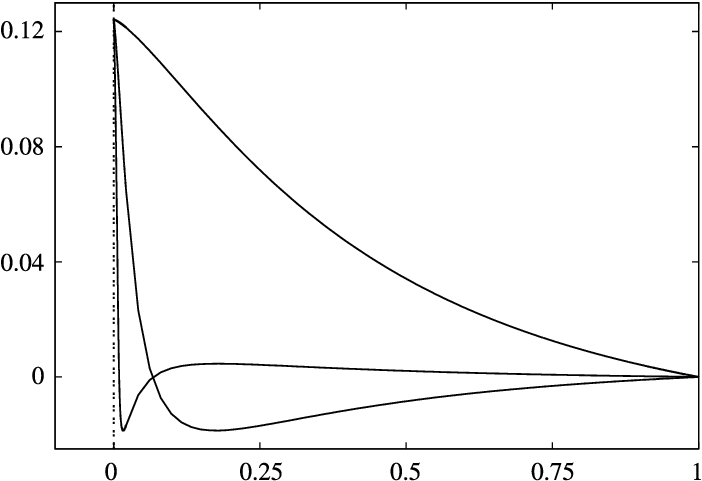}
			 \caption{A sequence of scalar field profiles starting with the same horizon data
in a scalarized model. Each solution possesses a different node number.}
			 \label{F2.23a}
			\end{figure}
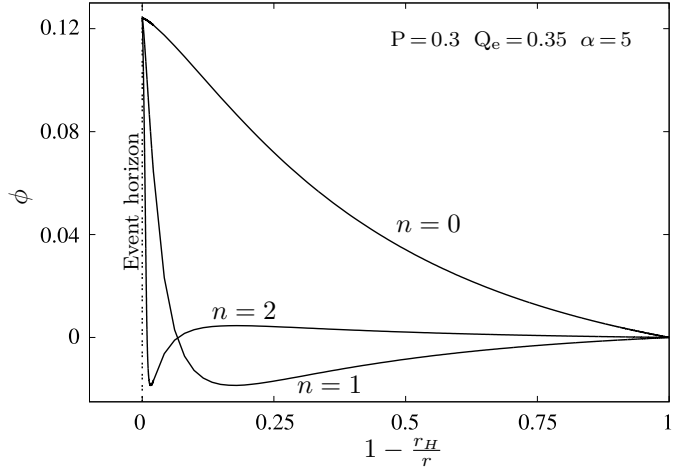

	The profile of the various functions resulting from the integration is not particularly enlightening, resembling those in the non-extremal case and shall not be shown here. However, we would like to point out a particular feature of the extremal scalarized BHs. There exists a (presumably) infinite family of solutions with the same horizon data as specified by $(\phi_0,\, r_H)$ $\big($or, equivalently, $(Q_e,\, P)\big)$, labelled by their node-number $n$. This is illustrated in Fig.~\ref{F2.23a}: the scalar field always starts at the same horizon value; however, the bulk profile is different. As expected for excited states, the mass of these solutions increases with $n$. We remark that no excited configurations were found in the dilatonic case, which always has $n=0\ze$.
%
		\subsection{An analytic approach: the attractor mechanism and entropy function}\label{S2.3.4}
%
	The numerical construction of the extremal BHs is a difficult numerical task. Let us now provide a different argument for the existence of the EMS extremal dyonic BHs: the existence of an exact solution describing a Robinson-Bertotti vacuum, namely an $AdS_2\times S^2$ spacetime. As for extremal RN BHs, we expect that this solution describes the neighbourhood of the event horizon of an extremal scalarized BH with non-zero magnetic and electric charges. As we shall see, both charges are mandatory for the Robinson-Bertotti vacuum to exist with a non-trivial scalar field.

	To search for the Robinson-Bertotti vacuum, we consider a new metric ansatz line element
			\begin{equation}\label{E2.3.68}
			 ds^2=v_0 (r) \left(-r^2dt^2+\frac{dr^2}{r^2}\right)+v_1(r)\big( d\theta^2+\sin^2\theta d\varphi^2\big) \ ,
			\end{equation}
	and the matter fields ansatze
			\begin{equation}\label{E2.3.69}
			 A_\mu =e\ze r dt+P\cos \theta\ze d \varphi\ , \qquad \phi=\phi_0 \ .
			\end{equation}
	The constant parameters $\{v_0,\, v_1;\, e,\, P,\, \phi_0\}$ satisfy a set of algebraic relations which result from the EMS equations \eqref{E1.2.2}-\eqref{E1.2.4}. However, instead of attempting to solve these, we shall, in what follows, determine these parameters by using the formalism proposed in\cite{sen2005black,astefanesei2006rotating,sen2008black}, thus by extremizing an {\it entropy function}.

	This approach also allows us to compute the BH entropy and show that the solutions exhibit an attractor-type behaviour. 
 
	Let  us consider the Lagrangian density of the model, as read off from \eqref{E2.1.1}, evaluated for the ansatz \eqref{E2.3.68}-\eqref{E2.3.69} and integrated over the angular coordinates,
			\begin{equation}\label{E2.3.70}
			 {\cal L} ^{\rm eff} _{EMS} = \frac{1}{2}\bigg[ v_0-v_1+f(\phi_0)\left(\frac{e^2 \ze v_1}{v_0}-\frac{P^2 \ze v_0}{v_1}\right)\bigg]\ .
			\end{equation}

	Then, following \cite{sen2005black,astefanesei2006rotating,sen2008black}, we define the entropy function ${\cal E}$ by taking the Legendre transform of the above integral with respect to a parameter $Q_e$,
			\begin{equation}\label{E2.3.71}
			 {\cal E}=  2\pi \big( e\ze Q_e -{\cal L}^{\rm eff} _{EMS} \big)\  ,
			\end{equation}
	where $Q_e={\cal E}_{,e}$ is the electric charge of the solutions. It follows as a consequence of the equations of motion that the constants $\{v_0,\, v_1\, ; e,\, \phi_0\}$ are solutions of the equations
			\begin{equation}\label{E2.3.72}
			 \frac{\partial {\cal E}}{\partial v_i}=0\,, \qquad \frac{\partial {\cal E}}{\partial \phi_0}=0\,,
  \qquad \frac{\partial {\cal E}}{\partial e }=0\, ,
			\end{equation}
	or, explicitly,
			\begin{align}\label{E2.3.73}
			 & \frac{\partial {\cal E}}{\partial v_0}  =  0\,\,\,\Rightarrow \,\,\	-1+\left(\frac{v_1}{v_0^2}\ze e^2+\frac{1}{v_1} P^2\right)f_i(\phi_0) =0 \ ,\\ \label{E2.3.74}
			 & \frac{\partial {\cal E}}{\partial v_1}  = 0\,\,\,\Rightarrow \,\,\, 1-\left(\frac{e^2}{v_0}+\frac{v_0}{v_1^2}\ze P^2\right)f_i(\phi_0)=0 \ ,\\\label{E2.3.75}
			 & \frac{\partial {\cal E}}{\partial \phi_0}  =  0\,\,\,\Rightarrow \,\,\, 	\big( P^2 v_0^2-e^2 v_1^2 \big)\hat{f}_i(\phi_0) = 0\ ,\\
			 & \frac{\partial {\cal E}}{\partial e}  = 0\,\,\,\Rightarrow \,\,\, Q_e = e\ze \frac{v_1}{v_0} f_i(\phi_0) \ . \label{E2.3.764}
			\end{align}
	The sum of~\eqref{E2.3.73} and \eqref{E2.3.74} leads to the generic relation 
			\begin{equation}\label{E2.3.77}
			 v_0=v_1\ .
			\end{equation}
	Thus, the ``radius'' of the $AdS_2$ is always equal with the one of $S^2$ in the metric \eqref{E2.3.68}. Then, the equation \eqref{E2.3.764} becomes 
			\begin{equation}\label{E2.3.78}
			 Q_e = e  f_i(\phi_0) \ .
			\end{equation} 

	Consequently, \eqref{E2.3.75} implies the existence of two different families of solutions:
			\begin{itemize}
			 \item[\textbf{a)}]  Eq.~\eqref{E2.3.75} is solved if $f_i(\phi)$ obeys $\hat{f}_i(\phi_0)=0$. Then, $e$ and $P$ are independent quantities and, from~\eqref{E2.3.73},
				\begin{equation}\label{E2.3.79}
				 v_0=v_1=\big( e^2+P^2\big) f_i(\phi_0) \ .
				\end{equation}
	This family of solutions is only possible in the scalarized case. In this case, $\hat{f}_i(\phi_0)=0$, with $\phi_0=0\ze$. Therefore, one obtains the near horizon geometry of the extremal RN BH, with a vanishing scalar field.

			\item[\textbf{b)}]  Eq.~\eqref{E2.3.75} is also solved if 
				\begin{equation}\label{E2.3.80}
				 e=P\ , \qquad \stackrel{\eqref{E2.3.764}}{\Rightarrow} \qquad Q_e= P f_i(\phi_0) \ ,
				\end{equation}
	and, from~\eqref{E2.3.73},
				\begin{equation}\label{E2.3.81}
				 v_0=v_1=2\ze P^2 f_i(\phi_0) \ .
				\end{equation}
	This family of solutions is possible for both the scalarized and dilatonic cases and demands both $Q_e,\, P$ to be non-vanishing. 
			\end{itemize}
	The scalarization mechanism is encoded in the existence of two different types of attractor solutions in the scalarized EMS models. This contrasts with the case of the dilatonic coupling, for which condition~\eqref{E2.3.80} is mandatory, and only one type of solution exists that requires both electric and magnetic charges to be present.

It is straightforward to check that in both cases, the entropy function, ${\cal E}$, evaluated at the critical attractor point is given by one-quarter of the area of angular sector in \eqref{E2.3.68},
			\begin{equation}\label{E2.3.82}
			 S=\pi v_1\ .
			\end{equation} 
	Finally, we remark that the correspondence of the above parameters with the ones in the near horizon expansion of the extremal BHs in Sec.~\ref{S2.3.3} is straightforward:
			\begin{equation}
			 v_1=r_H^2 \ , \qquad v_0=1/N_2 \ .
			\end{equation}
%
%
	\section{Axion EMS}\label{S2.4}
%
	In order to solve the strong CP problem, Peccei and Quinn introduced a pseudo-scalar known as the \textit{axion}\cite{peccei1977cp} (see also \cite{weinberg1978new,wilczek1978problem,kim2010axions}). It was later understood that besides offering a solution to the strong CP problem, the axion could have deep implications in cosmology, being a strong candidate for both cold and non-cold dark matter \cite{abbott1983cosmological,dine1983not,preskill1983cosmology,marsh2016axion}. It so happens that ultra-light axion fields arise naturally from compactifications in string theory \cite{conlon2006qcd,svrcek2006axions}. A series of experiments are being proposed and conducted in a quest to look for axionic imprints (see \cite{hook2018tasi,irastorza2018new,graham2015experimental}).
	
	The action that describes a real scalar field $\phi$, minimally coupled to Einstein's gravity, and non-minimally coupled to Maxwell's electromagnetism and to a Lorentz Chern-Simons term $\big($\textit{a.k.a.} Einstein-Maxwell-Axion (EMA)$\big)$ is\footnote{Observe that the axion is also a real scalar particle, however, to distinguish between the previous sections we will denote this model's scalar field as axionic field.}
		\begin{equation}
		 \mathcal{S}_{EMA}= \frac{1}{4}\int d ^ { 4 } x \sqrt { - g } \left[R - 2 \ze \phi_{,\mu} \phi ^{,\mu} - f(\phi)\ze F_{\mu \nu} F^{\mu \nu} - h ( \phi )\ze F_{\mu \nu} \Tilde{F}^{\mu \nu} \right],
		\end{equation}
	where $F_{\mu \nu}$ is the usual Maxwell tensor, and $\Tilde{F}^{\mu \nu}=\frac{1}{\sqrt{-g}}\epsilon^{\mu \nu \rho \sigma} F_{\rho \sigma}$ is its dual. The functions $h(\phi)$ and $f(\phi)$ are responsible for the non-minimal couplings between the scalar field and the source terms. Cases for which $h(\phi)=0$ are well studied in the previous Sec.~\ref{S2.1}-\ref{S2.3} (see also \cite{astefanesei2019einstein,herdeiro2018spontaneous,fernandes2019spontaneous}), we shall focus on the new cases for which $f(\phi)=1$ and $h(\phi) \neq 0\ze $. As usual, we will use the generic, spherically symmetric line element \eqref{E1.5.40}. Spherical symmetry and the presence of a magnetic charge impose a $4$-potential
		\begin{equation}
		 A_\mu =V(r) dt - P \cos{\theta} d\varphi,
		\end{equation}
	with $P$ the magnetic charge, and a solely radial dependent scalar field $\phi(r)$. The resulting effective Lagrangian from which the equations of motion may be derived comes as
		\begin{equation}\label{E2.4.90}
	     \mathcal{L}^{\rm eff} _{EMA}=-\frac{\sigma P^2\ze f}{2\ze r^2} + P\ze h\ze V' + \frac{1}{2\ze \sigma} r^2 f V'^{\, 2} + \frac{\sigma}{2} \left(1-N-rN'-r^2 N \phi'^{\, 2} \right),
		\end{equation}
	The solution is again under a first integral
		\begin{equation}
    	 V'=-\frac{Q_e+P\ze h}{r^2 f}\ze \sigma\ ,
		\end{equation}
	The equations of motion are
		\begin{align}\label{E2.4.92}
		 & \sigma '=-r\sigma \phi'^{\, 2}\ ,\\
		 & \left(\sigma r^2 N \phi' \right)' = -\frac{1}{2\sigma} r^2 \hat{f} V'^{\, 2} - P \hat{h} V' + \frac{\sigma P^2 \hat{f}}{2\ze r^2}\ ,\nonumber\\
		 & N'=-\frac{\frac{(P\ze h+Q_e)^2}{f}+P^2 f+r^2 \left(r^2 N \phi'^{\, 2}+N-1\right)}{r^3}\ .
		\end{align}
	To solve the set of ODEs we have to implement suitable boundary conditions for the desired functions and corresponding derivatives. We assume the existence of an event horizon at $r=r_H>0$ and that the solution possesses a power series expansion \eqref{E2.1.7}
		\begin{align}
   		& N  = -\frac{\big( Q_e+P\ze h(\phi_0)\big)^2 + P^2 f(\phi_0)^2 - r_H^2 f(\phi_0)}{r_H^3 f(\phi_0)} \left( r - r _ { H } \right) + \cdots \ ,\\
   		& \sigma = \sigma _ { 0 } -\sigma_0 \ze \phi _1 ^{\ze \ze 2}\ze \ze r_H\left( r - r _ { H } \right) + \cdots \ , \\ 
   		& \phi = \phi _ { 0 } -\frac{2P\big( Q_e+P\ze h(\phi_0)\big) f(\phi_0)\ze \hat{h}(\phi_0) + \hat{f}(\phi_0) \left[P^2f(\phi_0) - \big( Q_e+P\ze h(\phi_0)\ze)^2 \right]}{2r_H f(\phi_0) \left[\big( Q_e+P\ze h(\phi_0) \big)^2 - r_H^2f(\phi_0) + P^2f(\phi_0)^2 \right]} \left( r - r _ { H } \right) + \cdots\ ,\\
   		& V = -\frac{Q_e+P\ze h(\phi_0)}{r_H^2 f(\phi_0)}\sigma_0 \left( r - r _ { H } \right) + \cdots\ . 
		\end{align}
	which are fully determined via the two essential parameters $\phi_0$ and $\sigma_0\ze $. Also, of interest at the horizon, we have the Hawking temperature $T_H$, horizon area $A_H$, the energy density $\rho(r_H)$ and the Kretschmann scalar $K(r_H)$ given respectively by 
		\begin{align}
		 T_H & =\frac{1}{4\pi} N_1 \sigma _0\ ,\qquad	A_H=4\pi r_H^2\ , \qquad \rho(r_H) = 2\frac{ P^2 f(\phi_0)^2 + \big( Q_e+P h(\phi_0)\big)^2}{r_H^4 f(\phi_0)}\ ,\nonumber\\
		 K(r_H) & =\frac{4}{r_H^4}\left[ 3 - \frac{6Q_e^2}{r_H^2 f(\phi_0)} + \frac{5Q_e ^4}{r_H^4 f(\phi_0)^2} + \frac{10 P^2 Q_e ^2}{r_H^4} - \frac{6P^2 f(\phi_0)}{r_H^2} +\frac{5P^4 f(\phi_0)^2}{r_H^4} \right.\nonumber \\
		& \left. + \frac{10 P^4 h(\phi_0)}{r_H^4} - \frac{6P^2 h(\phi_0)^2}{r_H^2 f(\phi_0)} + \frac{5P^4 h(\phi_0)^4}{r_H^4 f(\phi_0)^2} + \right.\nonumber \\ 
		& \left. \frac{20PQ_e h(\phi_0)}{r_H^4 f(\phi_0)}\left( P^2 h(\phi_0) + \frac{Q_e^2 + P^2 h(\phi_0)^2 + 3 P Q_e h(\phi_0)}{f(\phi_0)} - \frac{3}{5}r_H^2 \right) \right]\ .
		\end{align}
	The asymptotic approximation of the solution in the far field takes the form
		\begin{align}
	 	 & N =1 - \frac{2M}{r} + \frac { Q_e ^ { 2 } + P^2 + Q _ { \phi } ^ { 2 } } { r^2 } + \cdots\ ,\qquad \phi = \frac { Q _ { \phi } } { r } + \frac{M Q_\phi}{r^2} + \cdots\ ,\nonumber \\
		 & V= \Psi_e + \frac { Q_e } { r } + \cdots\ , \qquad \sigma  = 1 + \frac { Q _ { \phi } ^ { 2 } } { 2 r ^ { 2 } } + \cdots\ ,\
		\end{align}
	The following results and definitions for our model will be useful later
		\begin{equation}\label{E2.4.99}
	     F_{\mu \nu}F^{\mu \nu} = \frac{Pf^2-\big(Q_e+P\ze h\big)^2}{f^2 \ze r^4}\ , \qquad F_{\mu \nu}\Tilde{F}^{\mu \nu} = \frac{P\big( Q_e+P\ze h\big)}{r^4 f}\ ,\qquad  q=\frac{\sqrt{Q_e^2+P^2}}{M}\ , 
		\end{equation}
			\subsubsection*{Physical relations}
	The Smarr law can be obtained via the integral mass formula, that for our model reads
				\begin{equation}
			     M=\frac{1}{2}T_H A_H - \frac{1}{16\pi} \int_V \big( 2\ze T^b_a - \textbf{T} \delta^b_a\big) k^a d\Sigma_b\ ,
				\end{equation}

	where $k^a$ is the time-like translational killing vector, and $\textbf{T}$ the trace of the energy-momentum tensor. The energy-momentum tensor is
				\begin{equation}
			     T_{\mu \nu} = \left[ f \left( F_{\mu \alpha} F_{\nu}^\alpha - \frac{1}{4} g_{\mu \nu} F_{\alpha \beta} F^{\alpha \beta} \right) +  \phi_{,\mu} \phi_{,\nu} - \frac{1}{2} g_{\mu \nu}\phi_{,\alpha}\phi^{,\alpha} \right]\ ,
				\end{equation}
	it turns out that the $F_{\mu \nu}\Tilde{F}^{\mu \nu}$ term (topological invariant) does not change $T_{\mu \nu}$. The resulting Smarr law
				\begin{equation}
			     M=\frac{1}{2}T_H A_H + \Psi_e Q_e + \Psi_m P + \Psi_A\ ,
				\end{equation}
	with
				\begin{equation}
				 \Psi_e = \int_{r_{H}}^{+\infty} dr\ \frac{Q_e+P\ze h}{r^2 f}\ze \sigma\ , \qquad \Psi_m =  \int_{r_{H}}^{+\infty} d r \  f \frac{P}{r^2}\ze \sigma \ ,
				\end{equation}
	the electrostatic and magnetostatic potential differences respectively. Let us introduce also the axion-like related electromagnetic energy
				\begin{equation}
    			 \Psi _A = \frac{1}{16 \pi} \int d^3 x \ze F_{\mu \nu} \Tilde{F}^{\mu \nu} f = - \int_{r_H}^{+\infty} dr\, V^\prime P \ze h \ .
				\end{equation}
	The solutions are also under a virial identity relation (see Chap.~\ref{C7} for further details)
				\begin{align}
				 \int _{r_H} ^{+\infty} dr\  \sigma\left[ \frac{2\ze r_H}{r}\left(1-\frac{m}{r}\right)-1 \right]r^2\phi'^{\, 2} = \int _{r_H} ^{+\infty} dr \ \frac{2\ze r_H -r}{r^3 \varepsilon _ \phi ^2 \ze\sigma }\left[ \big(Q_e -P \ze h \big)^2+P^2 \varepsilon _\phi ^2 \ze\sigma ^2 \right]\ .
				\end{align}
%
		\subsection{Onset of spontaneous scalarization}\label{S2.4.1}
%
	As discussed in the Sec.~\ref{S1.2} and \ref{S2.1}-\ref{S2.2}, in order for spontaneous scalarization (class \textbf{II.A}) to occur in a model, the coupling function (and respective derivatives) must satisfy a set of conditions. Let us then generalize such conditions to contain an axion-like coupling. For an EMA model to yield spontaneous scalarization, it must guarantee that:
			\begin{enumerate}
		     \item[\textbf{i)}] The Maxwell's theory must be recovered near infinity, hence, for an asymptotically vanishing scalar field profile: $f(0)=1\ze $. There is no condition imposed on the coupling function $h$, since the term $F_{\mu \nu}\Tilde{F}^{\mu \nu}$ is a topological invariant (this is not true for its derivatives).
    
    		\item[\textbf{ii)}] The system must accommodate a scalar free solution. The Klein-Gordon equation of motion is
				\begin{equation}\label{E2.4.101}
        		 \Box \ze \phi = \frac{\hat{f}F_{\mu \nu}F^{\mu \nu} + \hat{h} F_{\mu \nu}\Tilde{F}^{\mu \nu}}{4}\ ,
				\end{equation}
    from which, in order for a non-scalarized solution to exist, follows that $\hat{h}(0)=0= \hat{f}(0)$.
    
    		\item[\textbf{iii)}] Spontaneous scalarization occurs if the system is unstable against scalar perturbations $\delta \phi$. These obey (neglecting second order terms)
  	 			\begin{equation}
		        \big( \Box - \mu_{\rm eff}^2\big) \delta \phi = 0\ ,
		       \end{equation}
    with the effective mass $\mu_{\rm eff}^2<0$ given as
   			   \begin{align}\label{E2.4.103}
		        \mu_{\rm eff}^2= & \frac{\hat{\hat{f}}(0) F_{\mu \nu}F^{\mu \nu}|_{\phi=0} + \hat{\hat{h}}(0) F_{\mu \nu}\Tilde{F}^{\mu \nu}|_{\phi=0}}{4}\nonumber\\
		       \eqdef &\quad \frac{\hat{\hat{f}}(0) \left[P^2-\big( Q_e+P\ze h(0)\big)^2\right] + \hat{\hat{h}}(0) P\big( Q_e+P\ze h(0)\big)}{r^4}<0\ ,
			 \end{align}
   constraining the second derivatives of the coupling functions.
		\end{enumerate}

	Following a similar procedure as in Sec.~\ref{S1.2}, the spherical symmetry allows a spherical harmonic decomposition of the scalar field perturbation $\delta \phi$. The resulting linearized scalar field equation
			\begin{equation}
		     \frac{1}{ r ^ { 2 }\ze \sigma } \left( r ^ { 2 } N \ze\sigma\ze U _ { \ell } ' \right)' - \left[ \frac { \ell ( \ell + 1 ) } { r ^ { 2 } } + \mu _ { \mathrm { eff } } ^ { 2 } \right] U _ { \ell } = 0\ .
			\end{equation}
	Which is, again, an eigenvalue problem. The resulting solutions are the bifurcation points of the scalar free solution. Setting $\sigma=1$ and $N=1-\frac{2M}{r}+\frac{Q_e^2+P^2}{r^2}$ allows us to recover the dyonic RN metric (see Sec.~\ref{S2.3}). Then, a scalarized solution can be dynamically induced by a scalar perturbation of the background, as long as the scalar-free RN solution is in the unstable regime. 

\bigskip

	If we now carefully observe the condition \eqref{E2.4.103}, $\mu _{\rm eff}<0\ze $, one can observe that the standard axionic coupling~\cite{lee1991charged} with $f=1$ and $h_{A}=-\alpha\ze \phi$, has $\mu _{\rm eff} =0$ and hence cannot yield spontaneous scalarization (class \textbf{I}). In this regard, in the same spirit as the previous sections, let us introduce an axion-like coupling function $f=1$ and $h_{AL} =-\alpha \ze \phi ^2$ which has $\mu _{\rm eff}=-\alpha\ze P\ze Q_e\ze r^{-4}$ which is smaller than zero for positive charges (class \textbf{II.A}). Just as before, let us study both solutions and observe the differences between them and the previous EMS models.
%
		\subsection{Numerical solutions}\label{S2.4.2}
%

%
%
			\subsubsection*{Radial profiles}
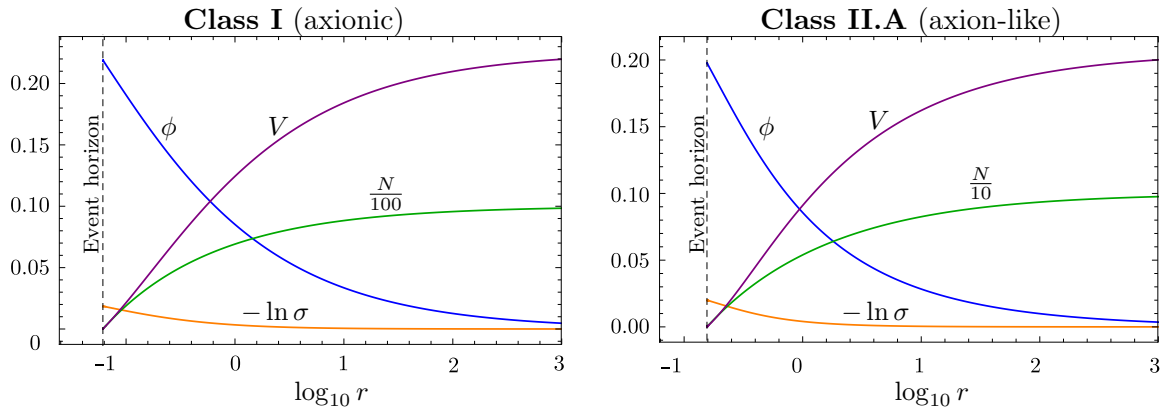
\begin{figure}[H]
			     \centering
			     \begin{picture}(0,0)
				 \put(72,130){{\bf Class \textbf{I}} (axionic)}
				 \put(288,130){{\bf Class \textbf{II.A}} (axion-like)}
				\end{picture}
			 \begin{picture}(0,0)
		  		 \put(110,-10){\small $\log _{10} r$}
		  		 \put(90,20){\small $-\ln \sigma $}
		  		 \put(100,90){\small $V$}
		  		 \put(137,65){\small $\frac{N}{100}$}
		  		 \put(60,90){\small $\phi$}
		  		\put(31,45){\begin{turn}{90}{$\scriptstyle{\rm Event\ horizon}$}\end{turn}}
	   			\end{picture}
			     \includegraphics[scale=0.8]{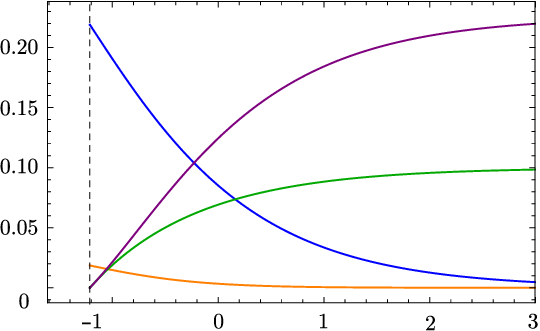}\hfill
			     			 \begin{picture}(0,0)
		  		 \put(110,-10){\small $\log _{10} r$}
		  		 \put(90,20){\small $-\ln \sigma $}
		  		 \put(100,92){\small $V$}
		  		 \put(137,69){\small $\frac{N}{10}$}
		  		 \put(59,90){\small $\phi$}
		  		\put(33,45){\begin{turn}{90}{$\scriptstyle{\rm Event\ horizon}$}\end{turn}}
	   			\end{picture}
    			 \includegraphics[scale=0.79]{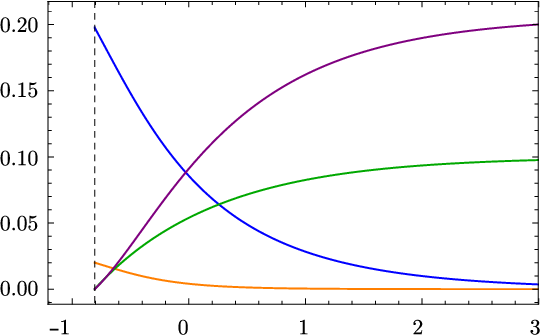}
    			 \caption{Scalarized BH radial functions for $\alpha=35\ze $, $Q_e=0.120\ze$, $P=0.012$ and $r_H=0.297\ze$. (Left panel) axionic coupling $h_A$; (right panel) axionic-like coupling $h_{AL}$.}
				 \label{F2.22}
				\end{figure}

	Some typical solutions of the various functions that define scalarized BHs obtained from numerical integration can be found in Fig.~\ref{F2.22} for $\alpha=35$ an axionic coupling $h_A$ (left panel) and an axion-like coupling $h_{AL}$ (right panel), while keeping $Q_e$, $P$ and $r_H$ constant.
\vspace{3mm}
	Just as before, the nodeless solutions have a monotonically decreasing scalar field ($\phi_0$ is the maximum of the scalar field). Data reveals that (as expected) larger $\alpha$ endow a higher level of scalarization. At least qualitatively, there is no noticeable difference between the profiles of both class \textbf{I} and \textbf{II.A} solutions.
			\subsubsection*{Domain of existence}
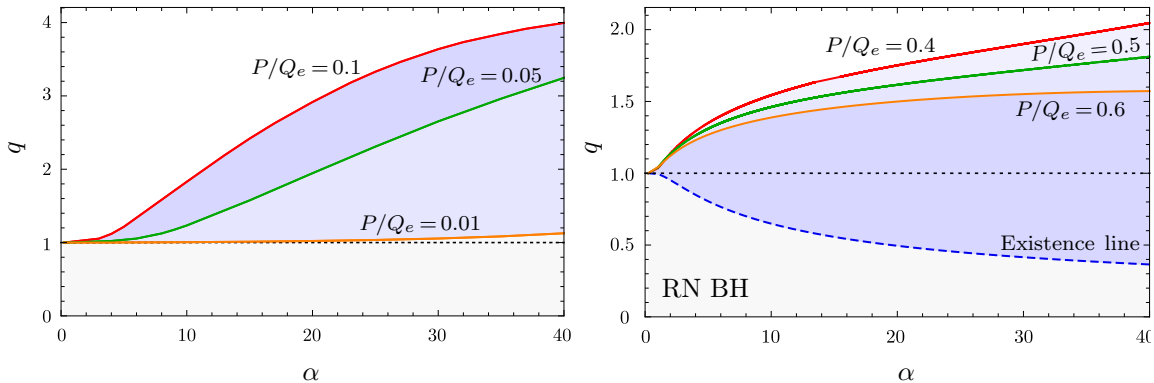
\begin{figure}[H]
				 \centering
				 \begin{picture}(0,0)
		  		 \put(105,-10){\small $\alpha$}
		  		\put(-5,75){\begin{turn}{90}{\small $q$}\end{turn}}
	   			\end{picture}	   		
	   		   		\begin{tikzpicture}[scale=0.5]
\node at (0,0) {\includegraphics[scale=0.8]{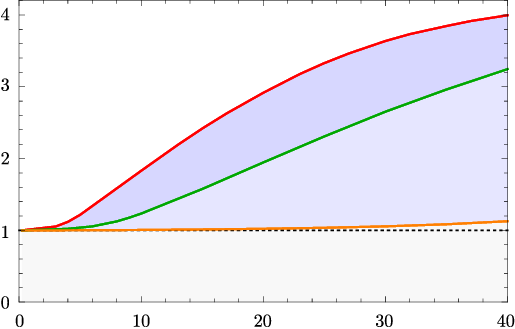}};
\node at (3,-1.3) { $\scriptstyle P/Q_e\, =\, 0.01$};
\node at (4.6,2.6){$\scriptstyle P/Q_e\, = \, 0.05$};
\node at (0,2.83) {$\scriptstyle P/Q_e \,  =\, 0.1$};
				\end{tikzpicture}
				 \begin{picture}(0,0)
		  		 \put(115,-10){\small $\alpha$}
		  		\put(-2,75){\begin{turn}{90}{\small $q$}\end{turn}}
	   			\end{picture}	   		
					\begin{tikzpicture}[scale=0.5]
\node at (0,0) {\includegraphics[scale=0.8]{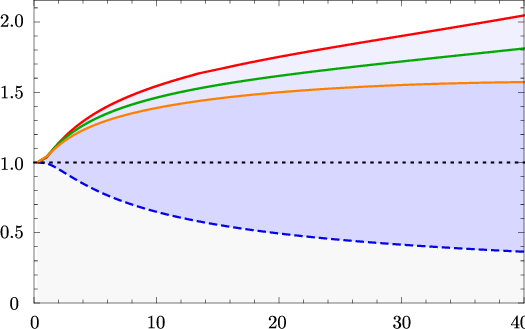}\hfill};
\node at (5,-1.8) { $\scriptstyle {\rm Existence\ line}$};
\node at (0,3.4){$\scriptstyle P/Q_e\, = \, 0.4$};
\node at (5,1.7){$\scriptstyle P/Q_e\, = \, 0.6$};
\node at (5.4,3.3){\begin{turn}{5}{$\scriptstyle P/Q_e\, = \, 0.5$}\end{turn}};
\node at (-4.6,-3) {\small RN BH};
				\end{tikzpicture}
				 \caption{Domain o existence (shadded blue regions) of scalarized solutions in the $(\alpha,\ze q)$-plane for: (left panel) $h_A$ and (right panel) $h_{AL}$. Both domains are bounded by a critical line (solid) at which solutions are singular, for each presented $P/Q_e$ value. The axion-like coupling (right panel) is bounded from bellow by the existence line (dashed blue).}
				 \label{F2.23}
				\end{figure}

	The domain of existence, in the $(\alpha,\ze q)$-plane, for scalarized solutions with an axionic coupling $h_{A}$ (left panel) and an axion-like coupling $h_{AL}$ (right panel) is presented in Fig.~\ref{F2.23}. In both cases, the domain of existence is delimited by a set of critical solutions that we call the \textit{critical line}. At the critical line, numerics suggest a divergence of the Kretschmann scalar and the temperature at the horizon, and vanishing of the horizon area (see Fig.~\ref{F2.30}), whilst $M$ and $Q_\phi$ remain finite and non-zero. At this critical line, all the solutions are overcharged $q>1$.
	
		Concerning the axionic coupling (class \textbf{I}) domain of existence, one observes that for higher magnetic charges, there is a broader domain of existence, which is opposite to the dyonic scalarized BHs case (see Sec.~\ref{S2.3}), for which extremal solutions are obtained for smaller values of $q$ for higher values of $P/Q_e$ along the same $\alpha=c^{\rm te}$ branche. This wider domain of existence is expected since the coupling $h\ze F_{\mu \nu} \Tilde{F}^{\mu \nu}$ is directly proportional to $P$: for higher values of $P/Q_e$ the scalar field couples more strongly to the source term.
		
		Regarding the axion-like coupling (class \textbf{II.A}), a different behaviour than the usual EMS model (see Sec.~\ref{S2.1}-\ref{S2.3}) is observed: numerics suggest that the temperature of the horizon (see Fig.~\ref{F2.30} right panel) tends to zero, while the Kretschmann scalar does not diverge, and the horizon area remains finite and non-zero (see Fig.~\ref{F2.30}), which is compatible with near-extremal BH solutions. 

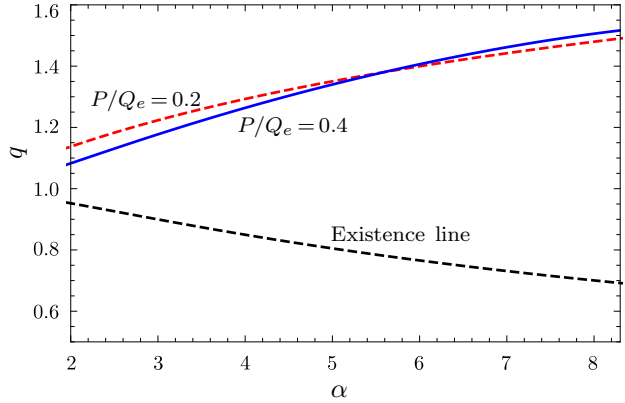
\begin{figure}[H]
				 \centering
				 \begin{picture}(0,0)
		  		 \put(115,-10){\small $\alpha$}
		  		 \put(115,49){$\scriptstyle {\rm Existence\ line}$}
		  		 \put(80,90){$\scriptstyle P/Q_e\, = \, 0.4$}
		  		 \put(25,100){$\scriptstyle P/Q_e\, = \, 0.2$}
		  		\put(-5,80){\begin{turn}{90}{\small $q$}\end{turn}}
	   			\end{picture}	   	
				 \includegraphics[width=.5\textwidth]{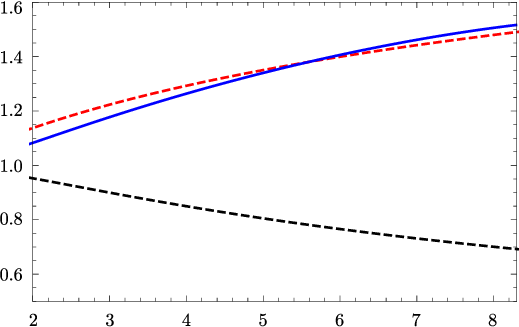}
				 \caption{Zoomed domain of existence for $P/Q_e = 0.2$ and $P/Q_e = 0.4\ze$. One observes that there is a region at which the bigger value of $P/Q_e$ allows greater overcharging and thus a wider domain of existence in that region.}
				 \label{F2.25}
				\end{figure}

\begin{figure}[H]
				 \centering
				 \begin{picture}(0,0)
		  		 \put(119,-10){\small $q$}
		  		 \put(25,30){$\scriptstyle \alpha = 3$}
		  		 \put(55,12){$\scriptstyle \alpha = 4$}
		  		 \put(80,18){$\scriptstyle \alpha = 5$}
		  		 \put(100,24){$\scriptstyle \alpha = 6$}
		  		 \put(115,32){$\scriptstyle \alpha = 7$}
		  		 \put(143,40){$\scriptstyle \alpha = 8$}
		  		 \put(160,47){$\scriptstyle \alpha = 10$}
		  		 \put(185,56){$\scriptstyle \alpha = 12$}
		  		 \put(200,69){$\scriptstyle \alpha = 15$}
		  		\put(-10,65){\begin{turn}{90}{\small $P/Q_e$}\end{turn}}
	   			\end{picture}	   	
				 \includegraphics[width=.5\textwidth]{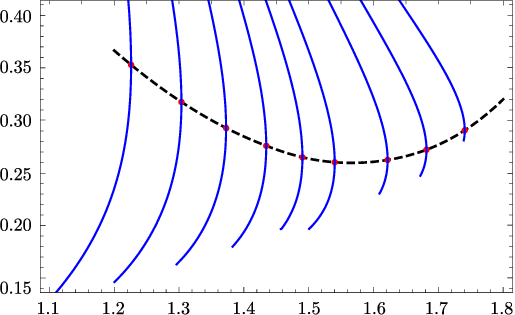}
				 \caption{$(P/Q_e,\,q)$-plane. The solid lines correspond to the extremal solution for each $P/Q_e$ value. The black curve is the curve that interpolates the optimum values of $P/Q_e$.}
				 \label{F2.27h}
				\end{figure}
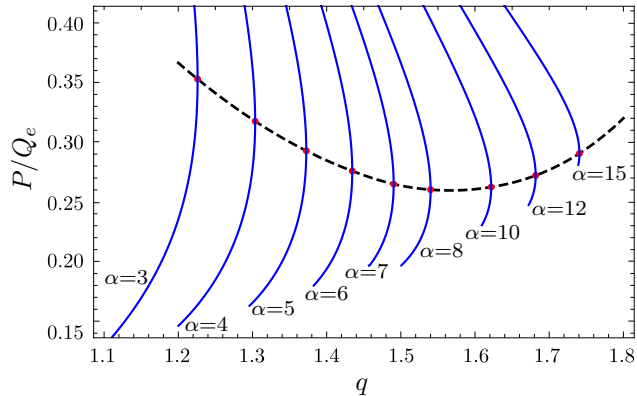

	A characteristic of extremal BH solutions is a vanishing horizon temperature, hence, in order to obtain the extremal BH solutions, a different near-horizon expansion must be done (see Sec.~\ref{S2.3}), and a double zero on the function $N$ has to be imposed, with the leading order terms being
				\begin{align}
			    & N=N_{2}\left(r-r_{H}\right)^{2}+\cdots\ , \nonumber\\
			    &\phi=\phi_{0}+\phi_{c}\left(r-r_{H}\right)^{k}+\cdots \ .
				\end{align}
	Taking $\phi_0$ and $r_H$ as the essential parameters one obtains
				\begin{equation}
			    P=r_H \, , \qquad Q_e=\alpha\ze \phi_0^2\ze r_H \ , \qquad N_{2}=\frac{1}{r_{H}^{2}} \ , \qquad k=\frac{1}{4}\left(1+\sqrt{1+16\ze \phi_{0}^{2}\ze \alpha^{2}}\right)\, ,
				\end{equation}
	A non-integer $k$ would imply that the derivative of all functions will diverge at some order as $r\to r_H$. Although everything is smooth to 2$^{\rm nd}$ order (in particular, the Kretschmann and Ricci scalar should be finite everywhere), the (suitable order) derivatives of the Riemann tensor (in an orthonormal frame) would diverge at the horizon, resulting in non-physical solutions. Thus, in order to obtain physical solutions, we arrive at another condition 
				\begin{equation}
			     \alpha = \frac{n\ze (2\ze n-1)}{2\ze \phi_0^2}\ , \qquad {\rm with}\qquad n \in \mathbb{N}\ .
				\end{equation}
	In Fig.~\ref{F2.23} (right panel), one can observe that there is a region of non-uniqueness where for the same charge to mass ratio $q<1$ RN BHs and scalarized BHs co-exist. Unlike in the axionic domain of existence (Fig.~\ref{F2.23} left panel), for higher $P/Q_e$ values, there is a narrower domain of existence. Such is unexpected since the coupling $h\ze F_{\mu \nu} \Tilde{F}^{\mu \nu}$ is directly proportional to $P$, which raises the question: is there any region of the domain of existence for which the domain of existence is wider for a greater $P/Q_e$ ratio? Indeed, there is, as presented in Fig.~\ref{F2.25}.
	For a region where $\alpha$ is small enough, the domain of existence is wider for higher values of $P/Q_e$. Numerics suggest this effect is related to the term proportional to $P\ze \hat{h}$ in the scalar field equation of motion. More details can be found in Fig.~\ref{F2.27h} that shows the $(P/Q_e ,\ze q)$-plane. The solid lines correspond to the extremal solution for each value of $P/Q_e$. One can observe that, for each $\alpha=c^{\rm te}$ line, there is an optimal value of $P/Q_e$ at which overcharging (and hence the domain of existence in that region) is maximum. The black curve is the curve that interpolates those optimal values of $P/Q_e$. Take the example of Fig.~\ref{F2.27h}: one can clearly see that for instance, for $\alpha=3$ there is a greater value of $q$ for $P/Q_e=0.4$ (greater magnetic charge) than for $P/Q_e=0.2\ze$.
		\subsubsection*{Perturbative stability}
	Let us also introduce a diagnosis analysis of perturbative stability against spherical perturbations that shall be applied to the solutions derived and discussed in the next section. 

\begin{figure}[H]
			 \centering
			 	\begin{picture}(0,0)
		  		 \put(100,-3){\small $\log _{10} \frac{r}{r_H}$}
		  		 \put(140,40){\small $\frac{P}{Q_e}=0.4$}
		  		\put(-7,78){\begin{turn}{90}{\small $U_\Omega$}\end{turn}}
	   			\end{picture}	  
			 \includegraphics[scale=0.82]{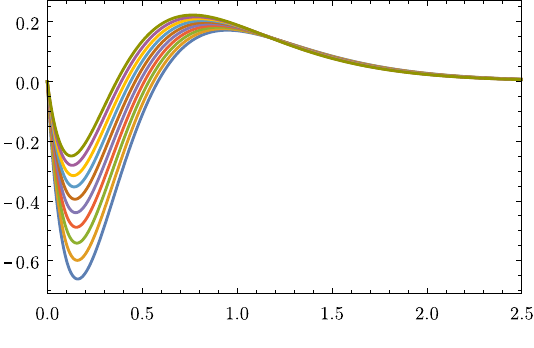}\hfill
			 	\begin{picture}(0,0)
		  		 \put(104,-3){\small $\log _{10} \frac{r}{r_H}$}
		  		 \put(140,40){\small $\frac{P}{Q_e}=0.5$}
		  		\put(-5,78){\begin{turn}{90}{\small $U_\Omega$}\end{turn}}
	   			\end{picture}
			 \includegraphics[scale=0.82]{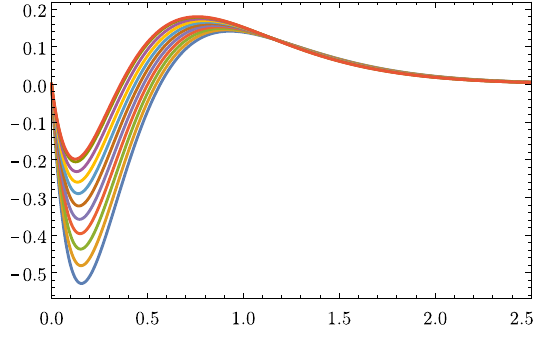}
			 \caption{Sequence of effective potentials, $U_\Omega$, for $\alpha=10$ and two values of $P/Q_e$ for an axion-like potential $h_{AL}$. There is always a region where the potential is negative. The deepest potentials occur for the largest $q$. The bottom (top) curves occur for $q=0.785\ (0.736)$ (left panel) or $q=0.746\ (0.701)$ (right panel).}
			 \label{F2.29}
			\end{figure}
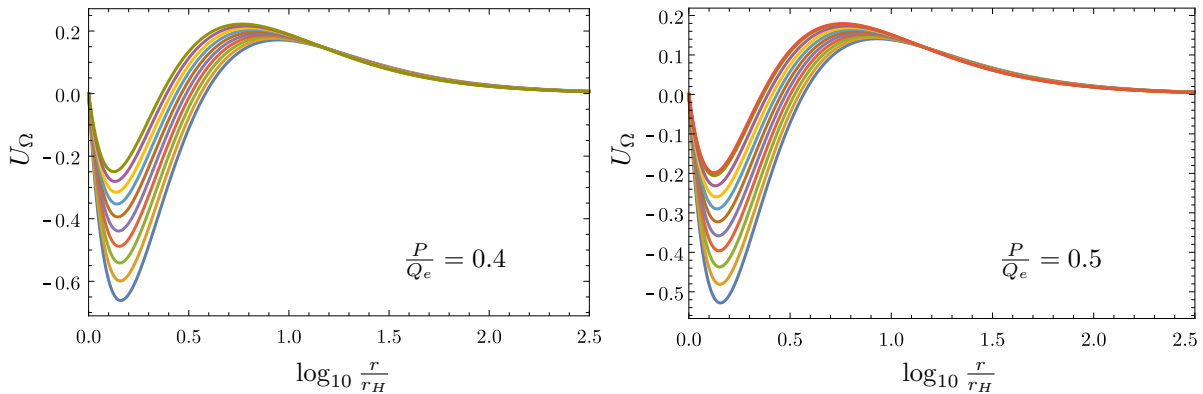

	Following the procedure in Sec.~\ref{S2.1.5} one obtains the effective potential that describes spherical perturbations $U_{\Omega}$:
			\begin{equation}
		     U_\Omega = U_0 + P\ze U_1 + P^2\ze U_2 \ ,
			\end{equation}
	with
			\begin{align}
             &U_0  = \frac{\sigma ^2N}{r^2} \left[ 1 - N -2\ze r^2\phi'^{\, 2} - \frac{Q_e ^2}{r^2 h} \mathcal{U} \right] \ , \nonumber\\
        	 &U_1 =\frac{\sigma^2N Q_e}{r^4 h} \left[ \hat{\hat{h}} + 4r\hat{f}\ze\phi' - 2f\ze \mathcal{U} \right] \ ,\nonumber \\
        	 &U_2 =\frac{\sigma ^2 N}{r^4} \left[ \frac{\hat{\hat{h}}}{2} +2r\phi' \hat{h} - h\big(1-2\ze r^2\phi'^{\, 2}\big) + \frac{1}{h} \left( \hat{\hat{f}} f - \hat{\hat{f}}^2 + 4r\phi'\hat{f}f - f^2 \mathcal{U}\right) \right] \ , \nonumber\\
 			&\mathcal{U}\equiv  1-2\ze r^2\phi'^{\, 2} + \frac{\hat{\hat{h}}}{2 h} + \left( 2\ze r\phi' -\frac{\hat{h}}{h}\right)\frac{\hat{h}}{h} \ .	
 			\end{align}
	An unstable mode would have $\Omega^2<0$, which for the asymptotic boundary conditions of our model is a bound state. It follows from a standard result in quantum mechanics (see $e.g.$~\cite{messiah1961quantum}). However, that \eqref{E2.1.24} has no bound states if $U_\Omega$ is everywhere larger than the lowest of its two asymptotic values, $i.e.$ if it is positive in our case. Thus an everywhere positive effective potential proofs mode stability against spherical perturbations.
			\begin{figure}[H]
			 \centering
			  \begin{picture}(0,0)
		  		 \put(95,-3){\small $\log _{10} \frac{r}{r_H}$}
		  		 \put(160,122){$\scriptstyle P\, =\, 0.045$}
		  		 \put(160,110){$\scriptstyle P\, =\, 0.048$}
		  		 \put(160,98){$\scriptstyle P\, =\, 0.052$}
		  		 \put(160,86){$\scriptstyle P\, =\, 0.055$}
		  		 \put(95,30){$\scriptstyle \alpha\, = \, 5\ \, r_H\, =\, 0.12\ \, Q_e\, = \, 0.12$}
		  		\put(-6,70){\begin{turn}{90}{\small $U_\Omega$}\end{turn}}
	   			\end{picture}	   	
			 \includegraphics[scale=0.8]{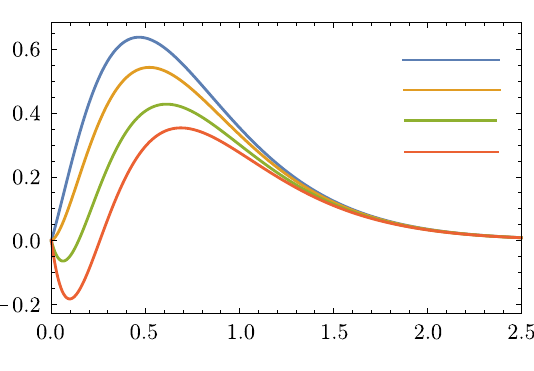}\hfill
			 			  \begin{picture}(0,0)
		  		 \put(95,-3){\small $\log _{10} \frac{r}{r_H}$}
		  		 \put(160,122){$\scriptstyle P\, =\, 0.045$}
		  		 \put(160,110){$\scriptstyle P\, =\, 0.048$}
		  		 \put(160,99){$\scriptstyle P\, =\, 0.052$}
		  		 \put(160,86){$\scriptstyle P\, =\, 0.055$}
		  		 \put(95,30){$\scriptstyle \alpha\, = \, 10\ \, r_H\, =\, 0.12\ \, Q_e\, = \, 0.12$}
		  		\put(-6,72){\begin{turn}{90}{\small $U_\Omega$}\end{turn}}
	   			\end{picture}	  
			 \includegraphics[scale=0.79]{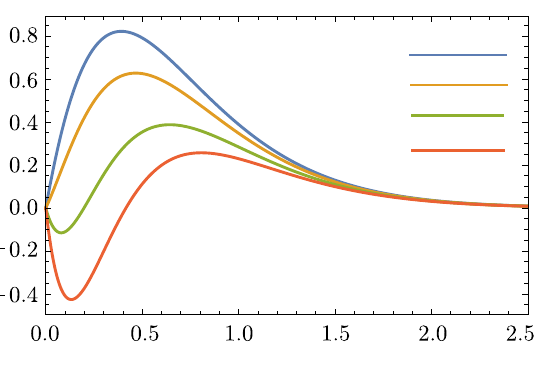}
			 \caption{Effective potential for spherical perturbations, $U_\Omega$, for BHs with axionic coupling $h_A$. The sample of solutions have $r_H=0.12$, $Q_e=0.12$ and $\alpha=5$ (left panel) or $\alpha=10$ (right panel). The potential is always positive until a critical $P/Q_e$ value is reached, beyond which a negative region appears.}
			 \label{F2.27}
			\end{figure}
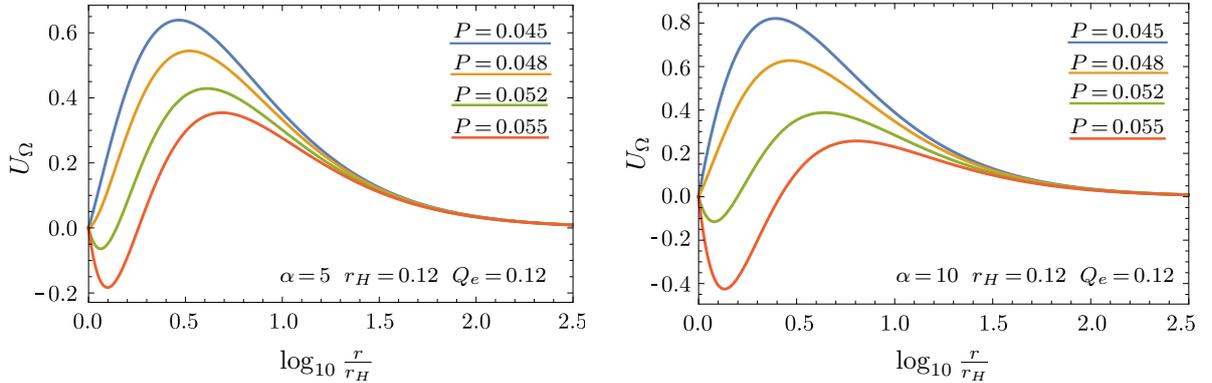	
	The effective potential for spherical perturbations $U_\Omega$, for a sample of BHs with axionic coupling $h_A$ (Fig.~\ref{F2.27}) and axionic-like coupling $h_{AL}$ (Fig.~\ref{F2.29}) is plotted. In both cases, it is not positive definite in all cases, but it is regular in the entire range $-\infty<x<+\infty$. For different $\alpha$ values, the potential always behaves similarly: for sufficiently small values of $P/Q_e$, the potential is always positive (and hence, free of instabilities), until $P/Q_e$ reaches a value at which the potential starts to have a negative region. The potential vanishes at the horizon and infinity.
	We recall that a region of negative potential does not imply instability. Thus, conclusions about linear stability require a study of quasi-normal modes, similarly to what was done in Sec.~\ref{S2.2} for the purely electrical case. Below, however, we shall present evidence, using fully non-linear numerical simulations, that the scalarized solutions are stable and form dynamically. 		
			\subsubsection*{Entropic preference}
\begin{figure}[H]
			\centering
				\begin{picture}(0,0)
						\put(82,135){\small $P/Q_e =0.5$}
				  		 \put(85,62){\small$\alpha =0$}		
				  		 \put(128,30){\small$\alpha =10$}		  		 
		  				 \put(100,100){\small$\alpha = 40$}
		  			     \put(125,12){$\scriptstyle {\rm Extremal}$}
				  		 \put(116,-10){\small $q$}
				  		\put(-8,68){\begin{turn}{90}{$a_H$}\end{turn}}
	   			\end{picture}
				 \includegraphics[scale=0.84]{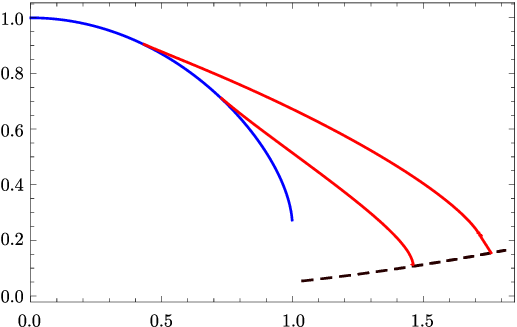}\hfill
				 		\begin{picture}(0,0)
						\put(82,135){\small $P/Q_e =0.6$}
				  		 \put(104,58){\small$\alpha =0$}		
				  		 \put(135,30){\small$\alpha =5$}		  		 
		  				 \put(100,105){\small$\alpha = 30$}
		  			     \put(125,13){$\scriptstyle {\rm Extremal}$}
				  		 \put(116,-8){\small $q$}
				  		 \put(148,68){\begin{turn}{-35}{\small $\alpha = 10$}\end{turn}}
				  		\put(-8,68){\begin{turn}{90}{$a_H$}\end{turn}}
	   			\end{picture}
				 \includegraphics[scale=0.84]{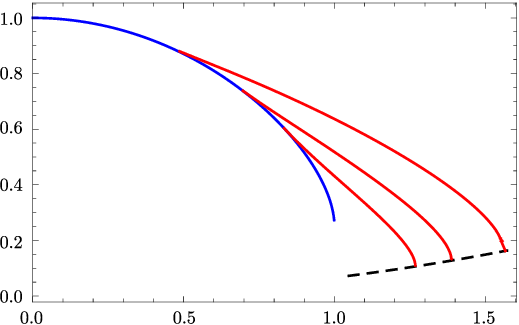}\vspace{5mm}\\			 	
				 		\begin{picture}(0,0)		  		 
		  				 \put(168,30){\small$\alpha = 1$}
				  		 \put(116,0){\small $q$}
				  		 \put(158,85){ $\alpha = 4$}
				  		 \put(60,100){ $\frac{P}{Q_e}=0.1$}
				  		\put(180,130){ $\alpha = 5$}
				  		\put(-8,78){\begin{turn}{90}{$t_H$}\end{turn}}
	   			\end{picture}
				 \includegraphics[scale=0.84]{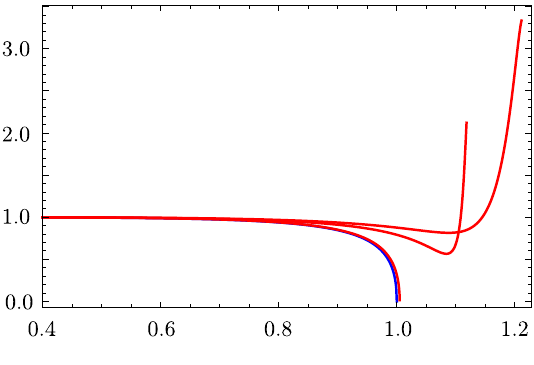}
				 \caption{(Top panel) $a_H$ \textit{vs.} $q$ for $P/Q_e = 0.5$ (left panel) and $P/Q_e =0.6$ (right panel). (Bottom panel) $t_H$ \textit{vs.} $q$ for $P/Q_e=0.1\ze $. All data for $h_{AL}$. The black line represents the sequence of non-scalarized BHs, while the blue dashed lines are sequences of (numerical data points representing) scalarized BHs for a given $\alpha$. The red line represents the entropy of the extremal BH solutions for the different $\alpha$ values.}
				 \label{F2.30}
				\end{figure}
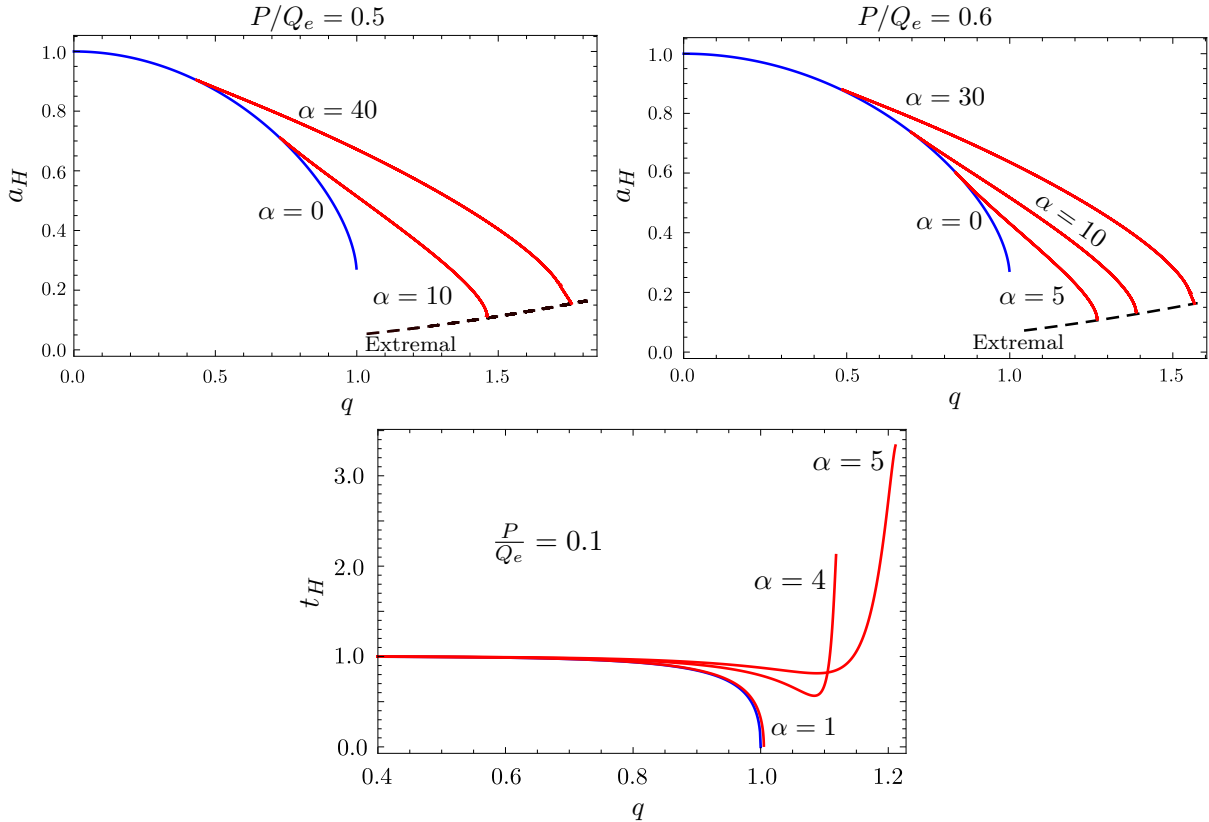

	In the current EMA model, the Bekenstein-Hawking BH entropy formula holds. Thus, the entropy analysis reduces to the analysis of the horizon area. It is convenient to use the already introduced reduced event horizon area $a_H$. Then, in the region where the RN BHs and scalarized BHs co-exist -- the non-uniqueness region --, for the same $q$ the scalarized solutions are always entropically preferred as seen in Fig.~\ref{F2.30}.
	
		The entropic preference, together with the instability of the scalar-free RN BHs against scalarization, suggest that the latter evolve towards the former when perturbed, at least if the evolutions are approximately conservative.

			\subsubsection*{Dynamical preference}
	 We have performed fully non-linear numerical evolutions to test this scenario\footnote{Dynamical evolution performed by Nicolas Sanchis-Gual.}, following our previous Sec.~\ref{S2.2}. The initial data is a dyonic RN BH, with ADM mass $M$, electric charge $Q_e$ and magnetic charge $P$.
	 
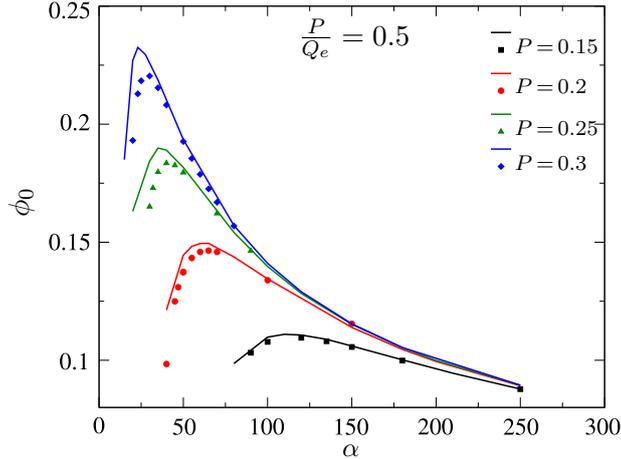
\begin{figure}[H]
				 \centering
				 		\begin{picture}(0,0)		  		 
				  		 \put(120,-5){\small $\alpha$}
						\put(181,103){ $\scriptstyle P\,=\, 0.3$}
				  		 \put(100,150){ $\frac{P}{Q_e}=0.5$}
						\put(181,116){ $\scriptstyle P\,=\, 0.25$}
				  		\put(181,132){ $\scriptstyle P\,=\, 0.2$}
				  		\put(181,147){ $\scriptstyle P\,=\, 0.15$}
				  		\put(-5,85){\begin{turn}{90}{$\phi_0$}\end{turn}}
	   			\end{picture}
				 \includegraphics[width=.5\textwidth]{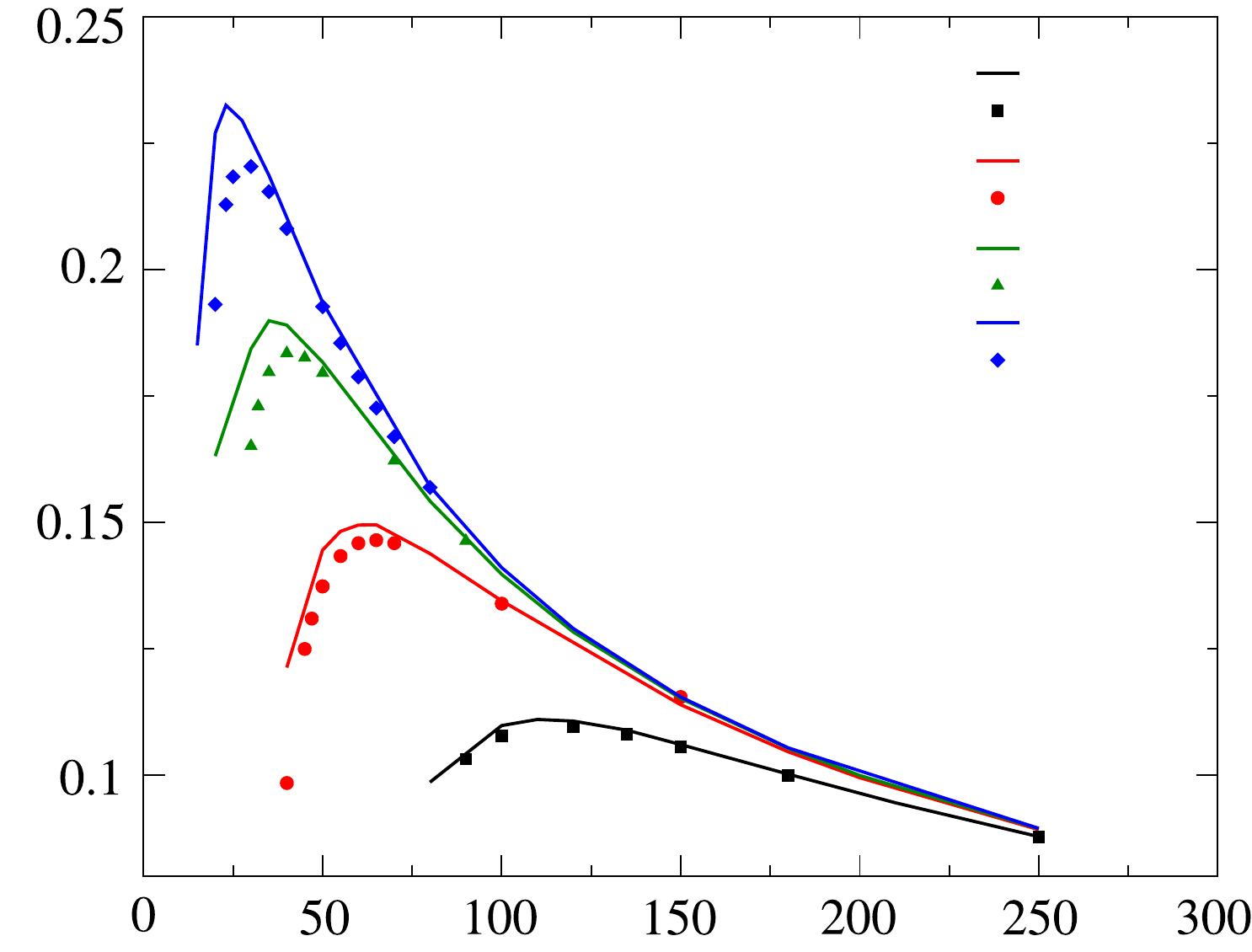}
				 \caption{Scalar field value at the horizon for different values of $Q_e$, while $P/Q_e=0.5\ze$, in the power-law coupling model. The solid lines are obtained from data of the scalarized BHs obtained as static solutions of the field equations. The individual points are obtained from the numerical evolutions, starting from a scalar-free RN BH with the same global charges $M,\ze Q_e,\ze P$. The agreement is better for the lower charges, showing that scalarization only redistributes the electric charge and energy between the horizon and the scalar hair, with minor leaking towards infinity. This leaking appears to become more significant for larger charges.}
				 \label{F2.31}
				\end{figure}

	 The numerical simulations show that the scalar perturbation triggers the spontaneous scalarization of the RN BH. The horizon electric charge decreases as the energy of the field increases, while the horizon magnetic charge remains unchanged until we reach equilibrium and a scalarized solution forms at the endpoint of the dynamical scalarization. The scalar cloud grows near the horizon and expands radially. The radial profile of the cloud decreases monotonically with increasing radii. In Fig.~\ref{F2.31}, we plot the scalar field value at the horizon $\phi_0$ for both the dynamical evolutions and the static solutions with the same $Q_e$ and $P$. We obtained a quite good agreement with the static solutions described above.
%

%
	\section{Einstein-Maxwell-Vector}\label{S2.5}
%
	When considering further generalizations of the spontaneous scalarization mechanism, it is natural to wonder if the spontaneously growing matter can be a vector or even a tensor. In this section, we consider the phenomenon of \emph{spontaneous vectorization}\footnote{An oral presentation about this section can be seen at~\cite{vectorVid}.}. 

	Vector fields and their role in extended theories of gravity have been discussed before~\cite{baum1970vector,gleiser2005linear}, and examples of BHs with vector hair have also been found~\cite{herdeiro2016kerr,fan2016black}. The phenomenon of vectorization was later considered in the extended Vector-Tensor-Gauss-Bonnet (eVTGB) theory~\cite{annulli2019electromagnetism, ramazanouglu2017spontaneous,ramazanouglu2019spontaneous} (the vector analogue to the eSTGB theory); in theories non-minimally coupled to matter~\cite{minamitsuji2020spontaneous,kase2020neutron}; and other theories of gravity~\cite{annulli2019electromagnetism,ramazanouglu2019generalized}. The main idea of this paper is to consider the EMS mechanism used in~\cite{herdeiro2018spontaneous} and extend it to an \textit{Einstein-Maxwell-Vector (EMV)} model. For that, let us consider the action:
	\begin{equation}\label{E2.4.131}
	\mathcal{S}_{EMV} = \frac{1}{4}\int d^4 x \sqrt{-g}\bigg[R - f(\textbf{B})F_{\mu\nu}F^{\mu\nu} - G_{\mu\nu}\bar{G}^{\mu\nu}\bigg] \ ,
	\end{equation}
	 where $G_{\mu\nu}= B_{\nu\, ,\mu}- B_{\mu\, , \nu}$ which represents the field strength of a (possibly complex) vector field $B_\mu$. While $A_\mu$ and $B_\mu$ are both vector fields, for nomenclature simplicity, from now on, we will refer to $A_\mu$ as Maxwell field and $B_\mu$ as vector field. 
	 
	The vector field is non-minimally coupled to the Maxwell term through the coupling function $f$, where $\textbf{B} = B ^\mu \bar{B}_\mu\ze$, and
	\begin{equation}\label{E2.4.132}
	f(0)=1 	\ ,
	\end{equation}
	so that we may recover Einstein-Maxwell when we have a trivial field $B^\mu$. Note that this is a straightforward generalization of the massless scalar case (Sec.~\ref{S2.1}), where a coupling $f(\phi)$ is considered.

\bigskip

	The stress-energy tensor for the model described by the action~\eqref{E2.4.131} is:
			\begin{align}\label{E2.3.133}
			 T_{\mu\nu} =& f\bigg(F_\mu^{\,\,\alpha}F_{\nu\alpha} - \frac{1}{4}g_{\mu\nu}F^{\alpha\beta}F_{\alpha\beta}\bigg) + \frac{1}{2}\bigg(G_\mu^{\,\,\alpha}\bar{G}_{\nu\alpha}+\bar{G}^{\alpha}_\mu G_{\nu\alpha}-\frac{1}{2}g_{\mu\nu}G^{\mu\nu}\bar{G}_{\mu\nu}\bigg) \nonumber \\
			 &+ \frac{1}{4}\hat{f} F^{\alpha\beta}F_{\alpha\beta}\big(B_\mu \bar{B}_\nu+ \bar{B}_\mu B_\nu\big) 
			 \ .
			\end{align}
	Note that, compared to the scalar case, the need to consider the scalar $\textbf{B}\equiv g^{\mu\nu}B_\mu \bar{B}_\nu$ in the coupling $f$ introduces the last term in \eqref{E2.3.133}. The fact that it can be negative allows the possibility of a violation of the weak energy condition.

	The massless vector field $B_\mu$ equations, which is described by a (massless) equation, and the Einstein equations come as:
			\begin{align}\label{E2.5.134}
			 \nabla _\mu G^{\mu\nu} &= \frac{1}{2}\hat{f}F_{\mu \nu} F^{\mu \nu} B^\nu \ , \\
			 R_{\mu\nu}&-\frac{1}{2}g_{\mu\nu}R = 2\ze T_{\mu\nu} \ .\label{E2.5.135}
			\end{align}
	The vector field $B^\mu$, while being massless, due to the interaction with the electromagnetic field, gains an effective mass $(\mu _{\rm eff})$
			\begin{equation}\label{E2.5.136}
			 \mu^2_{\rm eff} = \frac{1}{2}\hat{f}F_{\mu \nu}F^{\mu \nu} \ ,
			\end{equation}
	which, for certain forms of the coupling function and the electromagnetic field, can be negative. This translates into a \emph{tachyonic instability}, \textit{i.e.,} for an initial trivial configuration of $B_\mu$, corresponding to a RN spacetime, a small vector field perturbation, $\delta B_\mu$, grows exponentially and drives the system away from the RN solution. The result is a \textit{spontaneously vectorized RN BH} (VRN).

	For a purely electroestatic configuration $F_{\mu \nu}F^{\mu \nu}<0$: $\mu _{\rm eff} ^2 <0$ requires
			\begin{equation}
		 	\hat{f}>0 \ ,
			\end{equation}
	and the opposite sign for a purely magnetic configuration.	For a deeper study on the several possible coupling function solutions in the scalarized case, see Sec.~\ref{S1.2} (the same line of thought can be applied here). Both conditions are satisfied by a quadratic exponential coupling
			\begin{equation}
			f_B = e^{\alpha\ze \textbf{B}} \ .
			\end{equation}
	For this coupling, spontaneous vectorization of a purely electric RN BH occurs for $\alpha>0\ze$.
	
	Another important property of the model is that the equations of motion do not imply the Lorentz condition. If we take the divergence of \eqref{E2.5.134}
			\begin{equation}
			 \nabla^\mu B_\mu = - \frac{\nabla^\mu (\mu^2_{\rm eff})}{\mu^2_{\rm eff}}B_\mu \ ,
			\end{equation}	
	which, as we can see, does not correspond to the Lorentz condition since $\mu^2_{\rm eff}$ is now a function.

	The metric ansatz of a static, spherically symmetric spacetime is given by \eqref{E1.5.40}; the vector field inherits the metric spherical symmetry and, therefore, for its ansatz we consider \eqref{E1.5.44}; while the Maxwell field will only have an electrostatic component
			\begin{equation}\label{E2.4.140}
			A_\mu = V (r) dt \ ,
			\end{equation}

%
		\subsection{Vector theorems}\label{S2.5.1}
%

%
%
			\subsubsection*{No-vector-hair theorem}
	For this section, let us follow the work done by Herdeiro \textit{et al.}~\cite{herdeiro2016kerr} and generalize their results for our current model.\\
	
	\textbf{Theorem:} \textit{A spherically symmetric, static, asymptotically flat and electrically charged BH spacetime, regular on and outside the event horizon, which solves the Einstein-Maxwell complex-Proca field equations, and for which the massive Proca field inherits the spacetime spatial symmetries but can have a harmonic time dependence of the type $e^{-i\omega t}$ with $\omega\neq 0\ze$, cannot support a non-trivial, finite on and outside the horizon, Proca field.}\\
	
	To prove this argument, we will follow~\cite{pena1997collapsed}. If the vector field is given by \eqref{E1.5.44}, the Proca equations in the metric \eqref{E1.5.40} are
			\begin{align}\label{E2.4.141}
			 \left[\frac{r^2\big( B_t'-\omega B_r\big) }{\sigma}\right]' = \frac{\mu_{\rm eff}^2\, r^2 B_t}{\sigma N} \ ,\\
			 B_t' = \omega B_r\left(1-\frac{\mu_{\rm eff}^2\, \sigma^2N}{\omega^2}\right)\ .\label{E2.4.142}
			\end{align}
	For a BH solution, we assume the existence of an outermost horizon at $r=r_H>0\ze $, which requires $N(r_H)=0\ze $. Every $r>r_H$ surface will then be a time-like surface and $N'(r_H)>0\ze $. As the sign of $\sigma$ is irrelevant to the equations of motion, we can, without loss of generality, consider $\sigma(r_H)>0\ze $.
	
	 The proof comes as follows: consider that there is a small enough region close to the horizon, $r_H< r < r_1$, for which $\mu^2_{\rm eff}<0$ (which is always verified for a massless field). This means that
			\begin{equation}\label{E2.4.143}
			 1-\frac{\mu_{\rm eff}^2\ze \sigma^2N}{\omega^2}>0 \ ,
			\end{equation}
	is guaranteed in this region. Eq.~\eqref{E2.4.142} then implies that the sign of $B'_t$ is equal to the sign of $B_r\ze $. If we integrate \eqref{E2.4.141} in an interval $[r,r_c]	\subset\ ]r_H,r_1[$ and replace $B_t'$ by \eqref{E2.4.142}, we get
			\begin{equation}\label{E2.4.144}
			 B_r = -\frac{\omega}{r^2 \mu_{\rm eff}^2\sigma}\int_r^{r_c}dr\ze \frac{\mu_{\rm eff}^2 r^2 B_t}{\sigma N} \ ,
			\end{equation}
	which imposes that the sign of $B_r$ must be opposite to the sign of $B_t$.
	
	The theorem is now proven by contradiction. As we will see ahead, $B_t$ must be zero at the horizon. So, if $B_t'>0$ close to the horizon, then $B_t>0$ in this region. However, as we know from the considerations above, $B_t'$ has the same sign as $B_r$, implying that $B_r>0$ is the same sign as $B_t$, contradicting the equation above. The same reasoning applies if we consider $B_t'<0$, meaning that the only BH solution compatible with the conditions above is when $B_t=0=B_r\ze $: the Reissner-Nordstr\"om family of solutions.
	
	This same theorem can be generalized for the case where $\mu^2_{\rm eff}>0$ (for example, if we have a massive field\footnote{If the $B$ vector was massive (with mass $\mu _B$), $\mu^2_{\rm eff}$ would instead take the form $\mu^2_{\rm eff}= \mu_B^2 +\frac{1}{2}\hat{f}_B\ze \textbf{F}$, and $\alpha _{min}$ would be $\mu_B$-dependent (see Sec.~\ref{S2.2})} in the region $r_H<r<r_1$. As long as this region is small enough, we can always satisfy condition~\eqref{E2.4.143}. The fact that $N(r_H)=0$, implies that the \textit{lhs} of \eqref{E2.4.143} is very close to unity in this region. Since \eqref{E2.4.144} is independent of the sign of $\mu^2_{\rm eff}$, the rest of the theorem follows.
	
	Note that this theorem is not valid for $\omega=0\ze$. The latter imposes a solely $r$-dependent vector field. In that case, we can obtain the equation for $B_r$ from the Proca \eqref{E2.4.141}
			\begin{equation}\label{E2.4.145}
		 	 \nabla_t G^{tr} = 0 = \mu^2_{\rm eff}B^r \ .
			\end{equation}
	Since $\mu^2_{\rm eff}$ is assumed to be non-zero, we have that the radial component $B^r$ must vanish. Then, the only viable vector field ansatz is
			\begin{equation}\label{E2.4.146}
			 B_\mu = B_t(r)\ze dt \ .
			\end{equation}
		\subsubsection*{Flat spacetime electric no-go theorem}
	Let us now consider the real ansatz \eqref{E2.4.146} for the vector field. By assuming a purely electric field, given by \eqref{E2.4.140}, the electromagnetic equation of motion is
			\begin{equation}
			 \nabla_\mu(f_B F^{\mu\nu}) = 0 \Rightarrow V' = \frac{Q_e}{r^2 f_B} \ .
			\end{equation}
	The virial identity on flat spacetime is
			\begin{equation}\label{E2.4.148}
			 \int_{0}^{+\infty} dr\, \frac{1}{r^2} \left(r^4 B_t'^{\, 2}+\frac{Q_e^2}{f_B}\right) = 0 \ .
			\end{equation}
	Since both terms are always positive (for $f_B>0$), we find that the virial identity can only be respected for the trivial configuration $B'_t=0$ and $Q_e=0\ze$. When $Q_e=0\ze$, the effective mass term of $B_\mu$ vanishes, so $B_\mu$ gains gauge freedom and becomes a typical Maxwell field, allowing us to set $B_t =0\ze$.
	
	Alternatively, we can see this through a map to a scalar field, $\phi$. If we consider the effective action for this configuration
			\begin{equation}
			 \mathcal{S} = \frac{1}{4}\int d^4 x \bigg[- 2\frac{Q_e^2}{f(\textbf{B})\ze r^4} + 2B_t'^{\, 2}\bigg] \ ,
			\end{equation}
	and the mapping $B_t(r)\rightarrow i\phi(r)$, one recovers the effective action for the static, spherically symmetric EMS model~\eqref{E2.1.1} in flat spacetime ($R=0$)
			\begin{equation}
			 \mathcal{S} = \frac{1}{4}\int d^4 x \bigg[-2\frac{Q_e^2}{f(\phi) r^4} - 2 \phi'^{\, 2}	\bigg] \ .
			\end{equation}
	This means that the $B$ field with the ansatz \eqref{E2.4.146} acts as a \emph{ghost} scalar field.

%
		\subsection{Spontaneous vectorization}\label{S2.5.2}
%
%
%
			\subsubsection*{Bifurcation points}
	In the absence of backreaction, the EMV model can be seen as a RN BH that suffers a perturbation from a vector field $B_\mu$. In this case, the line element is the same as the RN BH, \eqref{E1.5.40} with
				\begin{equation}\label{E2.4.151}
				 \sigma = 1\ ,\qquad \mathrm{and}\qquad N=1-\frac{2M}{r}+\frac{Q_e^2}{r^2}\ , 
				\end{equation}
	where $M$ ($Q_e$) is the ADM mass (electric charge) of a RN BH. In this study we will consider the full model \eqref{E2.4.131}, however the coupling function needs to be linearly approximated in $\textbf{B}$ as $f_B= e^{\alpha\ze \textbf{B}} = 1+\alpha\ze \textbf{B}$.

	The Proca equation \eqref{E2.4.141} that describes a nodeless, massless vector field coupled to the Maxwell invariant and has the form \eqref{E2.4.146}, comes as:
				\begin{equation}\label{E2.4.152}
				 \frac{g^{rr}}{\sqrt{-g}} \Big[ \sqrt{-g}\, g^{rr} B_t '\Big]'+\alpha \frac{Q_e^2}{r^4} B_t =0\ , \qquad \mathrm{with} \qquad \mu _{\rm eff} ^2 = -\alpha \frac{Q_e ^2}{r^4}\ .
				\end{equation}
	A RN solution that supports spontaneous vectorization requires an effective mass $\mu _{\rm eff} ^2<0\ze$, and field equation reduces to an eigenvalue problem in $M$
				\begin{equation}\label{E2.4.153}
				 r^2 B_t ''+2\ze r B_t ' +\frac{\alpha\ze  Q_e^2 }{r (r-2 M)+Q_e^2} B_t = 0\ .
				\end{equation}
	While the $B^t$ equation is easy to deal with, due to the divergence at the horizon of $g^{tt}$, the value of $B^t (r_H)$ is not well defined. At the horizon, the physical vector field obeys $B_t=g_{tt}B^t(r_H)=0$ as well as at infinity $B_t (r\rightarrow +\infty) =0$. Close to the horizon, the vector field can be approximated as
				\begin{equation}\label{E2.4.154}
				 B_t = b_1 (r-r_H) -b_1 \frac{ r_H (\alpha -2) Q_e ^2}{2 \left(r_H ^2-Q_e^2 \right)} (r-r_H) ^2 + \cdots\ , \qquad \mathrm{with} \qquad M=\frac{Q_e^2+r_H ^2}{2 r_H}\ ,
				\end{equation}
	The field equation \eqref{E2.4.153} has an analytical solution that obeys the proper boundary conditions \eqref{E2.4.154} 
				\begin{equation}
				 B_t= z \, _2F_1\left[\frac{1}{4} \left(3-y\right);\,\frac{1}{4} \left(3+y\right);\, 2;\,-z\right] \ ,
				\end{equation}	
	with
				\begin{equation}
				 z = 4\ze Q_e^2\ze r_H \left(\frac{Q_e^2 r_H}{r^2}-\frac{Q_e^2}{r}-\frac{r_H ^2}{r}+r_H\right)\ ,\qquad \mathrm{and} \qquad y= \sqrt{4 \alpha-1}\ .
				\end{equation}
	Observe that $\alpha _{min} \geqslant \frac{1}{4}\ze$, which occurs for an extremal RN configuration (first bifurcation point), while $Q_e\rightarrow 0$ implies $\alpha \rightarrow +\infty\ze$. Observe that for each value of $\alpha$ and $Q_e$, \eqref{E2.4.154} yields a value of $M$ at which the vectorized solution bifurcates from the RN BH. The computation of all bifurcation points for a range of $\alpha$ gives the \textit{existence line}. 
			\subsubsection*{The full non-linear model}
	The set  of full non-linear field equations that result from the model \eqref{E2.4.131} with the ansatz \eqref{E1.5.40} and \eqref{E2.4.146} are
				\begin{align}
				  m' = \frac{N r^4  B_t'^{\, 2}-Q_e^2 \, e^{\frac{\alpha  B_t ^{2}}{N \sigma ^2}} \left(2\ze \alpha  \ze B_t^ {2}-N \sigma ^2\right)}{2 N r^2 \sigma ^2} \ ,\qquad \qquad \sigma ' = -\frac{B_t ^2\, \alpha \, Q_e^2 e^{\frac{\alpha  B_t ^2}{N \sigma ^2}}}{r^3 N^2 \sigma }\ ,\nonumber\\
				 V ' = -\frac{Q_e\, \sigma \,e^{\frac{\alpha  B_t ^2}{N \sigma ^2}}}{r^2}\ ,\qquad \qquad B_t'' = B_t ' \left(\frac{\sigma '}{\sigma }-\frac{2}{r}\right)- \frac{\alpha\,  Q_e ^2\, e^{\frac{\alpha  B_t ^2}{N \sigma ^2}}}{r^4 N}B_t \ .
				\end{align}
	Where the electric potential $V'$ is under a first integral that was used to simplify the other field equations. Close to the horizon, the metric functions and vector field can be approximated by a power series as
				\begin{align}
				 & m =  \frac{r_H}{2}+\frac{\frac{b_1 ^2 r_H^4}{\sigma _0 ^2}+Q_e ^2}{2 r_H ^2} (r-r_H)+\cdots\ ,\ \qquad \qquad \sigma = \sigma _0- \frac{b_1 ^2\ze r_H ^3 \ze\sigma _0 ^3\ze \alpha\ze  Q_e ^2}{\big[b_1 ^2 r_H ^4+\sigma _0 ^2 (Q_e ^2-r_H ^2)\big]^2}(r-r_H)+\cdots\ ,\nonumber \\
				 & V =  -\frac{Q_e \, \sigma _0}{r_H^2} (r-r_H) +\cdots \ , \qquad \qquad B_t  =  b_1 (r-r_H)+ b_2 (r-r_H)^2 +\cdots\ ,\nonumber\\ 
				 & b_2 = b_1 \left[ \frac{\alpha\ze  Q_e^2\, \sigma _0 ^4 \big( Q_e^2-r_H ^2\big)}{2\ze r_H \big( b_1 ^2 r_H ^4+\sigma _0 ^2 (Q_e^2-r_H ^2)\big)^2}-\frac{1}{r_H}\right]\ ,
				\end{align}
	with $b_1$ the value of the vector field derivative and $\sigma _0$ the value of the $\sigma$ function at the horizon. At infinity, we impose asymptotical flatness and the metric/field functions can be approximated by 
				\begin{equation}\label{E2.4.159}
				 m = M+\frac{Q_e^2+P_B ^2}{2r}+\cdots\ ,\qquad \sigma = 1- \frac{Q_e^2\alpha }{2 r^2}+\cdots\ ,\qquad	 V  =  \Psi _e -\frac{Q_e}{r}+\cdots \ ,\qquad B_t = \frac{P_B}{r} +\cdots\ ,
				\end{equation}
	with $\Psi _e$ the electrostatic potential difference at infinity and $P_B$ the ``vector charge`` obtained from the asymptotic decay. 
	
\bigskip

	Two horizon quantities of interest are the Hawking temperature and the horizon area \eqref{E2.1.8.5}. These, together with the horizon vector field derivative, $b_1$, and the horizon $\sigma$ value, $\sigma _0$, describe the relevant horizon data.

	The variation of the ADM Mass is described by the first law: $dM=T_HdS+\Psi _e\ze dQ_e$. The vectorized solutions obey the Smarr law
				\begin{equation}
				 M=\frac{1}{2}T_H S_H+ \Psi _e Q_e +M_B\ ,
				\end{equation}		
	where $M_B$ is the energy stored in the surrounding vector field, which can be computed through a Komar integral
				\begin{equation}\label{E2.4.162}
	 			 M_B = -\int d^3 x\ze  \sqrt{-g}\left( 2\ze T_t ^t-\textbf{T}\right)
	 	 = -4\pi \int _{r_H} ^{+\infty} dr \frac{Q_e ^2  \left(\alpha\ze  B_t ^2-N \sigma ^2\right)-r^4 N B_t'^{\, 2}}{f_B\, N r^2\ze \sigma }\ .
				\end{equation}
	In addition, the solutions satisfy the virial identity
				\begin{equation}\label{E2.4.163}
				 \int_{r_H}^{+\infty} dr\left[(r-r_H)\frac{\alpha\ze Q_e^2\, B_t^2}{2\ze r^3 N\, f_B \sigma} + (2\ze r_H-r) N^2\left(r^4 B_t'^{\, 2}+\frac{Q_e ^2\ze \sigma^2}{f_B} \right)\right]=0 \ .
				\end{equation}
	The generic vectorized solution is unknown in closed form, and a numerical approach is necessary. To solve the latter, we use the already established ODE numerical procedure in terms of the unknown parameters $b_1 $ and $\sigma _0$. In all the presented solutions, the virial identity gave an error of $\sim 10^{-8}$, while the Smarr law gave $\sim 10^{-4}$.

	At last, observe that the model possesses the scaling symmetry	$r\rightarrow \lambda\ze r$, $Q_e\rightarrow \lambda\ze Q_e\,$ where $\lambda >0$ is a constant. Under this scaling symmetry, all other quantities change accordingly, \textit{e.g.}, $M\rightarrow \lambda M$, while the coupling function $f $ remains unchanged. For the physical discussion let us use the reduced quantities \eqref{E2.1.15a}-(2.1.17).
			\subsubsection*{Light rings}
	One of the essential astrophysical properties of BHs is the presence of a \textit{light ring (LR)} -- since we are dealing with spherical symmetry, the LR is, in fact, a sphere: a photon sphere. To find the LR radius, $r_{LR}$, of a spherical spacetime, one must consider the null geodesics ($ds^2=0$) of the metric ansatz \eqref{E1.5.40} (the dot represents a derivative with respect to the proper time):
				\begin{equation}\label{E2.4.167}
				 \dot{r}^2 = \frac{E^2}{\sigma^2} - \frac{l^2N}{r^2} \ ,
				\end{equation}
	where $E$ and $L$ represent the energy and angular momentum of a photon along the geodesic, respectively. The LR is circular, implying $\dot{r}=0$ and $\ddot{r}=0$. The first condition relates the energy with the angular momentum of the photon $E=L\ze \sqrt{N}\sigma/ r$ while the second gives us the condition necessary to find $r_{LR}\ze$:
				\begin{equation}\label{E2.4.168}
				 \sigma\left(-2m'+\frac{2m}{r}\right)+2\left(1-\frac{2m}{r}\right)(r\sigma'-\sigma) = 0 \ .
				\end{equation}
	For the RN metric we have $\sigma=1$ and $m = M -Q_e^2/(2\ze r)$, giving us 
				\begin{equation}\label{E2.4.189}
				 r^{RN}_{LR} = \frac{3M\pm\sqrt{9M^2-8Q_e^2}}{2} \ .
				\end{equation}
	As demonstrated in \cite{cunha2017light,cunha2020stationary}, LR always come in pairs. For a BH, one of the LR is inside and the other outside the external horizon. 
%
		\subsection{Numerical results}\label{S2.5.3}

%
%
			\subsubsection*{Solutions profile}
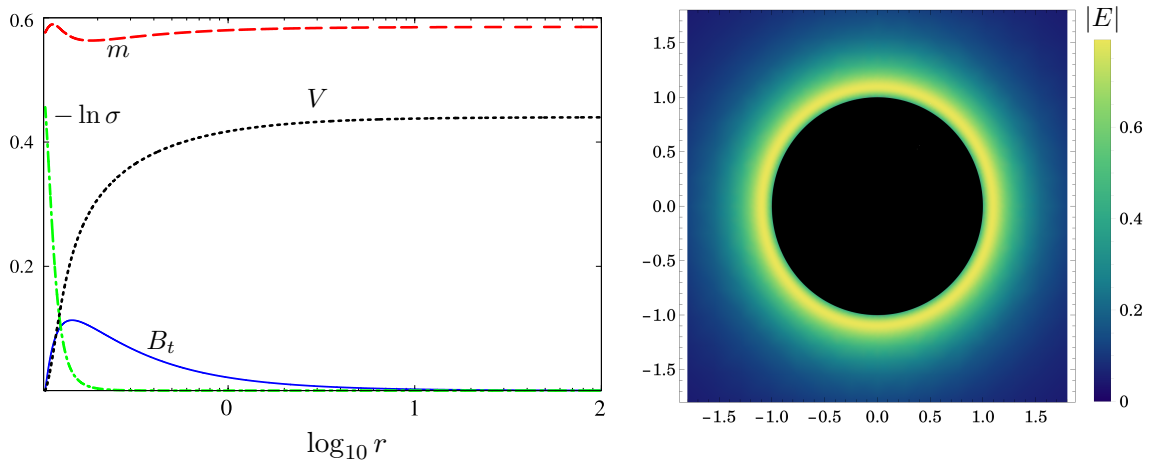
\begin{figure}[H]
				 \centering
					\begin{picture}(0,0)
				     \put(45,145){\small $m$}
	   				 \put(120,126){\small $V$}
	   				 \put(60,35){\small $B_t$}
	   				 \put(25,120){\small $-\ln \sigma$}
	   				 \put(120,-4){$\log _{10} r$}
	  				\end{picture}
			 	 \includegraphics[scale=0.65]{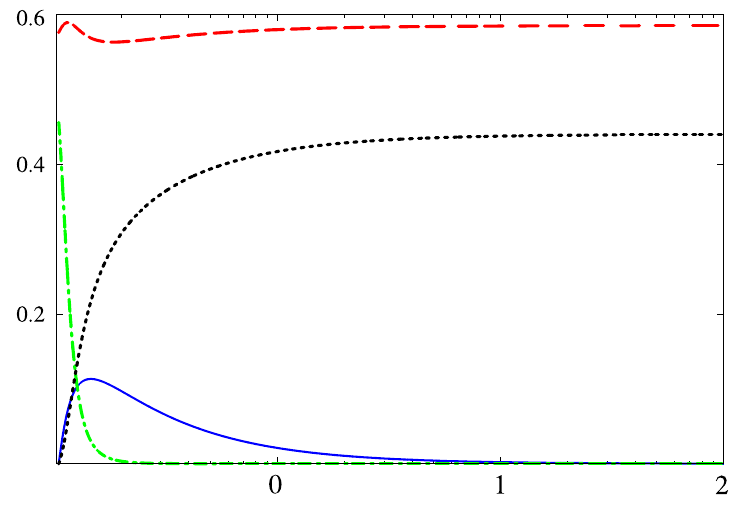}
					\begin{tikzpicture}[scale=0.5]
\node at (0,0) {\includegraphics[scale=0.111]{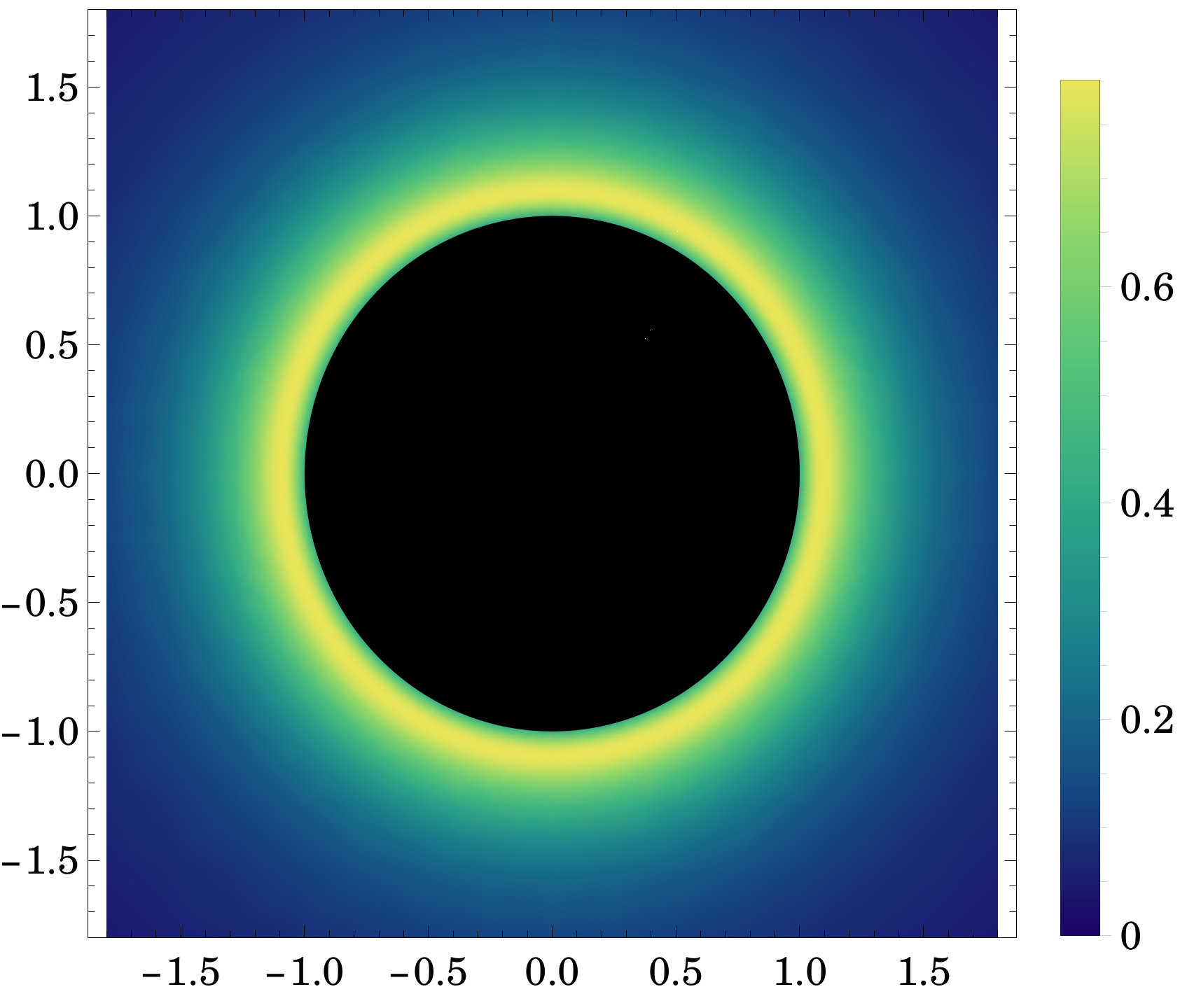}};
\node at (5.5,5.3) {\small $|E|$};
				\end{tikzpicture}
	 			\caption{(Left panel) graphical representation of a solution's functions profile and (right panel) density plot of the electric field strength along the equatorial plane for $\alpha =25$, $r_H=1.0$ and $Q_e=0.25\ze$.}
			  \label{F2.32}
			\end{figure}

	Let us start by studying the generic behaviour of the metric functions and the vector fields of a fundamental (nodeless) state VRN BH. In Fig.~\ref{F2.32} is represented the radial dependence of the various field functions for an illustrative solution with $\alpha =25$, $Q_e=0.25$, $r_H=1.0$, and a charge to mass ratio $q=0.427\ze $.
	A universal feature of the fundamental solutions is the existence of a bulge of $B_t$ around the event horizon. Since regularity imposes a null vector field at $r_H$ and infinity, the only non-trivial, nodeless vector field solutions possess a sharp increase very close to the horizon ($b_1 >0$), reaches a maximum and then ``slowly`` decays as $1/r$. 
	
	Numerical analysis shows that, for a fixed charge and mass, increasing $\alpha$ corresponds to an increase of the magnitude of the field $B_t$. The distance of the $B_t$'s maximum relative to the horizon increases slightly along with an increase in $\alpha$.

	Besides, due to the coupling with the Maxwell field, there will be a modulation of the electric potential that ultimately creates a non-monotonic electric field that can have important implications in the accretion disk formation (see Fig.~\ref{F2.32} right panel). In contrast, the electro-vacuum RN solution and the scalarized case are monotonically crescent, and the modulation associated with the latter is closer to damping.

	One other interesting characteristic of this model is a region with negative energy density. Observe the $m$ profile in Fig.~\ref{F2.32} (left panel). In the latter, there is a valley, which corresponds to the region where the $B_t$ reaches its peak. This can be easily understood by observing the Komar mass (mass associated with the external vector field): the negative energy density term in the stress-energy tensor \eqref{E2.3.133}, gives a negative contribution to the ADM mass, violating the weak energy condition.

	Regarding the $r_{LR}$, we show some values in Table~$2.4$. We can see that there is an increase of the LR radius when we increase the coupling constant $\alpha$ and that the LR radii of vectorized BHs are smaller than the corresponding RN black holes $r_{LR}<r_{LR} ^{RN}$.
				\begin{table}[H]
					\begin{center}
					 \caption{Light ring radii for four $\alpha $ values with $Q_e=0.25$ and $r_H=0.52\ze$.}
					 \vspace{2mm}
						\begin{tabular}{ c |cccc }
					  	 $\alpha$ & 6 & 8 & 10 & 12 \\ 
					 	 \hline
 						 $r_{LR}$ & $0.73$ & $0.77$ & $0.79$  & $0.80$ \\ 
 						 $r_{LR}/r^{RN}_{LR}$ & $0.77$ & $0.83$ & $0.84$ & $0.88$ 
						\end{tabular}
					\end{center}
										 \label{T2.4}
				\end{table}

			\subsubsection*{Domain of existence}
	In the same spirit as the scalarized BH solutions previously studied (\textit{cf.} Sec.~\ref{S2.1}-\ref{S2.3}), generating several solutions allows us to obtain a region of the domain of existence for the vectorized BH solutions. The latter is delimited by the existence line -- at which $b_1\rightarrow 0$ -- and a critical line -- with $b_1 \rightarrow +\infty\ze$.
	
\begin{figure}[H]
	 			 \centering
	 			 	\begin{picture}(0,0)
				     \put(45,150){$ \scriptstyle {\rm Extremal\ RN}$}
				     \put(145,133){$ \scriptstyle {\rm Existence\ line}$}
				     \put(155,20){$ \scriptstyle {\rm Critical\ line}$}
	   				 \put(30,20){$\scriptstyle {\rm RN\ BH}$}
	   				 \put(-5,82){\begin{turn}{90}{\small  $q$}\end{turn}}
	   				 \put(127,-4){$\alpha$}
	  				\end{picture}
				 \includegraphics[scale=0.66]{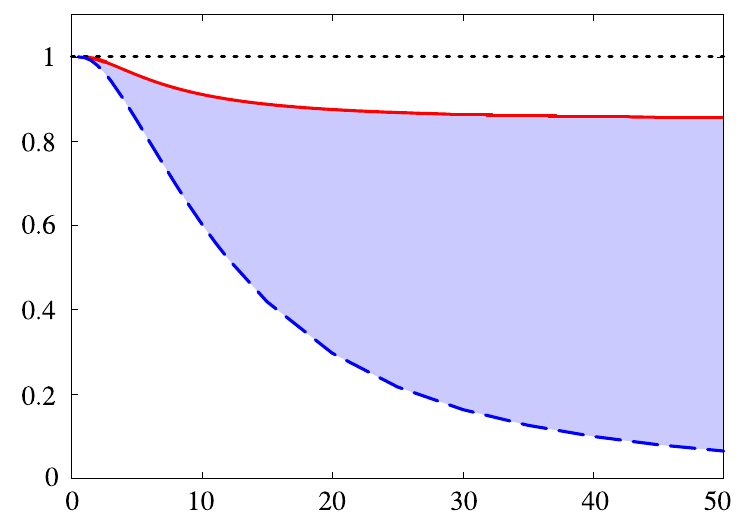}
	 			 \caption{Graphical representation of the domain of existence for a vectorized RN BH with exponential coupling $f_B$. The solid red line represents the bifurcation point for a given $\alpha$; the dashed blue line represents the critical solution with $r_H\to 0$, and the shaded blue region is the region where vectorized BHs can simultanoeusly exist with standard RN BHs.}
	 			 \label{F2.33}
				\end{figure}
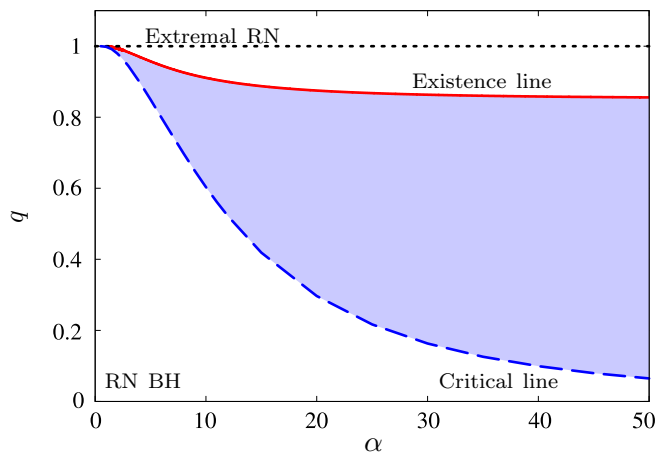

	Different than the scalar case, all the possible solutions are \textit{undercharged} ($q<1$), and in fact, the critical line always has a smaller $q$ than the existence line (see Fig.~\ref{F2.33}). For a fixed $\alpha $ value, one can go from the existence line to the critical line through an increase in $r_H$, meaning that solutions never become singular. Meanwhile, $b_1$ and $\sigma _0$ have a growing increase at the horizon and diverge at the critical line. We have also computed the Kretschmann and Ricci scalar \eqref{E2.1.9} and observed that the solution is everywhere regular along with the domain of existence, including the critical line. 

	Concerning the vector charge, $P_B$, one observes a monotonic increase along with the domain of existence for a fixed $\alpha$. While it starts at zero in the existence line, it grows to almost double $Q_e$ at the critical line.

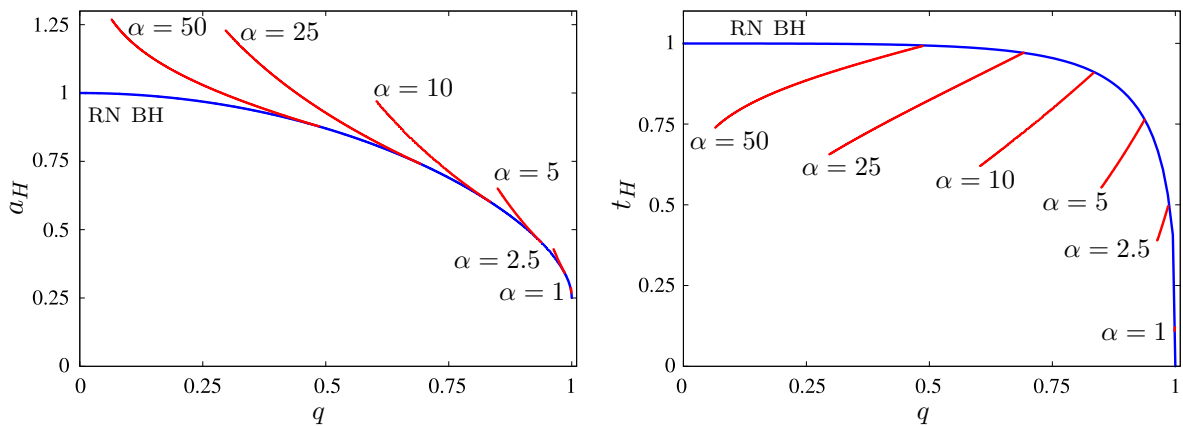
\begin{figure}[H]
	 			 \centering
	 			 \begin{picture}(0,0)
				  		 \put(28,105){$\scriptstyle{\rm RN\ BH}$}		
				  		 \put(135,115){\small$\alpha =10$}	
				  		 \put(180,83){\small$\alpha =5$}	
				  		 \put(165,50){\small$\alpha =2.5$}	
					     \put(182,38){\small$\alpha =1$}		  		 
		  				 \put(85,136){\small$\alpha = 25$}
						 \put(43,138){\small$\alpha = 50$}
				  		 \put(112,-7){\small $q$}
				  		\put(-1,70){\begin{turn}{90}{$a_H$}\end{turn}}
	   			\end{picture}
	 			 \includegraphics[scale=0.6]{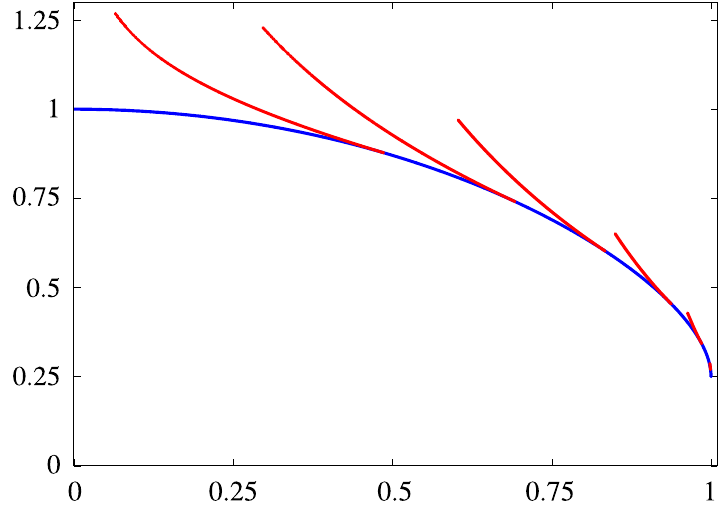}\hfill
	 			 	 			 \begin{picture}(0,0)
				  		 \put(28,95){\small$\alpha = 50$}		
				  		 \put(120,80){\small$\alpha =10$}	
				  		 \put(160,72){\small$\alpha =5$}	
				  		 \put(168,54){\small$\alpha =2.5$}	
					     \put(182,23){\small$\alpha =1$}		  		 
		  				 \put(70,85){\small$\alpha = 25$}
						 \put(43,138){$\scriptstyle{\rm RN\ BH}$}
				  		 \put(113,-7){\small $q$}
				  		\put(-1,72){\begin{turn}{90}{$t_H$}\end{turn}}
	   			\end{picture}
	 			 \includegraphics[scale=0.6]{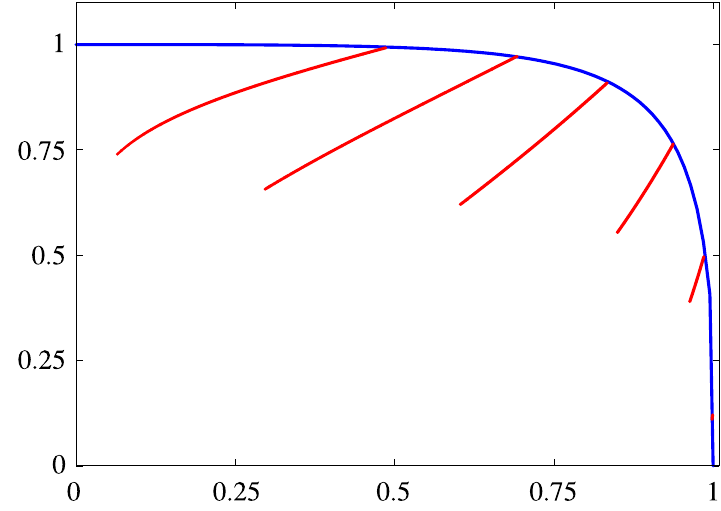}
	 			 \caption{(Left panel) Reduced area $a_H$ \textit{vs.} reduced charge $q$; (right panel) reduced horizon temperature $t_H$ \textit{vs.} $q$ for an EMV model with exponential coupling, $f_B$. The blue lines represent the non-vectorized RN BH, while the red lines represent the vectorized solutions for a series of couplings values $\alpha$.}
				 \label{F2.34}
				\end{figure}

	The domain of existence study shows that an increase in $\alpha$ implies a smaller value of the normalized electric charge for both the existence line and the critical line. However, the latter has a faster decrease in $q$, and hence the domain of existence broadens, tending to the Schwarzschild case for $\alpha \rightarrow+ \infty$ ($q=0.013$ for $\alpha =100$).
	In addition, to study the thermodynamical preference of vectorized solutions over an equivalent RN BH (see Fig.~\ref{F2.34}), we have computed the reduced area $a_H$ (left panel) and the reduced temperature $t_H$ (right panel). 
	
	From the thermodynamical study, we observe an entropic preference of the vectorized BHs in relation to an equivalent RN BH, which can be clearly understood by the fact that the ADM Mass of the vectorized BH is smaller than the mass contained in the central BH.
	Concerning the horizon temperature, Fig.~\ref{F2.34} (right panel), one observes a smaller horizon temperature for the vectorized solutions than an equivalent RN solution. In addition, the temperature decreases as one goes further from the existence line, however never reaching extremality ($t_H=0$).
	
	For a spontaneous vectorized configuration that allows two coupling constant values, the lower one will be entropically preferable. In comparison with the scalarized solutions Sec.~\ref{S2.1}, both solutions are entropically preferable over an equivalent RN BH. However, while in the scalarized case, a solution with a higher coupling constant has higher entropy, here, the opposite occurs. 	
%
	\section{Further remarks}\label{S2.6}
%
	In this chapter, we studied BH scalarization in the EMS model~\cite{herdeiro2018spontaneous} for six different choices of coupling function and performed several generalizations to the EMS model. We considered four different forms for the coupling function that can endow spontaneous scalarization of charged black holes, belonging to class \textbf{II.A}; one coupling that induces scalarized disconnected solutions, class \textbf{II.B}, and one dilatonic like coupling of class \textbf{I}. 	

	The examination of the static solutions allows two main conclusions concerning the purely electric EMS model with the six couplings. Firstly, for all cases studied, the scalarized solutions are entropically favoured over a comparable RN BH in the region where non-uniqueness holds. This creates a distinction with the BH scalarization in eSTGB model (see Sec.~\ref{C3}): for the same power-law coupling we have considered here, the scalarized BHs are not entropically favoured. In addition, the scalarized spherically symmetric, fundamental BH solutions are not necessarily perturbative stable. Thus, BH scalarization in the EMS and eSTGB models do not necessarily mimic one another for all couplings. 

	Secondly, the power-law, hyperbolic and exponential coupling are qualitatively very similar, albeit the exponential coupling maximizes differences concerning the RN case. On the other hand, the fractional coupling yields qualitative differences with the existence of a different type of boundary in the domain of existence, bounding the region where physical solutions exist, abiding the weak energy condition. This boundary is associated with the divergent behaviour of the coupling for a particular value of the scalar field. 

	Amongst the novel features, we have unveiled a new type of non-uniqueness amongst EMS models of class \textbf{II.B}. This is qualitatively different from what occurs in class \textbf{II.A} EMS models, but it exhibits a very curious analogy with a model which, a priori, seems completely unrelated. This analogy pertains to the five-dimensional vacuum Einstein gravity (see Appendix~\ref{O}). There are three different BH solutions for some regions of the domain of existence: the scalar-free (or bald) RN BH and the scalarized (or hairy) cold and hot BHs.

	Concerning the dynamical evolutions of class \textbf{II.A}, we have established that for small values of $q$, the evolutions of unstable RN BH lead to the formation of a scalarized BH with the same value of $q$, within numerical error. The evolution is essentially conservative. Such was explicitly observed for the exponential and power-law coupling. We expect the same behaviour to occur for the hyperbolic coupling. However, for sufficiently high values of $q$, scalarization decreases this value; thus a non-conservative process is taking over, expelling a non-negligible fraction of charge and energy from the BH with a dominance of the former. We have studied this in detail in the exponential coupling case but expect the same result to be observed in the power-law and hyperbolic coupling. For the case of the fractional coupling, we have only performed evolutions at large $q$ and in the region where RN BHs overlap with (physical) scalarized BHs. Scalarization was observed, and a decrease in the value of $q$ occurred. Finally, we have analyzed the evolution of unstable RN BHs under non-spherical perturbations and observed that a spherical scalarized BH emerges.

	Following establishing the main properties of the EMS model, we started the generalization of such a model by introducing a mass term to the scalar field. As in the case of other scalar-tensor theories, the mass term suppresses the effects of scalarization. We have done preliminary results of this model and observed that: $1)$ the existence line changes; $2)$ scalarization requires a larger $\alpha$ as compared to the scalar-free case; and $3)$ the mass term quenches the dispersion of the scalar field, which becomes more concentrated in the neighbourhood of the horizon.
	
	In an attempt to understand the effect of a magnetic charge in the EMS BHs (\textit{a.k.a.} dyon), we followed by generalizing the previous EMS model. In the well known dilatonic case, dyonic BHs have a regular extremal limit, whereas purely electrically (or magnetically) charged ones do not; the latter become singular, approaching a critical solution when endowed with the maximal possible charge for a given mass. Given the unique features of smooth extremal solutions, it is of interest to understand the status of these solutions in the generic EMS case since, for purely electric scalarized BHs, the maximal charge leads to a critical, rather than extremal, solutions~\cite{herdeiro2018spontaneous,fernandes2019spontaneous}. Here we have shown that for scalarized BHs, the conclusion is similar to dilatonic BHs (within a particular coupling regime) in this respect: dyonic BHs can have a regular extremal limit. Our analysis also allows constructing such dyonic extremal solutions for arbitrary coupling in the dilatonic case since solutions were only known (in closed analytic form) for some particular coupling values. Moreover, despite the defining difference in the two classes of solutions, Fig.~\ref{F2.1}-\ref{F2.4} and Fig.~\ref{F2.20} show that these two classes, for the illustrative families, present similar trends in the behaviour of physical quantities.

	As evidence for the existence of dyonic extremal scalarized BHs, we have used the fact that one expects such solutions to have a near-horizon geometry, which is a solution of the field equations. In both RN and Kerr extremal BHs (when $T_H=0$), the near-horizon geometry has an enhanced symmetry that contains an $AdS_2$ geometry (for Kerr, there exists a non-trivial fibration of $S^1$ on $AdS_2$ in the near-horizon geometry). It was proven in~\cite{sen2005black,astefanesei2006rotating} that the existence of $AdS_2$ factor is, in fact, at the basis of the attractor mechanism for extremal BHs rather than supersymmetry \cite{ferrara1995n,ferrara1996university}. In string theory, the attractor mechanism provides a non-renormalization theorem for the matching of statistical and thermodynamic entropies of extremal BHs \cite{dabholkar2007black} (see, also, Sec.~$5$ of \cite{astefanesei2008moduli}). Here, the attractor mechanism provides a clear and simple explanation of why the extremal limit is a naked singularity for solutions with a single charge and a smooth geometry for dyonic BHs. Besides enabling a partial analytical understanding of the extremal solutions, analyzing the near-horizon geometry provides an insight on how scalarization leaves a trace at the level of attractors, allowing two families of near-horizon geometries.

	The addition of an axionic (and axionic-like coupling) $h(\phi)F_{\mu \nu}\tilde{F}^{\mu \nu}$ to the EMS action, which has been previously studied in the context of BH spontaneous scalarization allows one to obtain a novel set of solutions. Depending on the choice of $h$, the model can accommodate BHs with axionic hair of class \textbf{I}, which do not reduce the hairless RN configuration; or solutions of class \textbf{II.A} with an axionic-like coupling that can generate spontaneously scalarized RN BHs. In this case, the latter may become unstable against scalar perturbations and spontaneously scalarized, which we have shown to occur dynamically in one illustrative example.
	
	At last, we considered replacing the scalar field with a vector field $B_\mu$. We first showed a no-hair theorem for a generic complex vector field ansatz, along with a flat spacetime no-go theorem extending the first. Notably, this no-hair theorem does not work for a vanishing field frequency $\omega = 0$, in which case we have a real vector field with only one component.
	
	We then attempted to construct the vectorized BH solutions with the last ansatz. An analytical study of the model showed that the classification established in Sec.~\ref{S1.2} for the scalar case is also valid in the presence of a vector field. We follow by considering a coupling of class \textbf{II.A}. The numerical results show a family of vectorized solutions bifurcating from the RN existence line, reaching a critical solution. Compared to the scalarized RN solutions, the $B_t (r)$ field's maximum is far from the horizon and slowly decays to zero at infinity. The difference occurs due to the imposition that both at the horizon and infinity, $B_t$ must be zero.

	A caveat to the vectorized solutions is that they violate the weak energy condition in a small region where the mass function $m(r)$ decreases. Since there is an additional contribution of the interaction term
to the energy in the vectorized case, a negative Komar mass outside the horizon emerges, creating a region with negative energy densities. 

	Another peculiar property of the vectorized solutions is that, while the scalarized solutions can be overcharged $Q_e/M >1$, the vectorized are always undercharged, $Q_e/M < 1$. A thermodynamical study tells us that, despite this property, the vectorized solutions are entropically preferred.

\clearpage\null\newpage

\chapter{Einstein-Maxwell-Scalar-Gauss-Bonnet}\label{C3}
%
	It is conceivable that deviations from GR occur \textit{only} for sufficiently large curvatures. An analogous realization of this idea was discussed in Ch.~\ref{C2} with the phenomenon of spontaneous scalarization of charged black holes. Recently, it gained a new guise in which the scalarization of the GR vacuum BH solutions becomes possible, in the context of extended scalar-tensor models that include the Gauss-Bonnet (GB) quadratic curvature invariant $R^2_{\rm GB} $, as first pointed out in~\cite{silva2018spontaneous,doneva2018new,antoniou2018evasion}. In the following, we shall dub the latter \textit{GB scalarization}.

	GB scalarization circumvents well-known no-hair theorems (see~\cite{herdeiro2015asymptotically} for a review) due to a certain class of non-minimal couplings between a real scalar field $\phi$ and the GB invariant. The phenomenon occurs for BHs in an appropriate mass range, defined by a dimensionful coupling in the model. Moreover, it can be triggered either if $R^2_{\rm GB}>0$ -- hereafter dubbed GB$^+$ scalarization~\cite{blazquez2018radial,doneva2021beyond,annulli2021clap,east2021dynamics,antoniou2021black, hod2019gauss,collodel2020spinning,herdeiro2019black,silva2019stability,brihaye2019hairy, minamitsuji2019scalarized,myung2019quasinormal,myung2019instability,doneva2018charged,witek2019black, brihaye2019spontaneous} --  or if $R^2_{\rm GB}<0$ -- hereafter dubbed GB$^-$ scalarization. For the Kerr family of GR, the latter only occurs for sufficiently fast-spinning BHs~\cite{dima2020spin,herdeiro2021spin,berti2021spin}, which justifies the terminology \textit{spin-induced scalarization}~\cite{dima2020spin}. By contrast, in the case of GB$^+$ scalarization, Kerr BHs can also scalarize, but rotation actually suppresses the effects of scalarization~\cite{cunha2019spontaneously,berti2021spin}.
	
\begin{figure}[H]
	 \centering
	 	  \begin{picture}(0,0)
						 \put(170,138){\begin{turn}{-42}{$\scriptstyle{\rm Extremal\ KN\ BHs}$}\end{turn}}
				  		 \put(132,-10){\small $j$}
				  		\put(-4,88){\begin{turn}{90}{$q$}\end{turn}}
	   			\end{picture}
	   			\begin{tikzpicture}[scale=0.5]
\node at (0,0) {\includegraphics[scale=0.70]{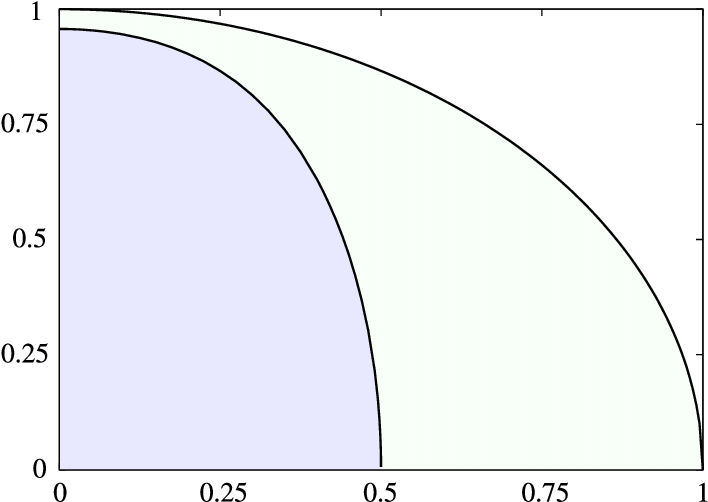}};
\node at (4.2,-4) {\small $R_{GB} ^2 (r_H,\ze 0) <0$};
\node at (-3,-3) {\small $R_{GB} ^2 (r_H,\ze 0) >0$};
				\end{tikzpicture}
	 \caption{GB invariant (in units of mass)  of a Kerr-Newman (KN) BH with mass $M$, angular momentum $J$ and charge $Q_e$, evaluated at the horizon's poles ($r=r_H$ and $\theta=0$), as a function of the dimensionless parameters $j=J/M^2$ and $q=Q_e/M$. Near extremality, $R^2_{GB}<0\ze$.}
	 \label{F3.1}
	\end{figure}
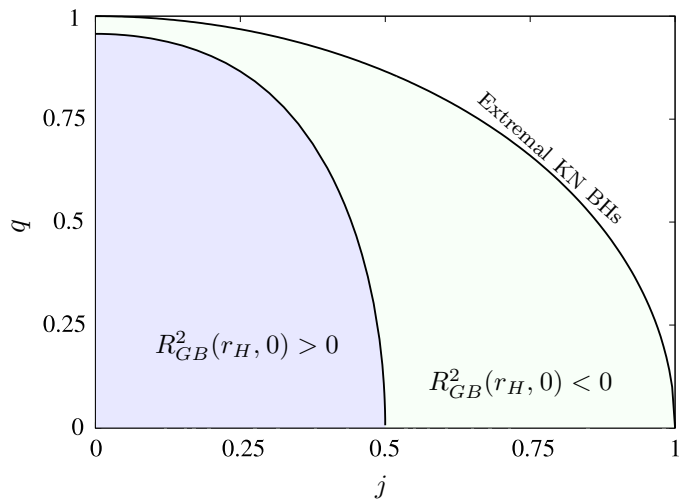

	Enlarging the model to include charged BHs, however, GB$^-$ scalarization ceases to rely solely on rotation. This can already be illustrated in electro-vacuum GR. The Kerr-Newman solution develops a negative GB invariant for either sufficiently large dimensionless angular momentum $j$ or sufficiently large dimensionless charge $q$. Thus, sufficiently near extremality, Kerr-Newman BHs develop regions with $R^2_{\rm GB}<0$ -- Fig.~\ref{F3.1}. One may expect that the boundary of the region with $R^2_{\rm GB}<0$ marks the onset of the solutions prone to GB$^-$ scalarization, as for the Kerr case~\cite{hod2020onset}. We shall confirm this expectation below, explicitly constructing some of the GB$^{-}$ scalarized Kerr-Newman solutions and comparing them with the corresponding GB$^{+}$ scalarized solutions.

	The previous paragraph's discussion shows that, within electro-vacuum BHs, GB$^-$ scalarization can be spin-induced or charge-induced (or both). Let us remark, however, that such GB charge-induced scalarization is different from the scalarization of charged BHs (\textit{cf.} Ch.~\ref{C2}), where the non-minimal coupling occurs between the scalar field and the Maxwell field, with the GB term (or any curvature corrections) being absent.  

	Additionally, the observations mentioned above on the GB$^-$ scalarization of Kerr-Newman BHs show this process occurs even for spherically symmetric, non-spinning Reissner-Nordstr\"om BHs~\cite{brihaye2019hairy}. Moreover, the negative GB invariant always occurs near the horizon. One may ask whether these features are generic. Is any charged BH model prone to GB$^-$ scalarization sufficiently close to the maximally allowed charge? Moreover, is the $R^2_{\rm GB}<0$  region supporting the scalarization always occurring in the immediate vicinity of the horizon? Interestingly, neither of these features is generic, as we shall illustrate by considering two alternative models of charged (spherical) BHs.
	
	\bigskip

	This chapter is based on the work publised in \cite{herdeiro2021aspects} and organized as follows. In Sec.~\ref{S3.1} we detail the Einstein-Maxwell-Scalar-Gauss-Bonnet model and its equations of motion. A general discussion on the tachyonic instability associated with scalarization is given in Sec.~\ref{S3.2} (see also Sec.~\ref{S1.2}), followed by the relevant physical quantities to describe the scalarized BHs. In Sec.~\ref{S3.3} we consider the GB$^\pm$ scalarization of the RN BH, first discussing the sign of the GB invariant for the electro-vacuum RN BH and then constructing both the linear scalar clouds and non-linear scalarized BHs. In Sec.~\ref{S3.4} we consider the GB$^\pm$ scalarization of the Kerr-Newman BH. After a brief discussion on the sign of the GB invariant, we discuss the construction of the solutions and provide a sample of numerical results, focusing on the GB$^-$ scalarization case. In Sec.~\ref{S3.5} we briefly consider Einstein-Maxwell-dilaton (Sec.~\ref{S3.5.1}) and Einstein-non-Abelian (Sec.~\ref{S3.5.2}) BHs, which yield two valuable lessons concerning GB$^\pm$ scalarization of charged BHs. We conclude with a brief discussion and final remarks in Sec.~\ref{S3.6}. 
%
	\section{The Einstein-Maxwell-Scalar-GB model}\label{S3.1}
%
	We wish to consider the Einstein-Maxwell-Scalar-GB (EMSGB) model, described by the following action
		\begin{equation}\label{E3.1.1}
		 \mathcal{S}_{\rm EMSGB}=\frac{1}{4} \int d^4 x\ze \sqrt{-g} 
\Big[ R -F_{\mu \nu}F^{\mu \nu}-2\ze \phi _{,\mu}\phi ^{,\mu} +\epsilon \ze \Lambda^2  f(\phi) R^2_{\rm GB} \Big] \ ,
		\end{equation}
	where $R$ is the Ricci scalar with respect to the spacetime metric $g_{\mu\nu}$, $R^2_{\rm GB}$  is the GB invariant
		\begin{equation}\label{E3.1.2}
		 R^2_{\rm GB} \equiv R_{\alpha\beta\mu\nu}R^{\alpha\beta\mu\nu}-4  R_{\alpha\beta} R^{\alpha\beta}+4R^2 \ ,
		\end{equation}  
	with $R_{\alpha\beta\mu\nu}$ the Riemann tensor, $R_{\alpha\beta}$ the Ricci tensor, $F_{\mu \nu}=A_{\nu\, ,\mu}-A_{\mu\, , \nu}$ is the Maxwell field strength tensor where $A_\mu$ is the $\textbf{U}(1)$ gauge potential, $f(\phi) $ is a coupling function of the real scalar field $\phi$ to the GB invariant, $\Lambda $ is a constant of the theory with dimension of length and $\epsilon=\pm 1$ is chosen for GB$^\epsilon$ scalarization.

	Varying the action~\eqref{E3.1.1} with respect to the metric tensor $g_{\mu \nu}$, gives the Einstein field equations,
		\begin{equation}\label{E3.1.3}
		 R_{\mu \nu}-\frac{1}{2}g_{\mu \nu} R =2\ze T_{\mu \nu}^{\rm (eff)}  \ .
		\end{equation}
	The effective energy-momentum tensor $T_{\mu \nu}^{\rm (eff)} $ has three distinct components:
		\begin{equation}\label{E3.1.4}
		 T_{\mu\nu}^{\rm (eff)}= T_{\mu\nu}^{\rm (S)}+ T_{ \mu\nu}^{\rm (M)}+ T_{ \mu\nu}^{\rm (GB)} \ ,
		\end{equation}
	consisting of the (pure) scalar and Maxwell parts, respectively,  
		\begin{equation}\label{E3.1.5}
		 T_{\mu \nu }^{\rm (S)}= \phi_{,\mu}\ze \phi_{,\nu} -\frac{1}{2}\ze g_{\mu\nu}\ze  \phi_{,\alpha}\ze\phi^{,\alpha} \ ,\qquad T_{\mu \nu }^{\rm (M)}= F_{\mu \alpha}\ze F_{\nu}^{\ \alpha}-\frac{1}{4}F_{\alpha\beta}F^{\alpha\beta} \ ,
		\end{equation}
	and a third contribution due to the Scalar-GB term in \eqref{E3.1.1}
		\begin{equation}\label{E3.1.6}
		 T_{\mu\nu}^{\rm (GB)}=  - 2\ze \epsilon\ze \Lambda^2 P_{\mu\gamma \nu \alpha}\nabla^\alpha \nabla^\gamma f \ ,
		\end{equation}
	where
		\begin{align}\label{E3.1.7}
		P_{\alpha\beta\mu\nu}& = -\frac14 \varepsilon_{\alpha\beta\rho\sigma} R^{\rho\sigma\gamma\delta} \varepsilon_{\mu\nu\gamma\delta} \nonumber\\
		 & = R_{\alpha\beta\mu\nu}+ g_{\alpha\nu}R_{\beta\mu} - g_{\alpha\mu} R_{\beta\nu} + g_{\beta\mu} R_{\alpha\nu}-g_{\beta\nu} R_{\alpha\mu} +\frac12 \left( g_{\alpha\mu}g_{\beta\nu} - g_{\alpha\nu}g_{\beta\mu}\right) R \ ,
		\end{align}  
	and $ \varepsilon_{\alpha\beta\rho\sigma}$ is the Levi-Civita tensor. The scalar field equation is 
		\begin{equation}\label{E3.1.8}
		 \nabla^2 \phi +\epsilon\ze \frac{\Lambda^2}{4}\hat{f} R^2_{\rm GB}=0 \ ,
		\end{equation} 
	while the (source-free) Maxwell equations have the usual form
		\begin{equation}\label{E3.1.9}
		 \nabla_\mu F^{\mu \nu}=0 \ .
		\end{equation} 

%
		\section{GB$^\epsilon$ scalarization of electro-vacuum solutions}\label{S3.2}
%
	Spontaneous scalarization manifests itself at the linear level as a tachyonic instability (\textit{cf.} Sec.~\ref{S1.2}). Let us assume that $\phi=0$ solves~\eqref{E3.1.8}, which will hold for a class of coupling functions. Then, the field equations reduce to those of electro-vacuum GR, and the corresponding solutions provide solutions of the full model~\eqref{E3.1.1} as well. In particular, the Kerr-Newman geometry will be a solution to this model. Thus, for concreteness, we shall refer to the scalarization of the Kerr-Newman solution in the following; but a similar discussion would hold for any GR electro-vacuum solution.

	Next, we consider scalar perturbations of the Kerr-Newman solution within the full model~\eqref{E3.1.1}. Assuming a small-$\phi$ expansion for the coupling function
			\begin{equation}\label{E3.1.10}
			 f(\phi)=f(0)+ \frac{1}{2} \hat{\hat{f}}(0) \ze \delta \phi^2+\mathcal{O}(\delta\phi^3) \ ,
			\end{equation}
	the linearized scalar field equation~\eqref{E3.1.8} around the Kerr-Newman solution becomes 
			\begin{equation}\label{E3.2.11}
			 (\Box-\mu_{\rm eff}^2)\ze \delta \phi =0 \ , \qquad {\rm where} \qquad  \mu_{\rm eff}^2=-\epsilon\ze\frac{ \Lambda^2}{4} \hat{\hat{f}}(0) R^2_{\rm GB} \  , 
			\end{equation} 
	where $\Box$ and $R^2_{\rm GB}$ are computed for the scalar-free Kerr-Newman solution.

	If $\mu_{\rm eff}^2$ is not strictly positive, the scalar field possesses a (spacetime dependent) tachyonic mass. Wherever this tachyonic mass is supported, such a region potentially supports a spacetime instability, precisely the GB scalarization. To simplify the discussion, we assume without any loss of generality that $\hat{\hat{f}}_{GB}(0)$ is strictly positive. Then the condition $\mu_{\rm eff}^2<0$ is equivalent to
			\begin{equation}\label{E3.2.12}
			 \epsilon\ze R^2_{\rm GB}>0~.
			\end{equation}
	When this condition is obeyed in some region(s) outside the BH horizon, GB$^\epsilon$ scalarization is triggered.

	If the Kerr-Newman BH reduces to a Schwarzschild BH of mass $M$, then
			\begin{equation}\label{E3.2.13}
			 \mu_{\rm eff}^2=  - \epsilon\ze \frac{\Lambda^2}{4}\hat{\hat{f}}(0) \frac{48 M^2}{r^6} \ , 
			\end{equation}
	and only GB$^{+}$ scalarization is possible.
%
		\subsection{Physical quantities of interest for scalarized BHs}\label{S3.2.1}
	When the above instability is present, there is also a different class of solutions for the model~\eqref{E3.1.1}, besides the electro-vacuum ones. These are the scalarized solutions. We are interested in the case of stationary BHs. These solutions possess three global charges: the mass $M$, the electric charge $Q_e$ and the angular momentum $J$. For the present solutions, there is also a ``scalar charge'' $Q_\phi$, which is not associated with a conservation law. There are also several relevant horizon quantities: the Hawking temperature $T_H$, the horizon area $A_H$, the entropy $S_H$ and horizon angular velocity $\Omega_H$. Unlike standard GR, the BH entropy is the sum of two terms,
			\begin{equation}\label{E3.2.14}
			 S_H=S_{\rm E}+S_{\rm SGB}\ ,\qquad{\rm with} \qquad 
S_{\rm E}=\frac{1}{4}A_H\ , \qquad  S_{\rm SGB}= \frac{\epsilon}{2}\Lambda^2 \int_{H} d^2 x \sqrt{h}\ze f\ze {\rm  R}^{(2)} \ ,
			\end{equation}
	where ${\rm  R}^{(2)}$ is the Ricci scalar of the induced horizon metric $h$. The solutions satisfy the Smarr law
			\begin{equation}\label{E3.2.15}
			 M=2\ze \Omega_H J+2\ze T_H S_H+\Psi_e\, Q_e+ M_\phi \ ,
			\end{equation}
	where $\Psi_e$ is the electrostatic potential and $M_\phi$ is a contribution of the scalar field
			\begin{equation}\label{E3.2.16}
			 M_\phi= \frac{1}{2} \int d^3 x \ze \, \sqrt{-g}\ze \phi _{,\mu} ^2\ .	
			\end{equation}
	Also, the solutions satisfy the first law of BH thermodynamics
			\begin{equation}\label{E3.2.17} 
			 dM=T_H dS_H +\Omega_H dJ +\Psi_e\, dQ_e \ ,
			\end{equation}
	in which there is no contribution from the scalar field.
	
	In this work, we shall focus on the quadratic coupling function,\footnote{For spherical symmetry, we have also explored the exponential coupling studied in~\cite{doneva2018new,doneva2018charged}
	\begin{equation}
	f_{GB} (\phi) = \frac{1-e^{-6\phi ^2}}{12}\ ,
	\end{equation}
 and observed that the behaviour is qualitatively similar.}
			\begin{equation}\label{E3.2.18}
			 f_{GB} (\phi)=\frac{\phi^2}{2} \ ,
			\end{equation}  
	which is the simplest choice of $f_{GB}$ that guarantees that $\phi=0$ satisfies the scalar equation~\eqref{E3.1.8}.

	It is helpful to observe that the equations of the model are invariant under the transformation
			\begin{equation}\label{E3.2.19}
			 r\to \lambda\ze r \ , \qquad \Lambda  \to \lambda\ze \Lambda \ ,
			\end{equation}
	with $r$ the radial coordinate and $\lambda>0$ an arbitrary positive constant. Only quantities invariant under \eqref{E3.2.19} ($e.g.$ $\frac{Q_e}{M}$ or $\frac{Q_e}{\Lambda}$) have a physical meaning. The reduced quantities \eqref{E2.1.15a}-$(2.1.17)$ still hold, to which we add the spin to mass ratio $j$
			\begin{equation}\label{E3.2.20}
			 j\equiv \frac{J}{M^2} \ ,\qquad \text{(spin to mass ratio)}
			\end{equation}
	which will be considered in what follows.
%
	\section{GB$^\epsilon$ scalarization of Reissner-Nordstr\"om BHs}\label{S3.3}
%
	Let us start by considering the spinless limit of the Kerr-Newman family, the RN BH. The corresponding metric and gauge field can be written from \eqref{E1.5.40} (see $e.g.$~\cite{townsend1997black})
	where 
		\begin{equation}\label{E3.3.21}
		 \sigma=1 \ , \qquad  N=1-\frac{2M}{r}+\frac{Q_e ^2}{r^2}\ , \qquad  V=\frac{Q_e}{r} \ .
		\end{equation}
	This BH possess an event horizon at
		\begin{equation}\label{E3.3.22}   
		 r=r_H=M+\sqrt{M^2-Q_e ^2} \ .
		\end{equation}
	Thus, $0\leqslant q\leqslant 1$ and $q=1$ for the extremal RN BH.
 
	The GB invariant of the RN metric reads
		\begin{equation}\label{E3.3.23}   
		 R^2_{\rm GB}   = \frac{8}{r^8}\Big[6M^2r^2-12\ze Q_e ^2Mr+5\ze Q_e^4 \Big] \ .
		\end{equation}
	For $Q_e\neq 0$ this always becomes negative for some region with $r>0$. The latter region, however, is cloaked by a horizon unless the largest root of the quadratic equation in (the square brackets in)~\eqref{E3.3.23} exceeds $r_H$. This condition is
		\begin{equation}\label{E3.3.24}
		 q\ze Q_e \left(1+\frac{1}{\sqrt{6}}\right)>r_H \ .
		\end{equation}
	Using~\eqref{E3.3.22} one can easily show that is possible for
		\begin{equation}\label{E3.3.25}   
		 q>q_{c}\simeq 0.957 \ . 
		\end{equation}
	Thus, for $q_c<q\leqslant 1$, a RN BH can undergo GB$^-$ scalarization.
%
		\subsection{The linear scalar clouds}\label{S3.3.1}
%
	At the onset of the tachyonic instability, the linearized scalar field equation \eqref{E3.2.11} on the RN background allows solutions known as \textit{scalar clouds}. These occur for a discrete set of RN solutions, each corresponding to a particular harmonic scalar field mode. To see this, we perform a harmonic decomposition of the scalar field as \eqref{E3.3.26}.

	For a RN BH background \eqref{E3.3.21}-\eqref{E3.3.23}, the linearized scalar equation \eqref{E3.2.11} becomes a radial equation
			\begin{equation}\label{E3.3.27a}
			 \Big( r^2 N\, U_\ell '\Big)'=\ell(\ell+1)\ze U_\ell-\epsilon\ze  \frac{\Lambda^2}{r^2} \left( \frac{12 M^2}{r^2}+\frac{10\ze Q_e^4}{r^4}-\frac{24\ze  MQ_e^2}{r^3}\right)~.
			\end{equation}
	This equation has the following asymptotic solutions: near the horizon
			\begin{equation}\label{E3.3.27}
			 U_\ell=u_0+\frac{r_H}{Q_e^2-r_H^2}\left[
-  \ell(\ell+1)+\epsilon \ze \frac{\Lambda^2}{r_H^2} \left(3-\frac{6\ze Q_e^2}{r_H^2}+\frac{Q_e^4}{r_H^4}\right)
\right]u_0(r-r_H)+\cdots \ , 
			\end{equation}
	where $u_0$ is an arbitrary non-zero constant which, in numerics, we set to $1$; and near spatial infinity
			\begin{equation}\label{E3.3.28}
			U_\ell = \frac{Q_\phi}{r^{\ell+1} } +  \cdots \ , 
			\end{equation}
	where $Q_\phi$ is the scalar charge for $\ell=0\ze$.
	
	Just as before (Sec.~\ref{C2}), solving \eqref{E3.3.27a} with the above asymptotic behaviours \eqref{E3.3.27}-\eqref{E3.3.28} can be viewed as an eigenvalue problem. In this chapter, we shall report results on nodeless spherically symmetric fundamental solutions only\footnote{Similar solutions are likely to exist for any other values of the quantum numbers,
some preliminary results being found for the $\ell=1$, $m=n=0$ case. An investigation of $Q_e=0$, $\ell=1$ static solutions has been reported in~\cite{collodel2020spinning}.}, $i.e.$ with $\ell=m=n=0\ze$.

\begin{figure}[H]
		\centering
			 	  \begin{picture}(0,0)
				  		 \put(112,-10){\small $q$}
				 		 \put(112,47){$\scriptstyle q$}
				  		 \put(30,20){\small $\epsilon = +1$}
				  		 \put(120,120){$\scriptstyle \epsilon\, =\, -1$}
				  		\put(-6,70){\begin{turn}{90}{\small $\Lambda /M$}\end{turn}}
				  		\put(37,85){\begin{turn}{90}{$\scriptstyle \Lambda /M$}\end{turn}}
	   			\end{picture}
		 \includegraphics[scale=0.60]{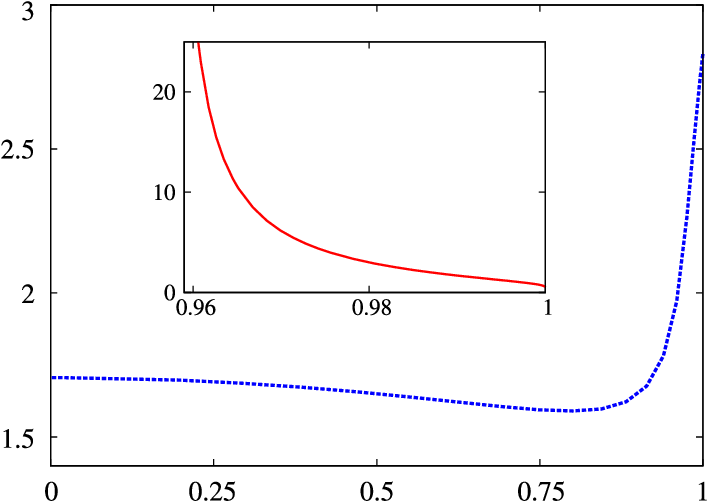}\hfill
		 			 	  \begin{picture}(0,0)
				  		 \put(100,-10){\small $\log _{10} \frac{r}{r_H}$}
				 		 \put(132,45){$\scriptstyle \log _{10} \frac{r}{r_H}$}
				  		 \put(45,18){\small $\epsilon = -1$}
				  		 \put(40,130){\small $\epsilon = +1$}
				  		 \put(148,115){$\scriptstyle q\, =\, 0.55$}
				  		 \put(148,68){$\scriptstyle \epsilon\, =\,+1$}
				  		 \put(165,25){\small $q=0.976$}
				  		\put(-6,76){\begin{turn}{90}{\small $\phi$}\end{turn}}
				  		\put(72,95){\begin{turn}{90}{$\scriptstyle \phi$}\end{turn}}
				  		\put(23,55){\begin{turn}{90}{$\scriptstyle {\rm Event\ horizon}$}\end{turn}}
	   			\end{picture}
		 \includegraphics[scale=0.61]{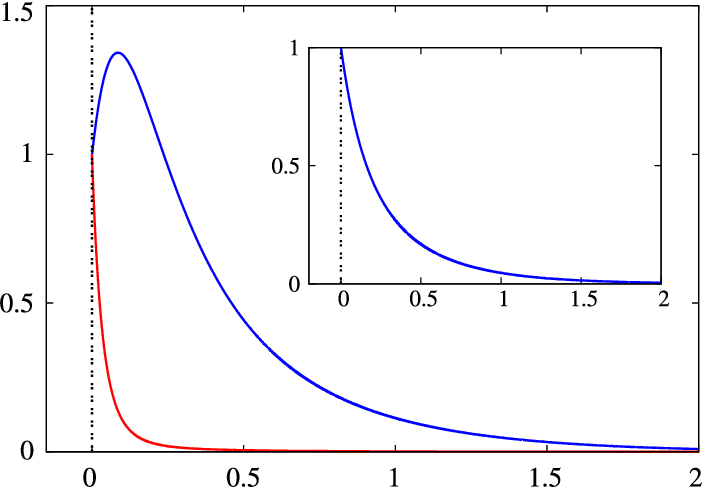}
		 \caption{(Left panel) dimensionless ratio $\Lambda/M$ of the set of  RN  solutions supporting the $\ell=m=n=0$ scalar cloud $vs.$ the reduced electric charge $q$ for $\epsilon=\pm 1$. (Right panel) typical radial profiles of the spherical, nodeless scalar clouds on a RN BH background.}
		 \label{F3.2}
		\end{figure}
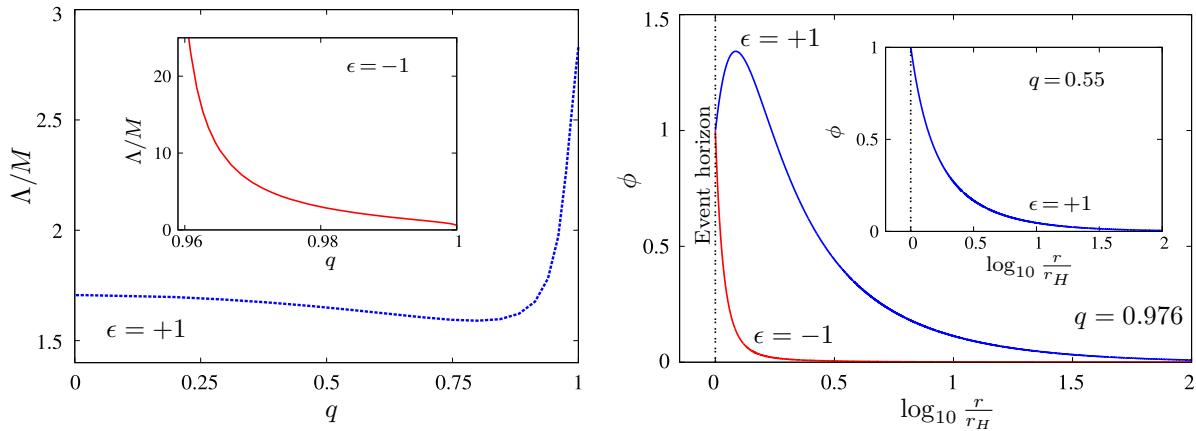

	For a given cloud's quantum numbers, taking $\Lambda$ as a fixed scale set in the action and fixing the reduced charged $q$, the radial equation has a solution for a specific dimensionless ratio $\Lambda/M$. For instance, for $\epsilon=+1$,  $\ell=m=n=0$ and $q=0$ the selected value is $\Lambda/M\sim 1.704$, corresponding to the initial point of the blue dashed curve in  Fig.~\ref{F3.2} (left panel). This is the zero mode of the GB$^+$ instability of Schwarzschild. It selects a mass scale. Smaller masses (larger $\Lambda/M$) describe BHs unstable against scalarization; larger masses  (smaller $\Lambda/M$) correspond to stable BHs.
	The variation of $\Lambda/M$ with increasing $q$ can be interpreted as follows. GB scalarization of Schwarzschild BHs may be attributed to a repulsive gravitational effect of the GB term, which only becomes dominant for sufficiently small BHs (in terms of $\Lambda$). Adding electric charge introduces two competing effects. On one hand, the electric charge provides a repulsive gravitational effect for RN BHs. Such facilitates scalarization, making it available for larger BHs (larger $M$, smaller $\Lambda/M$). On the other hand, the repulsive gravitational effect of the GB invariant, which is at the source of the scalarization phenomenon, becomes suppressed (and eventually the GB term even changes sign in some region) when increasing $q$. This suppresses scalarization, making it available only for smaller BHs (smaller $M$, larger $\Lambda/M$). The trend observed in the blue dashed curve in the main panel of  Fig.~\ref{F3.2} (left panel) suggests that for small $q$, the RN charge repulsion dominates and for large $q$, the GB charge suppression becomes dominant. In addition, the inset gives the behaviour for $\epsilon=-1$. In this case, as $q$ increases, in the allowed (large $q$) interval, the GB behaviour dominates, and due to the opposite sign, it provides an ever more significant repulsive contribution, thus facilitating GB$^-$ scalarization, which therefore is available for larger masses. We also remark that, for GB$^-$ scalarization, the ratio $\Lambda/M$ appears to diverge as $q \to q_{c}\ze$, while it stays finite as $q\to 1$.

	The profiles of typical scalar clouds are shown in Fig.~\ref{F3.2} (right panel).

	For $\epsilon=+1$ and moderate $q$, we recover the picture found in the Schwarzschild case, $Q_e=0\ze $: a monotonically decreasing profile starting with some finite value at the horizon (see the inset). This is also true for the $\epsilon=-1$ case (red curve in the main panel).

	Nevertheless, for $q>q_{c}$ and $\epsilon=+1$, a new qualitative behaviour emerges: the maximal value of the scalar cloud can be attained outside the horizon\footnote{This feature can be explained by studying \eqref{E3.3.27}~\cite{brihaye2019spontaneous}.} -- see the blue curve in the main panel.
%
		\subsection{The non-linear spherically symmetric scalarized BHs}\label{S3.3.2}
%
	The linear scalar clouds just discussed can be continued to the non-linear regime. Their backreaction originates scalarized BHs. 
We shall now discuss their construction for the case of the spherical, nodeless scalar clouds. 

	The ansatz to obtain the scalarized BH solutions is~\eqref{E1.5.40}, together with a radial scalar field $\phi\equiv \phi(r)$. The Maxwell equation~\eqref{E3.1.9} yields the first integral
			\begin{equation}\label{3.3.30}
			 V'=-\frac{Q_e\,\sigma}{r^2} \ .
			\end{equation}
	This introduces the electric charge measured at infinity, $Q_e$. The scalar field satisfies the equation
			\begin{equation}\label{E3.3.31}
			 \phi''+\left( \frac{2}{r}+\frac{N'}{N}+\frac{\sigma'}{\sigma}\right)\phi'- \frac{\epsilon\ze \Lambda^2}{r^2 N\sigma}\Big[
(3-5N)N'\sigma'+\sigma \big( (1-N)N''-N'^{\, 2})+2(1-N)N\sigma''\Big]\phi=0\ ,
			\end{equation}
	while the equations for the metric functions $N$ and $\sigma$ are too complex and shall not be displayed here.

	We are interested in BH solutions with an event horizon located at $r=r_H>0$. The equations of the model are subject to the following boundary conditions.
			\begin{align}\label{E3.3.32}
			 & N\big|_{r_H}=0\ , \quad \sigma\big|_{r_H}=\sigma_0\ , \quad \phi\big|_{r_H}=\phi_0\ ,\quad V\big|_{r_H}=0 \ ;\nonumber\\
			 & \quad N\big|_{+\infty}=1\ , \quad \sigma\big|_{+\infty}=1 \ ,  \quad \phi\big|_{+\infty}=0\ , \quad V\big|_{+\infty}=\Psi _e \ ,
			\end{align}
	where $\sigma_0$ and $\phi_0$ are constants fixed by numerics, and $\Psi_e$ is the electrostatic potential at infinity. The horizon data fix the Hawking temperature, the horizon area and the entropy of the solutions,
			\begin{equation}\label{E3.3.33}
			 T_H=\frac{\sigma_0\ze N'(r_H)}{4\pi}\ , \qquad A_H=4\pi r_H^2 \ , \qquad S=\pi r_H^2+\epsilon\ze \Lambda^2 f_{GB} (\phi_0) \ .
			\end{equation}
	A local solution compatible with this asymptotics can be constructed both at the horizon $\big($power series \eqref{E2.1.7}$\big)$ and infinity (power series in $1/r$). For example, the first terms in the far-field expression of the solutions read
			\begin{equation}\label{E3.3.34}
			 N= 1-\frac{2M}{r}+\frac{Q_e^2+Q_\phi^2}{r^2}+\cdots \ , \quad \phi=\frac{Q_\phi}{r}+\cdots \ , \quad V=\Psi_e-\frac{Q_e}{r}+\cdots\ , \quad \sigma=1-\frac{Q_\phi ^2}{2\ze r^2}+\cdots \ .
			\end{equation}
			\subsubsection*{Numerical results}
	With the details just laid out, we have numerically constructed the nonlinear continuation of the scalar clouds solving the full equations of the EMSGB model for both signs of $\epsilon$. 

\begin{figure}[H]
				 \centering
				 		 \begin{picture}(0,0)
				  		 \put(115,-10){\small $q$}
				  		 \put(55,36){\small $\epsilon = +1$}
				  		 \put(24,134){\small $Q_e = 0.0$}
				  		 \put(55,115){\small $\frac{Q_e}{\Lambda}=0.18$}
				  		 \put(98,133){\small $\frac{Q_e}{\Lambda}=0.34$}
				  		 \put(52,90){$\scriptstyle Q_e\, =\, 0$}	
				  		 \put(60,78){$\scriptstyle \frac{Q_e}{\Lambda}\, =\, 0.18$}	
				  		 \put(108,88){$\scriptstyle \frac{Q_e}{\Lambda}\, =\, 0.34$}	
				  		 \put(100,58){$\scriptstyle \frac{Q_e}{\Lambda}\, =\, 0.5$}		
				  		 \put(155,130){\small $\frac{Q_e}{\Lambda}=0.5$}
				  		 \put(112,38){$\scriptstyle \frac{Q_e}{\Lambda}\, =\, 0.52$}				  		 
				  		 \put(155,80){\small $\frac{Q_e}{\Lambda}=0.52$}
				  		\put(-6,74){\begin{turn}{90}{\small $t_H$}\end{turn}}
				  		\put(28,60){\begin{turn}{90}{$\scriptstyle  a_H$}\end{turn}}
				  		\put(100,18){$\scriptstyle q$}
	   			\end{picture}
			 	\includegraphics[scale=0.61]{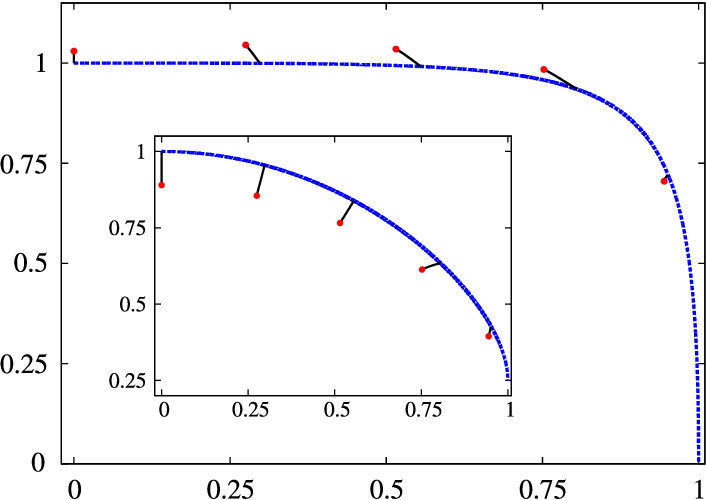}\hfill
			 					 		 \begin{picture}(0,0)
				  		 \put(115,-10){\small $q$}
				  		 \put(55,36){\small $\epsilon = -1$}
				  		 \put(60,136){\small $\frac{Q_e}{\Lambda}=0.22$}
				  		 \put(100,126){\small $\frac{Q_e}{\Lambda}=0.29$}
				  		 \put(155,120){\small $\frac{Q_e}{\Lambda}=0.36$}				  		 
				  		 \put(60,70){\begin{turn}{40}{$\scriptstyle \frac{Q_e}{\Lambda}\, =\, 0.22$}\end{turn}}
				  		 \put(166,97){\small $\frac{Q_e}{\Lambda}=0.43$}	  		 
				  		 \put(166,79){\small $\frac{Q_e}{\Lambda}=0.58$}
				  		 \put(170,65){\small $\frac{Q_e}{\Lambda}=0.72$}
				  		\put(-6,74){\begin{turn}{90}{\small $t_H$}\end{turn}}
				  		\put(26,60){\begin{turn}{90}{$\scriptstyle  a_H$}\end{turn}}
				  		\put(92,16){$\scriptstyle q$}
	   			\end{picture}
 				 \includegraphics[scale=0.61]{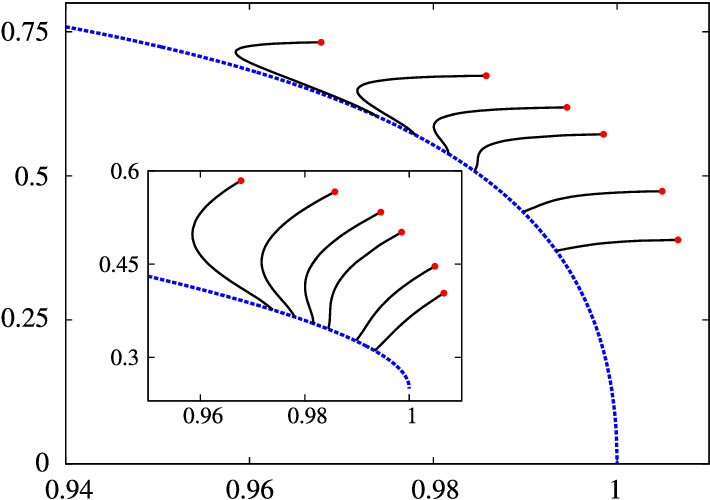} 
				 \caption{Reduced charge \textit{vs.} reduced temperature (main panels) and reduced charge \textit{vs.} reduced horizon area (insets) diagrams, in units set by the mass, for several illustrative families of GB$^\epsilon$ scalarized RN solutions, for both $\epsilon= +1$ (left panel) and $\epsilon= -1$ (right panel). The branches of scalarized solutions bifurcate from the electro-vacuum RN BHs (blue line) and terminate in critical configurations (red circles).}
				 \label{F3.3}
				\end{figure}
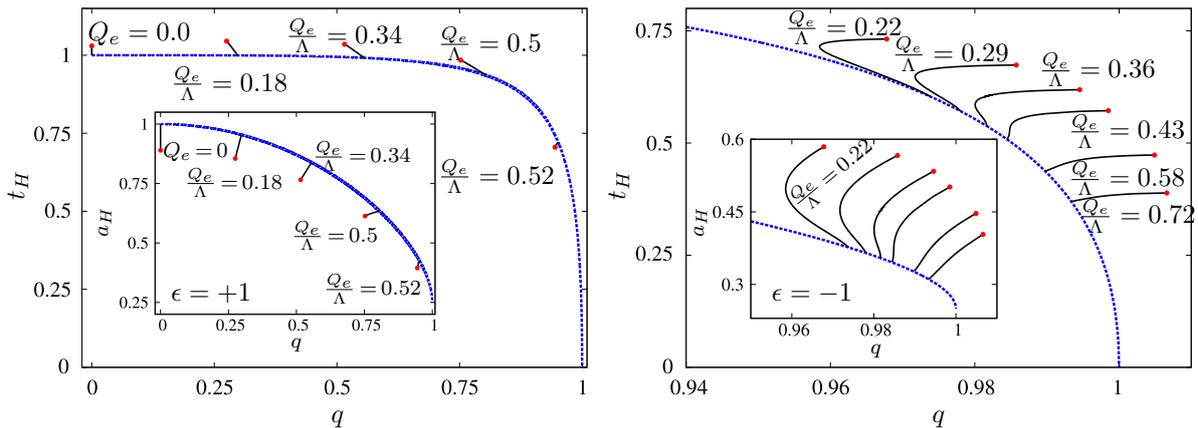

	Technically, the construction of the scalarized BHs is a one-parameter shooting problem in terms of the value of the scalar field at the horizon $\phi_0$. The input parameters are $r_H,\, Q_e$ and $\Lambda$. Fixing the length scale $\Lambda$ leads to a two-dimensional parameter space for the problem. The numerical results for several values of the ratio $Q_e/\Lambda$ are shown in Fig.~\ref{F3.3}.
	Fig.~\ref{F3.3} shows that for a given ratio $Q_e/\Lambda$ and both values of $\epsilon$, one finds a continuum of solutions that bifurcate from the corresponding RN BH supporting a scalar cloud with these parameters. This line has a finite extent, ending in a critical configuration where the numerical process fails to converge. A general explanation for this behaviour can be traced back to the radicand of a square root in the horizon expansion of the scalar field that vanishes as the critical set is approached. The latter is a generic feature of GB-Scalar models. An exception here is the $\epsilon=+1$ solutions emerging from RN BHs with $q>q_c$, in which case the critical configurations seem
to possess a curvature singularity for some radius outside the event horizon (see~\cite{brihaye2019spontaneous} for a discussion).
 
	From Fig.~\ref{F3.3} one may highlight two qualitatively different features when comparing $\epsilon=\pm 1$. First, scalarization reduces (increases) $a_H$  for $\epsilon=+1$ ($\epsilon=-1$). Secondly, ``overcharged'' solutions with $q>1$ exist for $\epsilon=-1$ only.
%
	\section{GB$^\epsilon$ scalarization of Kerr-Newmann BHs}\label{S3.4}
%
	Let us now address the GB$^\epsilon$ scalarization of the spinning, charged electro-vacuum Kerr-Newman BH. This is a solution of the model \eqref{E3.1.1}, with the coupling \eqref{E3.2.18}, together with a vanishing scalar field, $\phi=0$. This BH is described by its ADM mass $M$, total angular momentum per unit mass $a=J/M$ and electric charge $Q_e$. In Boyer-Lindquist coordinates it introduces two radial-dependent functions $\Sigma (r)$ and $\Delta (r)$ and reads (see $e.g.$~\cite{townsend1997black})
		\begin{equation}\label{E3.4.34}
		 ds^2 = - \frac{\Delta}{\Sigma} \big( dt -a \sin^2 \theta d\varphi \big)^2 +  \Sigma \bigg( \frac{dr^2}{\Delta} + d\theta^2 \bigg) + \frac{\sin^2 \theta}{\Sigma} \Big[ a dt - \left( \Sigma+a^2\sin^2\theta \right) d\varphi \Big]^2 \ ,
		\end{equation}
	and
		\begin{equation}\label{E3.4.35}
		 A_\mu = -\frac{Q_e r}{\Sigma} \left( dt - a \sin^2 \theta d\varphi \right) \ ,
		\end{equation}
	where 
		\begin{equation}\label{E3.4.36}
		 \Delta \equiv r^2 - 2Mr + a^2 + Q_e^2 \ , \qquad  \Sigma \equiv r^2 + a^2 \cos^2 \theta \ .
		\end{equation}
	The event horizon of this solution is located at
		\begin{equation}\label{E3.4.37}
		 r_H=M \left (1+\sqrt{1-j^2-q^2} \right).
		\end{equation}
	This implies the Kerr-Newman bound, $j^2+q^2 \leqslant 1$; an extremal BH saturates this bound. 
	
	The Kerr-Newman metric has a GB invariant
		\begin{align}\label{E3.4.38}
		 R^2_{\rm GB}  = &\frac{48 M^2}{\Sigma^3}\left[ 1-\frac{2a^2}{\Sigma^3}\big(3r^2-a^2 \cos^2\theta\big)^2\cos^2\theta
\right]\nonumber\\
		 & +\frac{8Q_e ^2}{\Sigma^6}\bigg\{ r^4(5Q_e ^2-12Mr +a^2\cos^2\theta \Big[2r^2(-19Q_e ^2+60 Mr)+5a^2(Q_e ^2-12Mr)cos^2\theta\Big] \bigg\}\ .
		\end{align}

\begin{figure}[h!]
		\centering
					 	 \begin{picture}(0,0)
						\put(18,87){\small $\frac{q}{j} = 3.07$}
						\put(17,67){\begin{turn}{10}{\small $\frac{q}{j} = 1.73$}	\end{turn}}		  		
				  		 \put(116,-10){\small $\theta$}
				  		\put(18,46){\small $\frac{q}{j} = 1$}
				  		 \put(70,36){\small $\frac{q}{j} = 0$}
				  		 \put(148,138){\small $q^2 + j^2 = 0.563$}				  		 
				  		\put(-8,60){\begin{turn}{90}{\small $R_{GB} ^2 (r_H, \theta)$}\end{turn}}
	   			\end{picture}
		\includegraphics[scale=0.63]{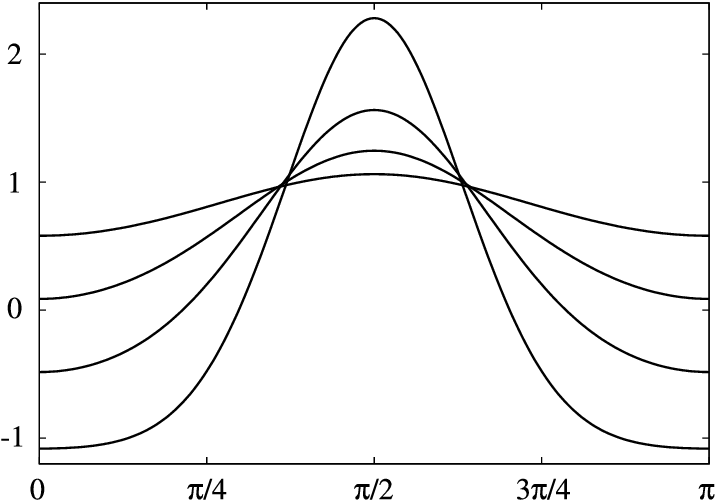}
		 \caption{The GB invariant at the horizon (in units of mass) as a function of the $\theta-$coordinate for several Kerr-Newman BHs.}
		 \label{F3.4}
		\end{figure}
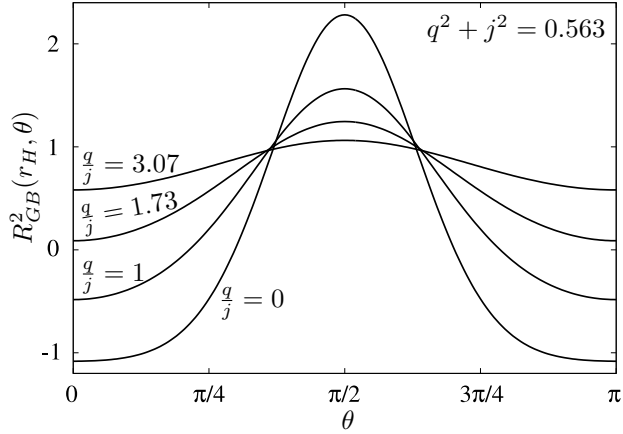

	We have studied\footnote{An alternative expression for \eqref{E3.4.38}, in terms of $P_1\equiv \big(1+\sqrt{1-j^2-q^2}\big)\frac{r}{r_H}$ and $P_2\equiv j \cos \theta$, is
		\begin{equation*}
		 R^2_{\rm GB} = \frac{48}{M^4 }\frac{1}{(P_1^2+P_2^2)^3}
\bigg\{ 1-\frac{2}{(P_1^2+P_2^2)^3} \Big[ P_2^2( 3P_1^2-P_2^2)^2 +q^2 \Big(	P_1(P_1^4-10P_1^2P_2^2+5P_2^4)-\frac{q^2}{12}\big(5P_1^4-38 P_1^2P_2^2+5P_2^4 \big)\Big)\Big]\bigg\}\ ,
		\end{equation*}
	a form which has been employed in our study.} the sign of this quantity as a function of the parameters $(j, q)$ -- Fig.~\ref{F3.1}, observing that the qualitative picture found in the Kerr ($q=0$) or RN ($j=0$) cases is still valid for a Kerr-Newman BH. Kerr-Newman BHs with $R^2_{\rm GB}<0$ have the potential to be scalarized for both signs of $\epsilon$. While $R^2_{\rm GB}$ is positive for large values of the radial coordinate, its sign close to the event horizon depends on the value of $(j,q)$. That is, for fixed $j$ (or $q$), the GB invariant $R^2_{\rm GB}$ always becomes negative in a region outside the horizon, for large enough values of $q$ (or $j$). In Fig. \ref{F3.1} we show the region in the $(j,q)$-domain where the GB invariant takes a negative sign at the poles of the horizon. In the presence of rotation, this region is located around the poles of the horizon, $\theta=0,\pi$ -- Fig. \ref{F3.4}.
%

%
		\subsection{Construction of the scalarized Kerr-Newman BHs}\label{S3.4.1}
%
	To construct the GB$^\epsilon$ scalarized Kerr-Newman BHs, we shall use numerical ansatz \eqref{E1.5.41}\footnote{Kerr-Newman BH can also be written in this coordinate system. The corresponding expressions in the Kerr limit can be found in~\cite{herdeiro2015construction}.} (see also \cite{cunha2019spontaneously,herdeiro2021spin}), supplemented with a non-zero gauge field \eqref{E1.5.45}.

	Setting $A_\varphi=V=0$ in \eqref{E1.5.45} results in the spontaneously scalarized Kerr BHs in~\cite{cunha2019spontaneously,herdeiro2021spin}, albeit with a different coupling function. The limit $F_W=A_\varphi=0$ results in the scalarized RN BHs~\cite{astefanesei2019einstein,fernandes2019spontaneous,herdeiro2018spontaneous} discussed above, albeit for a different radial coordinate.

	The general problem is solved subject to the following boundary conditions. Asymptotic flatness requires
			\begin{equation}\label{E3.4.39}
			 \lim_{r\rightarrow +\infty}{\mathcal{F}_i}=0\ , \qquad {\rm and} \qquad \lim_{r\rightarrow +\infty}{\phi} =\lim_{r\rightarrow +\infty}{A_\varphi}=0 \ , \qquad \lim_{r\rightarrow +\infty } V=\Psi_e  \ .
			\end{equation}
	Axial symmetry and regularity impose the following boundary conditions on the symmetry axis, $i.e.$ at $\theta=0,\,\pi$:
			\begin{equation}\label{E3.4.40}
			 \invbreve {\mathcal{F}}_i = \invbreve{\phi} =\invbreve{V}= A_\varphi=0 \ .
			\end{equation}
	Moreover, the absence of conical singularities implies also that $ F_1=F_2$ on the symmetry axis. 
 
	The event horizon is located at a constant $r=r_H>0$. Only non-extremal solutions can be studied within the metric ansatz \eqref{E1.5.41}. We introduce a new radial coordinate $x=\sqrt{r^2-r_H^2}\ze$, which simplifies the boundary conditions at the horizon and the numerical treatment of the problem.
	
	This results in the following boundary conditions at the horizon 
			\begin{equation}\label{E3.4.41}
			 F_{i\, ,x} \big|_{r=r_H}= \phi_{,x}\big|_{r=r_H} =  0 \ , \qquad F_W \big|_{r=r_H}=\Omega_H \ , \qquad A_{\varphi\, ,x} \big|_{r=r_H}=V_{,x} \big|_{r=r_H}=0 \ ,
			\end{equation}
	where the constant $\Omega_H>0$ is the horizon angular velocity. An approximate expansion of the solution compatible with these boundary conditions can easily be constructed.

	Specializing some of the aforementioned physical quantities of interest for the ansatz in use, we obtain that the following horizon data determine the Hawking temperature and the event horizon area,
			\begin{equation}\label{E3.4.42}
			 T_H=\frac{1}{4\pi r_H}e^{F_0^{(0)}(\theta)-F_1^{(0)}(\theta)} \ , \qquad A_H=2\pi r_H^2 \int_0^\pi d\theta \sin \theta~e^{F_1^{(0)}(\theta)+F_2^{(0)}(\theta)} \ ,
			\end{equation}
	with the near-horizon expansion $F_i=F_i^{(0)}(\theta)+x^2 F_i^{(2)}(\theta)+\cdots$, and $i=0,1,2$.
 
	The ADM mass $M$, angular momentum $J$, scalar ``charge'' $Q_\phi$, together with the magnetic dipole momentum $q_m$, the electrostatic potential $\Psi_e$ and  the electric charge $Q_e$ are read off from the far-field asymptotic of the metric and matter functions
			\begin{align}\label{E3.4.43}
			 & g_{tt} =-1+\frac{2M}{r}+\cdots\ ,\qquad g_{\varphi t} =-\frac{2J}{r}\sin^2\theta+\cdots\ ,\qquad \phi=-\frac{Q_\phi}{r}+\cdots \ ,\qquad\nonumber\\
			& A_\varphi=\frac{q_m \sin \theta }{r}+\cdots \ ,\qquad V=\Psi_e-\frac{Q_e}{r}+\cdots \ .
			\end{align}  
	We remark that both the metric functions and the scalar field are invariant $w.r.t.$ the transformation $\theta \to \pi-\theta$.
			\begin{figure}[H]
			 \centering
			 			\begin{picture}(0,0)
						\put(20,67){\small $Q_e =0$}	
						\put(65,137){\small $\frac{Q_e}{\Lambda} =0.12$}								  		
						\put(120,118){\small $\frac{Q_e}{\Lambda} =0.27$}	
						\put(165,93){\small $\frac{Q_e}{\Lambda} =0.42$}							
				  		 \put(116,-10){\small $q$}
				  		 \put(178,135){\small $\epsilon =-1$}
				  		 \put(25,20){\small $\Omega _H \Lambda = 0.469$}				  		 
				  		\put(-8,78){\begin{turn}{90}{\small $j$}\end{turn}}
	   			\end{picture}
			 \includegraphics[scale=0.62]{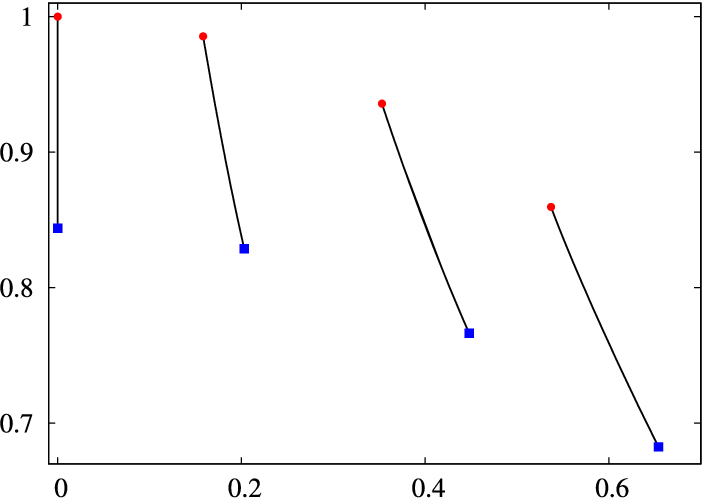}\hfill
			 			 \begin{picture}(0,0)
						\put(35,65){\small $\frac{Q_e}{\Lambda} =0.42$}	
						\put(65,135){\small $\frac{Q_e}{\Lambda} =0.24$}								  		
						\put(150,25){\small $\frac{Q_e}{\Lambda} =0.36$}	
						\put(165,93){\small $Q_e =0$}							
				  		 \put(125,-10){\small $j$}
				  		 \put(178,135){\small $\epsilon = +1$}
				  		 \put(28,20){\small $\Omega _H \Lambda = 0.469$}				  		 
				  		\put(-4,72){\begin{turn}{90}{\small $\Lambda /M $}\end{turn}}
	   			\end{picture}
 			 \includegraphics[scale=0.62]{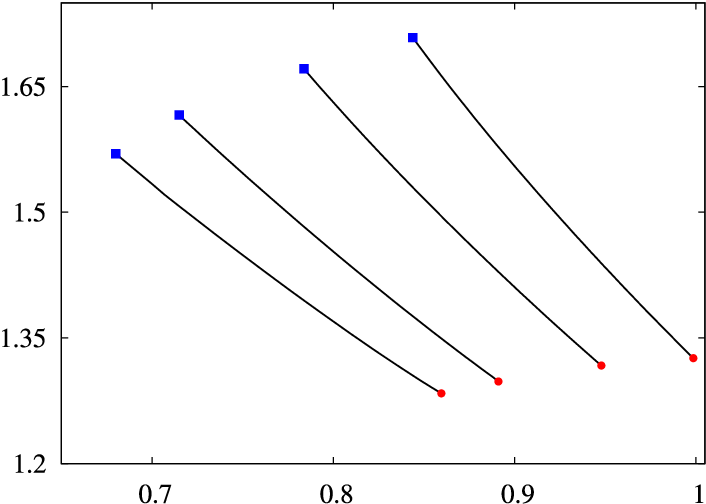}
			 \caption{Branches of GB$^{-}$ scalarized Kerr-Newman BHs in a $q$ $vs.$ $j$  plot (left panel) and $j$ $vs.$ $\Lambda/M$ plot (right panel). The branches are for a specific choice of $\Omega_H\Lambda=0.469$ and different choices of $Q_e/\Lambda$. The blue squares correspond to Kerr-Newman BHs with a vanishing
scalar field, while the red circles correspond to critical configurations.}
			 \label{F3.5}
			\end{figure}
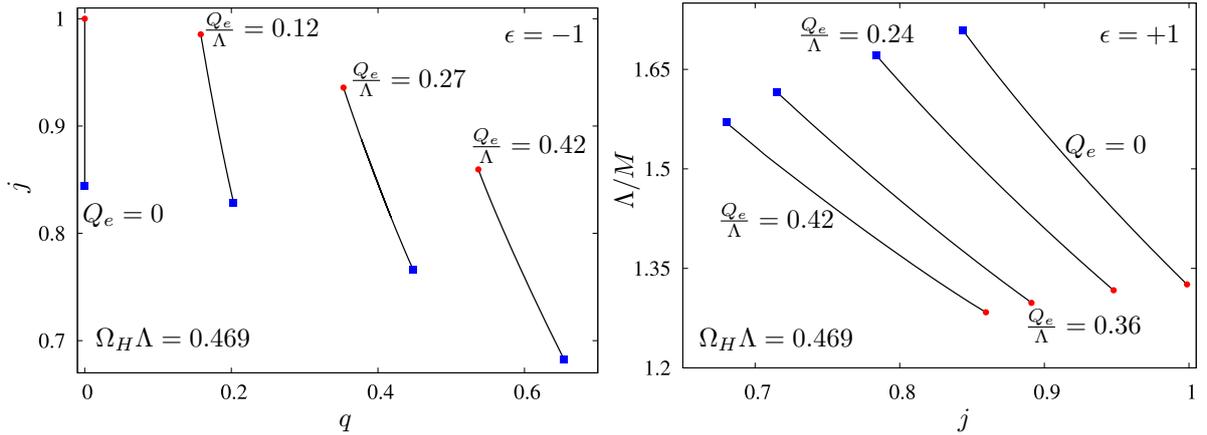
			\subsubsection*{Numerical results}
	With the setup just described, we have employed a numerical approach similar to the one in~\cite{herdeiro2021spin,herdeiro2016kerr} ($cf.$ Appendix~\ref{B}). The typical numerical error for the solutions so obtained and reported below are of the order of $10^{-3}$. 
 
	The scalarized Kerr-Newman solutions possess four independent charges: the three global charges shared with their electro-vacuum counterparts, $(M,\, J,\, Q_e)$, plus the scalar charge $Q_\phi$. In our approach, the input parameters are: the event horizon radius 
$r_H$, the horizon angular velocity $\Omega_H$, the asymptotic value of the electrostatic potential $\Psi _e$ (or the electric charge $Q_e$), together with the coupling constant $\Lambda$ (which specifies the theory). Therefore, after fixing the scale $\Lambda$, we are left with a three-dimensional parameter space. 

	Therefore, a complete scanning of such an ample parameter space of scalarized Kerr-Newman BHs is a time consuming task. Here we focus on illustrative solutions that already capture the generic behaviour. Moreover, although we have verified that spinning scalarized solutions exist for both signs of $\epsilon$, we shall focus on the results for the more novel case of spin/charge scalarization, $\epsilon =-1$. In practice, we have scanned the parameter space by varying both $r_H$ and $Q_e$ for several different values of $\Omega_H$. Alternatively, we have varied both $r_H$ and $\Omega_H$ for fixed values of $Q_e$. 

	Our numerical results suggest that the solutions share most of the properties of the scalarized Kerr BHs discussed in~\cite{berti2021spin,herdeiro2021spin}. In Fig.~\ref{F3.5} we display the reduced quantities $(q, j)$ (left panel) and $(j,\, M/\Lambda)$ (right panel) for solutions with $\Omega_H \Lambda= 0.469$ and illustrative values of the ratio $Q_e /\Lambda$. These scalarized solutions emerge from a Kerr-Newman BH supporting a zero-mode solution of the scalar equation -- a scalar cloud -- corresponding to the blue squares. Then, the sequence of solutions with constant $Q_e/\Lambda$ terminate at a critical configuration, as in the RN case reported before, corresponding to the red circles. The main trend observed in Fig.~\ref{F3.5} is that fixing $Q_e$ and $\Omega_H$ in units of $\Lambda$, the scalarized BHs have more mass (and thus smaller $q$ and $\Lambda/M$) and larger $j$. 
	
	Although our scanning of the full parameter space was limited, extrapolating the existing numerical data, we anticipate that the (three dimensional) domain of existence of $\epsilon=-1$ spinning, charged scalarized BHs is bounded by four sets of solutions: $i)$ the {\it existence surface}, which corresponds to the set of Kerr-Newman solutions  supporting scalar clouds; $ii)$ the set of {\it critical solutions}, which form  again a two dimensional surface;
$iii)$ the {\it static configurations}, $J=0$, which corresponds to the $\epsilon=-1$ scalarized RN solutions discussed in the previous Sec.~\ref{S3.3}; and $iv)$ the {\it neutral configurations}, $Q_e=0$, which were studied with the choice of the coupling function~\eqref{E3.2.18} in~\cite{berti2021spin}.

	As for the $Q_e=0$ case, the {\it existence surface} is universal for any expression of the coupling function allowing for scalarization. Concerning the set  $ii)$ ({\it critical solutions}) 
the numerical process fails to converge as it is approached, as in the static limit. The explanation for this behaviour can be traced back to the radicand of a square root in the horizon expansion of the solutions vanishes as the critical set is approached. Note also that the sets $ii)$-$iv)$ are not universal; they depend on the choice of the coupling function $f(\phi)$.
%
	\section{Lessons from alternative charged BHs}\label{S3.5}
%
	It is interesting to test the generality of some of the results concerning the interplay between the introduction of charge and the scalarization phenomenon. For this purpose, we shall be considering the more straightforward case of static, spherically symmetric, charged BHs in some alternative models rather than electro-vacuum.

	As a first observation, we remark that for a finite mass, asymptotically flat solution, the GB invariant $R^2_{\rm GB}$ is strictly positive for large enough $r$. As with the RN BH,
a sufficient condition for the occurrence of GB induced scalarization for $both$ signs of $\epsilon$
is that $R^2_{\rm GB}<0$ at the horizon. In general, however, 
the sign of the GB invariant at the horizon depends on the matter content. Indeed, for a generic spherically symmetric BH spacetime and using the metric ansatz \eqref{E1.5.40} with \eqref{E3.1.9}, a straightforward computation leads to the simple relation
		\begin{equation}\label{E3.5.44}
		 R^2_{\rm GB}\big |_{r=r_H}=\frac{12}{r_H^2}+16 \rho_{(H)}^2-\frac{16}{r_H^2}\left[2\rho_{(H)}+p_{\theta (H)}\right] \ ,
		\end{equation}
	where $r_H$ is the horizon radius, $\rho_{(H)}=-T_t^t(r_H)$
and $p_{\theta (H)}=T_\theta^\theta(r_H)$. One can easily see that, for a generic matter content, the above quantity has no definite sign.

	One may then ask the following two questions: $(1)$ is $R^2_{\rm GB}\big |_{r=r_H}<0$ close to the maximal charge for any charged BH model? $(2)$ is $R^2_{\rm GB}\big |_{r=r_H}<0$ a \textit{necessary} condition for GB$^-$ scalarization for any charged BH model?

	We will now show, by concrete illustrations, that both these questions have a $negative$ answer.
%
		\subsection{Einstein-Maxwell-dilaton BHs}\label{S3.5.1}
%
	To answer question $(1)$, we have investigated the sign of the GB invariant (together with its behaviour in the bulk) for the stringy generalization of the RN BH -- the Gibbons-Maeda-Garfinkle-Horowitz-Strominger (GMGHS) family of BHs~\cite{garfinkle1991charged,gibbons1988black}. In our context, these solutions are found for an  action of the form
			\begin{equation}\label{E3.5.45}
			 \mathcal{S}_{\rm EMdGB} = \frac{1}{4}\int d^4 x\ze \sqrt{-g} \left[R -2\ze\psi_{,\mu}\psi^{,\mu} -e^{\alpha\ze \psi }F_{\mu \nu}F^{\mu \nu}- 2\ze\phi_\mu \phi^\mu +\epsilon\ze\Lambda^2   f(\phi) R^2_{\rm GB} \right] \ ,
			\end{equation}
	describing an Einstein-Maxwell-dilaton-Scalar-GB model, which includes an extra scalar field (the dilaton $\psi$) with a non-minimally coupling with the Maxwell term, where $\alpha\geqslant 0$ is a constant of the theory. The GMGHS solution is found for $\phi=0$, with $f(\phi)$ satisfying the condition \eqref{E3.1.8}.
It is easy to prove that the behaviour found in the RN case ($\alpha=0$, $\psi=0$) is recovered for small enough values of $\alpha$. In that case, for large enough values of the electric charge, $R^2_{\rm GB}$ becomes negative in a region between the horizon and some maximal value of the radial coordinate. However, a direct computation shows that, for $\alpha>0.904$ (following the conventions used in~\cite{garfinkle1991charged}), $R^2_{\rm GB}$ is strictly positive at the horizon and also in bulk, irrespective of the value of the electric charge. Thus, as with Schwarzschild vacuum BHs, GB$^-$ scalarization of the GMGHS with large enough values of the dilaton coupling constant becomes impossible. This suggests that it would be interesting to check the status of GB$^-$ scalarization of the rotating counterpart of the GMGHS BH, the well known Kerr-Sen BH~\cite{sen1992rotating}.
%
		\subsection{Einstein-Yang-Mills BHs}\label{S3.5.2}
%
	To answer question $(2)$ above, we have investigated the sign of the GB invariant for the case of Einstein--Yang-Mills (EYM) BHs with \textit{SU}(2) non-Abelian hair (nA)~\cite{volkov1989non,volkov1999gravitating,bizon1990colored,kunzle1990spherically}. In our context, these solutions are found for an action of the form 
				\begin{equation}\label{E3.5.46}
				 \mathcal{S}_{\rm EYMSGB}=\frac{1}{4} \int d^4 x \ze\sqrt{-g}\left[R -F_{\mu \nu}^{(a)}F^{\mu \nu(a)}-2\ze\phi_{,\mu}\phi^{,\mu}+\epsilon\ze \Lambda^2  f(\phi) R^2_{\rm GB}\right] \ ,
				\end{equation} 
	with $F_{\mu \nu}^{(a)}$ the nA field strength and $a=1,2,3$. These so-called \textit{coloured} BHs are asymptotically flat and possess a \textit{single} global ``charge'' -- the ADM mass, despite the presence of a local magnetic field (see \cite{volkov1999gravitating,volkov2018hairy} for reviews). Another striking difference concerning their (magnetic) RN Abelian counterparts is the existence of a smooth solitonic limit \cite{bartnik1988particlelike}, obtained as the horizon size shrinks to zero. At the same time, there is no upper bound on their horizon size. However, the large EYM BHs are essentially Schwarzschild solutions;
the contribution of the YM fields to the total ADM mass becomes negligible, albeit these fields are still non-trivial.
 
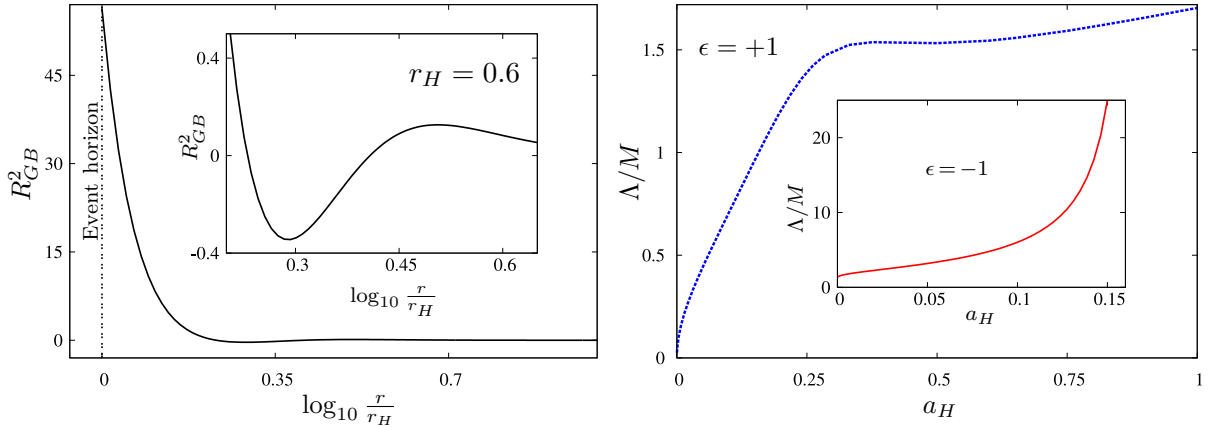
\begin{figure}[H]
					 \centering
					 	\begin{picture}(0,0)
				  		 \put(100,-10){\small $\log _{10} \frac{r}{r_H}$}
				 		 \put(118,32){$\scriptstyle \log _{10} \frac{r}{r_H}$}
				  		 \put(140,115){$r_H = 0.6$}
				  		\put(-10,70){\begin{turn}{90}{\small $R_{GB} ^2$}\end{turn}}
				  		\put(54,85){\begin{turn}{90}{$\scriptstyle R_{GB} ^2$}\end{turn}}
				  		\put(18,55){\begin{turn}{90}{$\scriptstyle {\rm Event\ horizon}$}\end{turn}}
	   			\end{picture}
					\includegraphics[scale=0.62]{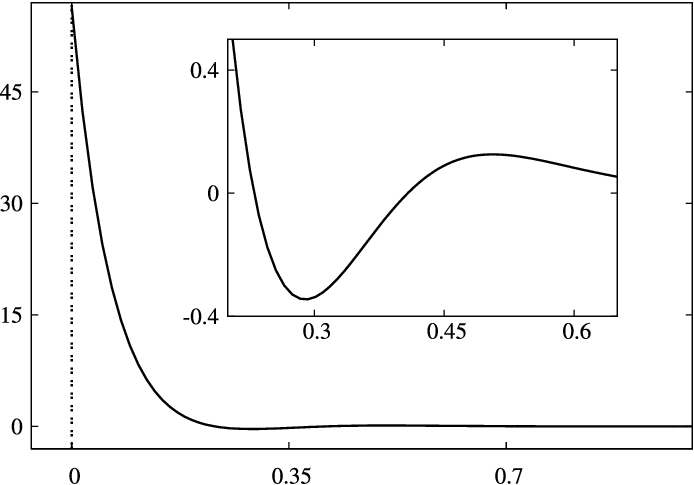}\hfill
						\begin{picture}(0,0)
				  		 \put(109,-10){\small $a_H$}
				  		 \put(25,125){\small $\epsilon = +1$}				  		 
				 		 \put(125,25){$\scriptstyle a_H$}
				  		 \put(110,80){$\scriptstyle \epsilon\, =\, -1$}
				  		\put(-5,70){\begin{turn}{90}{\small $\Lambda /M $}\end{turn}}
				  		\put(58,58){\begin{turn}{90}{$\scriptstyle \Lambda/M$}\end{turn}}
	   			\end{picture}
 				   \includegraphics[scale=0.62]{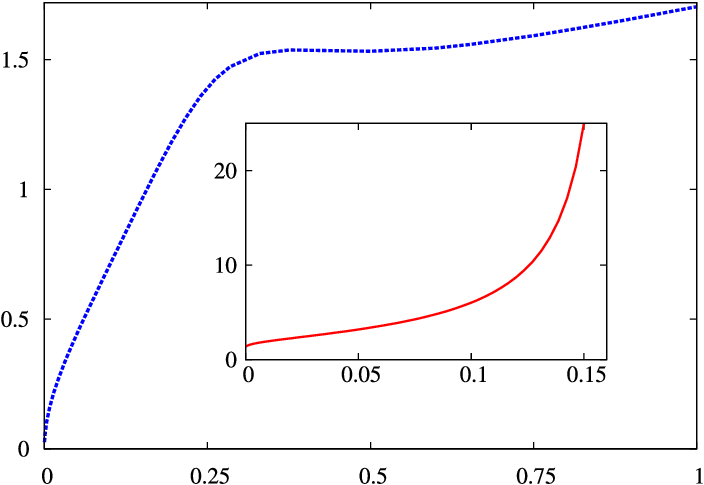}
					 \caption{(Left panel) GB invariant, $R^2_{GB}$, as a function of the radial coordinate for an EYM BH with horizon radius $r_H=0.6$. One notices a shell with $R^2_{GB}<0$, located outside the horizon.
(Right panel) $\Lambda/M$ of the set of EYM BHs supporting scalar clouds as a function of the reduced horizon area for $\epsilon=\pm 1$.}
					 \label{F3.6}
					\end{figure}

Contact with question $(2)$ above comes from observing that the GB invariant is always positive at the horizon for these solutions. However, $R^2_{\rm GB}$ may take negative values in a shell which does not touch the horizon, $i.e.$ for some range of the radial coordinate $r_H<r_1<r<r_2$ -- see Fig. \ref{F3.6} (left panel). This feature occurs for small enough BHs: using the metric form \eqref{E1.5.40} and the conventions in \cite{volkov1999gravitating}, we confirmed such shell is present for $0<r_H\leqslant 0.710\ze$, or, equivalently,
$0<a_H\leqslant 0.158\ze$.

	Given this qualitative difference with the RN case, one could ask whether GB$^\epsilon$ scalarization is still possible for both signs of $\epsilon$. The answer is positive, and we have constructed the corresponding  GB scalar clouds, $i.e.$ solved \eqref{E3.1.8} for a large set of EYM BH backgrounds and both values of $\epsilon$ -- see Fig.~\ref{F3.6} (right panel). As expected, $\epsilon=+1$ scalar clouds exist for all non-Abelian BHs. The value of the ratio $\Lambda/M\simeq 1.704$ corresponding to a Schwarzschild BH is approach asymptotically, as  $a_H\to 1$ ($i.e.$ large EYM BHs). Also, scalar clouds with $\epsilon=-1$ exist for {\it all} BHs with  $R^2_{\rm GB}<0$ in a shell outside the horizon. Nonlinear continuations of these scalar clouds should exist, but we did not construct them.

	This example makes clear that GB$^\epsilon$ scalarization of BHs is not necessarily supported and triggered near the event horizon. 

%
	\section{Further remarks}\label{S3.6}
%
	There is a well-known analogy between the spinning vacuum Kerr BHs and the electrically charged static RN solutions\footnote{Since the electric-magnetic duality is still valid for the (Abelian) models in this work, solutions possess a dual magnetic description.}. They possess many similar properties at the level of a thermodynamical description; in particular, both RN and Kerr BHs have an extremal limit with a finite horizon size.
In the context of this work, it is interesting to note that the GB invariant of a RN BH changes sign if $q$ is large enough as the Kerr one changes sign for sufficiently large $j$. Thus, it is natural to conjecture that the qualitative picture found concerning the GB$^\epsilon$ scalarization of Kerr BHs~\cite{berti2021spin,herdeiro2021spin,dima2020spin, cunha2019spontaneously} should be essentially recovered when replacing rotation by electric charge, with the existence of both $\epsilon=\pm 1$ scalarized solutions.

	This work confirms this conjecture and constructed the corresponding scalarized RN BHs. That is, we provide evidence for the following scenario: given an expression for the coupling function $f$, two classes of charged RN scalarized solutions may exist for the same global charges. The first one has 
	\begin{equation*}
	 \epsilon=+1\ ,
	\end{equation*}
	and can be viewed as a generalization of the $Q_e=0$ solutions in \cite{antoniou2018evasion,silva2018spontaneous, doneva2018charged}. The second has
	\begin{equation*}
	 \epsilon=- 1\ ,
	\end{equation*}
	and in this case, the condition $\mu_{\rm eff}^2<0$ is supported by a large enough charge to mass ratio $q>q_c=0.957$, which implies $R^2_{\rm GB}<0$ for some region outside the horizon.
	
	We have also presented a preliminary investigation of 
the spinning generalizations of the above BHs, $i.e.$, the scalarized Kerr-Newman BHs.

	In the last part of this chapter, we have addressed the generality of these results. First, we established that the GB invariant of the stringy dilatonic generalization of the RN BH becomes strictly positive for large enough values of the dilaton coupling constant. Also, we pointed out the possibility that the $\epsilon=-1$ BH scalarization may also appear in situations where $R^2_{\rm GB}$ is negative in a spherical $shell$ outside and disconnected from the horizon. This is the case of the coloured BHs in EYM theory. 
	
	In this chapter, to simplify the picture, we have assumed the absence of a self-interaction term for the scalar field in action \eqref{E3.1.1}. GB$^+$ scalarization of spherical BHs including such self-interactions is discussed $e.g.$ in \cite{doneva2019gauss,staykov2021axial,macedo2019self}. Our results could be generalized to include such self-interactions.

\clearpage\null\newpage	
	
%
\chapter{Black hole mimickers}\label{C4} 
%
	Bosonic stars (BSs) are speculative macroscopic Bose-Einstein condensates (\textit{cf.} Sec.~\ref{S1.3.2}). Such hypothetical ultra-light bosons could be part (or the whole) of the dark matter budget in  the Universe~\cite{suarez2014review,hui2017ultralight}. In particular,  compact and dynamically robust BSs  occur in a variety of different models~\cite{liebling2017dynamical}, thus being interesting for a diversity of theoretical and phenomenological strong gravity  studies -- see e.g.~\cite{kusmartsev1991stability,gleiser1988stability,gleiser1989gravitational,hawley2000boson ,palenzuela2008orbital,guzman2009three,diemer2013geodesic,cunha2015shadows,cunha2016chaotic,cao2016iron, zhou2017iron,vincent2016imaging,franchini2017constraining,cunha2017lensing,grandclement2017light ,sanchis2017numerical,cardoso2017tests,palenzuela2017gravitational,grould2017comparing,bezares2017final ,glampedakis2018well,croon2018gravitational,olivares2018tell,johnson2020constraining,calderon2020ultra ,siemonsen2021stability}. In fact, BSs have been suggested as possible BH mimickers~\cite{cardoso2019testing}. The purpose of this section is to assess this possibility in what concerns the BH shadow~\cite{falcke1999viewing,cunha2018shadows}, for equilibrium (or near-equilibrium) BSs.

	An essential feature of the paradigmatic BH model of general relativity, the Kerr BH~\cite{kerr1963gravitational}, is the existence of bound photon orbits (see $e.g.$ \cite{bardeen1972rotating}), which, in their simplest guise, are planar light rings (LRs). Furthermore, LRs have been shown to be a  generic feature of asymptotically flat stationary BHs, even beyond vacuum or beyond Einstein's theory~\cite{cunha2020stationary}. The existence of LRs around BHs impacts important strong gravity features, such as (the initial part of)  the  ringdown~\cite{cardoso2016gravitational} and the BH shadow~\cite{cunha2018shadows}. Thus, it has been generically assumed that in order to mimic these features, BSs should possess LRs. For the simplest models of BSs, where the bosonic field has no self-interactions, leading to the so-called \textit{mini}-BS, such ultra-compact solutions indeed exist, but only in particular regions of the parameter space, where the BSs have been shown to be  unstable~\cite{cunha2017lensing}. Nonetheless, one may wonder whether, in other models possessing self-interactions, ultra-compact BSs could arise for perturbatively stable BSs. However, as we show in this chapter, this is not the case for several examples of self-interactions. 

	Moreover, it has been shown that topologically trivial spacetime configurations, such as BSs,  develop not one but two LRs, when they become ultra-compact~\cite{cunha2017light}. Furthermore, if the bosonic matter obeys the null energy condition (which is the case for the standard bosonic fields considered), one of the LRs is stable. Such stable LRs have been argued to lead to a non-perturbative instability~\cite{keir2016slowly,cardoso2014light,benomio2018stable}. Little is known about the timescales of this putative instability, which, therefore, is not an unsurmountable obstacle \textit{per se} for ultra-compact  BSs to be dynamically robust. Nonetheless, together with the inability to have perturbatively stable ultra-compact BSs, this feature casts an additional shadow of doubt on the dynamical viability of ultra-compact BSs. 
	
	There is, however, a different possibility allowing a BS without LRs to mimic the appearance of a BH when lit by a surrounding accretion flow. If the source of light in the vicinity of the BS has the same morphology as it would have around a BH, the lensing of light,  and in particular a  similar central depression of the emission (the  shadow~\cite{falcke1999viewing}) would also be present~\cite{vincent2016imaging,grould2017comparing}. The key feature here is the cut-off in the emission due to the disk's inner edge, which is determined by the BH's innermost stable circular orbit (ISCO) for time-like geodesics. For a Schwarzschild BH of mass $M$, the ISCO is located at the areal  radius $r=6M$. For spherical BSs, there is no ISCO so that one could think that the disk, and the emission, continue to the centre: hence no shadow should be produced. In a recent work~\cite{olivares2018tell}, however, general relativistic magneto-hydrodynamic simulations were performed in static (scalar) BSs backgrounds, including general relativistic radiative transfer, observing qualitative similarities with BH spacetimes, in particular the central emission depression, in other words, an effective shadow. The key new feature observed in~\cite{olivares2018tell} is that despite the existence of stable time-like circular orbits in the BSs spacetime up to the centre,  the angular velocity of the orbits, $\Omega$, attains a  maximum at some areal radius ($R_{\Omega}$). This scale is observed to determine the inner edge of the accretion disk in the simulations in~\cite{olivares2018tell}. Under the assumptions therein, that the loss of angular momentum of the orbiting matter is driven  by  the  magneto-rotational  instability~\cite{balbus1991powerful} and that the radiation relevant for BH shadow observations is mostly due to synchroton emission. 

	Despite the qualitative similarity,  $i.e.$, the possibility of obtaining an effective shadow in a BS spacetime and in a realistic astrophysical environment (despite the absence of LRs or ISCO), the results in~\cite{olivares2018tell} raise two issues. Firstly, they show a quantitative difference between the BS and BH shadow for the cases analysed. Even by current observations, the two have different sizes for the same total mass. Secondly, and most importantly, the BSs in the analysis in~\cite{olivares2018tell} that display the new scale $R_{\Omega}$ are perturbatively unstable. Hence, one may wonder whether there are models in which spherical BSs can have a degenerate (effective) shadow with a comparable BH ($i.e.$ with the same  ADM  mass) in the perturbatively stable region. As we shall see below:  $(i)$ self-interactions of scalar bosonic fields can indeed yield perturbatively stable BS with the new scale $R_{\Omega}$. However, in this case, we could not get solutions with $R_{\Omega}=6M$. $(ii)$ for vector BSs [\textit{a.k.a.} \textit{Proca Stars} (PSs)], even without self-interactions, there are perturbatively stable stars with the new scale $R_{\Omega}$. We find a particular solution with $R_{\Omega}=6M$. Thus, models in which dynamically robust spherical BSs can have a degenerate (effective) shadow with a comparable Schwarzschild BH \textit{does exist}. {We confirm this possibility by explicitly studying the lensing of the aforementioned particular PS with  $R_{\Omega}=6M$, lit by an accretion disk with its inner edge at this special radius. The analysis, however,  clarifies that the degenerate shadow only occurs for some observation angles.}

	This chapter is based on the work published in \cite{herdeiro2021imitation} and is organised as follows. In Sec.~\ref{S4.1} we introduce the generic model for BSs and the field equations for both the scalar and vector cases. We compute the required asymptotic expansions at the star's centre and infinity. These are used to numerically compute the BSs solutions, of which we display the domain of existence in the following sections. In Sec.~\ref{S4.2} we derive the LR and time-like circular orbits' (TCOs) relations. These are then investigated for the scalar BSs cases in Sec.~\ref{S4.3} and the PSs in Sec.~\ref{S4.4}. We conclude with a summary of our main result and a discussion\footnote{An oral presentation about this Chap. can be seen at~\cite{ImitationVid}.}.
%
	\section{BS models}\label{S4.1}
%
	Let us quickly recall the model presented in Sec.~\ref{S1.3.2}. The Einstein-matter action, where the matter part describes a spin $s=0,\ze 1$ classical field minimally coupled to Einstein's gravity, reads
		\begin{equation}\label{E5.1.1}
		 \mathcal{S}_{s}=\frac{1}{4}\int d^4 x \sqrt{-g} \big[ R+4\mathcal{L}_s \big]\ ,
		\end{equation}
	with the matter Lagrangians for the spin-$0$ and spin-$1$ fields, respectively:
		\begin{equation}\label{E5.1.2}
		 \mathcal{L}_0 = -2\ze g^{\mu \nu} \big(\bar{\Phi} _{,\mu} \Phi _{,\nu}+\bar{\Phi} _{,\nu}\Phi _{,\mu} \big) - U_i(|\Phi |^2)\ , \qquad \qquad \mathcal{L}_1= - G_{\mu \nu}\bar{G} ^{\mu \nu}-U (\textbf{B})\ .
		\end{equation}
	The massive complex scalar field, $\Phi$, with mass $\mu _S$, is under a potential term $U_i(|\Phi |^2)$; the massive complex vector field, with mass $\mu _P$, has a $4$-potential $B^\alpha$ and is under a potential  $U (\textbf{B})$.  

	Variation of the action concerning the metric and matter fields leads to the following two sets of field equations, in the scalar and vector case, respectively
		\begin{align}
	 	 & E_{\mu \nu} =  \Big[\bar{\Phi} _{,\mu} \Phi _{,\nu}+\bar{\Phi} _{,\nu}\Phi _{,\mu}-g_{\mu \nu} \mathcal{L}_0 \Big]\ , \qquad \Box \Phi = \hat{U_i}\ \Phi \ ,\\
	 	 & E_{\mu \nu} = \left[\frac{1}{2} \big(G_{\mu \delta} \bar{G}_{\nu \gamma} +\bar{G}_{\mu \delta} G_{\nu \gamma} \big) g^{\delta \gamma}+\hat{U}\big( B_\mu \bar{B}_\nu +\bar{B}_\mu B_\nu - g_{\mu\nu}\mathcal{L}_1 \big)\right] \ , \qquad\frac{1}{2} \nabla_\mu G^{\mu \nu} = \hat{U} B^\nu\ ,
		\end{align}
	with $E_{\alpha \beta}$ the Einstein's tensor, $\Box$  $(\nabla)$ the covariant d'Alembertian (derivative) operator, $\hat{U_i}\equiv d U_i/d |\Phi |^2$ and $\hat{U}\equiv dU /d \textbf{B}$. 

	For the metric ansatz, we use the standard spherically symmetric solution \eqref{E1.5.40}; while the complex, spherically symmetric matter field ansatz \eqref{E1.5.44}-\eqref{E1.5.45} reads, for the scalar and vector cases, respectively:
		\begin{equation}\label{E5.1.5}
	 	 \Phi (r,t) = \phi (r) e^{-i \omega\ze t}\ ,\qquad \qquad B_\mu (r,t)=\big[ B_t(r) dt + \textit{i} B_r(r) dr \big] e^{-i\omega\ze t}\ ,
		\end{equation}
	where $\phi $ is  the scalar field amplitude and $B_t$ and $B_r$ are two real potentials that define the Proca ansatz. In both cases, $\omega $ is the field's frequency. In the Proca case, the field equation implies, for a Ricci-flat space, the Lorentz condition $\nabla _\mu (\hat{U}B^\mu) =0$, which is a dynamical condition, rather than a gauge choice.
	
	Both matter models possess a $\textbf{U}(1)$ global symmetry, under a global phase transformation: $\Phi \rightarrow \Phi e^{i\ze a}$ and $B_\mu \rightarrow B_\mu e^{i\ze a}$, where $a$ is a constant. This symmetry leads to a conserved Noether charge\footnote{Note that, in this case, the Noether charge is a true charge that is associated with a conserved Noether current and hence a Gauss law.}, $Q_s$, from the spatial integration of the time component of the conserved Noether current ($Q_S=\int _\Omega j_S ^t$), with
		\begin{equation}\label{E5.1.6}
		 j_0 ^\mu = -\textit{i}\ \big[ \bar{\Phi} \Phi^{,\mu} -\Phi \bar{\Phi}^{,\mu} \big]\ ,\qquad \qquad j_1 ^\mu = \frac{i}{2}\big[\bar{G} ^{\mu\nu} B_\nu -G^{\mu \nu}\bar{B}_\nu\big]\ ,
		\end{equation}
	where the subscripts in $j$ refer to the model $s=0,\, 1$. For all the other quantities (\textit{i.e.} $Q_s$ or $\mu _s$), we follow the notation $s=S\equiv \text{Scalar}$ and $s=P\equiv \text{Proca}$. For the vector model, we will focus on the following self-interactions potential:
		\begin{equation}\label{E5.1.7}
		 U=\frac{\mu _P ^2}{2}\ze \textbf{B}+\frac{\beta _P}{4} \textbf{B}^2\ .
		\end{equation}	 
	This model encompasses the mini-PSs solutions~\cite{brito2016proca} when the self-interaction coupling vanishes, $\beta _P =0$. 

	With the above setup, one obtains a system of coupled Einstein-matter ODEs. For each of the two models ($s=0,\, 1$) there are two ``essential'' Einstein equations:
		\begin{align}\label{E5.1.8}
		 & s=0:\qquad m' = r^2 \left[N \phi'^{\, 2}+\frac{\omega ^2 \phi ^2}{N \sigma ^2}+U_i \right]\ , \qquad
		\sigma ' = 2 \sigma r \Big[ \phi'^{\, 2}+\frac{\omega ^2 \phi^2}{N ^2 \sigma ^2}\Big]\ ,\\
		 & s=1:\qquad m' = r^2 \left[ \frac{(B_t'-\omega B_r)^2}{2\sigma ^2} + 	\left(\mu _P ^2 -\frac{3}{2}\beta_P \textbf{B}\right)\frac{B_t^2}{2 N \sigma^2}+\frac{U}{\textbf{B}}N\right] \ ,\nonumber\\
		 &\quad \ \qquad\qquad\sigma ' = 2r \sigma \hat{U} \left[ B_r^2+ \frac{B_t^2}{N^2 \sigma ^2}\right]\ .\label{E5.1.9}
		\end{align}
	To  close the system, the equations for the matter field functions are
		\begin{align}\label{E5.1.10}
		 & \phi '' = -\frac{2 \phi '}{r}-\frac{N' \phi '}{N}-\frac{\sigma ' \phi '}{\sigma}-\frac{\omega ^2 \phi}{N^2 \sigma ^2}+{\frac{\hat{U_i}}{N}}\, \phi\ ,\\
		& B_t ' = \omega B_r-2\frac{B_r \ze \sigma ^2 N}{\omega} \hat{U}\ , \qquad \qquad\left[\frac{r^2\big(\omega B_r - B_t'\big)}{\sigma}\right]'+\frac{2 \ze r^2 B_t}{N \sigma}\hat{U} =0\ .\label{E5.1.11} 
		\end{align}
			\subsubsection*{Asymptotic expansions and physical relations}
	In order to integrate the field equations \eqref{E5.1.8}-\eqref{E5.1.11} one must consider the asymptotic expansions \eqref{E2.1.7} with $r_H=0$. At the origin, the field equations can be approximated by a power series expansion in $r$ that guarantees $m (0)=0,\ \sigma (0) =\sigma _0 ,\ \phi (0)=\phi _0 , \ B_t(0)=b_0$ and $B_r(0)=0$
				\begin{align}\label{E1.5.12}
				 & s=0:\  m  = \frac{U_i \ze \sigma _0 ^2+\omega ^2 \phi _0 ^2}{3\ze \sigma _0} r^3 + \cdots\ ,	\quad \sigma  =  \sigma _ 0 + \frac{\omega ^2 \phi _0 ^2}{\sigma _0} r^2 + \cdots\ ,\nonumber\\
&\qquad \qquad \phi =  \phi _0 +\frac{\phi _0}{6} \left( \hat{U_i}-\frac{\omega ^2}{\sigma_ 0 ^2}\right) r^2 +\cdots \ ,\\ 
				& s=1:\ m =  \frac{2\ze b_0 ^2\ze \mu _P ^2\ze \sigma _0 ^2-3\ze b_0 ^4 \ze \beta _P }{ 12\ze\sigma_0 ^4} r^3 + \cdots\ , \qquad \sigma  =  \sigma _0 + \frac{ b_0 ^2\ze \mu _P ^2\ze \sigma _0 ^2+ b_0 ^4\ze \beta _P}{2\ze \sigma _0 ^3} r^2 + \cdots\ ,\nonumber\\
				&\, \ \quad \qquad B_t = b_0 -\frac{b_0 \left(b_0 ^2\ze \beta _P -\mu _P ^2\ze \sigma _0 ^2+\omega ^2\right)}{6\ze \sigma _0 ^2} r^2 +\cdots \ ,\qquad B_r = -\frac{b_0\ze \omega }{3\ze\sigma _0 ^2} r+\cdots \ . \label{E1.5.13}
				\end{align}
	At infinity, we impose asymptotic flatness and a finite ADM mass $M$: $m(+\infty ) =M,\ \sigma (+\infty) = 1,$  and $ \phi (+\infty) =0=B_t(+\infty)=B_r(+\infty)$. The values of $\sigma _0$ and $M$ are fixed by the numerics, while $\sigma (+\infty)$ fixes the following scaling symmetry of the system of equations: $\{ \sigma, \omega, b_0 \} \rightarrow \lambda \{ \sigma , \omega, b_0 \}$, with $\lambda >0$. An additional scaling symmetry, moreover, allows both mass's $\mu _P $ and $\mu _S $ to be set to unity 	$\big( \mu _P  = \mu _S =1 \big)$.

	The set of coupled ODE's are numerically integrated using a $(5)\,6^{\rm th}$ order adaptative step Runge-Kutta method, with a local error of $10^{-15}$ (see Appendix~\ref{A}). The boundary conditions are enforced using a shooting strategy, with a tolerance of $10^{-9}$ for the spatial asymptotic (at infinity) Scalar/Proca field decay value, while $m(r)\rightarrow M$ and $\sigma \rightarrow 1$. 
		
	Due to the lack of a surface, BSs do not have a well-defined radius. For the ``radius'' of BSs we will consider the areal radius of a spherical surface within which $99\% $ of all the mass is included; this radius is denoted $R_{99}$. The latter defines the BS compactness: $\mathcal{C}\equiv 2M_{99}/R_{99}$. This compactness is always smaller than unity, becoming unity for BHs.

	To test the numerical solutions, we have considered the so-called virial identities (see Ch.~\ref{C7} for the full treatment). These read for the $s=0,\,1$ cases, respectively,
				\begin{align}\label{E1.5.14}
				& s=0:\quad \int _0 ^{+\infty}dr\ r^2 \sigma \left[ \frac{\omega ^2 \phi ^2}{N \sigma ^2}\left(3-\frac{2 m}{rN}\right)-\phi'^{\, 2}-3\ze U_i\right]=0\ ,\\
				 & s=1:\quad \int _0 ^{+\infty} dr\ \frac{ r^2 }{2 N^3 \sigma ^3} \Bigg\{ B_t^4 (2-5 N) \beta _P+2 B_t^2 N \sigma ^2 \Big(N \left(3 B_r^2 N \beta _P +4\right)-1\Big)\Bigg.\nonumber\\
				& \qquad \quad \qquad \qquad\Bigg.+ N^3 \sigma ^2 \bigg[ 2 B_t' \left(B_t'-4\ze g\ze \omega \right)+B_r^2 \Big(6\ze \omega ^2-\sigma ^2 \big(B_r^2 N (N+2) \beta _P +4 N+2\big)\Big)\bigg]\Bigg\} =0\ .
				\label{E1.5.15}
				\end{align}
	The numerical accuracy can also be tested by the ADM mass expressions computed as a volume integral, which  can be compared to the value of $M$ computed from the mass function at infinity. The volume integrals read
				\begin{align}\label{E5.1.16}
				& s=0:\quad M= \int _0 ^{+\infty } dr\ r^2 \sigma \left[\frac{4\omega ^2 \phi ^2}{N\sigma ^2}-2 U_i\right]\ , \\
				& s=1:\quad M=  \int _0 ^{+\infty} dr\frac{ r^2}{4 N^2 \sigma ^4} \Big[-3 B_t^4 \beta _P +N \sigma ^2 \Big( 2 B_t^2  (B_r^2 N \beta _P+1)+2 N B_t'^{\, 2} \nonumber\\
				&\qquad\quad\qquad\qquad \qquad+N B_r^2 (B_r^2 N \beta _P +2)-2\ze \omega B_r \Big) \Big]\ .
				\end{align}
%
	\section{Light rings (LRs) and time-like circular orbits (TCOs)}\label{S4.2}	
%
	
	Let us now consider the basic equations to compute both LRs and TCOs in BSs spacetimes (see Sec.~\ref{S2.5.2} for a preliminary study). The radial geodesic equation for a particle around a BS, described by the metric \eqref{E1.5.40}, is
		\begin{equation}\label{E5.1.17}
		 \dot{r}^2 =\frac{E^2}{\sigma ^2}-\frac{L ^2 N}{r^2}+k N\ , 
		\end{equation}
	where $E,\, L$ represent the particle's energy and angular momentum. For null (time-like) geodesics $k=0$ $(k=-1)$.

			\subsubsection*{Light rings}
	Let us first consider null geodesics $(k=0)$. For a LR, $\dot{r}=0$, which relates $E$ and $L$, $E=\frac{L \sqrt{N} \sigma}{r}$. The LR is circular, which additionally imposes $\ddot{r}=0$. This gives the condition for the presence of a LR
		\begin{equation}\label{E5.1.18}
		 -r \sigma \left(\frac{-2 m'}{r}+\frac{2m}{r^2}\right)+2\left( 1-\frac{2m}{r}\right) \big( \sigma -r \sigma ' \big) =0\ .
		\end{equation}
	As shown in \cite{cunha2017light}, BSs' LRs always come in pairs -- one stable and one unstable -- corresponding to the two roots of \eqref{E5.1.18}. Here, we wish to find the first BS solution containing a LR; in other words, the first ultra-compact BS. Let $\omega _{LR}$ be the frequency of the first ultra-compact BS as we move along the (one dimensional) domain of existence, starting from the Newtonian limit ($cf.$ Secs.~\ref{S4.3} and~\ref{S4.4} below). The latter corresponds to a BS solution with two degenerate LRs~\cite{cunha2017light}.
			\subsubsection*{Time-like circular orbits}
		As argued in~\cite{olivares2018tell}, an accretion disk may have an inner edge even around BSs without an ISCO. This occurs if the angular velocity along TCOs attains a maximum at some radial distance. The corresponding areal radius is denoted $R_\Omega$.  This is computed by monitoring the angular frequency $\Omega$ at all radial points to obtain its maximum.
	
	In the following sections, we shall study the first ultra-compact BS and the existence of $R_\Omega$ for several different models. As an accuracy estimate, for all the computed BSs solutions, the virial identity and mass relations are obeyed within a factor of $~10^{-9}$.

	Let us now turn to TCOs ($k=-1$). They are described by the tangential $4$-velocity $u^\nu = \big( u^t, 0,0,u^\varphi \big)$ with the normalization condition $u^2=u^\nu u_\nu =-1$. The angular velocity $\Omega$ along these orbits is
	\begin{equation}\label{E324}
	\Omega = \frac{u^\varphi}{u^t} = \sqrt{\frac{\sigma}{2r}} \sqrt{\sigma N'+2 N \sigma '}\ .
	\end{equation}
%

%
	\section{Scalar boson stars}\label{S4.3}
%
	Many scalar BS models have been considered over the years -- see $e.g.$~\cite{schunck2003general}. Here, we shall divide our analysis into two different cases, depending on the choice of the self-interaction potential.

	The first case considers a polynomial self-interaction of the type:
		\begin{equation}\label{E5.3.21}
	 	 U_{\rm poly}=\mu_S ^2 \Phi ^2 + \beta_S\Phi ^4+ \gamma\ze \Phi ^6\ ,
		\end{equation}
	where $\gamma$ ($\beta_S$) is a coupling controlling the self-interaction of the sixth (fourth) order, thus, the potential is determined by two parameters, since we have already established that the mass can be set to unity, $\mu_S =1$.

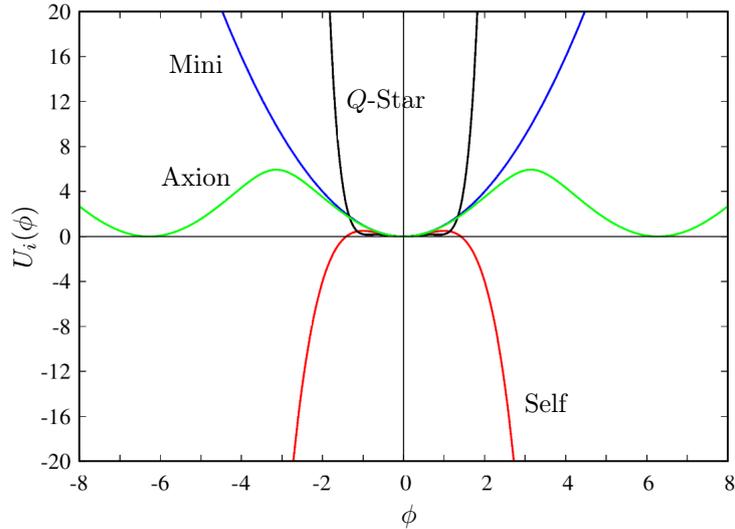
\begin{figure}[H]
			 \centering
	   			\begin{tikzpicture}[scale=0.5]
\node at (0,0) {\includegraphics[scale=0.25]{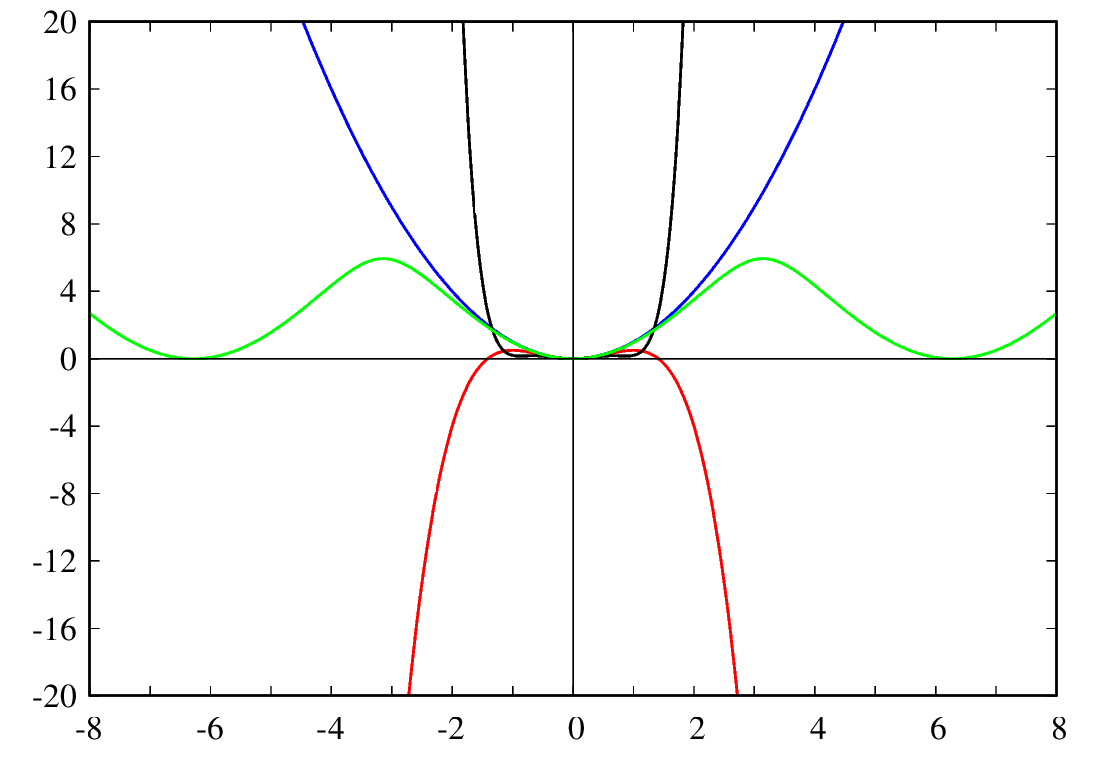}};
\node at (4.2,-4) {\small Self};
\node at (0.6,-7) {\small $\phi$};
\node at (-9.5,0.4) {\begin{turn}{90}{\small $U_i (\phi)$}\end{turn}};
\node at (-5,5) {\small Mini};
\node at (-5,2) {\small Axion};
\node at (-0,4) {\small $Q$-Star};
				\end{tikzpicture}
			 \caption{Graphical representation of the generic shape of the scalar field potentials.}
			 \label{F4.0}
    		\end{figure}

	There are three sub-cases of interest -- see Fig.~\ref{F4.0}. The most generic case occurs for $\gamma\neq 0 \neq \beta_S$ (black line). The latter BSs are known as $Q$-Stars~\cite{lynn1989q}, since they are self-gravitating generalizations of the flat spacetime $Q$-balls~\cite{coleman1985q}, with the constants $\mu_S,\, \beta_S$ and $\gamma$ subject to some conditions. For $\gamma=0$ and $\beta_S \neq 0$, one obtains the Colpi-Shapiro-Wassermann quartic scalar BSs~\cite{colpi1986boson} (red line for $\beta_S <0$). For $\gamma=\beta_S =0$ one recovers the scalar mini-BS~\cite{kaup1968klein,ruffini1969systems} (blue line).
	The second case considers the non-polynomial, axion-type potential (see\cite{guerra2019axion,delgado2020rotating})\footnote{This axion-type potential is also able to yield $Q$-ball solutions. A subject that the author of this thesis is currently pursuing.}
		\begin{equation}\label{E5.3.22}
		 U_{\rm axion}=\frac{2\ze \mu_S ^2 f_\alpha ^2}{\hbar B}\left[ 1-\sqrt{1-4B\sin ^2 \left( \frac{\Phi \sqrt{\hbar}}{2 f_\alpha}\right) }\ \right]\ ,
		\end{equation}
	The potential is characterized by two parameters: $f_\alpha $ and $\mu_S $ (see Fig.~\ref{F4.0} (green line) for a graphical representation). By expanding $U_{\rm axion} $ around the minimum $\Phi =0$
	\begin{equation}\label{E324n}
	U_{\rm axion} = \mu_S ^2 \Phi ^2 -\left(\frac{3B-1}{12}\right) \frac{\hbar\ze \mu _S ^2}{f_\alpha ^2}\Phi ^4 + \mathcal{O}(\Phi ^6) \ ,
\end{equation}		
%
%
%
%
		\subsection{The polynomial self-interaction: $\gamma=0$}\label{S4.3.1}
\begin{figure}[H]
			 \centering
		 		\begin{picture}(0,0)
			 	 \put(41.5,20){\small $1^{st}\ LR$}
			 	 \put(41.5,35){\small $\xi_{\rm min}$}
			 	 \put(41.5,49){\small $\xi _{\rm trans}$}
			 	 \put(115,-4){\small $\omega /\mu _S$}
			 	 \put(-4,74){\begin{turn}{90}{\small $M\ze \mu _S$}\end{turn}}
			 	 \put(136,82){${\scriptstyle \beta _S\, =\, 100}$}
			 	 \put(130,70){${\scriptstyle \beta _S\, =\, 0}$}
			 	 \put(114,60){${\scriptstyle \beta _S\, =\, -100}$}
				\end{picture}
			 \includegraphics[scale=0.61]{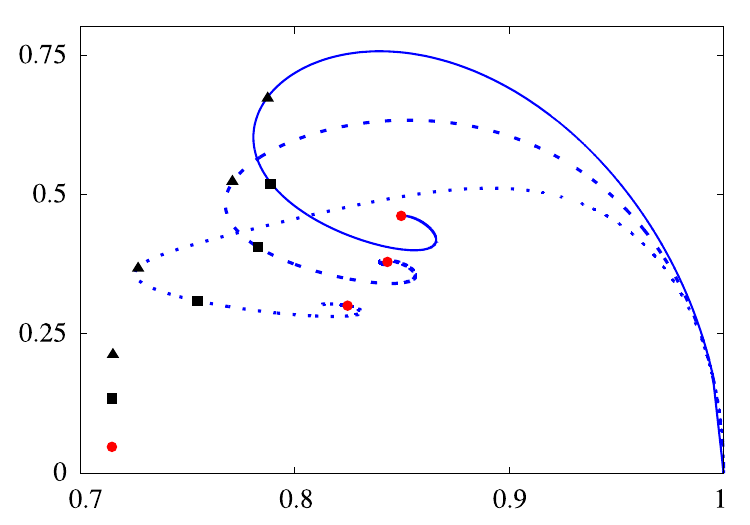}
		  		\begin{picture}(0,0)
				 \put(20,130){$\xi _{\rm min}$}
				 \put(75,48){$\chi(\xi _{\rm trans})$}
				 \put(105,-5){\small{$\beta _S$}}
				\end{picture}
			\includegraphics[scale=0.60]{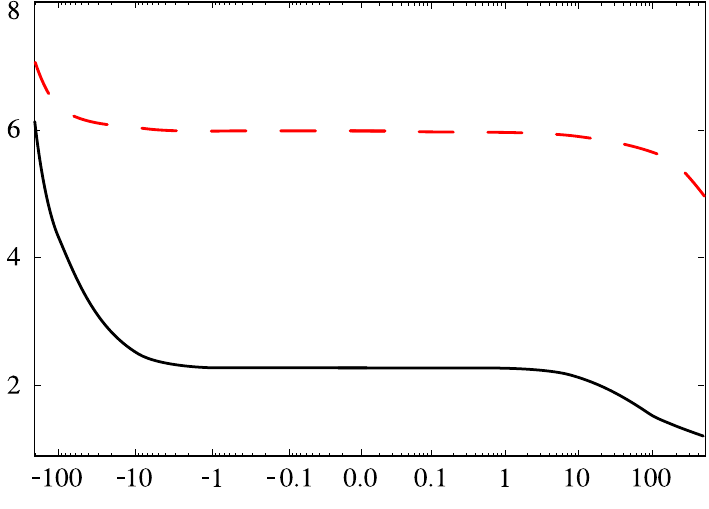}
	 	 	\caption{(Left panel) domain of existence of the self-interacting SBSs with the potential~\eqref{E5.3.21}, $\gamma=0$ and three different values of $\beta_S$: (solid) $\beta_S =100$; (dashed) $\beta _S = 0$ or mini-BSs; (dotted) $\beta_S =-100 $. (Right panel) $\xi_{\rm min}$ (dashed red line) and $\chi(\xi_{\rm trans})$ (solid black line) as a function of $\beta_S$.}
	 	 \label{F5.1}
		\end{figure}
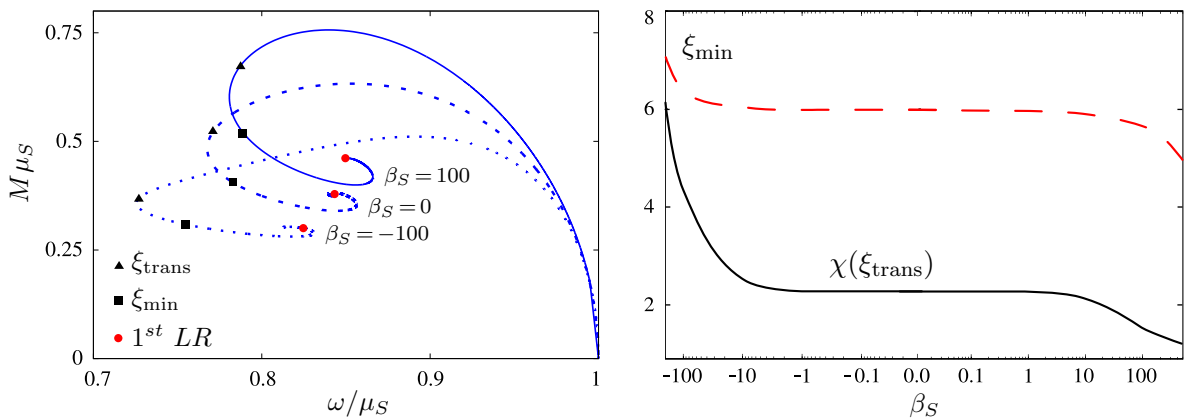

%
	First, let us consider the polynomial self-interaction with the quartic term only ($\gamma=0$). The domain of existence for three values of $\beta _S=\{-100,\, 0 ,\, 100\}$ can be observed in Fig.~\ref{F5.1} (left panel). The $\beta _S=0$ case corresponds to the standard mini-SBSs. This model exemplifies generic behaviours observed for SBSs with up to quartic self-interactions and PSs without self-interactions. In fact, in the spherical case, scalar and vector mini-BSs have been qualitatively similar in the generality of their physical and phenomenological properties studied in the literature so far, only with quantitative differences.\footnote{In the rotating case, however, a major dynamical difference was exhibited in~\cite{sanchis2019nonlinear}.} But in this section, a qualitative difference with phenomenological impact will be observed.
	The domain of existence of all studied SBSs in Fig.~\ref{F5.1} (left panel) shows a spiral behaviour starting at $\omega/\mu_S =1$  and $M \mu_S =0$,  corresponding to the Newtonian limit wherein the stars are less compact. Starting from this limit, the SBSs first increase (decrease) the ADM mass (frequency) until they reach a maximum mass ($M_{\rm max}$) at $\omega _{\rm crit}$. Then the mass decreases until the minimum frequency, which completes the first branch. After the back bending of the curve, there is a second branch. Several branches are observed, each ending on a back-bending of the curve, forming a spiral. The first branch, up to  $M_{\rm max}$, corresponds to the perturbatively stable SBSs solutions~\cite{gleiser1988stability,gleiser1989gravitational}. The latter is a common feature for spherical, fundamental SBSs solutions in the literature, also for PSs~\cite{brito2016proca}. However, we will see that a qualitatively different behaviour emerges in the axionic case discussed below. In all models herein, except the latter, we refer to the \textit{perturbatively stable} BSs as the solutions in the part of the first branch that lies between the Newtonian limit and the maximal mass solution. The remaining discussion focus on these ``standard'' models (they will be reconsidered in Sec.~\ref{S4.3.3} for the case of the axionic model).

	Moving along the spiral, the ADM mass ($M$) and frequency ($\omega$) undergo oscillations, as does the compactness -- see~\cite{cunha2017lensing} for a plot in the case of mini-SBSs. The field amplitude at the origin ($\phi_0$ for the scalar case and $b_0$ for Proca), on the other hand, grows monotonically. Therefore, the latter is a parameter that uniquely labels the solutions along the spiral. Thus, we define the ratio between the field amplitude at the origin ($\phi _0,\, b_0$) of a given solution along the spiral, denoted as $X$, and the field amplitude  at the origin of the maximal mass  solution,
			\begin{equation}\label{5.3.24}
	 		 \chi (X)\equiv \frac{\phi _0 ( X ) }{\phi _0 (M_{\rm max})} \ \ \ [{\rm{scalar}}] \qquad \textrm{or} \qquad \chi (X)\equiv \frac{b_0 ( X ) }{b_ 0 (M_{\rm max})} \ \ \ [{\rm{vector}}]\ ,
			\end{equation}
	as an indicator of how close the $X$ solution is from the perturbative stability crossing point (which is the maximal mass solution). In other words, $\chi(X)>1$ means the solution is \textit{perturbatively unstable}.

	Let us now turn our attention to the LRs and TCOs. In Fig.~\ref{F5.1} (left panel), the first ultra-compact SBS is denoted by a red circle for each of the three values of $\beta _S$ used. For $\beta _S =0$ this solution occurs in the third branch~\cite{cunha2017lensing}. Introducing self-interactions does not change, and the solution remains in the perturbatively unstable region with $\chi >1$.

	Consider now the TCOs. For small values of $\phi_0$, the areal radius of the maximum of $\Omega$, $R_\Omega$, is zero. Then, moving along the spiral, at a critical value of $\phi_0$, there is a \textit{transition}: $R_\Omega$ starts to move away from the origin. The transition solution, where $R_\Omega$ starts to depart from the centre, is denoted by a triangle on each  of the curves in Fig.~\ref{F5.1} (left panel). We can see this solution always has $\chi>1$: it is in the perturbatively unstable region.  

	As discussed in the introduction, $R_\Omega$ was seen to play the role of a BH ISCO in the simulations in~\cite{olivares2018tell}. However, the question arises, does it provide a similar scale for a BS and a Schwarzschild BH with the same ADM mass? To  analyse this possibility, we introduce the ratio
			\begin{equation}\label{E5.4.25}
			 \xi \equiv\frac{R_{\rm ISCO}}{R_\Omega} \  .
			\end{equation}

	For each value of $\beta_S$, we observe that there is a minimum value of $\xi$, denoted $\xi_{\rm min}$, which occurs for a solution with $\chi(\xi_{\rm min})>\chi(\xi_{\rm trans})>1$, where $\xi_{\rm trans}$ represents the transition solution wherein $R_\Omega$ starts to depart from the origin. The solution with $\xi_{\rm min}$  is denoted by a black square on each of the curves in Fig.~\ref{F5.1} (left panel).  

	To summarise, Fig.~\ref{F5.1} (left panel) shows that, in this model, SBS solutions with $R_\Omega\neq 0$ only occur after $M_{\rm max}$ and thus are perturbatively unstable. The transition solution responds to a positive (negative) coupling approaching (moving away) from the stable branch. We can observe from Fig.~\ref{F5.1} (right panel) that $\chi(\xi_{\rm trans})$ approaches unity for the largest positive values of $\beta _S$ but does not quite reach it within the values of $\beta _S$ explored. In this limit, the minimum value of $\xi_{\rm min}$ is still larger than $5\ze$.
	
	From the data presented, quartic self-interactions make SBSs have their transition solution closer to the stable region (than mini-SBSs) but not quite reaching it. The first ultra-light compact SBSs, on the other hand, does not approach the stable region noticeably.
%
			\subsection{The polynomial self-interaction:  $\gamma\neq 0$}\label{S4.3.2}
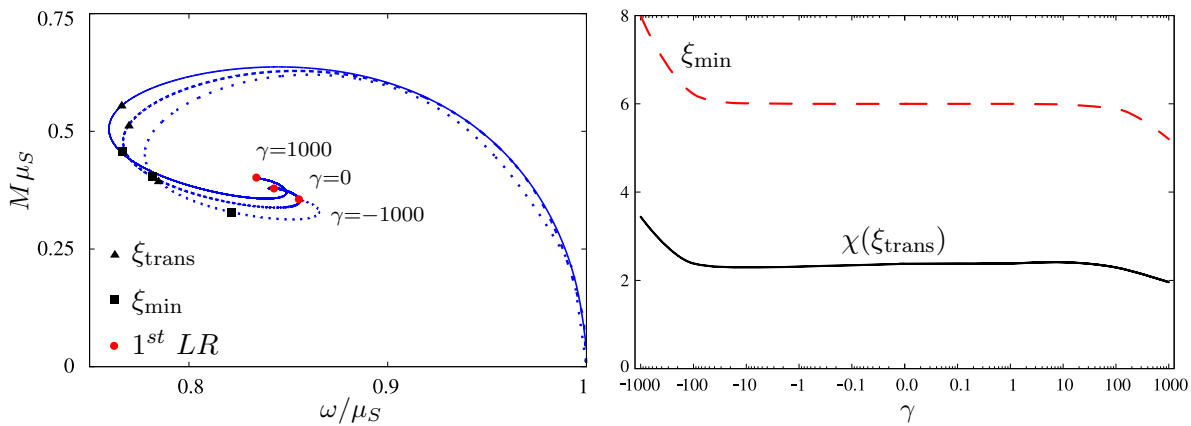
\begin{figure}[H]
				 \centering
		 			\begin{picture}(0,0)
		    		 \put(41.5,20){$1^{st}\ LR$}
					 \put(41.5,37){$\xi _{\rm min}$}
					 \put(41.5,55){$\xi _{\rm trans}$}
					 \put(88,94.5){${\scriptstyle \gamma =1000}$}
					 \put(108,84){${\scriptstyle\gamma =0 }$}
					 \put(116,70){${\scriptstyle\gamma =-1000}$}
					 \put(112,-4){\small $\omega /\mu _S$}
			 		 \put(-4,74){\begin{turn}{90}{\small $M\ze \mu _S$}\end{turn}}
					\end{picture}
		 		\includegraphics[scale=0.6]{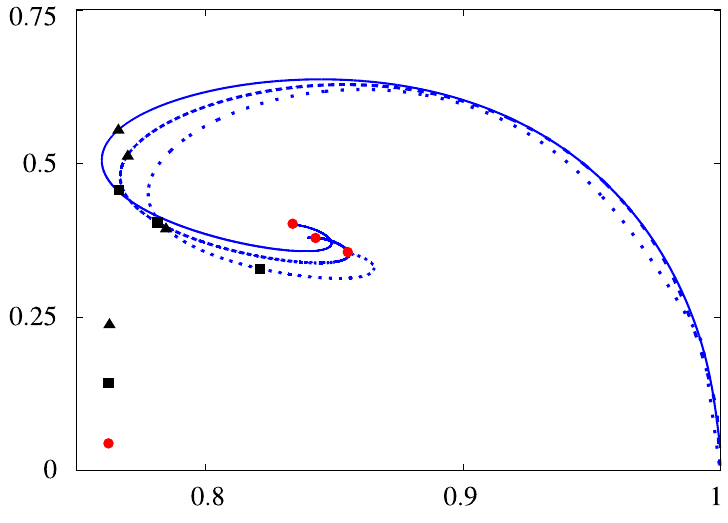}
		  			\begin{picture}(0,0)
					 \put(30,130){$\xi _{\rm min}$}
					 \put(90,60){$\chi(\xi _{\rm trans})$}
					 \put(112,-4){\small{$\gamma $}}
					\end{picture}
				\includegraphics[scale=0.6]{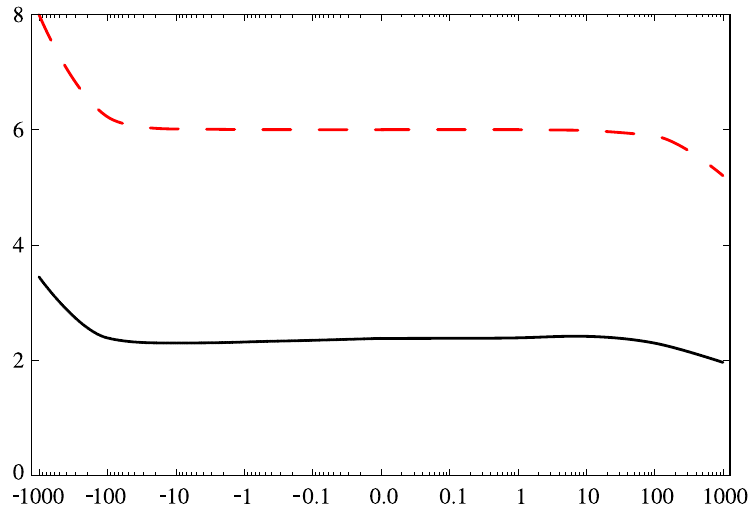}
	 	 		\caption{(Left panel) domain of existence of the self-interacting SBSs with the potential~\eqref{E5.3.21}, $\beta _S=1.0$ and three different values of $\gamma$: (solid) $\gamma  =1000$; (dashed) $\gamma = 0$; (dotted) $\gamma =-1000 $. (Right panel) $\xi_{\rm min}$ (dashed red line) and $\chi(\xi_{\rm trans})$ (solid black line) as a function of $\gamma$.}
			 \label{F5.2}
			\end{figure}

%
	We now consider the polynomial self-interaction with quartic and sextic terms ($\gamma\neq 0\neq \beta _S$). From the data presented before, and in an attempt to monitor the behaviour of the sextic coupling, we will consider a fixed $\beta_S=1.0\ze $. For each value of $\gamma$, we will have a family of solutions with an associated domain of existence, as represented in Fig.~\ref{F5.2} (left panel).
	The trends observed in Fig.~\ref{F5.2} are qualitatively similar to the ones seen in the previous Sec.~\ref{S4.3.1}, for $\gamma=0$ and varying $\beta _S$. In this case, however, both the first ultra-compact and the transition solutions respond to a positive (negative) coupling approaching (moving away) from the stable branch. Nevertheless, these solutions do not reach the perturbative stability region for the values of $\gamma$ explored.

	This analysis and the one in the previous subsection suggest that a simultaneous increase in both $\beta_S $ and $\gamma$ may bring $\xi _{\rm min}$ and $\chi (\xi _{\rm trans})$ closer to unity. To test this hypothesis, we have computed the domain of existence of a SBS with $\beta _S =100$ and $\gamma =1000$, corresponding to the two largest values that our code supported with trustable results. We obtained that the transition occurs closer to the maximum mass, $\chi (\xi _{\rm trans})=1.51$, but the $R_\Omega$ is still somewhat below the ISCO radius of the comparable BH: $\xi _{\rm min}=5.09$. Concerning the first ultra-compact solution, we obtained $\chi (LR)=6.58$, still far from the stability region. 
	
	From the data presented, the inclusion of a $6^{th}$ order self-interaction term does not lead to either ultra-compact SBSs or such stars with $R_\Omega\neq 0$ in the perturbatively stable region, although one observes a trend that could approach the stability region. We also observe that a $6^{th}$ self-interaction term has a minor impact on the overall SBS solutions than the quartic coupling, which is mainly associated with the fact that the scalar field is never larger than unity.
%
		\subsection{The axionic self-interaction}\label{S4.3.3}
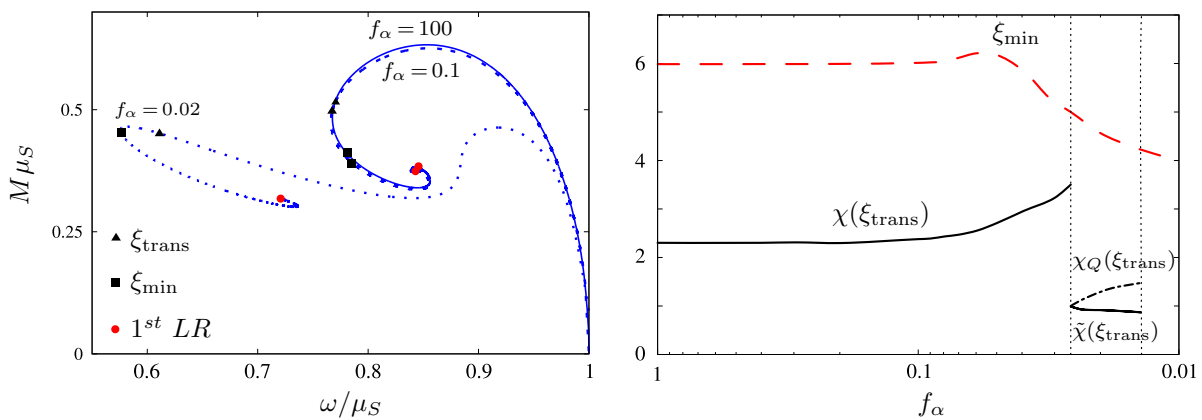
\begin{figure}[H]
				 \centering
					\begin{picture}(0,0)
		    		 \put(41,24){\small $1^{st}\ LR$}
					 \put(41,42){\small $\xi _{\rm min}$}
					 \put(41,58){\small $\xi _{\rm trans}$}
					 \put(35,108){\small ${\scriptstyle f_\alpha\, =\, 0.02}$}
					 \put(135,122){$\scriptstyle f_\alpha\, =\, 0.1$}
					 \put(130,138){$\scriptstyle f_\alpha\, =\, 100$}
					 \put(112,-4){\small $\omega /\mu _S$}
			 		 \put(-4,74){\begin{turn}{90}{\small $M\ze \mu _S$}\end{turn}}
					\end{picture}
				\includegraphics[scale=0.60]{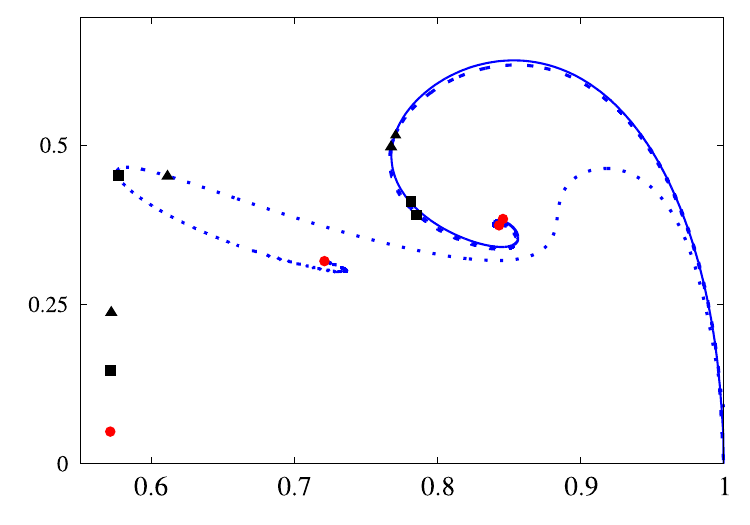}
				   \begin{picture}(0,0)
					\put(140,136){\small $\xi _{\rm min}$}
					\put(80,68){\small $\chi(\xi _{\rm trans})$}
					\put(170,24){$\scriptstyle \tilde{\chi} (\xi _{\rm trans})$}
					\put(170,51){$\scriptstyle \chi _{Q} (\xi _{\rm trans})$}
					\put(112,-4){\small{$f_\alpha $}}
				 \end{picture}
			   \includegraphics[scale=0.592]{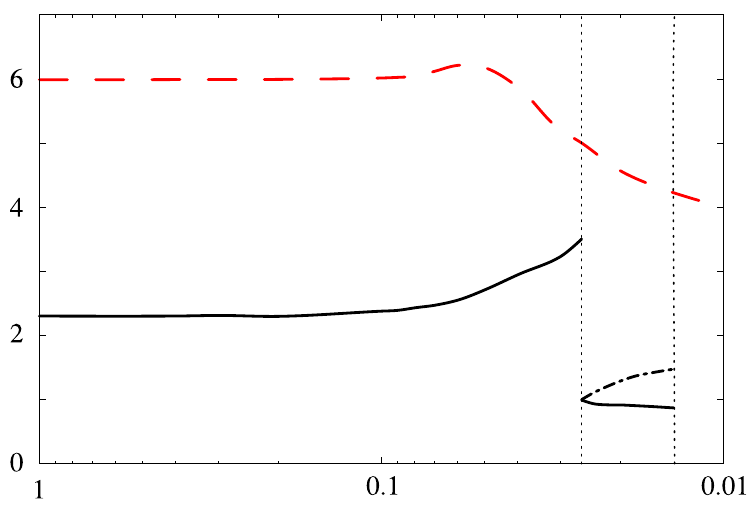}
	 		\caption{(Left panel) domain of existence of the self-interacting SBSs with the axionic potential~\eqref{E5.3.22} and three different values of $f_\alpha$: (solid) $f_\alpha =100$; (dashed) $f_\alpha =0.1$; (dotted) $f_\alpha =0.02 $. (Right panel) $\xi_{\rm min}$ (dashed red line), $\chi(\xi_{\rm trans})$  (solid black line) and $\chi _Q (\xi_{\rm trans})$ (dotted-dashed black line) as a function of $f_\alpha$. The left/right black vertical dotted lines define the value of $f_\alpha$ for which the second/third maximum first appears. The region to the right of the first vertical line encompasses BS solutions  with a relativistic stable branch.}
	 		 \label{F5.3}
			\end{figure}

%
	As our last example of spherical SBSs we use the axionic potential, first considered in~\cite{guerra2019axion}. The domain of existence for three different values of the coupling constant $f_\alpha=\{ 100,\, 0.1,\, 0.02\}$ is represented in Fig.~\ref{F5.3} (left panel). 
	For a large value of $f_\alpha$, one recovers the mini-SBSs. Then, as $f_\alpha$ decreases, qualitatively different properties start to emerge. Most notably, a second local maximum for the mass appears (see Fig.~\ref{F5.3} (right panel) left vertical line) that, for some value of $f_\alpha$ (close to $f_\alpha =0.02$), becomes the global maximum. This qualitatively different domain of existence impacts the stability region as we now discuss. However, before that, let us introduce two new quantities:
			\begin{equation}\label{E5.3.26}
	 		 \tilde{\chi} (\xi_{\rm trans}) =\frac{\phi _0 (\xi_{\rm trans})}{\phi _0 (2^{\rm nd} M_{\rm max})}\ , \qquad  \qquad \chi _Q(\xi_{\rm trans})=\frac{\phi _0 (\xi _{{\rm trans}})}{\phi _0 (Q=M)}\ ,
			\end{equation}
	where $\tilde{\chi}	(\xi_{\rm trans})$ $\big[\chi _Q(\xi_{\rm trans}) \big]$ gives a measure of the distance between the solutions where $R_\Omega$ depart from the origin and the second maximum (solution for which the Noether charge equals the ADM mass $Q_S\,\mu_S=M$).
	
\begin{figure}[H]
				 \centering
			 	 \includegraphics[scale=0.62]{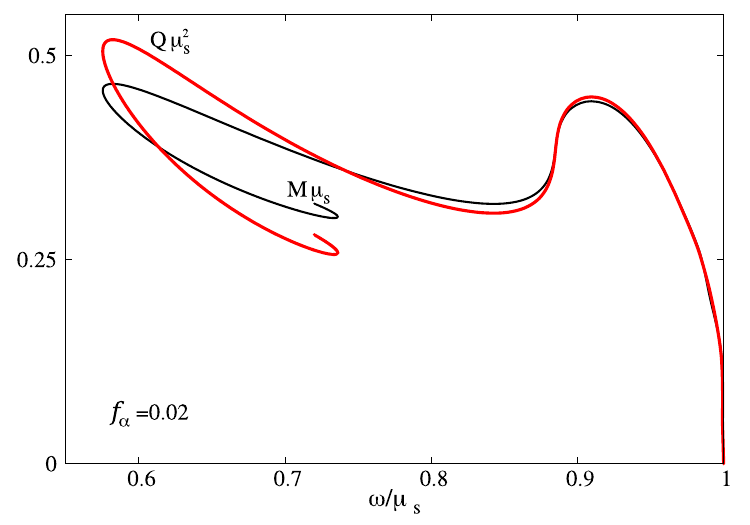}
    			 \includegraphics[scale=0.28]{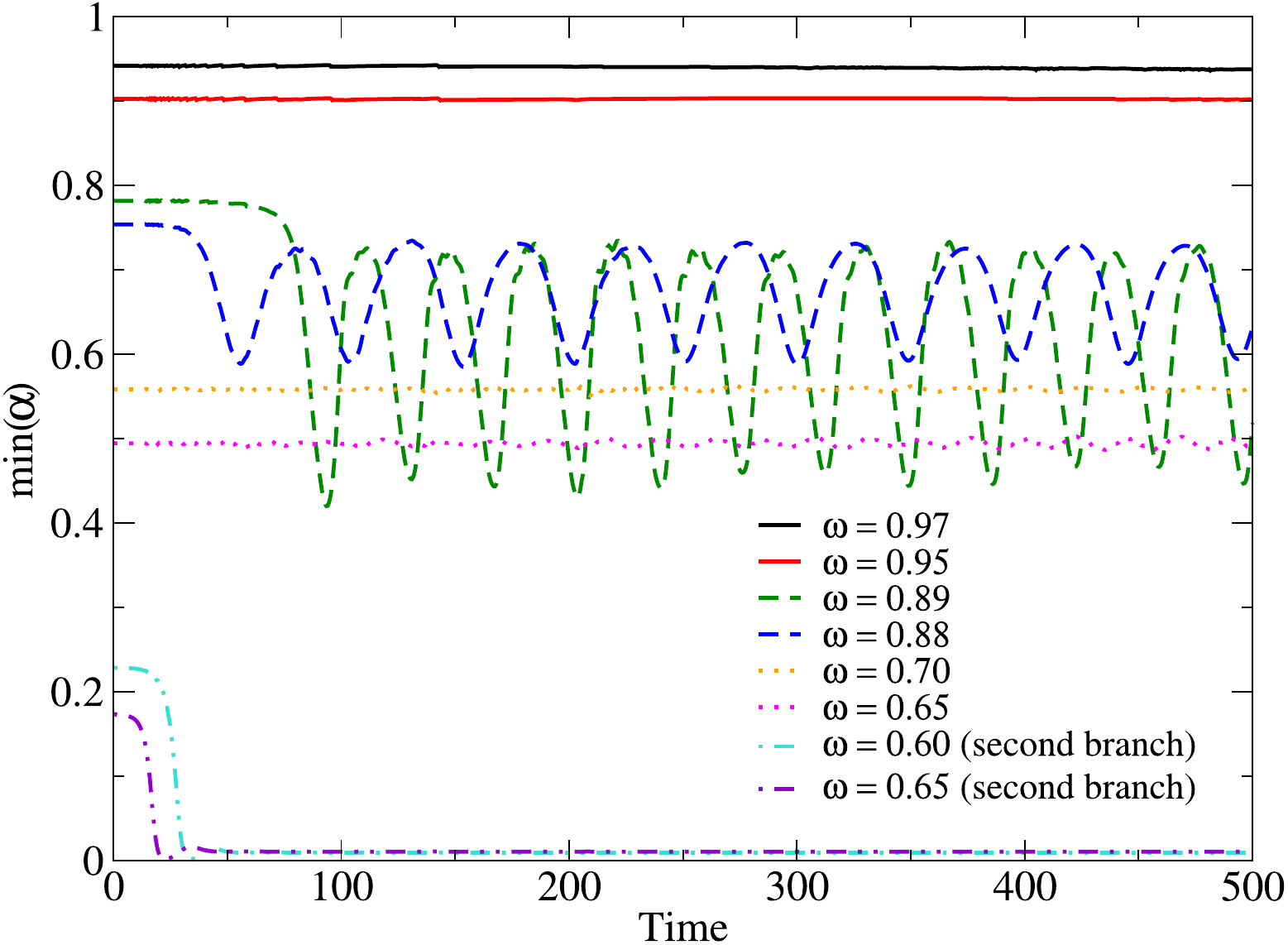}
				 \includegraphics[scale=0.28]{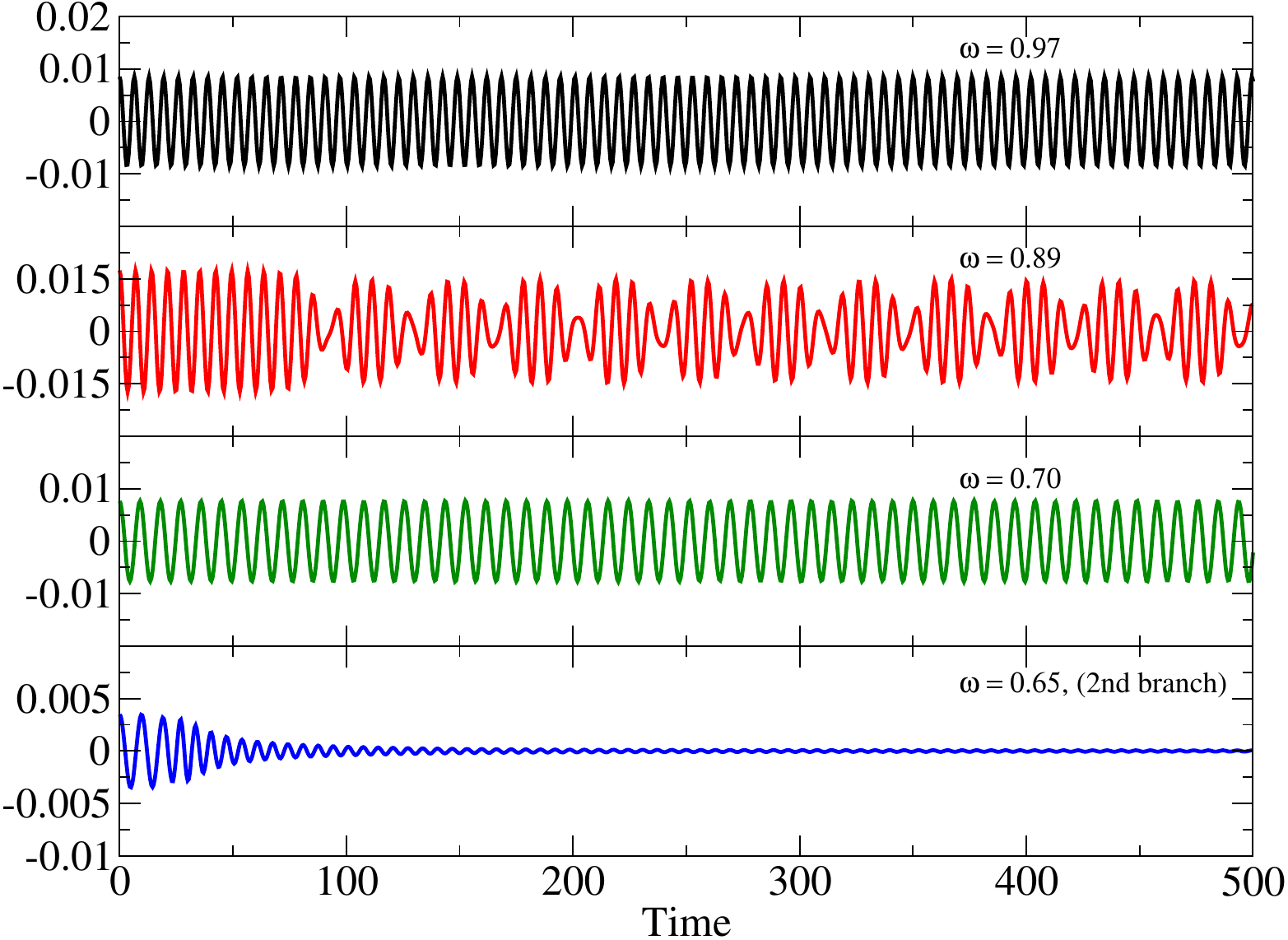}
                 \includegraphics[scale=0.28]{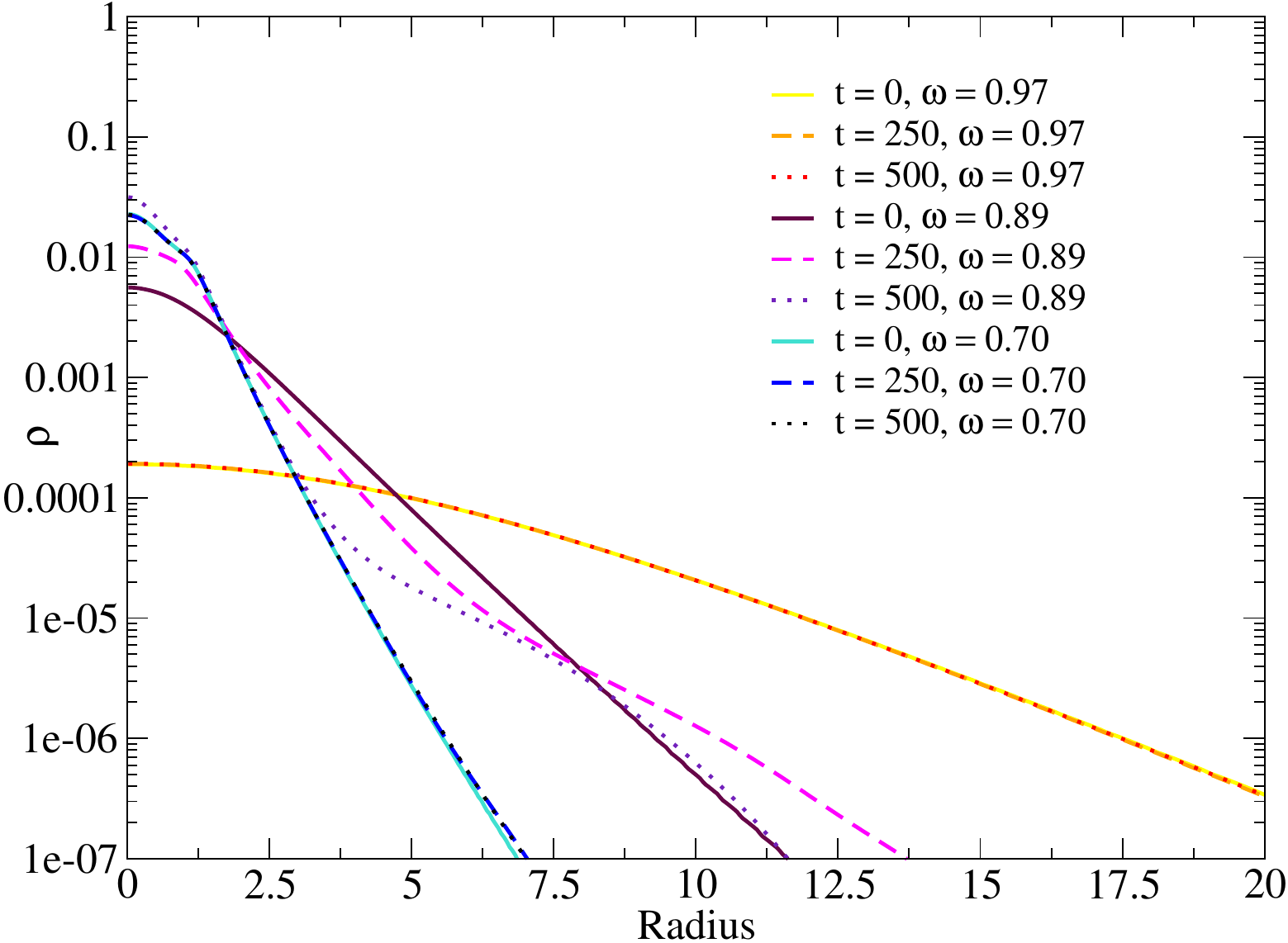}
	 		 	 \caption{(Top left panel) domain of existence of the self-interacting SBSs with the axionic potential~\eqref{E5.3.22} for both the ADM mass and Noether charge. Evolution of the minimum of the lapse (top right panel) and scalar field at an illustrative observation radius (bottom left panel) for several models. (Bottom right panel) evolution of the radial profile for three illustrative models.}
	 		 	 \label{F5.4}
				\end{figure}

	In all previous cases of SBSs, it has been argued that solutions in the $(M,\omega)$ domain of existence after the global maximum are unstable. The existence of such unstable modes originating at such a critical point of the ADM mass has been explicitly shown by perturbation theory studies in different works, $e.g.$~\cite{kusmartsev1991stability,gleiser1988stability,gleiser1989gravitational,brito2016proca}. However, could there be other stability regions further into the spiral? In~\cite{kleihaus2012stable} it was argued that catastrophy theory arguments suggest that BS models with a domain of existence akin to that for $ f_ \alpha=0.02$ in Fig.~\ref{F5.3}, have two stable branches: the \textit{Newtonian} one between the maximal frequency and the first (local) maximum of the mass, and the \textit{relativistic} one between the first local minimum of the mass and the second local (which can be global) maximum. The second region occurs to the right of the leftmost vertical dotted line in Fig.~\ref{F5.3} (right panel). To test this conclusion, we have performed fully non-linear numerical evolutions\footnote{Dynamical evolution performed by Nicolas Sanchis-Gual.}, using the same setup and code as described in~\cite{sanchis2016explosion,sanchis2016dynamical,escorihuela2017quasistationary,cunha2017lensing,herdeiro2018spontaneous}, of different SBSs corresponding to our axionic model~\eqref{E5.3.22}. The code uses spherical coordinates under the assumption of spherical symmetry employing the second-order Partially Implicit Runge-Kutta (PIRK) method developed by \cite{montero2012bssn,cordero2012partially,cordero2014partially}. Our results are exhibited in Fig.~\ref{F5.4}.

	The results in Fig.~\ref{F5.4} support that there are indeed two disjoint stable branches of solutions. In the top left panel, we show both the ADM mass and the Noether charge $Q_S$ for $f_\alpha\simeq 0.02$. The regions where $Q_S\,\mu_S<M$ $(\chi _Q <1)$ correspond to solutions with energy excess, which are not energetically stable. The first (top) branch of solutions between the minimum frequency and the crossing point between $M$ and $Q_S\, \mu_S$, corresponding to $\omega/\mu_S\in [0.582, 0.753]$ are stable. The aforementioned is the relativistic stable branch. On the other hand, the Newtonian stable branch corresponds  to  $\omega/\mu_S\in  [0.92, 1]$\footnote{In all the Newtonian branch, $Q_S\, \mu_S/M>1$ with the crossing point at $\omega/\mu_S\simeq 0.885$.}.  The stability in these branches is corroborated by the analysis in the top right panel, exhibiting the minimum of the lapse function $\alpha$ as a function of time, during the evolutions. Solutions in these stable branches have an approximately constant lapse, as illustrated by $\omega/\mu_S=0.97,0.95$ (Newtonian branch) and  $\omega/\mu_S=0.70,0.65$ (relativistic branch),\footnote{The small oscillations seen come from the interpolation, the different resolutions of the two grids in the two refinement levels used in the simulations and the outer boundary.} whereas the solutions in the frequency range in between exhibit large oscillations, as illustrated by $\omega/\mu_S=0.88, 0.89$. These solutions, however, do not decay into BHs. Indeed, two solutions in the second (bottom) branch of the left top panel, with $\omega/\mu_S=0.60, 0.65$, that collapse to BHs are also shown for comparison. The bottom left panel shows the scalar field extracted at an illustrative radius and corroborates the different qualitative behaviour between the stable and unstable branches. The bottom right panel exhibits the energy density of the SBSs, as a function of the radius, for the different models and for different times during the evolution. One can observe that: for  $\omega/\mu_S=0.97$ (Newtonian stable branch) and $\omega/\mu_S=0.7$ (relativistic stable branch), the radial profile does not change in time, but for $\omega/\mu_S=0.89$ the profile changes and the solution approaches the profile of the (more compact) solution in the relativistic stable branch. We conclude that the unstable models between the two stable branches do not collapse (even if perturbed) but rather migrate to the relativistic stable branch, where the SBSs are compacter.

	If we see now Fig.~\ref{F5.3} (right panel) we observe a region between $f_{\alpha} = [0.026, 0.016]$ where solutions can be both relativistic stable $ \tilde{\chi} (\xi_{\rm trans}) <1$ and energetically stable $\chi _Q (\xi_{\rm trans})>1$. Below $f_\alpha = 0.016$ the domain of existence becomes more complex -- with further local maximums -- and, while it could exhibit further stable regions, the numerical results are not precise enough to further explore that region.
	With the demonstrated evidence for a new stable branch, we can extract our main conclusion from Fig.~\ref{F5.3} (left panel): for $f_\alpha=0.02$ the transition point is already in the relativistic stable branch. Thus, unlike the previously analysed models, dynamically robust axionic BSs have $R_\Omega\neq 0$. However, we can see from Fig.~\ref{F5.3} (right panel) that the $\xi_{\rm min}$ is still not one, but further decreasing $f_\alpha$ exhibits a decreasing trend. 
%
			\subsection{Proca stars}\label{S5.3.4}
%
	Let us now consider the self-interacting PS model, with the potential~\eqref{E5.1.7}, first considered in~\cite{minamitsuji2018vector}. Setting the self-interactions coupling $\beta_P$ to zero, this model yields mini-PSs as solutions~\cite{herdeiro2017asymptotically,sanchis2017numerical,brito2016proca}. 

\begin{figure}[H]
		 	 \centering
		 		\begin{picture}(0,0)
				 \put(41,25){\small $1^{st}\ LR$}
				 \put(41,40){\small $\xi_{\rm min}$}
			 	 \put(41,55){\small $\xi^{2^{nd}} _{\rm trans}$}
			 	 \put(102,25){\small $\xi ^{1^{st}} _{\rm trans}$}
			 	 \put(102,41){\small $\xi =1$}
			 	 \put(155,63){${\scriptstyle \beta _P\, =\, 0.1}$}
			 	 \put(134,79){${\scriptstyle \beta _P\, =\, 0}$}
			 	 \put(120,95){${\scriptstyle \beta _P\, =\, -1}$}
			 	 	 \put(112,-4){\small $\omega /\mu _S$}
			 		 \put(-4,74){\begin{turn}{90}{\small $M\ze \mu _S$}\end{turn}}
				\end{picture}
		  	 \includegraphics[scale=0.61]{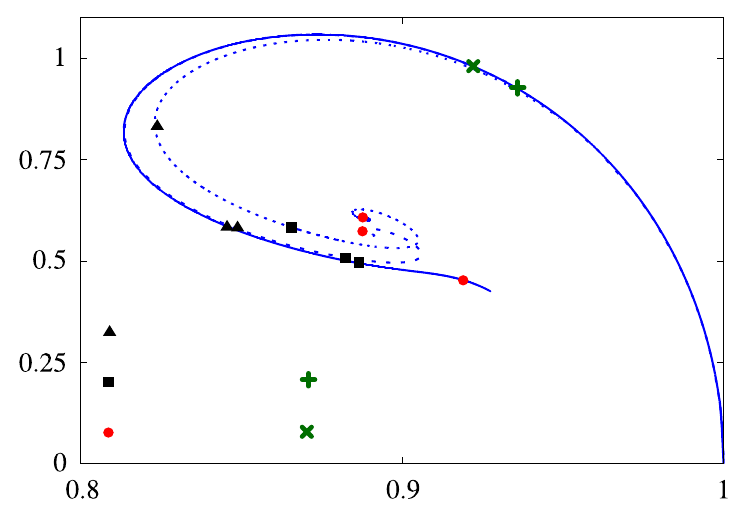}\hfill
		 	 \includegraphics[scale=0.59]{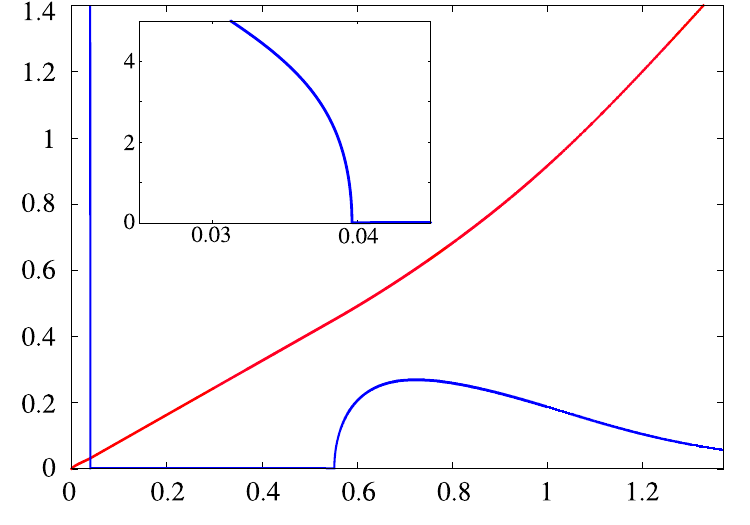}
		 		\begin{picture}(0,0)
		 	 	 \put(-50,70){\small $\beta _P = 0$}
			 	 \put(-150,124){\small $\scriptstyle R_\Omega$}
			 	 \put(-140,74){$\scriptstyle b_0$}
			 	 \put(-104,-4){\small $b_0$}
			 	 \put(-50,34){\small $R_\Omega$}
			 	 \put(-70,130){\small $\Omega _{\rm max}$}
				\end{picture}
	 	 	 \caption{(Left panel) domain of existence of the self-interacting PSs with three different values of the self-interaction coupling: (solid) $\beta _P =0.01$; (dashed) $\beta _P = 0$ or mini-PSs; (dotted) $\beta _P =-1.0\ze $. (Right panel) areal radius of the maximal angular velocity along TCOs, $R_\Omega$ (blue solid line)  and the corresponding value of the angular velocity $\Omega _{\rm max}$ (red solid line), as a function of the Proca field amplitude at the origin $b_0$ for a mini-PS.}
	 		 \label{F5.5}
			\end{figure}
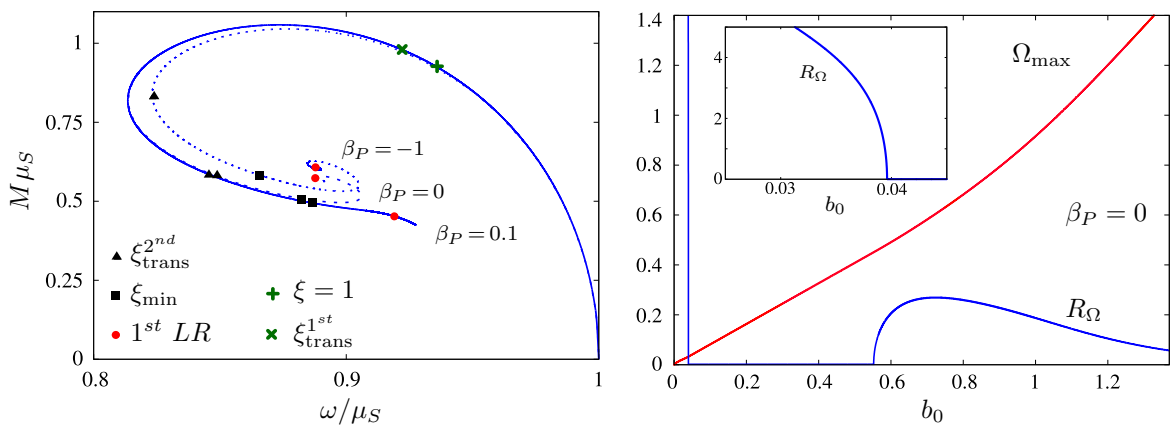

	The domain of existence of all studied PSs shows a spiral behaviour starting at $\omega/\mu_P =1$  and $M \mu_P =0$, corresponding to the less compact stars -- see Fig.~\ref{F5.5} (left panel). From this limit, the PSs first increase (decrease) the ADM mass (frequency) until they reach the maximum mass, $M_{\rm max}$, at $\omega _{\rm crit}$. Then the mass decreases until the minimum frequency, which completes the first branch. After the back bending of the curve, there is a second branch. The behaviour then depends on the coupling $\beta_P$. For $\beta_P\leqslant 0$, several branches are observed, each ending on a back bending of the curve, forming a   spiral. For $\beta _P >0$ the domain of existence is qualitatively different.\footnote{This may be understood~\cite{minamitsuji2018vector} from  the fact that for the fundamental PSs, $B_t$ must have a node; for $\beta _P>0$ a maximum of $b_0$ exists compatible with this requirement
			\begin{equation}
			 0<b_0\leqslant \frac{\mu _P\, \sigma _0}{\sqrt{\beta _P}}\ ,
			\end{equation}
	which was numerically confirmed. For $\beta _P <0$ there is also a critical value of $b_0$ beyond which solutions cease to exist; however it  appears to be less sensitive to the  coupling.} The first branch, up to  $M_{\rm max}$, corresponds to the perturbatively stable PSs solutions~\cite{brito2016proca}. 
	Let us now turn our attention to the LRs and TCOs. In Fig.~\ref{F5.5} (left panel), the first ultra-compact PS is denoted by a red circle for each of the three values of $\beta_P$ used. For $\beta _P=0$, this solution occurs at the beginning of the fourth branch~\cite{cunha2017lensing}. Introducing self-interactions, this solution remains in the perturbatively unstable region with $\chi >1$.

	Consider now the TCOs. Starting at $\omega/\mu_P =1$ and $M \mu_P =0$ ($b_0\rightarrow 0$), the areal radius of the maximum of $\Omega$, $R_\Omega$, is very large -- see the top inset in Fig.~\ref{F5.5} (right panel). Then, moving along the spiral, at a critical value of $b_0$, there is a first \textit{transition}: $R_\Omega\rightarrow 0$ (denoted in Fig.~\ref{F5.5} (left panel) as a green cross). This occurs for
		\begin{equation*}
	 	 \omega / \mu _P = 0.923 \ , \qquad  M \mu _P \approx 0.979 \ , \qquad \chi (\xi ^{1^{st}} _{\rm trans} )\approx 0.355 \ ,
		\end{equation*}
	well within the stable branch, and it is rather insensitive to the coupling $\beta_P$ in the models explored.

	Along the sequence of solutions between the Newtonian limit and the first transition, the ratio $\xi$~\eqref{E5.4.25} varies from a large value to zero. This means that a certain PS configuration in the \textit{perturbatively stable Newtonian branch} has $R_\Omega=6M$. This solution with $\xi =1$ (denoted in Fig.~\ref{F5.5} (left panel) as a green plus) has
		\begin{equation}\label{E5.3.28}
		 \omega / \mu _P \approx 0.936 \ , \qquad M \mu _P \approx 0.925\ , \qquad  \chi (\xi=1)\approx 0.248  \ .
		\end{equation}
	After the first transition, continuing along the spiral, $R_\Omega=0$ until a second transition occurs -- see the bottom inset in Fig.~\ref{F5.5} (right panel). The second transition solution, at which $R_\Omega$ moves away from the origin, is denoted by a triangle on each of the curves in Fig.~\ref{F5.5} (left panel). We can see this solution always has $\chi>1$: it is in the perturbatively unstable region, and it depends on $\beta_P$. The red curve in Fig.~\ref{F5.5} (right panel) shows that the maximal value of the angular velocity, $\Omega_{\rm max}$, increases monotonically with $b_0$.
			\begin{figure}[h!]
			 \centering
			  \includegraphics[scale=0.59]{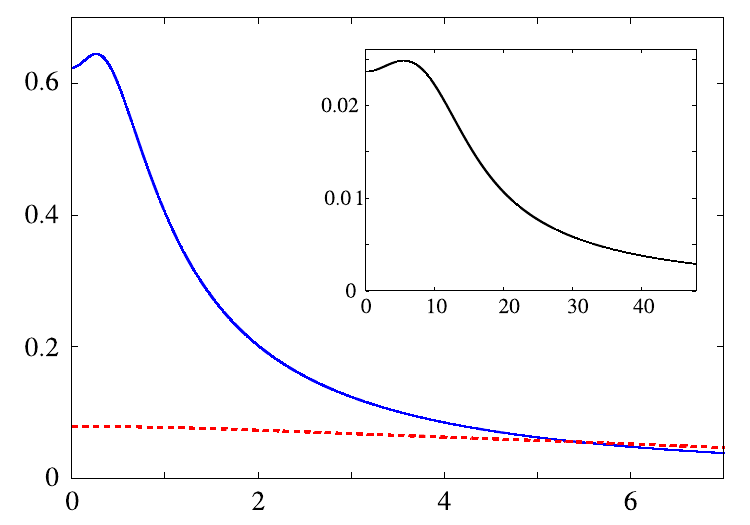}
		    	 \begin{picture}(0,0)
		     	  \put(-58,125){\small $\beta _P = 0$}
			 	  \put(-155,68){\small$\xi_{\rm min}$}
			 	  \put(-190,33){\small $M_{\rm max}$}
			 	  \put(-110,-5){\small $r$}
			 	  \put(-70,50){$\scriptstyle r$}
			 	  \put(-134,95){\begin{turn}{90}{$\scriptstyle \Omega $}\end{turn}}			 	  
			 	  \put(-220,78){\begin{turn}{90}{$\Omega $}\end{turn}}
			 	  \put(-70,100){$\scriptstyle \xi =1$}
				\end{picture}
					\begin{picture}(0,0)
				 \put(64,25){\small $-E_r$}
			 	 \put(45,112){\small $T_t ^t$}
			 	 \put(72,135){\small $j^t$}			 	
			 	 \put(150,30){$\scriptstyle B_r$}
			 	 \put(150,18){$\scriptstyle r$}
			 	 \put(138,40){$\scriptstyle B_t$}
			 	 \put(145,63){${\scriptstyle m}$}
			 	 \put(120,90){${\scriptstyle \sigma}$}
			 	 	 \put(114,-4){\small $r$}
				\end{picture}
			 \includegraphics[scale=0.59]{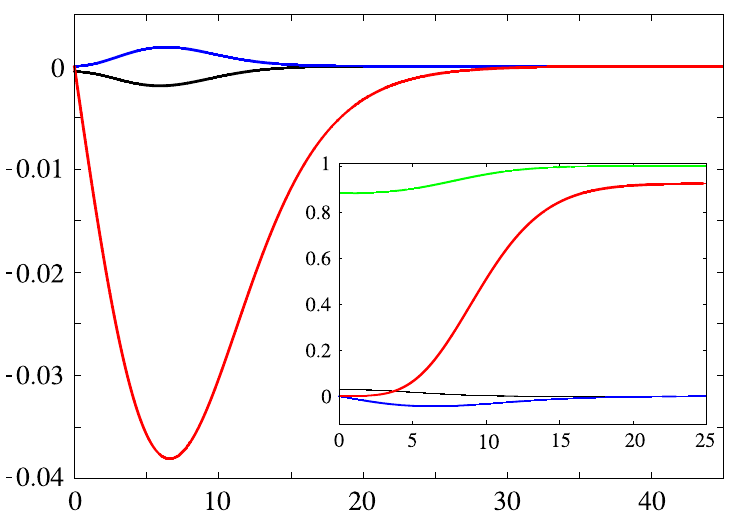}
	 	 	 \caption{(Left panel) $\Omega$ for the maximal mass mini-PS (red dashed line), for the solution with minimal $\xi$ (solid blue line) and (inset) for the $\xi=1$ solution (solid black line), for $\beta _P=0$. (Right panel) radial profiles of several quantities for the ``special'' solution  with $\xi=1$~\eqref{E5.3.28}: (main panel) energy density $T^t_t$ (see~\cite{brito2016proca} for the explicit expression), Noether charge density $j^t$~\eqref{E5.1.6} and ``electric field'' $E_r=| F_{rt}|$; (inset) metric functions~\eqref{E1.5.40} and Proca potential functions~\eqref{E5.1.7}.}
	 		 \label{F5.6}
			\end{figure}
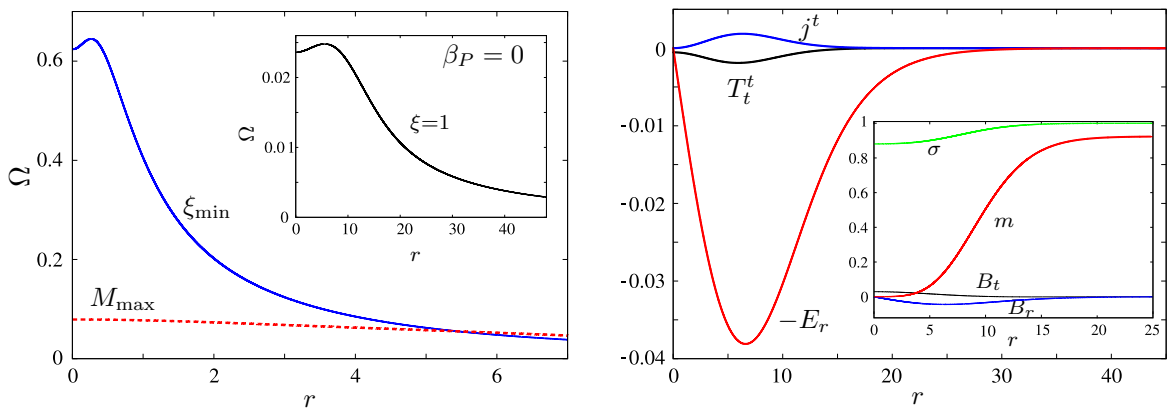
	In Fig.~\ref{F5.6} (left panel), we exhibit the radial profile of the angular velocity for two illustrative solutions: the red dashed line corresponds to the maximal mass solution, that has $\chi=1$, for which  $R_{\Omega}$ is still at the origin; the solid blue line is for a PS for which $R_{\Omega}$ is already away from the origin (after the second transition). 
	
	As in the scalar case, after the second transition, for each value of $\beta _P$, we observe that there is a minimum value of $\xi$, denoted $\xi_{\rm min}$, which occurs for a solution with $\chi(\xi_{\rm min})>\chi(\xi ^{2^{nd}} _{\rm trans})>1$,  where $\xi ^{2^{nd}} _{\rm trans}$ represents the second transition solution. The solution with $\xi_{\rm min}$  is denoted by a black square on each  of the curves in Fig.~\ref{F5.5} (left panel).  

	One may wonder why PSs allow for $R_\Omega\neq 0$ in the Newtonian stable branch, unlike the SBS. Whereas we do not have a final explanation, we would like to point out a special feature of PSs. Fig.~\ref{F5.6} (right panel) exhibits some physical quantities (main panel) as well as metric functions~\eqref{E1.5.40} and Proca potential functions~\eqref{E5.1.7} (inset) for the ``special'' solution with $\xi=1$~\eqref{E5.3.28}. One observes, in particular, that the energy density $T^t_t$ has a maximum away from the origin. The latter is a feature that had already been observed for PSs~\cite{brito2016proca,di2018dynamical} and it is more notorious precisely in the Newtonian branch. By contrast, for the SBSs, the maximum energy density is always at the origin.

\begin{figure}[h!]
			 \centering
			 \includegraphics[scale=0.28]{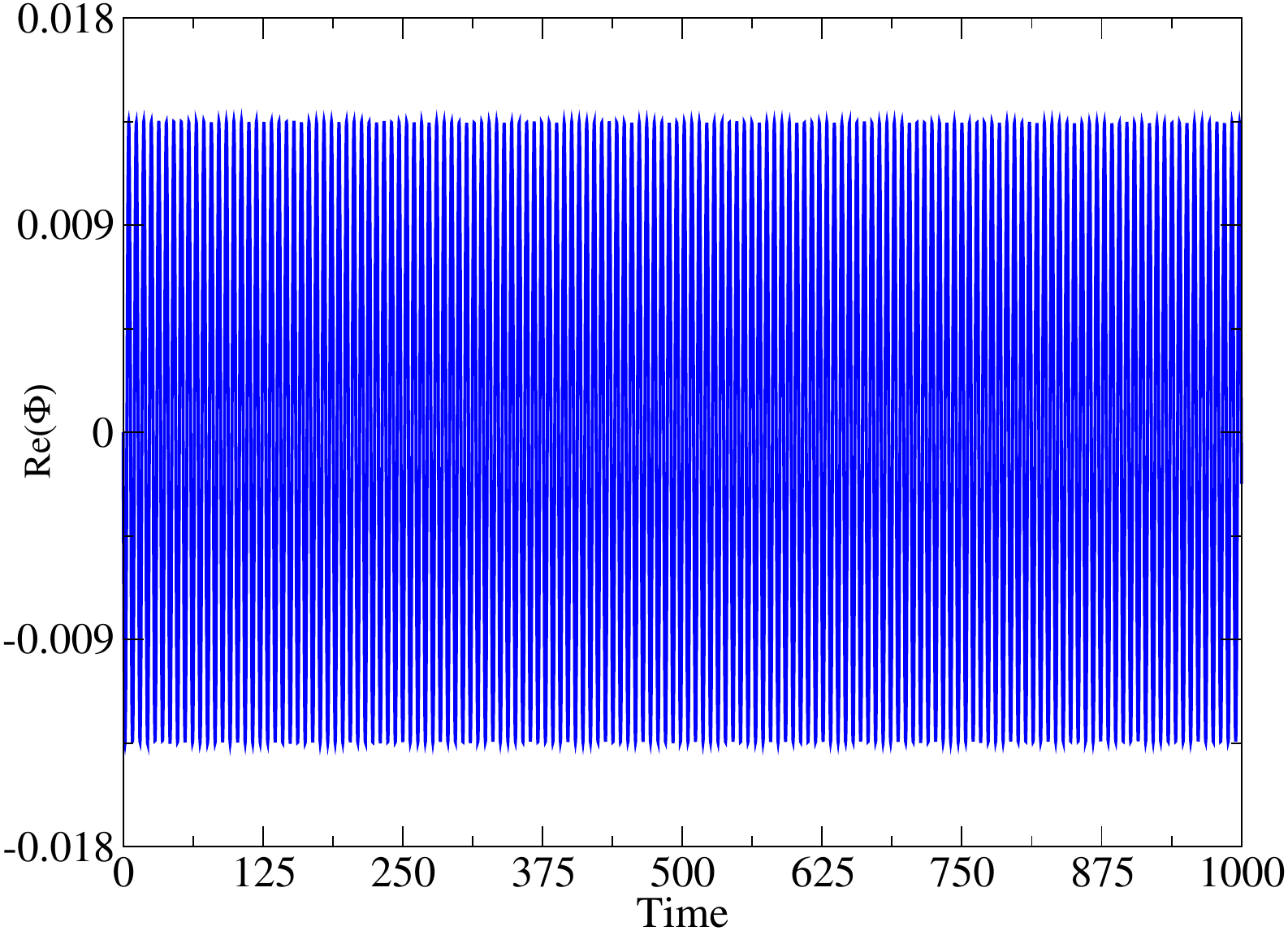}
   			 \includegraphics[scale=0.28]{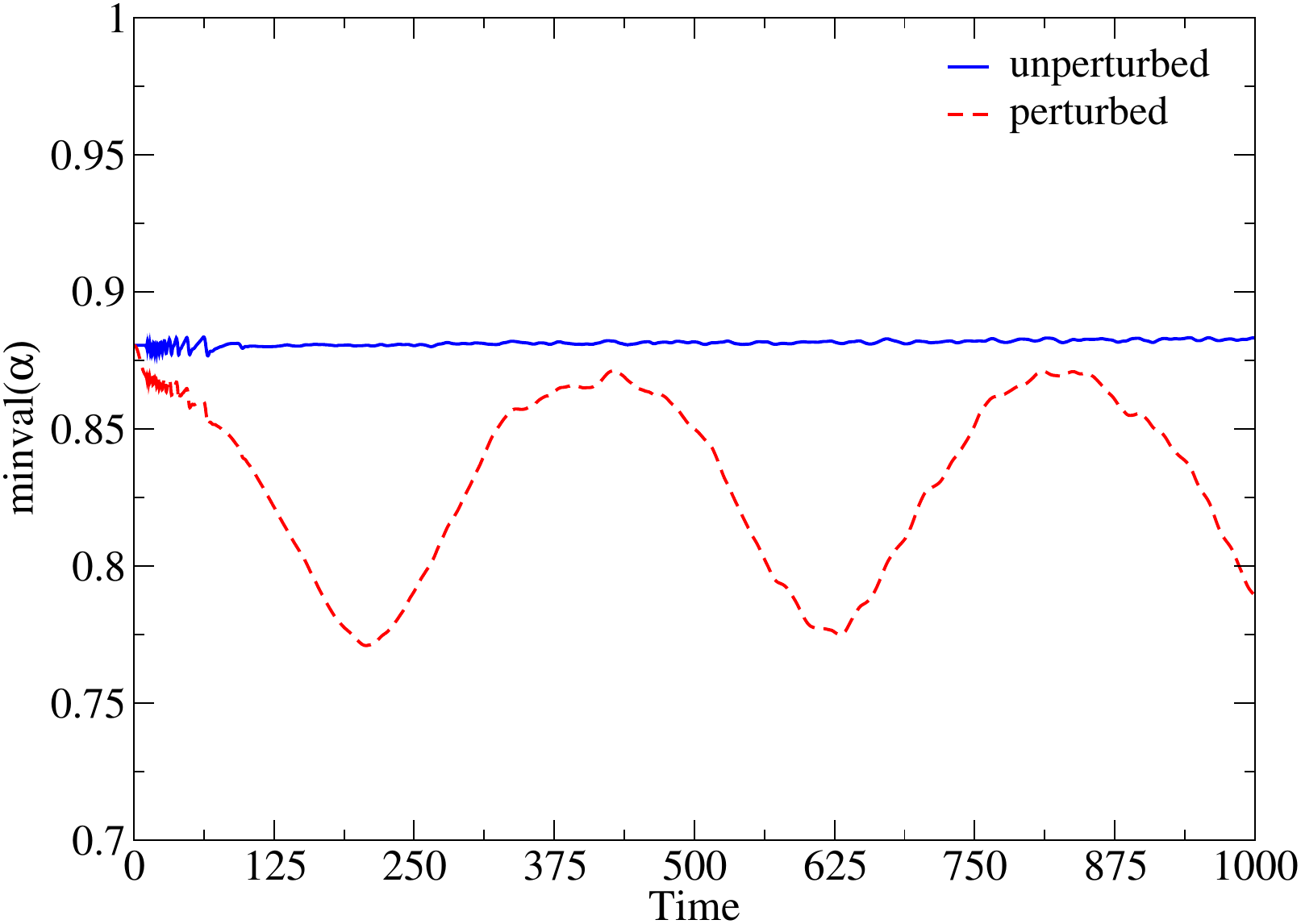}
			 \includegraphics[scale=0.3]{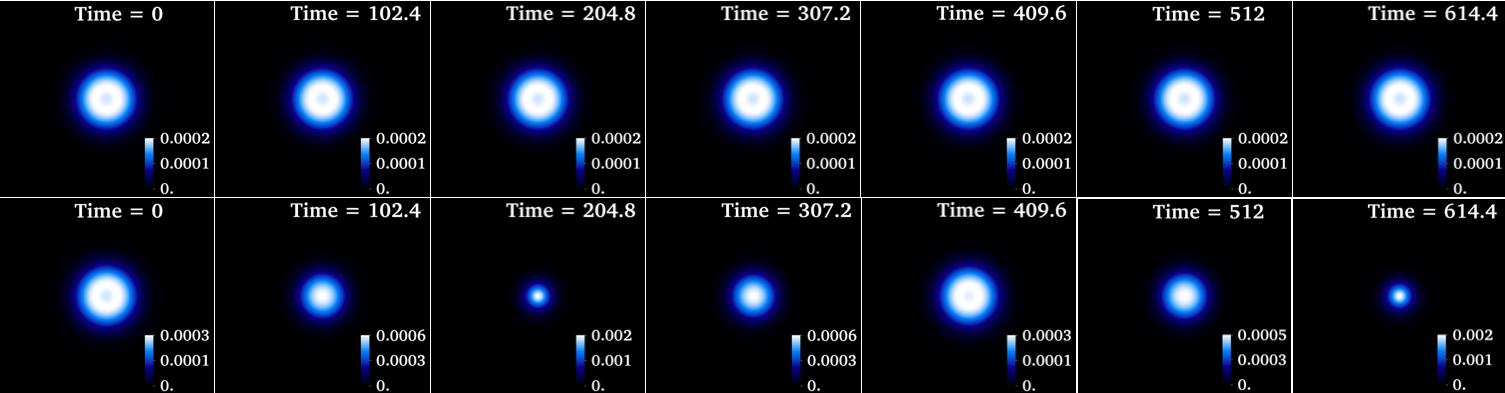}                     
	 		 \caption{Time evolution of the solution~\eqref{E5.3.28}: (top left panel) amplitude of the real part of the Proca scalar potential; (top right panel) minimum value of the lapse (also for a perturbation of the solution); (bottom panel) snapshots of the energy density, for both the unperturbed (top row) and perturbed (bottom row) evolutions.}
	 		 \label{F5.7}
			\end{figure}

	To summarise, PSs, while not being able to accommodate a LR in the perturbatively stable region, may have  $R_\Omega=R_{\rm ISCO}$ for dynamically stable stars, where the latter refers to a comparable ($i.e.$ same mass), Schwarzschild BH. In order to confirm the dynamical stability of the solution~\eqref{E5.3.28} we have evolved it using similar techniques to the ones described in Sec.~\ref{S4.2}. Here we have used the Einstein Toolkit to perform the dynamical evolutions\footnote{Dynamical evolution performed by Nicolas Sanchis-Gual.}~\cite{toolkit2012open,loffler2012einstein,zilhao2013introduction}, the Proca equations being solved with a modification of the \textsc{Proca} thorn \cite{witek2020canuda,zilhao2015nonlinear} for a complex Proca field; this setup has been used previously in~\cite{sanchis2017numerical,sanchis2019nonlinear}. The results are illustrated in Fig.~\ref{F5.7}. 

	The top left panel exhibits the time evolution of the amplitude of the real part of the Proca scalar potential of solution~\eqref{E5.3.28}; this is essentially the time component of the Proca potential, but see~\cite{sanchis2017numerical} for a precise definition. The maximal amplitude does not change. The top right panel shows the evolution of the minimum value of the lapse; this is essentially the time-time component of the metric, but see~\cite{sanchis2017numerical} for a precise definition. The two lines refer to the solution~\eqref{E5.3.28} and perturbation of this solution is obtained by multiplying the Proca field by $\times 1.05\ze $. The bottom panel shows snapshots of the time evolution of the energy density for both the unperturbed (top row) and perturbed (bottom row) evolutions. These evolutions confirm the expected stability of the solution~\eqref{E5.3.28}. One observes that the unperturbed solution is unaffected by the evolution, whereas the perturbed one oscillates but does not decay. 
	
	To conclude we remark that, in the PS's analysis $\big($and similarly in the scalar case \eqref{S4.2}$\big)$, we have allowed $\beta_P$ to take both positive and negative values as a means to see its impact on the LR and TCOs, even though negative values of $\beta_P$ lead to a self-interactions potential that is unbounded from below.
%
	\section{Lensing}\label{S4.4}
%
	Finally, we aim at confirming that the lensing of solution~\eqref{E5.3.28} lit by a thin accretion disk with its inner edge at $R_\Omega=6M$ indeed mimics the shadow of a mass $M$ Schwarzschild BH lit by a similar accretion disk. For this purpose, we have used an independent ray-tracing code to image both spacetimes' shadow and lensing\footnote{Lensing analysis performed by Pedro Cunha.}. This is the same code that was used in previous works, e.g.~\cite{cunha2018shadows,cunha2017lensing,cunha2016chaotic,cunha2015shadows,cunha2019spontaneously,cunha2017light,cunha2020stationary}, to numerically integrate the null geodesic equations $\ddot{x}^\mu + \Gamma^\nu_{\alpha \beta}\ze \dot{x}^\alpha \dot{x}^\beta = 0$. This procedure, $i.e.$, backwards ray-tracing, represents the propagation of light rays from the observer backwards in time towards the radiation source or the BH (if it exists).

	We consider a simplified astrophysical setup wherein the only radiation source is an opaque and thin accretion disk located on the equatorial plane around the central compact object. The disk has an inner edge with an areal radius $R_{\Omega}=6M$ in both spacetimes. For the PS, this radial location aims to mimic the inner edge of a stalled torus in the equatorial plane, inside which the Magneto-Rotational Instability (MRI) is essentially quenched~\cite{olivares2018tell}. To represent this system, we have imposed a luminosity profile for the disk, with a maximum at the disk edge and very fast decay as the radius increases.

The ray-tracing integration of a light ray stops when the photon reaches either: \textit{i)} the BH, \textit{ii)} the disk, or \textit{iii)} numerical infinity. Since the disk is the only light source assumed, photons that never intersect the disk via ray-tracing are shown as black pixels in the image. Black pixels thus include both photons that escape to numerical infinity and fall into the BH (forming the shadow).

\begin{figure}[h!]
		 \centering
		 \includegraphics[scale=0.20]{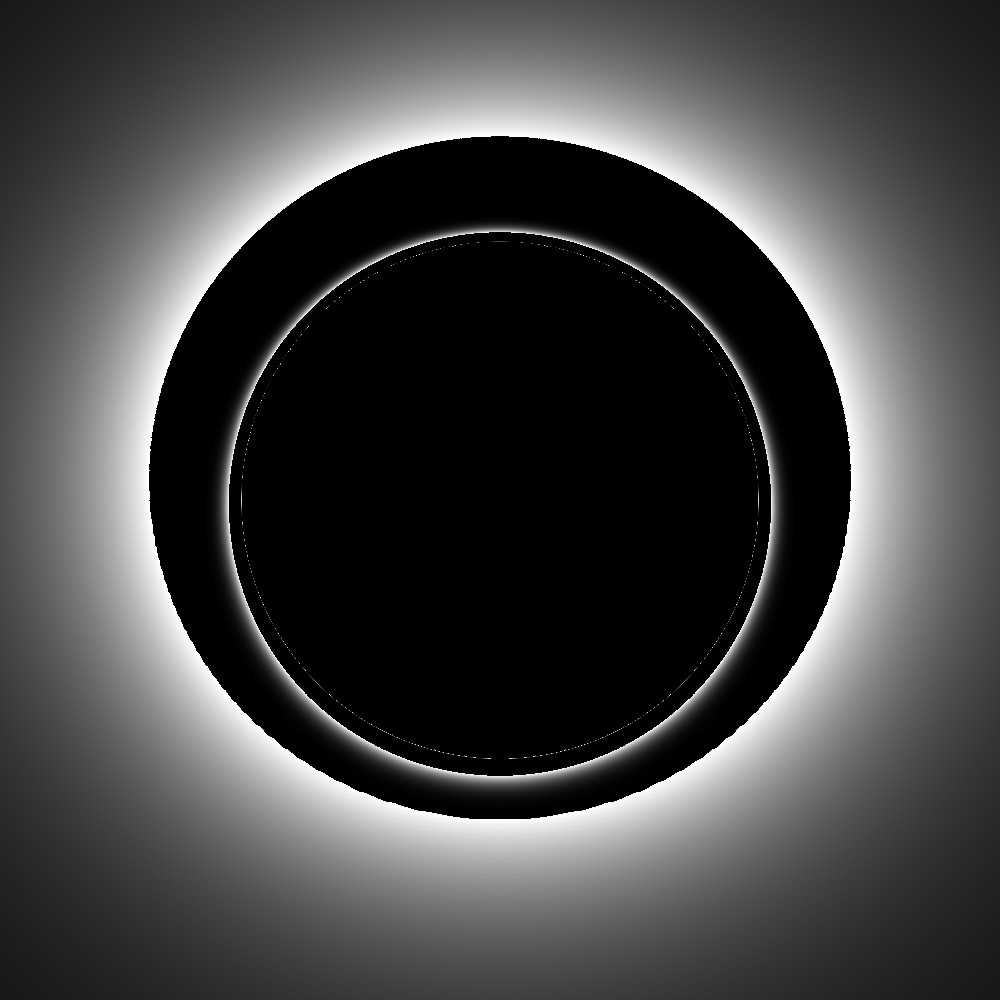}
		 \includegraphics[scale=0.20]{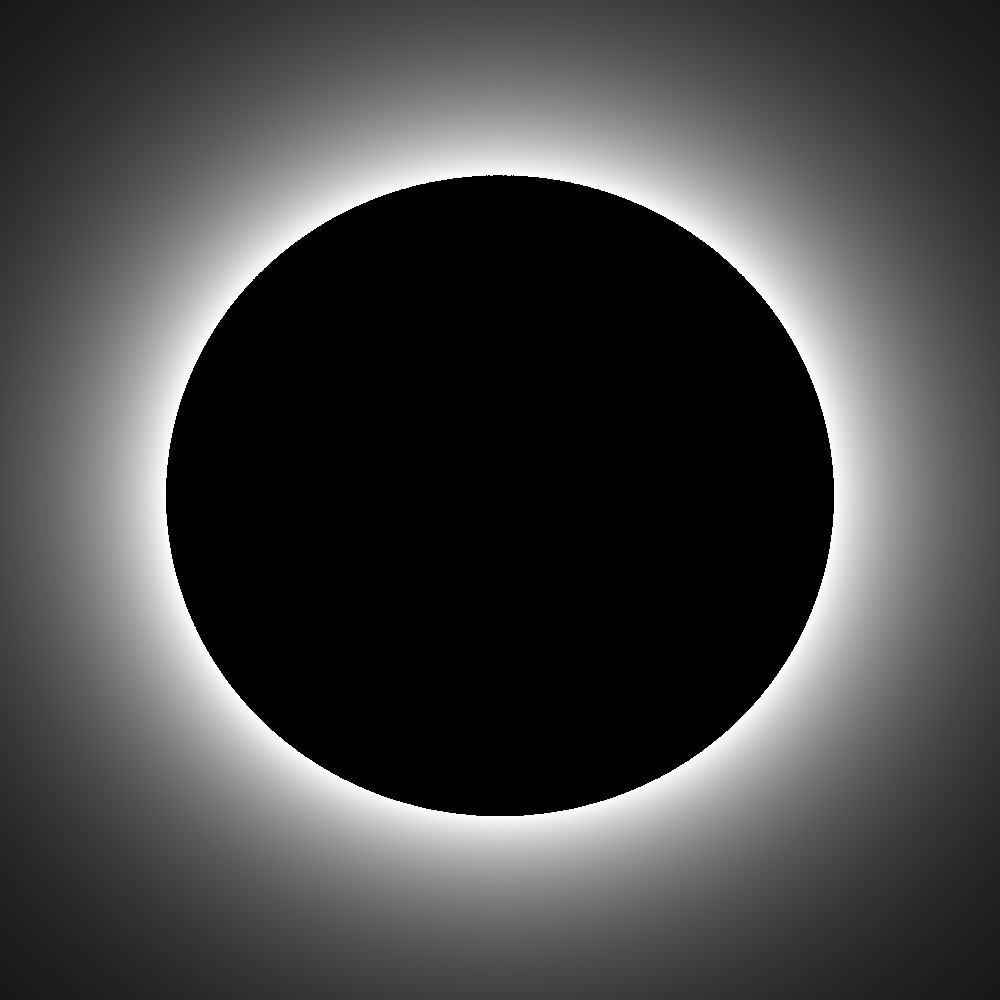}
		 \includegraphics[scale=0.20]{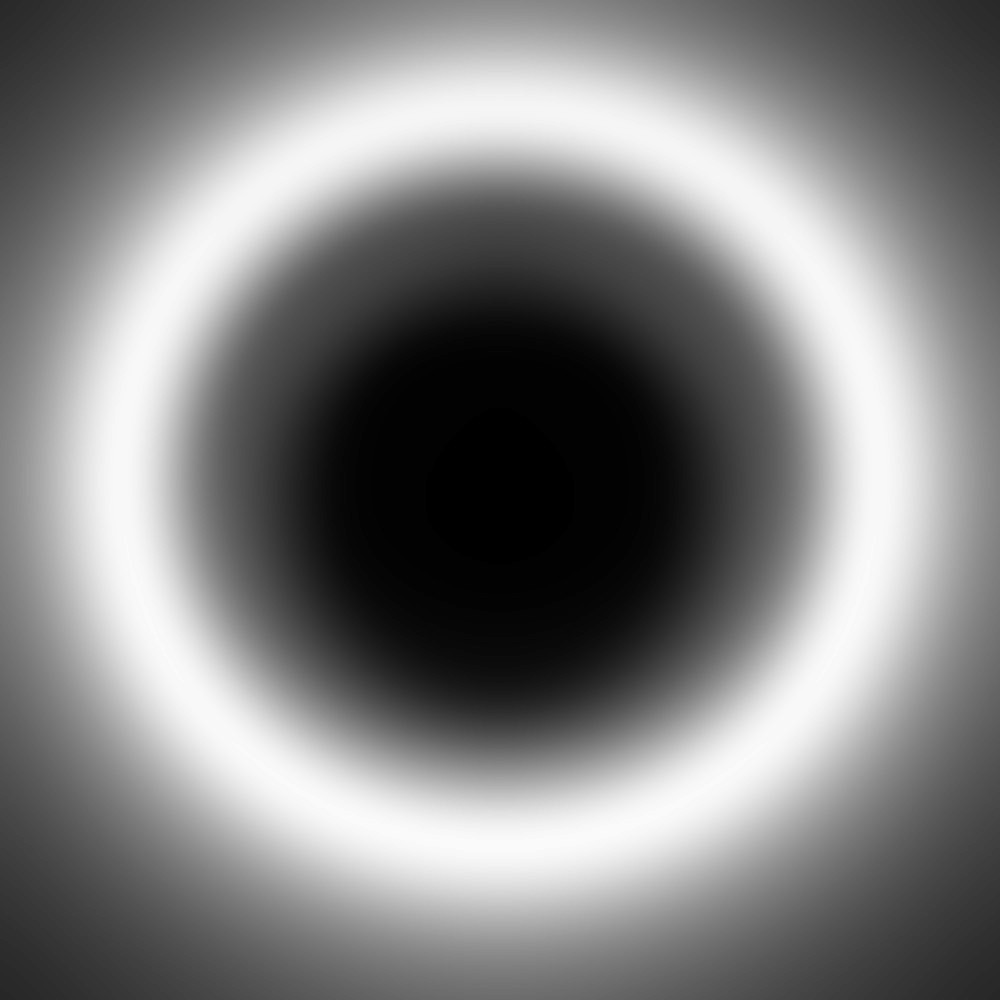}
		 \includegraphics[scale=0.20]{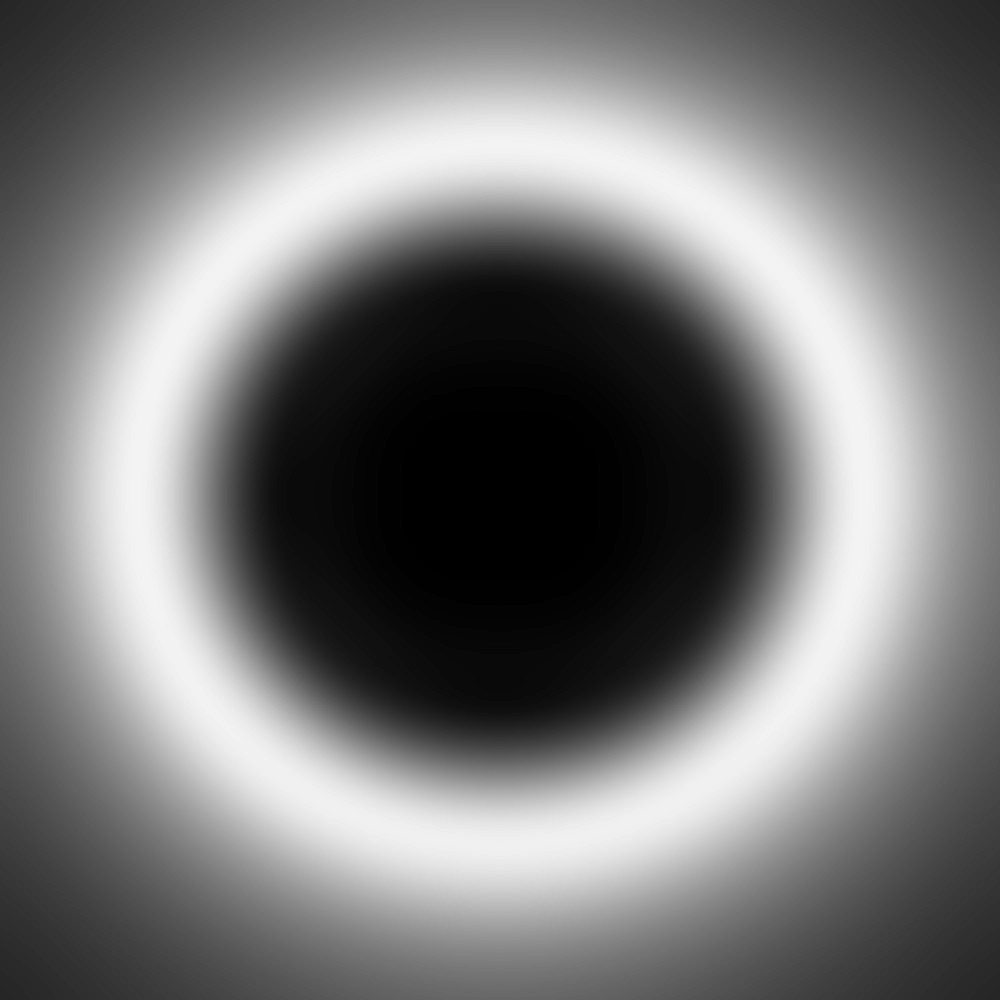}                   
	 	 \caption{Lensing at an observation angle of $\theta _O 17^o$ (almost polar): (top left) Schwarzschid; (top right) PS; (bottom left) Schwarzschid blurred (bottom right) PS blurred. }
	 		 \label{F5.8}
		\end{figure}

\begin{figure}[H]
		 \centering
		 \includegraphics[scale=0.20]{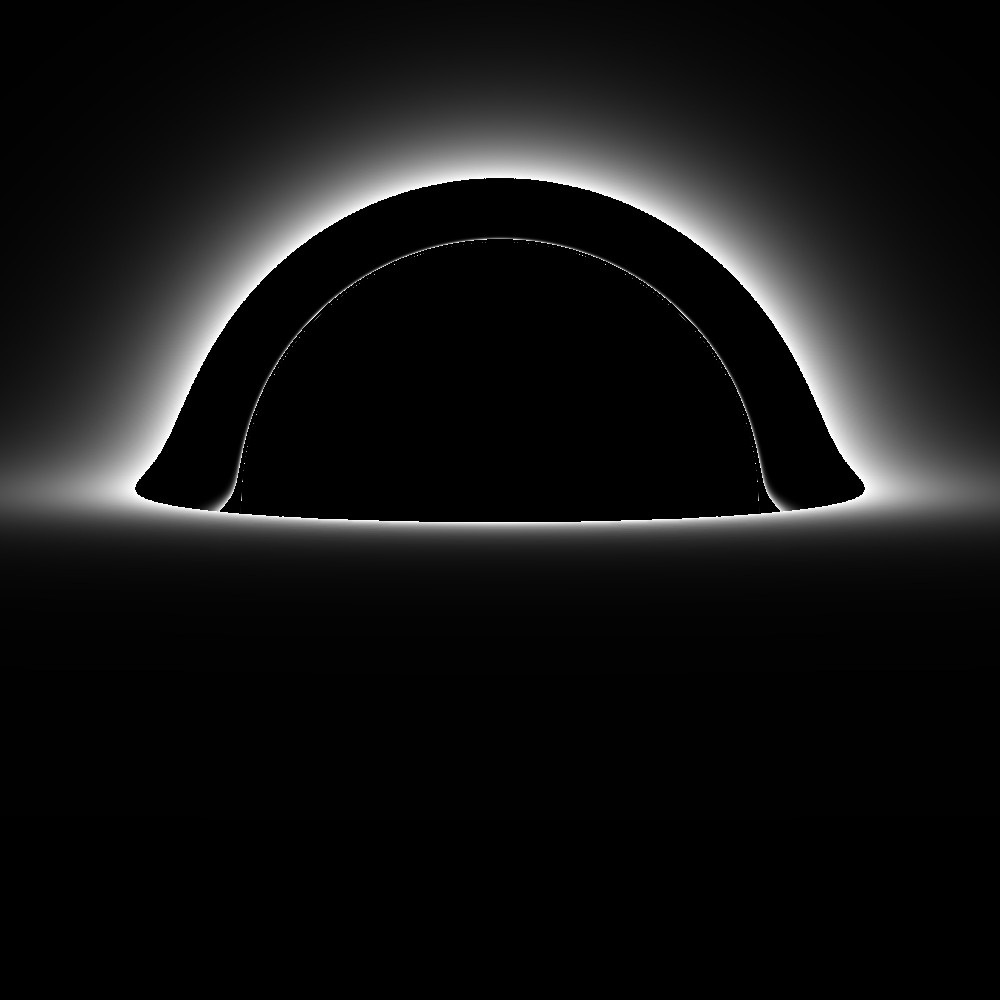}
		 \includegraphics[scale=0.20]{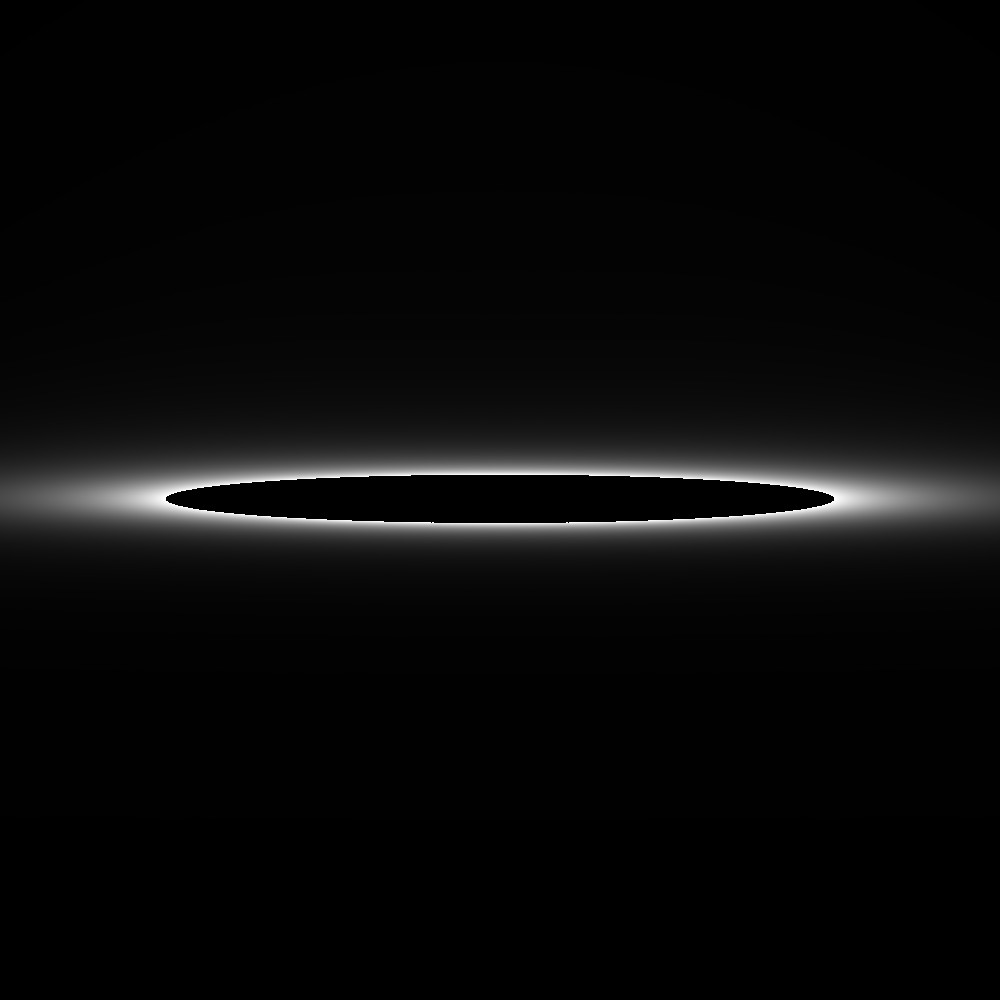}
		 \includegraphics[scale=0.20]{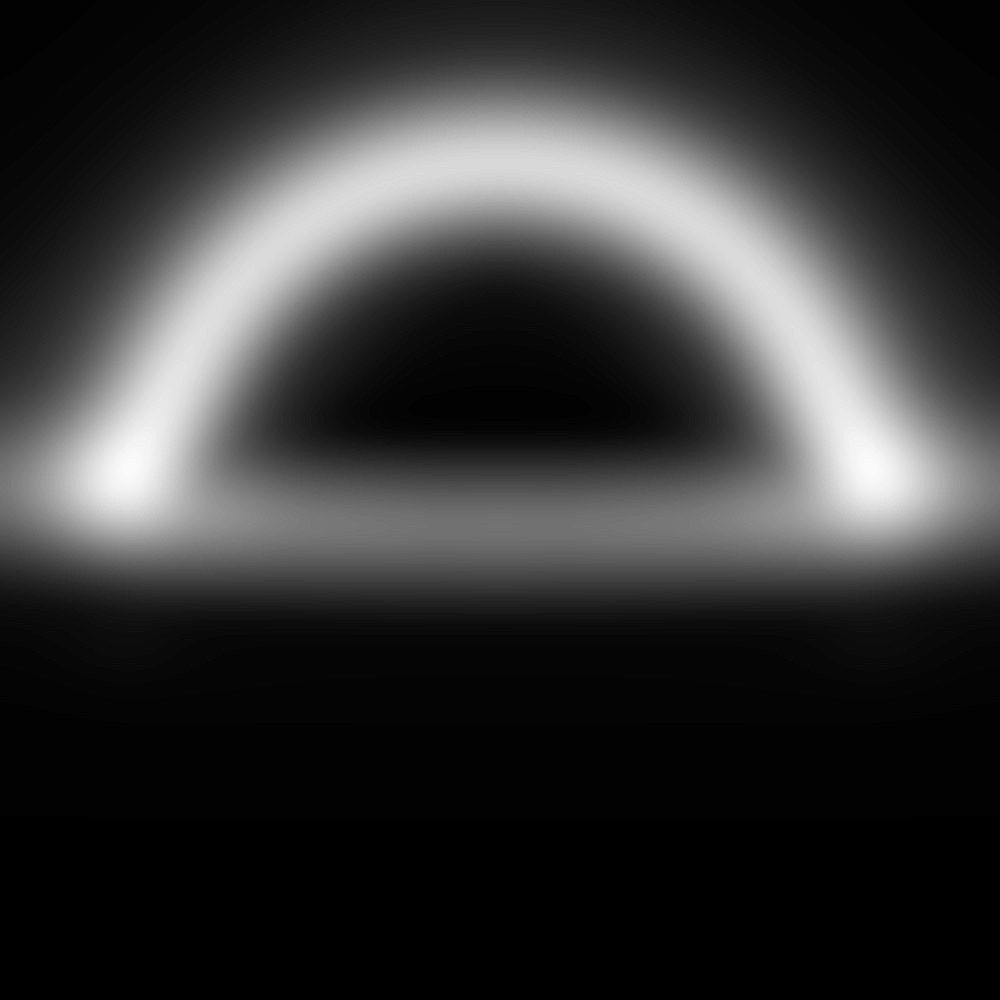}
		 \includegraphics[scale=0.20]{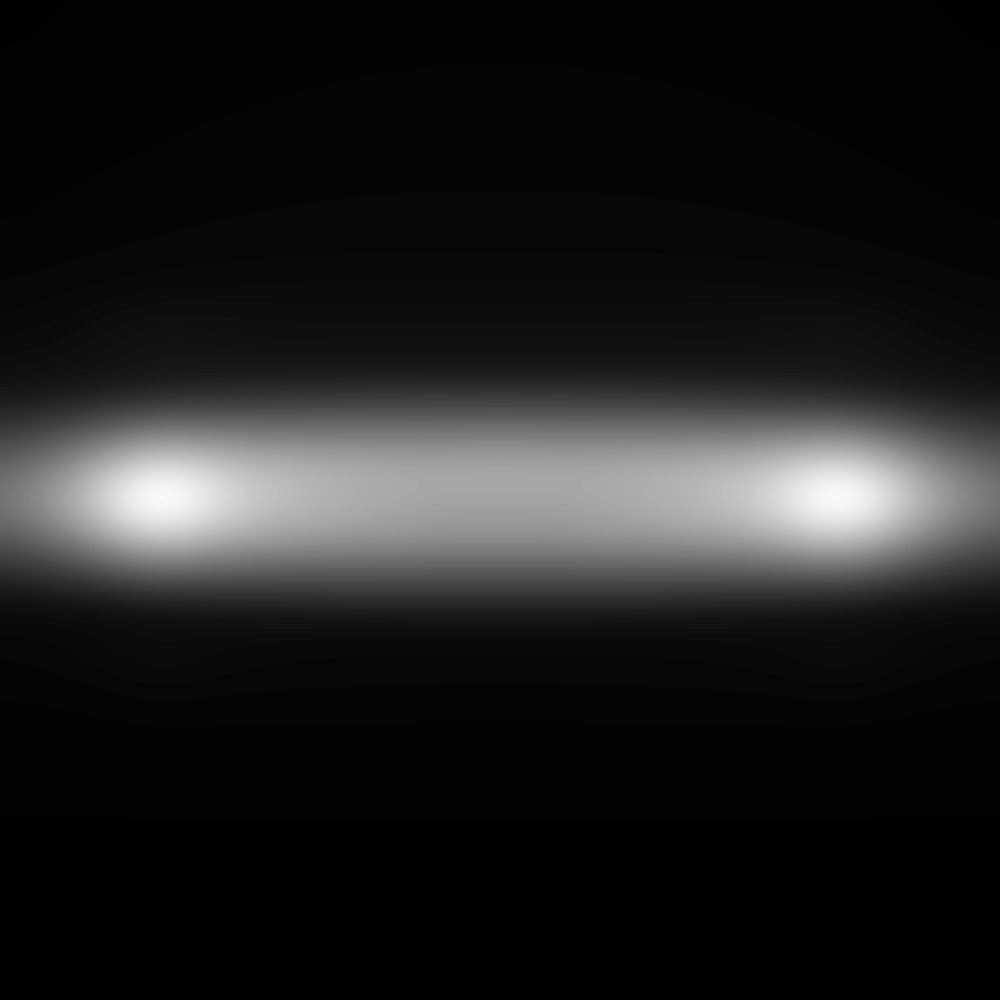}                   
	 	 \caption{Lensing at an observation angle of $\theta_o = 86^o$ (almost equatorial): (top left panel) Schwarzschid; (top right panel) PS; (bottom left panel) Schwarzschid blurred and (bottom right panel) PS blurred.}
	 		 \label{F5.9}
		\end{figure}

The lensed images are shown in Fig.~\ref{F5.8} and Fig.~\ref{F5.9}, and were obtained for an observer placed at an areal radius of $r_o=100M$ with a co-latitude angle $\theta_o=\{ 17 ^o , 86^o\}$, respectively. Local observation angles were locally discretized into a matrix $1000\times 1000$ of pixels, with both angles varying in the range $\pm\tan^{-1}(1/10)\simeq \pm 5.7 ^o\,$. This collection of pixels forms the displayed images.

	The most interesting case for degeneracy occurs for an observer close to the poles (Fig.~\ref{F5.8}). Concretely, the choice $\theta_o=17^o$, corresponds to the angle at which M87*, the target of the Event Horizon Telescope (EHT) 2017 run~\cite{event2019first}, was observed from Earth. For $\theta_o=17^o$ the images of the  Schwarzschild (top left panel) and PS (top right panel) look similar, although some finer additional lensing features are still visible in the Schwarzschild case. Whereas the central dark region in the PS case is due to the lack of source (disk), in the Schwarzschild BH image, there is a thin emission ring corresponding to a secondary image of the accretion disk, and indeed higher-order images at the very edge of the shadow. Such fine lensing details are absent in the PS ``shadow mimicker''. The latter's existence is only a consequence of the assumed absence of light coming from infinity.

	The potential similarity between the PS image and Schwarzschild one is further accentuated by considering that current EHT observations have a limited angular resolution of the order of the compact object itself. We can try to reproduce this effect by applying a Gaussian blurring filter to the images, which washes away smaller image details. The figures obtained after such a blurring procedure are shown in the bottom panels of Fig.~\ref{F5.8}, and indeed have an uncanny resemblance with each other, which illustrates how such a PS configuration might potentially mimic a Schwarzschild BH for electromagnetic channel observations.

	Let us now analyze a near-equatorial observation, choosing $\theta_o=86^o$. The corresponding images are shown in Fig.~\ref{F5.9}. In this case, the images of the Schwarzschild (top left panel) and PS (top right panel) are relatively different. In particular, the former resembles the now-familiar BH shape displayed in the prominent Hollywood movie {\it Interstellar}~\cite{james2015gravitational}, whereas the PS looks like what we might have naively expected: an accretion disk with a hole in it, as seen from the side. This is simple to interpret: such PS is still fairly Newtonian in some aspects; in particular, its gravitational potential is shallow, so the bending of light it produces is weak. Consequently, the accretion disk has an almost flat spacetime appearance, $i.e.$, a plane with a hole, whereas in the BH case, one sees the background of the disk raised due to considerable light bending. Again the bottom panels apply the same blurring as in Fig.~\ref{F5.8} and manifest that, even with limited resolution, under this almost equatorial observation, the two objects could be distinguished.
	
		Finally, this discussion aims only to be a proof of concept. The analysis presented herein has several caveats, such as assuming an idealized thin disk without a necessarily physical luminosity profile and not accounting for relativistic effects such as Doppler and gravitational redshifts. A complete GRMHD analysis and ray-tracing is required in the background of this PS to settle the question fully: to what degree can it imitate a BH observation? Nonetheless, the case built herein clearly confirms the degeneracy, but only under some observation conditions; under others, the different depth of the potential well impacts decisively in producing a different image.
%

%
	\section{Further remarks}\label{S5.5}
%
	The analysis we have presented in this chapter shows that models in which dynamically robust spherical BSs can have a degenerate (effective) shadow with a comparable Schwarzschild BH \textit{do exist}.

	In the case of SBSs, we have established that the most common models of SBSs cannot mimic fundamental phenomenological properties of BHs, such as the lensing of light or the accretion flow of matter if one imposes that these stars should be dynamically stable. Self-interaction cannot easily solve this issue, but as illustrated by the axionic model, appropriate self-interaction terms with sufficiently significant couplings may solve the issue. This is certainly an interesting possibility, which could be explored by approximation techniques for very large self-interactions like the ones suggested in~\cite{colpi1986boson}, rather than a full numerical approach.

	On the other hand, for PSs, we have found that the simplest model, without self-interactions, can mimic a comparable Schwarzschild BH, in the sense of having the new scale $R_\Omega=6M$, thus equal to the ISCO areal radius of the BH. As a word of caution, we remark that, despite the matching between $R_\Omega$ for the PS and the ISCO of the comparable Schwarzschild BH, the lensing in the different spacetime geometries leads to a slightly different shadow size (as a careful inspection of Fig.~\ref{F5.8} (top panels) reveals)\footnote{This is in the same spirit that in a BH spacetime lit from a faraway celestial sphere the shadow size is not determined by the areal radius of the LR but rather by the impact parameter of the LR photons.}. However, since along the Newtonian stable branch of PSs,  $R_\Omega$ varies from a considerable value down to zero, a precise shadow degeneracy will be achieved by a neighbouring solution of the special solution~\eqref{E5.3.28}. More importantly, our lensing analysis reveals the degeneracy only holds in certain degeneracy conditions. Interestingly, these include conditions similar to those for the M87* observations reported by the Event Horizon Telescope. It would be fascinating to perform general relativistic hydro-dynamical simulations on these PS backgrounds, similar to the ones in~\cite{olivares2018tell} for the scalar case, to confirm this degeneracy. 

	There are two key assumptions in the conclusions of the last paragraph: the BSs are near equilibrium and are spherical. Firstly, typical stars and BH candidates are spinning; do these results carry through to the spinning case? The answer is two-fold. Concerning the LRs,  several spinning BSs models have been discussed in the literature, and their LRs have been  computed -- see $e.g.$~ \cite{cunha2015shadows,cunha2016chaotic,grandclement2017light}. In all cases, LRs emerge beyond the first mass extremum. In the spinning case, however, perturbative stability computations showing that the mass extremum coincides with the crossing from stability to instability are absent. Moreover, it has been shown that spinning SBSs are unstable even in the region naively considered to be stable, whereas spinning PSs appear dynamically robust~\cite{sanchis2019nonlinear}. Concerning the TCOs, spinning BSs can have an ISCO, unlike the spherical case -- see $e.g.$~\cite{grandclement2014models,vincent2016imaging, franchini2017constraining,grould2017comparing,delgado2020rotating}. Nevertheless, it remains to see if there is any model in which the accretion flow really mimics that of a comparable BH (with the same mass and angular momentum) and which is,  moreover, dynamically robust.  

	Even if they cannot be ultra-compact, stable BSs are compact objects that can evolve in binaries. Recently, an intriguing degeneracy has been established for  GW190521~\cite{abbott2020gw190521}, showing that a collision of spinning PSs can fit the observed waveform with a slight statistical preference concerning the ``vanilla" binary BH  model~\cite{bustillo2021ultra}. Thus, even if BSs cannot simply imitate BHs in all of their phenomenologies, one cannot exclude that a population of BSs coexists with BHs, as part of the dark matter population,  in particular within a specific mass range, which would be determined by the mass of the ultra-light bosonic particle(s). In this sense, the existence of dynamically robust PSs that can imitate the BH lensing, as shown here, brings to the limelight the issue of degeneracy in lensing/shadow observations.
%
\chapter{Spectral decomposition}\label{C5}
%
	The newest results from gravitational waves detection by LIGO \cite{barish1999ligo} and the image obtained by the event horizon telescope \cite{event2019first} reveal a possible population of extremely compact objects, known as black holes. However, it is still uncertain if the observed BH candidates are depicted by general relativity, some alternative model of gravity, or even distinct compact objects without an event horizon. It is then time to emphasize or rule out possible candidates.

	In an attempt to explain the observed universe, it is tentative to connect the dark matter phenomena with the new observational data. Either through the study of ultra-compact objects (namely Boson Stars) or the study of alternative BHs models. 

	In general relativity, the paradigmatic BH (when in equilibrium) is the Kerr BH~\cite{kerr1963gravitational} (we are neglecting the presence of an electric charge due to its astrophysical irrelevance). The paradigm is based on the uniqueness theorems~\cite{robinson2009four,chrusciel2012stationary} and the no-hair conjectures~\cite{wheeler1971introducing}, which states that, after the gravitational collapse of a generic matter distribution, the exterior spacetime is solely characterized by the Kerr metric (see Sec.~\ref{S1.1}).

	Nevertheless, alternative BH models with additional degrees of freedom (\textit{a.k.a. hair}) are possible. In particular, we will focus on a new family of BHs with scalar hair, dubbed Kerr BHs with scalar hair (KBHsSH) \cite{herdeiro2015construction,herdeiro2014kerr}. These solutions continuously connect hairless Kerr BHs with spinning BSs. 

	There are still several questions about these hypothetical objects. In particular, while spinning BSs were shown to be perturbatively unstable \cite{sanchis2019nonlinear}, there is still an ongoing discussion about the stability of KBHsSH.

	In that regard, we propose the decomposition of KBHsSH into a \textit{spherical harmonic (SH) basis}\footnote{Work in developement with Nicola Franchini.}. The choice of basis arises from separating the radial and angular variables in spherical coordinates. Through this decomposition, we expect to obtain some insight into the structure and intricacies of these solutions.

	While the use of the SHs basis is a well-established technique to tackle differential equations (it transforms a set of coupled partial differential equations (PDEs) into coupled ordinary differential equations, see \cite{volkov2002spinning}), what we propose is an alternative procedure. The KBHsSH solutions are computed with a \textsc{fortran} code (see Appendix~\ref{B}) that solves \textit{exactly} (up to numerical error) the set of field equations, and the metric and scalar functions are, \textit{a posteriori}, decomposed into the SHs. Thus, this results in a kind of spectroscopic analysis to understand their structure, and hence we call this procedure a \textit{Spectral Decomposition (SD)}.

	As a first step in the analysis, we obtain an analytical formula to decompose the solution's functions into the SH basis and impose some restrictions/simplifications arising from the solutions symmetries. 

	Then, to test the method's accuracy, one studies how many SH are required to recover a decomposed function. In this regard, we shall divide the tests into two categories: global quantities (associated with an integral) and geometrical quantities. Concerning the former, we will examine parameters that globally define the scalar field, namely the scalar field mass contribution (Komar mass integral, $M_\Phi$) and the Noether charge $Q_S$. For the latter, we will consider two geometric properties of the spacetime: the radial position of the light rings ($r_{LR}$) and innermost stable circular orbits ($r_{ISCO}$). At last, we test our solution's decomposition with the virial identity. All these tests show that, in the worst case, the first four contributing SHs are enough to correctly describe the solutions with an accuracy better than $~10^{-4}$. Following the confirmation of the ability of the SD to correctly decompose a given solution, we finish by analyzing a set of solutions that continuously connect a hairless Kerr BH to a pure spinning SBS. We finish this chapter with some overall conclusions.
%
	\section{KBHsH model}\label{S6.1}
%
	The action that describes the Einstein-Klein-Gordon theory, with a massive complex scalar field $\Phi$, minimally coupled to gravity~\cite{herdeiro2015construction,herdeiro2014kerr} is 
		\begin{equation}\label{E6.1.1}
		 \mathcal{S}_{KBHsSH}=\frac{1}{4}\int d^4 x \sqrt{-g}\ze \Big[R -2\ze g^{\mu \nu} \big(\bar{\Phi}_{,\mu} \Phi_{,\nu}+\bar{\Phi}_{,\nu} \Phi_{,\mu}\big)- 2\ze\mu _S ^2 \bar{\Phi}\, \Phi\Big]\ ,
		\end{equation}
	where $\mu _S$ is the scalar field's mass. Observe that the action and relations for a KBHsSH are the same as the SBS configuration in Sec.~\ref{S4.2} when $U_i(\Phi)=\mu ^2 _S \Phi ^2$ and $r_H\neq 0$, still, since we are adding rotation and an event horizon, let us briefly present all the essential relations. Variation of the action with respect to the metric and scalar field yields the EKG field equations 
		\begin{align}\label{E6.1.2}
		 & R_{\mu \nu}-\frac{1}{2}g_{\mu \nu} R=2\,T_{\mu \nu} \ ,\\
		 & \Box \Phi  =  \mu _S ^2 \Phi\ ,\label{E6.1.3}
		\end{align}
	with the stress-energy tensor 
		\begin{equation}\label{E6.1.4}
		 T_{\mu \nu} = \bar{\Phi} _{,\mu } \Phi _{,\nu}+\Phi _{,\mu } \bar{\Phi} _{,\mu}-g_{\mu \nu}\left[\frac{1}{2} g^{\rho \sigma}\big(  \bar{\Phi} _{,\rho }  \Phi _{,\sigma}+ \Phi _{,\rho} \bar{\Phi} _{,\sigma}\big)+\mu _S ^2 |\Phi|^2\right] \ .
		\end{equation}
	The action \eqref{E6.1.1} is invariant under the global $\textbf{U}(1)$ transformation $\Phi \rightarrow e^{i\ze a} \Phi $, where $a $ is a constant. This induces a conserved scalar $4$-current: $j^\mu =-i (\bar{\Phi}\ze \Phi ^{,\mu}-\Phi\ze \bar{\Phi}  ^{,\mu})$ and, consequently, a conserved Noether charge:
		\begin{equation}\label{E6.1.5}
	 	 Q_S=\int _\Sigma d^3 x\, \sqrt{-g}\, j^t\ .
		\end{equation}
	For the metric ansatz, one chooses the standard metric ansatz compatible with axial symmetry~\eqref{E1.5.41}; while the complex scalar field ansatz contains both a harmonic time and azimuthal dependence \eqref{E1.5.44}
		\begin{equation}\label{E6.1.6}
		 \Phi (t, r, \theta, \varphi) =\phi (r,\theta ) e^{i(m\ze \varphi -\omega\ze t)}\ ,
		\end{equation}
	The full configuration is then described by five funcions of $(r,\ze\theta )$ that have to be decomposed: $\mathcal{F}_i$ ($i=0,\ze 1,\ze 2,\ze W$) and $\phi$.
	
	To numerically solve the field equations\footnote{The full set of equations of motion and boundary conditions can be seen in \cite{herdeiro2015construction}.}, we rely on a professional \textsc{fortran} solver, the \textsc{cadsol} program package~\cite{schauder1992cadsol,schonauer2001we} (see also Appendix~\ref{B}). The latter is designed to numerically solve a non-linear system of elliptic and parabolic PDEs through a finite difference method with mesh refinement and automatic control of the consistency order. With this package, one can numerically solve the field equations with a typical precision of $10^{-5}$.
%
	\section{Spectral decomposition}\label{S6.2}
%
	As already mentioned in Sec.~\ref{C5}, the decomposition of KBHsSH solutions in SH allows the separation of the radial and angular dependences. In addition, the scalar field ansatz \eqref{E6.1.6} reduces the usual SHs into \textit{associated Legendre function, $\mathcal{P} _\ell ^m (x)$}. These form a complete set, and thus, it is natural to expect that, by increasing the number of terms $k$ (\textit{a.k.a.} modes) in our expansion, one can obtain a series as close as desired to the exact solution. In this section, we shall present the generic decomposition procedure.

	Consider a generic function $X\equiv X(r,\ze\theta)$ that shall be decomposed into the SH basis. For a given $m\in \mathbb{Z}_0 $, such that $\ell\geqslant |m|$, the generic expansion in SHs is
		\begin{equation}\label{E6.2.7}
	 	 X (r,\theta ) =\sum _{\ell=m} ^{k} \textbf{x}^\ell (r) \, \sqrt{c_{\ell} ^m} \,  \mathcal{P}_\ell ^m (\cos \theta )\ ,
		\end{equation}		
	with $k$ the highest order term in the expansion -- full description should occur for $k\to +\infty$. The associated Legendre functions obey the orthogonality condition
		\begin{equation}\label{E6.2.8}
		 \int_{-1} ^{+1} dy \ze\ze\ze  \mathcal{P}_\ell ^m (y)\, \mathcal{P}_{\ell '} ^{m'} (y) = \frac{\delta _{\ell} ^{\ell'}}{c_{\ell} ^m}\delta _{m} ^{m'}\ , \qquad c_{\ell} ^m\equiv \frac{(2\ell+1)(\ell-m)!}{2(\ell+m)!} \ .
		\end{equation}

	To find the \textit{spectral functions} $\textbf{x}^\ell$ at a given position $r$ of each \textit{spectral mode $\ell$}, one as to perform the projection of $X$ into the basis: $\textbf{x}^\ell=  \mathcal{P}_\ell ^m\cdot X$. Here, the dot ($\cdot$) denotes a scalar product. Explicitly, this scalar product is given by:
		\begin{align}\label{E6.2.9}
		 \mathcal{P}_\ell ^m\cdot X & = \int _0 ^{\pi }d\theta\,  \sin \theta \  X \, \sqrt{c_{\ell} ^m} \,\mathcal{P}_\ell ^m (\cos \theta ) \nonumber \\
		 & =\sum _{\ell =m} ^{k} \textbf{x}^{\ell'} \int_0 ^{\pi} d\theta\, \sin \theta \,  \sqrt{c_{\ell'} ^m}\,  \mathcal{P}_{\ell '} ^m  (\cos \theta )\ze\ze\ze \sqrt{c_{\ell} ^m} \, \mathcal{P}_\ell ^m (\cos \theta )=\textbf{x}^\ell .
		\end{align}
	In the last step, we have used~\eqref{E6.2.8}. This gives us the formula for the spectral function $\textbf{x}^\ell$ such that~\eqref{E6.2.9} corresponds to the decomposition into SH of the generic function $X$ \footnote{A kind of spectral decomposition can also be performed for vectorial functions in vector spherical harmonics (VSH), such procedure can be seen in Appendix~\ref{F}.}.
%
		\subsection{Scalar field}\label{S6.2.1}
%
	Observe that \eqref{E6.1.3} is symmetric under reflections in the equatorial plane ($\theta =\pi /2$) \textit{i.e.,} if $\phi (r,\ze\theta)$ is a solution, so is $\phi (r,\ze\pi-\theta )$. In addition, due to the sole presence of odd powers of $\phi $ in the latter, $-\phi (r,\ze \theta ) $ is also a solution. Therefore, one has the freedom to choose the parity to be either $\phi (r,\ze\pi-\theta ) = \phi (r,\ze\theta )\ {\rm [even]\ or}\ \phi (r,\ze\pi -\theta )=-\phi (r,\ze\theta ) \ {\rm [odd]}$.
	
	In the case of $ \mathcal{P}^m _{\ell} (\cos \theta)$, they possess an even (odd) parity with respect to $\theta \rightarrow \pi -\theta $ for even (odd) values of $\ell +m$. As a result, it will be divided into modes that change sign under the reflection and modes that maintain it when the SD is performed.
	
	In this work we consider $\phi (r,\ze \theta ) =\phi (r,\ze\pi -\theta ) $ and $m=1$, resulting in the cancellation of all odd terms in the expansion. In other word, we focus on even parity solutions.
			\begin{equation}\label{E6.2.10}
		 	 \phi (r,\ze\theta ) =\sum _{n =0} ^{k} \textbf{f}^{\ze\ze 2n+1} (r) \, \sqrt{c_{2n +1} ^1} \,  \mathcal{P}_{2n +1} ^1 (\cos \theta )\ ,
			\end{equation}	 
	where $\textbf{f} ^{\, \ell} $ corresponds to the radial amplitude of the $\ell = 2n+1,\ n\in \mathbb{N} _0,$ spectral mode $\ell$, obtained through the projection
			\begin{equation}\label{E6.2.11}
			 \textbf{f}^{\, \ell} =\mathcal{P}_{\ell} ^1\cdot \phi = 2 \sqrt{c_\ell ^1 } \int _0 ^{\frac{\pi}{2}}d\theta\, \sin \theta \  \phi \ \mathcal{P}_\ell ^{1} (\cos \theta )\ . 
			\end{equation}
%

%
		\subsection{Metric functions}\label{S6.2.2}
%
	At last, let us decompose the metric functions $\mathcal{F}_i$ into the SH basis. Observe that the latter possess all the same symmetry and thus can be decomposed similarly. 

	The metric functions, $\mathcal{F}_i$, are axially symmetric and can be decomposed in terms of associated Legendre functions with $m=0$. Note that $\mathcal{F}_i$ are also symmetric under reflections along the equatorial plane, $\mathcal{F}_i (r,\ze\theta )=\mathcal{F}_i (r,\ze\pi-\theta )$. Again, due to the well-defined parity of $\mathcal{P}_{\ell} ^0 \equiv \mathcal{P}_\ell$, the odd terms (in $\ell$, with $m=0$) do not appear, and we only have to take into account the terms with even parity.

	The generic spectral expansion of $\mathcal{F}_i$ comes as
			\begin{equation}\label{E6.2.12}
	 		 \mathcal{F}_i (r,\ze\theta ) = \sum _{n =0} ^{k} \textbf{K}_{i} ^{2n} (r) \sqrt{c_{2n}}\, \mathcal{P}_{2n} (\cos \theta ) \ , 
		 	\end{equation}	 
	where $\textbf{K} _i ^{\ell} $ corresponds to the radial amplitude of the $\ell  = 2n,\ n\in \mathbb{N} _0,$ spectral mode obtained throught the projection
			\begin{equation}\label{E6.2.13}
			 \textbf{K}_i ^\ell = 2\, \sqrt{c_{\ell}} \int _0 ^{\frac{\pi}{2}}d\theta \, \sin \theta \  \mathcal{F}_i \,\mathcal{P}_\ell (\cos \theta )\ . 
			\end{equation}	
%

%
	\section{Test set}\label{S6.3}
%
	To numerically test the ability of the SD to correctly describe the solutions, let us consider five different tests. To solely test the scalar field decomposition, we compute the scalar field associated mass $M_\Phi $ and Noether charge $Q_S$ (while keeping the metric functions untouched). For the metric functions\footnote{Even though the metric functions are not explicitly $\Phi$-dependent, they have an implicit dependence.}, we will consider two geometric quantities: the LR and ISCO radii. The entire decomposed solution will be tested using the virial identity.
	
	The scalar field mass, $M_\Phi$ \eqref{E6.2.14}, and Noether charge, $Q_S$ \eqref{E6.2.15}, are given by the integrals~\cite{herdeiro2015construction}
		\begin{align}\label{E6.2.14}
		 & M_\Phi = 2\pi  \int _{r_H} ^{+\infty} dr \int _0 ^\pi d\theta\ r^2 \sin \theta\, e^{F_0 +2F_1 +F_2}\left[\mu _S ^2 -2\ze e^{-2F_2}\frac{\omega \big(\omega -m F_W\big)}{N}\right]\Phi ^2 \ ,\\
		& Q _S = 2\pi \int _{r_H} ^{+\infty} dr \int _0 ^\pi d\theta\ r^2 \sin \theta\, e^{-F_0 +2F_1 +F_2}\frac{m(\omega -m F_W)}{N}\ze\Phi ^2 \ .\label{E6.2.15}
		\end{align}
	The virial identity of the KBHsSH model comes as (see Ch.~\ref{C7} and \cite{herdeiro2021virial})
		\begin{align}\label{E6.2.16}
		 \int _0 ^\pi d\theta \int _{r_H} ^{+\infty} dr \ \Big[ I_R + I_\Phi+ I_U ^{[\Phi]}\Big]= I_{GHY} \  ,
		\end{align}
	where $I_R$, $I_\Phi$ and $I_U ^{[\Phi]}$ represent the Ricci, kinetic scalar field and self-interaction potential contributions to the virial identity. The $I_{GHY}$ term is the Gibbons-Hawking-York boundary term that contains the boundary contribution to the gravitational action. The detailed computation will be presented in Ch.~\ref{C7}.
%
		\subsection{Geodesic motion}\label{S6.3.1}
%
	Let us now generalize the work done in Sec.~\ref{S4.2}. Concerning the radii for which the ISCO and LR occurs, let us follow the work done in \cite{herdeiro2021imitation,delgado2021equatorial}.
	
	Consider a generic, asymptotically flat metric $\big( g_{\mu \nu} (r,\theta)\big)$ that is compatible with stationarity and axi-symmetry
		\begin{equation}\label{E6.3.17}
		 ds^2 = -g_{tt}\, dt^2 + g_{rr}\, dr^2 +g_{\theta \theta}\, d\theta ^2 + g_{\varphi \varphi}\, d\varphi ^2 + 2g_{t\varphi}\,dt d\varphi\ .
		\end{equation}
	The Lagrangian density that describes a particle moving in such geometry comes as
		\begin{equation}\label{E6.3.18}
		 \mathcal{L}=\frac{1}{2}\, g_{\mu \nu}\, \dot{x}^\mu \dot{x}^\nu ={\rm k}\ ,
		\end{equation}
	with ${\rm k}$ a parameter that distinguished between a time-like trajectory (${\rm k}=-1$) and a light-like trajectory (${\rm k}=0$). Assuming that the motion is planar and on the equatorial plane, $\theta =\pi /2$,
		\begin{equation}\label{E6.3.19}
		2\,\mathcal{L}=g_{tt}\, \dot{t} ^2 +2\ze g_{t\varphi}\, \dot{t}\dot{\varphi}+g_{rr}\,\dot{r}^2 + g_{\varphi \varphi}\, \dot{\varphi} ^2 = {\rm k}\ ,
		\end{equation}
	where the metric functions are now solely $r$-dependent. Due to the symmetry of the spacetime, one can define the energy, $E$, and angular momentum, $L$, of the particle as
		\begin{equation}\label{E6.3.20}
		 E =-g_{t\mu }\,\dot{x^\mu}=-g_{tt}\,\dot{t} -g_{t \varphi}\, \dot{\varphi}\ , \qquad \qquad L = g_{\varphi \mu}\, \dot{x}^\mu =g_{t\varphi }\, \dot{t}+g_{\varphi \varphi}\, \dot{\varphi} \ .
		\end{equation}
	This allows us to rewrite the Lagrangian as
		\begin{equation}\label{E6.3.21}
		 2\mathcal{L} = -\frac{g_{\varphi \varphi}\ze E^2 + 2\, g_{t \varphi}\ze E L +g_{tt} \ze L^2 }{g_{t\varphi }^2-g_{tt}\, g_{\varphi \varphi}} + g_{rr}\dot{r} ^2 ={\rm k}\ .
		\end{equation}
	Let us now introduce the particles effective potential $V_{\rm k} (r) \equiv g_{rr}\ze \dot{r} ^2$
		\begin{equation}\label{E6.3.22}
		 V_{\rm k} = {\rm k} + \frac{g_{\varphi \varphi}\ze E^2 + 2\, g_{t \varphi}\ze E L +g_{tt}\ze L^2 }{g_{t\varphi }^2-g_{tt}\,g_{\varphi \varphi}}\ ,
		\end{equation}
	where a particle that moves in a circular orbit has $V_{\rm k} = V_{\rm k} ' =0$.
			\subsubsection*{Time-like circular orbits}
	In the case of time-like particles, ${\rm k}=-1$, the condition $V_{-1}=V_{-1} ' = 0$, gives rise to two sets of algebraic equations for the energy and angular momentum of the particle, the latter's correspond to a co-rotating $\{ E_+ ,\ze L_+\} $, and a counter-rotating $\{ E_- ,\ze L_-\} $ orbit,
				\begin{equation}\label{E6.3.23}
				 E_{\pm} = -\frac{g_{tt}+g_{t\varphi}\ze \Omega _\pm}{\sqrt{-g_{tt}-2\, g_{t\varphi}-g_{\varphi\varphi}\ze \Omega _\pm ^2}}\ , \qquad \qquad L_\pm =\frac{g_{t\varphi}+g_{\varphi\varphi}\ze \Omega _\pm }{\sqrt{-g_{tt}-2\, g_{t\varphi}\ze \Omega _\pm -g_{\varphi\varphi}\ze\Omega _\pm ^2}}\ .
				\end{equation}
	where $\Omega _\pm$ is the angular frequency of the particle
				\begin{equation}\label{E6.3.24}
				 \Omega _\pm = \frac{1}{g_{\varphi\varphi} '} \Big[-g_{t,\varphi} ' \pm \sqrt{g_{t\varphi} '^{\, 2} -g_{tt}'\, g_{\varphi\varphi}'}\ \Big] \ .
				\end{equation}
	The stability of a particle's orbit can be analysed by the sign of $V_{-1} ''$. If it is positive (negative), the motion is unstable (stable). The ISCO corresponds to the last stable circular orbit, and hence it marks the threshold $V_{-1} '' =0$. The latter yields the condition for the radius ($r_{ISCO} ^\pm$) of the ISCOs
				\begin{equation}\label{E6.3.25}
				 g_{\varphi\varphi} ''\, E_\pm ^2 +2\, g_{t\varphi} ''\, E_\pm\ze L_\pm + g_{tt} ''\, L_\pm ^2 -(g_{t\varphi } ^2-g_{tt}\ze g_{\varphi\varphi})'' = 0\ .
				\end{equation}
			\subsubsection*{Light rings}
	In the case of light-like particles, ${\rm k}=0$, the first condition for the existence of a circular orbit, $V_0 =0$, yields one set of algebraic equations for the inverse impact parameter, $\eta = E/L $, for co-rotating, $\eta _+$, and counter-rotating $\eta _-$, orbits,
				\begin{equation}\label{E6.3.26}
				 \eta _\pm =\frac{1}{g_{\varphi\varphi}}\Big[-g_{t\varphi}\pm \sqrt{g_{t\varphi} ^2 -g_{tt}\ze g_{\varphi\varphi}}\ \Big]\ .
				\end{equation}
	The second condition $V_0 ' =0$, gives the condition for the presence of a LR
				\begin{equation}\label{E6.3.27}
				 g_{\varphi\varphi}\ze \eta _\pm ^2 + 2\ze g_{t\varphi }\ze \eta _\pm + g_{tt}  = 0\ .
				\end{equation}
	The radius at which the previous conditions holds corresponds to the LR radius $r_{LR} ^\pm$.

	With these two conditions set, we have three integral quantities to numerically test the ability of the SD to correctly describe the scalar and metric fields, two differential relations, and two linear relations.
%
	\section{Numerical tests}\label{S6.4}
%
%
		\subsection{Analytical Kerr BH decomposition}\label{S6.4.1}
%
	Let us start by testing the SD in a well known analytical solution. The analytical Kerr BH metric in Boyer-Lindquist coordinates introduces two radially dependent functions $\Sigma(r)$ and $\Delta(r)$ and comes as~\cite{townsend1997black}\footnote{Observe that the transformation between the numerical metric \eqref{E1.5.41} in Schwarzschild-like coordinates and the Kerr BH metric in Boyer-Lindquist coordinates \eqref{E6.4.28} is not direct. However, the Kerr solutions is also included see Sec.~\ref{S7.6}.}
			\begin{equation}\label{E6.4.28}
			 ds^2 = - \frac{\Delta}{\Sigma} \big( dt -a \sin^2 \theta d\varphi \big)^2 +  \Sigma \bigg( \frac{dr^2}{\Delta} + d\theta^2 \bigg) + \frac{\sin^2 \theta}{\Sigma} \Big[ a dt - \left( \Sigma+a^2\sin^2\theta \right) d\varphi \Big]^2 \ ,
			\end{equation}
	with 
			\begin{equation}\label{E6.4.29}
			 \Delta \equiv r^2 - 2Mr + a^2  \ , \qquad  \Sigma \equiv r^2 + a^2 \cos^2 \theta \ .
			\end{equation}
	The BH event horizon is located at $r_H =\big(M+\sqrt{M^2-a^2}\, \big)$.
	
	Careful examination of \eqref{E6.3.25} and \eqref{E6.3.27} shows that they only depend on the $t$ and $\varphi$ components of the metric, namely: $g_{tt},\, g_{\varphi \varphi}$ and $g_{t\varphi}$. Hence these are the only functions that need to be decomposed.
	
	For this solution, both the LRs and ISCOs radial coordinate have an analytical expression,
			\begin{equation}\label{E6.4.30}
	 		 r_{LR} ^\pm = r_S \bigg[1+\cos \left(\frac{2}{3} \arccos \left(\pm\frac{a}{M}\right)\right)\bigg]\ ,\qquad r_{ISCO} ^\pm =M \left[ 3+Z_2\pm\sqrt{(3-Z_1) (Z_1+2 Z_2+3)}\ \right]\ ,
			\end{equation}
	where $\chi = a/M$ and
			\begin{equation}\label{E6.4.31}
			 Z_1=\left(\sqrt[3]{1-\chi }+\sqrt[3]{\chi +1}\right) \sqrt[3]{1-\chi ^2}+1\ ,\qquad\qquad Z_2 = \sqrt{3\chi ^2 +Z_1 ^2}\ .
			\end{equation}
	To quantify the accuracy of our decomposition, let us recall the definition of the generic relative error, $err$:
			\begin{equation}\label{E6.4.32}
			 err = 1- \frac{X_{SD}}{X_{full}}\ ,
			\end{equation}
	where we compare the result obtained through the spectral decomposition $X_{SD}$ with the full solution $X_{full}$. Observe that $X$ can be either a sum of spectral modes (\textit{e.g.} $M_\Phi$) or a result obtained throught an evaluation of the sum of spectral modes (\textit{e.g.} $r_{LR}$).
	
\begin{figure}[H]
			 \centering
			 	\begin{picture}(0,0)
			 	 \put(45,120){\small $k=0$}
			 	 \put(110,155){$r_{LR}$}			 	
			 	 \put(154,30){\small$ k=2$}
			 	 \put(120,85){\small $k=1$}
			 	 \put(120,-6){\small $j$}
			 	 \put(2,80){\begin{turn}{90}{\small $err$}\end{turn}}
				\end{picture}
			 \includegraphics[scale=0.6]{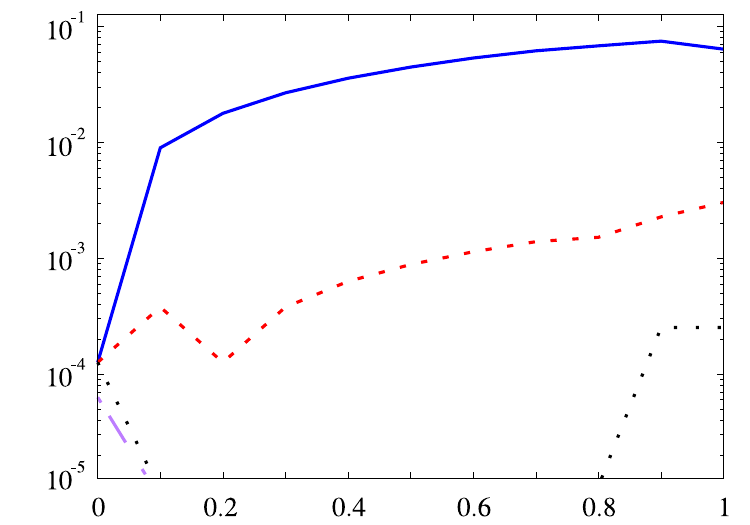}\hfill
			 \begin{picture}(0,0)
			 	 \put(58,135){\small $k=0$}
			 	 \put(110,155){$r_{ISCO}$}			 	
			 	 \put(154,40){\small$ k=3$}
			 	 \put(140,85){\small $k=2$}
			 	 \put(120,115){\small $k=1$}			 	 
			 	 \put(120,-6){\small $j$}
			 	 \put(2,80){\begin{turn}{90}{\small $err$}\end{turn}}
				\end{picture}
			 \includegraphics[scale=0.6]{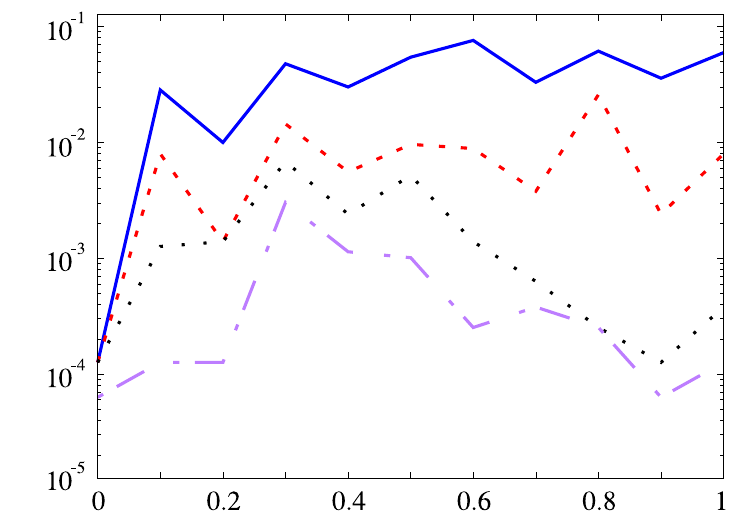}
			 \caption{Radial LR's (left panel) and ISCO's (right panel) $err$ convergence with the increase in the number of spectral modes: (solid blue) $k=0$; (dashed red) $k=1$; (dotted black) $k=2$, (dot-dashed purple) $k=3$.}
			 \label{F6.1}
    		\end{figure}
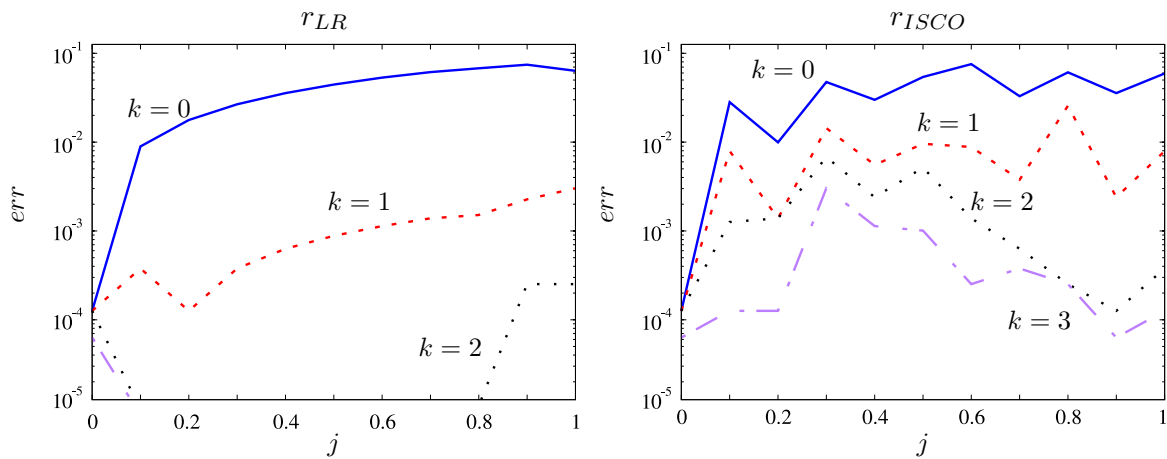

	Evaluation of $r_{LR}$ and $r_{ISCO}$ with both the analytical expression \eqref{E6.4.30} and the SD computation through \eqref{E6.3.27} and \eqref{E6.3.25} show a noticeable tendency towards decreasing the $err$ between them by increasing the number of spectral modes $k$ in the expansion -- see Fig.~\ref{F6.1}.
	\vspace{2mm}

	In Fig.~\ref{F6.1} one can observe that, while in both cases, the first four spectral modes are enough to correctly obtain the radial position with an  $err\sim 10^{-4}$, the ISCO's $err$ is much larger than the LR's.
	
	The cause for this difference lies in the radial conditions. While in the LR's case, the condition \eqref{E6.3.27} is only metric dependent, the ISCO's condition \eqref{E6.3.25} depends on its derivatives, which, when differentiating the recovered functions, introduces numerical noise that ends destroying the accuracy\footnote{This seems to be a numerical artefact since we are dealing with numerical data that is interpolated and derived several times.}. Still, the fact that the reconstructed functions can recover the geometric conditions is an excellent indicator of the power of such a technique.
%
		\subsection{Hairy black holes}\label{S6.4.2}
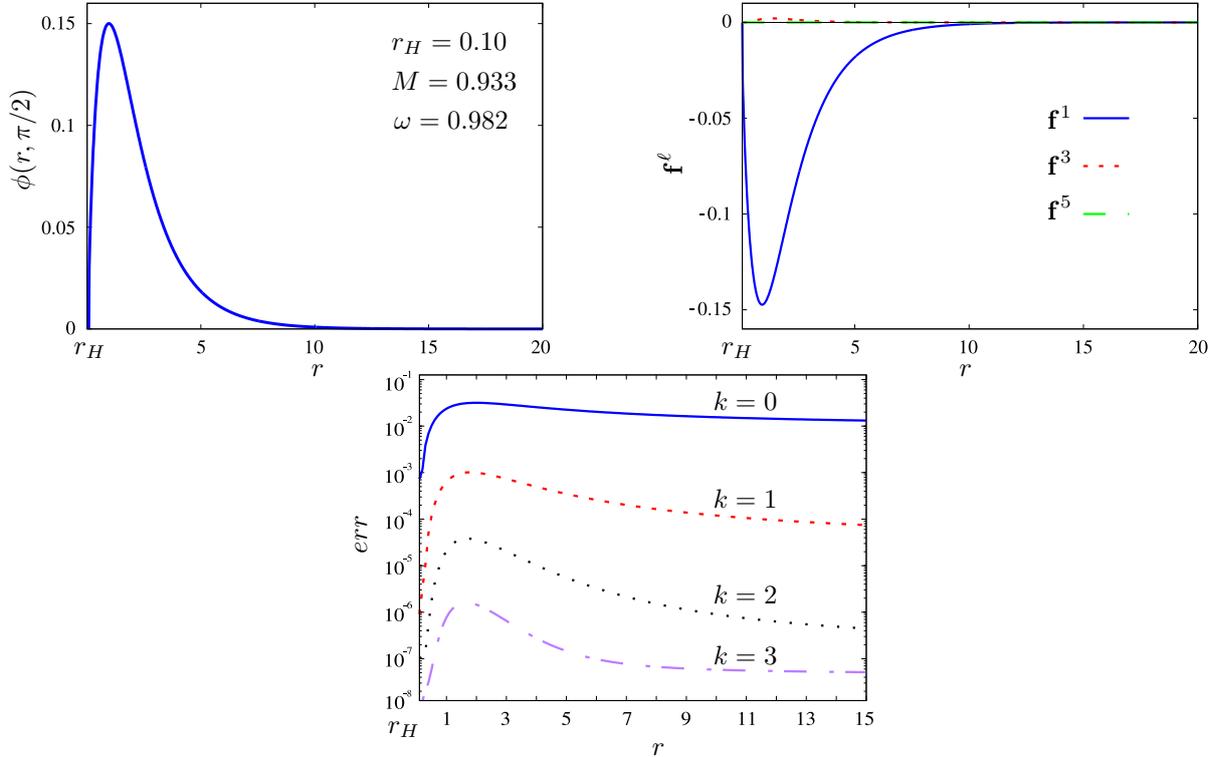
\begin{figure}[H]
			 \centering
			 	 \begin{picture}(0,0)
			 	 \put(135,115){\small $r_H = 0.10$}		
			 	 \put(135,100){\small $M = 0.933$}			 	 
			 	 \put(136,85){\small $\omega = 0.982$}	
			 	 \put(105,-8){\small $r$}
			 	 \put(16,0){\small $r_H$}
			 	 \put(-8,62){\begin{turn}{90}{\small $\phi(r,\pi /2)$}\end{turn}}
				\end{picture}
			 \includegraphics[scale=0.55]{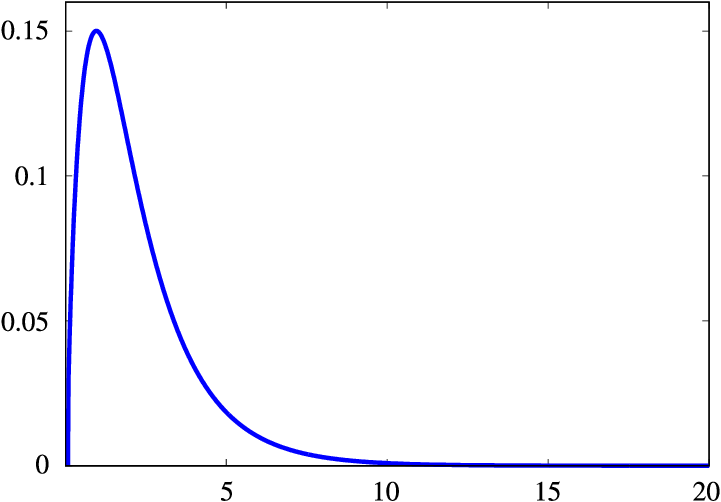}\hfill
			 	\begin{picture}(0,0)		 	 
			 	 \put(138,85){\small $\textbf{f}^{\,1} $}	
			 	 \put(138,68){\small $\textbf{f}^{\, 3} $}				 	 
			 	 \put(138,51){\small $\textbf{f}^{\, 5} $}
			 	 \put(105,-8){\small $r$}
			 	 \put(16,0){\small $r_H$}
			 	 \put(-8,68){\begin{turn}{90}{\small $\textbf{f}^{\, \ell}$}\end{turn}}
				\end{picture}
			 \includegraphics[scale=0.55]{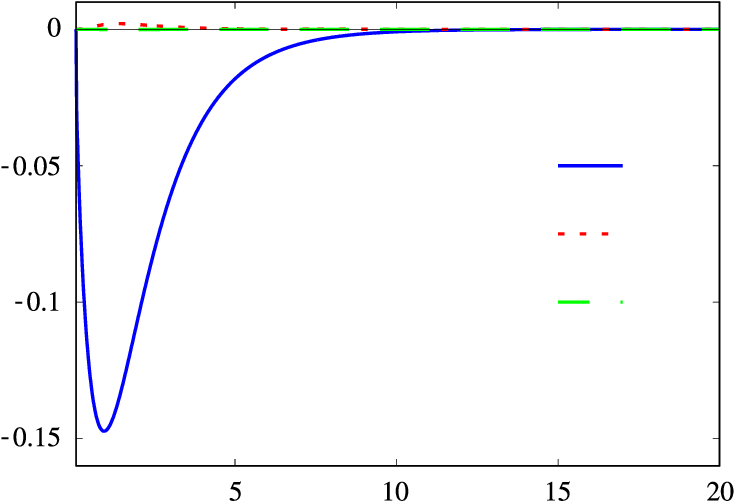}\vspace{1mm}\\
			 			 	\begin{picture}(0,0)		 	 
			 	 \put(138,122){\small $k=0$}	
			 	 \put(138,85){\small $k=1$}			 	 
			 	 \put(138,49){\small $k=2 $}
			 	 \put(138,26){\small $k=3 $}
			 	 \put(115,-8){\small $r$}
			 	 \put(16,0){\small $r_H$}
			 	 \put(4,68){\begin{turn}{90}{\small $err$}\end{turn}}
				\end{picture}
			 \includegraphics[scale=0.55]{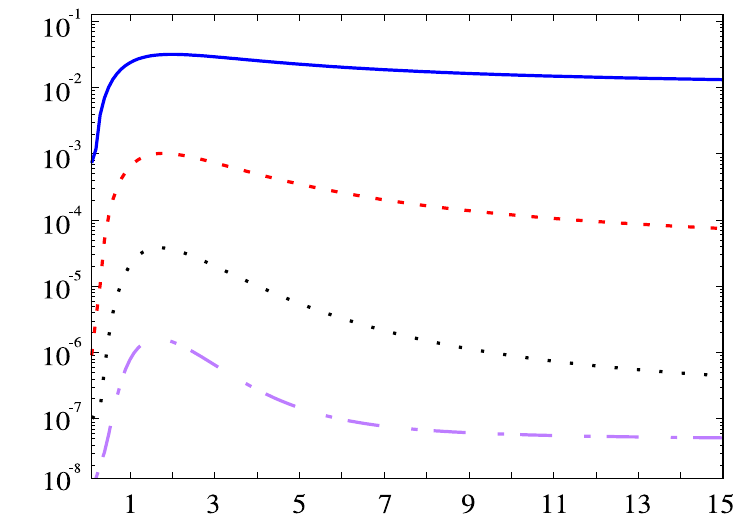}
			 \caption{(Left panel) radial profile of the numerically obtained $\phi (r,\frac{\pi}{2})$; (right panel) the first four spectral functions $\textbf{f}^{\ze 2\ell +1}$. (Bottom panel) $err$ of $\phi $ as a function of $(r,\ze \frac{\pi}{2})$ for (solid blue) $k=0$; (dashed red) $k=1$; (dotted black) $k=2$; (dot-dashed purple) $k=3$ modes.}
			 \label{F6.2}
    		\end{figure}

\begin{figure}[H]
			 \centering
			 	\begin{picture}(0,0)
			 	 \put(135,50){\small $r_H = 0.10$}		
			 	 \put(135,35){\small $M = 0.933$}			 	 
			 	 \put(136,20){\small $\omega = 0.982$}	
			 	 \put(105,-8){\small $r$}
			 	 \put(16,0){\small $r_H$}
			 	 \put(-8,55){\begin{turn}{90}{\small $F_0(r,\pi /2)$}\end{turn}}
				\end{picture}
			 \includegraphics[scale=0.55]{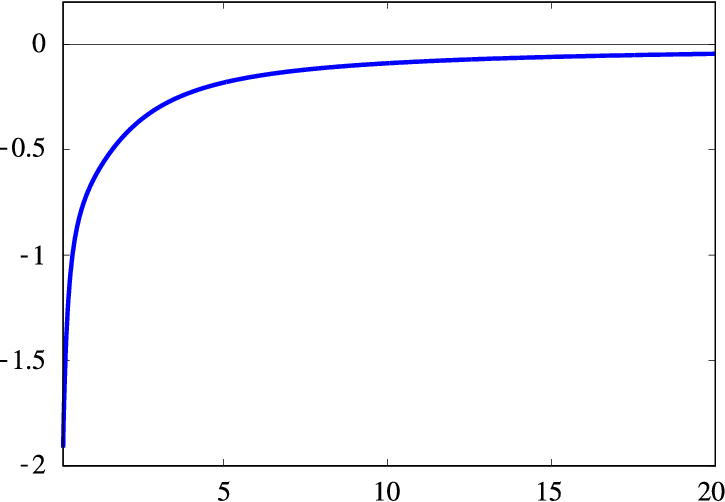}\hfill
			 			 	\begin{picture}(0,0)		 	 
			 	 \put(132,85){\small $\textbf{K}^{\,0} _0 $}	
			 	 \put(132,68){\small $\textbf{K}^{\, 2} _0 $}				 	 
			 	 \put(132,51){\small $\textbf{K}^{\, 4} _0 $}
			 	 \put(105,-8){\small $r$}
			 	 \put(16,0){\small $r_H$}
			 	 \put(-8,68){\begin{turn}{90}{\small $\textbf{K}^{\, \ell} _0$}\end{turn}}
				\end{picture}
			 \includegraphics[scale=0.55]{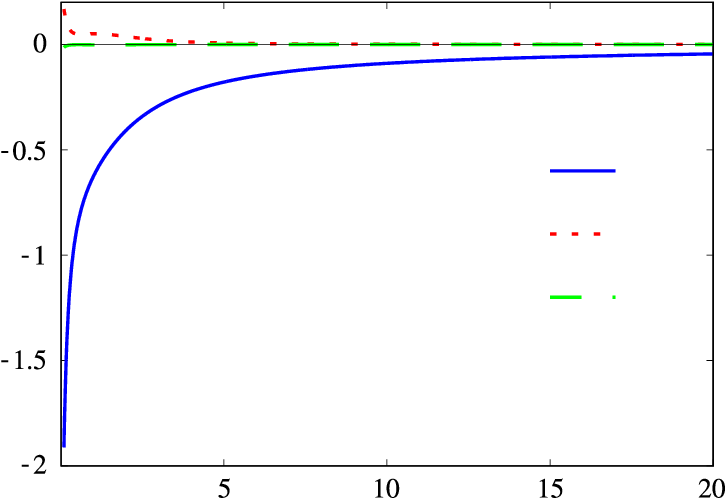}\vspace{1mm}\\
			 			 	\begin{picture}(0,0)		 	 
			 	 \put(138,120){\small $k=0$}	
			 	 \put(138,85){\small $k=1$}			 	 
			 	 \put(138,49){\small $k=2 $}
			 	 \put(138,26){\small $k=3 $}
			 	 \put(115,-8){\small $r$}
			 	 \put(16,0){\small $r_H$}
			 	 \put(4,68){\begin{turn}{90}{\small $err$}\end{turn}}
				\end{picture}
			 \includegraphics[scale=0.55]{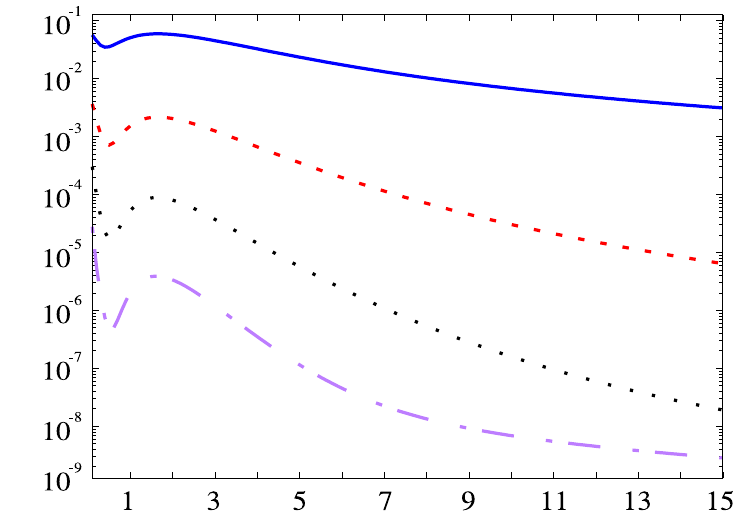}\\
			 \caption{(Left panel) radial profile of the numerically obtained $F_0 (r,\frac{\pi}{2})$; (right panel) the first four spectral functions $\textbf{K}_0 ^{2k}$. (Bottom panel) $err$ of $F_0 $ as a function of $(r,\ze \frac{\pi}{2})$ for (solid blue) $k=0$; (dashed red) $k =1$; (dotted black) $k=2$; (dot-dashed purple) $k=3$ modes.}
			 \label{F6.3}
    		\end{figure}
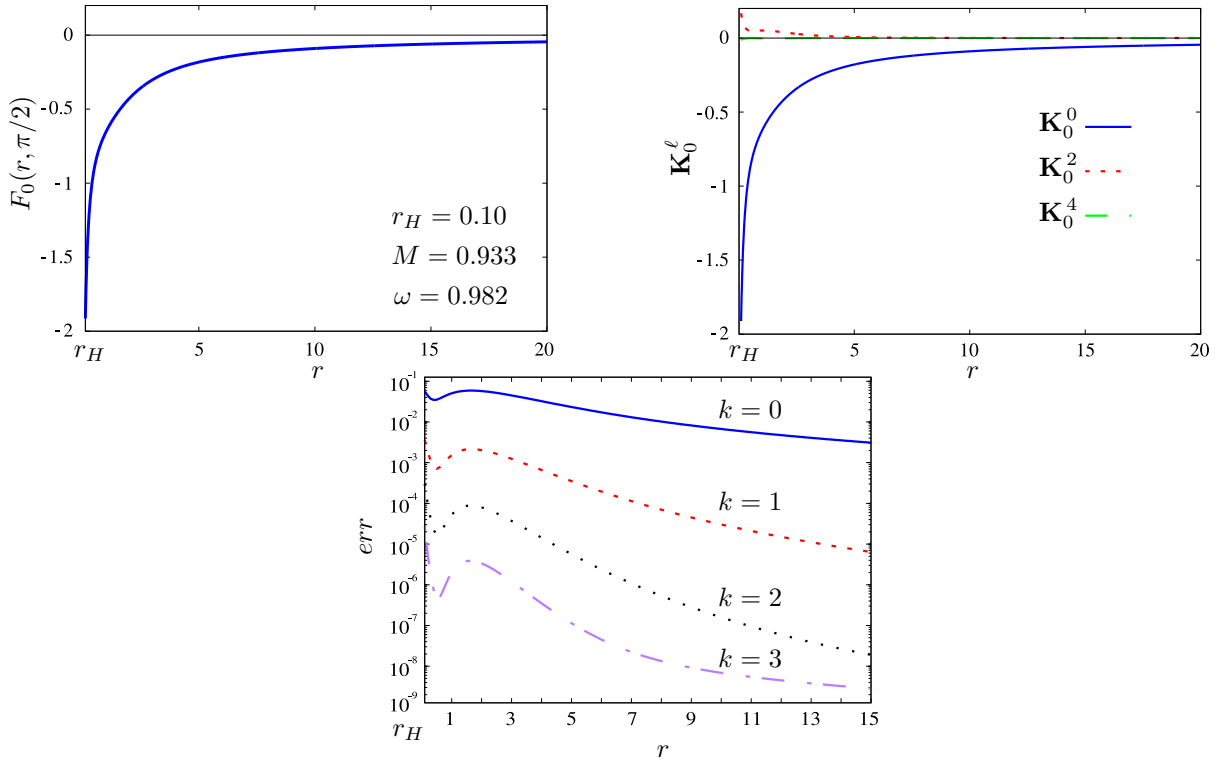

%
	At last, let us test our method in a more complex system. For this, consider five KBHsSH configurations \cite{Grav}:
			\begin{itemize}
			 \item[I)] A Kerr BH in the region of non-uniqueness with $r_H=0.07$, $\Omega _H =1.111$, $M=0.415$ and $J=0.172$,
			 \item[II)] A KBHsSH in the region of non-uniqueness with $r_H=0.20$, $\Omega _H =0.975$, $M=0.415$ and $J=0.172$,
		 	 \item[III)] A KBHsSH close to the main branch of spinning SBSs $r_H=0.10$, $\Omega _H =0.982$, $M=0.933$ and $J=0.739$,
			 \item[IV)] A KBHsSH in the region which is likely to be unstable $r_H=0.04$, $\Omega _H =0.680$, $M=0.975$ and $J=0.850$,
			 \item[V)] A typical spinning SBS belonging to the main branch of solutions with $r_H=0.00$, $\omega=0.800$, $M=1.310$ and $J=1.374$.
			\end{itemize}
	As a first, more qualitative analysis, consider the SD profile Fig.~\ref{F6.2} and \ref{F6.3} of scalar field and one exemplar metric function $F_0$ of solution Conf.~$III$.
	As one can observe, the first spectral mode $k=0$ ($\textbf{f}^{\, 1}$) has the most noticeable impact on the shape of the functions, followed by the second and third. In all cases, the fourth spectral mode is almost observationally negligible. In addition, from the $err$ plots, it is easy to observe that the region with the largest gradient is the hardest to describe. However, as one goes away from this region $err \rightarrow 0$ the solutions gets more well fitted by the expansion. 

	In a more quantitative analysis, let us look at the convergence of $M_\Phi$ and $Q_S$ with the increase in the number of spectral modes in the expansion, $k$, of $\Phi$, while keeping the metric functions intact.

			\begin{figure}[h!]
		 	 \centering
		 	 	\begin{picture}(0,0)		 	 
			 	 \put(52,138){\small $M_\Phi$}	
			 	 \put(85,138){\small $Q_S$}	
			 	 \put(118,139){\small $r_{LR}$}	
			 	 \put(150,139){\small $r_{ISCO}$}				 	 			 	 
			 	 \put(190,139){\small virial}	
			 	 \put(117,-8){\small $k$}
			 	 \put(-7,78){\begin{turn}{90}{\small $err$}\end{turn}}
				\end{picture}
			 \includegraphics[scale=0.62]{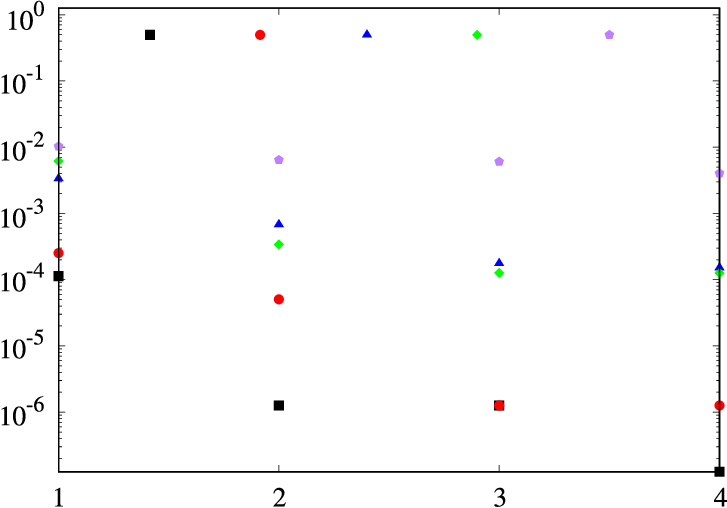}
			 \caption{Average of the five different KBHsSH configurations relative difference $err$ for several of quantities between the SD and full numerical solutions as a function of the number of spectral modes in the SD expansion $k$. Average of (black squares) scalar field mass, $M_\Phi$; (red dot) scalar charge $Q_S$; (blue triangle) LR radius, $r_{LR}$; (green rhombus) ISCO radius, $r_{ISCO}$ and (purple pentagon) virial identity. An increase in the number of the spectral modes is associated with a decrease of the $err$.}
			 \label{F6.7}
    		\end{figure}
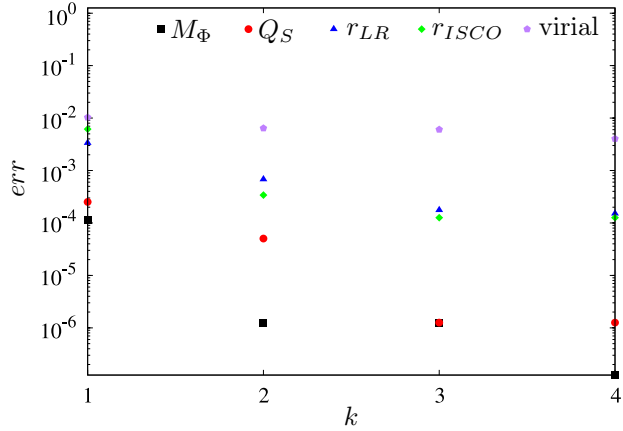
	As Fig.~\ref{F6.7} shows, both $M_\Phi$ (black squares) and $Q_S$ (red dots) tend to the respective numerical values with the increase in $k$. Taking only the first two spectral functions, $\textbf{f}^{\ze 1} $ and $\textbf{f}^{\ze 3}$, is already enough to obtain an $err$ smaller than $10^{-4}$. Full agreement with the numerical result is achieved with just four modes ($err\leqslant 10^{-6}$).
	
	Concerning the metric functions, just as before, the first four spectral modes, are enough to recover the $r_{LR}$ and $r_{ISCO}$ with an high accuracy $err\sim 10 ^{-4}$. With the $err$ associated to $r_{LR}$ always smaller than the one of $r_{ISCO}$, due to the derivatives. 
	
	For the overall decomposition test, \textit{a.k.a.} virial identity \eqref{E6.2.16}. The latter contains both $\Phi$ and $\mathcal{F}_i$ in an integral, and the $I_{GHY}$ obtained through an asymptotic decay (see Ch.~\ref{C7}), making it a very challenging SD test. 

	In Fig.~\ref{F6.7} (purple pentagon), one can easily see that $k=4$ is a suitable fit to the virial identity obtained by the pre-decomposed solution. In addition, we would like to point out that the virial identity is much more sensitive to the number of the spectral modes than any other test.

%
	\section{A curious behaviour}\label{S6.5}
%
	At last, let us try to apply the previous decomposition to the study of some solutions. Observe that KBHsSH continuously connects the hairless Kerr solution to the spinning SBS.
	
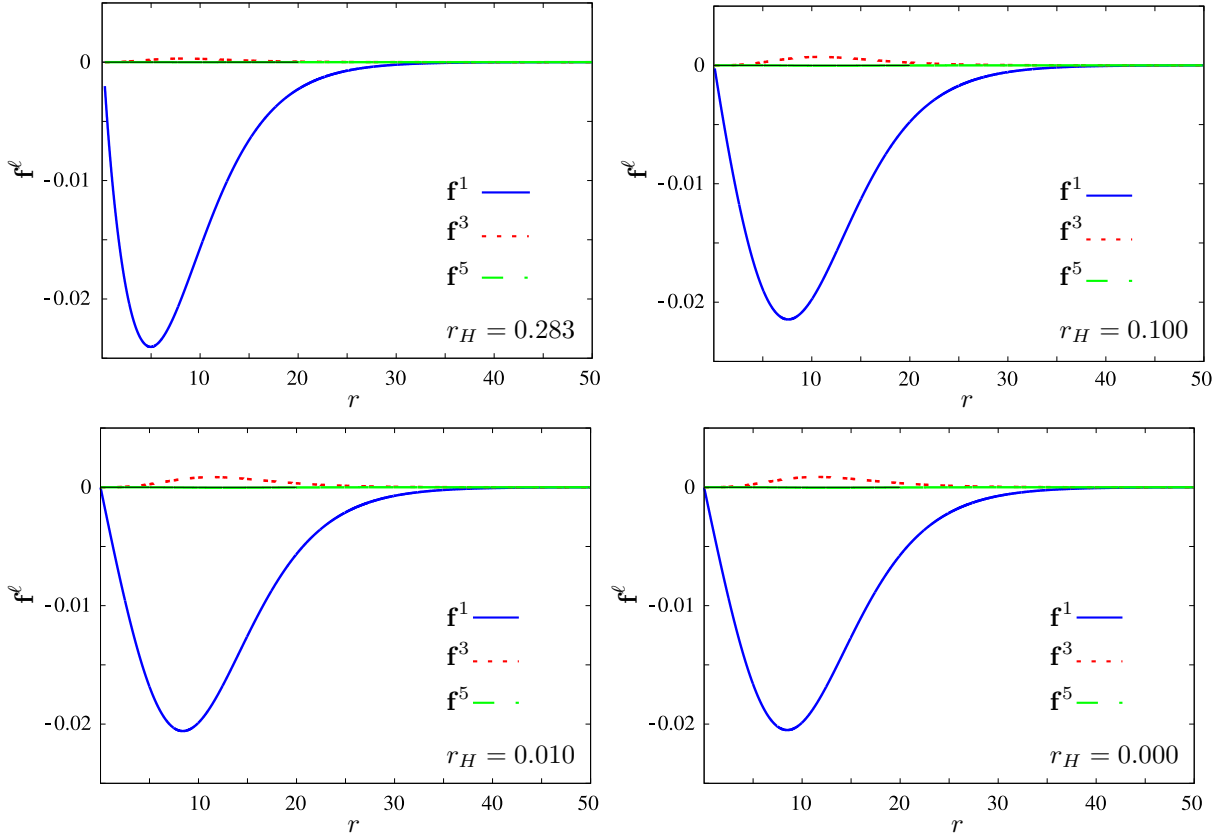
\begin{figure}[H]
		 \centering
		 	 	\begin{picture}(0,0)		 	 
			 	 \put(154,70){\small $\textbf{f}^{\ze\ze 1}$}	
			 	 \put(154,55){\small $\textbf{f}^{\ze\ze 3}$}	
			 	 \put(154,38){\small $\textbf{f}^{\ze\ze 5}$}		
			 	 \put(154,18){\small $r_H =0.283$}				 	 			 	 
			 	 \put(117,-8){\small $r$}
			 	 \put(-10,78){\begin{turn}{90}{\small $\textbf{f}^{\ze\ell}$}\end{turn}}
				\end{picture}
		 \includegraphics[scale=0.6]{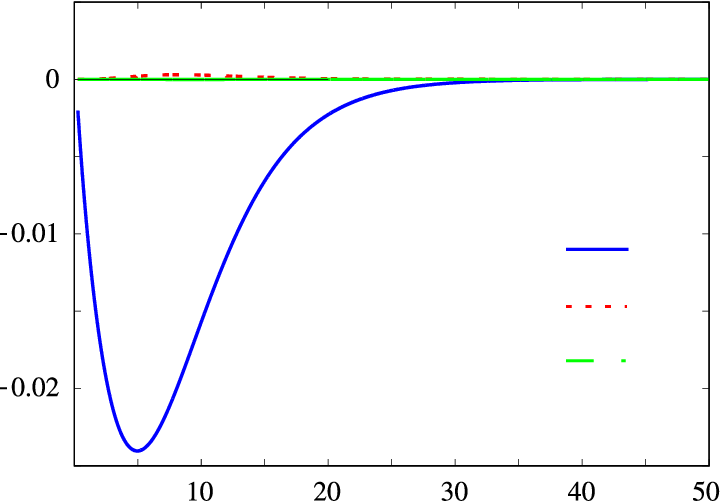}\hfill
		 		 	 	\begin{picture}(0,0)		 	 
			 	 \put(154,70){\small $\textbf{f}^{\ze\ze 1}$}	
			 	 \put(154,55){\small $\textbf{f}^{\ze\ze 3}$}	
			 	 \put(154,38){\small $\textbf{f}^{\ze\ze 5}$}		
			 	 \put(154,18){\small $r_H =0.100$}				 	 			 	 
			 	 \put(117,-8){\small $r$}
			 	 \put(-10,78){\begin{turn}{90}{\small $\textbf{f}^{\ze\ell}$}\end{turn}}
				\end{picture}
		 \includegraphics[scale=0.6]{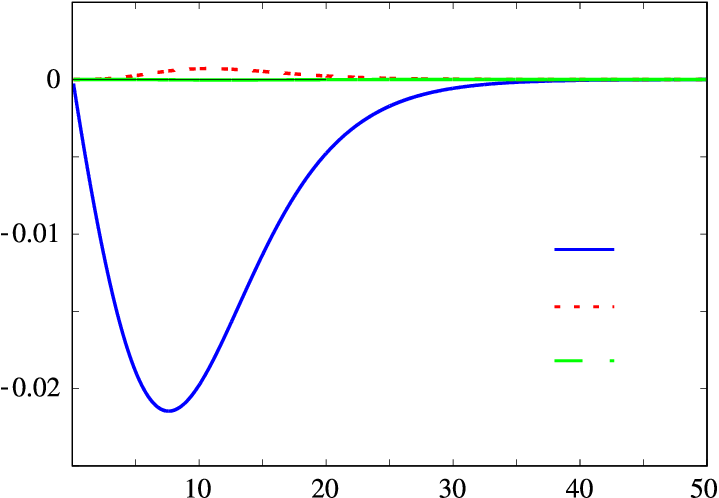}\vspace{5mm}\\
		 		 		 	 	\begin{picture}(0,0)		 	 
			 	 \put(154,70){\small $\textbf{f}^{\ze\ze 1}$}	
			 	 \put(154,55){\small $\textbf{f}^{\ze\ze 3}$}	
			 	 \put(154,38){\small $\textbf{f}^{\ze\ze 5}$}		
			 	 \put(154,18){\small $r_H =0.010$}				 	 			 	 
			 	 \put(117,-8){\small $r$}
			 	 \put(-10,78){\begin{turn}{90}{\small $\textbf{f}^{\ze\ell}$}\end{turn}}
				\end{picture}
		 \includegraphics[scale=0.6]{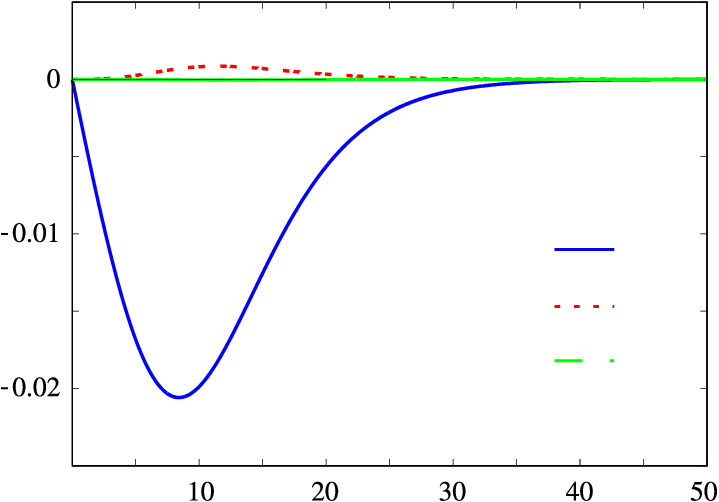}\hfill
		 		 \begin{picture}(0,0)		 	 
			 	 \put(154,70){\small $\textbf{f}^{\ze\ze 1}$}	
			 	 \put(154,55){\small $\textbf{f}^{\ze\ze 3}$}	
			 	 \put(154,38){\small $\textbf{f}^{\ze\ze 5}$}		
			 	 \put(154,18){\small $r_H =0.000$}				 	 			 	 
			 	 \put(117,-8){\small $r$}
			 	 \put(-10,78){\begin{turn}{90}{\small $\textbf{f}^{\ze\ell}$}\end{turn}}
			 	 				\end{picture}
		 \includegraphics[scale=0.6]{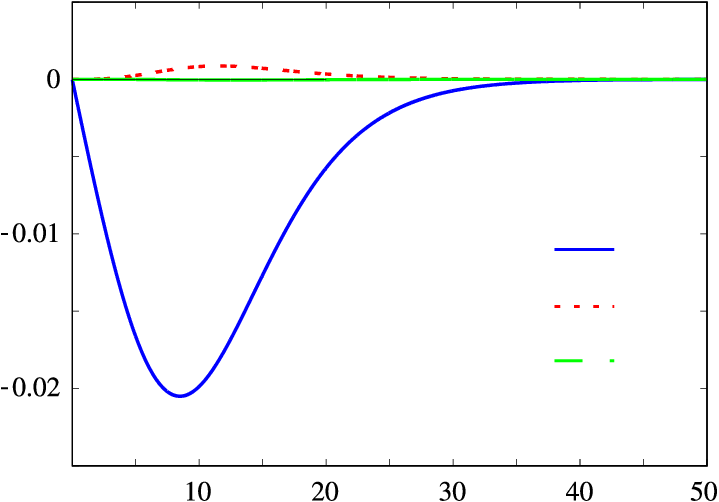}
		 \caption{First four spectral functions $\textbf{f} ^{\,\ell} $ of the scalar field surrounding a Kerr BH with $\omega = 0.96$. (Top left) configuration close to a Kerr BH $r_H=0.283$; (top right) configuration with $r_H=0.100$; (bottom left) configuration close to the SBS line $r_H=0.010$; (bottom right) a spinning SBS solution $r_H=0.000$. Full evolution can be seen at \cite{Evolution}.}
		 \label{F6.8}
    	\end{figure}

	In that regard, and as an example, let us solely consider the radial dependence of $\textbf{f}^{\,\ell}$ for Conf. $II-V$, see Fig.~\ref{F6.8}.
	
	From Fig.~\ref{F6.8}, we observe that as the impact of the scalar field increases -- from a perturbation of the Kerr background (Conf. $II$) to a pure spinning SBS (Conf. $V$) -- so does the impact of higher-order modes. 
	
	While the most significant contributor is always the first spectral mode, $k=0$, there is a visible increase in the contribution of the higher-order modes as we go from Kerr to a spinning SBS -- see the evolution from the top left panel to the bottom right panel for $\textbf{f}^{\,\ell}$ in Fig.~\ref{F6.8}. In particular, the growing impact of the second spectral function $\textbf{f}^{\, 3}$ can indicate a change of the solutions properties, becoming more like a spinning SBS than an hairy Kerr BH.
	
	Such a technique can be usefull in the study of stability, being an intermediate step in the constructuion of the machinery required to study the latter.
%
%
	\section{Further remarks}\label{S6.6}
%
	The goal of this chapter was to implement the decomposition in SH of gravitational functions. The latter separates the radial and angular dependences and, consequently, can give an essential insight into the structure of the solutions.

	To illustrate the procedure, we have decomposed the analytical Kerr and numerical Kerr BH with scalar hair solutions. While in the former, we only had to decompose the metric, in the latter, we also had to decompose the scalar field.

	The well-defined parity of each SH allows the separation of terms that contribute and terms that do not contribute to the decomposition. This property of the SH significantly simplifies the procedure.

	To test the ability of the SD to decompose a set of functions, we have compared the complete solution (pre-decomposition) with the decomposed solution in a series of tests. Namely, we tested the isolated scalar field decomposition (without decomposing the metric functions) with the Komar mass and Noether charge integrals; the metric decomposition with two geometric properties, the LRs and ISCOs radial position; and the decomposition of the general solution with the virial identity. 

	One has observed a difficulty of the SD to reproduce quantities computed through derivatives -- namely $r_{ISCO}$. The latter is mainly a numerical artefact. However, it shows a possible weakness of this method even though the SD procedure was able to correctly decompose the solution's functions (with a $err\sim 10^{-4}$) with just the first four terms of the expansion.

	Concerning the structure of each decomposed solution, one observes that the first spectral mode $k=0$ is always the leading contributor to the original function (with $\sim 95\% $ of the contribution), followed by the second and third spectral modes. The fourth mode is almost always negligible and only relevant for numerics. 

	At last, we have speculated the use of this technique as a tool to explore the stability of hairy black hole solutions in the future. Moreover, observe that a simple decomposition and analysis of the spectral functions can already indicate some distintions. 

\clearpage\null\newpage

%
\chapter{Virial identity}\label{C7}
%
	In this chapter\footnote{Part of this chapter is based on already published work \cite{herdeiro2021virial} (the spherically symmetric case) and on work on its way to be published (axially symmetric case, Sec.~\ref{S7.6} onward).}, we shall be interested in integral identities that are virial-like (and thus, following the literature, will be referred to as ``virial identities''), but in field theory rather than particle mechanics, obtained from \textit{scaling} arguments. The first example of such virial identities in field theory arose as a ``no-go'' theorem for \textit{solitons}. 

	The possible existence of soliton-type configurations (particle-like solutions inspired by \textit{solitary} wave solutions of the Korteweg-de-Vries equation~\cite{korteweg1895xli,whitham1965non,darrigol2005worlds,dauxois2006physics}) emerges as an interesting question in any non-linear field theory. The robustness against decay of the `shape' of such solutions is interpreted as a cancellation between non-linear and dispersive effects. In this context, \textit{Derrick's theorem}~\cite{derrick1964comments} was put forward in 1964 as a generic argument against the existence of stable, finite energy, time-independent solutions in a wide class of non-linear wave equations, in three or higher (spatial) dimensions -- see also~\cite{hobart1963instability,hobart1965non} for an earlier similar argument. This theorem results from a scaling argument; for a (3+1)-dimensional relativistic scalar field theory of a complex scalar field $\Phi$, with spatial gradient $\nabla \Phi$ and potential energy $U(\Phi)$, it results in the virial identity ($cf.$ Sec.~\ref{S7.2})
	\begin{equation}\label{E7.0.1}
	 \int d^3{\bf x} \left[\frac{(\nabla \Phi)^2}{3}+U(\Phi)\right] =0\ . \qquad \qquad {\rm {\bf [virial \  Derrick]}}
	\end{equation}
	Eq.~\eqref{E7.0.1} represents the prototypical virial identity in field theory. It has a simple interpretation: if the potential energy is non-negative, since $(\nabla \Phi)^2\geqslant 0$, then~\eqref{E7.0.1} can only be obeyed for a constant $\Phi=\Phi_0$ $\big($for which $U(\Phi_0)=0\big)$. Thus, there are no non-constant configurations, hence no solitons. 

	The usefulness of virial identities is not exhausted in establishing no-soliton theorems. In generic setups, which includes more general field theories (possibly also with gravity) and more general ansatze for the fields, virial/scaling identities serve to understand the balance between the different effects that allow the existence of solitonic or BH solutions (see $e.g.$ Sec.~\ref{S7.1}). In this sense, virial identities serve as a guide to construct new solutions. Additionally, as for solitons, they can also be used to establish no-go theorems for BHs with non-trivial matter fields, also known as ``no-hair'' theorems -- see $e.g.$~\cite{heusler1992scaling,heusler1996no,herdeiro2015asymptotically}. Furthermore, in the context of numerical solutions, virial identities serve as valid identities to test the accuracy of such numerical solutions -- see $e.g.$~\cite{fernandes2020einstein,herdeiro2018spontaneous,fernandes2019spontaneous}.

	Despite these (and other) exciting applications, the use of virial identities in the context of strong gravity has been mostly restricted to spherically symmetric solutions and a particular ``gauge'' (by which we mean coordinates \textit{plus} a parameterization) choice. The main goal of this chapter is to present a generic methodology for establishing virial identities for equilibrium, asymptotically flat, localized configurations using any ``gauge'' choice for the metric and matter fields. In doing so, we will unveil a key ingredient, hitherto neglected, that must be taken into account in relativistic gravity applications -- in general, there can be a non-trivial contribution from boundary terms. In the context of GR the appropriate boundary term is the Gibbons-Hawking-York (GHY) term~\cite{york1972role,gibbons1993action}, which must be considered in order to derive the correct virial identity.

	One can face the virial identity in a particular model, encompassing different fields as a ``word'' composed by different ``letters''. Computing the basic ``letters'' one can efficiently piece them together into the virial identity ``word'', for a model composed by the different fields analyzed here. After establishing a general methodology, we shall test the obtained virial identities, providing examples corresponding to different field theories and parameterization choices. Moreover, our analysis reveals a simpler ``gauge'' choice for which the gravitational part does not contribute. There is, therefore, a simple setup to compute virial identities in GR just by computing the contribution of the matter action, which can be safely used under the generic understanding presented here. 

	This chapter is organized as follows. We start in Sec.~\ref{S7.1} by considering the variational treatment in particle mechanics. This section serves two purposes. Firstly it builds a bridge between the scaling transformation that yields virial identities and the familiar standard variational treatment in Lagrangian mechanics. Secondly, it introduces the notion of \textit{effective action} (EA) that, in practice, is the central object used in building virial identities in field theory that yield a 1D EA (as in spherical symmetry). As we shall see, the virial identities obtained in this section $\big($\eqref{E7.1.13}, \eqref{E7.1.16}, \eqref{E7.1.19} and \eqref{E7.1.22}$\big)$ can then be used as general formulae for the subsequent problems found in field theory. In Sec.~\ref{S7.2.1}, we review Derrick's theorem as the paradigmatical illustration of a scaling argument and a virial identity. However, we also show how a change of ansatz leads to a way of circumventing Derrick's theorem allowing the existence of scalar field theory, flat spacetime solitons known as $Q$-balls~\cite{coleman1985q}. In Sec.~\ref{S7.2.2} we take a first look at GR. This section is meant as pedagogical, and the virial relations obtained therein are \textit{incomplete}. Our goal is to illustrate two points. Firstly, there are simpler ``gauge'' choices to compute virial identities. In the simplest parameterization, the Einstein-Hilbert (EH) action results in a scale-invariant EA; then, it does not contribute to the virial identity. Secondly, considering the case of electro-vacuum, we show that the (would be) virial identity derived solely from the EH-Maxwell action is incorrect, as the Reissner-Nordstr\"om does not obey it. The complete treatment is then introduced in Sec.~\ref{S7.2.3}, where we include the contribution of the GHY boundary term, and we provide the complete virial identities for the vacuum and electro-vacuum cases. In Sec.~\ref{S7.3}, we take advantage of the simplest ``gauge'' choice to compute the virial identity for various examples of field theories minimally coupled to Einstein's gravity by considering simply the contribution of the matter part. To emphasize the generic case, however, in Sec.~\ref{S7.4} we discuss the virial identities for electro-vacuum and (massive-complex) scalar-vacuum in isotropic coordinates, confirming the non-trivial contribution from the gravitational part that is mandatory in order for the virial identity to be obeyed by known solutions. We generalize the previous procedure to axial symmetry in Sec.~\ref{S7.6} and introduce a convenient gauge in Sec.~\ref{S7.6.1} (an exemple is given). We provide a discussion and our conclusions in Sec.~\ref{S7.8}.
%
	\section{Particle mechanics and effective actions}\label{S7.1}
%
	Some insight and valuable formulas that  will be used in the field theory case can be obtained by addressing first particle mechanics. Let us start with a recap of the elementary variational treatment. 
%
		\subsection{The standard variational treatment}\label{S7.1.1}
%
	Consider an action functional $\mathcal{S}$, depending on a set of $n$ generalized coordinates $q_j$ ($j=1\dots n$), their first derivatives in time, $\dot{q}_j$, and on the time coordinate $t$ itself (so that $\dot{q}_j = dq_j/dt$). The action is the time integral of the Lagrangian $L$:
			\begin{equation}\label{E7.1.2}
			 \mathcal{S}[q_j(t),\dot{q}_j(t),t] = \int_{t_i}^{t_f}L\left(q_j,\dot{q}_j,t\right)dt \ .
			\end{equation}
	In the standard variational problem one aims at finding the true \textit{path} of the particle in $\mathbb{R}^{n}$, which is a map
			\begin{align}\label{E7.1.3}
			 [t_i,t_f]\in \mathbb{R} & \rightarrow \mathbb{R}^{n} \nonumber \\
			 t  & \rightarrow q_j (t) \ ,
			\end{align}
	traveled as a function of (time) $t$. This path extremises the action functional. To compute it, one considers an arbitrary variation $\delta q_j(t)$ around a fiducial path, $q_j(t)$, where the endpoints are fixed, $\delta q_j(t_i)= \delta q_j(t_f)=0$. This generates a variation of the action $\delta \mathcal{S}$. Hamilton's principle (\textit{a.k.a.} principle of least action) selects the true path as the fiducial path if $\delta \mathcal{S}\Big|_{\delta q_j=0}=0$. 

	Explicitly, the variation (using the chain rule and integrating by parts) reads
			\begin{align}\label{E7.1.4}
	 		 \delta \mathcal{S} = \int_{t_i}^{t_f} \delta L dt &= \int_{t_i}^{t_f}\left( \frac{\partial L}{\partial \dot{q}_j}\delta\dot{q}_j +  \frac{\partial L}{\partial q_j}\delta q_j + \frac{\partial L}{\partial t}\delta t\right)dt\nonumber\\
			 &=\int_{t_i}^{t_f}\left( \frac{\partial L}{\partial \dot{q}_j}\frac{d}{dt}\delta q_j +  \frac{\partial L}{\partial q_j}\delta q_j\right)dt \\
	 &=\left[\frac{\partial L}{\partial \dot{q}_j}\delta q_j\right|^{t=t_f}_{t=t_i} + \int_{t_i}^{t_f}\left[ -\frac{d}{dt}\left(\frac{\partial L}{\partial \dot{q}_j}\right) +  \frac{\partial L}{\partial q_j}\right]\delta q_jdt  \ .\label{E7.1.5}
			\end{align}
	For arbitrary variations under fixed endpoints, the first term of the \textit{rhs} of the last equation vanishes, and the second term yields a set of \textit{differential} requirements for the true path, the \textit{Euler-Lagrange equations}
			\begin{equation}\label{E7.1.6}
			 \frac{d}{dt}\left(\frac{\partial L}{\partial \dot{q}_j}\right) = \frac{\partial L}{\partial q_j} \ .
			\end{equation}
%

%
		\subsection{A scaling transformation of an effective action}\label{S7.1.2}
	In the standard variational treatment~\eqref{E7.1.4} the term $(\partial L/\partial t) \delta t$ was dropped under the assumption that the Lagrangian has no explicit dependence on $t$. Moreover, arbitrary variations of the path were considered. We shall now consider a variation on the variational problem, where an explicit dependence on (the analogue of) $t$ is present, and it is a variation of this parameter that induces the variation of the ``path''. Instead of considering the \textit{path} traveled in time by a particle in $\mathbb{R}^{n}$, however, we shall consider the (spatial) \textit{profile} of a map:
			\begin{align}\label{E7.1.7}
			 [r_i,+\infty]\in \mathbb{R} & \rightarrow \mathbb{R}^{n} \nonumber \\
			 r  & \rightarrow q_j (r) \ ,
			\end{align}
	which is spanned as a function of a (spatial) coordinate $r$. Considering the field theory applications below, we choose the profile to start at $r=r_i$ and end at $r=+\infty$. There are infinitely many possible profiles, but the true one extremizes a certain \textit{effective action} (EA)
			\begin{equation}\label{E7.1.8}
			 \mathcal{S}^{\rm eff}[q_j(r),q'_j(r),r] = \int_{r_i}^{+\infty}\mathcal{L}\left( q_j,q'_j,r\right)dr \ .
			\end{equation}

	This EA does not have the physical dimensions of an action. Nevertheless, it plays the role of an action because it determines the true configurations through a variational principle. By the same token, we shall be referring to the integrand in~\eqref{E7.1.8} $\mathcal{L}$ as an effective Lagrangian. 

	We have considered arbitrary variations of a fiducial path $q_j(t)$ in the standard variational treatment. Now, we shall vary the independent parameter $r$ in a specific manner and consider the profile variation induced by the latter. Concretely, we consider a transformation $r\rightarrow \tilde{r}$ that \textit{scales} $r$ but keeps $r_i$ as a fixed point.  Thus

			\begin{equation}\label{E7.1.9}
			 r\rightarrow \tilde{r} = r_i + \lambda (r-r_i) \ ,
			\end{equation}
	where $\lambda$ is an arbitrary positive constant, such that $\tilde{r}=r_i$ for $r=r_i, \forall_\lambda$ (fixed point); the transformation trivializes for $\lambda=1$: $\tilde{r} = r $. The new profile induced by the scaling~\eqref{E7.1.9} is 
			\begin{equation}\label{E7.1.10}
			 q_j(r)\rightarrow q_{\lambda j}(r)=q_j( \tilde{r}) \ .
			\end{equation}
	The EA of the scaled profile becomes a \textit{function} of $\lambda$, denoted as $\mathcal{S}_\lambda^{\rm eff}$,
			\begin{align}\label{E7.1.11}
			 \mathcal{S}_\lambda ^{\rm eff} & = \int_{r_i}^{+\infty}\mathcal{L}_\lambda \left(q_j({r}),\frac{dq_j(r)}{dr},{r}\right)dr = \int_{r_i}^{+\infty}\mathcal{L}\left(q_j(\tilde{r}),\frac{dq_j(\tilde{r})}{dr},r\right)dr\nonumber\\
			 &= \int_{r_i}^{+\infty}\mathcal{L}\left(q_j(\tilde{r}),\, \lambda \frac{dq_j(\tilde{r})}{d\tilde{r}},\,\frac{\tilde{r}-r_i}{\lambda}+r_i\right)\frac{d\tilde{r}}{\lambda} \ .
			\end{align}
	The true profile obeys the \textit{stationarity condition}
			\begin{equation}\label{E7.1.12}
			 \frac{\partial \mathcal{S}^{\rm eff}_\lambda}{\partial \lambda} \bigg|_{\lambda=1}=0 \ ,
			\end{equation}
	which, from the last equality in~\eqref{E7.1.11} yields
			\begin{equation}\label{E7.1.13}
			 \int_{r_i}^{+\infty}\bigg[ \sum_j \frac{\partial \mathcal{L}}{\partial q'_j} q'_j -\mathcal{L} -\frac{\partial \mathcal{L}}{\partial r}(r-r_i)\bigg]dr = 0 \ .  \qquad \qquad {\rm {\bf [virial \  EA \ 1]}}
			\end{equation}
	Unlike the standard variational procedure, yielding a set of differential constraints, here we obtain an \textit{integral} constraint that should obey if the $q_j(r)$ are solutions of the Euler-Lagrange equations derived from~\eqref{E7.1.8}. Observe that the first two terms in the integrand of~\eqref{E7.1.13} combine into a ``Hamiltonian''
			\begin{equation}\label{E7.1.14}
			 \mathcal{H}\equiv  \sum_j \frac{\partial \mathcal{L}}{\partial q'_j} q'_j -\mathcal{L} \ .
			\end{equation}
%

		\subsection{Effective Lagrangians depending on second order derivatives}\label{S7.1.3}
%
	In field theory, we shall sometimes find effective Lagrangians depending \textit{also} on the second derivative of the profile functions $q''_j(r)=d^2q_j(r)/dr^2$. For instance, the EH Lagrangian $\big( cf.$~\eqref{E7.2.39} below$\big)$ depends on the second derivatives of the metric. In such cases, to consider the variational problem, the action~\eqref{E7.1.8} is replaced by the more general
			\begin{equation}\label{E7.1.15}
			 \mathcal{S}^{\rm eff}\big[ q_j(r),q'_j(r),q''_j(r),r\big ] = \int_{r_i}^{+\infty}\mathcal{L}\left( q_j,q'_j,q''_j,r\right)dr \ .
			\end{equation}
	Repeating the procedure of the previous subsection, \textit{mutatis mutandis}, we obtain the more general virial identity
			\begin{equation}\label{E7.1.16}
			 \int_{r_i}^{+\infty}\bigg[ \sum_j \frac{\partial \mathcal{L}}{\partial q'_j} q'_j+2 \sum_j \frac{\partial \mathcal{L}}{\partial q''_j} q''_j -\mathcal{L} -\frac{\partial \mathcal{L}}{\partial r}(r-r_i)\bigg] dr = 0 \ .  \qquad \qquad  {\rm {\bf [virial \ EA \ 2]}}
			\end{equation}	
%

%
		\subsection{Scalings affecting the integration limits}\label{S7.1.4}
%
	A further generalization is to consider a scaling that affects the integration limits. The simplest example is to replace~\eqref{E7.1.9} by
			\begin{equation}\label{E7.1.17}
			 r\rightarrow \tilde{r} =  \lambda r \ .
			\end{equation}
	This transformation impacts non-trivially on the lower limit of the action integral \eqref{E7.1.15}. To understand the corresponding contribution to the virial identity, we repeat the steps in \eqref{E7.1.11} (allowing, as in Sec.~\ref{S7.1.3}, a further $q''_j(r)$ dependence) to find
			\begin{equation}\label{E7.1.18}
			 \mathcal{S}^{\rm eff}_\lambda = \int_{\lambda r_i}^{+\infty}\mathcal{L}\left(q_j(\tilde{r}),\lambda \frac{dq_j(\tilde{r})}{d\tilde{r}},\lambda^2 \frac{d^2q_j(\tilde{r})}{d^2\tilde{r}},\frac{\tilde{r}}{\lambda}\right)\frac{d\tilde{r}}{\lambda} \ .
			\end{equation}
	Thus, the stationarity condition \eqref{E7.1.12} now yields an extra term:
			\begin{equation}\label{E7.1.19}
			 \int_{r_i}^{+\infty}\bigg[ \sum_j \frac{\partial \mathcal{L}}{\partial q'_j} q'_j+2 \sum_j \frac{\partial \mathcal{L}}{\partial q''_j} q''_j -\mathcal{L} -\frac{\partial \mathcal{L}}{\partial r}r\bigg]dr = r_i \ze \mathcal{L}(r_i) \ . \qquad \qquad  {\rm {\bf [virial \ EA \  3]}}
			\end{equation}	
%

%
		\subsection{Adding a total derivative to the effective Lagrangian}\label{S7.1.5}
	As a final discussion point, leading in fact to the formula that will be most used in the field theory applications below, we observe that, in some circumstances, there are boundary terms that can be added to the Lagrangian, which take the form of a total derivative. Consequently, these terms do not affect the bulk equations of motion. A total derivative can, however, affect the virial identity. Typically there can be a trade-off between considering a total derivative \textit{or} considering an effective Lagrangian with second-order derivatives (as in Sec.~\ref{S7.1.3}). The virial identities obtained using either perspective are equivalent (for an illustration, see Sec.~\ref{S7.3.1} below).

	To see the explicit form of the virial identity when a total derivative is present, consider an EA\footnote{When considering a total derivative we do not consider second derivatives in the effective Lagrangian, due to the trade-off between these two types of terms.}

		\begin{equation}\label{E7.1.20}
	 	 \mathcal{S}^{\rm eff}\big[ q_j(r),q'_j(r),r\big] = \int_{r_i}^{+\infty}\tilde{\mathcal{L}}\left( q_j,q'_j,r\right)dr \ ,
	 	\end{equation}
	where the new Lagrangian $\tilde{\mathcal{L}}$ contains a total derivative term
		\begin{equation}\label{E7.1.21}
		 \tilde{\mathcal{L}}\left(q_i,q'_i,r\right)=\mathcal{L}\left(q_i,q'_i,r\right)+\frac{d}{dr}f\left(q_i,q'_i,r\right) \ ,
		\end{equation}
	and $f$ is some function that depends on the same variables as the original effective Lagrangian $\mathcal{L}$, up to first derivatives. Performing the scaling~\eqref{E7.1.9}, the stationarity condition \eqref{E7.1.12} now yields
		\begin{equation}\label{E7.1.22}
		 \int_{r_i}^{+\infty}\bigg[ \sum_j \frac{\partial \mathcal{L}}{\partial q'_j} q'_j -\mathcal{L} -\frac{\partial \mathcal{L}}{\partial r}(r-r_i)\bigg]dr = \left[\frac{\partial f}{\partial r}(r-r_i) - \sum_i\frac{\partial f}{\partial q'_i} q'_i\right]^{+\infty}_{r_i} \ .  \qquad \qquad   {\rm {\bf [virial \ EA \  4]}}
		\end{equation}
	Eqs.~\eqref{E7.1.13}, \eqref{E7.1.16}, \eqref{E7.1.19} and \eqref{E7.1.22} provide useful relations that can be readily used in the context of EAs obtained from field theory models, as illustrated in the next sections.  
%
	\section{Spherical symmetry}\label{S7.2}
%

%
		\subsection{Flat spacetime field theory}\label{S7.2.1}
%
	Let us now address two flat spacetime relativistic (scalar) field theory examples. The mandatory first example is to review the original theorem by Derrick~\cite{derrick1964comments}, establishing the inexistence of solitons in a large class of non-linear field theories. We then consider a more generic ansatz for the scalar field configuration (allowing a harmonic time-dependence) and illustrate how the virial identity is compatible with the existence of solitons known as \textit{$Q$-balls}~\cite{coleman1985q}.
			\subsubsection*{Derrick's theorem}
	Consider the (possibly) non-linear Klein-Gordon equation, describing a real scalar test field on Minkowski spacetime:
				\begin{equation}\label{E7.2.23}
				 \Box\Phi=\frac{1}{2}\frac{dU (\Phi)}{d\Phi} \ ,
				\end{equation}
	where $U$ is a potential energy function. This can be derived from the following ``matter'' action:
				\begin{equation}\label{E7.2.24}
				 \mathcal{S}_{\rm m}^\Phi=\int d^4x\big[- \Phi _{,\mu} \Phi ^{,\mu} -U\big] \ .
				\end{equation}
	Splitting the spacetime coordinates $x^\mu=(t,\textbf{r})$ into temporal and spatial coordinates, the action may be rewritten as:
				\begin{equation}\label{E7.2.25}
				 \mathcal{S}_{\rm m}^\Phi=\int dt \big( S_0-S_1-S_2 \big) \ ,
				\end{equation}
	where
				\begin{equation}\label{E7.2.26}
				 S_0\equiv \int d^3x \, \dot{\Phi} ^2 \ , \qquad S_1\equiv \int d^3x\, (\nabla \Phi)^2 \ , \qquad S_2\equiv \int d^3x\, U \ ,
				\end{equation}
	and the integration is over the whole space. We will prove that no stable, time-independent, localized solutions exist for any potential energy. Time-independence implies $S_0=0$. By \textit{localized} we mean that $S_1$ and $S_2$ are finite. Due to the time-independence, we may consider the EA
				\begin{equation}\label{E7.2.27}
				 \mathcal{S}^{\rm eff}=S_1+S_2 \ .
				\end{equation}
	The existence of a localized solution, by Hamilton's principle, implies $\delta  \mathcal{S}^{\rm eff}=0$. Let the solution be $\Phi({\bf r})$; due to the time-independence, extremizing the EA is equivalent to extremizing the energy ($\delta  \mathcal{S}^{\rm eff} = \delta E$). The solution is stable if $\delta^2E\geqslant 0$.
	
	Let us define a scaled configuration $\Phi_\lambda({\bf r})=\Phi(\lambda\ze {\bf r})$, where  the radial coordinate suffers the dilation $r\rightarrow \tilde{r}=\lambda\ze r$. The energy of such scaled configuration is:
				\begin{equation}\label{E7.2.28}
				 E_\lambda= \int d^3x\left[ (\nabla \Phi_\lambda)^2 +U(\Phi_\lambda)\right] =\frac{S_1}{\lambda}+\frac{S_2}{\lambda^3} \ .
				\end{equation}
	Since, by assumption, the original configuration $\Phi({\bf r})$ (corresponding to $\lambda =1$) was a solution 
				\begin{equation}\label{E7.2.29}
				 \left(\frac{dE_\lambda}{d\lambda}\right)_{\lambda=1}=-S_1-3S_2=0 \ , \qquad \Leftrightarrow \qquad {S_2}=-\frac{S_1}{3} \ .
				\end{equation}
	Eq.~\eqref{E7.2.29} is Derrick's virial identity, \eqref{E7.0.1}. It relates the total ``kinetic'' and potential energy. As mentioned in the chapter's introduction (Ch.~\ref{C7}), inspection thereof is physically insightful: since the first term in the square bracket is everywhere positive, for positive definite potentials, there can be \textit{no solution}, regardless of being stable or not. On the other hand, 
				\begin{equation}\label{E7.2.30}
				 \left(\frac{d^2E_\lambda}{d\lambda^2}\right)_{\lambda=1}=2S_1+12S_2\stackrel{~\eqref{E7.2.29}}{=}-2S_1<0 \ ,
				\end{equation}
	since $S_1$ is manifestly positive. It follows that for \textit{any} $U$, even if it allows the existence of a solution (which may be the case for a non-positive $U$), the stretching of the hypothetical solution decreases its energy and thus, such a solution is unstable. These arguments illustrate how virial identities can establish \textit{no-go} theorems. A straightforward generalization to higher dimensions can be found in Appendix~\ref{I}.
			\subsubsection*{Circumventing Derrick's theorem: $Q$-balls}
	 In the original work~\cite{derrick1964comments}, Derrick observed that one way to circumvent the theorem would be to allow localized solutions that are periodic in time rather than time-independent. However, such configuration would not be static (or stationary) for a real scalar field. Various authors, starting with
Rosen~\cite{rosen1968particlelike}, considered a \textit{complex scalar field} $\Phi$, described by the matter action\footnote{Albeit still in flat spacetime, we allow the Minkowski metric $g$ to be written in curvilinear coordinates.}
				\begin{equation}\label{E7.2.31}
				 \mathcal{S}_{\rm m}^{\bar{\Phi}} =\int d^4x \sqrt{-g}  \bigg[-\frac{1}{2}g^{\mu\nu}(\Phi_{,\mu}\bar{\Phi}_{,\nu} +\bar{\Phi}_{,\mu}\Phi_{,\nu})-U(|\Phi|)\bigg] \ ,
				\end{equation}
	with a harmonic time-dependence:
				\begin{equation}\label{E7.2.32}
				 \Phi(t,r)=\phi(r) e^{-i\omega t} \ ,
				\end{equation}
	which guarantees a time-independent energy-momentum tensor. Moreover, there is a global symmetry and a conserved scalar Noether charge. Then, for some classes of potentials (yielding non-linear models), localized stable solutions exist, which are known, following Coleman~\cite{coleman1985q}, as $Q$-balls (since the Noether charge is typically labelled $Q$).

	Let us derive a virial identity for spherical solutions in this model to analyze how it is compatible with spherical $Q$-balls. We use the standard spatial spherical coordinates for the Minkowski background: $(t,\,r,\,\theta,\,\phi)$. Due to the spherical symmetry, the action is $(\theta,\, \varphi)$-independent and these terms can be integrated right away. Repeating Derrick's argument, we now have that $\mathcal{S}_{\rm m}^{\bar{\Phi}} =- \int dt\ze  \mathcal{S}^{\rm eff}$, where the EA $\mathcal{S}^{\rm eff}$ is written as:
				\begin{equation}\label{E7.2.33}
				 \mathcal{S}^{\rm eff}= \int_0^{+\infty}  dr \, r^2 \big[ - \omega^2\phi^2+\phi'^{\, 2} +U(|\phi|)\big]\equiv S_0+S_1+S_2 \ .
				\end{equation}
	Consider, again, a scaled configuration $ \phi_\lambda(r)=\phi(\lambda\ze r)$. Its EA is 
				\begin{equation}\label{E7.2.34}
				 \mathcal{S}^{\rm eff}_ \lambda=  \int_0^\infty dr  \, r^2 \left[ -\omega^2\phi_\lambda^2+\phi_\lambda'^{\, 2} +U(|\phi_\lambda|)\right]=\frac{S_0+S_2}{\lambda^3} + \frac{S_1}{\lambda}\ .
				\end{equation}
	Thus
				\begin{equation}\label{E7.2.35}
	 			 \left(\frac{d\mathcal{S}^{\rm eff}_ \lambda}{d\lambda}\right)_{\lambda=1}=0 \  \qquad \Leftrightarrow \qquad {S_0+S_2}=-\frac{S_1}{3} \ ,
				\end{equation}
	or, explicitly, 
				\begin{equation}\label{E7.2.36}
	 			 \int_0^{+\infty}  dr \,  r^2 \big[- \omega^2\phi^2+\frac{1}{3}\phi'^{\, 2} +U(|\phi|)\big] = 0 \ . \qquad \qquad {\rm {\bf [virial  \ {\rm Q-}balls]}}
				\end{equation}
	One observes that the harmonic time-dependence yields a term with the opposite sign $(-\omega^2\phi^2)$, so the obstruction raised by Derrick's theorem does not necessarily apply. However, the existence of solutions depends on the choice of the potential. If one chooses the potential to be solely a mass term $U(|\phi|)=\mu^2\phi^2$, then~\eqref{E7.2.36} becomes:
				\begin{equation}\label{E7.2.37}
				 \int_0^{+\infty}  dr \,  r^2 \left[\big( \mu^2- \omega^2\big) \phi^2+\frac{1}{3}\phi'^{\, 2} \right] = 0 \ ,		
				\end{equation}
	and for bound states, which obey $\omega<\mu$, one immediately concludes the inexistence of solutions. In other words, the virial identity~\eqref{E7.2.36} implies that the scalar field must have self-interactions, even with the harmonic time-dependence, in order to yield solitonic solutions. Indeed, $Q$-balls are constructed taking an everywhere positive potential with self-interactions, and for which $U-\omega^2\phi^2<0$ in some spatial regions.

	Finally, let us remark how~\eqref{E7.2.36} can be readily obtained from applying the virial identity formulas for the EAs in Sec.~\ref{S7.1}. Comparing~\eqref{E7.2.33} with~\eqref{E7.1.8} one identifies $r_i=0$ and the effective Lagrangian
				\begin{equation}\label{E7.2.38}
				 \mathcal{L}(\phi,\phi',r)=r^2 \big[- \omega^2\phi^2+\phi'^{\, 2} +U\big]\ .
				\end{equation}
	Then, applying~\eqref{E7.1.13}, a one line computation yields~\eqref{E7.2.36}.

%
		\subsection{GR in spherical symmetry -- an incomplete treatment}\label{S7.2.2}
%
	We now consider Einstein's gravity. When deriving solutions of the field equations, one considers the EH action 
			\begin{equation}\label{E7.2.39}
			 \mathcal{S}_{\rm EH}=\frac{1}{4}\int d^4x \sqrt{-g} R  \ ,
			\end{equation}
	where $R$ is the Ricci scalar of the spacetime metric $g_{\mu\nu}$ with determinant $g$. In this way one neglects possible boundary terms. As such, in this section, we shall be considering models with total action
			\begin{equation}\label{E7.2.40}
			 \mathcal{S}=\mathcal{S}_{\rm EH}+ \mathcal{S}_{\rm m} \ ,
			\end{equation}
	where $\mathcal{S}_{\rm m}$ is some matter/fields action. This treatment will turn out to be incomplete. To be clear, the (would be) virial identities derived in this section are incomplete (and will be completed in the next section). The purpose of this section is twofold. Firstly, it serves as a pedagogical introduction to the need for the GHY boundary term in deriving the correct virial identities. Secondly, it serves as an illustration of how the virial identity derived for any such model depends on the choice of $\mathcal{S}_{\rm m}$ \textit{and} on the parameterization chosen for the metric. We shall now investigate such ``gauge'' choices, starting with the simplest possible case: a spherically symmetric spacetime in vacuum GR. 
			\subsubsection*{Vacuum: $\sigma-N$ parameterization in Schwarzschild coordinates}
	An often used ansatz for a spherically symmetric metric spacetime is \eqref{E1.5.40}
				\begin{equation}\label{E7.2.41}
				 ds^2 = -\sigma^2N dt^2 + \frac{dr^2}{N}+ r^2\big( d\theta^2+\sin^2\theta d\varphi^2\big) \ .
				\end{equation}
	This ansatz uses Schwarzschild-like coordinates, where $r$ is the areal radius, together with parameterizing functions $\sigma$ and $N$\footnote{We have already widely used this ansatz in all the previous sections (first introduced in Sec.~\ref{S1.6}).}. The EH action can then be re-expressed in terms of an EA $\mathcal{S}_{\rm EH}=\int dt\ze \mathcal{S}^{\rm eff}$, where
				\begin{equation}\label{E7.2.42}
				 \mathcal{S}^{\rm eff} = \int_{r_i} ^{+\infty} dr\  \sigma \ze r^2 R = -\int_{r_i} ^{+\infty} \Big\{ r \big[3 r N' \sigma '+2 N \left(r \sigma '' + 2 \sigma '\right)\big]+ \sigma \left(r^2 N''+4 r N'+2 N-2\right)\Big\} dr \ .
				\end{equation}
	A distinctive feature is that this action depends on the second derivatives of $\sigma,\, N$. The second derivative terms can be collected into a \textit{total derivative}, such that this EA is cast in the form~\eqref{E7.1.20} with
				\begin{equation}\label{E7.2.43}
				 \mathcal{L}(\sigma, N; \sigma',N'; r)=-2\ze \sigma \left(-1+N+rN'\right) \ , \qquad	 f(\sigma, N; \sigma',N'; r) = -2\ze r^2N\sigma'-r^2N'\sigma \ .
				\end{equation}
	Admitting the existence of an event horizon, we take $r_i$ in~\eqref{E7.1.20} to be $r_i=r_H$, such that $N(r_H)=0$. Then, the virial identity is readily obtained from~\eqref{E7.1.11}, yielding 
				\begin{equation}\label{E7.2.44}
				 2\int_{r_H}^{+\infty}\sigma \left[N-1+(r-r_H)N'\right]dr= \bigg[\left(2rN\sigma'+rN'\sigma\right)(2r_H-r)\bigg]^{+\infty}_{r_H}  \ . 
				\end{equation}
	A test on this identity is provided by the Schwarzschild solution, 
				\begin{equation}\label{E7.2.45}
				 N = 1-\frac{2M}{r} \ , \qquad\qquad \sigma = 1 \ ,
				\end{equation}
	with $M=c^{\rm te}$. Indeed, for these choices, both sides of~\eqref{E7.2.45} give $-4M$. Thus, the total derivative term in the EA, albeit not contributing to the equations of motion, gives a non-trivial contribution to the virial identity~\eqref{E7.2.44}. 

	Alternatively, we could have faced the EA~\eqref{E7.2.39} as being of the type of~\eqref{E7.1.16} with an effective Lagrangian depending also on second derivatives:
				\begin{equation}\label{E7.2.46}
				 \mathcal{L}(\sigma, N; \sigma',N';\sigma'',N''; r)= -  r \left[3 r N' \sigma '+2 N \left(r \sigma '' + 2 \sigma '\right)\right]- \sigma \left(r^2 N''+4 r N'+2 N-2\right) \ .
				\end{equation}
	Then, applying~\eqref{E7.1.16} yields an identity that is equivalent to~\eqref{E7.2.44}. This illustrates the equivalence observed between the virial identities~\eqref{E7.1.16} and~\eqref{E7.1.22} in concrete examples.

	Let us emphasise that, despite the non-trivial check provided by the Schwarzschild solution, the (would be) virial identity~\eqref{E7.2.44} is \textit{incomplete}. The correct version will be given below in \eqref{E7.2.64}.
				\subsubsection*{Vacuum: $\sigma-m$ parameterization in Schwarzschild coordinates}
	Virial identities depend not only on the choice of coordinates but also on metric functions. Such is sharply illustrated by reconsidering the metric ansatz of the previous subsection~\eqref{E7.2.41} but with a seemingly innocuous modification: taking as the parameterizing function the Misner-Sharp mass $m(r)$ function~\cite{misner1973gravitation}, instead of $N$, given by
					\begin{equation}\label{E7.2.47}
					 N = 1-\frac{2\ze m}{r} \ .
					\end{equation}
	In this case, the EA can be written as
					\begin{equation}\label{E7.2.48}
	 \mathcal{S}^{\rm eff} = 4\int _{r_i} ^{+\infty} \sigma m'dr + \int_{r_i} ^{+\infty}  \frac{d}{dr}\Big[2\sigma'r(2m-r)+2\sigma(m'r-m)\Big]dr \ .
					\end{equation}
	This EA is again of the form~\eqref{E7.1.20} with
					\begin{equation}\label{E7.2.49}
					 \mathcal{L}(\sigma, m; \sigma',m'; r)=4\ze\sigma  m' \ , \qquad	 f(\sigma, m; \sigma',m'; r) = 2\ze \sigma'r\ze (2\ze m-r)+2\ze \sigma(m'r-m) \ .
					\end{equation}
	Again, admitting the existence of an event horizon, we take $r_i$ in~\eqref{E7.1.20} to be $r_i=r_H$, such that $2\ze m(r_H)=r_H$ and applying~\eqref{E7.1.22} yields 
					\begin{equation}\label{E7.2.50}
					 \Big[-2\ze\sigma'(r^2+2\ze m\ze r_H-2\ze r\ze r_H)-2\ze \sigma\ze  m'\ze r_H\Big]_{r_H}^{+\infty} = 0 \ . 
					\end{equation}
	For Schwarzschild, $m = M$ and $\sigma=1$, and the identity is trivially satisfied.

	The peculiar feature of the (would be) virial identity~\eqref{E7.2.50} is the absence of the integral term; only the boundary term contributes. Which is a consequence of the EH action for this ansatz being invariant (up to a boundary term) under the scaling transformation~\eqref{E7.1.17}, which is manifest from the fact that the integrand (plus integration measure) of the first term in~\eqref{E7.2.48} is $\sigma\ze \frac{dm}{dr}dr$. Therefore, we learn, by example, that an appropriate choice of parameterization functions can simplify the virial identities by trivializing some terms. Thus, in spherical symmetry, the metric gauge~\eqref{E7.2.41} with the $\sigma - m$ parameterization functions~\eqref{E7.2.47} is the most straightforward choice for computing virial identities, which we shall therefore use in (most of) the following cases.

	Again, we emphasize that, despite the check of the Schwarzschild solution (which now is more trivial), the (would be) virial identity~\eqref{E7.2.50} is \textit{incomplete}. The correct version will be given below in \eqref{E7.2.67}.
			\subsubsection*{Electro-vacuum: an inconsistency}
	Our final example of this section will clarify that there is one key ingredient missing in the computation of virial identities for GR. We now consider spherically symmetric solutions in electro-vacuum. The action is~\eqref{E7.2.40} with 
				\begin{equation}\label{E7.2.51}
				 \mathcal{S}_{\rm m}^{\rm Maxwell}=-\frac{1}{4}\int d^4x \sqrt{-g} F_{\mu\nu}F^{\mu\nu} \ .
				\end{equation}
	Following the conclusion at the end of the last subsection we take the metric gauge~\eqref{E7.2.41} with the $\sigma - m$ parameterization functions~\eqref{E7.2.47}, and the ansatz for gauge potential 
				\begin{equation}\label{E7.2.52}
				 A_\mu = -V(r)dt \ .
				\end{equation}
	Defining the EA as $\mathcal{S}_{\rm EH}+\mathcal{S}_{\rm m}^{\rm Maxwell}=\int dt\, \mathcal{S}^{\rm eff}$, we find that the EA is again of the form~\eqref{E7.1.20} with
				\begin{equation}\label{E7.2.53}
				 \mathcal{L}(\sigma, m,V; \sigma',m',V'; r)=4\ze \sigma \ze m'+\frac{2\ze r^2 V'^{\, 2}}{\sigma} \ , \qquad	 f(\sigma, m; \sigma',m'; r) = 2\ze \sigma'\ze r\ze (2\ze m-r)+2\ze \sigma(m'r-m) \ .
				\end{equation}
	The difference with~\eqref{E7.2.49} is the extra term depending on $V'^{\, 2}$ in the effective Lagrangian. Applying~\eqref{E7.1.22}, the new identity becomes
				\begin{equation}\label{E7.2.54}
				 \int_{r_H}^{+\infty}\frac{r V'^{\, 2}}{\sigma}(2\ze r_H-r)=	 \Big[-\sigma'(r^2+2\ze m\ze r_H-2\ze r\ze r_H)-\sigma m' r_H\Big]_{r_H}^{+\infty } \ . 
				\end{equation}
	If \eqref{E7.2.54} were the correct virial identity, the RN solution, which has 
				\begin{equation}\label{E7.2.55}
				 m = M-\frac{Q_e ^2}{2r} \ , \qquad \qquad \sigma = 1 \ , \qquad \qquad V=-\frac{Q_e}{r} \ ,
				\end{equation}
	should verify it. However, whereas the \textit{lhs} of~\eqref{E7.2.54} vanishes, the \textit{rhs} gives
				\begin{equation}\label{E7.2.56}
				 -  m' r_H\Big|_{r_H}^{+\infty} = \frac{Q^2}{2\ze r_H}\neq 0 \ .
				\end{equation}
	The fact that \eqref{E7.2.54} is not satisfied with the RN solution means this is not the correct virial identity for the electro-vacuum model. 
	
	In the next section, we propose that the boundary term of the gravitational action is mandatory in the correct treatment of virial identities in GR. This boundary term is the GHY term. As we shall see, the contribution of such a term for the vacuum case turns out to be trivial for the Schwarzschild solution with the parameterizations discussed in this section. The aforementioned explains the accidental (and thus misleading) check provided by the Schwarzschild solution to the incomplete vacuum GR virial identities~\eqref{E7.2.44} and~\eqref{E7.2.50}. However, in the electro-vacuum case, the boundary term contributes the incomplete virial identity~\eqref{E7.2.54} which is non-trivial for the RN solution and which precisely makes it verify the correct virial identity, given below in \eqref{E7.2.68}.
%

%
		\subsection{GR in spherical symmetry - adding the missing GHY term}\label{S7.2.3}
%
	The GHY~\cite{york1972role,gibbons1993action,hawking1996gravitational,brown1993microcanonical} term is a surface term that is necessary for GR to have a well posed variational principle in a manifold with a boundary. In a BH spacetime case (such as the Schwarzschild spacetimes), there are boundaries at the horizon and at spatial infinity that, in principle, need to be considered. 

The complete gravitational action on a manifold $\mathcal{M}$, including the boundary term, is
			\begin{equation}\label{E7.2.57}
			 \mathcal{S}_{grav}= \mathcal{S}_{EH}+\mathcal{S}_{GHY} =\frac{1}{4} \int_\mathcal{M} d^4x \sqrt{-g}R + \frac{1}{2}\int_{\partial\mathcal{M}} d^3x \sqrt{-\gamma}\big( K-K_0\big)  \ ,
			\end{equation}
	where $K =\nabla_\mu n^\mu$ is the extrinsic curvature of the boundary $\partial\mathcal{M}$ with normal $n^\mu$, and $\gamma$ is the associated 3-metric of the boundary. The additional $K_0$ term corresponds to the extrinsic curvature in flat spacetime (the background metric), necessary to obtain a finite result.

	The GHY boundary term will give an extra total derivative to the EA. This will remain consistent with the vacuum case and fix the issue raised in the electro-vacuum case. In this section, we will compute it in the spherical case (the axial case will be dealt with in Sec.~\ref{S7.5}), under the parametrizations we have considered in Sec.~\ref{S7.2}.
			\subsubsection*{Vacuum: $\sigma-N$ parameterization in Schwarzschild coordinates}
	We consider the metric ansatz \eqref{E7.2.41} again. Assume the spacetime has a spherical surface boundary at a specific radius $r$ (like the spatial sections of the event horizon). Thus, the normal vector is $n = \sqrt{N}\,\partial_r$. Then 
				\begin{align}
	 			 \sqrt{-\gamma} &= \sigma\sqrt{N} r^2\sin\theta  \ ,\\
				 K&= \nabla_\mu n^\mu = \partial_r n^r + \frac{2}{r}n^r +\frac{\sigma'}{\sigma}n^r = \frac{1}{2}\frac{N'}{\sqrt N} + \left(\frac{2}{r}+\frac{\sigma'}{\sigma}\right)\sqrt{N} \ ,\\	
				 K_0&= \frac{2}{r} \ ,\\
				 \sqrt{-\gamma}\big( K-K_0\big) &= \left[\frac{r^2}{2}\sigma N'  + 2\ze r\ze \sigma \big( N-\sqrt{N}\,\big) +  r^2\sigma'N\right]\sin\theta \ .
				\end{align}
	Defining as before an EA contribution for the GHY term, $ \mathcal{S}_{grav}=(4\pi)^{-1}\int dt\, \mathcal{S}^{\rm eff}$, we obtain an EA as in~\eqref{E7.1.20} with an \textit{extra} total derivative, defined by 
				\begin{equation}\label{E7.2.62}
				 f^{GHY} = r^2\sigma N' + 4\ze r\ze \sigma \big( N-\sqrt{N}\, \big) + 2\ze r^2\sigma'N \ .
				\end{equation}
	Comparing with~\eqref{E7.2.46}, the old $f$ cancels out completely. This removes the second derivatives from the complete EA (precisely the goal of the boundary term in this case), which remains of the form~\eqref{E7.1.20} with
				\begin{equation}\label{E7.2.63}
				 \mathcal{L}(\sigma, N; \sigma',N'; r)=-2\ze \sigma \left(-1+N+rN'\right) \ , \qquad	 f(\sigma, N; \sigma',N'; r) =  4\ze r\ze \sigma\big( N-\sqrt{N}\,\big)  \ .
				\end{equation}
	Then, the virial identity obtained from~\eqref{E7.1.22} is
				\begin{equation}\label{E7.2.64}
				 2\int_{r_H}^{+\infty}\sigma \Big[N-1+(r-r_H)N'\Big]dr= \bigg[4\ze \sigma\ze\big( N-\sqrt{N}\,\big) (r-r_H)\bigg]^{+\infty}_{r_H}  \ . \qquad  {\rm {\bf [Virial \  vacuum \ GR \ \sigma-N ]}}
				\end{equation}
	This is the complete virial identity for vacuum GR in the $\sigma-N$ parameterization $\big($correcting \eqref{E7.2.44}$\big)$. One can check that the Schwarzschild solution~\eqref{E7.2.45} still obeys it. The \textit{lhs} remains unchanged whereas the \textit{rhs} still gives $-4M$ (which now comes from the limit at $r=+\infty$).
			\subsubsection*{Vacuum: $\sigma-m$ parameterization in Schwarzschild coordinates}
	For the $\sigma-m$ parameterization, on the other hand, where $N$ is replaced by $m$ via~\eqref{E7.2.47}, the extra total derivative from the GHY boundary term is
				\begin{align}
				 f^{GHY}  &= 2\ze r\ze \sigma'(r-2\ze m)-2\ze \sigma \left[m'\ze r+2\ze r\ze  \sqrt{1-\frac{2\ze  m}{r}}-2\ze r+3 m\right] \ .\
				\end{align}
	Adding this contribution to old $f$ in~\eqref{E7.2.53} cancels out the second derivatives in the complete EA, which remains of the form~\eqref{E7.1.20} with
				\begin{equation}\label{E7.2.66}
				 \mathcal{L}(\sigma, m; \sigma',m'; r)=4 \ze \sigma\ze  m' \ , \qquad	 f(\sigma, m; \sigma',m'; r) = -4\ze \sigma\ze \left[ \sqrt{r^2-{2\ze  m\ze r}}-r+2 \ze m\right]  \ .
				\end{equation}
	The virial identity obtained from~\eqref{E7.1.22} is then
				\begin{equation}\label{E7.2.67}
				 \bigg[-4\ze \sigma\left(\frac{r-m}{\sqrt{r^2-2\ze m\ze r}}-1\right) (r-r_H)\bigg]^{+\infty}_{r_H}=0  \ . \qquad  {\rm {\bf [Virial \  vacuum \ GR \ \sigma-m ]}}
				\end{equation}
	One can check that for the Schwarzschild solution ($\sigma=1$, $m=M=c^{\rm te}$) this is obeyed (considering carefully the $r=+\infty$ limit). Thus, this is the complete virial identity for vacuum GR in the $\sigma-m$ parameterization $\big($correcting~\eqref{E7.2.50}$\big)$. 

			\subsubsection*{Electro-vacuum: solving the inconsistency}
	From the results in the previous section we can straightforwardly put together the virial identity for the electro-vacuum case to be
				\begin{align}\label{E7.2.68}
				 &\int_{r_H}^{+\infty}\frac{r V'^{\, 2}}{\sigma}(2\ze r_H-r)=	 \bigg[-2\ze \sigma\left(\frac{r-m}{\sqrt{r^2-2\ze m\ze r}}-1\right) (r-r_H)\bigg]^{+\infty}_{r_H} \ .\nonumber \\
&\qquad \qquad\qquad \qquad \qquad\qquad\qquad\qquad \qquad {\rm {\bf [Virial \   electro-vacuum \ GR \ \sigma-m ]}}
				\end{align}
	It is now simple to check that the RN solution~\eqref{E7.2.55} verifies this virial identity (both \textit{lhs} and \textit{rhs} vanish).
                                                                                                                                                        
%
	\section{GR in spherical symmetry ($\sigma-m$ parameterization): illustrations}\label{S7.3}
%
	Being in control of the correct methodology, we shall now compute the virial identity for different models. We shall always use the metric ansatz~\eqref{E7.2.41} with the $\sigma-m$ parameterization~\eqref{E7.2.47}. The gravitational part of the action is given by $ \mathcal{S}_{grav}$, \eqref{E7.2.57}. The corresponding contribution to the virial identity is~\eqref{E7.2.67}. This boundary term does not contribute to the matter models to be considered here. This is a consequence of the behaviour of $m$ and $\sigma$ at infinity and at the origin/horizon, depending on whether we consider solitonic solutions or BHs. At infinity, these models have the asymptotic behaviour
			\begin{equation}\label{E7.2.69}
			 \sigma=1+\mathcal{O}\left(\frac{1}{r}\right) \ , \qquad m=M+\mathcal{O}\left(\frac{1}{r}\right) \ .
			\end{equation}
	A careful analysis of the $r\rightarrow +\infty$ limit of~\eqref{E7.2.67} shows it does not contribute. For the lower limit of~\eqref{E7.2.67}, the models we consider have the following behaviour close to the horizon
			\begin{equation}\label{E7.2.70}
			 \sigma=\sigma_H+\mathcal{O}\left(r-r_H\right)\ , \qquad m=\frac{r_H}{2}+\mathcal{O}\left(r-r_H\right) \ ,
			\end{equation}
	we can see that the limit will be proportional to $(r-r_H)^{1/2}$, rendering the horizon 
contribution zero; for solitons, at the origin,
			\begin{equation}\label{E7.2.71}
			 \sigma=\sigma_0+\mathcal{O}\left(r^{n_1}\right) \ , \qquad m=\mathcal{O}\left({r^{n_2}}\right) \ , 
			\end{equation}
	where $n_1$ and $n_2$ are model dependent but typically greater than $1$ (for example, $n_2=3$ for all models discussed in this section). The latter implies that the $r=0$ contribution also vanishes. Thus, the virial identity comes solely from the matter action. This illustrates how the correct choice of parameterizing functions simplifies the computation of virial identities.

	In all cases in this section, we end up with an EA of the type~\eqref{E7.1.8} with an effective Lagrangian
			\begin{equation}\label{E7.2.72}
			 \mathcal{L}(\sigma,m,X;\sigma',m',X';r) \ ,
			\end{equation}
	where $X$ denotes collectively the parameterizing functions coming from the matter sector. The corresponding virial identity is then computed from~\eqref{E7.1.13}.
	
	For all models discussed in this section, we have solved the field equations numerically and evaluated the displayed virial identities for a large sample of solutions in each case. Although the relative errors depend on the values of various input parameters, they are typical of order $10^{-5}$ or smaller. An explicit illustration of this sort of numerical checking is provided in Sec.~\ref{S7.4.3}.

%
		\subsection{Solitonic solutions}\label{S7.3.1}
%
Let us start by considering solitonic solutions, thus without an event horizon ($r_i=r_H=0$).
		\subsubsection*{Scalar boson stars}
	Scalar boson stars~\cite{kaup1968klein,merafina1989systems,ruffini1969systems} are self-gravitating lumps of a complex, massive scalar field -- see also~\cite{schunck2003general,liebling2017dynamical,herdeiro2017asymptotically,herdeiro2020asymptotically} and Ch.~\ref{C4} and \ref{C5}. They mimic $Q$-balls in their harmonic time-dependence. In spherical symmetry, they are described by the same scalar field ansatz as $Q$-balls~\eqref{E7.2.31}. Nevertheless, unlike the latter, they do not require a self-interacting scalar field; GR provides the necessary non-linearities.  

	Consider the action that describes the self-gravitating complex scalar field, using the ansatz \eqref{E7.2.31} in a model with a self-interactions potential $U(\Phi)$
			\begin{equation}\label{E7.3.73}
			 \mathcal{S}= \mathcal{S}_{grav}+\mathcal{S}_{\rm m}^{\bar{\Phi}} \ ,	 
			\end{equation}
	where the latter action is explicitly given by~\eqref{E7.2.31}. The resulting effective matter Lagrangian is,
			\begin{equation}
			 \mathcal{L} (\sigma, m, \phi; \sigma',m', \phi '; r) = r^2  \sigma \left[ \frac{r \ze \omega ^2 \phi ^2}{(r-2\ze m) \sigma ^2}-\left(1-\frac{2\ze m}{r}\right) \phi'^{\,2} +U(|\phi|)\right] \ .
			\end{equation}
	Then, the virial identity reads
			\begin{equation}\label{E7.3.75}
			 \int _0 ^{+\infty} dr \,  r ^2 \sigma \left[-\frac{r\ze \omega^2 \phi ^2}{\sigma ^2}\frac{3\ze r-8\ze m}{(r-2\ze m )^2}+\phi'^{\,2} +  3   \, U \right] = 0  \ . \qquad {\rm {\bf [virial  \ scalar \ boson \ stars]}}
			\end{equation}
	For $m=0$, $\sigma=1$, this reduces to the $Q$-balls virial identity~\eqref{E7.2.29}. Eq.~\eqref{E7.3.75} allows an immediate conclusion: if $\omega=0$ and the potential $U$ is everywhere non-negative, the identity can never be respected, leading to a \textit{no-go} theorem~\cite{heusler1996no}. Thus gravity is not enough to circumvent Derrick's theorem; even with gravity, a finite oscillation frequency $\omega$ is necessary to have self-gravitating scalar solitons (with a time-independent spacetime). We will see in Sec.~\ref{S7.3.2} a special case: a matter model for which no solitons exist in flat spacetime but where the coupling to Einstein's gravity makes them possible.
		\subsubsection*{Dirac stars}
	 Einstein's gravity minimally coupled with spin $\frac{1}{2}$ fields allows the existence of self-gravitating solitons~\cite{finster1999particlelike}. These solitons are also known as \textit{Dirac Stars} -- see also~\cite{herdeiro2020asymptotically,dolan2015bound,herdeiro2017asymptotically}. The corresponding action is 
			\begin{equation}
			 \mathcal{S} = \mathcal{S}_{grav}-\frac{i}{4} \int d^4 x \sqrt{-g} \Bigg[ 2\Big(\psi ^{[A]} \overline{\hat{\slashed D}} \overline{\psi}^{[A]}-\overline{\psi} ^{[A]} \hat{\slashed D} \psi ^{[A]} \Big)+U(|\Psi|)\Bigg]\ ,  
			\end{equation}
	where $\psi$ is a Dirac $4$-spinor, with four complex components, the index $[A]$ corresponds to the number of copies of the Lagrangian. For a spherically symmetric configuration, one should consider, at least, two spinors with an equal potential $U(\Psi)$; a single spinner will necessarily make the soliton rotate, yielding a stationary axially symmetric spacetime~\cite{herdeiro2019asymptotically}, rather than a spherical, static spacetime. The ``slashed'' derivative is $\hat{\slashed D}\equiv \gamma ^\mu \hat{D} _\mu$, where $\gamma ^\mu$ are the curved space gamma matrices and $\hat{D}=\partial _\mu +\Gamma _\mu $ is the spinorial covariant derivative, with $\Gamma _\mu $ being the spin connection matrices.

	For the Dirac field, the matter ansatz introduces two real functions, $h(r)$ and $j(r)$ where $z (r) \equiv (1+i) h +(1-i) j$ 
			\begin{align}
			 &&
			 \psi^{[1]} = \begin{pmatrix} 
			 \cos(\frac{\theta}{2})\, z
			 \\ 
			 i \sin(\frac{\theta}{2})\, \bar z
			 \\ 
			 -i \cos(\frac{\theta}{2})\, \bar z
			 \\ 
			 -\sin(\frac{\theta}{2})\,  z
			 \end{pmatrix}
			 e^{i\big(\frac{1}{2}\varphi-\omega\ze t\big) } \ , \qquad
			 \label{down}
			 \psi^{[2]} = \begin{pmatrix} 
			 i \sin(\frac{\theta}{2})\, z
			 \\ 
			 \cos(\frac{\theta}{2}) \,\bar z
			 \\ 
			 \sin(\frac{\theta}{2}) \,\bar z
		 	 \\ 
			 i \cos(\frac{\theta}{2}) \,  z
			 \end{pmatrix}
			 e^{i\big( -\frac{1}{2}\varphi-\omega\ze t\big) } \ ,
			\end{align}
	and $\Psi =i \overline{\psi}^{[A]}\psi^{[A]} = 4(h^2 - j^2 )$. The effective matter Lagrangian is
			\begin{align}
			 \mathcal{L} (\sigma, m, h,j; \sigma',m', h',j'; r) = r^2 \sigma  \left[ \sqrt{1-\frac{2 m}{r}} \left(j h'-h j' \right)-\frac{\omega  \left(h ^2+j^2\right)}{\sqrt{1-\frac{2 m}{r}} \sigma }+\frac{2 hj }{r}+\frac{U}{4}\right]\ .
			\end{align}
	The resulting virial identity\footnote{Here and in some other cases below, the identity is expressed in terms of $N$, rather than $m$, for compactness, although the computation is made with the  $\sigma-m$ parameterization.}
			\begin{align}\label{E7.3.79}
			 &\int _0 ^{+\infty} dr\ \frac{r^2 \sigma }{\sqrt{N} } \Bigg[(3 N+1) \left(j h'-h j'\right)+\frac{\omega  \left(h^2+j^2\right)}{\sigma }\left(\frac{1}{N}-7\right)+\left(\frac{ 8 hj}{r}+\frac{3}{2}  U\right) \sqrt{N} \Bigg]= 0 \ . \nonumber\\
			 &\qquad\qquad \qquad\qquad\qquad\qquad\qquad\qquad \qquad\qquad\qquad\qquad \quad {\rm {\bf [virial \ Dirac \ stars]}}
			\end{align}
	Unlike the scalar case, this identity does not provide any clear indication for the mechanism allowing the existence of solutions. However, in the flat spacetime limit, \eqref{E7.3.79} reduces to
			\begin{equation}
			 \int _0 ^{+\infty} dr\ r^2 \bigg[\big( j h'-hj'\big)+\frac{2 h j}{r}-\frac{3 }{2} \omega (h^2 +j ^2)+\frac{3}{8}U\bigg] = 0 \ ,
			\end{equation}
	which can be further simplified through the field equations to yield
			\begin{equation}
			 \int _0 ^{+\infty } dr \ r^2 U =\int _0 ^{+\infty} dr\ r^2 \Big[4\ze \omega \big( h^2+j^2\big) \Big]\ .
			\end{equation}
	Then, one observes that for strictly positive potential, $U>0$, the solutions are supported by the harmonic time-dependence, with $\omega>0$.
			\subsubsection*{Vector boson stars (Proca stars)}
	Spherical vector boson stars, \textit{a.k.a. Proca stars}~\cite{brito2016proca} (see also~\cite{landea2016charged,duarte2016asymptotically,herdeiro2017asymptotically,minamitsuji2018vector, herdeiro2021imitation}), can be found in GR minimally coupled to complex, massive vector fields. The model's action comes as
			\begin{equation}
			 \mathcal{S}= \mathcal{S}_{grav}-\frac{1}{4} \int d^4 x \sqrt{-g} \Big[ G_{\mu \nu} \bar{G}^{\mu \nu} +U(\textbf{B})\Big]\ .
			\end{equation}
	where the complex vector field's ansatz is 
			\begin{equation}
			B_\mu =\big[ B_t(r) dt+i B_r(r) dr \big] e^{-i\omega\ze t} \ .
			\end{equation}
	The vector field  is under a self-interacting potential $U$, where $\textbf{B}\equiv B_\mu \bar{B}^{\mu}$. One obtains the effective matter Lagrangian
			\begin{equation}\label{E7.3.84}
			 \mathcal{L} (\sigma, m, B_t,B_r; \sigma',m', B_t ',B_r '; r) = \frac{r^2}{\sigma}\left[ - \left(B_t '-\omega  B_r \right)^2 +\sigma ^2 U\right]\ .
			\end{equation}
	The resulting virial identity is 
			\begin{align}\label{E7.3.85}
			 &\int _0 ^{+\infty} dr \ \frac{r^2}{\sigma} \bigg[-\big( \omega B_r - B_t '\big) \big(3\ze\omega B_r - B_t '\big)+3\ze \sigma ^2 U+ \frac{1-N}{N^2} \hat{U} \big(\sigma ^2 N^2 B_r ^2 + B_t ^2\big) \bigg] =0\ . \nonumber\\
			 &\qquad\qquad \qquad\qquad\qquad\qquad\qquad\qquad \qquad\qquad\qquad\qquad \quad{\rm {\bf [virial \ Proca \ stars]}}
			\end{align}
	This identity reduces to the one in~\cite{brito2016proca} for a massive, free complex vector field. In the absence of self-interactions, the above relation can be used to rule out non-gravitating solutions.
			\subsubsection*{Einstein-Maxwell-Scalar (EMS) solitons} \label{PLEMS}
	The EMS model is described by the action (already presented in Ch.~\ref{C2})
				\begin{equation}\label{E7.3.86}
				 \mathcal{S} _{EMS}= \mathcal{S}_{grav}+ \frac{1}{4} \int d^4 x \sqrt{-g} \Big[-2\ze \phi_{,\mu} \phi ^{,\mu} - f(\phi ) F_{\mu\nu}F^{\mu \nu} - U(\phi )\Big] .
				\end{equation}
	In this model $F_{\mu \nu}$ is the Maxwell tensor and $\phi $ is a real scalar field that is non-minimally coupled to the Maxwell term through the coupling function $f $. Moreover, we admit a self-interactions potential $U$ for the scalar field. Particle-like soliton configurations were found in \cite{herdeiro2020class} (see also~\cite{herdeiro2019inexistence}). These configurations have a scalar field that depends only on the radial coordinate, $\phi \equiv \phi(r)$.

	For an electric $4$-vector potential, $A_\mu = V dt$, the resulting effective matter Lagrangian is
				\begin{equation}\label{E7.3.87}
				 \mathcal{L} (\sigma, m, \phi ; \sigma',m', \phi '; r) =r^2 \sigma \left[f\ze\frac{2V'^{\,2}}{ \sigma ^2}- \left(1-\frac{2 m}{r}\right)   \phi '^{\,2} -U\right]\ .
				\end{equation}
	A first integral is obtained from the field equations, that simplifies the EA, namely,
				\begin{equation}\label{E7.3.88}
				 V' =-\frac{Q_e}{r^2 \varepsilon_\phi}\ ,
				\end{equation}
	where we recall that $\varepsilon _\phi = f\ze \sigma ^{-1}$ can be thought as a relative electric permittivity that is caused by the non-minimal coupling between the scalar and Maxwell fields.	
	
	Replacing the first integral into the Maxwell term, the resulting virial identity is
				\begin{equation}\label{E7.3.89}
				 \int _0 ^{+\infty} dr \left[ r ^2 \sigma\ze \phi'^{\,2} +3 r ^2 \sigma \, U - 2\ze \frac{Q_e ^2}{r^2 \varepsilon _\phi}  \right] = 0 \ . \qquad {\rm {\bf [virial \ EMS \ solitons]}}
				\end{equation}
	The virial identity informs us that particle-like solution can be supported by the electric charge or a negative potential.

%
		\subsection{Black holes}\label{S7.3.2}
%
	As already mentioned in the chapter's introduction (Ch.~\ref{C7}), virial theorems can be used to establish no-hair theorems for BHs (see~\cite{herdeiro2015asymptotically} for a review). Heusler and Straumann obtained virial identities with that goal in \cite{heusler1996no} and \cite{heusler1992scaling} for the Einstein-Klein-Gordon model (that we shall refer to as \textit{scalar-vacuum} -- Sec.~\ref{S7.3.2}) and Einstein-Yang-Mills model (Sec.~\ref{S7.3.2}). In order to consider BHs, in this subsection we take $r_i=r_H\neq 0$.
			\subsubsection*{No scalar hair theorem}
	The virial identity obtained for the model defined by~\eqref{E7.3.73} can be generalized to include a putative horizon scale $r_H$. Using a scalar field ansatz with a harmonic time-dependence~\eqref{E7.2.31} one obtains\footnote{We remark that there is a factor of $1/2$ difference as compared to (46) in~\cite{herdeiro2015asymptotically}, which comes from a different action normalization.}
				\begin{align}\label{E7.3.90}
				 &\int_{r_H}^{+\infty }dr~
	 			 \bigg\{\sigma\left[\left( \frac{2\ze r_H}{r}\left(1-\frac{m}{r}\right)-1
\right)r^2\phi'^{\, 2}+\left(\frac{2\ze r_H}{r}-3\right)r^2U\right]+\nonumber\\
	 			 & \frac{1}{\sigma}\bigg[\frac{3\ze (r-r_H)(r-2\ze m)+r(r_H-2\ze m)}{(r-2\ze m)^2}\bigg]\omega^2r^2\phi^2 
\bigg\} =0 \ .\qquad {\rm {\bf [virial \ scalar \ vacuum]}}
				\end{align}
	Putting $r_H=0$ we recover~\eqref{E7.3.75}. On the other hand, putting $\omega=0$ one keeps only the second line. For this special case, inspection shows that the pre-factor of $U$ and the first term (in the second line) are negative for $r>r_H$. This establishes a no-hair theorem for this model with $\omega=0$~\cite{heusler1996no}. This virial identity is not enough, however, to establish a no-hair theorem for $\omega\neq 0$, albeit such theorem can be established using other methods~\cite{pena1997collapsed,graham2014stationary}.
			\subsubsection*{EMS BHs}
	Let us reconsider the EMS model \cite{blazquez2021quasinormal,blazquez2020einstein,fernandes2019spontaneous,herdeiro2018spontaneous} described by the action \eqref{E7.3.86} (see also Ch.~\ref{C2} for a full treatment), but now taking into account the presence of an event horizon. Then, the virial identity reads
			\begin{equation}\label{E7.3.91}
			 \int _{r_H} ^{+\infty} dr\  \bigg[I_\Phi (0, r_H)+ I_U ^{[\Phi]} (r_H)-f I_M (r_H) \bigg]\ =0 \  . \qquad {\rm {\bf [virial \ EMS \ BHs]}}
			\end{equation}
	where the scalar terms are	
			\begin{align}\label{E7.3.92}
			 & I_\Phi(\omega,r_H) = \frac{1}{\sigma}\bigg[\frac{3\ze (r-r_H)(r-2\ze m)+r(r_H-2\ze m)}{(r-2\ze m)^2}\bigg]\omega^2\ze r^2\phi^2 + 
			 \sigma\left( \frac{2\ze r_H}{r}\left(1-\frac{m}{r}\right)-1
\right)r^2\phi'^{\, 2} \ ,\\
			& I_U ^{[\Phi ]} (r_H) = r\ze U (2\ze r_H-3\ze r)\sigma \ ,\label{E7.3.93}
			\end{align} 
	whereas the Maxwell term reads
			\begin{equation}
		 	 I_M (r_H) = 2\frac{ (2\ze r_H-r) Q_e ^2 }{\varepsilon _\phi ^2 r^3 \sigma } \ .
			\end{equation}
	As expected~\eqref{E7.3.91} reduces to \eqref{E7.3.89} when $r_H=0$. The identity~\eqref{E7.3.91}  tells us that a non-trivial scalar hair requires a non-zero electric charge. Indeed, as mentioned in the previous subsection,  $I_\Phi (0, r_H)<0$ outside the horizon; furthermore (since $\sigma>0$) for a non-negative potential  $I_U ^{[\Phi]}$ is non-positive outside the horizon; thus the positive contribution must come from the Maxwell term. Observe that when $Q_e=0$, and replacing $I_\Phi (0, r_H)\rightarrow I_\Phi (\omega, r_H)$, then~\eqref{E7.3.91} becomes~\eqref{E7.3.90}.
			\subsubsection*{Einstein-Maxwell-Axion (EMA) BHs}
	The EMA model with an axionic-like coupling~\cite{fernandes2019charged} (see also Sec.~\ref{S2.4}) is described by the action 
				\begin{equation}
				 \mathcal{S}_{EMA} = \mathcal{S}_{grav}+ \frac{1}{4} \int d^4 x \sqrt{-g} \bigg[-2\ze \phi_{,\mu}\phi ^{,\mu} -f(\phi ) F_{\mu\nu}F^{\mu \nu} - h(\phi) F_{\mu \nu} \tilde{F}^{\mu \nu}- U(\phi)\bigg] ,
				\end{equation}
	where $\tilde{F}^{\mu \nu} = \frac{\epsilon ^{\mu \nu \rho \delta} F_{\rho \delta}}{2\sqrt{-g}} $, $\epsilon ^{\mu \nu \rho \delta}$ is the Levi-Civita tensor density and $h(\phi)$ is an additional coupling function. 
				\begin{equation}
				 \mathcal{L} (\sigma, m, \phi ; \sigma',m', \phi '; r) = 4\ze f \left(\frac{r^2 V'^{\,2}}{\sigma }-\frac{P^2 \sigma }{r^2}\right)-r (r-2\ze m) \sigma\ze \phi '^{\,2}+2 P\ze h\ze V' - r^2 \sigma\ze U\ .
				\end{equation}
	A first integral can be obtained from the equations of motion
				\begin{equation}
				 V' = -\frac{Q_e+P \ze h}{r^2 \varepsilon _\phi}\ .
				\end{equation}
	Then, the virial identity for this model is 
				\begin{equation}
				 \int _0 ^{+\infty} dr \ \bigg[ I_\Phi (0,r_H) +I_U ^{[\phi]} (r_H)-f\ze I_M (r_H,\ze h) \bigg] =0 \ ,  \qquad {\rm {\bf [Virial \ EMA \ BHs]}}
				\end{equation}
	where
				\begin{equation}
				 I_M (r_H,\, h) = \frac{4}{r^3 \varepsilon _ \phi ^2 \sigma } \big( 2\ze r_H -r\big) \big[ (Q_e -P\ze h )^2+P^2 \varepsilon _\phi ^2 \ze\sigma ^2 \big]\ .
				\end{equation}

			\subsubsection*{Einstein-Maxwell-Vector (EMV) BHs}
	The EMV model \cite{oliveira2021spontaneous,fan2016black} (see also Sec.~\ref{S2.5}) is described by the action
				\begin{equation}\label{E7.3.95}
				 \mathcal{S} _{EMV} = \mathcal{S}_{grav}+\frac{1}{4} \int d^4 x \sqrt{-g} \Big[-G_{\mu \nu}G^{\mu \nu} - f(\textbf{B} ) F_{\mu\nu}F^{\mu \nu} - U(\textbf{B})\Big]\ ,
				\end{equation}
	where $B_\mu $ is a real vector field that is non-minimally coupled to the Maxwell term $F_{\mu \nu} F^{\mu \nu}$ through the coupling function $f $, for which self-interactions (and a mass term) are described by the potential $U$. For the vector field we consider, following \cite{oliveira2021spontaneous} and Sec.~\ref{S2.5}, a time-independent vector field  ansatz, $B_\mu= B_t (r) dt$. Assuming a purely electric field, the effective matter Lagrangian becomes
				\begin{equation}\label{E7.3.96}
				 \mathcal{L} (\sigma, m, B_t ; \sigma',m', B_t '; r) = \frac{r^2}{\sigma} \big[ -B_t'^{\,2}-f\ze V'^{\,2}-\sigma ^2\ze U\big] \ .
				\end{equation}
	Then, using the electromagnetic equation of motion to obtain a first integral (the charge $Q_e$),
				\begin{equation}
			 	 \nabla_\mu(f F^{\mu\nu}) = 0 \Rightarrow V' =- \frac{Q_e\ze \sigma }{r^2 f} \ ,
				\end{equation}
	the corresponding virial identity becomes
				\begin{align}
				 \int _{r_H} ^{+ \infty} dr  \ \, \bigg[ \frac{r-r_H}{r}\frac{N-1}{\sigma N^2} \frac{ Q_e ^2 B_t ^2}{ r^2f ^2 }\hat{f}+\frac{r\ze (2\ze r_H -r)}{\sigma}\left( B_t'^{\, 2}+\frac{Q_e ^2 \sigma ^2}{r^4f}\right)  - I_U^{[B]} (r_H) \bigg]=0\ , \nonumber\\
				 {\rm {\bf [virial \ EMV \ BHs]}}
				\end{align}
	where $I_U ^{[B]}$ corresponds to the contribution from the potential of the vector field
				\begin{equation}
			 	 I_U ^{[B]} = r^2\sigma\Big[3\ze U-\frac{r-r_H}{r}\frac{N-1}{\sigma N^2}\hat{U}B_t^2\Big] \ .
				\end{equation}
	For flat spacetime and $U=0$ this reduces to
				\begin{equation}\label{E7.3.100}
				 \int_{0}^{+\infty} dr\frac{1}{r^2} \left(r^4 B_t'^{\, 2}+\frac{Q_e^2}{f}\right) = 0 \ .
				\end{equation}
	If $f>0$, the virial identity~\eqref{E7.3.100} informs us that only the trivial configuration $B'_t=0$ and $Q_e=0$ is possible. In this case, of course, $B_\mu$ also became a gauge field (since the mass term vanished).
			\subsubsection*{Einstein-Yang-Mills (EYM) BHs and solitons}
	Yang-Mills theories \cite{yang1954conservation} (see also Sec.~\ref{S3.5.2} for a brief discussion) are gauge theories based on non-Abelian Lie groups. These theories are at the core of the standard model of particle physics. Minimally coupling these ``matter'' models to Einstein's gravity leads to EYM theories described by the action
				\begin{equation}\label{E7.3.101}
				 \mathcal{S} _{EYM}=\mathcal{S}_{grav}-\frac{1}{2} \int d^4 x\  \sqrt{-g}\, {\rm Tr}(F^2)\ .
				\end{equation}
	As an illustration of the role of virial identities in EYM models, let us follow the work done by Heusler~\cite{heusler1996no}. One considers the purely magnetic $\textbf{SU}(2)$ configuration with the gauge potential $1$-form $A$ 
				\begin{equation}
				 A=[p(r)-1](\tau _\varphi\ze d\theta -\tau_\theta\sin \theta\ze d\varphi )\ .
				\end{equation}
	The usual basis of $\textbf{SU}(2)$ is denoted as $(\tau_r,\tau_\theta,\tau_\varphi)$~\cite{bartnik1988particlelike}; also $\tau _\theta \equiv \tau _{r\, ,\theta} $, $\tau _\varphi \sin \theta \equiv \tau _{r\, ,\varphi}$ and $\tau _r \equiv (2 i |\overrightarrow{r}|)^{-1} (\overrightarrow{r}, \overrightarrow{\delta})$; $p(r)$ is an unkown radial function, determined by solving the field equations. The effective matter Lagrangian is
				\begin{equation}\label{E7.3.103}
				 \mathcal{L} (\sigma, m, p ; \sigma',m', p'; r) = \sigma \left[\frac{1}{2}\left(1-\frac{2\ze m}{r}\right)p'^{\,2} +\frac{(1-p^2)^2}{4r^2}\right]\ .
				\end{equation}
	The virial identity in the presence of an event horizon is
				\begin{equation}
				 \int _{r_H} ^{+\infty} dr \ I_{YM} (r_H) =0\ , \qquad\qquad {\rm {\bf [Virial \ EYM]}}
				\end{equation}
	where the Yang-Mills term is
				\begin{equation}\label{E7.3.105}
				 I_{YM} = \frac{\sigma}{2} \left\{\left[1+\frac{2 m}{r}\left(\frac{r_H}{r}-2\right)\right]N p'^{\, 2}+\left[1-\frac{2r_H}{r}\frac{(1-p^2)^2}{2r^2}\right]\right\}\ .
				\end{equation}
	In the presence of a horizon, the virial identity does not exclude the existence of BHs with hair. In fact these BHs exist~\cite{volkov1989non,volkov1990black,bizon1990colored,kunzle1990spherically} and were an influential counter-example to the no-hair conjecture~\cite{bizon1994gravitating,volkov1999gravitating}. The same occurs when $r_H\rightarrow 0$: the virial identity allows the existence of self-gravitating solitonic objects. These solitons exist, as first pointed out by Barnik and Mckinnon~\cite{bartnik1988particlelike}. However, in the absence of gravity
	\begin{equation}
	 \int _0 ^{+\infty} dr \Bigg[ \frac{p'^{\, 2}}{2}+\frac{(1-p^2)^2}{4r^2}\Bigg] =0 \ ,
	\end{equation}
	which shows that no flat spacetime Yang-Mills solitons exist. So, in this case, the coupling of the Yang-Mills source to Einstein's gravity is enough to allow particle-like solutions, which are forbidden in flat spacetime.
			\subsubsection*{Einstein-Maxwell-gauged scalar (EMgS) BHs}
	A \textit{gauged} complex scalar field minimally coupled to both the electromagnetic field and Einstein's gravity is described by the action 
				\begin{equation}\label{E7.3.107}
				 \mathcal{S} _{EMgS}= \mathcal{S}_{grav}+\frac{1}{4} \int \sqrt{-g} \bigg[-F_{\mu \nu } F^{\mu \nu} -4 g^{\mu\nu} D_{(\mu} \Phi \bar{D}_{\nu)} \bar{\Phi} - U(|\Phi|)\bigg] \ ,
				\end{equation}
 	where $D_\mu = \partial_\mu -ieA_\mu$ is the covariant gauge derivative. In this case, the global $\textbf{U}(1)$ symmetry of the scalar field is gauged. Charged (gauged) boson stars in this model have been discussed in~\cite{jetzer1989charged,pugliese2013charged}. Hairy BHs in this class of models (with self-interactions) are also possible and discussed in~\cite{herdeiro2020spherical,hong2020spherically}.

 	For a purely electric spherical configuration~\eqref{E7.2.52} and a scalar field with a harmonic time-dependence~\eqref{E7.2.31}, we get the following effective matter Lagrangian
 				\begin{equation}
				 \mathcal{L} (\sigma, m, \phi, V; \sigma',m', \phi ',V'; r) = r^2  \sigma \left[
			\left(1-\frac{2 m}{r}\right) \phi'^{\, 2} +U-\frac{ (\omega-e V)^2 \phi^2}{(1-\frac{2\ze m}{r})\sigma^2}-\frac{V'^{\, 2}}{2\ze\sigma^2}\right]\ .
 				\end{equation}
	Then the corresponding virial identity for BH solutions reads \cite{herdeiro2020spherical}
				\begin{align}
				 & \int_{r_H}^{+\infty} dr~  r^2 \left\{\left(1-\frac{2\ze r_H}{r}\right)\frac{V'^{\, 2}}{2\ze \sigma}+\left[
3-\frac{2r_H}{r}\left(1-\frac{3\ze m}{r}\right)-\frac{8\ze m}{r}\right]\frac{(\omega-e V)^2 \phi^2}{N^2 \sigma}
\right\}\nonumber \\
				 & = \int_{r_H}^{+\infty} dr~ r^2 \sigma \left\{\left[1-\frac{2\ze r_H}{r}\left(1-\frac{m}{r}\right) 
		 \right]\phi'^{\, 2} + \left(3-\frac{2\ze r_H}{r} \right)U
\right\}~, \nonumber\\
&\qquad\qquad \qquad\qquad\qquad\qquad \qquad\qquad\qquad\qquad \qquad\qquad {\rm {\bf [Virial \ EMgS]}}		\label{E7.3.109}
				\end{align}
	which reduces to \eqref{E7.3.75} for $e=V=0$ case. One notices that both factors in front of the scalar quantities on the $lhs$ have a fixed, positive sign, such that all this integral is strictly positive
(here we assume $U>0$). Therefore no solutions with $\phi \neq 0$ can exist for $V=0$ (no Maxwell field) and $\omega=0$. Also, the factors in front of the Maxwell quantities on the $rhs$ are indefinite (although they become positive asymptotically). Thus, for $V\neq 0$ and/or $\omega\neq 0$ a solution becomes possible (but not guaranteed). 
			\subsubsection*{Einstein-Maxwell-Gauss-Bonnet (EMGB)}
	The model that describes a charged, non-minimally coupled to the Gauss-Bonnet curvature term, real scalar field (see Ch.~\ref{C3}) that surrounds an electrically charged BH is described by the action,
				\begin{equation}\label{E7.3.110}
				 \mathcal{S}_{EMGB}=\mathcal{S}_{grav}+ \frac{1}{4} \int d^4 x \sqrt{-g} \Bigg[-F_{\mu \nu}F^{\mu \nu}+f(\phi) R_{GB} ^2-2\, \phi _{,\mu}\, \phi^{,\mu} +U(\phi)\Bigg]\ .
				\end{equation}
	Observe that, in spherical symmetry, the Gauss-Bonnet term is always the same since it only depends on metric functions. So, one can compute it \textit{a priori}.

	For a purely electric spherical configuration~\eqref{E7.2.52} and a scalar field with a harmonic time-dependence~\eqref{E7.2.31}, we get the following effective matter Lagrangian
				\begin{align}\label{E7.3.111}
				 \mathcal{L} & (\sigma, m, \phi ; \sigma',m', \phi '; r) = r^2 \sigma \bigg[\frac{2V'^{\, 2}}{ \sigma ^2}- \left(1-\frac{2 m}{r}\right)   \phi '^{\, 2} -U\nonumber\\
				 &+f\frac{4}{r^4}\Big( (-3+5 N) N' \sigma ' +\sigma \big(N^{' 2}+(-1+N)N''\big) +2(-1+N)N \sigma ''\Big)\bigg]\ .
				\end{align}
	The virial identity in the presence of an event horizon is
				\begin{equation}
				 \int _{r_H} ^{+\infty} dr \ \Big[I_\Phi (0,\, r_H)+ I_U ^{[\Phi]} (r_H)-I_M(r_H) +f\ze I_{GB} (r_H)\Big] =0\ ,  \qquad\qquad {\rm {\bf [Virial \ EMsGB]}}
				\end{equation}
	with the $I_{GB}$ given as
				\begin{align}\label{E7.3.114}
				 I_{GB} = \frac{4}{r^3}\Bigg[& r^2 N' \Big(2 \big(r_H-2 r\big) \sigma '+\big(2 r_H-3 r\big) \sigma  N'\Big)-r (N-1) \Big(2 r \big(2 r- r_H\big) \sigma ''\nonumber \\
				 &+\sigma ' \big(5 r (3 r-2 r_H) N'+2 r_H\big)+\sigma  \big(r (3 r-2 r_H) N''+4 r_H N'\big)\Big)\nonumber\\
				 &+(N-1)^2 \Big(r \big(2 r (2 r_H-3 r) \sigma ''-5 r_H \sigma '\big)+2 r_H \sigma \Big)\Bigg]\ ,
				\end{align}
	In the absence of an horizon, one can obtain a particle like solution \cite{kleihaus2020properties,kleihaus2020particle}
%
	\section{GR in spherical symmetry and isotropic coordinates}\label{S7.4}
%
	An alternative coordinate system to deal with spherical spacetimes, often useful, is given by \textit{isotropic} coordinates -- see $e.g.$~\cite{townsend1997black}. In isotropic coordinates, the radial coordinate is not the areal radius. This section will compute the virial identity in isotropic coordinates for two cases: electro-vacuum and (massive, complex) scalar vacuum. We shall see that the correct virial identities, including a non-trivial contribution from the GHY boundary term, are obeyed by known solutions of these models (the RN BH and SBS). Further confirming that the GHY term is indeed required to construct the virial identity in a generic coordinate system and parameterization.
	
%
		\subsection{A general result}\label{S7.4.1}
%
	Let us consider a general model, described by the action $\mathcal{S}= \mathcal{S}_{grav}+\mathcal{S}_{\rm m}$, where $\mathcal{S}_{grav}$ also includes the GHY boundary term, while $\mathcal{S}_{\rm m}$ is the matter field(s) action (with the presence of first-order derivatives, only). As for the line element, we consider a general form  in terms of two functions $f_0 (r)$ and $f_1 (r)$
			\begin{equation}\label{E7.4.114}
			 ds^2=-f_0^2 dt^2+ f_1^2\big[dr^2+r^2 (d\theta^2+\sin^2\theta d\varphi^2)\big] \ .
			\end{equation}
	The computation of the gravity effective action is very similar to the case of Schwarzschild coordinates. Although the bulk action $\sqrt{-g}\ze R$ depends again on the second derivatives of the metric functions $f_0$ and $f_1$, which can be collected into a \textit{total derivative}, such that this EA is cast in the form~\eqref{E7.1.20} with
			\begin{equation}\label{E7.4.115}
			 \int dr  f_0\ze f_1^3\ze r^2 R  = \int dr \left[ 2r^2 \left(2f_0'f_1'+\frac{f_0f_1'^{\, 2}}{f_1}\right)+ f' \right] \  ,~~{\rm with}~~ f=-2\ze r^2\big( f_1 f_0'+2 f_0 f_1'\ze)~.
			\end{equation}
	Again, we assume that the spacetime boundary is a spherical surface at some radius $r$, with a normal vector $n=1/f_1\partial_r$. Then one finds\footnote{Note that, in computing $K_0$, one considers a (flat) background metric with a two-sphere of radius $r f_1$.}
			\begin{align}
			 \sqrt{-\gamma} &=  f_0f_1^2 r^2\sin\theta  \ ,\\
			 K&= \nabla_\mu n^\mu =\frac{1}{f_1}\left(\frac{2}{r}+\frac{f_0'}{f_0}\right)+\frac{2f_1'}{f_1^2} \ ,	\\	
			 K_0&= \frac{2}{r f_1} \ ,\\
			 \sqrt{-\gamma}(K-K_0) & =  r^2\big( f_1 f_0'+2f_0f_1' \big) \sin\theta \ .
			\end{align}
	One can easily see that, different from the case of Schwarzschild-like coordinates, the contribution of the GHY boundary term cancels out {\it completely} the total derivative in the gravity bulk action \eqref{E7.4.115}. Then one finds the following gravity effective Lagrangian
			\begin{equation}
			 \mathcal{L} (f_0,f_1; f_0',f_1'; r) =  2\ze r^2 \left( 2 f_0'f_1'+\frac{f_0f_1'^{\, 2}}{f_1}\right)\ .	
			\end{equation}
	When adding the EA for the matter sector of the model, the result \eqref{E7.1.22} implies the following 
 form of the generic virial identity  
			\begin{equation}\label{E7.4.121}
			 {\cal V}_{grav}+{\cal V}_m =0 \ ,   \qquad \qquad {\rm {\bf [virial \ isotropic \ general]}}
			\end{equation} 
	with the gravity contribution
			\begin{equation}\label{E7.4.122}
			 {\cal V}_{grav}=-2\int_{r_i}^{+\infty} dr\left[ r(r-r_i)f_1 \left(2f_0'+\frac{f_0 f_1' }{f_1}
\right)\right] ,
			\end{equation}
	${\cal V}_m $ being the matter contribution (as resulting from \eqref{E7.1.22}, in terms of matter field(s) effective Lagrangian $ \mathcal{L}_m$).

%
		\subsection{Electro-vacuum}\label{S7.4.2}
%
	As the most straightforward application of the above results, let us consider the electro-vacuum case, with the Maxwell action as given by~\eqref{E7.2.51}. The electric field is again purely electric, with $A_\mu=V dt$, while the Maxwell equations can be integrated to give
			\begin{equation}\label{E7.4.123}
			 V'=\frac{Q_e}{r^2}\frac{f_0}{f_1} \ ,
			\end{equation}
	with $Q_e$ the electric charge.
	
	The contribution ${\cal V}_m$ of the Maxwell field to the virial \eqref{E7.4.121} is computed from~\eqref{E7.1.22} (with  $\mathcal{L}_M=2\ze r^2 f_1 V'^{\, 2}/f_0$). After using \eqref{E7.4.123} the final result reads
			\begin{equation}\label{E7.4.124}
			 \int_{r_H}^{+\infty} dr\left\{ \frac{f_0\ze Q_e^2}{f_1 r^3}+rf_1'\left( 2f_0'+\frac{f_0f_1'}{f_1}\right)
\right\}(r-2\ze r_H)=0 \ .   \qquad {\rm {\bf [virial \ electro-vacuum \ isotropic]}}
			\end{equation}
	After replacing the expression of the RN solution
			\begin{equation}
		 	 f_0=\frac{1-\frac{r_H^2}{r^2}}{1+\frac{M}{r}+\frac{r_H^2}{r^2}}\ , \qquad  
		 	 f_1 =1+\frac{M}{r}+\frac{r_H^2}{r^2}\ ,  \qquad {\rm where} \qquad r_H^2=\frac{M^2 -Q^2}{4}\ ,
			\end{equation}
	the identity~\eqref{E7.4.124} simplifies to
			\begin{equation}
			 \int_{r_H}^{+\infty} dr\ze \frac{4\ze r_H^2}{r^3}(r-2\ze r_H)=4\ze r_H^2\left( \frac{r_H}{r^2}-\frac{1}{r}\right)\bigg|^{+\infty} _{r_H}=0 \ .  
			\end{equation}
	Hence confirming that the RN solution obeys the identity~\eqref{E7.4.124}. Had we not included the GHY contribution, however, there would be an extra contribution to the identity coming from $f=-4\ze r^2f_0f_1'-2\ze r^2f_1f_0' $ in~\eqref{E7.4.121}. Then, from~\eqref{E7.4.122}, this would give the extra contribution to the virial identity \eqref{E7.4.121}
			\begin{equation}
			 \left[\frac{\partial f}{\partial r}(r-r_i) - \sum_i\frac{\partial f}{\partial q'_i} q'_i\right]^{+\infty}_{r_H}=-\Big[2\ze r(r-2\ze r_H)(2f_0f_1'+f_1f_0')\Big]^{+\infty}_{r_H}=2(M-2\ze r_H) \ .
			\end{equation}
	The non-vanishing for $Q_e \neq 0$ means that a virial identity derived solely from the EH plus Maxwell actions is not obeyed by the RN solution (albeit, accidentally, it is obeyed by the Schwarzschild solution as in the discussion of Sec.~\ref{S7.2}). The correct identity must be derived from the full gravitational action, including the GHY boundary term. Moreover, using isotropic coordinates, the contribution of the gravitational action to~\eqref{E7.4.124} is non-vanishing (and both the EH and GHY terms must be considered), unlike the special ``gauge'' discussed in Sec.~\ref{S7.3}. 
%
		\subsection{(Massive-complex) scalar vacuum}\label{S7.4.3}
%
	As a second illustration, let us reconsider the SBS already discussed in Sec.~\ref{S7.3}. The action is given by~\eqref{E7.3.73}, and the scalar field ansatz by~\eqref{E7.2.31}. In order to test the virial identity for concrete solutions, we take the most straightforward choice for the potential, with a mass term only, $U(|\phi|) = {\mu^2} \phi ^2$. Employing the metric ansatz \eqref{E7.4.114} again results in the scalar field effective Lagrangian
			\begin{equation}
			 \mathcal{L}_{\bar{\Phi}}= r^2 f_0f_1^3 \left[ \frac{\phi'^{\, 2}}{f_1^2}+\left(\mu^2 -\frac{\omega ^2}{f_0^2}\right)\phi^2 \right] \ .
			\end{equation}
	In the absence of an event horizon, the scaling of the radial coordinate is simply $r\rightarrow \tilde{r}= \lambda\ze r$. Then, following the standard procedure, we obtain the simple expression for the scalar field contribution to the  virial identity \eqref{E7.4.121}
			\begin{equation}\label{E7.4.129}
			 {\cal V}_m= 4\int _{0} ^{+\infty} dr ~r^2 f_0f_1\left[ \phi'^{\, 2}+3f_1^2\left(\mu^2-\frac{\omega ^2}{f_0^2}\right)\phi^2\right] \ .
			\end{equation}
	Then, the whole virial identity~\eqref{E7.4.121} reads
			\begin{equation}
			 \int_{0}^{+\infty} dr\left\{ r^2f_1 \left( 2f_0'+\frac{f_0 f_1' }{f_1}-2 f_0  \left[  \phi'^{\, 2}+3f_1^2\left(\mu^2-\frac{\omega ^2}{f_0^2}\right)\phi^2\right]\right)\right\}=0 \ .   
\qquad {\rm {\bf [virial \ BS \ isotropic]}}
			\end{equation}
			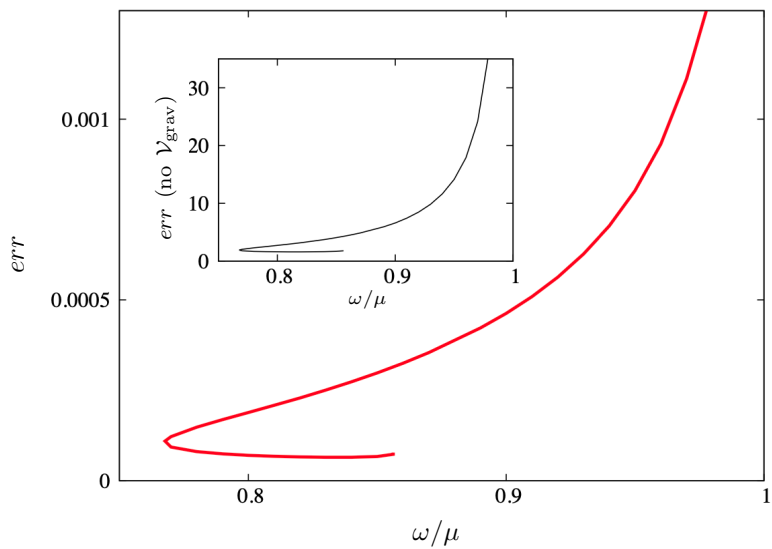
\begin{figure}[h!]
				\centering
	   			\begin{tikzpicture}[scale=0.5]
\node at (0,0) {\includegraphics[scale=0.25]{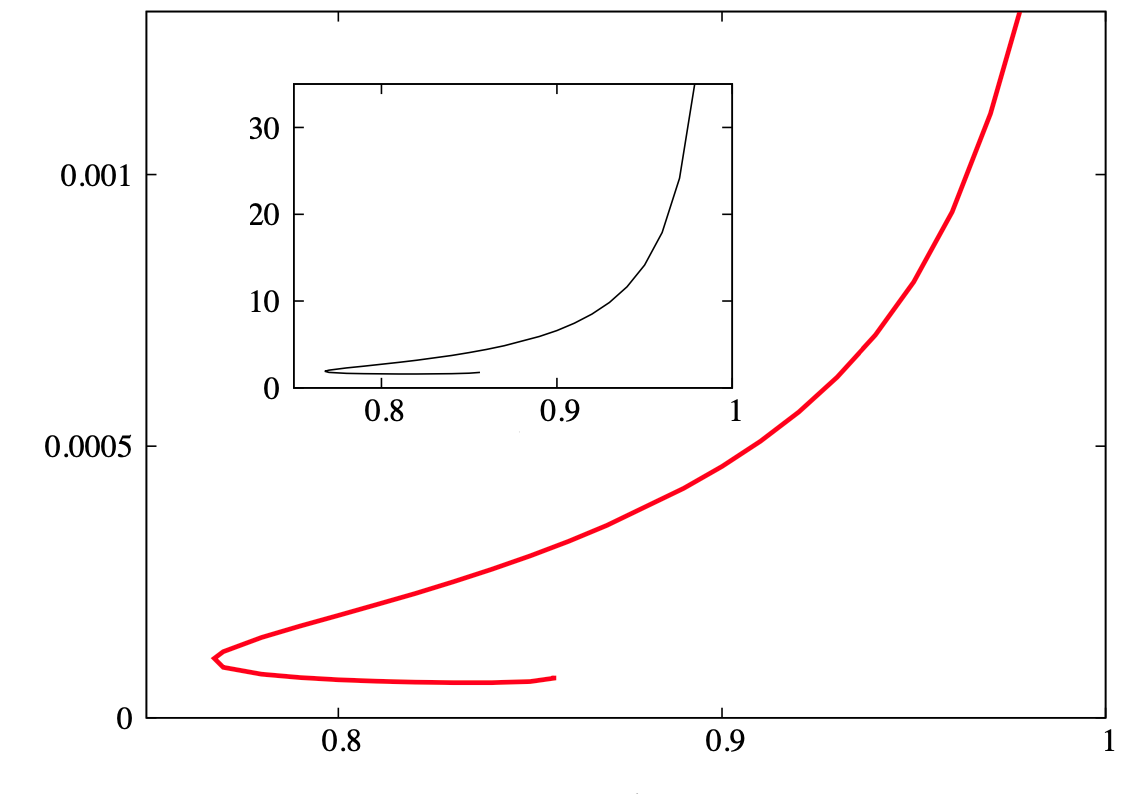}};
\node at (1,-7) {\small $\omega / \mu $};
\node at (-0.8,-0.8) {\small $\scriptstyle \omega / \mu $};
\node at (-10,0.4) {\begin{turn}{90}{\small $err$}\end{turn}};
\node at (-6.1,2.7) {\begin{turn}{90}{$\scriptstyle err\ {\rm (no\ {\cal V}_{grav})}$}\end{turn}};
				\end{tikzpicture}
				 \caption{Relative error $err$~$(1.5.48)$ for the virial identity satisfied by numerical SBS in isotropic coordinates is shown as a function of the ratio between the field frequency and field's mass. The inset shows the same relative error but without including the boundary term in ${\cal V}_{grav}$.}
				 \label{F7.1}
			\end{figure}
	Unlike the electro-vacuum case, no exact solutions are known for a SBS. As seen in Fig.~\ref{F7.1}, $err$ is never zero for a numerical solution\footnote{In constructing the SBSs in isotropic coordinates, we have used the approach described in \cite{herdeiro2015asymptotically} (and in particular, the same solver and the same grid choice). The $err$ increase with $\omega \to \mu$ can be attributed to the delocalization of the solutions in this limit, with $\phi\to 0$ and $(f_1,\ze f_0)\to 1$.}, and takes values compatible with other error estimates. The natural interpretation of this result is that the virial relation \eqref{E7.4.121} also holds for SBS in isotropic coordinates.

	As for the role of the GHY term, an analogous computation to one of the previous subsection yields (taking into account the asymptotic behaviour of the BSs) an extra $-2M$ contribution to the gravity part in the virial identity. Such is fundamental for the solutions to obey the virial identity. 
In Fig. \ref{F7.1} (inset), we show the same relative error as in the main panel, but where ${\cal V}_{grav}$ does not include the contribution from the GHY boundary term. One observes that the error becomes order unity or larger, in this case.
%
%
%
	\section{Axial symmetry}\label{S7.5}
%
	In a Sec.~\ref{S7.2}-\ref{S7.4} (see also \cite{herdeiro2021virial}), hereafter referred to as spherical section, we have presented an introduction to virial identities in field theory. 
	
	The present section seeks to address virial identities that apply for higher dimensional EAs, in particular having in mind axially symmetric (rather than spherical) equilibrium configurations. As we shall see, the latter case leads to 2-dimensional (2D) rather than 1D EAs. We shall propose a concrete methodology to apply the scaling procedure and obtain the corresponding virial identity for an $n$-dimensional ($nD$)  EA where one of the integration variables can be suitably identified with a radial coordinate. Although the procedure is technically more involved in practical applications, it is a simple and natural generalization of the treatment in the spherical section for 1D EAs. In particular, a similar procedure can be applied to EAs that depend on all three spatial variables in suitable coordinates or even on any number of spatial variables (if, for instance, one considers field theories in arbitrary dimensions). Thus, the discussion of axially symmetric solutions presented here, albeit quite relevant on its own, can also be faced as an illustration of the general case of EAs depending on more than one variable.

	The proposed method to deal with scalings of $nD$ EAs results from the understanding (from the spherical section) that this scaling is a way to obtain a particular variation of a fiducial solution. Since stationary configurations obey the Euler-Lagrange equations of the EA under arbitrary variations, the integral identity obtained from any particular variation should obey the solutions of the Euler-Lagrange equation. Thus, the scaling chosen amounts to whatever transformation can be dealt with and transformed into a helpful expression. Also, as discussed in the spherical section, the scaling can be done using different choices for the metric ansatz and thus different radial coordinates. Thus, we should understand from the outset that the virial identities obtained here via scaling transformations of an $nD$ EA need not be the most generic identities obtainable from such action. One cannot exclude that scalings involving several coordinates in independent or dependent ways yield exciting identities.
%
		\subsection{The $n$-dimensional effective action}\label{S7.5.1}
%
	Consider an $n$-dimensional EA
			\begin{equation}\label{E7.5.131}
			 \mathcal{S}^{\rm eff}[q_j(r,\theta_\alpha);q'_j(r,\theta_\alpha),\partial_\alpha{q}_j(r,\theta_\alpha);r,\theta_\alpha] = \int\dots\int \prod_{\alpha=1}^{n-1} d\theta_\alpha \int_{r_i}^{+\infty}\hat{\mathcal{L}}(q_j;q'_j,\partial_\alpha{q}_j;r,\theta_\alpha)\ze dr \ ,
			\end{equation}
	where $q_j$ ($j=1\dots \mathcal{N}$) are a set of $\mathcal{N}$ functions,  $\{\theta_\alpha\}$ are a set of $n-1$ variables on which the EA depends and the derivatives are denoted ${q_j'}(r,\theta_\alpha)\equiv \partial_r q_j(r,\theta_\alpha)$ and $\partial_\alpha{q_j}(r,\theta_\alpha)\equiv \partial_{\theta_\alpha} q_j(r,\theta_\alpha)$. The effective Lagrangian is allowed to include a total \textit{radial} derivative 
			\begin{equation}\label{E7.5.132}
\hat{\mathcal{L}}(q_j;q'_j,\partial_\alpha{q}_j;r,\theta_\alpha)={\mathcal{L}}(q_j;q'_j,\partial_\alpha{q}_j;r,\theta_\alpha)+\frac{d}{dr}f(q_j;q'_j,\partial_\alpha{q}_j;r,\theta_\alpha) \ .
			\end{equation}
	To obtain a virial identity, we now have more freedom in the variables we may scale. Here, we shall focus on the simplest case in which a single scaling in a single coordinate (the radial coordinate) is performed. Consequently, the methodology closely mimics that of the spherical case. In this spirit, we consider the scaling transformation~\eqref{E7.1.9} which varies a fiducial configuration $q_j(r,\theta_\alpha)$ into  $q_{\lambda j}(r,\theta_\alpha)=q_j(r_i+\lambda(r-r_i),\theta_\alpha)$.  Following the discussion of Sec.~\ref{S7.2}, we find, from the stationarity condition~\eqref{E7.1.12} a virial identity
			\begin{align}\label{E7.5.133}
			 \int\dots\int \prod_{\alpha=1}^{n-1} d\theta_\alpha  \left\{\int_{r_i}^{+\infty}\left[ \frac{\partial \mathcal{L}}{\partial r}(r-r_i) -\sum_i \frac{\partial \mathcal{L}}{\partial q'_i} q'_i +\mathcal{L}\right]dr -  \left[\frac{\partial f}{\partial r}(r-r_i) - \sum_i\frac{\partial f}{\partial q'_i} q'_i\right]^{+\infty}_{r_i} \right\}=0 \ . \nonumber  \\ 
			 {\rm {\bf [virial \ nD \ EA]}}
			\end{align}
	Such is no more no less than the integral in the $\theta_\alpha$ coordinates of the virial identity for 1D EAs~\eqref{E7.1.22}. In the following, \eqref{E7.5.133} will be the formula that shall be used in field theory applications. For the specific case of stationary and axially symmetric spacetimes, $n=2$ and the (only) $\theta$ integral in~\eqref{E7.5.133}  is, for the standard polar coordinate, between $0$ and $\pi$.\footnote{In this case, for ease of notation, we make $\theta_1\rightarrow \theta$.} 
%
		\subsection{Flat spacetime field theory}\label{E7.5.2}
%
	Let us start by discussing field theories in flat spacetime. The aforementioned is free of the complexities of the gravitational case that shall be encountered in the following sections. As such, consider the $4$-dimensional Minkowski spacetime in standard spherical coordinates, for which the line element reads.
			\begin{equation}
			 ds^2 = -dt ^2 + dr^2 +r^2 \big( d\theta ^2 + \sin ^2 \theta d\varphi ^2 \big)\ . 
			\end{equation}
			\subsubsection*{Ungauged spinning $Q$-balls}
	Consider a complex scalar field model with a (yet unspecified) potential, described by the action
				\begin{equation}\label{E7.5.135}
	 \mathcal{S}_{\rm m}^{\bar{\Phi}} =\int d^4x \sqrt{-g}  \bigg[-\frac{1}{2}g^{\mu\nu}(\Phi _{,\mu}\bar{\Phi}_{,\nu} + 	\bar{\Phi}_{,\mu}\Phi_{,\nu})-U(|\Phi|)\bigg] \ .
				\end{equation}
	Such a model allows spinning solitons known as spinning $Q$-balls~\cite{volkov2002spinning,kleihaus2005rotating} if one considers the scalar ansatz \eqref{E1.5.44}
				\begin{equation}\label{E7.5.141}
				 \Phi (t,r,\theta,\varphi ) \equiv \phi (r,\theta ) e^{i(m\ze \varphi- \omega\ze t )}\ ,
				\end{equation}
	Defining the EA as $\mathcal{S}_{\rm m}^{\bar{\Phi}} =-\frac{1}{2} \int dt\, \mathcal{S}^{\rm eff}$,\footnote{The factor of $1/2$ is arbitrary and chosen for convenience.} we obtain an EA of the type~\eqref{E7.5.131} with $r_i=0$, $n=2$, $f=0$ and the effective Lagrangian
				\begin{equation}
				 \mathcal{L}(\phi; \phi',\invbreve{\phi};\, r, \theta)=r^2\sin\theta\bigg[-\omega^2\phi^2+\phi'^{\, 2}+\frac{\invbreve{\phi} ^2}{r^2}+\frac{m^2\phi^2}{r^2\sin^2\theta}+U(|\phi |)\bigg] \ .
				\end{equation}
	Then, applying~\eqref{E7.5.133} we obtain the virial identity\footnote{All $I_Y$ virial identity contributions are $(r,\ze\theta)$-dependent, however, for notation simplicity, these will be omitted.}
				\begin{equation}\label{E7.5.138}
				 \int _0 ^{\pi} d\theta\ \int _0 ^{+\infty} dr \Big[ I_\Phi (\omega , m) +I_U ^{[\Phi]} \Big] =0\ , \qquad\qquad {\rm {\bf [virial \ spinning \ Q-balls]}}	
				\end{equation}
	where
				\begin{align}
				 I_\Phi (\omega , m) &\equiv r^2 \sin \theta \bigg[-3\ze \omega ^2 \phi ^2 +\phi '^{\, 2}+\frac{\invbreve{\phi}^2}{r^2}+\frac{m^2\phi ^2}{r^2 \sin ^2 \theta }\bigg] \ ,\\
				 I_U^{[\Phi]}  &\equiv  r^2 \sin \theta \, 3\ze U\ .
				\end{align}
	It can be easily checked that for spherical solutions, since $\invbreve{\phi}=0$, and $m=0$, then~\eqref{E7.5.138} reduces to the virial identity for $Q$-balls obtained in the spherical section. 

	The identity~\eqref{E7.5.138} shows that even in axi-symmetry, the $\omega^2$ term is the only negative term for non-negative potentials. If the potential is just a mass term, $U=\mu^2\phi^2$, the bound state condition $\omega^2<\mu^2$ implies that the integrand is everywhere positive, ruling out non-trivial solutions, just like in the static case. Thus, the existence of spinning $Q$-balls for non-negative potentials also requires self-interactions such that $U-\omega^2\phi^2$ becomes negative in some spacetime regions; rotation \textit{per se} cannot support such solitons without self-interactions.

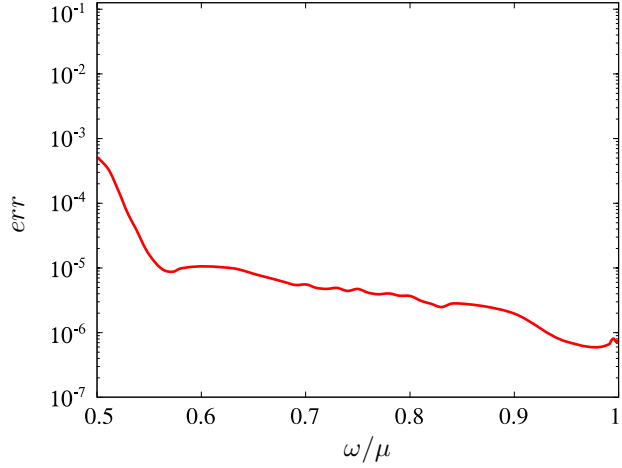
\begin{figure}[H]
			 	 \centering
			 	  \begin{picture}(0,0)		 	  	 			 	 			 	 
			 	 \put(126,-8){\small $\omega /\mu$}
			 	 \put(2,76){\begin{turn}{90}{\small $err$}\end{turn}}
			 	 \end{picture}
			 	 \includegraphics[scale=0.65]{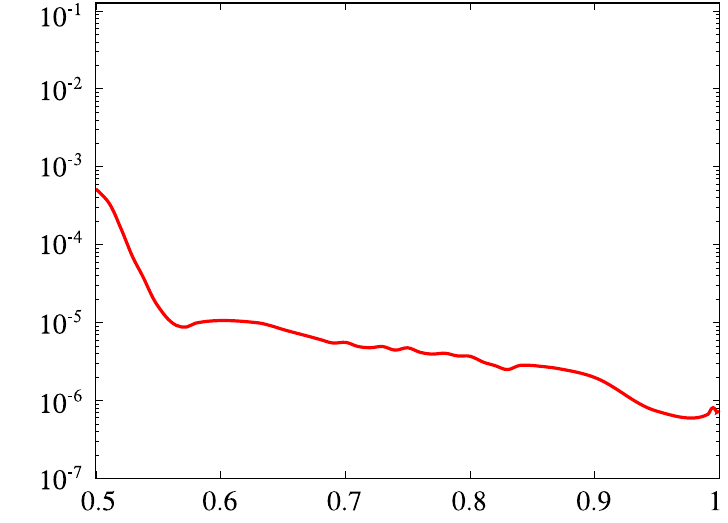}
			 	 \caption{Relative error between the two terms in the virial identity~\eqref{E7.5.138} for an ungauged, spinning $Q$-balls model with $U= \mu ^2 \Phi ^2 +\alpha \Phi ^4 +\beta \Phi ^6$ and $\mu = 1.0,\ \alpha= -1.8$ and $\beta = 1.0\ze$.}
				 \label{F7.2}
    			\end{figure}

	Since there are no known closed-form solutions for spinning Q-balls, we can numerically test the relation \eqref{E7.5.138}. To do this, let us redefine the integral of the virial component $Y$:
				\begin{equation}
				 \mathcal{V}_Y=\int _0 ^{\pi} d\theta\ \int _0 ^{+\infty} dr I_Y (r,\ze\theta) \ .
				\end{equation}
	An illustration of this relative error is given in Fig.~\ref{F7.2}.
			\subsubsection*{Gauged spinning $Q$-balls}
\begin{figure}[H]
				 \centering
				 		 		 \begin{picture}(0,0)		 	 
			 	 \put(195,80){\small $\omega = 1.0$}	
			 	 \put(174,95){\small $\omega = 0.9$}	
			 	 \put(55,96){\small $\omega = 0.6$}				 	 
			 	 \put(125,97){\small $\omega = 0.8$}				 	 			 	 			 	 
			 	 \put(130,-8){\small $e$}
			 	 \put(2,78){\begin{turn}{90}{\small $err$}\end{turn}}
			 	 				\end{picture}
				 \includegraphics[scale=0.65]{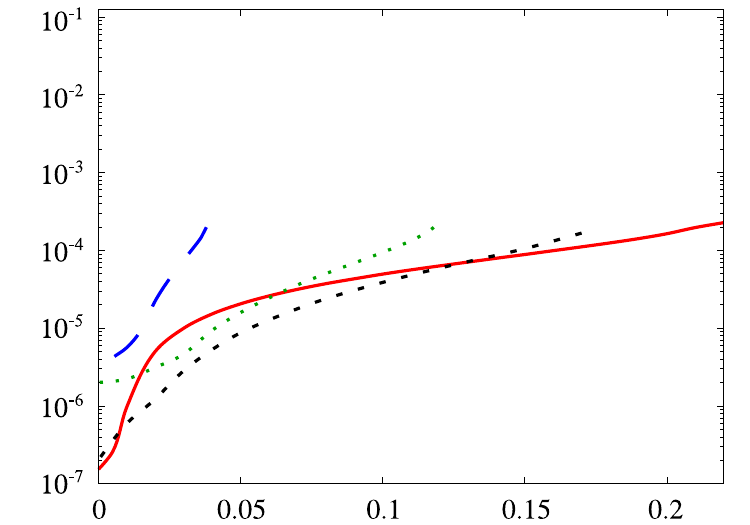}
				 \caption{Relative error between the three terms in the virial identity~\eqref{E7.5.146} for a gauged spinning $Q$-ball model with $U= \mu ^2 \Phi ^2 +\alpha \Phi ^4 +\beta \Phi ^6$ and $\mu = 1.0,\ \alpha= -1.8$ and $\beta = 1.0\ze$.}
				 \label{F7.3}
	    		\end{figure}
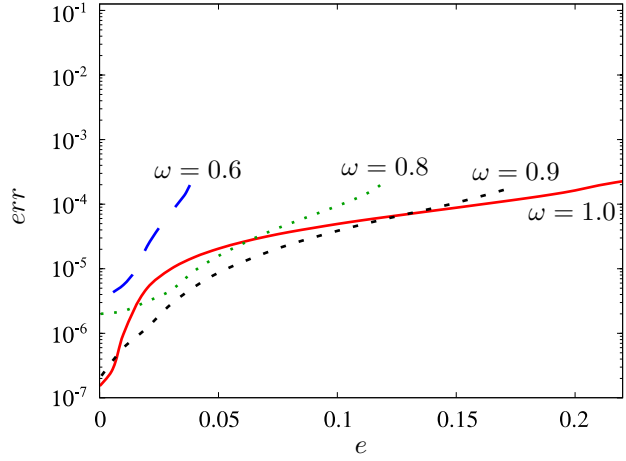

	As a generalization of the previous model, one can gauge the scalar field via the gauge covariant derivative $D_\mu =\partial _\mu-i e A_\mu $ which introduces the gauge coupling constant $e$. One can thus consider the Maxwell-gauged scalar field theory
				\begin{equation}\label{E7.5.143}
				 \mathcal{S} _ {m} ^{\bar{\Phi}_e}=\frac{1}{4} \int d^4 x \sqrt{-g} \bigg[-F_{\mu \nu } F^{\mu \nu} -4\ze g^{\mu\nu} D_{(\mu} \Phi \bar{D}_{\nu)} \bar{\Phi} - U(|\Phi|)\bigg] \ .
				\end{equation}
	Under an ansatz that keeps the scalar field form~\eqref{E1.5.44}, we consider a $4$-vector potential of the form \eqref{E1.5.45} (with $F_W =0$)
				\begin{equation}
				 A_\mu = V(r,\ze\theta) dt+A_\varphi (r,\ze\theta)d\varphi \ .
				\end{equation}	 
	The rotation of the solitons induces the magnetic part of this ansatz. Then, there are indeed gauged spinning $Q$-ball solutions \cite{radu2008stationary}. Repeating the previous subsection, we obtain an effective Lagrangian
				\begin{align}
				 \mathcal{L} (\phi,V,A_\varphi; \phi',V',A_\varphi',\invbreve{\phi},\invbreve{V},\invbreve{A}_\varphi; r, \theta) =  &\frac{\sin \theta}{2 r^2} \Bigg[ 2\ze r^2 \phi ^2 \bigg( \frac{(m-A_\varphi e)^2}{\sin^2\theta}-r^2\big( q V + \omega\big)^2\bigg)\nonumber\\
				 +& \frac{A_\varphi'^{\, 2} r^2 + \invbreve{A}_\varphi^2}{\sin^2\theta}-r^2\big( r^2 V'^{\, 2}+\invbreve{V}^2\big)+2\ze r^2\big( r^2 \phi'^{\, 2}+ \invbreve{\phi}^2\big) +2\ze r^4 U(\phi) \Bigg]
				\end{align}
	Then, applying~\eqref{E7.5.133} we obtain the virial identity
				\begin{equation}\label{E7.5.146}
				 \int _0 ^{+\infty} dr \int _0 ^\pi d\theta \ \Bigg[ I_\Phi (\omega, m ) + I_U^{[\Phi]} +I_e (\omega, m ,e )\Bigg]=0\ , \qquad    {\rm {\bf [virial\ gauged \ spinning \ Q-balls]}}	
				\end{equation}	 
	where the novel gauged term $I_e(\omega,\ze m ,\ze e )$ (with respect to the ungauged model) is given as
				\begin{align}
				 I_e(r,\ze\theta)= \phi ^2 \left[3 e  V r^2\sin\theta \big( e V+2\ze \omega \big)- \frac{q A_\varphi}{\sin\theta}  \big( e A_\varphi -2\ze m\big) \right]+\big( r^2 V'^{\, 2} +  \invbreve{V}^2\big)\frac{\sin\theta}{2}+\frac{r^2 A'^{\, 2}_\varphi + \invbreve{A}_\varphi^2}{2\ze r^2\sin\theta}\ .
				\end{align}
	In the current case, the virial identity \eqref{E7.5.146} contains three components, and the relative error is not as straightforward as before. However, one can note that for this case, both $\mathcal{V}_\Phi $ and $\mathcal{V}_U ^{[\Phi]}$ are always positive and $\mathcal{V}_e$ is always negative. So the equality $\mathcal{V}_\Phi + \mathcal{V}_U ^{[\Phi]} = \mathcal{V}_e$ must hold and the relative error of this relation can be given by
				\begin{equation}\label{E7.5.148}
				 err=1+\frac{\mathcal{V}_\Phi +\mathcal{V}_U ^{[\Phi]}}{|\mathcal{V}_e|}\ ,
				\end{equation}
	which would vanish for an infinity accuracy solution (see Fig.~\ref{F7.3}).
	To summarize, these examples with spinning $Q$-balls (ungauged and gauged) confirm that the virial identity for such axially symmetric cases is the $\theta$ integral of the virial identity that would be obtained regarding the EA as 1D only.
%
		\subsection{Schwarzschild-like coordinates in axial symmetry}\label{S7.5.3}
%
	The GHY~\cite{york1972role,gibbons1993action,hawking1996gravitational,brown1993microcanonical} term is a surface term that is necessary for GR to have a well posed variational principle in a manifold with a boundary. For a BH spacetime ($e.g.$ Schwarzschild spacetimes), there are boundaries at the horizon and at spatial infinity that, in principle, need to be considered. 

The complete gravitational action on a manifold $\mathcal{M}$, including the boundary term, is
			\begin{equation}\label{E6.5.153}
			 \mathcal{S}_{grav}= \mathcal{S}_{EH}+\mathcal{S}_{GHY} =\frac{1}{4} \int_\mathcal{M} d^4x \sqrt{-g}R + \frac{1}{2}\int_{\partial\mathcal{M}} d^3x \sqrt{-\gamma}(K-K_0)  \ ,
			\end{equation}
	where $K =\nabla_\mu n^\mu$ is the extrinsic curvature of the boundary $\partial\mathcal{M}$ with normal $n^\mu$, and $\gamma$ is the associated 3-metric of the boundary. The additional $K_0$ term corresponds to the extrinsic curvature in flat spacetime, necessary to obtain a finite result.

	The GHY boundary term yields an extra total derivative to the EA. In this section, we will compute it for axially symmetric spacetimes \eqref{E1.5.41}, under the following stationary and axially symmetric ansatz:
			\begin{equation}\label{E7.5.149}
			 ds^2 = -e^{2F_0}H dt^2+ e^{2F_1}\left(\frac{dr^2}{H}+r^2d\theta^2\right) + e^{2F_2}r^2\sin^2\theta\big( d\varphi-F_Wdt\big) ^2, \quad \mathrm{with}\;\;\; H=1-\frac{r_H}{r} \ .
			\end{equation}
	The aforementioned widely used ansatz introduces four parametrizing functions $\mathcal{F}_i(r,\ze\theta)$. Moreover, well known analytic solutions, such as the Kerr metric, can also be put in this form, as we shall see in the following subsection.

	Derrick's argument follows through the scaling of the radial component $r\rightarrow r_\lambda= r_H+ \lambda(r-r_H)$ and the metric/matter function.

			\subsubsection*{Vacuum: the Kerr solution}
	Let us start by considering vacuum GR. Then, the complete action to be considered is~\eqref{E6.5.153}. The virial identity for an axially symmetric vacuum configuration under the parameterization~\eqref{E7.5.149} is the sum of the Einstein-Hilbert action plus the GHY term
				\begin{equation}\label{E7.5.151}
				 \int _0 ^\pi d\theta \int _{r_H} ^{+\infty} dr \  I_R(r_H)= I_{GHY} \ . \qquad \qquad {\rm {\bf [Virial \ vacuum \ GR]}} 
				\end{equation}
	Concerning the GHY term contribution $I_{GHY}$, using the normal vector $n^\mu\partial_\mu= \sqrt{N}e^{-F_1}\partial_r$ and $K_0=\frac{2}{r}$, we find a surface contribution akin to the extra right term in \eqref{E7.5.133} with
				\begin{equation}\label{E7.5.152}
				 f^{GHY} = \sin \theta \, e^{F_0+F_2} \left[2 r (r-r_H)( F_0'+ F_1' + F_2') -4 r \sqrt{1-\frac{r_H}{r}} e^{F_1}+4 r-3 r_H\right]\ ,
				\end{equation}
	which gives us
				\begin{align} \label{E7.5.153}
			 	 I_{GHY} &= \int_0^\pi d\theta\left[\frac{\partial f^{GHY}}{\partial r}(r-r_i) - \sum_i\frac{\partial f^{GHY}}{\partial q'_i} q'_i\right]^{+\infty}_{r_H}\nonumber \\
			     &= 2\int_0^\pi d\theta\sin\theta  e^{F_0+F_2} \left[(r-r_H)^2( F_0'+F_1'+F_2')-(2 r-r_H) \sqrt{1-\frac{r_H}{r}} e^{F_1}+2(r-r_H)\right]_{r_H}^{+\infty}  \ .
				\end{align}
	Which can be simplified by considering appropriate boundary conditions for a BH. Due to asymptotic flatness, we assume the following behaviour for large $r$
				\begin{equation}\label{E7.5.154}
				 F_0= \frac{c_t}{r} +\cdots \ , \qquad	F_1= -\frac{c_t}{r} + \cdots \ ,\qquad  F_2= -\frac{c_t}{r} + \cdots  \ ,\qquad F_W= -\frac{c_t}{r^3} + \cdots  \ ,
				\end{equation}
	where $c_t$ is a solution dependent constant\footnote{We can write the metric for Kerr spacetime in these coordinates, $cf.$ \eqref{E7.5.155}-\eqref{E84}. In this case $c_t = r_H/2 - M$, where $M$ is the ADM mass of the Kerr BH.}. At the horizon we assume 
				\begin{equation}\label{E7.5.155}
				 \mathcal{F}_i(r,\ze\theta)\Big|_{r_H} = \mathcal{F}_{iH}(\theta) \ , \qquad	 \mathcal{F}_i'\Big|_{r_H} = 0 \ .
				\end{equation}
	Then, the GHY term simplifies to
				\begin{equation}\label{E7.5.156}
				 I_{GHY} = 2\int_0^\pi d\theta\, c_t\sin\theta = 4\ze c_t \ .
				\end{equation}
	Now we consider the EH term contribution. We obtain:
				\begin{align}\label{E7.5.157}
				  I_R(r_H) = \frac{ e^{F_2-F_0}\sin \theta}{2}  \Bigg\{ & r^2 \sin ^2 \theta  e^{2 F_2} \left(r \big(3r-4r_H\big) F_W'^{\, 2}+ 3\invbreve{F}_W^2\right)\nonumber\\
				 &-4 e^{2 F_0} \Bigg[ (r-r_H) \Bigg( (r-r_H) \Big(F_0 '' +F_1 '' +F_2'^{\, 2}+F_2''\Big)\nonumber\\
				 &+F_0' \Big((r-r_H) \big(F_0'+F_2'\big)+2\Big)+F_1'+3 F_2'\Bigg)+\invbreve{F_0} \left(\invbreve{F_1}+\cot \theta \right)+\invbreve{F}_0^2+\invbreve{\invbreve{F}}_0\nonumber\\
				 &+\invbreve{\invbreve{F}}_1+\invbreve{\invbreve{F}}_2+\invbreve{F}_2 \Big( \invbreve{F}_2+2 \cot \theta \Big)\Bigg]\Bigg\} \ .
				\end{align}
	This illustrates how the interpretation of such identities will become much more difficult in the axially symmetric case. Nonetheless, we can check that known solutions obey the corresponding identity. In the vacuum case, the only BH solution is the Kerr metric~\cite{kerr1963gravitational}. This solution can be cast in the form~\eqref{E7.5.149} with~\cite{herdeiro2015construction}
				\begin{align}\label{7.5.157}
				 F_0& = -F_2 \ ,\\
				 F_1& = \frac{1}{2} \ln \left[\frac{c_t\ze (c_t-r_H) \cos ^2\theta }{r^2}+\left(1-\frac{c_t}{r}\right)^2\right] \ ,\\
				 F_2& = \frac{1}{2}\ln \left[ e^{-2 F_1} \left(\frac{c_t\ze (r_H-c_t) \sin ^2 \theta  \left(1-\frac{r_H}{r}\right)}{r^2}+\left(\frac{c_t (c_t-r_H)}{r^2}+\left(1-\frac{c_t}{r}\right)^2\right)^2\right)\right]\ ,\\
				 F_W &= \frac{\left(1-\frac{c_t}{r}\right) \sqrt{c_t \ze (c_t-r_H)} (r_H-2 c_t) e^{-2 (F_1+F_2)}}{r^3}\ .\label{E84}
				\end{align}
	With asymptotic form~\eqref{E7.5.153}.	

	After replacing the metric functions and corresponding derivatives, the $I_R$ term is even in the $\theta \in [0,\ze\pi ]$ interval, this allows us to perform the angular integration of \eqref{E7.5.156} between $\theta : 0 \rightarrow \pi/2 $. To further simplify the computations, let us perform a coordinate transformation $x\equiv \sin \theta $. This means that $\cos \theta = \sqrt{1-x^2}$ and $d\theta = \frac{dx}{\sqrt{1-x^2}}$. Then,  the $\theta$-integration is performed from $x: 0\rightarrow 1$ and we end up with a radial integral.

	The integral could be performed analytically (using algebraic manipulation software) and yields the same analytical result as \eqref{E7.5.155}. Thus, the Kerr solution verifies the virial identity \eqref{E7.5.151}.
			\subsubsection*{(Massive, complex) scalar-vacuum: mini spinning boson stars}
	We consider a complex, massive scalar field minimally coupled to Einstein's gravity as a second example. The action of the model is
				\begin{equation}
				 \mathcal{S}=  \mathcal{S}_{grav}+	 \mathcal{S}_{\rm m}^{\bar{\Phi}}\ ,
				\end{equation}
	where the two terms are given by~\eqref{E7.2.57} and~\eqref{E7.5.135}. This model admits as solutions a family of axially symmetric, asymptotic flat solitons, known as spinning SBS -- see $e.g.$~\cite{schunck1998rotating,schunck2003general,liebling2017dynamical,herdeiro2019asymptotically}. For such solutions $r_H=0$ and $H=1$. There are different families of such solutions depending on $U(|\Phi|)$. Here we shall use a simple mass term potential $U=\mu^2|\Phi|^2$ and the corresponding solutions are dubbed \textit{mini (spinning) SBS}.

	Using the results from the previous subsection and obtaining the additional matter sector contribution, we get
				\begin{equation}
				 \int _0 ^\pi d\theta \int _0 ^{+\infty} dr \ \Big[ I_R(0) + I_\Phi(\omega, m)+ I_U ^{[\Phi]}\Big]= I_{GHY} \ . \qquad {\rm {\bf [virial\  (mini) \ spinning \ SBS]}} \ .
				\end{equation}
	We once again use the boundary conditions \eqref{E7.5.153}, obtaining $I_ {GHY} =4\ze c_t$ and
				\begin{align}
				 I_\Phi(\omega, m) &=e^{F_2-F_0} \Big[-e^{2F_0} \left(r^2 \phi'^{\, 2}+\invbreve{\phi} ^2\right)+3\ze e^{2 F_1} r^2 \phi ^2 (\omega -m F_W)^2\Big]\sin \theta - e^{2 F_1+F_0-F_2} \frac{m^2 \phi ^2}{\sin\theta} \ ,\\
				 I_U ^{[\Phi]} & =  -3\ze \mu ^2\ze r^2 \phi ^2 \sin \theta\, e^{F_2+2F_1+F_0} \ .
				\end{align}
			\subsubsection*{Secondary relations}
	Observe that $f^{GHY}$ in \eqref{E7.5.151} and, as a consequence $I_{GHY}$, are $F_W$-independentent. In addition, let us consider the metric \eqref{E7.5.149} where $W\equiv \frac{F_W}{r}$. 
	 
	 Since the $f^{GHY}$ term is $F_W$-independent, such transformation should keep it invariant. On the other hand, the Ricci contribution will change to accommodate the new metric function while keeping the remaining function's contribution untouched.  
				\begin{align}
				 &W\ {\rm contribution\ to\ I_R:\ } \frac{1}{2} \sin ^3 \theta \, e^{3F_2-F_0} \left[\left(W-r W'\right) \left(W-(r-2 r_H) W'\right)+\invbreve{W}^2\right]\ ,\\
				 &{\rm F_W\ contribution\ to\ I_R:\ } \frac{1}{2} r^2 \sin ^3 \theta \, e^{3 F_2-F_0} \left[r (3 r-4 r_H) F_W'^{\, 2}+3 \invbreve{F}_W^2\right]\ .
				\end{align}		 
	 Therefore, since we have never explicitly stated the shape of the $F_W$ function in our virial identity calculations, it is expected that one could reverse the transformation, $F_W\equiv W r$, and recover the initial relation. However, there is now a new term that is not present on the initial relation \eqref{E7.5.156}
				\begin{equation}
				 F_W=W r {\rm\ contribution\ to\ I_R\ :\ } \frac{1}{2} r^2 \sin ^3 \theta e^{3 F_2-F_0} \left[F_W' \left(r (r-2 r_H) F_W'-2 r_H F_W\right)+\invbreve{F}_W^2\right]\ ,
				\end{equation}		 
	and should therefore give a null contribution to the virial identity
				\begin{equation}
			 	 \int _0 ^\pi d\theta \int _{r_H} ^{+\infty} dr\ r^2 \sin ^3 \theta\, e^{3 F_2-F_0} \left[F_W' \left(r (r-r_H) F_W'+r_H F_W\right)+\invbreve{F}_W^2\right] =0\ .
				\end{equation}
	This creates a new condition that the $F_W$ function should obey. Algebraic computation of the Kerr metric can be performed and obeys the previous condition.
	
	In the absence of an event horizon and the presence of a scalar field, \textit{i.e.} BS, one obtains the following relation
			\begin{equation}
			 \int _0 ^\pi d\theta \int _{0} ^{+\infty} dr\ r^2 \sin \theta\,  e^{F_2-F_0} \left[2\ze m\ze e^{2 F_1}  F_W (m F_W-\omega )\phi^2+\sin ^2 \theta e^{2 F_2} \left(r^2 F_W'^{\, 2}+\invbreve{F}_W^2\right)\right] =0\ ,
			\end{equation}
	which, for $m>0$ implies that $m\ze F_W < \omega$. Observe that such secondary relations do not seem to introduce new information; however, they can be used to simplify and/or clarify the information already included in the complete virial identity. 
%
	\section{Axial symmetry: the convenient gauge}\label{S7.6}
%
	The metric ansatz used in the last Sec.~\ref{S7.5} gives rise to a non-trivial virial identities contribution from the gravitational part of the action, as seen in the explicit examples of Sec.~\ref{S7.5.3}. As discussed in the spherical section, there is a convenient gauge choice (the $\sigma-m$ parameterization ~\eqref{E1.5.40} therein) that trivializes the gravitational action contribution, making the whole virial identity originate from the matter action. This section discusses an ansatz for the axially symmetric case that produces the same effect: it trivializes the contribution from the gravitational part of the action.  

	Consider the following ansatz:
		\begin{equation}\label{E7.6.170}
		 ds^2 =-F_0^2 dt^2 + F_1^2 dr^2 + (r-r_H) ^2  F_1^2 d\theta ^2+F_2^2\big( d\varphi - F_W dt\big)^2 \ ,
		\end{equation}	 
	where $\mathcal{F}_i$ are the four parameterizing functions, while $r_H$ corresponds to the horizon's radius, which in these coordinates will always be a topological sphere.

	The remarkable property of the ansatz \eqref{E7.6.170} is that, after scaling the metric functions and respective derivatives (\textit{e.g} $F_W'\rightarrow F_{W\lambda}'/\lambda$), the integral term associated with the Ricci scalar becomes
		\begin{equation}\label{E7.6.171}
		 \sqrt{-g_\lambda}\, R _\lambda = \frac{1}{\lambda}\sqrt{-g}\, R\ ,
		\end{equation}
	and hence, the transformations $dr \rightarrow \lambda\ze dr$ will keep the action invariant under scaling transformations. Thus, the EH action for this ansatz becomes scale-invariant; the EH action does not contribute to the virial identity ($I_R =0$).
	
	Let us now justify that the GHY boundary term also vanishes for this ansatz. In the metric parameterization \eqref{E7.6.170}, using the normal vector $n^\mu\partial_\mu= \frac{1}{F_1}\partial_r$ and $K_0=\frac{2}{(r-r_H)F_1}$ the extra total derivative from the GHY boundary term is
		\begin{align}
		 f^{GHY}=& \frac{F_0^2 F_2(r-r_H)}{\sqrt{F_0^2 + F_2^2 F_W ^2}}\Bigg[\frac{F'_0}{F_0} + \frac{F'_1}{F_1} + \frac{F'_2}{F_2} - \frac{1}{r- r_H} + \frac{
 F_2^2 F_W ^2}{F_0^2} \bigg(\frac{F'_1}{F_1} + 2 \frac{F'_2}{F_2} - \frac{1}{r - r_H} + \frac{F_W'}{F_W}\bigg)\Bigg]\ .
		\end{align}
	The resulting GHY contribution from $f^{GHY}$ to the virial identity is
		\begin{align}
		 I_{GHY}=\int _0 ^\pi d\theta \ 0\,\Big|_{r_H} ^{+\infty} \ = 0\ .
		\end{align}
	Which is metric function independent. The GHY term is automatically null due to the form of the metric ansatz, meaning that this result can be used for any solution that obeys the convenient ansatz.
%
		\subsection{Examples using the convenient ``gauge'' -- black holes and solitons}\label{S7.6.1}

%
%
			\subsubsection*{BHs with scalar hair}
	The Kerr BHs with scalar hair (KBHsSH) \cite{herdeiro2015construction,herdeiro2014kerr} (see also Ch.~\ref{C5}) can be seen as a generalization of the spinning SBS with the inclusion of an event horizon $r_H \neq 0$ at their centre. Consider the action that describes a self-gravitating complex scalar field, described by the ansatz \eqref{E7.6.170} in a model with a self-interaction potential $U(\Phi)$ 
				\begin{equation}\label{E7.7.174}
				 \mathcal{S} = \mathcal{S}_{grav} + \mathcal{S}_m ^{\bar{\Phi}}\ ,
				\end{equation}
	 with the complex scalar field described by the ansatz~\eqref{E7.5.141}. The resulting effective Lagrangian is
				\begin{align}
				 \mathcal{L} (\phi; \phi ', \invbreve{\phi};r,\theta) = &(r-r_H) F_1 ^2 F_2 \sqrt{F_0^2+ F_2^2 F_W ^2} \bigg[\frac{m^2 F_0^2-\omega F_2^2(\omega-2\ze m F_W)}{F_2^2(F_0^2+ F_2^2 W^2)}\ze \phi ^2\nonumber\\
				 &+\frac{1}{F_1^2}\bigg(\phi '^{\, 2} +\frac{\invbreve{\phi}^2}{(r-r_H)^2}\bigg)+U\bigg]\ .
				\end{align}
	
	Then, the virial identity reads
				\begin{equation}\label{E7.6.176}
				 \int _{r_H} ^{+\infty} dr \ \int _0 ^{\pi} d\theta \Big[ I_\Phi (\omega, m, r_H) + I_U ^{[\Phi] } (r_H) \Big] =0 \ ,\qquad \qquad {\rm {\bf [virial\ KBHsSH]}}
				\end{equation}
	where the virial terms are
				\begin{align}
				 & I_\Phi (\omega , m,r_H)  = \frac{2\ze (r-r_H) F_1 ^2 }{F_2 \sqrt{ F_0 ^2 +F_2 ^2 F_W ^2}} \bigg[m^2 F_0^2-\omega F_2^2\big( \omega-2\ze m F_W\big)\bigg] \phi ^2\ ,\\
				 & I_U ^{[\Phi]} (r_H) = 2 \big( r-r_H\big) F_1 ^2 F_2 \sqrt{F_0^2+ F_2^2 F_W^2}\, U \ .
				\end{align}

	The SBS virial identity is obtained if we set $r_H=0$ in \eqref{E7.6.176}. To test relation \eqref{E7.6.176}, we use three rotating scalar hairy Kerr BH configurations and two rotating SBS configurations, one with a simple mass potential $U=\mu ^2 \phi ^2 $ and one with an axionic coupling \eqref{E5.3.22}. In all cases, we obtained an absolute (relative) difference for the virial identity of $\sim 10 ^{-2}$ $(\sim 10^{-2})$.

\bigskip

	We have also computed the virial identity for Kerr BHs with vector hair and Kerr BHs in the EMS model, however, the relations are too complicate and besides being used as a numerical test, we were not able to obtain any insight into the solutions.
%
	\section{Further remarks}\label{S7.8}
%
	This chapter aims to present a primer for a clear and efficient understanding of virial identities in non-linear field theories, particularly in relativistic gravity. As explained in Sec.~\ref{S1.5}, virial identities result from a specific type of variational principle obtained from an EA. Thus, they should be obeyed by the solutions of the Euler-Lagrange equations obtained from that EA, which extremise \textit{any} variation. Nonetheless, virial identities are integral identities that \textit{appear} independent from the field equations. Thus, their analysis provides different insights and checks than those provided by the analysis of the (differential) field equations.

	Considering an appropriate ansatz in any non-linear field theory in spherical symmetry leads to an EA in the radial variable. Then, \eqref{E7.1.13}, \eqref{E7.1.16}, \eqref{E7.1.19} and \eqref{E7.1.22} provide a straightforward way to compute the virial identity. Nevertheless, the EA must contain all terms necessary to define the model completely. In the case of non-linear field theories for which the original action contains second derivatives of the fundamental variables, the well-posedness of the field equations in manifolds with boundaries requires the introduction of boundary terms. Whereas the latter is irrelevant for many analyses (such as computing the bulk solutions of the field equations), such boundary terms can, and in general, will contribute to virial identities. This is the case of GR, for which the EH action has second-order derivatives of the metric and the complete gravitational action~\eqref{E7.2.57} needs the GHY boundary term. We have shown that this term must be considered to derive the correct virial identities in GR.

	Nonetheless, there is a special ``gauge'' choice $\big($corresponding to the $\sigma-m$ parametrization in Schwarzschild coordinates~\eqref{E1.5.40} with~\eqref{E7.2.55}$\big)$ where one can get away with neglecting the boundary term and indeed the whole gravitational action for the virial identity. This occurs because the EH action for this ``gauge'' choice leads to a scale-invariant EA, and the GHY boundary term does not contribute, at least for the boundary conditions that apply to asymptotically flat regular solitons or BHs. In this context, it is important to stress that the scaling transformation leading to virial identities is \textit{not} a diffeomorphism; the EA results from the integral of scaled configurations, which is not simply a coordinate transformation in the integral. Thus, in general, the EH action will contribute to virial identities. However, it turns out that there is a nice ``gauge'' choice for which it does not, facilitating thus the computation of virial identities. 

	The generalization to higher-dimensional actions was also performed with a special focus on axial symmetry. The same treatment as in the spherical symmetry was performed. We introduced a ``special gauge'' that simplifies the virial identity computation in axial symmetry. Still, the systems' higher complexity prevents one from taking conclusions from the computed relations. An interesting future question that we hope to consider is the case of modified gravity, for which the boundary term needs to be appropriately modified.  
%
\chapter{Conclusion}\label{C8}
%
	In this thesis, we studied several hypothetical objects made and/or surrounded by bosonic fields. We developed a numerical code to construct such solutions, developed the machinery to study their properties and established the complete procedure to obtain the virial identity of a given model that allows one to permit or rule out the existence of such solutions. Except for the numerical procedure -- that will be developed in Appendix~\ref{A} and \ref{B}, all these topics were introduced in Ch.~\ref{C1}.
	
	Chapters \ref{C2} and \ref{C3} were dedicate to scalarized BHs. In Ch.~\ref{C2} charged black holes with a scalar field non-minimally coupled to the Maxwell invariant were studied. Such objects were discussed within the Einstein-Maxwell-Matter model and can be categorized in two classes (presented in Ch.~\ref{S1.2}): the dilatonic solutions (\textit{a.k.a.} class \textbf{I}) and the scalarized solutions (\textit{a.k.a.} class \textbf{II}). While the formers do not reduce to the Reissner-Nordstr\"om solutions, the latter do. These can be further divided into two subclasses: scalarized connected (class \textbf{II.A}) -- that endow spontaneous scalarization -- and scalarized disconnected (class \textbf{II.B}). 
	
	In the first section, we studied and compared the different properties of all the solutions and observed that all of the models can be overcharged and entropically prefered. Concerning the class \textbf{II.A}, the spontaneously scalarized solutions, since such solutions can be obtained from a perturbation of the electro-vacuum BH configuration, we performed a dynamical evolution and perturbative stability analysis. Solutions are perturbatively stable and dynamically favourable. Some further generalizations of the EMS model were performed in the same spirit as before. 
	
	Chapter \ref{C3} is dedicated to the scalarization phenomena in an extended-scalar-tensor theory. In the latter, the scalar field is non-minimally coupled to the Gauss-Bonnet invariant and endows scalarization of black holes. We constructed such scalarized solutions for Reissner-Nordstr\"om and Kerr-Newman black holes for both signs of the coupling. 

	Even though its astrophysical relevance is still under debate, the relative simplicity of the Einstein-Maxwell-Matter model allowed us to perform several dynamical and perturbative studies for several coupling functions and possible generalizations. The simplicity and flexibility of such models make them one crucial tool to study more complex models, such as the Gauss-Bonnet gravity, which are more computationally demanding. However, the latter also gave us essential insight into the properties of the spontaneous scalarization phenomena. In the end, the balance between a simpler model, which allowed several generalizations and a more complex model allowed us to have a better insight into the scalarization phenomena. A strategy that we certainly will implement in future studies.

	We followed the previous two chapters with two additional chapters dedicated to boson stars. In Ch.~\ref{C3} we constructed scalar and vector boson star solutions and observed that, while dynamically robust scalar boson stars can not easily mimic the shadow of a spherically symmetric black hole, vector boson stars can do it even in the simplest model. The constructed shadow mimics the one of a Schwarzschild black hole but only in one of the possible observation configurations.
	
	In the next chapter (Ch.~\ref{C4}), we studied spinning boson stars containing a Kerr black hole at its centre. Such solutions are much more complex, and a deep study cannot be easily performed. We introduced and developed the decomposition into the spherical harmonics basis of all the metric and scalar functions to circumvent such a problem. We observed that four modes are already enough to correctly describe the set of solutions. In addition, we have also decomposed a set of solutions that continuously connect between a hairless Kerr black hole and a spinning scalar boson star. We observed a change in the spherical harmonics structure of the solutions. While it is still premature to make any conclusion about the latter behaviour, it gives us a good indication of the potential power of such a technique, making it an essential instrument for future, more exciting studies.
	
	We finished this thesis (Ch.~\ref{C7}) with the study of Derrick's argument to construct the virial identity of a general relativistic model. We observed that applying the argument to a gravitational action is not enough to obtain the correct (complete) identity, and an additional term, the Gibbons-Hawking-York term, is necessary. The latter is a non-dynamical term computed at the boundaries and rescales the action to have a finite value. Several models in both spherical and axial symmetry are studied. In both cases, a ``convenient'' ansatz that significantly reduces the computational difficulty of the identity is introduced.

	Virial identities are of extreme importance, not only as a check of the solutions (in particular the numerically obtained ones) but also to establish no-go and no-hair theorems. The latter has been mainly performed under spherical symmetry -- due to a convenient metric gauge. However, the study performed in this thesis allowed us to extend such studies to the more astrophysically relevant axial symmetry. While being much more complex and, for now, not allowing us to obtain any relevant insight, it has the potential to develop further and maybe establish new theorems. The complete study of such identities is then fundamental to further pursue and rule out new objects. 
	
	Bosonic fields are a vibrant and flexible topic to study; they can easily describe some of the observed unknown phenomena of the Universe and introduce additional degrees of freedom to black holes. Their existence in nature (besides the standard model particles) is still unknown, but the inflation phenomena, dark matter and dark energy give an exciting possibility for their occurrence in nature.
	
	This thesis has always considered asymptotically flat solutions for BHs and BSs. However, such objects can also occur in non-asymptotically flat (Anti)-de Sitter spacetimes. Even though some studies have already been performed (\textit{e.g.}\cite{brihaye2020black,duarte2016asymptotically,kichakova2014spinning}), however much is still to be done. In addition, with all the new gravitational wave data, and some without a good ``vanilla'' explanation, it would be fascinating to perform full non-linear dynamical simulations of collisions/mergers of both scalarized black holes and boson stars. In addition, we expect to perform further developments concerning the stability of the Kerr black holes with scalar hair through spectral decomposition. As a future project, we intend to further extend the virial identity computation to alternative theories of gravity. A unified picture of the virial identity is under development.
%
\appendix

\clearpage\null\newpage

%
\chapter{ODE Solver}\label{A}
%
	In mathematics, an ordinary differential equation (ODE)~\cite{zill2012first} is an equation that relates one or more functions of an independent variable and the respective derivatives of those functions. The term ordinary is used to distinguish between the more general partial differential equation, which may contain more than one independent variable\footnote{An oral presentation about the workings of the ODE solver applied to spherically symmetric SBSs can be seen at \cite{ODEVid}.}.
	
	The mathematical description of a change uses differentials and derivatives. Various differentials, derivatives and functions, become related via equations such that a differential equation is a result that describes dynamically changing phenomena, evolution and variation. ODEs arise in many contexts of mathematics, social and natural sciences.
	
	Let $F$ be a function of $x$ and $y$ and derivatives of $y$. Then an equation of the form 
	\begin{equation}
	 F\big( x,\ze y,\ze y',...,\ze y^{(n-1)}\big) = y^{(n)}\ , 
	\end{equation}
	is called an explicit ODE of order $n$, while an implicit ODE of order $n$ takes the form
	\begin{equation}
	F\big( x,\ze y,\ze y',...,\ze y^{(n)}\big) =0\ .
	\end{equation}

	One can expand $y$ using a Taylor's series. This would give a reasonably good approximation to the function, especially if the point is close enough to some known starting point, and we take enough terms.
	
	However, one of the drawbacks of the Taylor's expansion is the need to differentiate the function once for each new term one wants to calculate. Such can be troublesome for complicated functions and not work well in computational modelling.
	
	In an attempt to provide a method of approximating a function without differentiating the original equation, in $1900$, Carl Runge and Wilhelm Kutta proposed a new method to simulate as many steps of Taylor's expansion while only evaluating the original function.
	
	The idea of Runge and Kutta was to obtain a method that only uses first derivatives calculated explicitly in the adjacent points of integration. 
	
	The Runge-Kutta methods propagate a solution over an interval by combining the information from several Euler-style steps (each involving one evaluation of the \textit{rhs}) and then using the information obtained to match a Taylor series expansion up to some higher order. 
	
	Runge-Kutta succeeds virtually always, but it is not usually the fastest, except when evaluating $F$ is cheap and moderate accuracy ($\leqslant 10 ^{-5}$) is required. 
%
	\section{Integrator: Runge-Kutta method}\label{A1}
%
	In numerical analysis, the Runge-Kutta methods~\cite{runge1895numeric,ascher1998computer,butcher1963coefficients,kutta1901post} (RK) are a family of implicit and explicit iterative methods, which include the well-known routine called the Euler method -- widely used in the temporal discretization of ODEs.
	
	Consider the initial value problem:
		\begin{equation}\label{A0}
		 y' = f(x,y)\ ,\qquad \qquad y(x_0)=a\ ,
		\end{equation}
	where $y$ is an unknown function of the $x$-independent coordinate, which we would like to approximate. The first relation gives the rate at which $y$ changes -- and that such a rate depends both on $x$ and $y$. The second relation is known as a initial value and states that: at the initial point $x_0$, the corresponding value of $y$ is $a$. The function $f$ and the initial conditions $x_0$ and $a$ are known.
	
	As with all the explicit interactive integration methods, the idea is to obtain an approximation of the following point $y_{n+1}$ from the current, known, point $y_n$. 

	In general, the family of explicit Runge-Kutta methods of order $s$ is given by
		\begin{equation}\label{A5}
		 y_{n+1} = y_n + h \sum _{i=1} ^s b_i k_i\ ,
		\end{equation}
	where $k_i$ are the intermediate integration increments and $h$ is the integration step. The weighted average of the increments estimates the slope specified by the function $f$. With
		\begin{align}
		 & k_1 = f\big( x_n, y_n\big)\ ,\\
		 & k_2 = f\big( x_n + c_2h, y_n + h(a_{21}k_1)\big)\ ,\\
		 & k_3 = f\big( x_n + c_3h, y_n + h(a_{31}k_1+a_{32}k_2)\big)\ ,\\
		 & .\nonumber\\
		 & .\nonumber\\
		 & k_s = f\big( x_n + c_s h, y_n + h(a_{s1}k_1+a_{s2}k_2+...+a_{s\, s-1}k_{s-1})\big)\ .
		\end{align}
	The $k_1$ ($k_s$) correspondes to the slope at the begginning (end) of the interval, while $k_2$ to $k_{s-1}$ correspond to the slope at the midpoints of the interval.
	
	While averaging of $s^{th}$ slopes, greater weight is given to the slopes at the midpoint. The RK method of order $s$ has a local truncation error of $\mathcal{O}(h^{s+1})$, while the total accumulated error of $\mathcal{O}(h^s)$.

\vspace{3mm}
	 \textbf{Iteration tasks:}
	
		\begin{itemize}
	 	 \item[1)] Impose the set of initial conditions and initiallize the integration vectors: $y_0=a$ at the integration point $x_0$.
		 \item[2)] Calculate, sequentially, the slopes $k_i$ by evaluating the integration function $f(x)$ at the intermediate points.
		 \item[3)] Calculate the next step of the function $y_{n+1}$ throught \eqref{A5}.
	 	 \item[4)] Update the independent variable $x_{n+1}=x_n + h$. 
		 \item[5)] Repeat the process until one reaches the desired value of the independent variable $x$.
		\end{itemize}		

	A given method is especified by an integer $s$, the coefficients $a_{ij}$ (for $1\leqslant j < i \leqslant s$), $b_i$ (for $i=1,2,...,s$) and $c_i$ (for $i=2,3,...,s$). The matrix $[a_{ij}]$ is called the Runge-Kutta matrix, while the $b_i$ and $c_i$ are known as the weights and the nodes. These data are usually arranged in a mnemonic device known as a Butcher tableou (after John C. Butcher).
		\begin{figure}[H]
		 \centering
	 	 \includegraphics[scale=1.0]{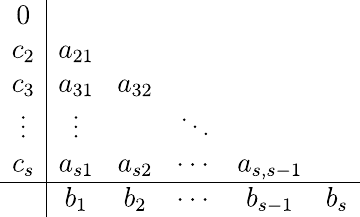}
	 	 \caption{Butcher tableau for a generic RK method.}
	 	 \label{F}
		\end{figure}
	
	A Taylor series expansion shows that the RK method is consistent if and only if
		\begin{equation}
		 \sum _{i=1} ^s b_i =1\ .
		\end{equation}
%
	\section{Adaptative step-size Runge-Kutta methods}\label{A2}
%
	Adaptative methods are designed to produce an estimate of the local truncation error of a single Runge-Kutta step~\cite{hairer1993solving,press1992adaptive}. A good ODE integrator should exert some adaptative control over its progress, making frequent adjustments to its step-size. Usually, an adaptative step-size control aims to reach a desired accuracy in the solution with minimum computational effort. Basically, several small steps should tiptoe through rought regions, while a few great strides should cover large monotonous regions. 
	
	The resulting gains in efficiency can sometimes be factors of ten, hundred or more. In addition, accuracy can be demanded not directly in the solution itself, but in some related conserved quantity that can be monitored.
	
	Implementation of an adaptative step-size control requires that the stepping algorithm signal information about its performance through an estimate of its truncation error. Observe that the calculation of this information will add to the computational overhead, however, the gains in speed will generally be repaid.
	
	There are several strategies to implement an adaptative step-size. Nevertheless, we will consider the \textit{Embedded Runge-Kutta formulas}. Initially invented by Fehlberg. An interesting fact about RK formulas is that for orders $s>4$, it is required to computation more intermediate steps than $s$ (though never more than $s+2$). 
	
	Fehlberg discovered a fifth and fourth-order method based on the same six intermediate slopes. The difference between the two approximations of $y(x+h)$ can then be used to estimate the truncation error to adjust the step-size. Since Fehlberg's original formula, several other embedded RK formulas have been found.
	
	There is, however, the danger that, by using the same evaluation points to advance the function and to estimate the error, an erroneous evaluation was performed (unlike \textit{e.g.} step-doubling, where error estimate is based on independent function evaluations). Still, experience has shown that this concern is not a problem in practice. Accordingly, embedded RK formulas, which are roughly a factor of two more efficient, have superseded algorithms based on step-doubling.
	
	Let us consider a RK method of order $s$ and $s+1$, both containing $M$ intermediate steps. The next integration point $n+1$ obtained by evaluating both methods 
		\begin{align}\label{A1}
		 y^s _{n+1} =& y_n +h\times (c_1 k_1+c_2 k_2 +...+c_Mk_M) +\mathcal{O}(h^{s+1})\ ,\\
		 y^{s+1} _{n+1} =& y_n +h\times (c_1 ^* k_1+c_2^* k_2 +...+c_M ^* k_M) +\mathcal{O}(h^{s+1})\ ,\label{A2}
		\end{align}
	and so, one can introduce the local error estimate as
		\begin{equation}\label{A6}
		 \Delta \equiv y_{n+1} ^{s+1} - y_{n+1} ^{s} = \sum_{i=1} ^{M} (c_i ^*-c_i)k_i\ . 
		\end{equation}
	Having an estimative of the error at a given point $n+1$, we need to consider keeping it within desired bounds; one has to find a way to relate $\Delta$ with $h$. Acording to \eqref{A1}-\eqref{A6}, $\Delta$ scales as $h^{s+1}$. If we take a step $h_1$ and produce an error $\Delta _1$, therefore, the step $h_0$ that would have given some other value $\Delta _0$ is already estimated as
		\begin{equation}\label{A3}
		 h_0 = h_1 \left| \frac{\Delta_0}{\Delta_1}\right| ^{\frac{1}{s+1}}\ .
		\end{equation}
	Henceforth, we will let $\Delta_0$ denote the desired accuracy. Then \eqref{A3} is used in two ways: 
		\begin{itemize}
	 	 \item If $\Delta_1$ is larger than $\Delta_0$ in magnitude, the equation tells how much to decrease the step-size when we retry the present (failed) step. 
		 \item Else, if $\Delta _1$ is smaller than $\Delta _0$, the equation tells how much we can safely increase the step-size for the next step. 
		\end{itemize}
	Local extrapolation consists in accepting the $s+1$ order value $y_{n+1} ^{s+1}$, even though the error estimate actually applies to the $s$ order value $y_{n+1} ^s$.
	
	The current notations hide that $\Delta _0$ is a vector of desired accuracies, one for each ODE in the set. In general, our accuracy requirement will be that all equations are within their respective allowed errors. We  rescale the step-size according to the needs of the ``worst-offender'' equation.
	
\vspace{3mm}	
	\textbf{Iteration tasks:}
	
		\begin{itemize}
		 \item[1)] Compute all the simultaneous intermediate slopes $k_i$ for a given integration point $x_n$.
		 \item[2)] Calculate the two different order integrated points $y_{n+1} ^{s}$ and $y_{n+1} ^{s+1}$ with the same intermediate slopes $k_i$ but different weihted averages.
		 \item[3)] Compute the local error estimate \eqref{A6}.
	 	 \item[4)] If the local error estimate is larger than the desired accuracy, correct the integration step throught \eqref{A3} and repeat the process.
		 \item[5)] Else, update all the integration dependent quantities ($x_{n+1},\, y_{n+1}=y_{n+1}^{s+1},\, h,\, ...$) with the higher order method, and redo the entire process for the next integration step.
		\end{itemize}	
	In our work, we used an adaptive step-size embedded $5(6)$-RK method (\textit{a.k.a.} Runge-Kutta-Verner method~\cite{pfister2015literature,fawzi2016embedded,macdonald2003constructing,press1989integration, balac2013embedded}). The corresponding Butcher tableu follows.
		\begin{table}[H]
			\begin{center}
			 \caption{Butcher Tableau for a $5(6)$ Runge-Kutta-Verner method \cite{verner1978explicit}.}
			 \vspace{2mm}
				\begin{tabular}{ c |c ccccccc }
			  	 $c_i$ & $a_{ij}$ &  &  &  &  &  &  & \\ 
				 $\frac{1}{18}$ & $\frac{1}{18}$ &  &  &  &  &  &  & \\ 
				 $\frac{1}{6}$ & $-\frac{1}{12}$ & $\frac{1}{4}$  &  &  &  &  &  & \\ 
				 $\frac{2}{9}$ & $-\frac{2}{81}$ & $\frac{4}{27}$ & $\frac{8}{81}$ &  &  &  &  & \\ 
				 $\frac{2}{3}$ & $\frac{40}{81}$ & $-\frac{4}{11}$ & $-\frac{56}{11}$ & $\frac{54}{11}$  &  &  &  & \\ 
				 $1$ & $-\frac{369}{73}$ & $\frac{72}{73}$ & $\frac{5380}{219}$ & $-\frac{12285}{584}$  & $\frac{2695}{1752}$ &  &  & \\ 
				 $\frac{8}{9}$ & $-\frac{8716}{891}$ & $\frac{656}{297}$ & $\frac{39520}{891}$ & $-\frac{416}{11}$ & $\frac{52}{27}$  & $0$ &  & \\ 
				 $1$ & $\frac{3015}{256}$ & $-\frac{9}{4}$ & $-\frac{4219}{78}$ & $\frac{5985}{128}$ & $-\frac{539}{384}$ & $0$ & $\frac{693}{3328}$ & \\
				 \hline
				 $b_j ^5$ & $\frac{3}{80}$ & $0$ & $\frac{4}{25}$ & $\frac{234}{1120}$ & $ \frac{77}{160}$ & $\frac{73}{700}$ & $0$ & $0$\\
				 $b_j ^6$ & $\frac{57}{640}$ & $0$ & $-\frac{16}{65}$ & $\frac{1377}{2240}$ & $ \frac{121}{320}$ & $0$ & $\frac{891}{8320}$ & $\frac{2}{35}$			
				\end{tabular}
			\end{center}
		\end{table}
%
	\section{Parallelization of the integration}
%
	Parallel computing~\cite{almasi1994highly,kumar2008parallel} is a type of computation in which many calculations or processes are carried out simultaneously. If two computations are independent of each other (and respective outputs), they can be done simultaneously instead of one after the other. As one can imagine, for computation with many calculations, even two processes in parallel can already decrease the computation time almost in half. Depending on the computation hardware and the integration structure, one can decrease the computational time by more than $10$ times\footnote{Observe that while parallel computing is a potent tool, it adds complexity to the code. For some more basic computation, a well-performed optimization is preferable.}.
	
	We decided to allocate $4$ threads to each running code. In other words, we will parallelize four computations at a time. To obtain the maximum out of our parallelization, one should parallelize the procedure containing the most computations. In our case, this is the computation of all the intermediate slopes.
	
	However, if one carefully observes the integration strategy that we are using (Sec.~\ref{A1}-\ref{A2}), one can observe that a given slope $k_i$ is dependent on the previous step $k_{i-1}$, making it impossible to parallelize the slope computation. The same occurs for the integration computation.
	
	Note that one could estimate the optimal step-size of the integrator at each integration point. However, this requires several step-size trials, which are all independent from each other. On the other hand, the computation for a given step-size can be done independently. Hence, being a possible point to tackle and parallelize. 
	
	In our current case, since we aim at using $4$ threads, we will use one thread to compute with the previous step-size ($h$), one with the double of the previous step-size ($2\times h$), one with half ($h/2$) and one with a quarter ($h/4$). Observe that such a strategy can be further generalized to contain more threads and hence perform more trials of the step-size simultaneously. However, we have observed that increasing the parallelization after $4$ reaches the point of diminishing returns.
	
	Since we assume that the solutions are well behaved and have the complicated/easy regions in a well-defined region, we will also update the step-size for the next point with the current, accurate one. This will help us accurately describe the initial complicated region with a small step-size and quickly cover the constant decaying tail with a large step-size.
	
\vspace{3mm}	
	\textbf{Iteration tasks:}
	
		\begin{itemize}
		 \item[1)] With $h$ as the initial step-size, compute the additional integration step-sizes.
		 \item[2)] Compute, in parallel, the next integration point for each integration step-size and correspondent local error estimate.
		 \item[3)] If there are solutions with a local error estimate smaller than the desired accuracy, propagate the solution with the higher step-size and update all the required quantities.
		 \item[4)] Else, use the smallest step-size as the new $h$ and repeat the process.
		\end{itemize}	
%
%
	\section{Shooting strategy}
%
	In numerical analysis, the shooting is a method for solving boundary value problem~\cite{ziegel1987numerical}. It transforms the boundary value problem into several initial value problems with different initial conditions. The latter's are used to find the solution that satisfies the boundary conditions of the boundary value problem. In layman's terms, one ``shoots'' out trajectories in different directions from one boundary until one finds the trajectory that ``hits'' the other boundary condition.
	
	Suppose one wants to solve the boundary-value problem \eqref{A0}, with an additional boundary condition at some point $x_f$ such that $y(x_f)=y_f$. 
	
	We know that a $x_0$ such that $y(x_0=u)=a$ solves the boundary conditions problem, but we do not know the precise value of $u$.
	
	The shooting method is a process of solving the initial value problem for many different values of $x_0$ until one finds the right solution $x_0=u$ that satisfies the desired boundary conditions. Typically, one does so numerically. The solution(s) correspond to root(s) of
		\begin{equation}
		 F(u)=y(u)-y_1\ .
		\end{equation}
	To systematically vary the shooting parameter $x_0$ and find the root, one can employ standard root-finding algorithms like the bisection method (Sec.~\ref{A.4.1}) or the Secant strategy (Sec.~\ref{A.4.2}), the two strategies that we have implemented.
	
	Roots of $F$ and solutions to the boundary value problem are equivalent. If $u$ is a root of $F$, then $y(u)$ solves the boundary value problem. Conversely, if the boundary value problem has a solution $y(u)$, it is also the unique solution $y(x_0)$ of the initial value problem where $u=x_0$, so $u$ is a root of $F$.
%
		\subsection{Bisection method}\label{A.4.1}
%
	The bisection method~\cite{ziegel1987numerical,corliss1977root,la2002numerical} is a root-finding method that can be applied to any continuous functions. The method consists on the bisection of an interval where the continuous function changes sign (this method relies on Bolzano's intermediate value theorem).
	
	The method is applicable for numerically solving the equation $F(x)=0$ for the real variable $x$, and $F$ a continuous function in the interval $[a,b]$ with $F(a)$ and $F(b)$ having opposite signs. In this case, by the intermediate value theorem, the continuous function $F$ must have, at least, one root in this interval $F(c)=0$.
	
	At each step, the method divides the interval into two parts/halves by computing the midpoint $c=(a+b)/2$ of the interval and the value of the function $F(c)$ at that point. Unless $c$ itself is a root, there are two possibilities: either $F(a)$ and $F(c)$ have oppositive signs and bracket a root, or $F(c)$ and $F(b)$ have opposite signs and bracket a root. The method selects the sub-interval that is guaranteed to bracket a root. In this way, an interval that contains a root of $F$ is reduced in width by $50\%$ at each step. The process is continued until the interval is sufficiently small. 
		
	Explicitly, if $F(a)$ and $F(c)$ have opposite signs, then the method sets $c$ as the new value $b$, and if $F(b)$ and $F(c)$ have opposite signs, then the method sets $c$ as the new $a$. It guarantees that the extremes of the intervals have the opposite sign and hence a root in between.
	
\vspace{3mm}	
	\textbf{Iteration tasks:}
			\begin{itemize}
			 \item[1)] Stablish the interval $[a,b]$ such that a root is guaranteed to exist in it.
			 \item[2)] Calculate the midpoint of the interval $c=\frac{a+b}{2}$.
			 \item[3)] Calculate the function value at the midpoint $F(c)$.
			 \item[4)] If the convergence is satisfactory, return $c$ and stop the iteration.
			 \item[5)] Examine the sign of $F(c)$ and replace either $\big( a,F(a)\big)$ or $\big( b,F(b)\big)$ with $\big( c,F(c)\big)$, so that there is a zero crossing within the new interval.
			\end{itemize}		
	The method is guaranteed to converge to a $F$'s root if the latter is a continuous function on the interval $[a,\ze b]$ and $F(a)$ and $F(b)$ have opposite signs. The absolute error is halved at each step, so the method converges linearly. Specifically, if $c_1=\frac{a+b}{2}$ is the midpoint of the initial interval, and $c_n$ is the midpoint of the interval in the $n^{th}$ step, then the difference between $c_n$ and a solution $c$ is bounded by
			\begin{equation}
			 |c_n -c| \leqslant \frac{|b-a|}{2^n}\ .
			\end{equation}
	Despite the bisection method being optimal concerning a worst-case performance under absolute error criteria, it is sub-optimal for average performance under standard assumptions and asymptotic performance. Better performance can be achieved with the Secant strategy (discussed next Sec.~\ref{A.4.2}). However, the robustness and guaranteed convergence of the method is a welcome tradeoff for the more difficult numerical problems, which is why we use it. 
%
		\subsection{Secant method}\label{A.4.2}
%
	In numerical analysis, the secant method~\cite{aziz2014numerical,papakonstantinou2013origin,avriel2003nonlinear} is a root-finding algorithm that uses a succession of roots of secant lines to better approximate a root of a function $F$. The secant method can be thought of as a finite-difference approximation of Newton's method. However, the secant method predates Newton's method by over $300$ years.
	
	The method is applicable to solve the same type of problems as the bisection method, namely, solving the $F(x)=0$ equation. However, the secant method does not require the root to remain bracketed and hence does not always converge. 
	
	The recurrence relation defines the secant method
			\begin{equation}\label{A4}
			 x_{n+1} = x_{n} -F(x_n)\frac{x_n - x_{n-1}}{F(x_n)-F(x_{n-1})}\ .
			\end{equation}
	As one can see, the secant method requires two initial values $x_0$ and $x_1$, which should, ideally, be chosen to lie close to the real root.

\vspace{3mm}
	\textbf{Iteration tasks:}
	
			\begin{itemize}
			 \item[1)] Stablish the two initial guesses $x_0$ and $x_1$ such that they are close to a root of $F(x)$.
			 \item[2)] Calculate the function value at $F(x_{n-1})$.
			 \item[3)] Calculate the function value at $F(x_{n})$.
			 \item[4)] If the convergence is satisfactory, return $x_n$ and stop the iteration.
			 \item[5)] Else, compute the new guess $x_{n+1}$ with the recurrence relation \eqref{A4}, so that one can obtain a better guess for the next trial.
			\end{itemize}
	The iterates $x_n$ of the secant method, converge a $F$'s root when the initial values $x_0$ and $x_1$ are sufficiently close to it. The order of convergence is $\approx 1.618$ (the golden ratio). In other words, the convergence is super-linear but not quadratic. 
	
	If the two initial guesses are not ``close enough'' to the root, then there is no guarantee that the secant method converges. In addition, the presence of a difference in the denominator of \eqref{A4} opens the possibility of a numerical division by zero and still not finding a root with the desired accuracy. 
	
	In general, we use the secant method to obtain the roots of the solutions. If the latter cannot obtain or converge to the solution, we will then utilize the Bisection method.
	
\clearpage\null\newpage	
	
%
\chapter{PDE Solver}\label{B}
%
	In mathematics, a PDE can be viewed as a generalization of the ODEs to include more independent variables.	The order of a PDE is the order of the highest derivative involved. These are used to mathematically formulate and thus aid the solution of, physical and other problems involving functions of several variables.
	
	The dependence on more than one independent variable makes PDEs a much more complicated problem to solve. Due to that, it would be impracticable to create our code. Hence, to solve a system of PDEs and impose the proper boundary conditions, we will use the professional CADSOL/FIDISOL program package. 
	
	The FIDISOL/CADSOL (Finite Difference Solver/Cartesian Arbitrary DOmain Solver)~\cite{gross1993fidisol,schonauer2001we,schauder1992cadsol} is a program package designed to numerically solve PDEs with arbitrary boundary conditions. The numerical method is the finite difference method with fixed step-size and restricted to a rectangular domain. The package is written in \textsc{Fortran} programing language.

%
	\section{Finite difference methods}\label{C.1}
%
	In mathematical analysis, the finite difference methods (FDM)~\cite{gautschi1997numerical,grossmann2007numerical,iserles2009first} are a class of numerical techniques developed to solve differential equations through the approximation of each derivative with finite differences. For this, the integration domain is discretized, and the value of the solution at these discrete points is approximated by solving algebraic equations containing finite differences and values from nearby points.
	
	 The FDM converts PDEs, which may be non-linear, into linear equations systems that matrix algebra techniques can solve.
	 
	 Consider a well-behaved function $f(x)$ whose derivatives are approximated. At an unknown point $x=x_0+h$, one can obtain an approximation to the actual value expanding in a Taylor series around the known point $x=x_0$
	\begin{equation}
	f(x_0+h) = f(x_0) +\frac{f'(x_0)}{1!}h+\frac{f^{(2)} (x_0)}{2!}h^2+...+\frac{f^{(n)}(x_0)}{n!}h^n+\mathcal{O}(h^{n+1})\ ,
\end{equation}	  
	where $n!$ denotes the factorial of the Taylor polynomial degree $n$. As an example of the procedure, let us compute the finite difference expression for the first derivative of $f$ by first truncating the Taylor polynomial:
	\begin{equation}
	f(x_0+h)=f(x_0)+f'(x_0)h+\mathcal{O}(h^2)\ .
	\end{equation}
	Assuming that the higher-order contributions $\mathcal{O}(h^2)$ are sufficiently small, one can rearrange the expression in order to separate $f'(x_0)$ and obtain the approximation of the first derivative of $f$
	\begin{equation}
	f'(x_0) \approx \frac{f(x_0+h)-f(x_0)}{h}\ ,
	\end{equation}
 	which can easily be transformed into the recurrence relation that computes the derivative at a given point $x_n$ from the difference between the values of the function at the point $x_{n+1}$ and $x_n$
	\begin{equation}
	f'(x_n) = \frac{f(x_{n+1})-f(x_n)}{h}\ .
	\end{equation}
	The two primary sources of error in a FDM are the round-off error -- the loss of precision due to computer rounding of decimal quantities -- and truncation error -- the loss of precision from the non-consideration of all the Taylor series terms. 
	
	The expression for the local truncation error of a given method with order $n$ is given by
	\begin{equation}
	\mathcal{O}\big( h^{(n+1)}\big) =\frac{f^{(n+1)}}{(n+1)!}h^{n+1}\ .
	\end{equation}
%
%
	\section{Newton-Raphson method}\label{S1}
%
	In numerical mathematics, Newton's method~\cite{anton2014calculo,gil2007numerical,suli2003introduction}, also known as the Newton-Raphson method, is a root-finding algorithm that produces successively better approximations to the roots of a real-valued function. The latter must be differentiable $f'$, and and initial guess close to the root is required. 
	
	If the function satisfies the previous assumptions, the new, better guess $x_1$ to the function's root comes as
		\begin{equation}
	 	 x_1 = x_0 - \frac{f(x_0)}{f'(x_0)}\ ,
		\end{equation}
	Geometrically, $(x_1,\ze 0)$ is the intersection of the $x$-axis and the tangent to the graph of $f$ at $\big( x_0,\ze f(x_0)\big)$: the improved guess is the unique root of the linear approximation at the initial point. The resulting recurrence relation is
		\begin{equation}
		 x_{n+1} = x_n - \frac{f(x_n)}{f'(x_n)}\ .
		\end{equation}
	In principle, the method is quadratic in convergence. However, if the root sought has a multiplicity greater than one, the convergence is linear.
	
	One may also use Newton's method to solve systems of $k$ equations, which amounts to finding the (simultaneous) zeros of $k$ continuously differentiable functions $f:\, \mathbb{R} \to \mathbb{R} $. This is equivalent to finding the roots of a single vector-valued function $F:\, \mathbb{R}^k\to \mathbb{R} ^k$. In the formulation given above, the scalars $x_n$ are replaced by vectors $\textbf{x}_n$ and instead of dividing the function $f(x_n)$ by its derivatives $f'(x_n)$ one has to left multiply the function $F(\textbf{x}_n)$ by the inverse of its $k\times k$ Jacobian matrix $J_F (\textbf{x}_n)$. This results in the expression
		\begin{equation}
		 \textbf{x}_{n+1}=\textbf{x}_n -J_F (\textbf{x}_n) ^{-1} F(\textbf{x}_n)\ .
		\end{equation}	 
	Rather than computing the inverse of the Jacobian matrix (matrix with all the computed derivatives of $F$ \textit{w.r.t.} the independent variables), one may save time and increase numerical stability by solving the system of linear equations
		\begin{equation}
		 J_F(\textbf{x}_n) (\textbf{x}_{n+1}-\textbf{x}_n)=-F(\textbf{x}_n) \ ,
		\end{equation}
	For the unknown $\textbf{x}_{n+1}-\textbf{x}_n$.
	
	The Newton's-Raphson method is a powerful technique to compute roots of a function -- moreover, it can also be used as a shooting strategy--; however, it is plagued by some difficulties:
		\begin{itemize}
		 \item \textbf{Difficulty in calculating the derivative of a function:} Newton's method requires that the derivative can be calculated directly. An analytical expression for the derivative may not be easily obtainable (especially in the shooting application).
		 \item \textbf{Failure to converge:} Newton's method is only guaranteed to converge if certain conditions are satisfied. 
		 \item \textbf{Invertebility of the Jacobian matrix:} The Jacobian of all functions must be provided, which can be a taxative procedure. In addition, the computation of the inverse of the Jacobian matrix is a problem by itself.
		\end{itemize}
%

%
	\section{FIDISOL/CADSOL professional package}\label{S1}
%
	Since the package is based on a FDM and a Newton's-Raphson method, it expects several key points:

	\textbf{Program usage:}
		\begin{itemize}
		 \item The system of PDEs must be written in the form:
			\begin{equation}
			 F(x,y,f; \partial _i f;\partial _{ij} f) = 0\ ,
			\end{equation}
	where $f$ is the set of functions that we want to find an approximation to, and $i,\ze j = (x,\ze y)$ are the generic independent variables. 
	
		 \item The Jacobian $J_F$ of all functions must be provided. This can easily be computed through the calculus of the derivatives of $F$ concerning $f$ and its derivatives. 
		 \item Provide an initial guess close enough to the root we want to obtain. 

		 \item Provide a mesh for the independent variables $(x,y)$ with $N_x \times N_y$ points.
	
 		 \item The FDM requires an initial solutions profile.
	
		 \item Provide the set of proper boundary conditions.
	
		\end{itemize}
	Luckily, after some iterations, the solver should converge and obtain a numerical approximation to the function $f$. 
%
\chapter{Exact solutions with a linear coupling}\label{C}
%

%
	\section{Purely electric  BHs}\label{C.1}
%
	Purely electric dilatonic solutions of~\eqref{E2.1.1} with the dilatonic coupling~\eqref{E1.2.6} where first considered by Gibbons and Maeda~\cite{gibbons1986black,gibbons1988black} and Garfinkle, Horowitz and Strominger~\cite{garfinkle1991charged}. The BH solution has the line element~\eqref{E1.5.40} with $\sigma (r)=1$ and
		\begin{equation}\label{EC.1.1}
		 N(r) =\left(1-\frac{r_+}{r} \right)\left(1-\frac{r_-}{r}\right)^{\frac{1-\alpha^2}{1+\alpha^2}}\ , \qquad r=r \left(1-\frac{r_-}{r}\right)^{\frac{ \alpha^2}{1+\alpha^2}}\ , 
		\end{equation}
	together with the Maxwell potential and dilaton field\footnote{Following the conventions in the work, we fix $\phi(+\infty)=0$ for all solutions in the Appendix.}
		\begin{equation}
		 A=\frac{Q_e}{r}dt \ , \qquad e^{2\ze\alpha\ze \phi}=\left(1-\frac{r_-}{r}\right)^{\frac{2\alpha}{1+\alpha^2}}\ .
		\end{equation}
	The two free parameters $r_{+}$ and $r_{-}$ (with $r_-<r_+$) are related to the ADM mass, $M$, and (total) electric charge, $Q_e$, by
		\begin{equation}   
		 M = \frac{1}{2}\left[r_+ +\left(\frac{1-\alpha^2}{1+\alpha^2}\right)r_-\right] \ , \qquad Q_e=\left(  	\frac{r_-\,r_+}{1+\alpha^2}\right)^{\frac{1}{2}} \ .
		\end{equation}
	For all $\alpha $, the surface $r= r_H=r_+$ is the location of the (outer) event horizon, with
		\begin{equation} 
		 A_H=4\pi r_+^2\left(1-\frac{r_-}{r_+} \right)^{\frac{2\ze \alpha^2}{1+\alpha^2}},~~
		 T_H=\frac{1}{4\pi}\frac{1}{r_+-r_-}\left(1-\frac{r_-}{r_+}\right)^{\frac{2}{1+\alpha^2}} \ .
		\end{equation}
	The extremal limit, which corresponds to the coincidence limit $r_- = r_+$, results in a singular solution (as can be seen $e.g.$ by evaluating the Kretschmann scalar). In this limit, the event horizon area goes to zero for $\alpha \neq 0\ze$. However, the Hawking temperature only goes to zero in the extremal limit for $\alpha<1$, while for $\alpha=1$, it approaches a constant, and for $\alpha>1$, it diverges.

	The reduced quantities~\eqref{E2.1.15a}-\eqref{E2.1.15c} have the following exact expressions:
		\begin{align}
		& q=\frac{2\sqrt{(1+\alpha^2)x}}{1+\alpha^2(1-x)+x}\ ,\qquad a_H=\frac{(1+\alpha^2)^2(1-x)^{\frac{2\alpha^2}{1+\alpha^2}}}{(1+\alpha^2(1-x)+x)^2}\ ,\nonumber\\
		 & t_H=\frac{(1-x)^{\frac{1-\alpha^2}{1+\alpha^2}}(1+\alpha^2(1-x)+x)}{1+\alpha^2}  \ ,
		\end{align}
	where $0\leqslant x\leqslant 1$ is a parameter.
	
%
	\section{Dyonic BHs}
%

%
		\subsection{$\alpha=1$}
%
	A dyonic dilatonic BH solution of \eqref{E2.1.1} with the dilatonic coupling~\eqref{E1.2.6} and $\alpha=1$, was found in~\cite{kallosh1992supersymmetry}, and extensively discussed in the literature, since it can be embedded in ${\cal N}=4$ supergravity. Taking the form \eqref{E1.5.40} with $\sigma =1$ and
			\begin{equation}\label{EC.1.6}
			 \phi= \frac{1}{2} \ln \frac{r+Q_\phi}{r-Q_\phi}\ ,\qquad  N= \frac{(r-r_{+})(r-r_{-})}{r^{2}-Q_\phi^{2}}\ ,\qquad r^2=  r^{2}-Q_\phi^{2} \ ,
			\end{equation}
	where
			\begin{equation}\label{EC.1.7}
			 r_{\pm}=M\pm \sqrt{M^{2}+Q_\phi- {Q_e}^{2}- {P}^{2}}\ ,
			\end{equation}
	with the outer horizon at $r_H=r_+$, while $M,\, Q_e$ and $P$ are the mass and electric and magnetic charges. $Q_\phi$ corresponds to the scalar charge, which, however, is not an independent parameter (the hair is secondary):
			\begin{equation}\label{EC.1.8}
			 Q_\phi =\frac{ {P}^{2}- {Q_e}^{2}}{2M}\ .
			\end{equation}
	The extremal limit of the above solution corresponds to $r_+=r_-$, in which case one finds two relation between the charges
			\begin{equation} 
			 0 = M^{2}+Q_\phi^{2}- {Q_e }^{2}- {P}^{2} \ \Longrightarrow  \ \ (M+Q_\phi)^2 -2 {P}^{2}=0 \qquad {\rm and} \qquad (M-Q_\phi)^2 -2 \ze {Q_e}^{2}=0 \ .
			\end{equation}  
	The horizon area and Hawking temperature of the solutions are 
			\begin{equation}\label{EC.1.10}
			 A_H=4 \pi \big( 2 M r_+ -P^2-Q_e^2\big)\ , \qquad T_H=\frac{1}{2\pi} \frac{r_+-M}{2M r_+-P^2-Q_e ^2} \ .
			\end{equation}
	The expression of the reduced quantities is  more involved in this case:
			\begin{equation}\label{EC.1.11}  
			 a_H=\frac{1}{4} (2x-q^2)\ , \qquad t_H=\frac{4(x-1)}{2x-q^2} \ ,
			\end{equation}
	with $x$ a parameter expressed in terms of $q$ as a solution of the equation (where $k=\frac{P}{Q_e}$)
			\begin{equation}\label{EC.1.12}  
			 q^4-\frac{4(1+k^2)^2}{(1-k^2)^2}\big( q^2+x(x-2)\big)=0 \ .
			\end{equation}
%

%
		\subsection{$\alpha=\sqrt{3}$}
%
	A dyonic dilatonic BH solution of \eqref{E2.1.1} with the dilatonic coupling~\eqref{E1.2.6} and $\alpha=\sqrt{3}$, was found in~\cite{gibbons1986black,dobiasch1982stationary}. This case arises from a suitable Kaluza-Klein reduction of a five-dimensional vacuum BH. In the extremal limit, one obtains a non-BPS BH that can be embedded in ${\cal N}=2$ supergravity.

	The generic solution can be written again in the form \eqref{E1.5.40} with $\sigma =1$ and
			\begin{equation}\label{EC.2.}
 			 N =\frac{(r-r_{+})(r-r_{-})}{\sqrt{AB}}\ , \qquad r^{2}= \sqrt{AB} \qquad {\rm and} \qquad  e^{\frac{4\ze \phi}{\sqrt{3}}} = \frac{A}{B} \ ,
			\end{equation}
	where
			\begin{equation}
			 A= (r-r_{A_{+}})(r-r_{A_{-}})\ , \qquad B= (r-r_{B_{+}})(r-r_{B_{-}}) \ .
			\end{equation}
	In the above relations one defines
			\begin{equation}
			 r_{\pm}=M\pm\sqrt{M^{2}+Q_\phi^{2}-{P}^{2}-{Q_e}^{2}} \ ,
			\end{equation}
	where, again, the outer horizon is at $r_H=r_+$, and
			\begin{equation}
			 r_{A_{\pm}}=\frac{1}{\sqrt{3}}Q_\phi\pm {P}\sqrt{\frac{2Q_\phi}{Q_\phi-\sqrt{3}M}} \ , \qquad r_{B_{\pm}}=-\frac{1}{\sqrt{3}}Q_\phi\pm {Q_e}\sqrt{\frac{2Q_\phi}{Q_\phi+\sqrt{3}M}} \ .
			\end{equation}
	The solution possesses again three parameters $M,\, Q_e$ and $P$ which fix the scalar charge $Q_\phi$ via the equation
			\begin{equation}
			 \frac{2}{\sqrt{3}}Q_\phi=\frac{ {Q_e }^{2}}{\sqrt{3}M+Q_\phi}- \frac{ {P}^{2}}{\sqrt{3}M-Q_\phi} \ ,
			\end{equation}
	while the horizon area and the Hawking temperature are given by
			\begin{align}
			 & A_H=4\pi\sqrt{(r_{+}-r_{A_{+}})(r_{+}-r_{A_{-}})(r_{+}-r_{B_{+}})(r_{+}-r_{B_{-}})}\ , \\
			 & T_H=\frac{1}{4\pi}\frac{r_+-r_-}{\sqrt{(r_{+}-r_{A_{+}})(r_{+}-r_{A_{-}})(r_{+}-r_{B_{+}})(r_{+}-r_{B_{-}})}}\ .	
			\end{align} 
	The corresponding expressions for $a_H$ and $t_H$ as a function of $q$ (and the ratio $P/Q_e$) can be derived directly from the above relations; however, they are too complicated to include here.
	
\clearpage\null\newpage	
	
%
\chapter{Five dimensional vacuum Einstein gravity analogy}\label{O}
%
	An influential result in BH physics was the discovery of the vacuum black ring in five-dimensional Einstein's gravity~\cite{emparan2001supergravity,emparan2006black}. Black rings come in two types (fat and thin) and co-exist with the Myers-Perry BH~\cite{myers1986black} in five-dimensional vacuum gravity, being distinguished by their event horizon topology. They carry two physical parameters, mass $M$ and angular momentum $J$, but a set of reduced quantities typically characterises them: respectively, the reduced angular momentum, horizon area and temperature:
	\begin{equation}\label{E320}
	 j=\frac{3}{4}\sqrt{\frac{3 \pi}{2 }}\frac{J}{M^{3/2}}\ ,\qquad \quad
	 a_H=\frac{3}{16} \sqrt{\frac{3}{\pi}} \frac{A_H}{M^{3/2}}\ ,\qquad \quad
	 t_H=4 \sqrt{\frac{\pi}{3}} T_H \sqrt{M}\ . 
	\end{equation}
	The overall factors in the above expressions are taken to agree with the usual conventions in the literature~\cite{emparan2001supergravity,emparan2006black}. 

\begin{figure}[h!]
	 \centering
 		 \begin{picture}(0,0)		 	 	
			 	 \put(174,38){\small BRs}	
			 	 \put(50,96){\small MP BHs}				 	 				 	 			 	 			 	 
			 	 \put(112,-8){\small $q$}
			 	 \put(0,78){\begin{turn}{90}{\small $a_H$}\end{turn}}
			\end{picture}	 
	 \includegraphics[scale=0.6]{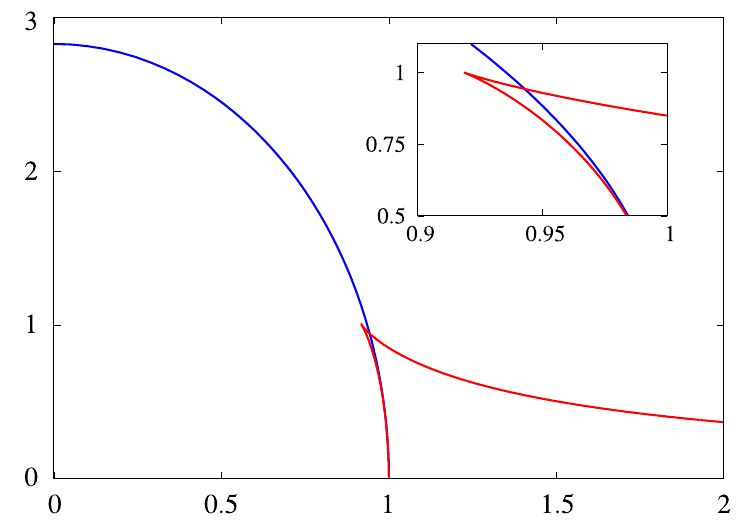}
	 \includegraphics[scale=0.6]{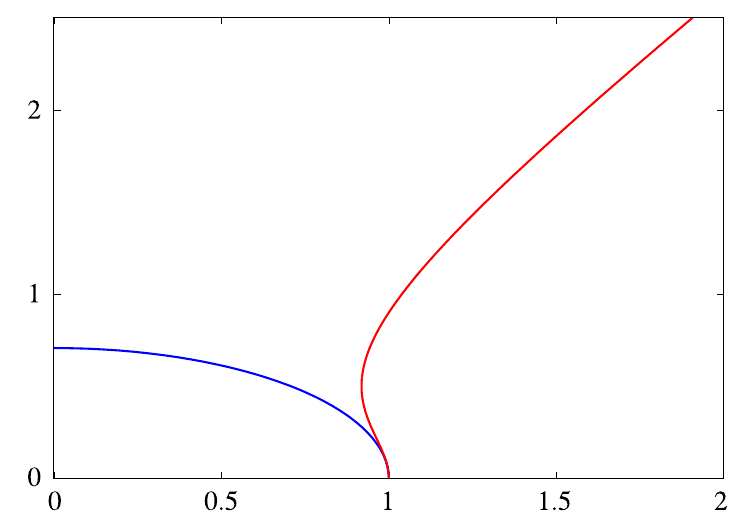}
	  		 \begin{picture}(0,0)		 	 	
			 	 \put(174,125){\small BRs}	
			 	 \put(24,72){\small MP BHs}				 	 				 	 			 	 			 	 
			 	 \put(114,10){\small $q$}
			 	 \put(0,90){\begin{turn}{90}{\small $t_H$}\end{turn}}
			\end{picture}	 
	 \caption{(Left panel) branches of five-dimensional vacuum gravity black rings (red curves) and Myers-Perry BHs (blue curve) in a reduced area $vs.$ reduced angular momentum diagram. (Right panel) reduced temperature $vs.$ reduced angular momentum for the same solutions. Black rings are colder when they are fat (in the first branch) and hotter when they are thin (in the second branch).}
	 \label{F22}
	\end{figure}
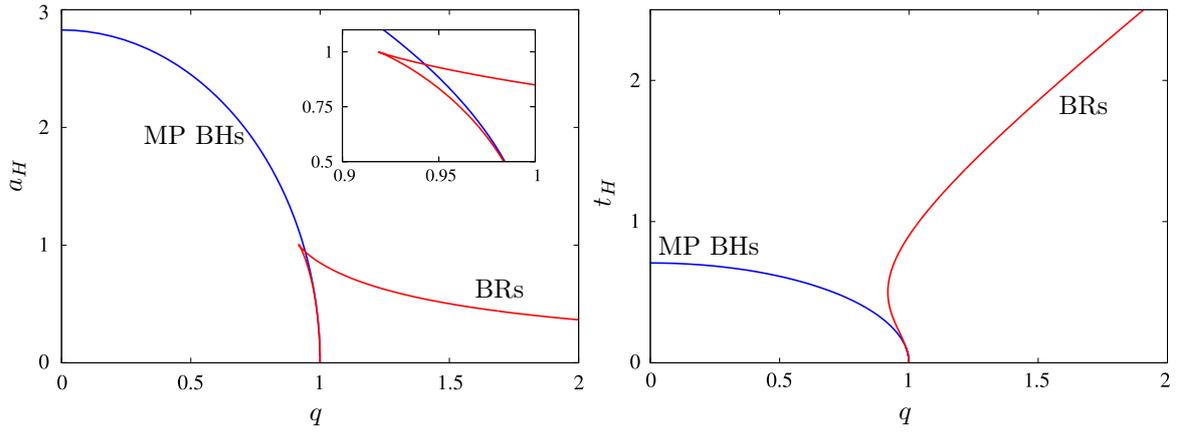

	In Figure~\ref{F22} we exhibit the BHs of vacuum five-dimensional gravity in a reduced area (left panel) and a reduced temperature (right panel) $vs.$ reduced angular momentum plot.
	The parallelism with Fig.~\ref{F22} (left panel) and~\ref{F2.11} is uncanny, with [Myer-Perry BHs, fat rings, thin rings] playing the role of [RN, cold scalarized, hot scalarized] BHs and the reduced angular momentum being mapped to the reduced charge.  

	In particular, one observes that the Myers-Perry (RN) BHs exist for a finite range of $0\leqslant j \leqslant 1$ ($0\leqslant q \leqslant 1$), where the solution with $j=0$ ($q=0$) corresponds to the Tangherlini (Schwarzschild) BH. Black rings (scalarized BHs), on the other hand, can be overrotating (overcharged). At $j=1$, the fat black rings and the Myers-Perry BHs degenerate to the same (singular) extremal solution. On the other hand, the cold scalarized BHs connect with the (regular) extremal RN BH at $q=1$. 

	Fat black rings become thinner, with lower $j$ and larger $a_H$ until a bifurcation point, where they become thin black rings. Cold scalarized BHs become hotter, with lower $q$ and larger $a_H$ until a bifurcation point, where they become hot scalarized BHs. Thin black rings (hot scalarized BHs) become overspinning (overcharged). At the moment, however, no additional weight can be given to this curious analogy. 

%
\chapter{Vector decomposition}\label{F}
%
	In mathematics, vector spherical harmonics (VSH) extend the scalar spherical harmonics (SH). Following the convention of Barrera \textit{et al.} \cite{barrera1985vector}, for a certain spherical harmonic $Y_\ell ^m (\theta, \phi )$, the three VSH are defined has:
	\begin{equation}
	 \textbf{Y}_\ell ^m = e_r \  Y_\ell ^m \ , \qquad \qquad \Psi _\ell ^m = r\ \nabla  Y_\ell ^m\ ,\qquad \qquad \chi _\ell ^m = \vec{r}\times \nabla Y_\ell ^m \ . 
	\end{equation}
	In the current convention $e_r$ is the radial unit vector and $\vec{r}$ corresponds to a radial vector with norm equal to the radius: $\vec{r}=(r,0,0)$. The radial factors are included to guarantee the dimension consistency of the VSH. The angular dependency of VSH and the usual SH are intentionally omitted for simplicity. 
	
	Similar to the SH case, the main interest in the VSH basis is the separation of the radial and angular dependences of a vector field. In the most generic case, a vector field $\overrightarrow{W}$ admits the expansion in vectorial multipoles 
	\begin{equation}
	 \overrightarrow{W}=\sum _{\ell=0} ^{+ \infty} \sum _{m=-\ell} ^\ell \left[ ^r W_\ell ^m (r) \textbf{Y} _\ell ^m  + ^{(1)}W_\ell ^m (r) \Psi _\ell ^m + ^{(2)} W_\ell ^m (r)\chi _\ell ^m \right]\ .
	\end{equation}
	The labels in the $W$ components reflect that $^r W _\ell ^m$ is the radial component of $\overrightarrow{W}$, while $^{(1)} W_\ell ^m$ and $^{(2)} W_\ell ^m$ are two transverse components with respect to $e_r$.
	
	VSH are symmetric and orthogonal. With this properties it is possible to obtain a formula for each vectorial multipolar function:
	\begin{align}
	 ^r W _\ell ^m  &= \int d\Omega _2\, \overrightarrow{W}\cdot \bar{\textbf{Y}}_\ell ^m \ ,\\
	  ^{(1)} W_\ell ^m &=\frac{1}{\ell(\ell+1)} \int d\Omega _2\, \overrightarrow{W}\cdot \bar{\Psi}_\ell ^m \ ,\\
	  ^{(2)} W_\ell ^m  &=\frac{1}{\ell(\ell+1)} \int d\Omega _2\, \overrightarrow{W}\cdot \bar{\chi} _\ell ^m\ .
	\end{align}
	With this method, one can numerically decompose any vector field in the VSH basis and isolate the radial and angular dependences of the field. Observe that the previous procedure is laborious but generic.

\clearpage\null\newpage

%
\chapter{Derrick's theorem in higher dimensions}\label{I}
%
	Consider the $D=n+1$ dimensional flat spacetime with the metric
	\begin{equation}
	 ds_D^2 = -dt^2 + \sum_{i}^{n}dx_i^2 \ .
	\end{equation}
	The scalar field action is now
	\begin{equation}
	 \mathcal{S}_D=\int dt \int d^{n} x\left[-\Phi_{,\ze\mathcal{M}}\ze \bar{\Phi}^{,\ze\mathcal{M}} -U(\Phi)\right] \ .
	\end{equation}
	where the index $\mathcal{M}$ takes values between $0$ and $n$. By following the same arguments as above, we obtain
	\begin{equation}
	 \mathcal{S}_D=-\int dt\, E^D = -\int dt \big(I_1^D + I_2^D\big) \ ,
	\end{equation}
	where
	\begin{equation}
	 I_1^D\equiv \int d^{n}x \, (\nabla_{n} \Phi)^2 \ , \qquad I_2^D\equiv \int d^{n}x\, U \ ,
	\end{equation}
	with $\nabla_{n}$ being the $n$-dimensional spatial gradient. Assuming once again the same 1-parameter family of configurations $\Phi_\lambda({\bf r})=\Phi(\lambda\ze {\bf r})$ and extremizing the energy in the same way, we obtain the following virial identity
	\begin{align}\label{EG.0.5}
	 \left(\frac{dE^D_\lambda}{d\lambda}\right)_{\lambda=1}&= (-n+1)I_1^D-(n+1)I_2^D =0 \ .\qquad {\rm {\bf [virial \ Derrick \ higher \ D]}}
	\end{align}
	Moreover, the stability condition is, using the virial identity,
	\begin{align}
	 \left(\frac{d^2E^D_\lambda}{d\lambda^2}\right)_{\lambda=1}&= n(-n+1)I_1^D -(n+1)(-n-2)I_2^D = 2(-n+1)I_1^D \ .
	\end{align}
	We see that for any $n>1$, we always have that any solution to the Klein-Gordon equation is unstable. At the same time, the virial identity \eqref{EG.0.5} shows that both of the terms involved have the same sign for $n>1$ and a positive definite potential, meaning that, in such case, there are no solutions regardless of stability.

\clearpage\null\newpage

\chapter{List of Publications}\label{J}
%
This thesis is based on a number of publications by the author. These publications are:
	\begin{itemize}
		\item Carlos A.R. Herdeiro, Alexandre M. Pombo, and Eugen Radu. \textbf{Asymptotically flat
scalar, Dirac and Proca stars: discrete vs. continuous families of solutions.} \textit{Physics Letters B, 773:654-662, 2017.}

		\item Pedro G.S. Fernandes, Carlos A.R. Herdeiro, Alexandre M. Pombo, Eugen Radu, and Nicolas Sanchis-Gual.\textbf{ Spontaneous scalarization of charged black holes: coupling dependence and dynamical features.} \textit{Classical and Quantum Gravity, 36(13): 134002, 2019.}
		
		\item Dumitru Astefanesei, Carlos Herdeiro, A. Pombo, and E. Radu. \textbf{Einstein-Maxwell-scalar black holes: classes of solutions, dyons and extremality.} \textit{Journal of High Energy Physics, 2019(10):1-27, 2019.}
		
		\item Pedro G.S. Fernandes, Carlos A.R. Herdeiro, Alexandre M. Pombo, Eugen Radu, and Nicolas Sanchis-Gual. \textbf{Charged black holes with axionic-type couplings: Classes of solutions and dynamical scalarization.} \textit{Physical Review D, 100(8):084045, 2019.}
		
		\item Jose Luis Bl\'azquez-Salcedo, Carlos A.R. Herdeiro, Jutta Kunz, Alexandre M. Pombo, and Eugen Radu. \textbf{Einstein-Maxwell-scalar black holes: the hot, the cold and the bald.} \textit{Physics Letters B, 806:135493, 2020.}
		
		\item Jose Luis Bl\'azquez-Salcedo, Carlos A.R. Herdeiro, Sarah Kahlen, Jutta Kunz, Alexandre M. Pombo, and Eugen Radu.\textbf{ Quasinormal modes of hot, cold and bald Einstein-Maxwell-scalar black holes.} \textit{The European Physical Journal C, 81(2):1-16, 2021.}
		
		\item Carlos A.R. Herdeiro, Alexandre M. Pombo, Eugen Radu, Pedro V.P. Cunha, and Nicolas Sanchis-Gual. \textbf{The imitation game: Proca stars that can mimic the Schwarzschild shadow.} \textit{Journal of Cosmology and Astroparticle Physics, 2021(04):051, 2021.}
		
		\item Jo\~ao M. S. Oliveira and Alexandre M. Pombo. \textbf{Spontaneous vectorization of electrically charged black holes.} \textit{Physical Review D, 103(4):044004, 2021.}
		
		\item Carlos A. R. Herdeiro, Jo\~ao M. S. Oliveira, Alexandre M. Pombo, and Eugen Radu. \textbf{Virial identities in relativistic gravity: 1D effective actions and the role of boundary terms.} \textit{Physical Review D, 104(10):104051, 2021.}
		
		\item Carlos A. R. Herdeiro, Alexandre M. Pombo, and Eugen Radu.\textbf{Aspects of Gauss-Bonnet scalarization of charged black holes.} \textit{Universe 7 (12), 483, 2021}
		
	\end{itemize}
%
  \bibliographystyle{unsrt}
  \bibliography{ThesisBib}


\end{document}